# Computational investigation of single herbal drugs in Ayurveda for diabetes and obesity using knowledge graph and network pharmacology

Priyotosh Sil[a,b,1], Rahul Tiwari[a,b,1], Vasavi Garisetti[a], Shanmuga Priya Baskaran[a,b], Fenita Hephzibah Dhanaseelan[c], Smita Srivastava[c], Areejit Samal[a,b,*]

[a] *The Institute of Mathematical Sciences (IMSc), Chennai 600113, India*

[b] *Homi Bhabha National Institute (HBNI), Mumbai 400094, India*

[c] *Department of Biotechnology, Bhupat and Jyoti Mehta School of Biosciences, Indian Institute of Technology Madras, Chennai, 600036, India*

[1]Priyotosh Sil and Rahul Tiwari contributed equally to this work and should be considered as Joint-First authors

[*]Corresponding author: asamal@imsc.res.in

## Abstract

Metabolic diseases such as type 2 diabetes and obesity represent a rapidly escalating global health burden, yet current therapeutic strategies largely target isolated symptoms or single molecular pathways. To address this limitation, we developed an integrated computational pipeline leveraging knowledge graph, pathway analysis and network pharmacology to elucidate the multi-target mechanisms of Ayurvedic Single Herbal Drugs (SHDs). SHDs associated with diabetes and obesity were curated from the Ayurvedic Pharmacopoeia of India (API), followed by phytochemical identification using IMPPAT database, yielding a shortlist of 11 SHDs and their 188 phytochemicals after drug-likeness and bioavailability filtering. Subsequently, molecular targets of the phytochemicals in SHDs, disease-associated genes and therapeutic targets of FDA-approved drugs, were curated via integration of data from several databases. Pathway enrichment analysis revealed significant functional overlap between SHD-associated and disease-associated pathways. Furthermore, all curated data were embedded into a Neo4j-based knowledge graph, enabling SHD–disease intersection analysis that prioritized key disease-relevant targets, including PTPN1, GLP1R, and DPP4. Also, the SHD – Target – FDA-approved drug profile elucidated the molecular and mechanistic aspects of the SHDs as a phytochemical cocktail, and is in alignment with the clinically studied synergistic FDA-approved drug combinations. Network pharmacology based protein–protein interaction analysis identified PPARG as another central regulator. Using a quantitative framework, we identified phytochemical pairs within SHDs, which were structurally dissimilar and target-wise distinct, yet acted on shared or different disease-associated pathways, indicating complementary and potentially synergistic interactions. Molecular docking analysis of two selected druggable targets (PPARG and DPP4) identified putative lead phytochemicals, including Chitraline, Isovitexin, and Pakistanine, for DPP4, and Sulfurein, Sesamin, and Pterosupin for PPARG. Overall, this systematic computational framework integrating knowledge graph and network pharmacology provides a robust approach for mechanistically dissecting SHDs and bridges traditional Ayurvedic knowledge with modern molecular and systems-level biomedical research.

**Keywords:** Ayurvedic Single Herbal Drugs; Diabetes; Obesity; Knowledge Graph; Network Pharmacology; Synergistic phytochemical combinations

## 1. Introduction

Metabolic disorders have emerged as a defining global health crisis of the 21st century, driven by rapid urbanization and profound lifestyle transitions. Sedentary behavior, increased consumption of energy-dense diets, and heightened psychosocial stress have contributed to a sharp rise in obesity, type 2 diabetes mellitus, and related metabolic complications. The IDF Diabetes Atlas (2025) estimates that 589 million adults worldwide live with diabetes, a number projected to reach 853 million by 2050[1], while the World Obesity Atlas (2025) projects adult obesity will increase from 524 million in 2010 to nearly 1.13 billion by 2030[2]. Globally, diabetes alone accounted for nearly 3.4 million deaths in 2024, equivalent to one death every nine seconds[1]. Taken together, these trends underscore the urgent need for effective, multi-targeted, and mechanism-driven therapeutic strategies capable of addressing the complex metabolic, inflammatory, and hormonal dysregulation underlying metabolic disorders.

Diabetes mellitus and obesity exhibit deeply interconnected pathophysiology and are often described collectively as *diabesity*, reflecting their shared metabolic, hormonal, and inflammatory foundations[3]. Diabetes comprises a group of metabolic disorders marked by progressive disruption of metabolic homeostasis. Type 1 diabetes results from immune-mediated destruction of pancreatic β-cells leading to absolute insulin deficiency[4], whereas type 2 diabetes primarily arises from insulin resistance accompanied by relative insulin insufficiency[5]. Obesity is a chronic and multifactorial condition influenced by genetic, environmental and lifestyle factors, representing a profound dysregulation of the adipose–insulin axis[6,7]. Current therapeutic approaches including insulin sensitizers, incretin-based therapies, α-glucosidase inhibitors and centrally acting appetite modulators remain largely palliative, and are frequently constrained by safety and tolerability concerns[8–10], underscoring the persistent gap between symptomatic management and durable metabolic homeostasis.

To address this predominantly symptomatic management of metabolic disorders, it is important to treat such diseases via a holistic interconnected network approach rather than as isolated symptoms, a perspective in which Ayurveda excels. As a traditional system of Indian medicine, Ayurveda emphasizes on regaining the balance of body systems with the help of herbal formulations that can modulate disease processes at multiple biological levels. Importantly, the inherent multi-compound and multi-target nature of herbal formulations mirror the networked architecture of complex metabolic disorders by enabling simultaneous modulation of interconnected pathological submodules. In Ayurveda, both diabetes and obesity are well characterized. Diabetes is described as *Madhumeha*, *Prameha*, commonly associated with *kapha* imbalance and excessive *kleda*, with therapeutic strategies aimed at restoring metabolic stability (*sthiratva*) through herbal interventions, dietary regulation (*ahara*), and lifestyle modification (*vihara*)[11]. Obesity is defined as excessive accumulation of *meda* (adipose tissue) and *mamsa* (muscle), leading to body flaccidity, and is classified as a *Santarpanottha Vikara*, a disorder arising from over-nutrition and linked to metabolic dysfunction and cardiovascular risk[12]. Ayurvedic principles, rooted in their relevance to metabolic regulation, are being actively investigated, and associated formulations are increasingly being explored for the management of both diabetes and obesity, highlighting their potential to address the systemic nature of metabolic disorders[13–15].

Computational approaches[16–19] including network pharmacology have gained substantial momentum as complementary strategies for elucidating the mechanistic basis of herbal formulations used in traditional Chinese and Indian systems of medicine. In the context of traditional Chinese medicine, network pharmacology is both well established and extensively used. For instance, Pan *et al.*[20] investigated the multi-target mechanisms of Huanglian decoction in type 2 diabetes mellitus. In the context of traditional Indian medicine, network-based studies have also explored both single- and poly-herbal Ayurvedic interventions for treatment of metabolic disorders[21,22]. Despite these advances, published *in silico* studies in Ayurveda are largely limited to small-scale and formulation-specific analyses that examine disease associations from a network perspective. Therefore, a comprehensive computational pipeline that systematically integrates traditional knowledge with diverse available biological datasets is needed to uncover the molecular mechanisms underlying Ayurvedic herbal formulations for chronic disorders.

Knowledge graphs offer a unified framework for systematic integration of heterogenous biomedical datasets, enabling their effective use in computational drug discovery[23]. A Knowledge Graph (KG) is essentially a directed graph wherein nodes denote distinct biological or conceptual entities, and edges specify the type and nature of their interactions. In recent times, KGs have been increasingly employed to encapsulate the diverse information in biological databases. Notably, the traditional Chinese medicine database Herb 2.0[24] has implemented a KG-based representation to organize the vast information on herbal formulations. Subsequent studies have employed graph embedding and deep learning for link prediction between previously unconnected biological entities in KG for herbal formulations used in traditional Chinese medicine (see e.g., Duan *et al.*[25]). In contrast to traditional Chinese medicine, there is an unmet need to encapsulate diverse information on Ayurvedic herbal formulations, their phytoconstituents, molecular targets, and linked diseases, into a large-scale KG to enable computational efforts toward traditional Indian knowledge based drug discovery and repurposing. In particular, coupling KG-based representations of traditional Ayurvedic knowledge with established network pharmacology paradigms holds significant promise for unravelling deeper mechanistic insights into herbal drug action than conventional network pharmacology approaches alone.

Building on the holistic principles of Ayurveda, knowledge graphs, and network pharmacology, this computational study investigates how Single Herbal Drugs (SHDs) for type 2 diabetes and obesity can function as natural multi-target interventions, leveraging their phytochemical diversity to modulate complex disease networks in a manner analogous to modern combination therapies. Our study begins by examining the pathway-level interactions of the diseases and SHDs, providing insights into their overlapping mechanisms. By employing a KG-based framework, this study systematically integrates heterogenous data from diverse sources to uncover hidden patterns, offering insights not only into the molecular-level actions of SHDs but also into their broader network-level mechanisms through SHD–Target–Drug associations. Furthermore, we present a quantitative framework to assess complementarity of phytochemicals in disease modulation, enabling systematic assessment of their combined actions. In doing so, this study bridges traditional Ayurvedic wisdom with modern biomedical science, providing a scalable, systematic, computational pipeline (**Figure 1**) that can be applied to other target-modulated diseases to reveal underlying mechanisms and therapeutic potential.

## 2. Methods

### 2.1 Compilation of single herbal drugs for both diabetes and obesity from the Ayurvedic Pharmacopoeia of India

Information on Ayurvedic single drugs was obtained from the Ayurvedic Pharmacopoeia of India (API; https://pcimh.gov.in/show_content.php?lang=1&level=1&ls_id=56&lid=54). Notably, computational investigation in this work is restricted to plant-derived single drugs, which are hereafter referred to as 'Single Herbal Drugs' (SHDs). The API provides botanical identity and the plant part(s) used in each SHD. SHDs were selected based on explicit therapeutic indications for both diabetes and obesity. Specifically, only those SHDs were retained for which the 'therapeutic uses' section of the API explicitly mentioned *diabetes mellitus* or its traditional Ayurvedic equivalents (e.g., *Madhumeha*, *Prameha*), as well as *obesity* or corresponding Ayurvedic terms (e.g., *Medoroga*, *Sthoulya*), which align with the clinical features of diabetes and obesity in modern biomedical terminology.

### 2.2 Phytochemical retrieval and screening for drug-likeness and bioavailability

For each SHD, phytochemical constituents were retrieved from the Indian Medicinal Plants, Phytochemistry and Therapeutics (IMPPAT)[26,27] database using the basic search module, with searches restricted to the plant part(s) specified in the API. In addition, phytochemicals reported as Thin-Layer Chromatography (TLC) data in the API were included when a valid chemical identifier (e.g., PubChem CID or ChemSpider ID) or a resolved chemical structure could be obtained from published literature. Drug-likeness of all compiled phytochemicals was evaluated using Lipinski's Rule of Five[28] as implemented in the SwissADME[29] webserver, retaining compounds with no more than one rule violation. Predicted oral bioavailability was also assessed using SwissADME[29], and compounds with a bioavailability score $\geq 0.50$ were selected. The final phytochemical set for each SHD comprised compounds satisfying both drug-likeness and bioavailability criteria. For downstream analyses, only SHDs containing at least 10 phytochemicals were retained (**Figure 1**).

### 2.3 Identification of clinical trials for herbs constituting SHDs for both diabetes and obesity

Clinical trials related to the selected SHDs were retrieved from the International Clinical Trials Registry Platform (ICTRP)[30]. Each corresponding herb was queried using the advanced search option to obtain a comprehensive list of associated clinical trials. Retrieved trials were screened for relevance to the diseases of interest, namely diabetes mellitus and obesity. Relevance was assessed by examining the *Condition* field for terms including 'Diabetes', 'Type 2 Diabetes', 'Hyperglycaemia', and 'Obesity'. The *Intervention* field was subsequently reviewed to verify the plant part(s) used in each trial. Trials specifying plant parts different from those used in the SHDs were excluded, whereas trials that either did not specify a plant part or explicitly mentioned the relevant plant part were retained.

**2.4 Phytochemical classification and chemical similarity analysis**

Phytochemicals retrieved from the IMPPAT[27] database were annotated using the ClassyFire[23] classification scheme provided by IMPPAT, which assigns hierarchical chemical taxonomy categories including kingdom, superclass, class, and subclass. For phytochemicals not available in IMPPAT, ClassyFire based chemical classification was performed using the classyfireR[31] package.

To evaluate structural relationships among phytochemicals derived from the SHDs, a Chemical Similarity Network (CSN) was constructed. Structure Data Format (SDF) files for each phytochemical were obtained from IMPPAT when available; for compounds not present in IMPPAT, structures were retrieved from PubChem or manually drawn using the PubChem Sketcher[32]. Pairwise chemical similarity among phytochemicals was quantified by computing the Tanimoto coefficient (Tc) based on Extended Connectivity Fingerprints with diameter 4 (ECFP4)[33] as implemented in RDKit[34]. A network was then generated in which nodes represent phytochemicals and edges correspond to pairwise similarities weighted by the corresponding Tc values. Edges with $Tc < 0.5$ were removed, and the resulting high similarity CSN was visualized using Cytoscape[35].

**2.5 Retrieval of human protein targets of phytochemicals**

Phytochemicals that remained after drug-likeness and bioavailability filtering were further examined to identify their known human protein targets. Target associations were collected by querying multiple databases, namely NPASS[36], BindingDB[37], and ChEMBL[38], which compile experimentally derived bioactivity and interaction data linking small molecules to their protein targets. Searches across all databases were restricted to human protein targets. For data obtained from ChEMBL[38], additional quality control measures were applied. Entries containing annotations such as 'not active', 'not determined', 'inactive', 'inconclusive', 'undetermined', 'not evaluated', and 'no data' in the *comment* field were excluded. Records with non-empty entries in the *data validity comment* field were also removed. No additional filtering criteria were applied to data retrieved from NPASS[36] or BindingDB[37].

## 2.6 Retrieval of disease-associated targets for diabetes and obesity

High-confidence disease-associated target sets for type 2 diabetes and obesity were assembled through a structured and integrative workflow comprising two complementary components. First, therapeutic targets of United States Food and Drug Administration (US FDA) approved drugs (referred to as FDA-approved drugs) were curated to capture well established, clinically validated disease targets. Second, disease-associated genes were systematically collected from multiple independent databases to ensure broad and robust coverage.

### 2.6.1 Approved drugs and their therapeutic targets

A systematic, stepwise strategy was employed to curate approved drugs relevant to type 2 diabetes and obesity, along with their associated therapeutic targets. Initially, DrugBank[39] was queried to identify candidate drugs linked to these conditions by screening the *Indication* field using disease relevant keywords, including 'diabetes', 'diabetic', 'hyperglycemia', 'insulin', 'glycaemic', 'obesity', and 'weight loss'. Each retrieved drug was subsequently subjected to manual curation to ensure high-confidence and clinical relevance. Drugs were retained only if they were approved, not withdrawn from the market, and classified as either small molecules or peptides based on the *Modality* information. In addition, reported indications and summary descriptions were carefully reviewed to exclude ambiguous or non-specific entries. This multi-level verification resulted in a refined set of drugs with clear therapeutic relevance to type 2 diabetes or obesity. Thereafter, therapeutic target information for the curated drugs was compiled. Target annotations were initially retrieved from the Therapeutic Target Database (TTD)[40] and further supplemented using the comprehensive review by Santos *et al.*[41], which integrates validated therapeutic targets from DrugCentral[42], ChEMBL[43], and canSAR[44]. For drugs included in the Santos *et al.*[41] dataset, all reported targets were directly incorporated. For drugs not covered by this resource or TTD[40], target information was manually curated from DrugCentral[42], retaining only those targets explicitly marked with a 'Green Tick' in the *Mechanism of Action* field.

### 2.6.2 Systematic retrieval of disease-associated genes

In parallel, disease-associated genes were curated using a multi-database integration strategy. Gene-disease associations were first retrieved from DisGeNET[45] by selecting disease-specific Concept Unique Identifiers (CUIs) corresponding to type 2 diabetes mellitus (CUI: C0011854) and obesity (CUI: C0028754). Associated gene sets were filtered to retain high-confidence entries with a Gene-Disease Association (GDA) score $\geq 0.4$, an Evidence Index (EI) $> 0.5$, and gene type restricted to protein coding genes. Additional curated associations were incorporated by using the datasets from the DISEASES database[46]. Human-specific text-mining, curated knowledge, and experimental datasets were obtained, and genes with relevance scores $\geq 3$ were retained. Obesity-associated genes were extracted using Disease Ontology identifier DOID:9970, while type 2 diabetes-associated genes were identified using DOID:9352. Ensembl protein identifiers were converted to Entrez Gene IDs using the *mygene* package[47], and unresolved mappings were manually curated using NCBI Gene resources[48] to ensure identifier consistency. Gene-disease associations were further retrieved from the Comparative Toxicogenomics Database (CTD)[49] using disease-specific Medical Subject Headings (MeSH) identifiers (MESH:D003924 for type 2 diabetes, MESH:D009765 for obesity). The resulting gene sets were refined using the SynGO platform[50] to retain only human-derived entries. Furthermore, GeneCards[51] was queried using disease-specific search terms (‘Type 2 diabetes’ and ‘Obesity’). Retrieved genes were filtered using a relevance score threshold of $\geq 15$, and only protein-coding genes were retained.

### 2.6.3 Integration and consensus-based selection of disease-associated targets

Finally, disease-associated genes obtained from all sources were integrated with the therapeutic targets of FDA-approved drugs curated earlier. To construct a high-confidence disease target set, a consensus-based filtering strategy was applied. All therapeutic targets associated with FDA-approved drugs were retained by default. Remaining candidate genes were included only if supported by at least two independent databases among DisGeNet[45], DISEASES[46], CTD[49] and GeneCards[51]. This integrative approach ensured a robust and biologically meaningful compilation of disease-associated targets for downstream analyses.

## 2.7 Pathway enrichment analysis

Biological pathway enrichment was performed to characterize functional processes associated with the different gene sets, including protein targets of phytochemicals derived from the SHDs as well as disease-associated gene sets. Enrichment analysis was conducted using the KEGG 2021 Human pathway collection implemented in the Enrichr platform[52]. For each input gene set, Enrichr generated a ranked list of KEGG pathways based on statistical enrichment derived from the overlap between pathway gene members and the input genes.

To improve robustness and reduce potential bias arising from small pathway sizes, only KEGG pathways comprising more than 50 genes were retained for further analysis. Pathways with a p-value less than 0.01 were considered significantly enriched. Each significant pathway was subsequently assigned to its corresponding hierarchical category using annotations from the KEGG BRITE database[53]. Pathways belonging to high-level disease categories not directly relevant to the pathophysiology of type 2 diabetes or obesity, such as cancer-related, neurodegenerative, or infectious disease pathways, were excluded from downstream analyses. For visualization and comparative interpretation, the top 20 enriched KEGG pathways were selected for each gene set.

### 2.8 Construction of a Knowledge Graph

A Knowledge Graph (KG) for SHDs in Ayurveda for diabetes and obesity was constructed using Neo4j[54], a graph database platform. Data obtained from the preceding curation and integration steps were transformed into a standardized triplet-based format comprising a source identifier, source node type, relationship, relationship properties, target identifier, and target node type.

The KG comprised multiple node categories, including the final set of SHDs (represented by their traditional Ayurvedic Sanskrit names and corresponding botanical names), the filtered set of phytochemicals satisfying drug-likeness and bioavailability criteria, a target set encompassing gene/protein targets either known to interact with phytochemicals in the graph or identified as disease-associated genes, two metabolic diseases (type 2 diabetes and obesity), and FDA-approved drugs for these conditions (**Figure 2**). Relationships between nodes were defined based on their biological and pharmacological relevance. Specifically, SHDs were connected to phytochemicals through the ‘contains_chemical’ relationship, with associated properties capturing contextual information such as plant parts used. Phytochemicals were linked to targets via the ‘interacts_with’ relationship, targets were associated with diseases through 'corresponds_to', approved drugs were connected to diseases using 'treats', and approved drugs were linked to targets via the 'affects' relationship. This triplet-based representation enabled systematic data ingestion and efficient retrieval within the KG. An interactive HTML-based visualization of the KG was generated using PyVis[55] to facilitate interactive exploration and visual inspection.

### 2.9 Knowledge Graph based analysis

The constructed KG was subsequently used for integrative analyses. First, the distribution of phytochemicals across SHDs was examined to determine how many SHDs are linked to each phytochemical. Using the SHD-phytochemical and phytochemical-target relationships encoded in the KG, indirect SHD-target associations were derived. For each SHD, the union of all known targets connected through its constituent phytochemicals was considered as the corresponding SHD-specific target set. This representation further enabled evaluation of the overlap of targets between each pair of SHDs as well as ranking of protein targets based on the number of SHDs known to modulate them. Pairwise similarity between two SHDs (say, a and b) based on their target sets was quantified using the Jaccard similarity index, defined as $J(A, B) = |A \cap B| \,/\, |A \cup B|$, where $A$ and $B$ represent the respective SHD-specific target set

In addition, overlap analyses were performed to assess the shared molecular landscape between type 2 diabetes and obesity, as well as between disease-associated gene sets and the SHD-specific target sets. Specifically overlap were computed between: (i) genes associated with type 2 diabetes and obesity, and (ii) protein targets of individual SHDs and disease-associated genes in different conditions. These overlap analyses were accompanied by statistical significance testing using a right-tailed hypergeometric test to evaluate whether the observed intersections exceeded random expectation, with all human protein-coding genes annotated in Ensembl[62] used as the background universe.

Further, the KG was leveraged to explore potential multi-target and combinatorial relevance of SHDs. For each SHD, targets associated with its constituent phytochemicals were compared with known therapeutic targets of FDA-approved drugs for type 2 diabetes and obesity. Overlap between these target sets was used to identify cases where phytochemicals from an SHD are associated with the same protein targets as FDA-approved drugs. Such overlaps provide insight into shared molecular mechanisms. In particular, this framework enabled identification of SHDs whose phytochemical constituents collectively target protein sets that are jointly modulated by clinically used drug combinations. These patterns are consistent with a potential multi-compound and multi-target mode of action of SHDs, conceptually analogous to combination pharmacotherapy.

**2.10 Network pharmacology analysis**

Network pharmacology analysis was performed to prioritize key proteins within overlapping target sets of SHDs and disease-associated genes. Protein–protein interaction (PPI) networks were constructed using the STRING (version 12.0)[57] database, restricting interactions to human proteins and applying a confidence score threshold of 0.4. For each SHD–disease pair, a PPI network was generated from the intersection of the SHD-specific target set and the corresponding disease-associated gene set. Only networks containing a Largest Connected Component (LCC) with at least ten nodes were retained for further analysis.

Topological analysis of each retained network was performed using the CytoHubba[16] plugin in Cytoscape[35]. Five complementary centrality-based algorithms namely, Maximal Clique Centrality (MCC), Maximum Neighborhood Component (MNC), degree centrality, betweenness centrality, and closeness centrality, were applied independently. For each algorithm, the ten top-ranked proteins were identified. To derive a high-confidence hub set for a given SHD–disease network, the intersection of the top-ranked protein sets across all five algorithms was computed, retaining only proteins consistently ranked among the top nodes by all methods. This procedure yielded a consensus hub protein set for each SHD–disease pair. Subsequently, for each disease condition, consensus hub sets obtained across all relevant SHDs were combined, and proteins were prioritized based on the frequency with which they appeared across multiple SHD-specific networks. This frequency-based aggregation enabled ranking of proteins according to their recurrent centrality across SHD–disease contexts.

### 2.11 Molecular docking analysis

Molecular docking was carried out using AutoDock Vina version 1.2.3[58]. The target protein structures were prepared by removal of water molecules and ions not involved in binding, and addition of hydrogen atoms. The prepared target protein structures and ligand 3D conformers were converted to PDBQT file formats using Open Babel[59] version 3.1.0. For each target protein, docking grids were centred on the co-crystallized reference biological ligands and important residues from the literature. Subsequently, protein-ligand docking was performed using AutoDock Vina by setting the exhaustiveness parameter to 12. For each ligand, multiple poses were generated, and thereafter, PoseBusters[60] was used to screen the docking poses for steric clashes and ligand geometry strain issues. The best docked pose passing the PoseBusters[60] validity checks and with the best binding affinity was retained. The corresponding protein-ligand complex for the selected pose was generated and further analysed using in-house Python scripts.

## 3. Results and discussion

### 3.1 Ayurvedic single herbal drugs indicated for both diabetes and obesity and their phytochemical profiles

A total of 24 Ayurvedic Single Herbal Drugs (SHDs) were initially identified from the Ayurvedic Pharmacopoeia of India (API) as being indicated for both diabetes and obesity, based on explicit references to these conditions or corresponding Ayurvedic nosological terms (Methods). The complete list of these SHDs, including the corresponding plant species, plant part(s) used, family, API volume and page references, and IUCN status, is provided in **Supplementary Table S1**. The identified SHDs are documented for a broad spectrum of therapeutic applications, and **Supplementary Table S1** lists the comprehensive therapeutic indications for the SHDs as per the API. Among the initially identified SHDs, Arjuna (*Terminalia arjuna*; bark) appeared twice, in API Volume II and Volume VIII. After removing this duplication, 23 unique SHDs explicitly associated with the concurrent management of both diabetes and obesity in Ayurveda were retained for subsequent analyses.

Phytochemical screening of the retained SHDs was subsequently performed using the IMPPAT[26,27] database and the API (Methods). This search led to reported phytochemicals for 20 of the 23 SHDs, while the remaining 3 SHDs lacked phytochemical information and were excluded from further analysis. The retained SHDs are listed in **Supplementary Table S2**. Phytochemicals obtained exclusively from the API were retained only when they could be unambiguously mapped to a recognized database such as PubChem, or when their chemical structures could be reliably confirmed from the literature. The consolidated phytochemical dataset of each SHD was then subjected to drug-likeness and oral bioavailability filtering (Methods). The number of phytochemicals before and after this filtration step is reported in **Supplementary Table S2**. This multi-step screening markedly reduced the phytochemical pool associated with each SHD. To ensure robust downstream analyses, a minimum threshold of ten phytochemicals per SHD was applied, resulting in 11 SHDs that met this criterion (**Table 1**). All subsequent analyses in this study were therefore conducted on these 11 SHDs.

In a preliminary assessment of clinical relevance, 7 of the 11 SHDs were found to be associated with at least one registered clinical trial related to either diabetes or obesity (Methods). Notably, *Gymnema sylvestre* (Meṣaśṛṅgī) exhibited the strongest clinical evidence base, with 22 registered trials for diabetes (predominantly, type 2 diabetes) and 3 trials for obesity. These findings underscore the translational and clinical relevance of these SHDs for the studied conditions. Detailed clinical trial information is provided in **Supplementary Table S3**.

While clinical trial evidence supports the therapeutic relevance of these herbs, their observed clinical effects are ultimately mediated by the combined action of their constituent phytochemicals. Following all filtration steps, a total of 188 phytochemicals were retained across 11 SHDs for downstream analyses. The number of compounds per SHD is provide in the column ‘Filtered set of phytochemicals’ of **Table 1**. These compounds were subjected to ClassyFire[31] classification (Methods), which successfully categorized 187 phytochemicals within the hierarchical chemical taxonomy. SHD-wise classification of all compounds at the levels of kingdom, superclass, and class is presented in **Supplementary Table S4**. As shown in **Figure 3a**, the most abundant chemical superclass is lipid and lipid-like molecules (42.2%), followed by polypropanoids and polyketides (21.9%).

To examine the structural diversity, a Chemical Similarity Network (CSN) was constructed using all 188 phytochemicals (Methods). The CSN comprised 22 connected components (each containing two or more nodes) and 51 isolated nodes, indicating both shared and distinct chemical structures within the dataset (**Figure 3b**). Interestingly, in most cases, connected components consisted primarily of chemicals from the same ClassyFire[31] superclass, suggesting that structural similarity within the CSN largely aligns with superclass-level chemical classification.

### 3.2 Enriched pathways of SHD-associated targets and their overlap with disease pathways

To better elucidate the potential molecular mechanisms underlying the therapeutic relevance of the selected 11 SHDs for type 2 diabetes and obesity, pathway enrichment analyses were performed for both SHD-associated targets and disease-associated genes (Methods).

### 3.2.1 SHD-associated targets and their pathway enrichment analysis

A curated set of human protein targets associated with the filtered phytochemicals of the 11 SHDs was assembled from NPASS[36], BindingDB[37], and ChEMBL[43] (Methods). Of the 188 phytochemicals, target information could be retrieved for 103 phytochemicals, yielding a compiled set of 759 unique human protein targets. **Supplementary Table S5** lists, for each SHD, the phytochemicals with at least one known associated target, the corresponding protein targets, and the source databases supporting each phytochemical–target association. For each SHD, **Table 1** summarizes the number of phytochemicals with at least one known human target and the corresponding number of unique protein targets. These SHD-associated targets were subsequently subjected to pathway enrichment analysis (Methods). The top 20 enriched KEGG pathways for each SHD are shown in **Supplementary Figures S1-S11**. Across the 11 SHDs, we observed a recurrent enrichment of broad categories of signal transduction, metabolic regulation, lipid metabolism, and inflammation-related pathways.

### 3.2.2 Disease-associated genes and their pathway enrichment analysis

High-confidence disease-associated gene sets were compiled for type 2 diabetes and obesity based on targets for FDA-approved drugs and evidence from multiple databases (Methods). For type 2 diabetes, 24 drugs yielded 63 unique therapeutic targets, and after integration with multi-database evidence, 279 disease-associated genes were retained (**Supplementary Table S6**). For obesity, eight drugs corresponded to ten unique therapeutic targets, with 253 genes retained following consensus filtering (**Supplementary Table S7**). KEGG pathway enrichment analysis was subsequently performed on these gene sets (Methods), and the top 20 enriched pathways are shown in **Supplementary Figures S12-S13**.

Comparison of the top 20 enriched KEGG pathways for type 2 diabetes and obesity yielded a consolidated union of 25 pathways (**Figure 4**), of which 15 were shared between the two diseases. This substantial overlap indicates a high degree of functional convergence between the two conditions despite their partially distinct genetic triggers.

Among the shared pathways, the Non-Alcoholic Fatty Liver Disease (NAFLD) pathway emerged prominently, consistent with its established role as a hepatic manifestation of metabolic syndrome characterized by dysregulated lipid metabolism and fatty acid accumulation[61]. This was complemented by the enrichment of the Adipocytokine signaling pathway, which mediates systematic inflammation via adipokines such as leptin and resistin that impair insulin sensitivity and promote progression from obesity to metabolic disease[62]. The concurrent enrichment of AMPK signaling pathway alongside NAFLD further underscores the central role of impaired cellular energy sensing in the metabolic coexistence of type 2 diabetes and obesity[63,64]. At the level of proximal disease drivers, co-enrichment of Insulin Resistance, Insulin signaling, and Glucagon signaling pathways, highlight bi-hormonal dysregulation as a core feature of both diseases[65,66]. Additionally, enrichment of the AGE-RAGE signaling pathway aligns with evidence that chronic hyperglycemia driven inflammatory feedback perpetuates insulin resistance and contributes to metabolic complications[67,68].

### 3.2.3 Overlap between SHD- and disease-enriched pathways

Building on this shared disease pathway landscape, we next compared the top enriched pathways of the 11 SHDs with those of type 2 diabetes and obesity. Recall that the union of the top 20 enriched pathways for the two diseases comprised 25 pathways, of which 15 were shared between them (**Figure 4**). When these 25 disease-associated pathways were compared with the top 20 enriched pathways of each SHD, the average overlap exceeded 8 pathways per

SHD. For the 15 pathways common to both diseases, the average overlap was ~ 6 pathways per SHD.

Among the 15 pathways shared between the two diseases, those that appeared in the top 20 enriched pathways of more than 50% of the SHDs were Lipid and atherosclerosis (8 SHDs), FoxO signaling (8 SHDs), NAFLD (7 SHDs), Longevity regulating (7 SHDs), Insulin Resistance (7 SHDs), Insulin signaling (6 SHDs) and Adipocytokine signaling (6 SHDs). Among obesity-specific pathways, cAMP signaling (7 SHDs) and Neuroactive ligand-receptor interaction (6 SHDs) were most recurrent (**Figure 4**). These patterns indicate convergence of most SHDs on core metabolic, lipid-related, stress response, and neuroendocrine signaling pathways.

At the level of individual SHDs, distinct but informative patterns emerged. For instance, *Aegle marmelos* showed preferential enrichment for pathways associated with secondary complications and lipid dysregulation, including NAFLD[69], Diabetic cardiomyopathy[70], and Lipid and Atherosclerosis[71] (**Figure 4**). This pathway footprint indicates that *Aegle marmelos* engages processes linked to lipid metabolism and metabolic complications, in addition to pathways related to glycemic control. By contrast, *Terminalia arjuna* displayed a different functional profile, characterized by enrichment of Insulin resistance and Insulin signaling pathways alongside AGE-RAGE signaling and Lipid and Atherosclerosis[72,73]. This distribution suggests concomitant engagement of mechanisms governing insulin responsiveness and vascular stress pathways implicated in diabetes-associated cardiovascular risk.

Overall, this pathway overlap analysis reveals that most SHDs consistently intersect with key metabolic and inflammatory pathways implicated in both type 2 diabetes and obesity.

### 3.3 Systems-level knowledge graph analysis of SHD–disease interactions

While pathway enrichment analysis revealed functional overlap between disease-associated and SHD-associated pathways, it did not resolve how individual protein targets are interconnected at the molecular level. To obtain this entity-level perspective, we leveraged an integrated Knowledge Graph (KG; Methods), enabling a systems-level analysis of relationships among SHDs, phytochemicals, protein targets, FDA-approved drugs, and disease-associated genes within a unified framework.

The resulting KG comprised a total of 1332 nodes, including 11 SHDs, 188 phytochemicals, 1099 unique human protein targets (integrating both phytochemical-associated and disease-associated targets), 32 FDA-approved drugs for diabetes or obesity, and 2 disease nodes (type 2 diabetes and obesity). These entities were connected by 3973 curated relationships, consisting of 224 SHD–phytochemical edges, 3082 phytochemical–target interactions, 532 target–disease associations, 103 drug–target interactions and 32 drug–disease links. This Ayurvedic KG constructed for the 11 SHDs and implemented in Neo4j[54] can be accessed via our GitHub repository at: https://asamallab.github.io/SHDKG-DiabetesObesity/

### 3.3.1 Shared phytochemicals and targets among SHDs along with disease gene overlap

The KG implemented in Neo4j[54] was queried to assess phytochemical sharing among SHDs, revealing limited overlap in their phytochemical composition. Notably, the most widely distributed compound was β-sitosterol, present in 7 of 11 SHDs, and only 22 of the 188 phytochemicals were documented in at least two SHDs. These findings indicate substantial heterogeneity in the known phytochemical profiles of the SHDs, with the majority of compounds being herb-specific in our curated dataset. **Supplementary Table S8** lists all 188 phytochemicals, the number of SHDs in which each occurs, and the corresponding SHDs. Nevertheless, the CSN presented in **Figure 3b** indicates that, despite limited direct sharing, many compounds across different SHDs exhibit commonality at the level of chemical superclass and structural features.

To evaluate functional similarity among SHDs at the protein level, we compared target overlap for all SHD pairs using the Jaccard similarity index (Methods). This analysis revealed that only one SHD pair exhibited a markedly high similarity, namely *Commiphora wightii* and *Tecomella undulata* with a Jaccard similarity of 0.7281, followed by the pair *Commiphora wightii* and *Gymnema sylvestre* with a similarity of 0.5030. The overall mean Jaccard similarity across all 55 SHD pairs was 0.2617 (**Supplementary Figure S14**), indicating generally low to moderate target overlap among most SHDs. Notably, when considering only phytochemicals with known targets, the phytochemical-level mean Jaccard similarity (0.0427) was substantially lower than the mean target-level similarity, suggesting greater convergence among SHDs at the protein level.

Using the KG, we assessed gene-level overlap between type 2 diabetes and obesity. Our curated dataset contained 279 type 2 diabetes-associated genes (**Supplementary Table S6**) and 253 obesity-associated genes (**Supplementary Table S7**), of which 62 genes were shared between the two conditions. This overlap was highly significant (p=$2.11 \times 10^{-59}$; Methods). This gene level convergence aligns with the pathway enrichment results presented in **Figure 4**, wherein 15 of the top 20 enriched pathways were commonly enriched in both diseases. Together, the concordance between shared genes and shared pathways supports a close biological relationship between type 2 diabetes and obesity, consistent with their well-documented clinical comorbidity and mechanistic interdependence[74].

### 3.3.2 Overlap between targets of SHDs and disease-associated genes, and prioritization of recurrent targets

Utilising the KG, we next quantified the alignment between SHD-associated targets and disease-associated genes for all 11 SHDs and both diseases, type 2 diabetes and obesity. For each SHD–disease pair, we evaluated the overlap between SHD-associated targets and disease gene sets. We observed that all SHD–disease pairs exhibited non-zero and statistically significant overlap ($p < 0.05$; **Supplementary Table S9**), indicating that each SHD shares disease-relevant molecular targets beyond random expectation.

To identify disease targets recurrently modulated by multiple SHDs, we focused on genes targeted by at least eight out of 11 SHDs in each condition. This criterion yielded six high-frequency targets for type 2 diabetes and seven for obesity (**Figure 5**). Among these, PTPN1 (Protein tyrosine phosphatase non-receptor type 1) targeted by nine SHDs, and GLP1R (Glucagon like peptide 1 receptor) targeted by eight SHDs, emerged as shared high-frequency targets for both diseases. Among the non-shared high-frequency targets, DPP4 (Dipeptidyl peptidase-4), which was targeted by nine SHDs, was specific to type 2 diabetes and is a therapeutic target of established FDA-approved drugs (**Supplementary Table S6**).

The biological relevance of these recurrent targets is supported by their well-established roles in metabolic regulation. PTPN1 is a critical regulator in both type 2 diabetes and obesity[75,76]. In peripheral tissues, PTPN1 attenuates insulin signaling by dephosphorylating the insulin receptor and IRS1 (Insulin receptor substrate 1), thereby promoting systemic insulin resistance[77]. In the hypothalamus, PTPN1 dampens leptin signaling via JAK2 (Janus kinase 2), linking impaired satiety to adiposity[78]. GLP1R is a clinically-validated metabolic target, and its pharmacological agonism is a widely-used therapeutic strategy for concurrent management of type 2 diabetes and obesity[79]. Its activation leverages the incretin effect to enhance glucose dependent insulin secretion and suppress glucagon release, thereby improving glycemic control[80], while also reducing appetite and slowing gastric emptying to promote sustained weight loss[81]. DPP4 regulates endogenous incretin activity by degrading GLP1 and GIP (Glucose-dependent insulinotropic polypeptide)[82]. Pharmacological inhibition of DPP4 preserves physiologic incretin levels, enhancing glucose-dependent insulin secretion and suppressing glucagon release[83]. Notably, DPP4 inhibitors are characterized by a weight-neutral clinical profile[84].

In sum, the findings indicate that, despite low phytochemical and target overlap, several SHDs converge on core metabolic control nodes in type 2 diabetes and obesity.

### 3.3.3 Molecular basis of phytochemical cocktails in SHDs for type 2 diabetes and obesity using knowledge graph

Target based analyses, as performed above, indicated that SHDs converge on a significant set of disease-associated targets, suggesting coordinated therapeutic potential, although the underlying molecular mechanisms remain incompletely understood. As complex mixtures of phytochemicals, SHDs characteristically exert multi-target and multifaceted effects. To explore these properties, we examined the overlap between targets modulated by SHD-derived phytochemicals and those targeted by FDA-approved drugs in the context of type 2 diabetes and obesity.

Leveraging the KG, we identified shared targets between SHD-derived phytochemicals and FDA-approved drugs (**Figure 6**), revealing functional "cocktails" of pharmacological actions in which individual SHDs recapitulate combination drug-like target coverage at the network level. Notably, 9 of the 11 SHDs exhibited drug-combination like target coverage in type 2 diabetes, some of which corresponds to combinations of FDA-approved drugs studied and/or used clinically for this condition (**Table 2**). A comparable pattern was also observed for obesity.

To further elucidate how these network-level interactions may translate into therapeutic effects, we examined the potential mechanism of action of *Aegle marmelos* as a representative SHD. Querying the KG for type 2 diabetes-associated targets in conjunction with *Aegle marmelos* revealed a functional drug-like cocktail comprising metformin (a glucose-lowering and insulin-sensitizing agent acting through INSR and oxidative phosphorylation related genes[85]), DPP4 inhibitors (sitagliptin, vildagliptin, saxagliptin, alogliptin, and linagliptin)[86], miglitol (which attenuates postprandial glucose excursions by inhibiting carbohydrate digestion via α-glucosidase and maltase-glucoamylase)[87], pioglitazone (which reduces insulin resistance through modulation of PPARG)[88] and GLP1R-associated therapies, including semaglutide, tirzepatide, and pramlintide[89] (**Supplementary Table S10**). These drugs target distinct yet complementary facets of type 2 diabetes pathophysiology.

At the molecular level, *Aegle marmelos* modulates key targets involved in glucose homeostasis, insulin sensitivity and incretin signaling, including INSR, PPARG, DPP4, GLP1R, and α-glucosidase related enzymes, collectively recapitulating the target coverage of clinically studied drug combinations. This multi-target profile mirrors combination strategies evaluated in clinical practice, including Metformin–Sitagliptin (Janumet)[90], Metformin–Pioglitazone[91], Semaglutide–Metformin[92] as well as experimentally explored pairings such as Sitagliptin–Miglitol[93], which has been shown to exert complementary effects on incretin regulation in controlled studies (**Table 2**).

These findings suggest that *Aegle marmelos* operates as a natural "polypill", integrating multiple therapeutic axes within a single SHD. Importantly, this multi-target, network-level strategy is not unique to *Aegle marmelos*. Most of the other SHDs similarly exhibit coordinated modulation of complementary pathways, effectively recapitulating the pharmacological actions of multiple drugs within an SHD (**Figure 6**; **Table 2**; **Supplementary Table S10**). Together, these results highlight the broader polypharmacological potential of SHDs to mirror multi-target pharmacological strategies for managing complex metabolic disorders such as type 2 diabetes and obesity.

### 3.4 Network pharmacology analysis of SHD–disease target overlap

Building on the KG-based overlap analysis which identified disease-relevant targets shared across multiple SHDs, we next examined these intersections from a network topological perspective. While the KG-based analysis prioritized targets based on their frequency of occurrence across SHDs, it did not capture their relative importance within the underlying protein–protein interaction (PPI) network. To obtain this complementary view, we performed network pharmacology based analysis to identify highly central ('hub') proteins within the overlapping SHD–disease target space. We constructed separate STRING[57] based PPI networks for each SHD–disease pair (Methods). This approach enabled identification of topologically central proteins that recur across multiple SHD–disease contexts.

Of the 11 selected SHDs, 7 SHD–type 2 diabetes pairs yielded PPI networks with a Largest Connected Component (LCC) containing at least ten nodes and were retained for network pharmacology analysis. The four herbs namely, *Berberis aristata*, *Diospyros malabarica*, *Tecomella undulata* and *Tectona grandis*, were excluded due to LCC with fewer than ten nodes. For obesity, 8 of the 11 SHDs met this criterion, while three herbs namely, *Berberis aristata*, *Diospyros malabarica* and *Tectona grandis* were excluded.

For each of the 15 SHD–disease pairs, topological analysis was performed on the LCC of the corresponding PPI network using five centrality based algorithms (Methods). A consensus hub set was defined for each network as the intersection of the top ten proteins ranked by all five measures. While this analysis was conducted for all SHDs, representative LCC visualizations for *Aegle marmelos* and *Terminalia arjuna* are shown in **Supplementary Figure S15** for illustration. **Supplementary Table S11** lists the ten top-ranked proteins for each centrality measure, along with the consensus hub sets for all considered SHD–disease pairs.

To identify recurrent key proteins across SHDs, consensus hub sets were aggregated separately for each disease. This yielded 21 unique consensus hub proteins for type 2 diabetes and 20 for obesity (**Supplementary Tables S12-S13**). Six proteins (PPARG, PPARA, AKT1, STAT3, GLP1R and ACE) appeared as consensus hubs in at least one network for both diseases. When ranked by the number of SHD–disease networks in which they appeared, PPARG emerged as the most recurrent common hub (present in 6 PPIs for diabetes and 7 PPIs for obesity) (**Supplementary Tables S12-S13**). Notably, PPARG is also a therapeutic target for a FDA-approved drug for type 2 diabetes (**Supplementary Table S6**). In the target overlap analysis reported in an earlier section and summarized in **Figure 5**, targets were restricted to those modulated by eight or more SHDs. PPARG fell just below this threshold, being targeted by seven SHDs. However, it consistently emerged as a consensus hub across all seven corresponding SHD–obesity and six SHD–type 2 diabetes PPI networks, further underscoring its central role in the disease-relevant interaction landscape.

The recurrence of common consensus hubs across multiple SHD–disease networks suggests that several SHDs may converge on a shared set of topologically central proteins. Further, the identification of PPARG as a consensus hub in networks for both disease conditions, and as a therapeutic target for an approved drug, supports the biological relevance of the network-based prioritization and indicates that the PPI-centred approach reveals targets of clinical relevance.

### 3.5 Complementarity of phytochemical combinations in SHDs

SHDs comprise multiple phytochemicals and are therefore expected to act through coordinated, multi-component mechanisms. A central mechanistic question is whether these constituents act redundantly on the same targets or complementarily on the disease-relevant networks. Prior studies indicate that synergistic effects may arise from compounds acting on distinct targets within the same disease context or through modulation of different related pathways[94,95]. Further, higher structural similarity between two chemicals has been shown to negatively correlate with synergistic potential of the pair combination, highlighting the role of chemical diversity in combinatorial efficacy[96].

To examine this, we analyzed phytochemical pairs at structural, target, and pathway levels to identify chemically dissimilar combinations with minimal target overlap, configurations consistent with complementary network-level actions. For each SHD, phytochemicals targeting at least one disease-associated protein were retained. Pairwise target similarity (Jaccard) was computed using only the subset of targets overlapping the disease-associated gene set, and structural similarity was measured using the Tc (Methods). Pathway-level similarity was similarly calculated by restricting analysis to pathways enriched in disease-associated genes (top 25 pathways shown in **Figure 4**), ensuring that both target- and pathway-level comparisons reflect disease-relevant interactions rather than global target overlap. Across 10 of the 11 SHDs, most phytochemical pairs exhibited low target similarity (Jaccard $< 0.25$) and low structural similarity (Tc $< 0.5$). These patterns support predominantly complementary, rather than redundant, modes of action within SHDs. The structural, target-level, and pathway-level similarity score of phytochemical pairs for all SHDs are provided in **Supplementary Table S14**.

Here, we discuss *Aegle marmelos* and *Terminalia arjuna* as representative examples. In both SHDs, more than 85% of phytochemical pairs (with at least one disease-associated target) satisfied the stringent complementarity criteria (target-level similarity $< 0.25$; Tc $< 0.5$), highlighting a prevalence of structurally diverse compounds engaging distinct targets (**Figure 7**; **Supplementary Table S14**). In *Aegle marmelos*, β-sitosterol–marmin (Tc = 0.08; target similarity = 0) exemplifies this pattern combining DPP4 modulation with the modulation of mitochondrial complexes I and IV, a multi-mechanistic configuration analogous to combination strategies such as metformin–sitagliptin therapy[90]. A similar complementary profile was observed for β-sitosterol–umbelliferone (Tc = 0.03; target similarity = 0), resembling dual-mechanism approaches such as sitagliptin–miglitol treatment[93]. A comparable trend was evident in *Terminalia arjuna*, where β-sitosterol–ellagic acid (Tc = 0.02; target

similarity = 0) and β-sitosterol–baicalein (Tc = 0.02; target similarity = 0.14) showed complementary engagement of DPP4- and GAA-associated mechanisms[93].

These target-level complementarities were further reinforced at the pathway level. Several structurally and target-wise distinct pairs converged on shared biological pathways; for example, catechol–ellagic acid displayed high pathway similarity (0.80), and β-sitosterol–ellagic acid showed moderate pathway similarity (0.44) under similar conditions. Such convergence suggests that chemically diverse phytochemicals modulate common disease-relevant processes through different molecular routes. Collectively, these findings demonstrate that our integrative framework captures complementary phytochemical interactions across structural, target, and pathway dimensions, offering a systems-level perspective on SHD mechanisms. The mechanistic parallels with established drug-combination strategies further support the biological plausibility of these interactions. While experimental validation is required to confirm synergy, the observed patterns are consistent with potential synergy arising from distributed, multi-target modulation within disease-relevant networks.

### 3.6 Molecular docking analysis of DPP4 and PPARG

Based on the KG-based overlap analysis and the network pharmacology results, we carried out docking studies to assess phytochemical binding to the prioritized targets (Methods). We focused on DPP4, which is a type 2 diabetes specific target recurrently modulated by multiple SHDs, and PPARG, the most recurrent common consensus hub across type 2 diabetes and obesity networks. We performed molecular docking for chemical-target pairs that were not present in our curated list, in order to evaluate the potential binding for experimentally unexplored phytochemicals.

For docking, we used the crystal structures of DPP4 (PDB id: 1X70) and PPARG (PDB id: 5Y2O). The co-crystallized ligands, sitagliptin and pioglitazone for DPP4 and PPARG, respectively, were used as biological references for defining the binding region. For DPP4, the pocket was defined around residues central to inhibitor anchoring and S1/S2 subsite occupancy (Glu205, Glu206, Ser209, Phe357, Arg358, Tyr547, Tyr662)[97]. For PPARG, the pocket comprised of the activation pocket and surrounding hydrophobic cavity important for agonist binding (Cys285, Ser289, His323, Tyr327, Phe363, Met364, His449, Tyr473, Phe282, Ile281, Leu330, Leu453, Leu469)[98]. For each docked phytochemical, the interacting protein residues in the docked pose were compared against the above binding pocket residue sets. Compounds were retained only if they interacted with at least 4 of the 7 DPP4 pocket residues or at least 8 of the 13 PPARG pocket residues. All the compounds that passed this criterion were prioritized and ranked based on the binding affinity to identify the top candidates for each target (**Supplementary Tables S15-S16**).

From the docking analyses, it was observed that in DPP4 (1X70), the top-ranked phytochemicals showing predicted affinities stronger than the biological reference sitagliptin (-8.871 kcal/mol), include Chitraline (CAS_77754-91-7; -9.756 kcal/mol), Isovitexin (CID:162350; -9.699 kcal/mol) and Pakistanine (CID:193239; -9.278 kcal/mol). These ligands occupied the same functional pocket as the reference and made key site contacts. Chitraline interacted with Ser209, Phe357, Arg358, Tyr547 and Tyr662, while Isovitexin and Pakistanine additionally involved Glu206, consistent with interacting with the anchoring region. Sitagliptin showed contacts spanning the full residue set (Glu205, Glu206, Ser209, Phe357, Arg358, Tyr547, Tyr662), supporting that the docked phytochemicals are present in the inhibitor cavity (**Figure 8a**).

In PPARG (5Y2O), the top-ranked phytochemicals showing predicted affinities comparable to the biological reference Pioglitazone (CID:4829; -8.842 kcal/mol), include Sulfurein (CID:10071442; -8.844 kcal/mol), followed by Sesamin (CID:5204; -8.708 kcal/mol) and Pterosupin (CID:133775; -8.242 kcal/mol). These ligands occupied the same ligand binding cavity as the reference and made key pocket contacts. Pioglitazone engaged with the activation-pocket interacting with Cys285, Ser289, His323, Tyr327, Phe363, Met364, His449 and Tyr473, together with hydrophobic packing residues such as Phe282, Leu330 and Leu453. Sulfurein primarily made contacts with Cys285, Ser289, Tyr327, Phe363, Met364, His449, Ile281 and Leu330, indicating stable binding within the pocket. The other top-ranked ligands also mapped to the same cavity and showed contacts with core binding residues supporting consistent presence within the 5Y2O binding site (**Figure 8b**).

In sum, these compounds show promising *in silico* binding to DPP4 and PPARG and may be considered as putative natural product leads.

**4. Conclusions**

This study presents an integrative framework to elucidate the molecular mechanisms of Ayurvedic Single Herbal Drugs (SHDs) for metabolic disorders by combining pathway analysis, knowledge graph based analysis, and network pharmacology (**Figure 1**). Pathway enrichment analysis revealed functional convergence between SHD-associated and disease-associated pathways, providing a biological context for downstream analyses. Guided by these pathway-level insights, a biomedical knowledge graph enabled structured SHD–disease intersection analysis, leading to the prioritization of key disease-relevant targets including PTPN1, GLP1R, and DPP4. Extension of this framework into an SHD – Target – FDA-approved drug network facilitated mechanistic interpretation of the effect of phytochemical cocktails underlying the multi-target action of SHDs. Further, network pharmacology analysis identified PPARG as a central regulator linking insulin signaling and adipose tissue function. We further applied a quantitative framework integrating chemical structural similarity, target profiles, and disease-enriched pathway associations to assess phytochemical complementarity in disease modulation. This analysis revealed recurring multi-level patterns in which structurally diverse phytochemicals with minimal target overlap modulate either shared or distinct disease-relevant pathways, consistent with complementary and potentially synergistic interactions across SHDs. Lastly, molecular docking analyses led to identification of additional promising phytochemical leads for therapeutic targets, PPARG and DPP4, of existing FDA-approved drugs. While constrained by incomplete SHD–phytochemical and phytochemical–target annotations, future integration of experimentally validated phytochemical interactions and functional validation studies will further strengthen the translational relevance of this framework and its ability to bridge traditional medicinal knowledge with modern molecular biomedical research.

**Acknowledgement**

Areejit Samal would like to acknowledge funding from the Department of Atomic Energy (DAE), Government of India via Apex project to The Institute of Mathematical Sciences (IMSc) Chennai. Fenita Hephzibah Dhanaseelan has received a fellowship under the Prime Minister's Research Fellowship (PMRF) scheme, Ministry of Education, Government of India.

**Author contribution statement**

**Priyotosh Sil:** Conceptualization, Data Curation, Formal Analysis, Methodology, Visualization, Writing; **Rahul Tiwari:** Conceptualization, Data Curation, Formal Analysis, Methodology, Visualization, Writing; **Vasavi Garisetti:** Data Curation, Formal Analysis, Methodology, Visualization, Writing; **Shanmuga Priya Baskaran:** Data Curation, Formal Analysis, Visualization; **Fenita Hephzibah Dhanaseelan:** Formal Analysis; **Smita Srivastava:** Conceptualization, Formal Analysis, Writing; **Areejit Samal:** Conceptualization, Supervision, Formal Analysis, Methodology, Writing.

**Declaration of competing interest**

The authors declare that they have no known competing financial interests or personal relationships that could have appeared to influence the work reported in this paper.

# Tables

**Table 1: Summary of filtered phytochemicals and associated human protein targets for the selected SHDs indicated for the management of diabetes and obesity in Ayurveda.** For each of the 11 SHDs, the table reports the number of phytochemicals retained after applying bioavailability (≥ 0.50) and drug-likeness (≤ 1 Lipinski's rule violation) criteria. It further provides the number of filtered phytochemicals with experimentally supported or curated human protein targets, along with the corresponding count of unique human protein targets.

| **Traditional name of SHD in API** | **Plant name** | **Part(s)** | **Filtered set of phytochemicals** | **Phytochemicals with known human target(s)** | **Total number of human targets** |
|---|---|---|---|---|---|
| Bilva | *Aegle marmelos* | Bark | 11 | 10 | 268 |
| Dāruharidrā | *Berberis aristata* | Stem | 10 | 5 | 55 |
| Palāśaḥ | *Butea monosperma* | Seed | 28 | 14 | 226 |
| Guggulu | *Commiphora wightii* | Plant exudate | 21 | 11 | 195 |
| Tinduka | *Diospyros malabarica* | Fruit | 15 | 7 | 111 |
| Nāhī | *Enicostema axillare* | Whole plant | 15 | 7 | 360 |
| Meṣaśṛṅgī | *Gymnema sylvestre* | Leaf | 55 | 33 | 292 |
| Asana | *Pterocarpus marsupium* | Wood | 21 | 15 | 238 |
| Rohitaka | *Tecomella undulata* | Bark | 14 | 7 | 161 |
| Śāka | *Tectona grandis* | Wood | 14 | 8 | 83 |
| Arjuna | *Terminalia arjuna* | Bark | 20 | 12 | 388 |

**Table 2: Comparison of FDA-approved drug combinations and mimicking SHDs.** The table summarizes clinically approved drug combinations and associated evidence from published studies, alongside SHDs identified to mimic their multi-target mechanisms based on shared molecular targets.

| **Clinically studied FDA-approved drug combinations** | **Known studies** | **SHDs mimicking FDA-approved drug combinations** |
|---|---|---|
| Metformin + Sitagliptin (Janumet) | Charbonnel *et al.*[90] | *Enicostema axillare* \| *Terminalia arjuna* \| *Aegle marmelos* |
| Metformin + Pioglitazone | Ragazzi *et al.*[91] | *Enicostema axillare* \| *Terminalia arjuna* \| *Aegle marmelos* |
| Metformin + Miglitol | Chiasson *et al.*[99] | *Enicostema axillare* \| *Terminalia arjuna* \| *Aegle marmelos* |
| Sitagliptin + Pioglitazone | Yoon *et al.*[100] | *Enicostema axillare* \| *Aegle marmelos* \| *Terminalia arjuna* \| *Pterocarpus marsupium* \| *Butea monosperma* \| *Commiphora wightii* |
| Sitagliptin + Miglitol | Aoki *et al.*[93] | *Enicostema axillare* \| *Pterocarpus marsupium* \| *Terminalia arjuna* \| *Aegle marmelos* |
| Semaglutide + Metformin | Ahrén *et al.*[92] | *Enicostema axillare* \| *Aegle marmelos* |
| Semaglutide + Pioglitazone | Papaetis *et al.*[101] | *Enicostema axillare* \| *Aegle marmelos* \| *Pterocarpus marsupium* \| *Butea monosperma* \| *Commiphora wightii* \| *Gymnema sylvestre* |

# Figures

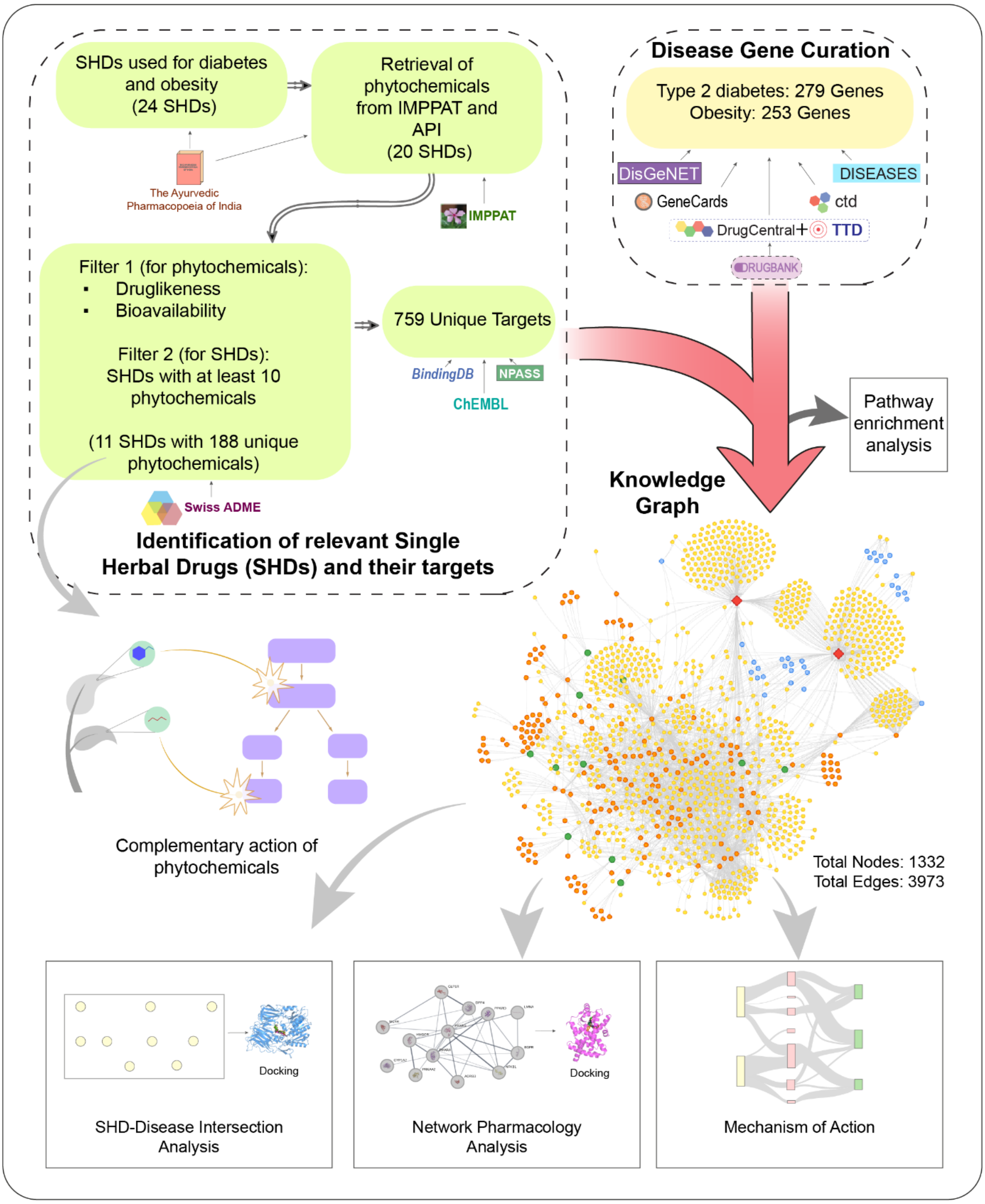

**Figure 1. Schematic overview of the computational workflow leveraging knowledge graph and network pharmacology to elucidate the molecular mechanisms of Ayurvedic single herbal drugs (SHDs) for diabetes and obesity.**

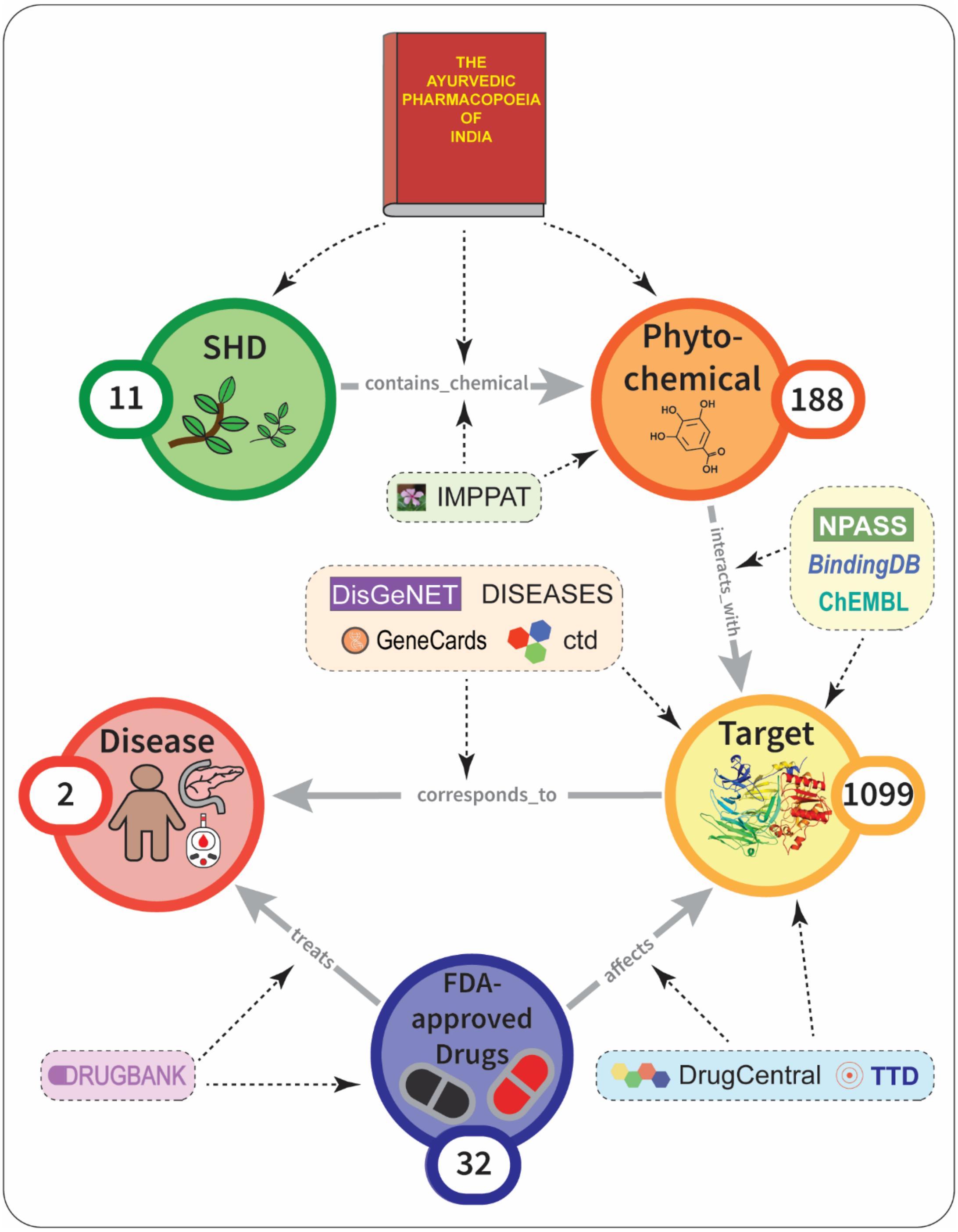

**Figure 2. Schematic representation of different associations present in the knowledge graph constructed for the single herbal drugs (SHDs) in Ayurveda indicated for the management of both diabetes and obesity.**

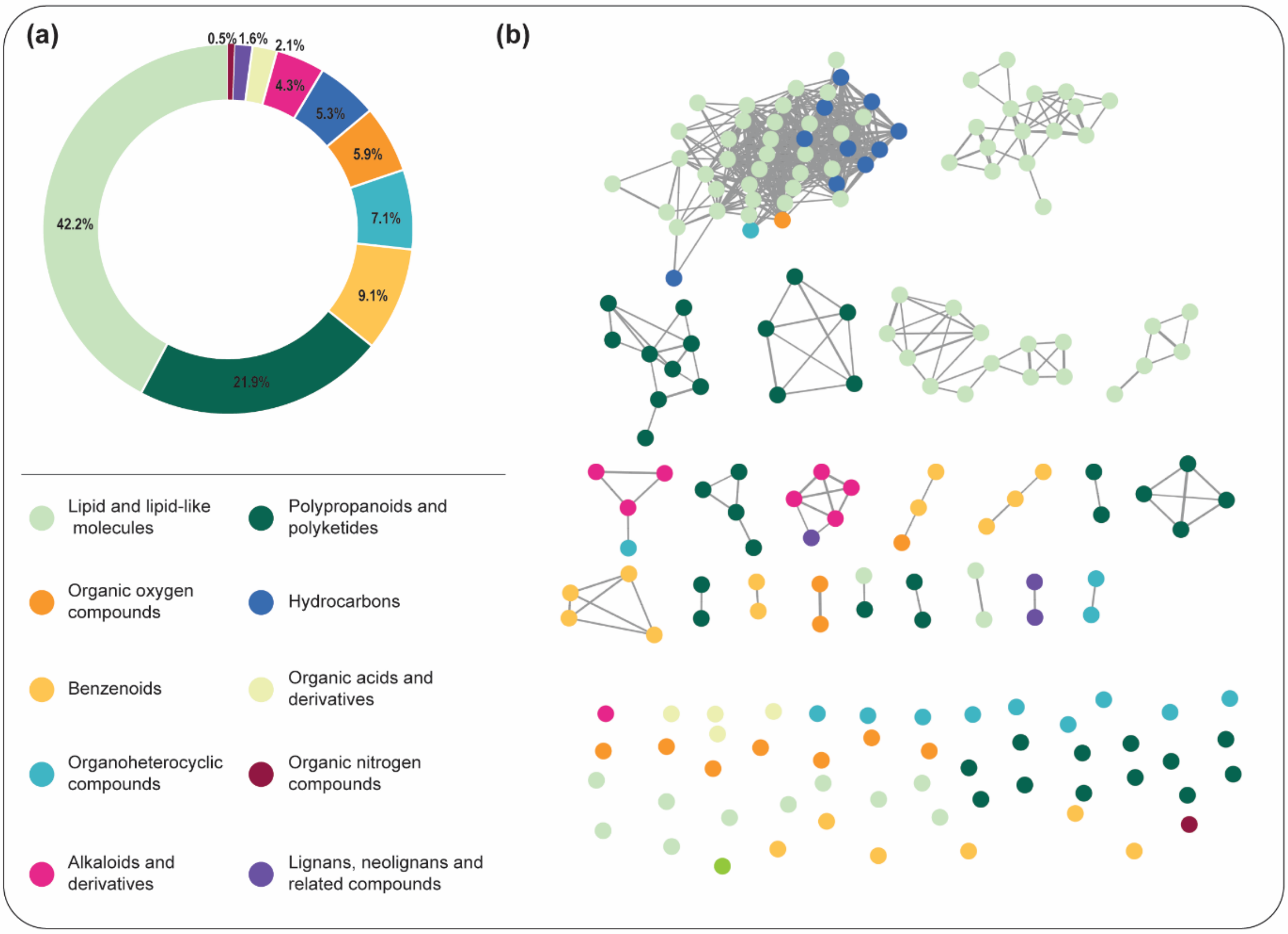

**Figure 3. Chemical classification and structural similarity for 188 phytochemicals from 11 SHDs. (a)** ClassyFire-based classification of phytochemicals at the superclass level, showing the relative distribution of chemical classes. Lipid and lipid-like molecules dominate the dataset of 188 phytochemicals followed by polypropanoids and polyketides. **(b)** Chemical similarity network constructed for all 188 phytochemicals using a pairwise Tanimoto coefficient threshold of $\geq 0.50$ (Methods). Nodes are colored according to their ClassyFire superclass. While many compounds remain isolated, connected components mostly consist of phytochemicals from the same ClassyFire superclass.

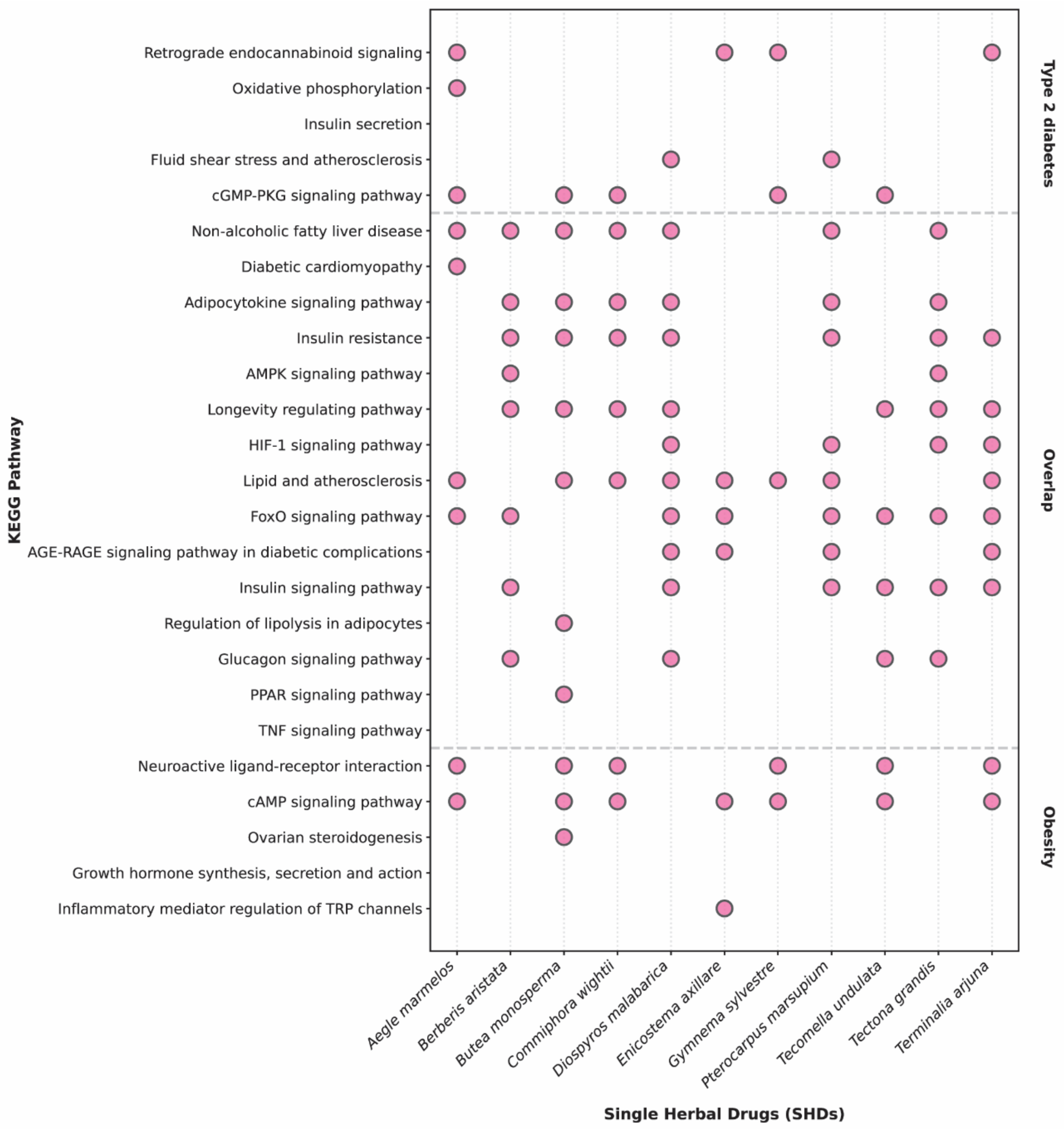

**Figure 4. Overlap between SHD-associated and disease-associated enriched pathways.** The y-axis lists the 25 KEGG pathways comprising the union of the top 20 enriched pathways for type 2 diabetes and obesity based on curated disease-associated gene sets. Pathways shared between both diseases are separated from those specific to type 2 diabetes or obesity by horizontal dashed lines. The x-axis represents the 11 SHDs, with a dot indicating that a given pathway also appears among the top 20 enriched pathways for that SHD. The figure shows substantial pathway-level convergence between SHDs and both disease conditions.

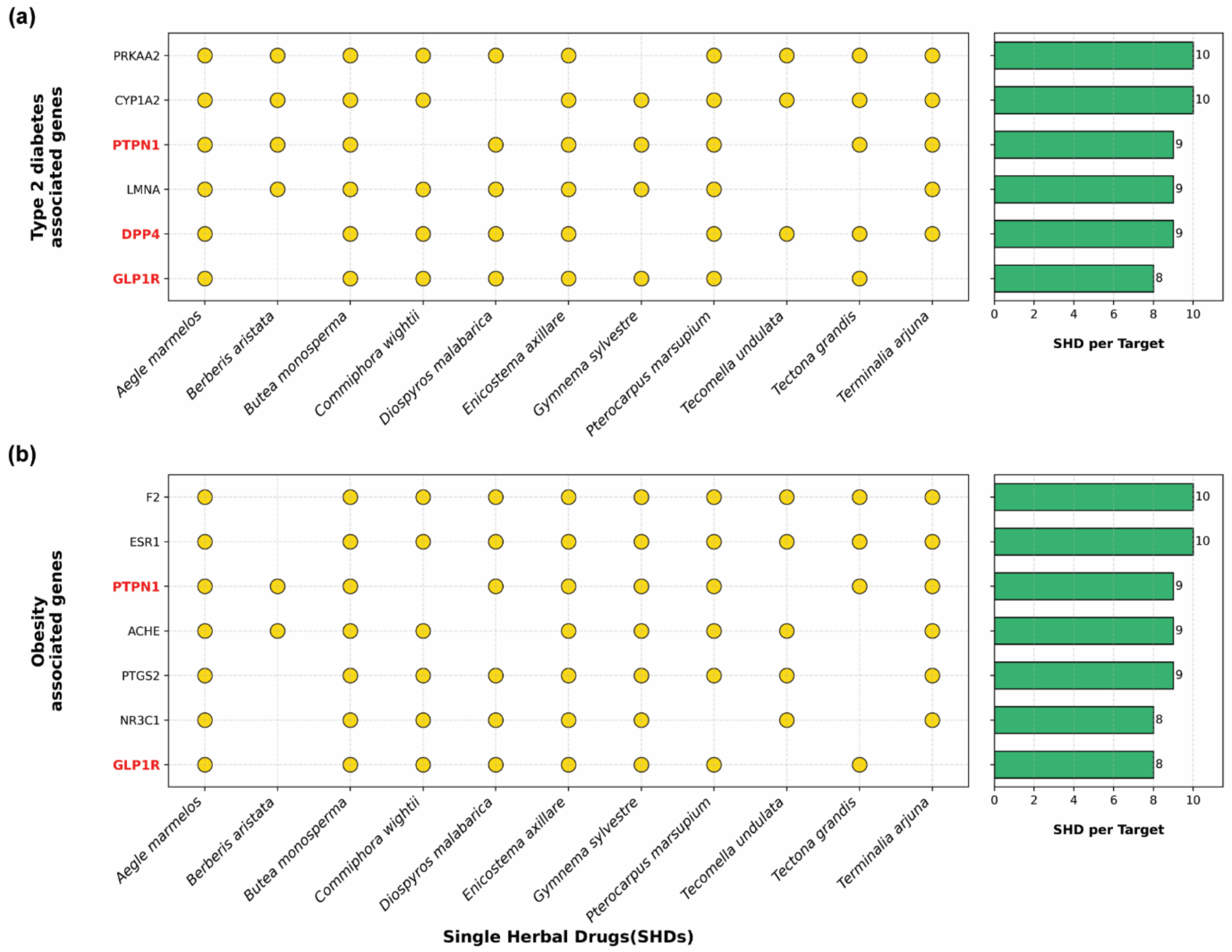

**Figure 5. Disease-associated targets recurrently modulated by multiple SHDs. (a)** Type 2 diabetes, and **(b)** obesity associated targets modulated by at least eight of the 11 SHDs. In each panel, the x-axis represents SHDs and the y-axis lists the disease-associated targets. A dot indicates that a given SHD modulates the corresponding target. The horizontal bar plots display the number of SHDs modulating each target in the curated dataset. Targets that are shared between both diseases and/or are therapeutic targets of FDA-approved drugs are highlighted in red. The figure illustrates that multiple SHDs converge on a common set of core metabolic control nodes in type 2 diabetes and obesity.

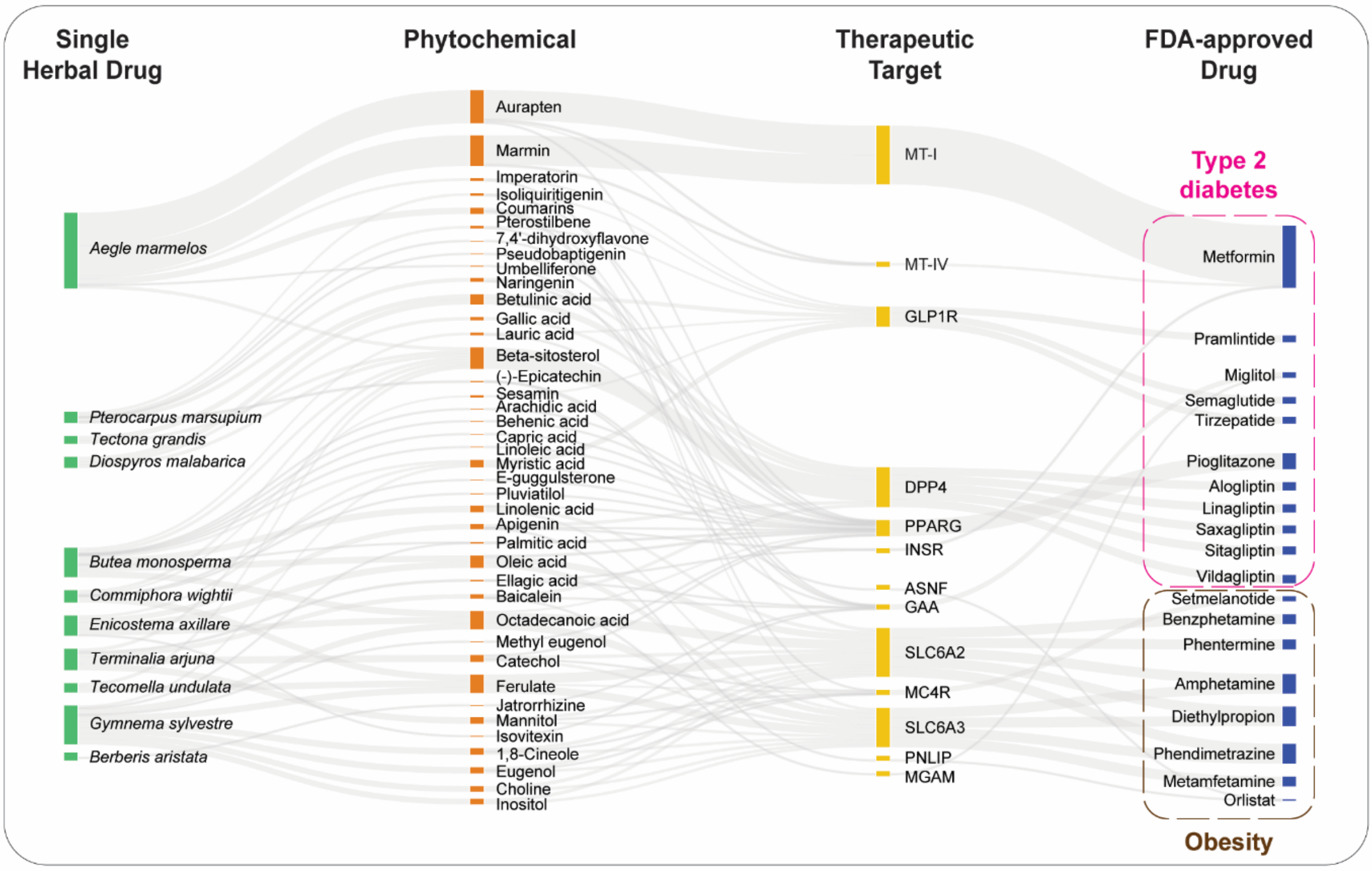

**Figure 6. Sankey plot illustrating the connectivity between Single Herbal Drugs (SHDs), their phytochemicals, molecular targets, and FDA-approved drugs.** SHDs are mapped to FDA-approved drugs through shared molecular targets to highlight potential mechanistic similarities between the actions of SHDs and combinations of known FDA-approved drugs. For metformin, multiple genes associated with mitochondrial complex I (e.g., NDUF and MT-ND family genes) are grouped and represented as MT-I in the Sankey plot. Genes associated with mitochondrial complex IV (e.g., COX family genes) are represented as MT-IV (see **Supplementary Table S13**).

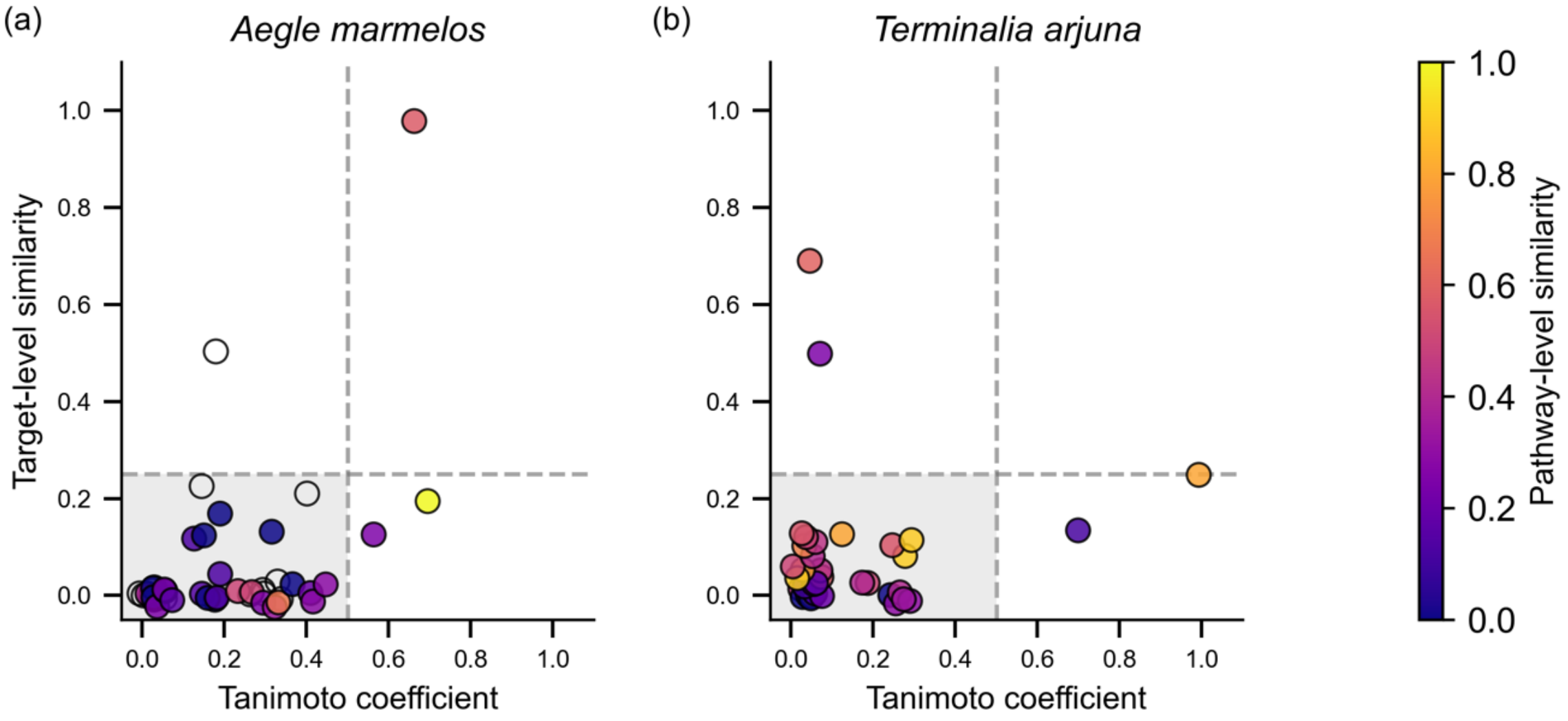

**Figure 7. Multi-level similarity profiles of phytochemical pairs from (a) *Aegle marmelos* and (b) *Terminalia arjuna*.** Each panel displays phytochemical pairs positioned according to chemical similarity and target-level Jaccard similarity, considering only pairs in which both compounds have at least one known disease-associated protein target. Vertical and horizontal dashed lines indicate a Tanimoto coefficient (Tc) of 0.5 and a target-level similarity of 0.25, respectively. Point color represents pathway-level Jaccard similarity, while hollow markers denote chemical pairs for which pathway-level similarity could not be computed because disease-overlapping targets of at least one phytochemical were not represented among the top enriched disease pathways considered (25 pathways in **Figure 3**). The shaded grey region highlights pairs with low structural and target-level similarity, reflecting chemically diverse compounds that act on distinct but potentially related biological mechanisms. A small visual jitter was applied to point positions to improve visibility of overlapping data points.

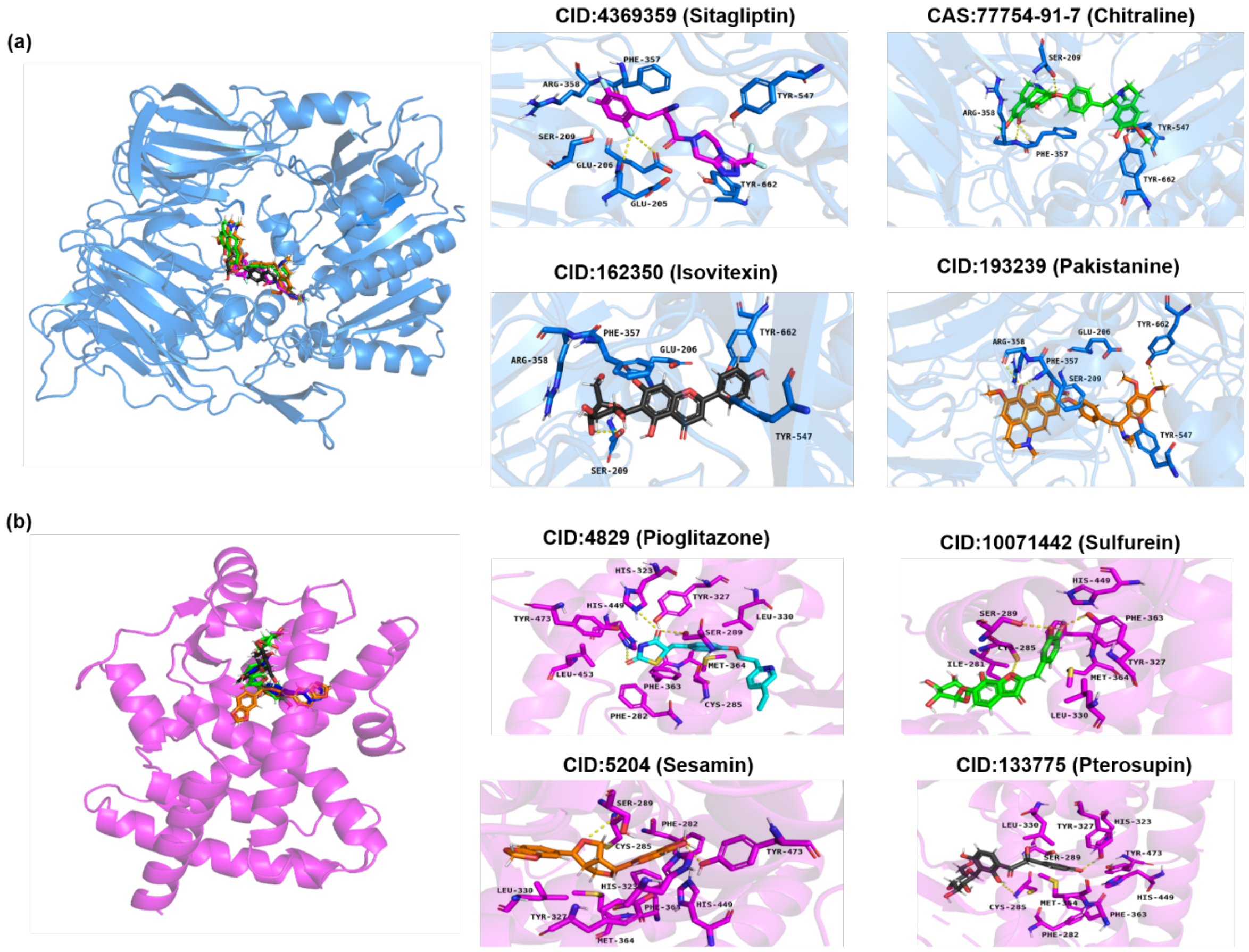

**Figure 8. Cartoon representations and binding site interaction views for the best-ranked docked complexes of (a) DPP4 (PDB: 1X70), and (b) PPARG (PDB: 5Y2O).** For each target, the left panel shows the overall protein cartoon with the docked ligands superposed in the binding pocket. The right panels show enlarged views of the binding site for the biological reference co-crystallized ligand and selected docked compounds. Ligands are shown as sticks (colored by element), and interacting protein residues are shown as sticks with residue labels. Dashed lines indicate predicted hydrogen bond interactions, while nearby hydrophobic pocket residues are displayed to highlight stabilizing contacts within the binding cavity.

# Supplementary Figures S1-S15

## for

# Computational investigation of single herbal drugs in Ayurveda for diabetes and obesity using knowledge graph and network pharmacology

Priyotosh Sil[a,b,1], Rahul Tiwari[a,b,1], Vasavi Garisetti[a], Shanmuga Priya Baskaran[a,b], Fenita Hephzibah Dhanaseelan[c], Smita Srivastava[c], Areejit Samal[a,b,*]

[a] *The Institute of Mathematical Sciences (IMSc), Chennai 600113, India*

[b] *Homi Bhabha National Institute (HBNI), Mumbai 400094, India*

[c] *Department of Biotechnology, Bhupat and Jyoti Mehta School of Biosciences, Indian Institute of Technology Madras, Chennai, 600036, India*

[1]Priyotosh Sil and Rahul Tiwari contributed equally to this work and should be considered as Joint-First authors

[*]Corresponding author: asamal@imsc.res.in

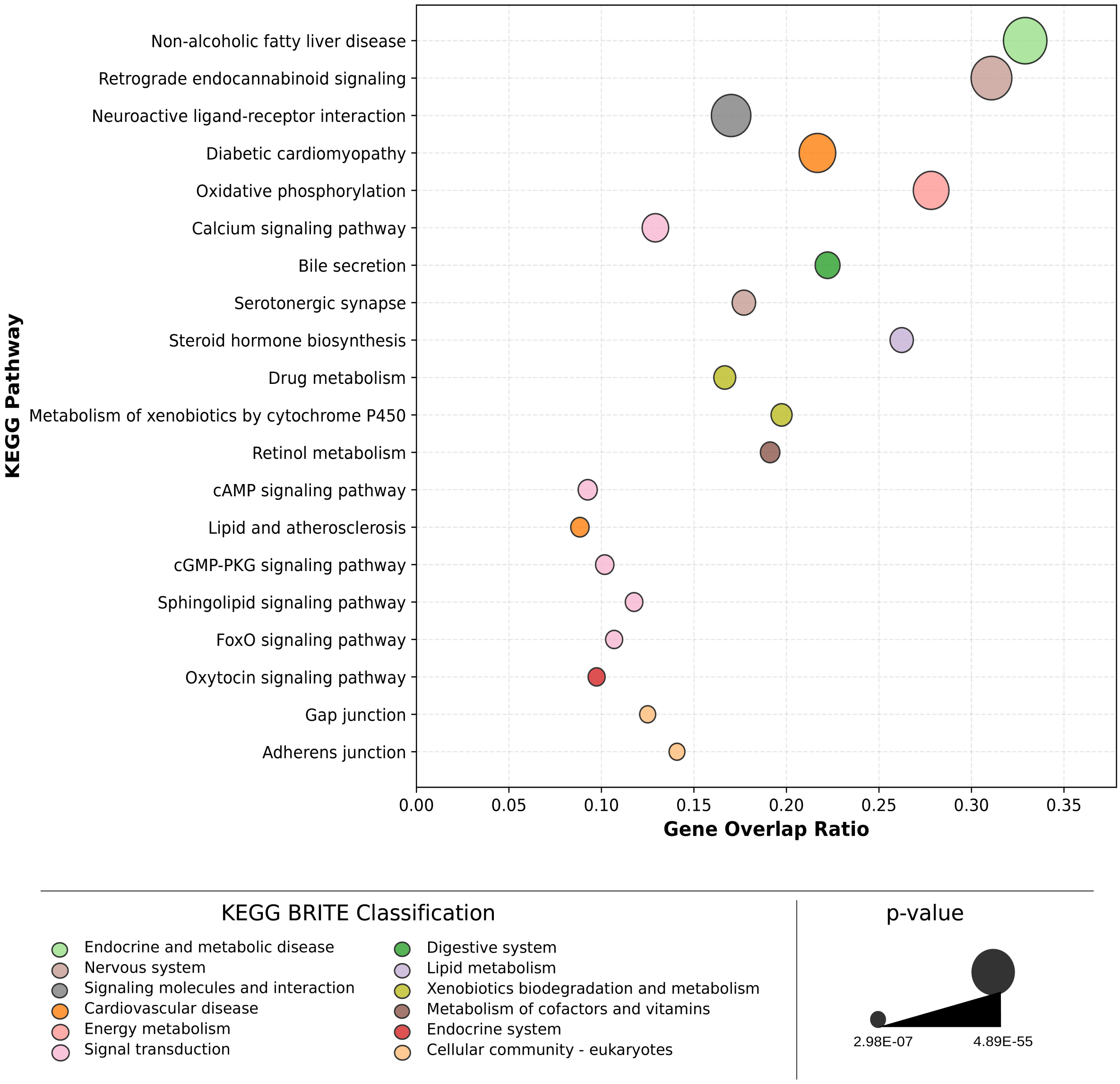

**Figure S1. Pathway enrichment analysis of human protein targets associated with phytochemicals from the SHD *Aegle marmelos*.** The bubble plot illustrates the 20 most significantly enriched KEGG pathways ($p < 0.01$) for this SHD. Bubble size reflects enrichment significance, whereas the position along the x-axis represents the fraction of pathway genes that overlap with the human protein targets associated *Aegle marmelos*. Bubble colors denote KEGG BRITE categories at Level 2.

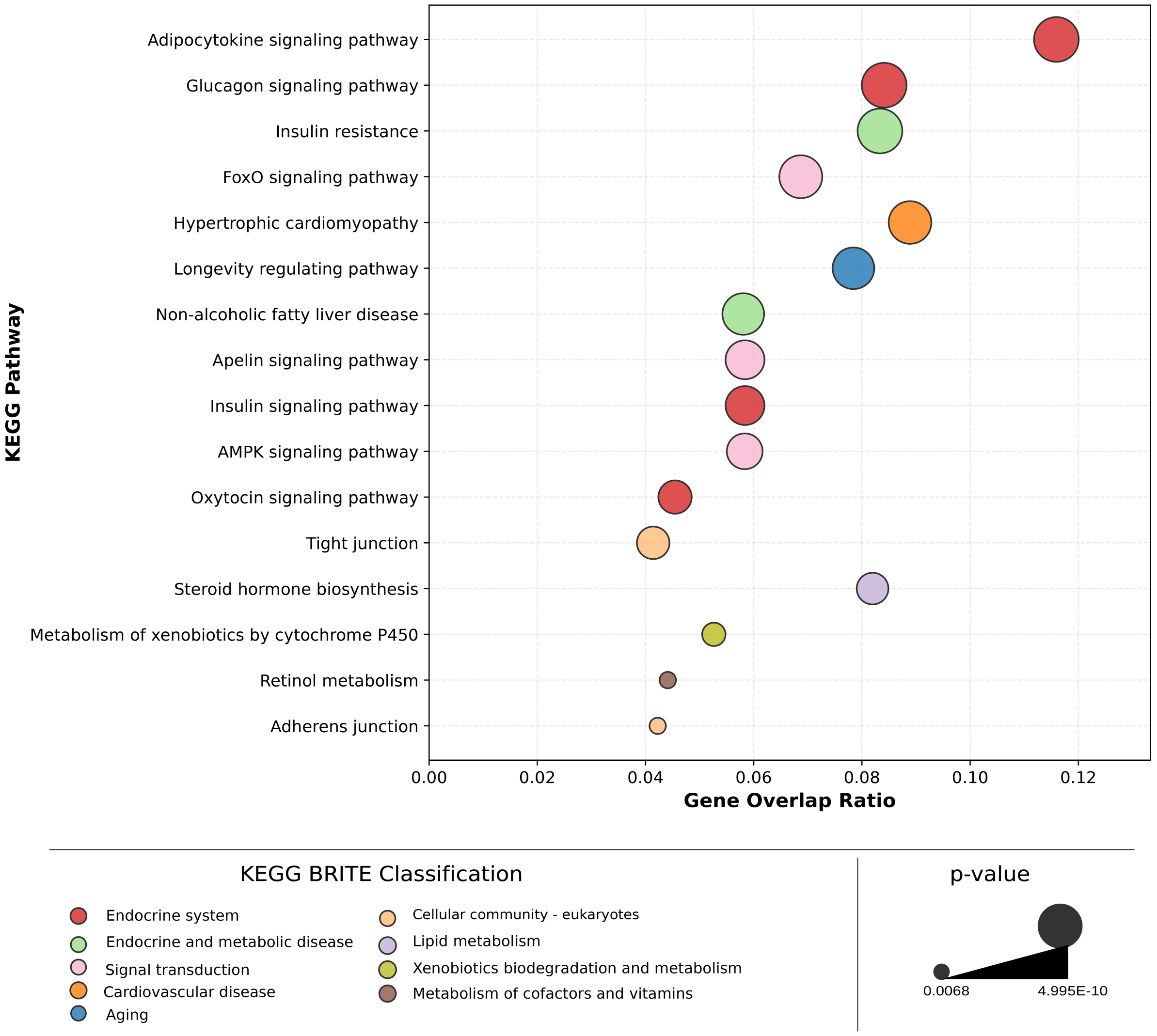

**Figure S2. Pathway enrichment analysis of human protein targets associated with phytochemicals from the SHD *Berberis aristata*.** The bubble plot illustrates the 16 significantly enriched KEGG pathways ($p < 0.01$) for this SHD; no additional pathway met this significance threshold. Bubble size reflects enrichment significance, whereas the position along the x-axis represents the fraction of pathway genes that overlap with the human protein targets associated with *Berberis aristata*. Bubble colors denote KEGG BRITE categories at Level 2.

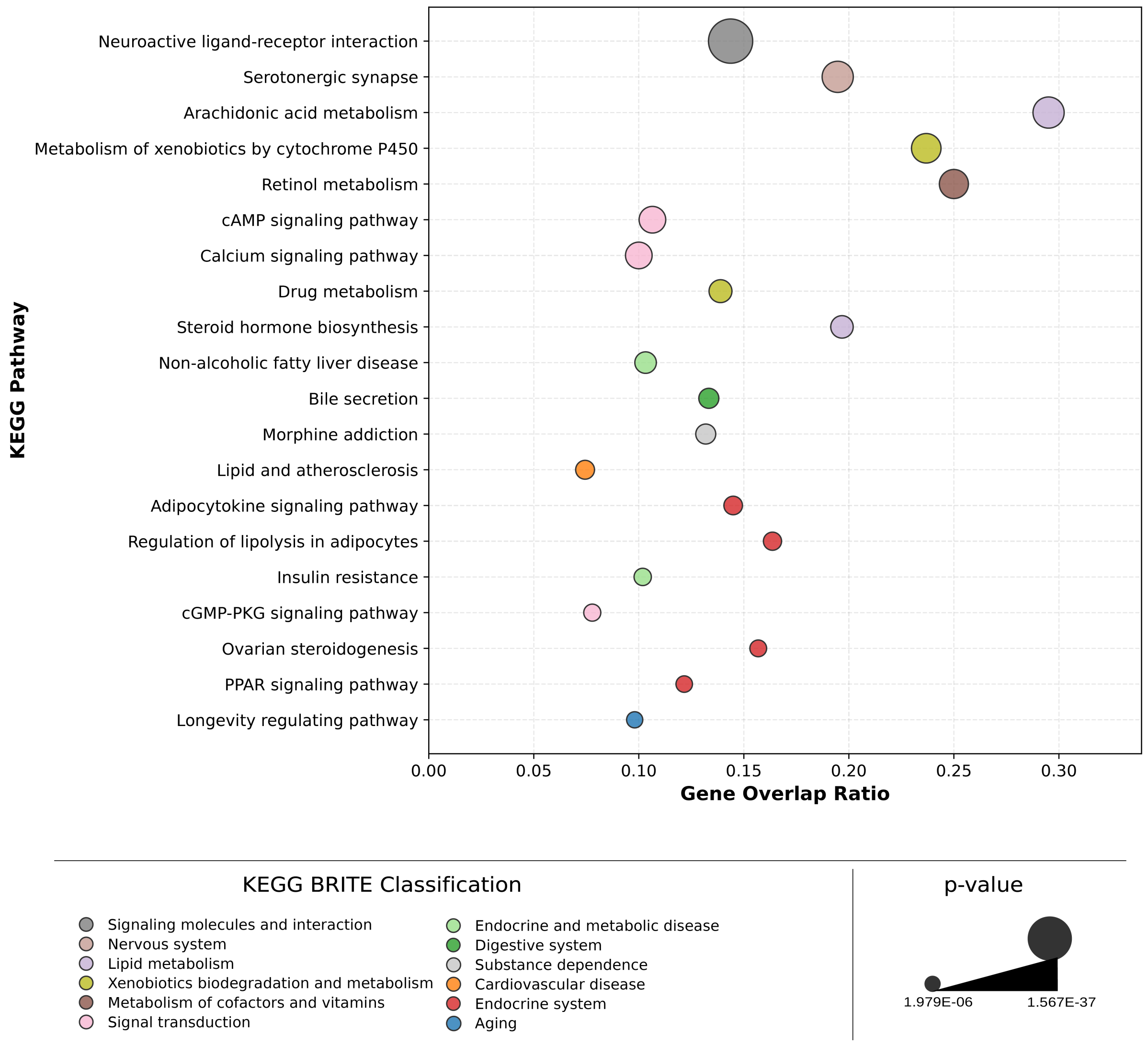

**Figure S3. Pathway enrichment analysis of human protein targets associated with phytochemicals from the SHD *Butea monosperma*.** The bubble plot illustrates the 20 most significantly enriched KEGG pathways ($p < 0.01$) for this SHD. Bubble size reflects enrichment significance, whereas the position along the x-axis represents the fraction of pathway genes that overlap with the human protein targets associated with *Butea monosperma*. Bubble colors denote KEGG BRITE categories at Level 2.

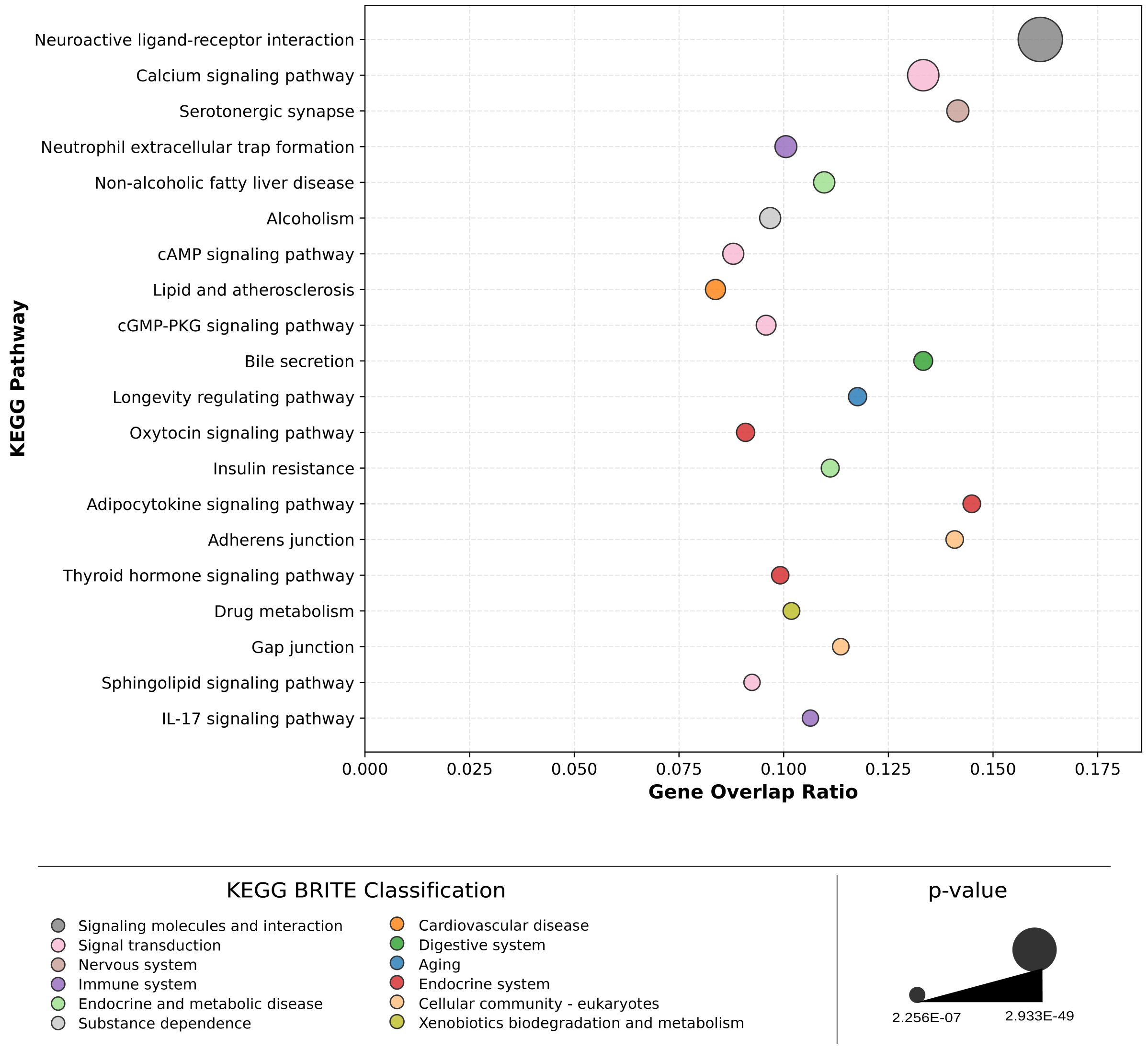

**Figure S4. Pathway enrichment analysis of human protein targets associated with phytochemicals from the SHD *Commiphora wightii.*** The bubble plot illustrates the 20 most significantly enriched KEGG pathways ($p < 0.01$) for this SHD. Bubble size reflects enrichment significance, whereas the position along the x-axis represents the fraction of pathway genes that overlap with the human protein targets associated with *Commiphora wightii*. Bubble colors denote KEGG BRITE categories at Level 2.

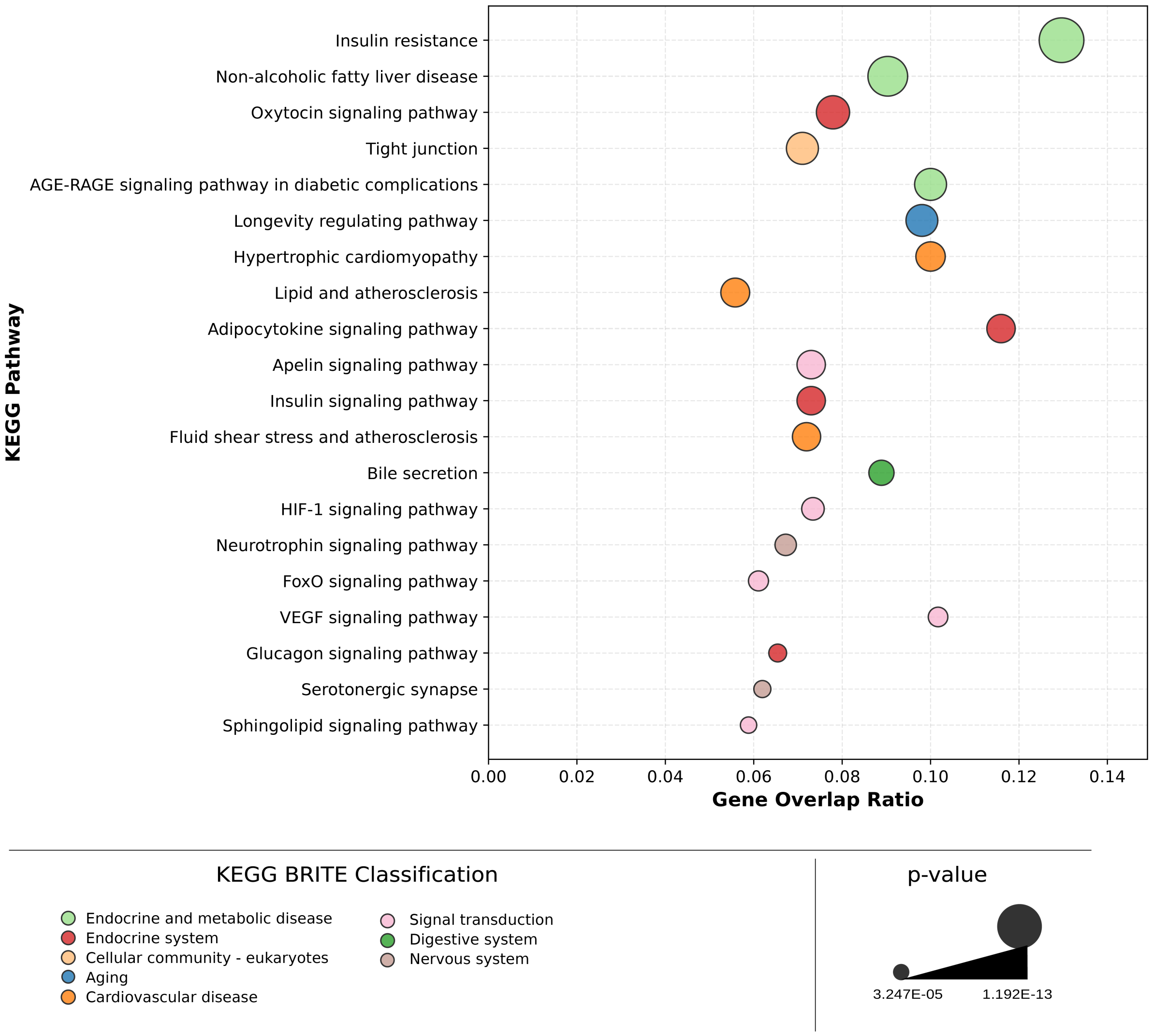

**Figure S5: Pathway enrichment analysis of human protein targets associated with phytochemicals from the SHD *Diospyros malabarica.*** The bubble plot illustrates the 20 most significantly enriched KEGG pathways ($p < 0.01$) for this SHD. Bubble size reflects enrichment significance, whereas the position along the x-axis represents the fraction of pathway genes that overlap with the human protein targets associated with *Diospyros malabarica*. Bubble colors denote KEGG BRITE categories at Level 2.

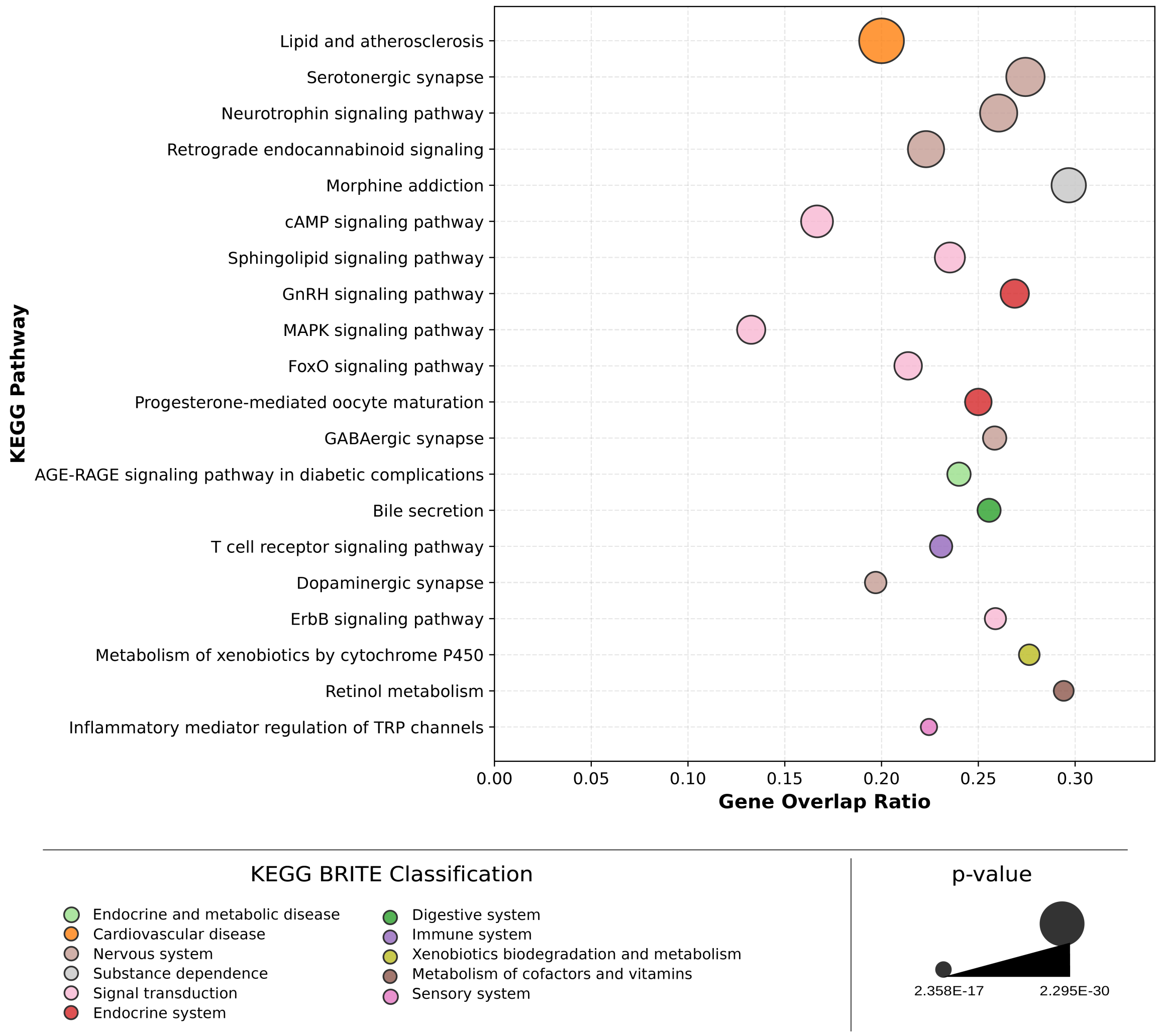

**Figure S6. Pathway enrichment analysis of human protein targets associated with phytochemicals from the SHD *Enicostemma axillare.*** The bubble plot illustrates the 20 most significantly enriched KEGG pathways ($p < 0.01$) for this SHD. Bubble size reflects enrichment significance, whereas the position along the x-axis represents the fraction of pathway genes that overlap with the human protein targets associated with *Enicostemma axillare*. Bubble colors denote KEGG BRITE categories at Level 2.

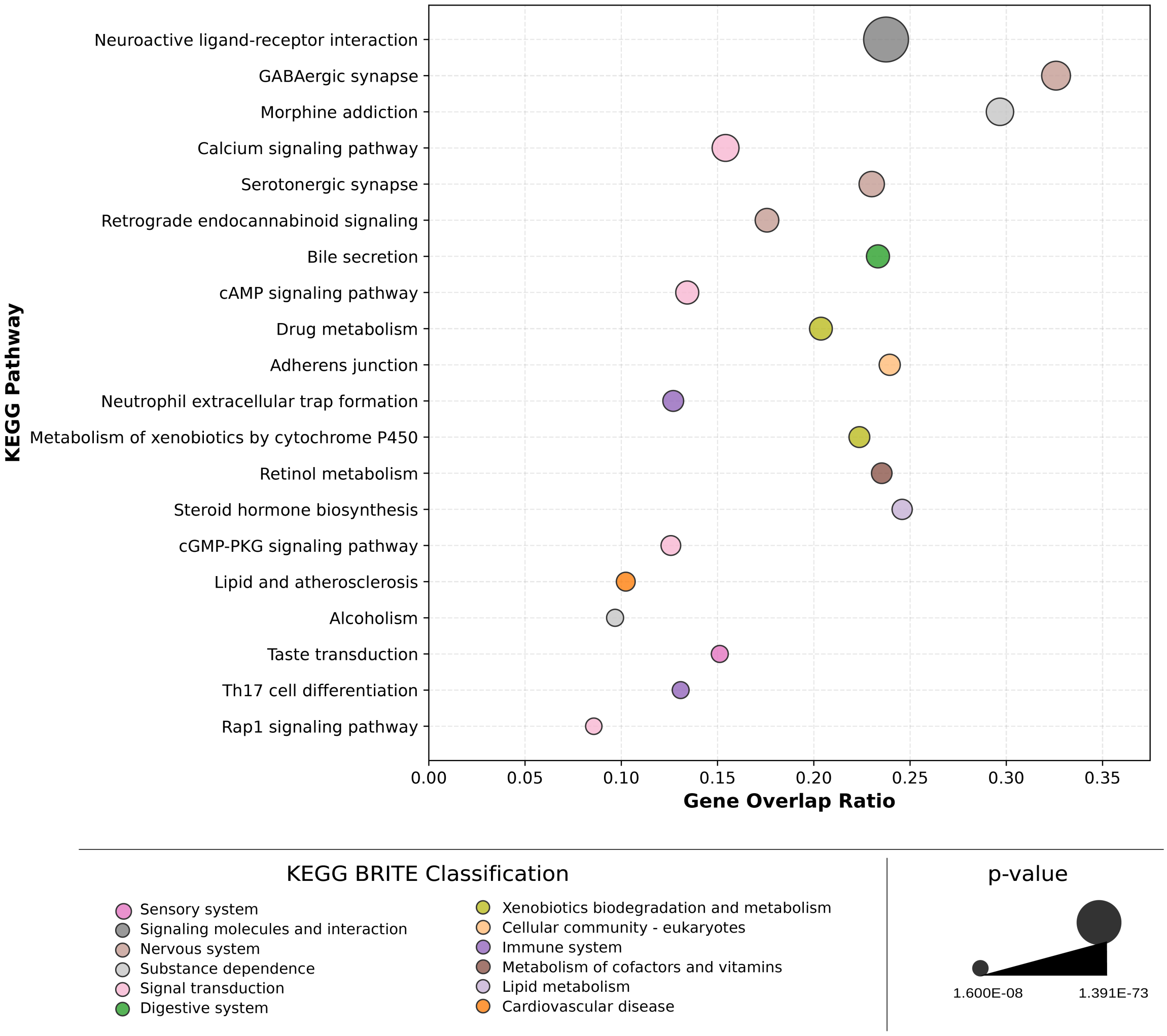

**Figure S7: Pathway enrichment analysis of human protein targets associated with phytochemicals from the SHD *Gymnema sylvestre.*** The bubble plot illustrates the 20 most significantly enriched KEGG pathways ($p < 0.01$) for this SHD. Bubble size reflects enrichment significance, whereas the position along the x-axis represents the fraction of pathway genes that overlap with the human protein targets associated with *Gymnema sylvestre*. Bubble colors denote KEGG BRITE categories at Level 2.

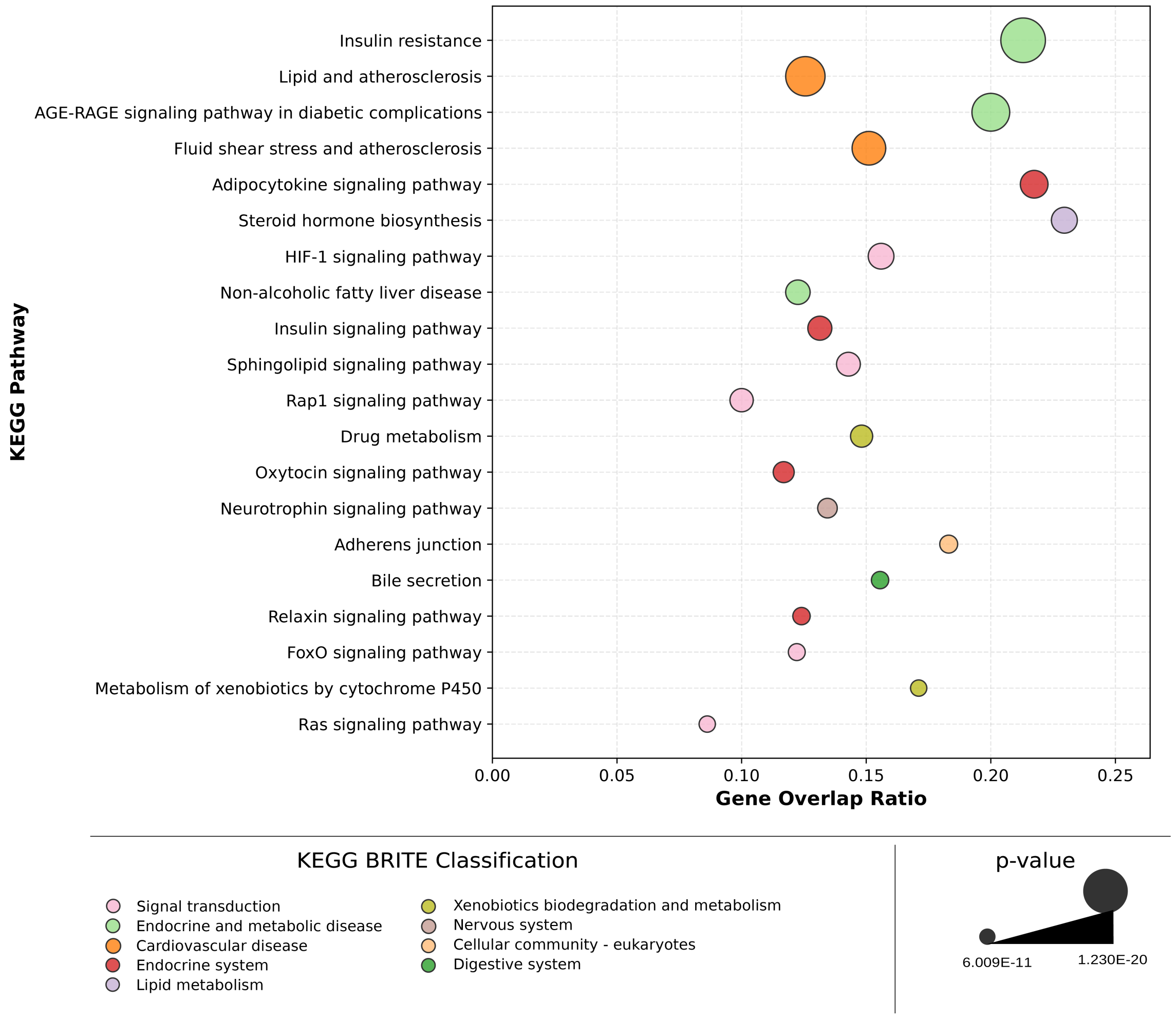

**Figure S8. Pathway enrichment analysis of human protein targets associated with phytochemicals from the SHD *Pterocarpus marsupium*.** The bubble plot illustrates the 20 most significantly enriched KEGG pathways ($p < 0.01$) for this SHD. Bubble size reflects enrichment significance, whereas the position along the x-axis represents the fraction of pathway genes that overlap with the human protein targets associated with *Pterocarpus marsupium*. Bubble colors denote KEGG BRITE categories at Level 2.

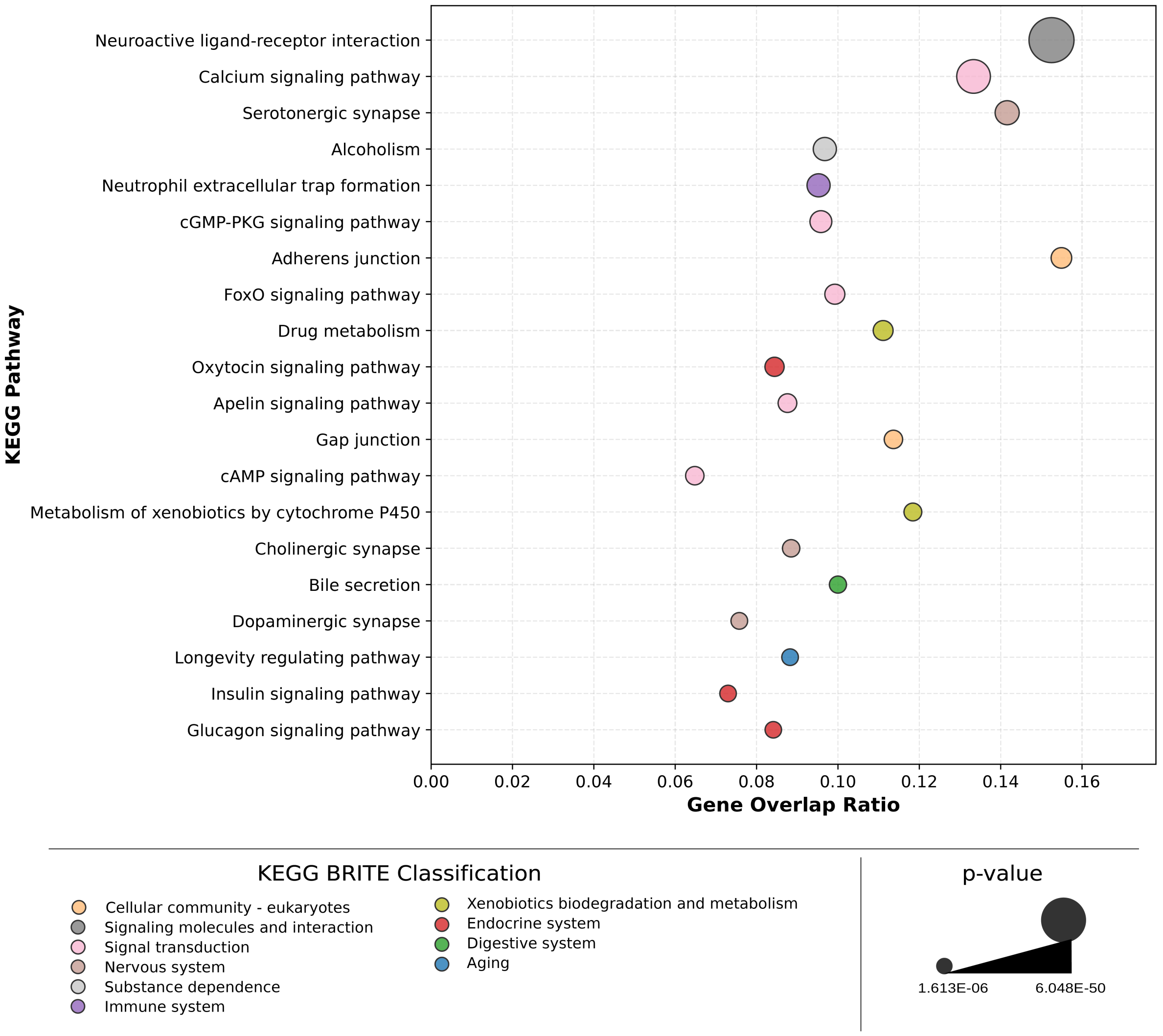

**Figure S9: Pathway enrichment analysis of human protein targets associated with phytochemicals from the SHD *Tecomella undulata*.** The bubble plot illustrates the 20 most significantly enriched KEGG pathways ($p < 0.01$) for this SHD. Bubble size reflects enrichment significance, whereas the position along the x-axis represents the fraction of pathway genes that overlap with the human protein targets associated with *Tecomella undulata*. Bubble colors denote KEGG BRITE categories at Level 2**.**

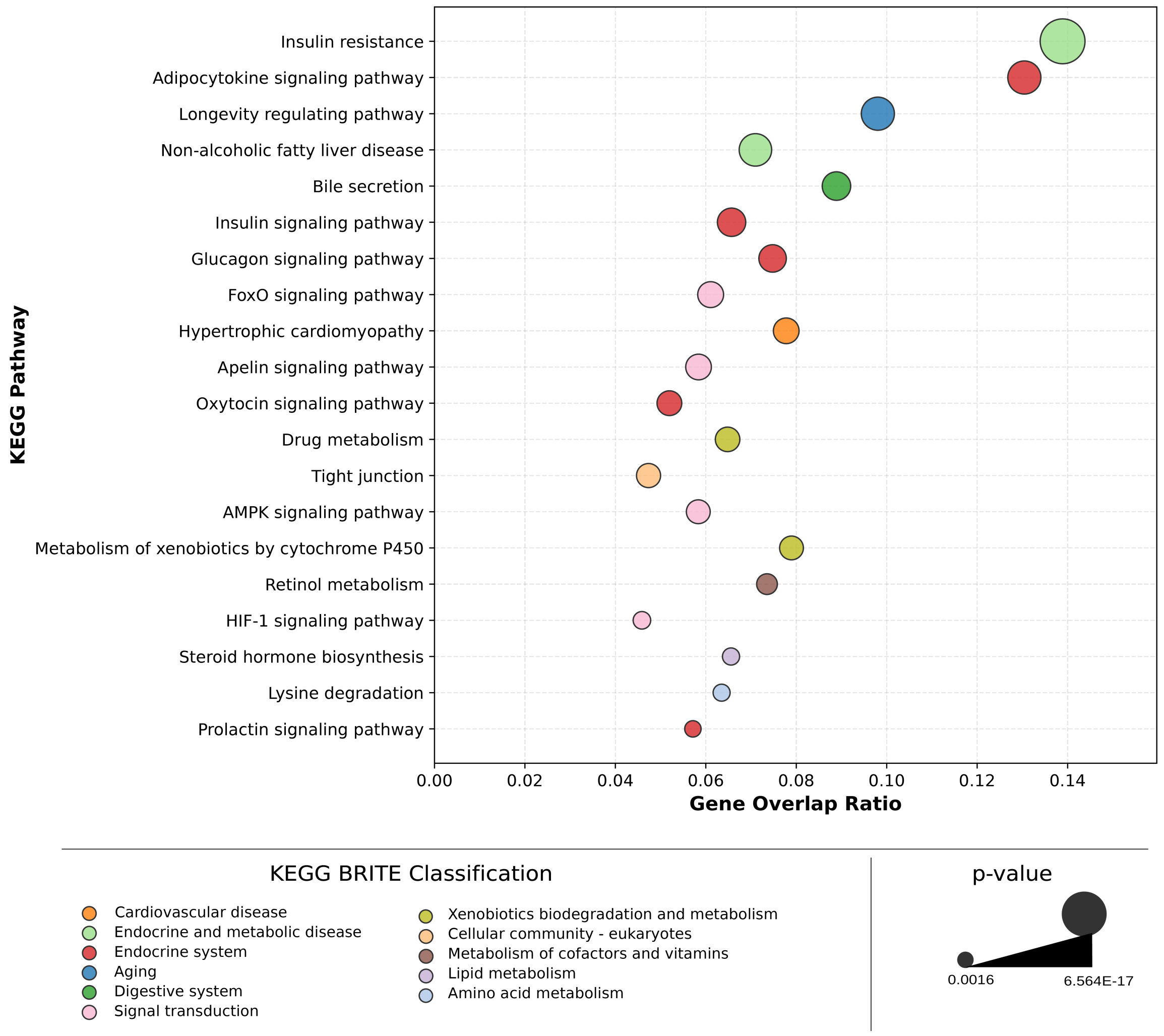

**Figure S10: Pathway enrichment analysis of human protein targets associated with phytochemicals from the SHD *Tectona grandis.*** The bubble plot illustrates the 20 most significantly enriched KEGG pathways ($p < 0.01$) for this SHD. Bubble size reflects enrichment significance, whereas the position along the x-axis represents the fraction of pathway genes that overlap with the human protein targets associated with *Tectona grandis*. Bubble colors denote KEGG BRITE categories at Level 2**.**

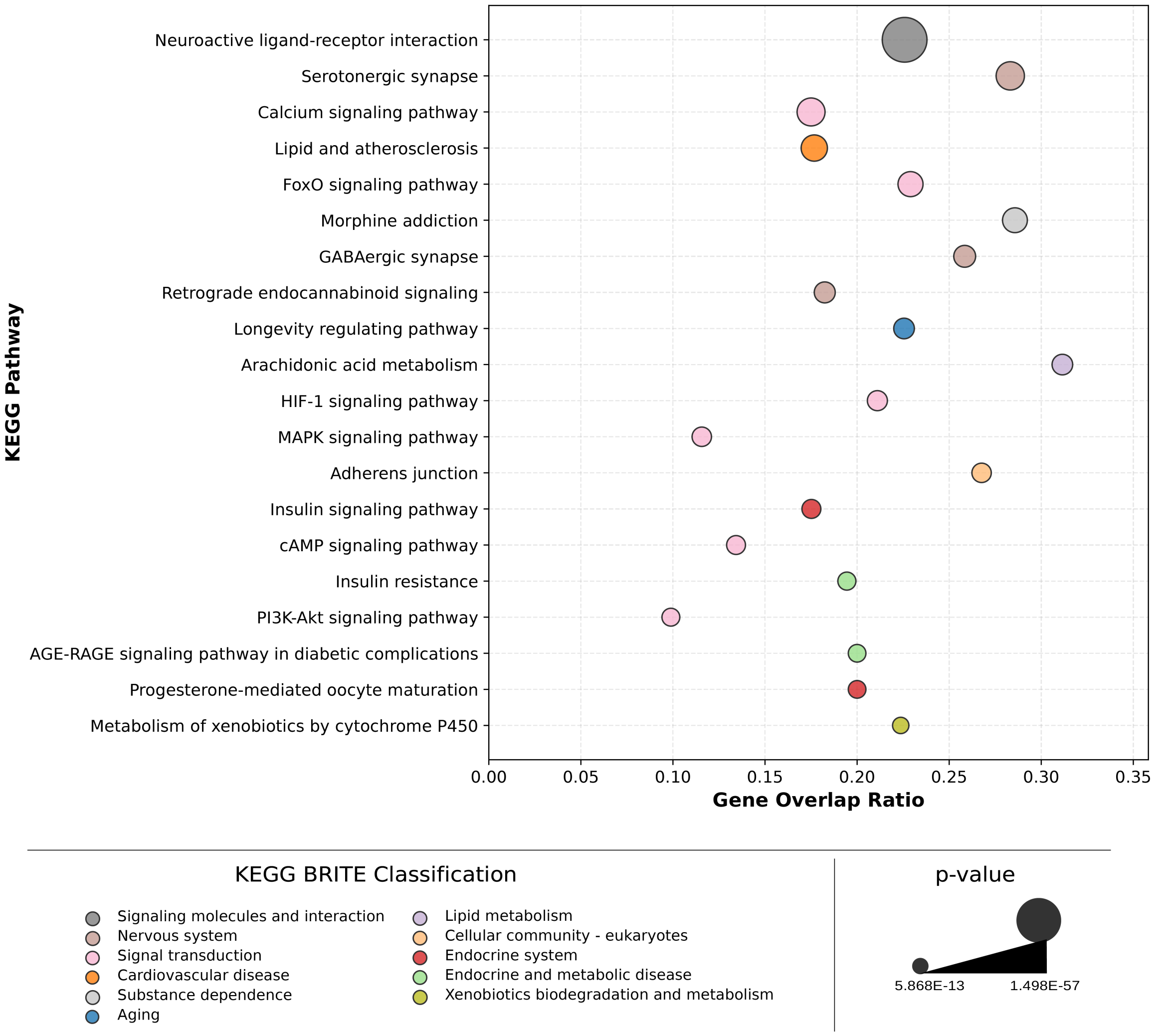

**Figure S11: Pathway enrichment analysis of human protein targets associated with phytochemicals from the SHD *Terminalia arjuna.*** The bubble plot illustrates the 20 most significantly enriched KEGG pathways ($p < 0.01$) for this SHD. Bubble size reflects enrichment significance, whereas the position along the x-axis represents the fraction of pathway genes that overlap with the human protein targets associated with *Terminalia arjuna*. Bubble colors denote KEGG BRITE categories at Level 2**.**

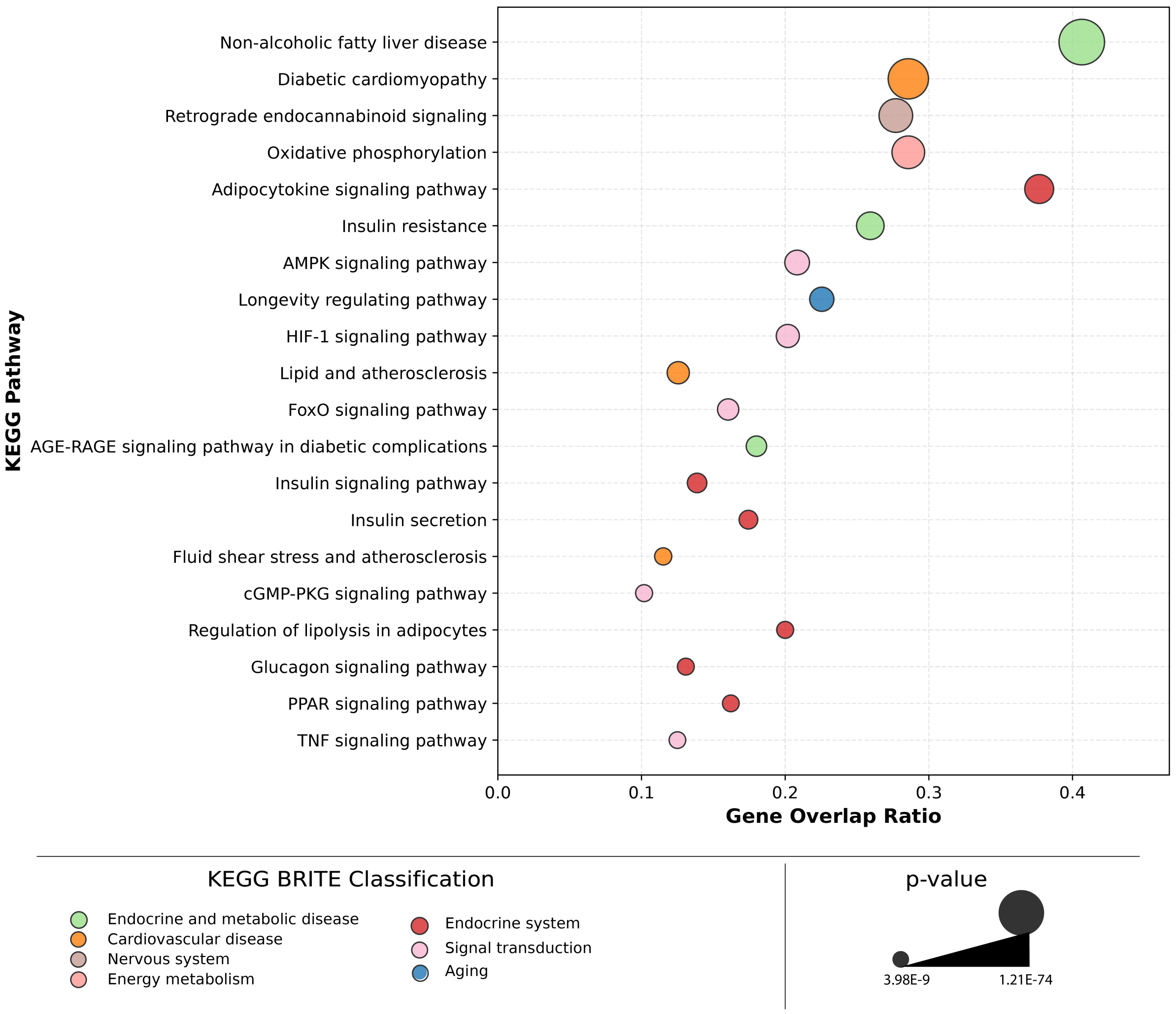

**Figure S12: Pathway enrichment analysis of curated genes associated with type 2 diabetes.** The bubble plot illustrates the 20 most significantly enriched KEGG pathways ($p < 0.01$) for the genes associated with type 2 diabetes. Bubble size reflects enrichment significance, whereas the position along the x-axis represents the fraction of pathway genes that overlap with the genes associated type 2 diabetes. Bubble colors denote KEGG BRITE categories at Level 2.

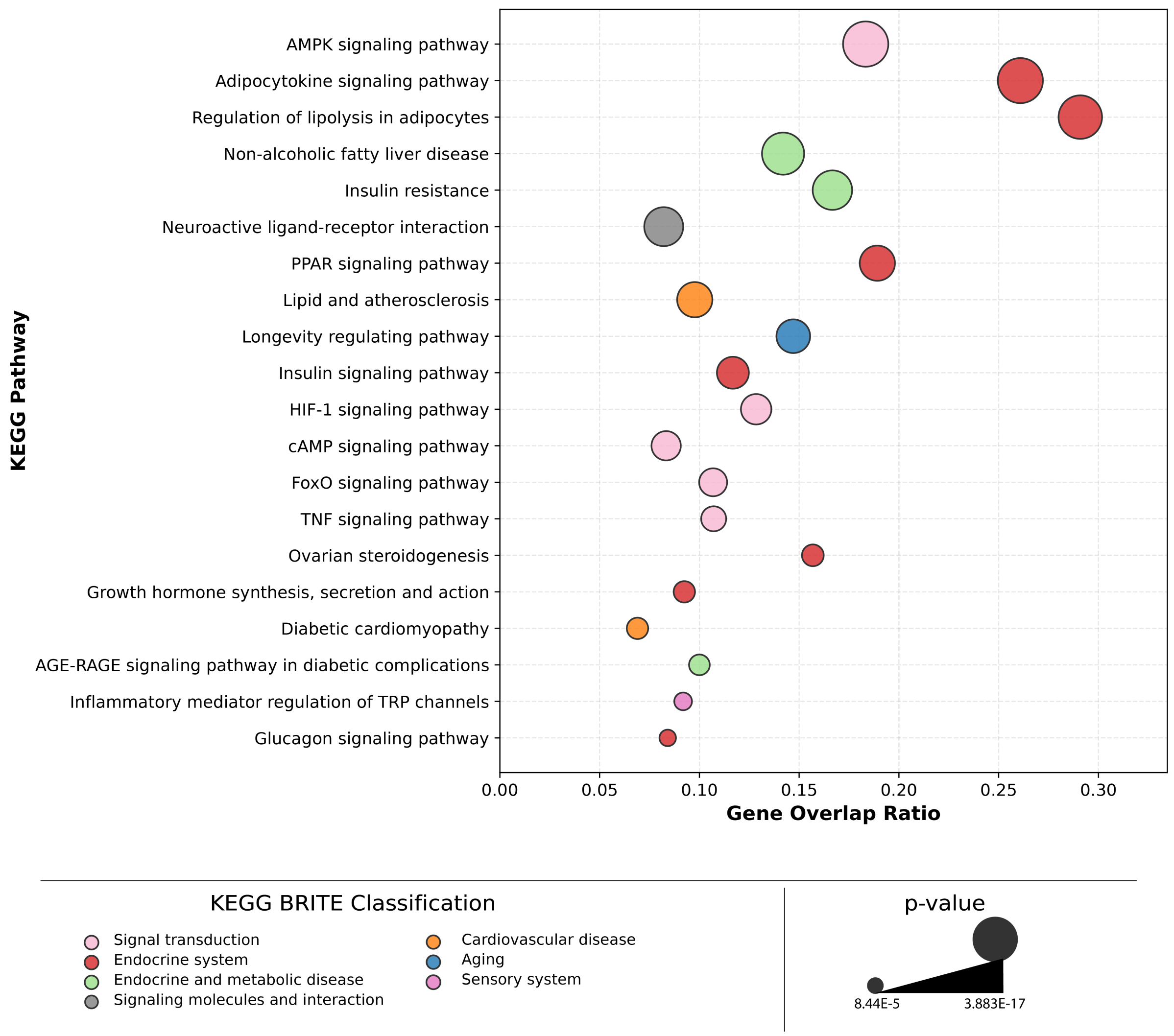

**Figure S13: Pathway enrichment analysis of curated genes associated with obesity.** The bubble plot illustrates the 20 most significantly enriched KEGG pathways ($p < 0.01$) for the genes associated with obesity. Bubble size reflects enrichment significance, whereas the position along the x-axis represents the fraction of pathway genes that overlap with the genes associated obesity. Bubble colors denote KEGG BRITE categories at Level 2.

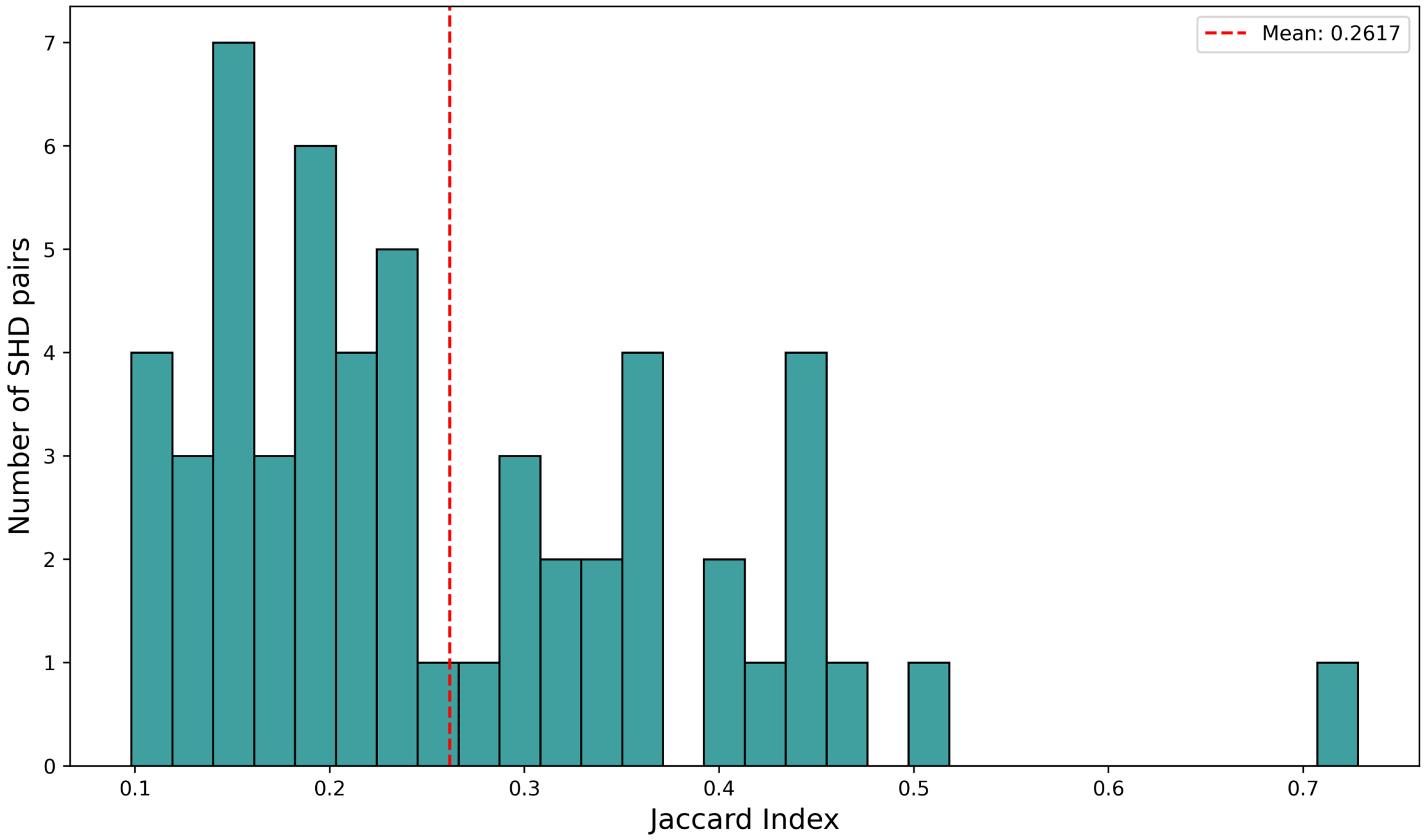

**Figure S14. Distribution of target-level similarity for SHD pairs using Jaccard index values.** The x-axis in this plot correspond to the Jaccard index values. Based on 11 SHDs, a total of 55 pairwise comparisons were considered. The histogram indicates the frequency, i.e., the number of SHD pairs for a given Jaccard index. The vertical dashed red line indicates the mean of the distribution. This illustrates that the pairwise target overlap among SHDs is low to moderate.

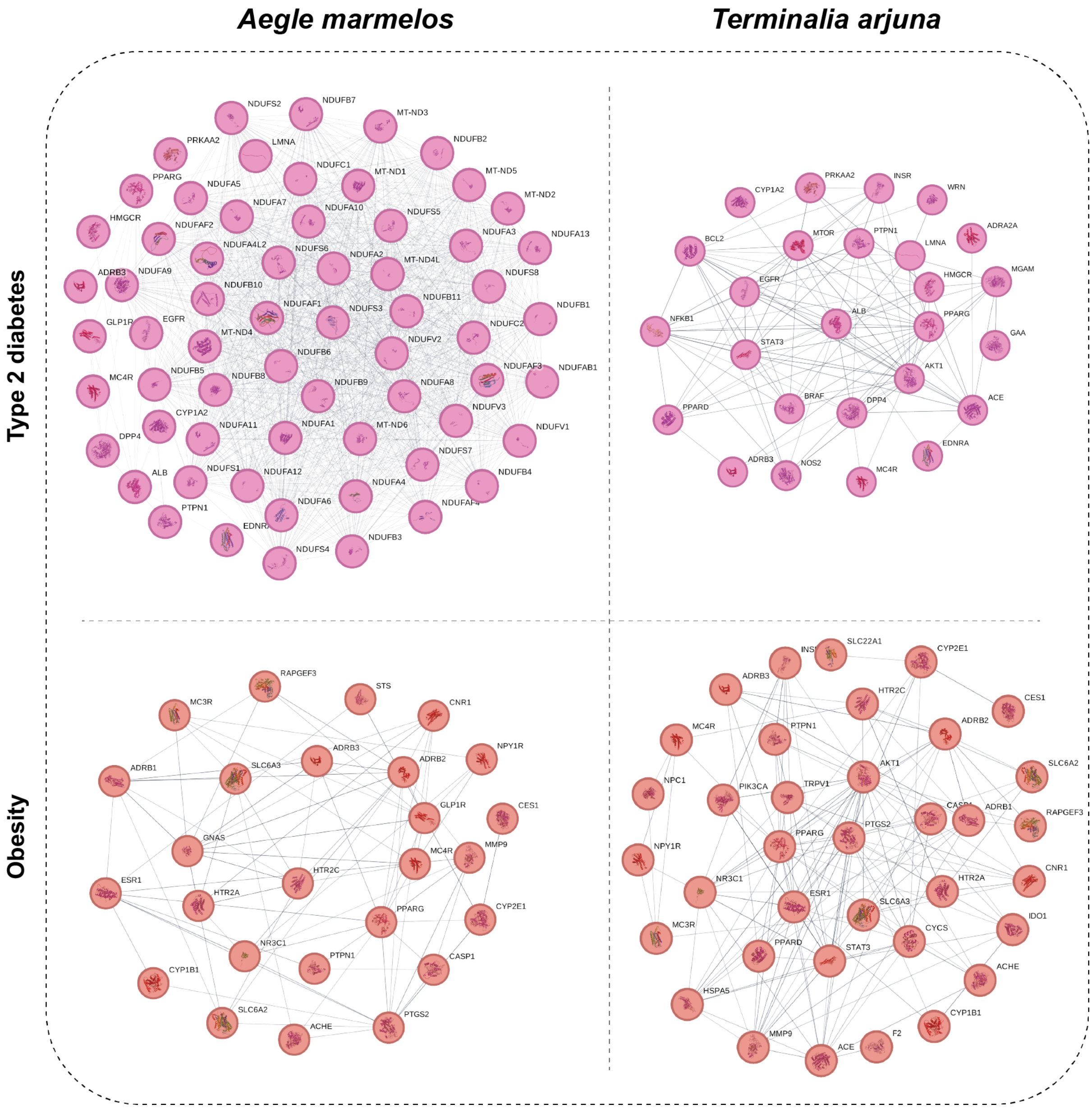

**Figure S15: Largest Connected Components (LCCs) of the protein-protein interaction (PPI) networks for SHD–disease overlaps for *Aegle marmelos* and *Terminalia arjuna*.** LCCs of PPI networks constructed from targets shared between disease-associated proteins and those modulated by *Aegle marmelos* and *Terminalia arjuna*. The top panel shows LCCs based on overlaps with type 2 diabetes–associated targets, while the bottom panel shows LCCs derived from overlaps with obesity-associated targets.

# Supplementary Tables S1-S16
# for
# Computational investigation of single herbal drugs in Ayurveda for diabetes and obesity using knowledge graph and network pharmacology

Priyotosh Sil[a,b,1], Rahul Tiwari[a,b,1], Vasavi Garisetti[a], Shanmuga Priya Baskaran[a,b], Fenita Hephzibah Dhanaseelan[c], Smita Srivastava[c], Areejit Samal[a,b,*]

[a] *The Institute of Mathematical Sciences (IMSc), Chennai 600113, India*

[b] *Homi Bhabha National Institute (HBNI), Mumbai 400094, India*

[c] *Department of Biotechnology, Bhupat and Jyoti Mehta School of Biosciences, Indian Institute of Technology Madras, Chennai, 600036, India*

[1]*Priyotosh Sil and Rahul Tiwari contributed equally to this work and should be considered as Joint-First authors*

**Corresponding author: asamal@imsc.res.in*

**Table S1: Ayurvedic Single Herbal Drugs (SHDs) documented in the Ayurvedic Pharmacopoeia of India (API) that are indicated for the management of diabetes and obesity.** The table lists 24 SHDs and reports, for each entry, the traditional Ayurvedic name, corresponding botanical name, synonymous name, plant part used, and therapeutic indications as described in the API, along with the relevant API volume and page reference. Additional information includes the plant family, IUCN conservation status, and the availability of associated phytochemicals as recorded in the IMPPAT database and/or the API.

| Serial Number | Traditional name from API | Plant name from API | Synonymous Name | Plant part(s) used from API | Therapeutic uses from API | API Volume | Page number | Family | IUCN status | Availability of phytochemical data (IMPPAT / API) |
|---|---|---|---|---|---|---|---|---|---|---|
| **1** | Guggulu | *Commiphora wightii* (Arn.) Bhand, Syn.*Balsamodend* | - | Exudate | Āmavāta, Kuṣṭha, Prameha, Vātavyādi, Granthi, Śopha,Gaṇḍamālā, Medoroga | API Vol 1 | 43 | Burseraceae | Critically Endangered | Yes |
| **2** | Asana | *Pterocarpus marsupium* | - | Heart-wood | Kṛmiroga, Kuṣṭha, Prameha, Pāṇḍu, Medodoṣa | API Vol 1 | 12 | Fabaceae | Near Threatened | Yes |
| **3** | Yava | *Hordeum vulgare* Linn | - | Dried fruits | Śvāsa, Kāsa, Medoroga, Pīnasa, Prameha, Tṛṣṇā, Urustambha,Kanharoga, Tvagroga | API Vol 2 | 175-176 | Poaceae | NA | Yes |
| **4** | Arjuna | *Terminalia arjuna* W.& A. | - | Stem bark | Medoroga, Vraṇa, Hṛdroga, Kṣatakṣaya, Prameha, Tṛṣā, Vyaṅga | API Vol 2 | 18 | Malvaceae | Least Concern | Yes |
| **5** | Asana | *Pterocarpus marsupium* Roxb. | - | Dried stem bark | Pāṇḍu, Prameha, Medodoṣa, Kuṣṭha, Kṛmiroga, Śvitra,Madhumeha, Sthoulya | API Vol 3 | 20 | Fabaceae | Near Threatened | Yes |
| **6** | Śāka | *Tectona grandis* Linn. f. | - | Dried heart wood | Kuṣṭha, Raktapitta, Mūtraroga, Pāṇḍu, Prameha, Medoroga, Dāha, Śrama, Tṛṣṇā, Kṛmiroga, Garbhasrāva, Garbhapātana | API Vol 3 | 174-175 | Lamiaceae | Endangered | Yes |
| **7** | Bilva | *Aegle marmelos* Corr. | - | Dried stem bark | Chardi, Vātavyādhi, Śūla, Śotha, Atīsāra, Raktātisāra, Kukṣiśūla, Āmaśūla, Arśa, Medoroga, Grahaṇīroga, Madhumeha, Pravāhikā | API Vol 4 | 12-13 | Rutaceae | Near Threatened | Yes |
| **8** | Muṇḍītikā | *Sphaeranthus indicus* Linn. | - | Dried whole plant | Apau, Mūtrakṛcchra, Kṛmiroga, Vātarakta, Pāṇḍu, Yoni Roga, Āmātisāra, Kāsa, Ślīpada, Apasmāra, Plīhāroga, Medoroga, Gudaroga, Prameha, Chardi | API Vol 4 | 70-72 | Asteraceae | Least Concern | Yes |
| **9** | Sarja | *Vateria indica* Linn. | - | Exudate | Pāṇḍu, Karṇa Roga, Prameha, Kuṣṭha, Bādhirya, Vraṇa, Atīsāra, Kaṇḍū, Visphoṭa, Medoroga, Grahaṇī, Vātarakta, Kṣudraroga, Lippa, Mānasa Roga, Mūṣika Viṣa, Vidradhi, Dagdhaka, Yoni Roga, Rakta Doṣa, Kṛmiroga | API Vol 4 | 106-107 | Dipterocarpaceae | Vulnerable | No |
| **10** | Kadaraḥ | *Acacia suma* Buch.-Ham. | - | Dried pieces of heart wood | Madhumeha, Mukharoga, Udarda, Kaṇḍū, Medodoṣa, Vraṇa, Pāṇḍu, Kuṣṭha, Śvitra, Raktadoṣa | API Vol 5 | 54-55 | Fabaceae | NA | Yes |
| **11** | Palāśaḥ | *Butea monosperma* (Lam.) Kuntze | - | Seed | Kṛmi, Vraṇa, Gulma, Gudajaroga, Arśa, Raktavikāra, Vātarakta, Udararoga, Kāsa, Kaṇḍū, Tvagroga, Prameha, Yonidoṣa, Sukradoṣa, Mūtrakṛcchra, Kuṣṭha, Pāmā, Dadru, Dāha, Plīhāroga, Atīsāra, Netraśukra, Śūla, Medoroga, Pāṇḍu, Aśmarī, Vṛścika-viṣa | API Vol 5 | 127 | Fabaceae | Least Concern | Yes |
| **12** | Tūnī | *Cedrela toona* Roxb. | *Toona hexandra* | Stem bark | Bāla Pravāhikā, Vraṇa, Dāha, Yoniroga, Kaṇḍū, Kuṣṭha, Gaṇḍamālā, Raktavikāra, Raktapitta, Śvetakuṣṭha, Prameha, Viṣavikāra, Medovikāra | API Vol 5 | 179-180 | Meliaceae | Least Concern | Yes |
| **13** | Viralā | *Diospyros exsculpta* Buch. - Ham. | - | Dried stem bark | Udarda, Prameha, Raktapitta, Aruci, Atīsāra, Vibandha, Pittaroga, Karṇasrāva, Vraṇa, Agnidagdha Vraṇa, Atidagdha Vraṇa, Bhagna, Tṛṣṇā, Dāha, Yoniroga, Medoroga | API Vol 5 | 195-196 | Ebenaceae | NA | Yes |
| **14** | Tiniśaḥ | *Ougeinia oojeinensis* (Roxb.) Hochr. | - | Wood | Śotha, Kuṣṭha, Atīsāra, Raktātisāra, Pravāhikā, Raktavikāra, Raktapitta, Prameha, Śvitra, Vraṇa, Kṛmi, Pāṇḍuroga, Medoroga, Dāha | API Vol 5 | 172-173 | Fabaceae | Least Concern | Yes |
| **15** | Kapītana | *Thespesia populnea* (L.) Soland. ex Correa | - | Stem bark | Raktapitta, Prameha, Raktavikāra, Yoniroga, Dāha, Tṛṣṇā, Medoroga, Vraṇa, Śotha, Tvagroga, Bālavisarpa, Pāmā, Kaṇḍū, Dadru | API Vol 5 | 63-64 | Meliaceae | Least Concern | No |
| **16** | Dhava | *Anogeissus latifolia* | *Terminalia anogeissiana* | Dried fruits | Aśmarī (Calculus), Arśa (Piles), Mūtrakṛcchra (Dysuria), Medoroga (Obesity), Pāṇḍu (Anaemia), Prameha (Metabolic disorder), Raktavīkāra (Disorders of blood), Upadaṁśa (Syphilis / soft chancre) | API Vol 6 | 34-35 | Combretaceae | NA | Yes |
| **17** | Dhava | *Anogeissus latifolia* | *Terminalia anogeissiana* | Dried stem bark | Aśmarī (Calculus), Arśa (Piles), Karṇasrāva (Otorrhoea), Kuṣṭha (Leprosy / diseases of skin), Mūtrakṛcchra (Dysuria), Medoroga (Obesity), Pāṇḍu (Anaemia), Prameha (Metabolic disorder), Raktavīkāra (Disorders of blood), Upadaṁśa (Syphilis / soft chancre), Visarpa (Erysepales) | API Vol 6 | 36-37 | Combretaceae | NA | Yes |
| **18** | Tinduka | *Diospyros peregrina* Gurke | *Diospyros malabarica* | Ripe fruits | Pakva phala- Āsmarī (Calculus), Aruci (Tastelessness), Kapharoga (Disease due to Kapha dosa), Prameha (Metabolic disorder), Raktadoṣa (Disorders of blood) Apakva phala- Atisīra (Diarrhoea), Bhagna (Fractures), Dāha (Burning sensation), Kuṣṭha (Leprosy / diseases of skin), Śotha (Oedema), Medoroga (Obesity), Pravīhikā (Dysentery), Raktapitta (Bleeding disorder), Udarda (Urticaria), Vraṇa (Ulcer) | API Vol 6 | 173-175 | Ebenaceae | Least Concern | Yes |
| **19** | Nāhī | *Enicostemma axillare* (Lam.) | - | Whole plant | Kṛmi (Worm infestation), Śotha (Oedema), Madhumeha (Diabetes mellitus), Medoroga (Obesity), Prameha (Metabolic disorder), Raktavīkāra (Disorders of blood), Tvak roga (Skin diseases), Viṣamajvara (Intermittent fever), Vibandha (Constipation), Yakrtdaurbalya (Poor function of liver) | API Vol 6 | 152 | Gentianaceae | Least Concern | Yes |
| **20** | Vajrānna | *Pennisetum typhoides* | - | Leaf bases | Prameha (Metabolic disorder), Śaitya (Milk), Santarpaṇjanya roga, Sthaulya (Obesity) | API Vol 6 | 185-186 | Poaceae | NA | No |
| **21** | Rohitaka | *Tecomella undulata* (Sm.) | - | Dried stem barks | Gulma (Abdominal lump), Kṛmi (Helminthiasis), Kāmalā (Jaundice), Karnaroga (Disease of ear), Kuṣṭha (Leprosy / diseases of skin), Medoroga (Obesity), Netraroga (Diseases of the eye), Plihodara (Splenomegaly), Prameha (Metabolic disorder), Raktadoṣa (Disorders of blood), Raktavīkāra (Disorders of blood), Śūla (Pain / Colic), Svetapradara (Leucorrhoea), Vibandha (Constipation), Vraṇa (Ulcer), Yakṛt (Liver) | API Vol 6 | 170 | Bignoniaceae | Endangered | Yes |
| **22** | Arjuna | *Terminalia arjuna* | - | Stem bark | Hṛdroga (heart disease); Kṣatakṣaya (emaciation due to injury), Medoroga(obesity), Prameha (increased frequency and turbidity of urine); Vraṇa (ulcer); Tṛṣā (Thirst);Vyaṅga (dark shade on face due to stress and excessive excercise /localized hyper pigmentation of skin) | API Vol 8 | 13-17 | Combretaceae | Least Concern | Yes |
| **23** | Dāruharidrā | *Berberis aristata* DC. | - | Dried stem | Āmātisāra (diarrhoea due to indigestion), Kaṇḍū (pruritis), Kapharoga (disease due to kapha doṣa), Karṇaroga (disease of ear), Medoroga (obesity), Mukharoga (disease of mouth), Netraroga (disease of eye), Prameha (increased frequency and turbidity of urine), Ūrustambha (stiffness in thigh muscles), Vraṇa (wound) | API Vol 9 | 16 | Berberidaceae | Least Concern | Yes |
| **24** | Meṣaśṛṅgī | *Gymnema sylvestre* | - | Dried leaf | Gulma (abdominal lump), Kuṣṭha (disease of skin), Prameha(increased frequency and turbidity of urine),Sthaulya (obesity), Śiraḥśūla (headache), Vidradhi (abscess) | API Vol 9 | 46 | Apocynaceae | NA | Yes |

**Table S2: Phytochemical and target profiling of Ayurvedic Single Herbal Drugs (SHDs) indicated for the management of diabetes and obesity.** This table summarizes 20 SHDs (for which phytochemical information is available in IMPPAT and/or the API), reporting the number of phytochemicals identified before and after drug-likeness and bioavailability screening. A minimum cut-off of ten retained phytochemicals was applied to select SHDs for downstream analysis. The filtered phytochemicals were subsequently queried against multiple databases, and the corresponding number of associated human molecular targets identified for each SHD is reported.

| Traditional name in API | Plant name | Part | Chemicals before drug-likeness and bioavailability analysis | Chemicals after drug-likeness and bioavailability analysis (with known mapping or structure) | Contains at least 10 phytochemicals | Phytochemicals with known human targets | Total number of human targets |
|---|---|---|---|---|---|---|---|
| Kadaraḥ | *Acacia suma* | Wood | 5 | 5 | No | - | - |
| Bilva | *Aegle marmelos* | Bark | 11 | 11 | Yes | 10 | 268 |
| Dāruharidrā | *Berberis aristata* | Stem | 10 | 10 | Yes | 5 | 55 |
| Palāśaḥ | *Butea monosperma* | Seed | 31 | 28 | Yes | 14 | 226 |
| Guggulu | *Commiphora wightii* | Plant exudate | 21 | 21 | Yes | 11 | 195 |
| Viralā | *Diospyros exsculpta* | Bark | 5 | 5 | No | - | - |
| Tinduka | *Diospyros malabarica* | Fruit | 15 | 15 | Yes | 7 | 111 |
| Nāhī | *Enicostema axillare* | Whole Plant | 19 | 15 | Yes | 7 | 360 |
| Meṣaśṛṅgī | *Gymnema sylvestre* | Leaf | 78 | 55 | Yes | 33 | 292 |
| Yava | *Hordeum vulgare* | Fruit | 3 | 2 | No | - | - |
| Tiniśaḥ | *Ougeinia oojeinensis* | Wood | 2 | 2 | No | - | - |
| Asana | *Pterocarpus marsupium* | Wood | 21 | 21 | Yes | 15 | 238 |
| Asana | *Pterocarpus marsupium* | Bark | 5 | 4 | No | - | - |
| Muṇḍītikā | *Sphaeranthus indicus* | Whole Plant | 6 | 6 | No | - | - |
| Rohitaka | *Tecomella undulata* | Bark | 17 | 14 | Yes | 7 | 161 |
| Śāka | *Tectona grandis* | Wood | 14 | 14 | Yes | 8 | 83 |
| Dhava | *Terminalia anogeissiana* | Bark | 10 | 8 | No | - | - |
| Dhava | *Terminalia anogeissiana* | Fruit | 3 | 3 | No | - | - |
| Arjuna | *Terminalia arjuna* | Bark | 25 | 20 | Yes | 12 | 388 |
| Tūnī | *Toona hexandra* | Bark | 2 | 2 | No | - | - |

**Table S3: Clinical trials associated with selected Single Herbal Drugs (SHDs) in relation to diabetes and obesity.** This table summarizes registered clinical trials involving the selected SHDs, retrieved from the World Health Organization International Clinical Trials Registry Platform (WHO-ICTRP). For each SHD for which clinical trial data was available, the table reports the corresponding trial identification numbers and the disease condition(s) evaluated, with a focus on diabetes and obesity.

| Plant name | Conditions with trials | Type 2 diabetes trials count | Obesity trials count | Trial IDs (Type 2 diabetes) | Trial IDs (Obesity) |
|---|---|---|---|---|---|
| *Aegle marmelos* | Diabetes | 3 | 0 | CTRI/2020/07/026465 \| CTRI/2017/09/009765 \| CTRI/2017/09/009704 | |
| *Berberis aristata* | Diabetes | 5 | 0 | IRCT20230202057310N3 \| CTRI/2019/08/020948 \| TCTR20180509002 \| CTRI/2024/06/068301 \| IRCT201207191774N5 | |
| *Butea monosperma* | Diabetes | 1 | 0 | CTRI/2020/01/023027 | |
| *Enicostema axillare* | Diabetes | 3 | 0 | CTRI/2017/09/009765 \| CTRI/2017/09/009704 \| CTRI/2009/091/000605 | |
| *Gymnema sylvestre* | Diabetes \| Obesity | 22 | 3 | CTRI/2025/04/084530 \| CTRI/2025/03/082396 \| CTRI/2025/02/080905 \| CTRI/2024/07/071682 \| NCT06426966 \| CTRI/2024/02/063330 \| CTRI/2023/02/049426 \| CTRI/2022/10/046705 \| ACTRN12622001023741 \| CTRI/2021/04/033094 \| NCT04745780 \| CTRI/2020/01/022662 \| CTRI/2018/07/014903 \| CTRI/2018/03/012483 \| CTRI/2017/11/010528 \| CTRI/2017/09/009765 \| CTRI/2017/09/009704 \| SLCTR/2015/008 \| CTRI/2015/03/005660 \| CTRI/2014/12/005281 \| CTRI/2009/091/000605 \| CTRI/2010/091/001244 | ACTRN12624000685516 \| NCT06426966 \| ACTRN12622001023741 |
| *Pterocarpus marsupium* | Diabetes | 8 | 0 | CTRI/2018/03/012483 \| CTRI/2017/09/009765 \| CTRI/2017/09/009704 \| CTRI/2015/01/005471 \| CTRI/2024/10/075883 \| CTRI/2020/01/022662 \| CTRI/2015/04/005719 \| CTRI/2008/091/000286 | |
| *Terminalia arjuna* | Diabetes | 3 | 0 | TCTR20241031003 \| CTRI/2015/04/005719 \| CTRI/2015/02/005565 | |

**Table S4: ClassyFire-based chemical classification of phytochemicals associated with the selected Single Herbal Drugs (SHDs).** This table presents the chemical classification of the phytochemicals obtained after drug-likeness and bioavailability analysis across the selected SHDs. Each phytochemical is mapped to its corresponding IMPPAT identifier and an external chemical identifier (e.g., PubChem CID, ChemSpider ID, or CAS number), along with its source. Chemical classification was performed using the ClassyFire framework, categorizing each compound according to its hierarchical taxonomy, including kingdom, superclass, class, and subclass.

| Plant name | Part | IMPPAT identifier | External chemical identifier | Chemical name | Source | Kingdom | Superclass | Class | Subclass |
|---|---|---|---|---|---|---|---|---|---|
| *Pterocarpus marsupium* | Wood | IMPHY012473 | CID_259846 | Lupeol | IMPPAT | Organic compounds | Lipids and lipid-like molecules | Prenol lipids | Triterpenoids |
| *Pterocarpus marsupium* | Wood | IMPHY014908 | CID_72276 | (-)-Epicatechin | IMPPAT | Organic compounds | Phenylpropanoids and polyketides | Flavonoids | Flavans |
| *Pterocarpus marsupium* | Wood | IMPHY010550 | CID_439246 | Naringenin | IMPPAT | Organic compounds | Phenylpropanoids and polyketides | Flavonoids | Flavans |
| *Pterocarpus marsupium* | Wood | IMPHY011826 | CID_10494 | Oleanolic acid | IMPPAT | Organic compounds | Lipids and lipid-like molecules | Prenol lipids | Triterpenoids |
| *Pterocarpus marsupium* | Wood | IMPHY011542 | CID_91457 | Beta Eudesmol | IMPPAT | Organic compounds | Lipids and lipid-like molecules | Prenol lipids | Sesquiterpenoids |
| *Pterocarpus marsupium* | Wood | IMPHY005430 | CID_5281611 | 5-deoxykaempferol | IMPPAT | Organic compounds | Phenylpropanoids and polyketides | Flavonoids | Flavones |
| *Pterocarpus marsupium* | Wood | IMPHY006489 | CID_638278 | Isoliquiritigenin | IMPPAT | Organic compounds | Phenylpropanoids and polyketides | Linear 1,3-diarylpropanoids | Chalcones and dihydrochalcones |
| *Pterocarpus marsupium* | Wood | IMPHY009356 | CID_126 | P-Hydroxybenzaldehyde | IMPPAT | Organic compounds | Organic oxygen compounds | Organooxygen compounds | Carbonyl compounds |
| *Pterocarpus marsupium* | Wood | IMPHY001869 | CID_114829 | Liquiritigenin | IMPPAT | Organic compounds | Phenylpropanoids and polyketides | Flavonoids | Flavans |
| *Pterocarpus marsupium* | Wood | IMPHY000908 | CID_442410 | Garbanzol | IMPPAT | Organic compounds | Phenylpropanoids and polyketides | Flavonoids | Flavans |
| *Pterocarpus marsupium* | Wood | IMPHY004580 | CID_5281805 | Pseudobaptigenin | IMPPAT | Organic compounds | Phenylpropanoids and polyketides | Isoflavonoids | Isoflav-2-enes |
| *Pterocarpus marsupium* | Wood | IMPHY010592 | CID_5282073 | 7,4'-dihydroxyflavone | IMPPAT | Organic compounds | Phenylpropanoids and polyketides | Flavonoids | Flavones |
| *Pterocarpus marsupium* | Wood | IMPHY006487 | CID_638088 | Stilbenes | IMPPAT | Organic compounds | Phenylpropanoids and polyketides | Stilbenes | |
| *Pterocarpus marsupium* | Wood | IMPHY011492 | CID_688857 | 7-hydroxyflavanone | IMPPAT | Organic compounds | Phenylpropanoids and polyketides | Flavonoids | Flavans |
| *Pterocarpus marsupium* | Wood | IMPHY004573 | CID_5281727 | Pterostilbene | IMPPAT | Organic compounds | Phenylpropanoids and polyketides | Stilbenes | |
| *Pterocarpus marsupium* | Wood | IMPHY009579 | CID_134369 | Marsupsin | IMPPAT | Organic compounds | Phenylpropanoids and polyketides | Aurone flavonoids | Auronols |
| *Pterocarpus marsupium* | Wood | IMPHY009654 | CID_133775 | Pterosupin | IMPPAT | Organic compounds | Phenylpropanoids and polyketides | Linear 1,3-diarylpropanoids | Chalcones and dihydrochalcones |
| *Pterocarpus marsupium* | Wood | IMPHY005318 | CID_185124 | Propterol | IMPPAT | Organic compounds | Phenylpropanoids and polyketides | Linear 1,3-diarylpropanoids | Cinnamylphenols |
| *Pterocarpus marsupium* | Wood | IMPHY011467 | CHEMSPIDER_10306372 | Propterol B | IMPPAT | Organic compounds | Phenylpropanoids and polyketides | Linear 1,3-diarylpropanoids | Cinnamylphenols |
| *Pterocarpus marsupium* | Wood | IMPHY006496 | CID_6437266 | Parsupol | IMPPAT | Organic compounds | Lipids and lipid-like molecules | Prenol lipids | Monoterpenoids |
| *Pterocarpus marsupium* | Wood | IMPHY013393 | CID_44257328 | 7-hydroxy-5,4'-dimethoxy-8-methylisoflavone-7-rhamnoside | IMPPAT | Organic compounds | Phenylpropanoids and polyketides | Isoflavonoids | Isoflavonoid O-glycosides |
| *Commiphora wightii* | Plant exudate | IMPHY014836 | CID_222284 | Beta-sitosterol | IMPPAT | Organic compounds | Lipids and lipid-like molecules | Steroids and steroid derivatives | Stigmastanes and derivatives |
| *Commiphora wightii* | Plant exudate | IMPHY014831 | CID_5281515 | Caryophyllene | IMPPAT | Organic compounds | Lipids and lipid-like molecules | Prenol lipids | Sesquiterpenoids |
| *Commiphora wightii* | Plant exudate | IMPHY006558 | CID_68972 | n-Triacontanol | IMPPAT | Organic compounds | Lipids and lipid-like molecules | Fatty Acyls | Fatty alcohols |
| *Commiphora wightii* | Plant exudate | IMPHY006300 | CID_5997 | Cholesterol | IMPPAT | Organic compounds | Lipids and lipid-like molecules | Steroids and steroid derivatives | Cholestane steroids |
| *Commiphora wightii* | Plant exudate | IMPHY014842 | CID_5280794 | Stigmasterol | IMPPAT | Organic compounds | Lipids and lipid-like molecules | Steroids and steroid derivatives | Stigmastanes and derivatives |
| *Commiphora wightii* | Plant exudate | IMPHY003485 | CID_31253 | Myrcene | IMPPAT | Organic compounds | Lipids and lipid-like molecules | Prenol lipids | Monoterpenoids |
| *Commiphora wightii* | Plant exudate | IMPHY014895 | CID_5204 | Sesamin | IMPPAT | Organic compounds | Lignans, neolignans and related compounds | Furanoid lignans | |
| *Commiphora wightii* | Plant exudate | IMPHY003893 | CID_90472510 | Mukulol | IMPPAT | Organic compounds | Lipids and lipid-like molecules | Prenol lipids | Diterpenoids |
| *Commiphora wightii* | Plant exudate | IMPHY011802 | CID_445858 | Ferulate | IMPPAT | Organic compounds | Phenylpropanoids and polyketides | Cinnamic acids and derivatives | Hydroxycinnamic acids and derivatives |
| *Commiphora wightii* | Plant exudate | IMPHY010189 | CID_11747713 | Cembrene | IMPPAT | Organic compounds | Lipids and lipid-like molecules | Prenol lipids | Diterpenoids |
| *Commiphora wightii* | Plant exudate | IMPHY005588 | CID_5281384 | Cembrene A | IMPPAT | Organic compounds | Lipids and lipid-like molecules | Prenol lipids | Diterpenoids |
| *Commiphora wightii* | Plant exudate | IMPHY008907 | CID_101297673 | Guggulsterol I | IMPPAT | Organic compounds | Lipids and lipid-like molecules | Steroids and steroid derivatives | Bile acids, alcohols and derivatives |
| *Commiphora wightii* | Plant exudate | IMPHY008908 | CID_101297674 | Guggulsterol II | IMPPAT | Organic compounds | Lipids and lipid-like molecules | Steroids and steroid derivatives | Cholestane steroids |
| *Commiphora wightii* | Plant exudate | IMPHY008746 | CID_101297675 | Guggulsterol III | IMPPAT | Organic compounds | Lipids and lipid-like molecules | Steroids and steroid derivatives | Cholestane steroids |
| *Commiphora wightii* | Plant exudate | IMPHY002189 | CID_101750 | Alpha-camphorene | IMPPAT | Organic compounds | Lipids and lipid-like molecules | Prenol lipids | Sesquiterpenoids |
| *Commiphora wightii* | Plant exudate | IMPHY003974 | CID_161109 | 20 alpha hydroxy-4-pregnen-3-one | IMPPAT | Organic compounds | Lipids and lipid-like molecules | Steroids and steroid derivatives | Pregnane steroids |
| *Commiphora wightii* | Plant exudate | IMPHY004048 | CID_92747 | 20 beta hydroxy -4-pregnen-3-one | IMPPAT | Organic compounds | Lipids and lipid-like molecules | Steroids and steroid derivatives | Pregnane steroids |
| *Commiphora wightii* | Plant exudate | IMPHY003654 | CID_6439929 | E-guggulsterone | IMPPAT | Organic compounds | Lipids and lipid-like molecules | Steroids and steroid derivatives | Androstane steroids |
| *Commiphora wightii* | Plant exudate | IMPHY006624 | CID_70695727 | Pluviatilol | IMPPAT | Organic compounds | Lignans, neolignans and related compounds | Furanoid lignans | |
| *Commiphora wightii* | Plant exudate | IMPHY004031 | CID_122173119 | Z-guggulsterone | IMPPAT | Organic compounds | Lipids and lipid-like molecules | Steroids and steroid derivatives | Androstane steroids |
| *Commiphora wightii* | Plant exudate | IMPHY003895 | CID_5368823 | Allylcembrol | IMPPAT | Organic compounds | Lipids and lipid-like molecules | Prenol lipids | Diterpenoids |
| *Terminalia arjuna* | Bark | IMPHY014836 | CID_222284 | Beta-sitosterol | IMPPAT | Organic compounds | Lipids and lipid-like molecules | Steroids and steroid derivatives | Stigmastanes and derivatives |
| *Terminalia arjuna* | Bark | IMPHY007450 | CID_971 | Oxalic acid | IMPPAT | Organic compounds | Organic acids and derivatives | Carboxylic acids and derivatives | Dicarboxylic acids and derivatives |
| *Terminalia arjuna* | Bark | IMPHY011688 | CID_91472 | Friedelin | IMPPAT | Organic compounds | Lipids and lipid-like molecules | Prenol lipids | Triterpenoids |
| *Terminalia arjuna* | Bark | IMPHY008910 | CID_12410 | Hentriacontane | IMPPAT | Organic compounds | Hydrocarbons | Saturated hydrocarbons | Alkanes |
| *Terminalia arjuna* | Bark | IMPHY011735 | CID_65084 | (+)-Gallocatechol | IMPPAT | Organic compounds | Phenylpropanoids and polyketides | Flavonoids | Flavans |
| *Terminalia arjuna* | Bark | IMPHY005537 | CID_5281855 | Ellagic acid | IMPPAT | Organic compounds | Phenylpropanoids and polyketides | Tannins | Hydrolyzable tannins |
| *Terminalia arjuna* | Bark | IMPHY004079 | CID_289 | Catechol | IMPPAT | Organic compounds | Benzenoids | Phenols | Benzenediols |
| *Terminalia arjuna* | Bark | IMPHY011729 | CID_6251 | Mannitol | IMPPAT | Organic compounds | Organic oxygen compounds | Organooxygen compounds | Carbohydrates and carbohydrate conjugates |
| *Terminalia arjuna* | Bark | IMPHY011737 | CID_72277 | Epigallocatechol | IMPPAT | Organic compounds | Phenylpropanoids and polyketides | Flavonoids | Flavans |
| *Terminalia arjuna* | Bark | IMPHY011826 | CID_10494 | Oleanolic acid | IMPPAT | Organic compounds | Lipids and lipid-like molecules | Prenol lipids | Triterpenoids |
| *Terminalia arjuna* | Bark | IMPHY011627 | CID_73641 | Arjunolic acid | IMPPAT | Organic compounds | Lipids and lipid-like molecules | Prenol lipids | Triterpenoids |
| *Terminalia arjuna* | Bark | IMPHY011885 | CID_440835 | Leucodelphinidin | IMPPAT | Organic compounds | Phenylpropanoids and polyketides | Flavonoids | Flavans |
| *Terminalia arjuna* | Bark | IMPHY012649 | CID_622032 | Tomentosic acid | IMPPAT | Organic compounds | Lipids and lipid-like molecules | Prenol lipids | Triterpenoids |
| *Terminalia arjuna* | Bark | IMPHY001431 | CID_15385516 | Arjunic acid | IMPPAT | Organic compounds | Lipids and lipid-like molecules | Prenol lipids | Triterpenoids |
| *Terminalia arjuna* | Bark | IMPHY002542 | CID_14034812 | Cerasidin | IMPPAT | Organic compounds | Phenylpropanoids and polyketides | Linear 1,3-diarylpropanoids | Chalcones and dihydrochalcones |
| *Terminalia arjuna* | Bark | IMPHY005607 | CID_5281605 | Baicalein | IMPPAT | Organic compounds | Phenylpropanoids and polyketides | Flavonoids | Flavones |
| *Terminalia arjuna* | Bark | IMPHY012650 | CID_12444386 | Arjungenin | IMPPAT | Organic compounds | Lipids and lipid-like molecules | Prenol lipids | Triterpenoids |
| *Terminalia arjuna* | Bark | IMPHY008675 | CID_132568257 | Terminic acid | IMPPAT | Organic compounds | Lipids and lipid-like molecules | Prenol lipids | Triterpenoids |
| *Terminalia arjuna* | Bark | IMPHY002418 | CAS_82178-34-5 | Arjunolone | IMPPAT | Organic compounds | Phenylpropanoids and polyketides | Flavonoids | O-methylated flavonoids |
| *Terminalia arjuna* | Bark | IMPHY005947 | CID_5365034 | Myristyl oleate | IMPPAT | Organic compounds | Lipids and lipid-like molecules | Fatty Acyls | Fatty acid esters |
| *Tectona grandis* | Wood | IMPHY014836 | CID_222284 | Beta-sitosterol | IMPPAT | Organic compounds | Lipids and lipid-like molecules | Steroids and steroid derivatives | Stigmastanes and derivatives |
| *Tectona grandis* | Wood | IMPHY012003 | CID_64971 | Betulinic acid | IMPPAT | Organic compounds | Lipids and lipid-like molecules | Prenol lipids | Triterpenoids |
| *Tectona grandis* | Wood | IMPHY003002 | CID_3884 | Lapachol | IMPPAT | Organic compounds | Benzenoids | Naphthalenes | Naphthoquinones |
| *Tectona grandis* | Wood | IMPHY007191 | CID_6773 | Tectoquinone | IMPPAT | Organic compounds | Benzenoids | Anthracenes | Anthraquinones |
| *Tectona grandis* | Wood | IMPHY009853 | CID_931 | Naphthalene | IMPPAT | Organic compounds | Benzenoids | Naphthalenes | |
| *Tectona grandis* | Wood | IMPHY007192 | CID_6780 | Anthraquinones | IMPPAT | Organic compounds | Benzenoids | Anthracenes | Anthraquinones |
| *Tectona grandis* | Wood | IMPHY004090 | CID_3037329 | Dehydrotectol | IMPPAT | Organic compounds | Organoheterocyclic compounds | Naphthopyrans | |
| *Tectona grandis* | Wood | IMPHY006054 | CID_72734 | Dehydro-alpha-lapachone | IMPPAT | Organic compounds | Organoheterocyclic compounds | Naphthopyrans | Naphthopyranones |
| *Tectona grandis* | Wood | IMPHY007262 | CID_67030 | Anthraquinone-2-carboxylic acid | IMPPAT | Organic compounds | Benzenoids | Anthracenes | Anthracenecarboxylic acids and derivatives |
| *Tectona grandis* | Wood | IMPHY003260 | CID_3574508 | Tecomaquinone I | IMPPAT | Organic compounds | Phenylpropanoids and polyketides | Isochromanequinones | Benzoisochromanequinones |
| *Tectona grandis* | Wood | IMPHY003235 | CID_344310 | Anthraquinone-2-carboxaldehyde | IMPPAT | Organic compounds | Benzenoids | Anthracenes | Anthraquinones |
| *Tectona grandis* | Wood | IMPHY005026 | CID_5318245 | 5-hydroxyalpachol | IMPPAT | Organic compounds | Benzenoids | Naphthalenes | Naphthoquinones |
| *Tectona grandis* | Wood | IMPHY014299 | CID_14283285 | 9,10-dimethoxy-2-methyl-1,4-anthraquinone | IMPPAT | Organic compounds | Benzenoids | Anthracenes | Anthraquinones |
| *Tectona grandis* | Wood | IMPHY014278 | CID_13970503 | 1-hydroxy-5-methoxy-2-methyl-9,10-anthraquinone | IMPPAT | Organic compounds | Benzenoids | Anthracenes | Anthraquinones |
| *Aegle marmelos* | Bark | IMPHY014836 | CID_222284 | Beta-sitosterol | IMPPAT | Organic compounds | Lipids and lipid-like molecules | Steroids and steroid derivatives | Stigmastanes and derivatives |
| *Aegle marmelos* | Bark | IMPHY012473 | CID_259846 | Lupeol | IMPPAT | Organic compounds | Lipids and lipid-like molecules | Prenol lipids | Triterpenoids |
| *Aegle marmelos* | Bark | IMPHY003490 | CID_323 | Coumarins | IMPPAT | Organic compounds | Phenylpropanoids and polyketides | Coumarins and derivatives | |
| *Aegle marmelos* | Bark | IMPHY001552 | CID_1550607 | Aurapten | IMPPAT | Organic compounds | Lipids and lipid-like molecules | Prenol lipids | Terpene lactones |
| *Aegle marmelos* | Bark | IMPHY011661 | CID_334704 | Marmesin | IMPPAT | Organic compounds | Phenylpropanoids and polyketides | Coumarins and derivatives | Furanocoumarins |
| *Aegle marmelos* | Bark | IMPHY006258 | CID_6450230 | Marmin | IMPPAT | Organic compounds | Phenylpropanoids and polyketides | Coumarins and derivatives | |
| *Aegle marmelos* | Bark | IMPHY007265 | CID_6760 | Skimmianine | IMPPAT | Organic compounds | Organoheterocyclic compounds | Quinolines and derivatives | Furanoquinolines |
| *Aegle marmelos* | Bark | IMPHY005587 | CID_5281426 | Umbelliferone | IMPPAT | Organic compounds | Phenylpropanoids and polyketides | Coumarins and derivatives | Hydroxycoumarins |
| *Aegle marmelos* | Bark | IMPHY008913 | CID_10212 | Imperatorin | IMPPAT | Organic compounds | Phenylpropanoids and polyketides | Coumarins and derivatives | Furanocoumarins |
| *Aegle marmelos* | Bark | IMPHY011596 | CID_107936 | Fagarine | IMPPAT | Organic compounds | Organoheterocyclic compounds | Quinolines and derivatives | Furanoquinolines |
| *Aegle marmelos* | Bark | IMPHY012643 | CID_600671 | Aegelinol | IMPPAT | Organic compounds | Phenylpropanoids and polyketides | Coumarins and derivatives | Pyranocoumarins |
| *Butea monosperma* | Seed | IMPHY014836 | CID_222284 | Beta-sitosterol | IMPPAT | Organic compounds | Lipids and lipid-like molecules | Steroids and steroid derivatives | Stigmastanes and derivatives |
| *Butea monosperma* | Seed | IMPHY000060 | CID_11005 | Myristic acid | IMPPAT | Organic compounds | Lipids and lipid-like molecules | Fatty Acyls | Fatty acids and conjugates |
| *Butea monosperma* | Seed | IMPHY007327 | CID_985 | Palmitic acid | IMPPAT | Organic compounds | Lipids and lipid-like molecules | Fatty Acyls | Fatty acids and conjugates |
| *Butea monosperma* | Seed | IMPHY004631 | CID_5281 | Octadecanoic acid | IMPPAT | Organic compounds | Lipids and lipid-like molecules | Fatty Acyls | Fatty acids and conjugates |
| *Butea monosperma* | Seed | IMPHY011797 | CID_445639 | Oleic acid | IMPPAT | Organic compounds | Lipids and lipid-like molecules | Fatty Acyls | Fatty acids and conjugates |
| *Butea monosperma* | Seed | IMPHY012723 | CID_5280934 | Linolenic acid | IMPPAT | Organic compounds | Lipids and lipid-like molecules | Fatty Acyls | Lineolic acids and derivatives |
| *Butea monosperma* | Seed | IMPHY014990 | CID_5280450 | Linoleic acid | IMPPAT | Organic compounds | Lipids and lipid-like molecules | Fatty Acyls | Lineolic acids and derivatives |
| *Butea monosperma* | Seed | IMPHY015072 | CID_12309055 | Beta-Sitosterol-beta-D-glucoside | IMPPAT | Organic compounds | Lipids and lipid-like molecules | Steroids and steroid derivatives | Stigmastanes and derivatives |
| *Butea monosperma* | Seed | IMPHY014838 | CID_5742590 | Beta-sitosterol glucoside | IMPPAT | Organic compounds | Lipids and lipid-like molecules | Steroids and steroid derivatives | Stigmastanes and derivatives |
| *Butea monosperma* | Seed | IMPHY011394 | CID_10467 | Arachidic acid | IMPPAT | Organic compounds | Lipids and lipid-like molecules | Fatty Acyls | Fatty acids and conjugates |
| *Butea monosperma* | Seed | IMPHY007212 | CID_8215 | Behenic acid | IMPPAT | Organic compounds | Lipids and lipid-like molecules | Fatty Acyls | Fatty acids and conjugates |
| *Butea monosperma* | Seed | IMPHY011619 | CID_73170 | Alpha-amyrin | IMPPAT | Organic compounds | Lipids and lipid-like molecules | Prenol lipids | Triterpenoids |
| *Butea monosperma* | Seed | IMPHY000165 | CID_11197 | Lignoceric acid | IMPPAT | Organic compounds | Lipids and lipid-like molecules | Fatty Acyls | Fatty acids and conjugates |
| *Butea monosperma* | Seed | IMPHY003104 | CID_2969 | Capric acid | IMPPAT | Organic compounds | Lipids and lipid-like molecules | Fatty Acyls | Fatty acids and conjugates |
| *Butea monosperma* | Seed | IMPHY003301 | CID_379 | Caprylic acid | IMPPAT | Organic compounds | Lipids and lipid-like molecules | Fatty Acyls | Fatty acids and conjugates |
| *Butea monosperma* | Seed | IMPHY003016 | CID_3893 | Lauric acid | IMPPAT | Organic compounds | Lipids and lipid-like molecules | Fatty Acyls | Fatty acids and conjugates |
| *Butea monosperma* | Seed | IMPHY009826 | CID_102515444 | Phosphatidylethanolamine | IMPPAT | Organic compounds | Lipids and lipid-like molecules | Glycerophospholipids | Glycerophosphoethanolamines |
| *Butea monosperma* | Seed | IMPHY005085 | CID_16898 | n-Heneicosanoic acid | IMPPAT | Organic compounds | Lipids and lipid-like molecules | Fatty Acyls | Fatty acids and conjugates |
| *Butea monosperma* | Seed | IMPHY010212 | CID_10071442 | Sulfurein | IMPPAT | Organic compounds | Phenylpropanoids and polyketides | Flavonoids | Flavonoid glycosides |
| *Butea monosperma* | Seed | IMPHY009395 | CID_12309899 | Isocoreopsin | IMPPAT | Organic compounds | Phenylpropanoids and polyketides | Flavonoids | Flavonoid glycosides |
| *Butea monosperma* | Seed | IMPHY002064 | CID_12303942 | Coreopsin | IMPPAT | Organic compounds | Phenylpropanoids and polyketides | Flavonoids | Flavonoid glycosides |
| *Butea monosperma* | Seed | IMPHY003013 | CID_42607822 | Isomonospermoside | IMPPAT | Organic compounds | Phenylpropanoids and polyketides | Flavonoids | Flavonoid glycosides |
| *Butea monosperma* | Seed | IMPHY003036 | CID_42607524 | Monospermoside | IMPPAT | Organic compounds | Phenylpropanoids and polyketides | Flavonoids | Flavonoid glycosides |
| *Butea monosperma* | Seed | IMPHY006875 | CID_198727 | Palasonin | IMPPAT | Organic compounds | Organoheterocyclic compounds | Furofurans | |
| *Butea monosperma* | Seed | IMPHY014052 | CID_5312784 | 15-hydroxypentacosanoic acid | IMPPAT | Organic compounds | Lipids and lipid-like molecules | Fatty Acyls | Fatty acids and conjugates |
| *Butea monosperma* | Seed | IMPHY000570 | CID_10790 | Aleuritic acid | IMPPAT | Organic compounds | Lipids and lipid-like molecules | Fatty Acyls | Fatty acids and conjugates |
| *Butea monosperma* | Seed | IMPHY013481 | CID_49831545 | Delta-lactone | IMPPAT | Organic compounds | Organoheterocyclic compounds | Lactones | Delta valerolactones |
| *Butea monosperma* | Seed | IMPHY008663 | CID_5311264 | Lysophosphatidylcholine | IMPPAT | Organic compounds | Lipids and lipid-like molecules | Glycerophospholipids | Glycerophosphocholines |
| *Diospyros malabarica* | Fruit | IMPHY014836 | CID_222284 | Beta-sitosterol | IMPPAT | Organic compounds | Lipids and lipid-like molecules | Steroids and steroid derivatives | Stigmastanes and derivatives |
| *Diospyros malabarica* | Fruit | IMPHY013522 | CID_71597391 | Triterpenoids | IMPPAT | Organic compounds | Lipids and lipid-like molecules | Prenol lipids | Triterpenoids |
| *Diospyros malabarica* | Fruit | IMPHY015072 | CID_12309055 | Beta-Sitosterol-beta-D-glucoside | IMPPAT | Organic compounds | Lipids and lipid-like molecules | Steroids and steroid derivatives | Stigmastanes and derivatives |
| *Diospyros malabarica* | Fruit | IMPHY014838 | CID_5742590 | Beta-sitosterol glucoside | IMPPAT | Organic compounds | Lipids and lipid-like molecules | Steroids and steroid derivatives | Stigmastanes and derivatives |
| *Diospyros malabarica* | Fruit | IMPHY012223 | CID_73145 | Beta amyrin | IMPPAT | Organic compounds | Lipids and lipid-like molecules | Prenol lipids | Triterpenoids |
| *Diospyros malabarica* | Fruit | IMPHY012473 | CID_259846 | Lupeol | IMPPAT | Organic compounds | Lipids and lipid-like molecules | Prenol lipids | Triterpenoids |
| *Diospyros malabarica* | Fruit | IMPHY012021 | CID_370 | Gallic acid | IMPPAT | Organic compounds | Benzenoids | Benzene and substituted derivatives | Benzoic acids and derivatives |
| *Diospyros malabarica* | Fruit | IMPHY007273 | CID_68171 | N-hexacosanol | IMPPAT | Organic compounds | Lipids and lipid-like molecules | Fatty Acyls | Fatty alcohols |
| *Diospyros malabarica* | Fruit | IMPHY012003 | CID_64971 | Betulinic acid | IMPPAT | Organic compounds | Lipids and lipid-like molecules | Prenol lipids | Triterpenoids |
| *Diospyros malabarica* | Fruit | IMPHY004271 | CID_72326 | Betulin | IMPPAT | Organic compounds | Lipids and lipid-like molecules | Prenol lipids | Triterpenoids |
| *Diospyros malabarica* | Fruit | IMPHY009359 | CID_12407 | Hexacosane | IMPPAT | Organic compounds | Hydrocarbons | Saturated hydrocarbons | Alkanes |
| *Diospyros malabarica* | Fruit | IMPHY007426 | CID_92785 | Taraxerone | IMPPAT | Organic compounds | Lipids and lipid-like molecules | Prenol lipids | Sesterterpenoids |
| *Diospyros malabarica* | Fruit | IMPHY008639 | CID_22296838 | Marsformosanone | IMPPAT | Organic compounds | Lipids and lipid-like molecules | Prenol lipids | Triterpenoids |
| *Diospyros malabarica* | Fruit | IMPHY006812 | CID_7092583 | Peregrinol | IMPPAT | Organic compounds | Lipids and lipid-like molecules | Prenol lipids | Diterpenoids |
| *Diospyros malabarica* | Fruit | IMPHY002547 | CID_14033983 | Pongaflavone | IMPPAT | Organic compounds | Phenylpropanoids and polyketides | Flavonoids | Pyranoflavonoids |
| *Berberis aristata* | Stem | IMPHY005665 | CID_2353 | Berberine | IMPPAT | Organic compounds | Alkaloids and derivatives | Protoberberine alkaloids and derivatives | |
| *Berberis aristata* | Stem | IMPHY005330 | CID_19009 | Palmatine | IMPPAT | Organic compounds | Alkaloids and derivatives | Protoberberine alkaloids and derivatives | |
| *Berberis aristata* | Stem | IMPHY007190 | CID_72323 | Jatrorrhizine | IMPPAT | Organic compounds | Alkaloids and derivatives | Protoberberine alkaloids and derivatives | |
| *Berberis aristata* | Stem | IMPHY000342 | CID_11066 | Oxyberberine | IMPPAT | Organic compounds | Organoheterocyclic compounds | Isoquinolines and derivatives | Isoquinolones and derivatives |
| *Berberis aristata* | Stem | IMPHY003578 | CID_442333 | Oxyacanthine | IMPPAT | Organic compounds | Lignans, neolignans and related compounds | | |
| *Berberis aristata* | Stem | IMPHY008157 | CID_630739 | Karachine | IMPPAT | Organic compounds | Alkaloids and derivatives | Protoberberine alkaloids and derivatives | |
| *Berberis aristata* | Stem | IMPHY005761 | CID_193239 | Pakistanine | IMPPAT | Organic compounds | Alkaloids and derivatives | Aporphines | |
| *Berberis aristata* | Stem | IMPHY001751 | CID_156697 | Kalashine | IMPPAT | Organic compounds | Alkaloids and derivatives | Aporphines | |
| *Berberis aristata* | Stem | IMPHY002379 | CAS_77754-91-7 | Chitraline | IMPPAT | Organic compounds | Alkaloids and derivatives | Aporphines | |
| *Berberis aristata* | Stem | IMPHY003363 | CID_181478 | 1 - O - methyl pakistanine | IMPPAT | Organic compounds | Alkaloids and derivatives | Aporphines | |
| *Gymnema sylvestre* | Leaf | IMPHY004235 | CID_94715 | Glucuronic acid | IMPPAT | Organic compounds | Organic oxygen compounds | Organooxygen compounds | Carbohydrates and carbohydrate conjugates |
| *Gymnema sylvestre* | Leaf | IMPHY014919 | CID_439215 | Galacturonic acid | IMPPAT | Organic compounds | Organic oxygen compounds | Organooxygen compounds | Carbohydrates and carbohydrate conjugates |
| *Gymnema sylvestre* | Leaf | IMPHY009624 | CID_12366 | Ethyl palmitate | IMPPAT | Organic compounds | Lipids and lipid-like molecules | Fatty Acyls | Fatty acid esters |
| *Gymnema sylvestre* | Leaf | IMPHY000060 | CID_11005 | Myristic acid | IMPPAT | Organic compounds | Lipids and lipid-like molecules | Fatty Acyls | Fatty acids and conjugates |
| *Gymnema sylvestre* | Leaf | IMPHY007327 | CID_985 | Palmitic acid | IMPPAT | Organic compounds | Lipids and lipid-like molecules | Fatty Acyls | Fatty acids and conjugates |
| *Gymnema sylvestre* | Leaf | IMPHY004631 | CID_5281 | Octadecanoic acid | IMPPAT | Organic compounds | Lipids and lipid-like molecules | Fatty Acyls | Fatty acids and conjugates |
| *Gymnema sylvestre* | Leaf | IMPHY009355 | CID_12592 | Tetracosane | IMPPAT | Organic compounds | Hydrocarbons | Saturated hydrocarbons | Alkanes |
| *Gymnema sylvestre* | Leaf | IMPHY007620 | CID_957 | Octanol | IMPPAT | Organic compounds | Lipids and lipid-like molecules | Fatty Acyls | Fatty alcohols |
| *Gymnema sylvestre* | Leaf | IMPHY007100 | CID_8193 | Dodecanol | IMPPAT | Organic compounds | Lipids and lipid-like molecules | Fatty Acyls | Fatty alcohols |
| *Gymnema sylvestre* | Leaf | IMPHY014842 | CID_5280794 | Stigmasterol | IMPPAT | Organic compounds | Lipids and lipid-like molecules | Steroids and steroid derivatives | Stigmastanes and derivatives |
| *Gymnema sylvestre* | Leaf | IMPHY012223 | CID_73145 | Beta amyrin | IMPPAT | Organic compounds | Lipids and lipid-like molecules | Prenol lipids | Triterpenoids |
| *Gymnema sylvestre* | Leaf | IMPHY012473 | CID_259846 | Lupeol | IMPPAT | Organic compounds | Lipids and lipid-like molecules | Prenol lipids | Triterpenoids |
| *Gymnema sylvestre* | Leaf | IMPHY004055 | CID_305 | Choline | IMPPAT | Organic compounds | Organic nitrogen compounds | Organonitrogen compounds | Quaternary ammonium salts |
| *Gymnema sylvestre* | Leaf | IMPHY003963 | CID_247 | Betaine | IMPPAT | Organic compounds | Organic acids and derivatives | Carboxylic acids and derivatives | Amino acids, peptides, and analogues |
| *Gymnema sylvestre* | Leaf | IMPHY010072 | CID_2758 | 1,8-Cineole | IMPPAT | Organic compounds | Organoheterocyclic compounds | Oxanes | |
| *Gymnema sylvestre* | Leaf | IMPHY010080 | CID_6918391 | b-Elemene | IMPPAT | Organic compounds | Lipids and lipid-like molecules | Prenol lipids | Sesquiterpenoids |
| *Gymnema sylvestre* | Leaf | IMPHY008910 | CID_12410 | Hentriacontane | IMPPAT | Organic compounds | Hydrocarbons | Saturated hydrocarbons | Alkanes |
| *Gymnema sylvestre* | Leaf | IMPHY003999 | CID_439655 | Tartaric acid | IMPPAT | Organic compounds | Organic oxygen compounds | Organooxygen compounds | Carbohydrates and carbohydrate conjugates |
| *Gymnema sylvestre* | Leaf | IMPHY002667 | CID_13849 | Pentadecanoic acid | IMPPAT | Organic compounds | Lipids and lipid-like molecules | Fatty Acyls | Fatty acids and conjugates |
| *Gymnema sylvestre* | Leaf | IMPHY009482 | CID_12409 | n-Nonacosane | IMPPAT | Organic compounds | Hydrocarbons | Saturated hydrocarbons | Alkanes |
| *Gymnema sylvestre* | Leaf | IMPHY009413 | CID_12535 | n-Triacontane | IMPPAT | Organic compounds | Hydrocarbons | Saturated hydrocarbons | Alkanes |
| *Gymnema sylvestre* | Leaf | IMPHY003536 | CID_3314 | Eugenol | IMPPAT | Organic compounds | Benzenoids | Phenols | Methoxyphenols |
| *Gymnema sylvestre* | Leaf | IMPHY006696 | CID_7127 | Methyl eugenol | IMPPAT | Organic compounds | Benzenoids | Benzene and substituted derivatives | Methoxybenzenes |
| *Gymnema sylvestre* | Leaf | IMPHY002983 | CID_2682 | 1-Hexadecanol | IMPPAT | Organic compounds | Lipids and lipid-like molecules | Fatty Acyls | Fatty alcohols |
| *Gymnema sylvestre* | Leaf | IMPHY007039 | CID_7410 | Acetophenone | IMPPAT | Organic compounds | Organic oxygen compounds | Organooxygen compounds | Carbonyl compounds |
| *Gymnema sylvestre* | Leaf | IMPHY006315 | CID_61303 | 2-Pentadecanone | IMPPAT | Organic compounds | Organic oxygen compounds | Organooxygen compounds | Carbonyl compounds |
| *Gymnema sylvestre* | Leaf | IMPHY006898 | CID_8221 | Octadecanol | IMPPAT | Organic compounds | Lipids and lipid-like molecules | Fatty Acyls | Fatty alcohols |
| *Gymnema sylvestre* | Leaf | IMPHY012712 | CID_5280435 | Phytol | IMPPAT | Organic compounds | Lipids and lipid-like molecules | Prenol lipids | Diterpenoids |
| *Gymnema sylvestre* | Leaf | IMPHY001135 | CID_10408 | Hexahydrofarnesyl acetone | IMPPAT | Organic compounds | Lipids and lipid-like molecules | Prenol lipids | Sesquiterpenoids |
| *Gymnema sylvestre* | Leaf | IMPHY009377 | CID_12406 | Pentacosane | IMPPAT | Organic compounds | Hydrocarbons | Saturated hydrocarbons | Alkanes |
| *Gymnema sylvestre* | Leaf | IMPHY001896 | CID_11636 | n-Heptacosane | IMPPAT | Organic compounds | Hydrocarbons | Saturated hydrocarbons | Alkanes |
| *Gymnema sylvestre* | Leaf | IMPHY006971 | CID_8181 | Methyl palmitate | IMPPAT | Organic compounds | Lipids and lipid-like molecules | Fatty Acyls | Fatty acid esters |
| *Gymnema sylvestre* | Leaf | IMPHY009485 | CID_12413 | Pentatriacontane | IMPPAT | Organic compounds | Hydrocarbons | Saturated hydrocarbons | Alkanes |
| *Gymnema sylvestre* | Leaf | IMPHY009483 | CID_12411 | Tritriacontane | IMPPAT | Organic compounds | Hydrocarbons | Saturated hydrocarbons | Alkanes |
| *Gymnema sylvestre* | Leaf | IMPHY012018 | CID_119 | Gamma-butyric acid | IMPPAT | Organic compounds | Organic acids and derivatives | Carboxylic acids and derivatives | Amino acids, peptides, and analogues |
| *Gymnema sylvestre* | Leaf | IMPHY010081 | CID_264 | Butyric acid | IMPPAT | Organic compounds | Lipids and lipid-like molecules | Fatty Acyls | Fatty acids and conjugates |
| *Gymnema sylvestre* | Leaf | IMPHY011802 | CID_445858 | Ferulate | IMPPAT | Organic compounds | Phenylpropanoids and polyketides | Cinnamic acids and derivatives | Hydroxycinnamic acids and derivatives |
| *Gymnema sylvestre* | Leaf | IMPHY007182 | CID_8209 | Tetradecanol | IMPPAT | Organic compounds | Lipids and lipid-like molecules | Fatty Acyls | Fatty alcohols |
| *Gymnema sylvestre* | Leaf | IMPHY011044 | CID_5352845 | 2-Dodecenol | IMPPAT | Organic compounds | Lipids and lipid-like molecules | Fatty Acyls | Fatty alcohols |
| *Gymnema sylvestre* | Leaf | IMPHY006220 | CID_643915 | Angelic acid | IMPPAT | Organic compounds | Lipids and lipid-like molecules | Fatty Acyls | Fatty acids and conjugates |
| *Gymnema sylvestre* | Leaf | IMPHY006981 | CID_798 | Indole | IMPPAT | Organic compounds | Organoheterocyclic compounds | Indoles and derivatives | Indoles |
| *Gymnema sylvestre* | Leaf | IMPHY011805 | CID_441437 | D-quercitol | IMPPAT | Organic compounds | Organic oxygen compounds | Organooxygen compounds | Alcohols and polyols |
| *Gymnema sylvestre* | Leaf | IMPHY003695 | CID_9548706 | Germacrene A | IMPPAT | Organic compounds | Lipids and lipid-like molecules | Prenol lipids | Sesquiterpenoids |
| *Gymnema sylvestre* | Leaf | IMPHY009367 | CID_12397 | Pentadecanol | IMPPAT | Organic compounds | Lipids and lipid-like molecules | Fatty Acyls | Fatty alcohols |
| *Gymnema sylvestre* | Leaf | IMPHY008252 | CID_892 | Inositol | IMPPAT | Organic compounds | Organic oxygen compounds | Organooxygen compounds | Alcohols and polyols |
| *Gymnema sylvestre* | Leaf | IMPHY006959 | CID_785 | Hydroquinone | IMPPAT | Organic compounds | Benzenoids | Phenols | Benzenediols |
| *Gymnema sylvestre* | Leaf | IMPHY001808 | CID_114522 | Ketones | IMPPAT | Organic compounds | Lipids and lipid-like molecules | Fatty Acyls | Fatty alcohol esters |
| *Gymnema sylvestre* | Leaf | IMPHY012767 | CID_5283384 | 9,12,15-Octadecatrienal | IMPPAT | Organic compounds | Lipids and lipid-like molecules | Fatty Acyls | Fatty aldehydes |
| *Gymnema sylvestre* | Leaf | IMPHY009893 | CID_9015 | p-Guaiacol | IMPPAT | Organic compounds | Benzenoids | Phenols | Methoxyphenols |
| *Gymnema sylvestre* | Leaf | IMPHY001213 | CID_10290861 | Conduritol A | IMPPAT | Organic compounds | Organic oxygen compounds | Organooxygen compounds | Alcohols and polyols |
| *Gymnema sylvestre* | Leaf | IMPHY014911 | CID_8123 | Ethyl octadec-9-enoate | IMPPAT | Organic compounds | Lipids and lipid-like molecules | Fatty Acyls | Fatty acid esters |
| *Gymnema sylvestre* | Leaf | IMPHY001589 | CID_15560302 | Gymnestrogenin | IMPPAT | Organic compounds | Lipids and lipid-like molecules | Prenol lipids | Triterpenoids |
| *Gymnema sylvestre* | Leaf | IMPHY012014 | CID_12101 | m-Ethyl phenol | IMPPAT | Organic compounds | Benzenoids | Phenols | 1-hydroxy-4-unsubstituted benzenoids |
| *Gymnema sylvestre* | Leaf | IMPHY009717 | CID_6365430 | Tetradecadiene | IMPPAT | Organic compounds | Hydrocarbons | Unsaturated hydrocarbons | Olefins |
| *Gymnema sylvestre* | Leaf | | CID_284 | Formic Acid | API | Organic compounds | Organic acids and derivatives | Carboxylic acids and derivatives | Carboxylic acids |
| *Enicostema axillare* | Whole Plant | IMPHY000060 | CID_11005 | Myristic acid | IMPPAT | Organic compounds | Lipids and lipid-like molecules | Fatty Acyls | Fatty acids and conjugates |
| *Enicostema axillare* | Whole Plant | IMPHY001896 | CID_11636 | n-heptacosane | IMPPAT | Organic compounds | Hydrocarbons | Saturated hydrocarbons | Alkanes |
| *Enicostema axillare* | Whole Plant | IMPHY009372 | CID_124034 | Swertisin | IMPPAT | Organic compounds | Phenylpropanoids and polyketides | Flavonoids | Flavonoid glycosides |
| *Enicostema axillare* | Whole Plant | IMPHY009482 | CID_12409 | n-nonacosane | IMPPAT | Organic compounds | Hydrocarbons | Saturated hydrocarbons | Alkanes |
| *Enicostema axillare* | Whole Plant | IMPHY008689 | CID_162350 | Isovitexin | IMPPAT | Organic compounds | Phenylpropanoids and polyketides | Flavonoids | Flavonoid glycosides |
| *Enicostema axillare* | Whole Plant | IMPHY005316 | CID_191120 | Erythrocentaurin | IMPPAT | Organic compounds | Organoheterocyclic compounds | Benzopyrans | 2-benzopyrans |
| *Enicostema axillare* | Whole Plant | IMPHY003257 | CID_354616 | Gentianine | IMPPAT | Organic compounds | Organoheterocyclic compounds | Pyranopyridines | |
| *Enicostema axillare* | Whole Plant | IMPHY011797 | CID_445639 | Oleic acid | IMPPAT | Organic compounds | Lipids and lipid-like molecules | Fatty Acyls | Fatty acids and conjugates |
| *Enicostema axillare* | Whole Plant | IMPHY004661 | CID_5280443 | Apigenin | IMPPAT | Organic compounds | Phenylpropanoids and polyketides | Flavonoids | Flavones |
| *Enicostema axillare* | Whole Plant | IMPHY004631 | CID_5281 | Octadecanoic acid | IMPPAT | Organic compounds | Lipids and lipid-like molecules | Fatty Acyls | Fatty acids and conjugates |
| *Enicostema axillare* | Whole Plant | IMPHY005642 | CID_5281564 | Enicoflavin | IMPPAT | Organic compounds | Organoheterocyclic compounds | Lactones | Delta valerolactones |
| *Enicostema axillare* | Whole Plant | IMPHY005435 | CID_5281617 | Genkwanin | IMPPAT | Organic compounds | Phenylpropanoids and polyketides | Flavonoids | O-methylated flavonoids |
| *Enicostema axillare* | Whole Plant | IMPHY007273 | CID_68171 | N-hexacosanol | IMPPAT | Organic compounds | Lipids and lipid-like molecules | Fatty Acyls | Fatty alcohols |
| *Enicostema axillare* | Whole Plant | IMPHY004271 | CID_72326 | Betulin | IMPPAT | Organic compounds | Lipids and lipid-like molecules | Prenol lipids | Triterpenoids |
| *Enicostema axillare* | Whole Plant | IMPHY000526 | CAS_22108-77-6 | Gentiocrucine | IMPPAT | Organic compounds | Organoheterocyclic compounds | Lactones | Delta valerolactones |
| *Tecomella undulata* | Bark | IMPHY014836 | CID_222284 | Beta-sitosterol | IMPPAT | Organic compounds | Lipids and lipid-like molecules | Steroids and steroid derivatives | Stigmastanes and derivatives |
| *Tecomella undulata* | Bark | IMPHY006558 | CID_68972 | n-Triacontanol | IMPPAT | Organic compounds | Lipids and lipid-like molecules | Fatty Acyls | Fatty alcohols |
| *Tecomella undulata* | Bark | IMPHY015072 | CID_12309055 | Beta-Sitosterol-beta-D-glucoside | IMPPAT | Organic compounds | Lipids and lipid-like molecules | Steroids and steroid derivatives | Stigmastanes and derivatives |
| *Tecomella undulata* | Bark | IMPHY014838 | CID_5742590 | Beta-sitosterol glucoside | IMPPAT | Organic compounds | Lipids and lipid-like molecules | Steroids and steroid derivatives | Stigmastanes and derivatives |
| *Tecomella undulata* | Bark | IMPHY006972 | CID_68406 | n-Octacosanol | IMPPAT | Organic compounds | Lipids and lipid-like molecules | Fatty Acyls | Fatty alcohols |
| *Tecomella undulata* | Bark | IMPHY009482 | CID_12409 | n-Nonacosane | IMPPAT | Organic compounds | Hydrocarbons | Saturated hydrocarbons | Alkanes |
| *Tecomella undulata* | Bark | IMPHY009413 | CID_12535 | n-Triacontane | IMPPAT | Organic compounds | Hydrocarbons | Saturated hydrocarbons | Alkanes |
| *Tecomella undulata* | Bark | IMPHY003002 | CID_3884 | Lapachol | IMPPAT | Organic compounds | Benzenoids | Naphthalenes | Naphthoquinones |
| *Tecomella undulata* | Bark | IMPHY001896 | CID_11636 | n-Heptacosane | IMPPAT | Organic compounds | Hydrocarbons | Saturated hydrocarbons | Alkanes |
| *Tecomella undulata* | Bark | IMPHY011802 | CID_445858 | Ferulate | IMPPAT | Organic compounds | Phenylpropanoids and polyketides | Cinnamic acids and derivatives | Hydroxycinnamic acids and derivatives |
| *Tecomella undulata* | Bark | IMPHY006695 | CID_7121 | Veratric acid | IMPPAT | Organic compounds | Benzenoids | Benzene and substituted derivatives | Benzoic acids and derivatives |
| *Tecomella undulata* | Bark | IMPHY004090 | CID_3037329 | Dehydrotectol | IMPPAT | Organic compounds | Organoheterocyclic compounds | Naphthopyrans | |
| *Tecomella undulata* | Bark | IMPHY004462 | CID_5321494 | Undulatosides A | IMPPAT | Organic compounds | Organic oxygen compounds | Organooxygen compounds | Carbohydrates and carbohydrate conjugates |
| *Tecomella undulata* | Bark | | | Tecomin | API | | | | |

**Table S5: Plant–phytochemical–target associations derived from multiple target identification sources.** This table compiles associations linking plants to their constituent phytochemicals and the corresponding human molecular targets. For each entry, the table lists the plant source, phytochemical, associated target(s), and the database sources from which the target information was obtained.

| Plant name | Part | PubChem CID | Entrez Gene ID | Source database(s) |
|---|---|---|---|---|
| *Pterocarpus marsupium* | Wood | CID_259846 | 1066 | ChEMBL \| NPASS |
| *Pterocarpus marsupium* | Wood | CID_259846 | 387 | ChEMBL |
| *Pterocarpus marsupium* | Wood | CID_259846 | 5770 | BindingDB \| ChEMBL \| NPASS |
| *Pterocarpus marsupium* | Wood | CID_259846 | 5879 | ChEMBL |
| *Pterocarpus marsupium* | Wood | CID_259846 | 7153 | ChEMBL \| NPASS |
| *Pterocarpus marsupium* | Wood | CID_259846 | 8824 | BindingDB \| ChEMBL \| NPASS |
| *Pterocarpus marsupium* | Wood | CID_259846 | 998 | ChEMBL |
| *Pterocarpus marsupium* | Wood | CID_72276 | 10073 | NPASS |
| *Pterocarpus marsupium* | Wood | CID_72276 | 10599 | ChEMBL \| NPASS |
| *Pterocarpus marsupium* | Wood | CID_72276 | 10919 | ChEMBL \| NPASS |
| *Pterocarpus marsupium* | Wood | CID_72276 | 11201 | ChEMBL \| NPASS |
| *Pterocarpus marsupium* | Wood | CID_72276 | 11309 | ChEMBL |
| *Pterocarpus marsupium* | Wood | CID_72276 | 118429 | NPASS |
| *Pterocarpus marsupium* | Wood | CID_72276 | 1432 | BindingDB |
| *Pterocarpus marsupium* | Wood | CID_72276 | 1514 | BindingDB \| NPASS |
| *Pterocarpus marsupium* | Wood | CID_72276 | 1576 | NPASS |
| *Pterocarpus marsupium* | Wood | CID_72276 | 213 | ChEMBL |
| *Pterocarpus marsupium* | Wood | CID_72276 | 2147 | ChEMBL \| NPASS |
| *Pterocarpus marsupium* | Wood | CID_72276 | 2194 | ChEMBL \| NPASS |
| *Pterocarpus marsupium* | Wood | CID_72276 | 23621 | BindingDB \| ChEMBL |
| *Pterocarpus marsupium* | Wood | CID_72276 | 248 | BindingDB \| NPASS |
| *Pterocarpus marsupium* | Wood | CID_72276 | 249 | BindingDB \| ChEMBL \| NPASS |
| *Pterocarpus marsupium* | Wood | CID_72276 | 251 | BindingDB \| ChEMBL \| NPASS |
| *Pterocarpus marsupium* | Wood | CID_72276 | 2548 | ChEMBL \| NPASS |
| *Pterocarpus marsupium* | Wood | CID_72276 | 2581 | NPASS |
| *Pterocarpus marsupium* | Wood | CID_72276 | 276 | ChEMBL |
| *Pterocarpus marsupium* | Wood | CID_72276 | 277 | BindingDB |
| *Pterocarpus marsupium* | Wood | CID_72276 | 28234 | ChEMBL \| NPASS |
| *Pterocarpus marsupium* | Wood | CID_72276 | 29127 | NPASS |
| *Pterocarpus marsupium* | Wood | CID_72276 | 29994 | NPASS |
| *Pterocarpus marsupium* | Wood | CID_72276 | 3028 | NPASS |
| *Pterocarpus marsupium* | Wood | CID_72276 | 328 | NPASS |
| *Pterocarpus marsupium* | Wood | CID_72276 | 351 | BindingDB \| ChEMBL |
| *Pterocarpus marsupium* | Wood | CID_72276 | 3837 | NPASS |
| *Pterocarpus marsupium* | Wood | CID_72276 | 390245 | ChEMBL \| NPASS |
| *Pterocarpus marsupium* | Wood | CID_72276 | 4088 | NPASS |
| *Pterocarpus marsupium* | Wood | CID_72276 | 4137 | NPASS |
| *Pterocarpus marsupium* | Wood | CID_72276 | 4297 | NPASS |
| *Pterocarpus marsupium* | Wood | CID_72276 | 4780 | NPASS |
| *Pterocarpus marsupium* | Wood | CID_72276 | 4790 | NPASS |
| *Pterocarpus marsupium* | Wood | CID_72276 | 4846 | ChEMBL |
| *Pterocarpus marsupium* | Wood | CID_72276 | 51426 | ChEMBL \| NPASS |
| *Pterocarpus marsupium* | Wood | CID_72276 | 5226 | BindingDB \| NPASS |
| *Pterocarpus marsupium* | Wood | CID_72276 | 5315 | ChEMBL |
| *Pterocarpus marsupium* | Wood | CID_72276 | 55775 | NPASS |
| *Pterocarpus marsupium* | Wood | CID_72276 | 5602 | BindingDB |
| *Pterocarpus marsupium* | Wood | CID_72276 | 5893 | ChEMBL \| NPASS |
| *Pterocarpus marsupium* | Wood | CID_72276 | 5901 | NPASS |
| *Pterocarpus marsupium* | Wood | CID_72276 | 596 | ChEMBL \| NPASS |
| *Pterocarpus marsupium* | Wood | CID_72276 | 6035 | ChEMBL \| NPASS |
| *Pterocarpus marsupium* | Wood | CID_72276 | 6622 | ChEMBL \| NPASS |
| *Pterocarpus marsupium* | Wood | CID_72276 | 7253 | NPASS |
| *Pterocarpus marsupium* | Wood | CID_72276 | 7498 | BindingDB \| ChEMBL \| NPASS |
| *Pterocarpus marsupium* | Wood | CID_72276 | 759 | BindingDB |
| *Pterocarpus marsupium* | Wood | CID_72276 | 760 | BindingDB |
| *Pterocarpus marsupium* | Wood | CID_72276 | 762 | BindingDB |
| *Pterocarpus marsupium* | Wood | CID_439246 | 10013 | ChEMBL |
| *Pterocarpus marsupium* | Wood | CID_439246 | 10411 | NPASS |
| *Pterocarpus marsupium* | Wood | CID_439246 | 10599 | ChEMBL \| NPASS |
| *Pterocarpus marsupium* | Wood | CID_439246 | 11309 | ChEMBL \| NPASS |
| *Pterocarpus marsupium* | Wood | CID_439246 | 1244 | ChEMBL \| NPASS |
| *Pterocarpus marsupium* | Wood | CID_439246 | 1543 | BindingDB \| ChEMBL \| NPASS |
| *Pterocarpus marsupium* | Wood | CID_439246 | 1544 | BindingDB \| ChEMBL \| NPASS |
| *Pterocarpus marsupium* | Wood | CID_439246 | 1545 | BindingDB \| ChEMBL \| NPASS |
| *Pterocarpus marsupium* | Wood | CID_439246 | 1557 | ChEMBL \| NPASS |
| *Pterocarpus marsupium* | Wood | CID_439246 | 1559 | NPASS |
| *Pterocarpus marsupium* | Wood | CID_439246 | 1576 | ChEMBL \| NPASS |
| *Pterocarpus marsupium* | Wood | CID_439246 | 1588 | BindingDB \| ChEMBL \| NPASS |
| *Pterocarpus marsupium* | Wood | CID_439246 | 1803 | BindingDB \| ChEMBL \| NPASS |
| *Pterocarpus marsupium* | Wood | CID_439246 | 2099 | ChEMBL \| NPASS |
| *Pterocarpus marsupium* | Wood | CID_439246 | 2100 | ChEMBL \| NPASS |
| *Pterocarpus marsupium* | Wood | CID_439246 | 2203 | ChEMBL \| NPASS |
| *Pterocarpus marsupium* | Wood | CID_439246 | 231 | ChEMBL \| NPASS |
| *Pterocarpus marsupium* | Wood | CID_439246 | 2322 | ChEMBL \| NPASS |
| *Pterocarpus marsupium* | Wood | CID_439246 | 253430 | ChEMBL \| NPASS |
| *Pterocarpus marsupium* | Wood | CID_439246 | 25939 | ChEMBL |
| *Pterocarpus marsupium* | Wood | CID_439246 | 2739 | ChEMBL \| NPASS |
| *Pterocarpus marsupium* | Wood | CID_439246 | 276 | ChEMBL |
| *Pterocarpus marsupium* | Wood | CID_439246 | 277 | BindingDB |
| *Pterocarpus marsupium* | Wood | CID_439246 | 28234 | ChEMBL \| NPASS |
| *Pterocarpus marsupium* | Wood | CID_439246 | 2932 | BindingDB \| ChEMBL \| NPASS |
| *Pterocarpus marsupium* | Wood | CID_439246 | 3028 | NPASS |
| *Pterocarpus marsupium* | Wood | CID_439246 | 3292 | BindingDB \| ChEMBL \| NPASS |
| *Pterocarpus marsupium* | Wood | CID_439246 | 3294 | BindingDB \| ChEMBL \| NPASS |
| *Pterocarpus marsupium* | Wood | CID_439246 | 3417 | ChEMBL \| NPASS |
| *Pterocarpus marsupium* | Wood | CID_439246 | 351 | ChEMBL |
| *Pterocarpus marsupium* | Wood | CID_439246 | 4128 | NPASS |
| *Pterocarpus marsupium* | Wood | CID_439246 | 4129 | NPASS |
| *Pterocarpus marsupium* | Wood | CID_439246 | 43 | BindingDB \| ChEMBL \| NPASS |
| *Pterocarpus marsupium* | Wood | CID_439246 | 4363 | ChEMBL \| NPASS |
| *Pterocarpus marsupium* | Wood | CID_439246 | 4780 | ChEMBL \| NPASS |
| *Pterocarpus marsupium* | Wood | CID_439246 | 4985 | ChEMBL \| NPASS |
| *Pterocarpus marsupium* | Wood | CID_439246 | 4986 | ChEMBL \| NPASS |
| *Pterocarpus marsupium* | Wood | CID_439246 | 4988 | ChEMBL \| NPASS |
| *Pterocarpus marsupium* | Wood | CID_439246 | 51053 | NPASS |
| *Pterocarpus marsupium* | Wood | CID_439246 | 51447 | ChEMBL \| NPASS |
| *Pterocarpus marsupium* | Wood | CID_439246 | 5243 | ChEMBL \| NPASS |
| *Pterocarpus marsupium* | Wood | CID_439246 | 54205 | ChEMBL \| NPASS |
| *Pterocarpus marsupium* | Wood | CID_439246 | 5444 | ChEMBL \| NPASS |
| *Pterocarpus marsupium* | Wood | CID_439246 | 54575 | ChEMBL \| NPASS |
| *Pterocarpus marsupium* | Wood | CID_439246 | 54576 | ChEMBL \| NPASS |
| *Pterocarpus marsupium* | Wood | CID_439246 | 5465 | ChEMBL \| NPASS |
| *Pterocarpus marsupium* | Wood | CID_439246 | 54657 | ChEMBL \| NPASS |
| *Pterocarpus marsupium* | Wood | CID_439246 | 54658 | ChEMBL \| NPASS |
| *Pterocarpus marsupium* | Wood | CID_439246 | 54659 | ChEMBL \| NPASS |
| *Pterocarpus marsupium* | Wood | CID_439246 | 5467 | ChEMBL \| NPASS |
| *Pterocarpus marsupium* | Wood | CID_439246 | 5468 | ChEMBL \| NPASS |
| *Pterocarpus marsupium* | Wood | CID_439246 | 55775 | NPASS |
| *Pterocarpus marsupium* | Wood | CID_439246 | 5693 | BindingDB \| ChEMBL \| NPASS |
| *Pterocarpus marsupium* | Wood | CID_439246 | 57016 | ChEMBL |
| *Pterocarpus marsupium* | Wood | CID_439246 | 5743 | ChEMBL \| NPASS |
| *Pterocarpus marsupium* | Wood | CID_439246 | 5745 | NPASS |
| *Pterocarpus marsupium* | Wood | CID_439246 | 5770 | BindingDB \| ChEMBL \| NPASS |
| *Pterocarpus marsupium* | Wood | CID_439246 | 590 | ChEMBL \| NPASS |
| *Pterocarpus marsupium* | Wood | CID_439246 | 6462 | BindingDB \| ChEMBL \| NPASS |
| *Pterocarpus marsupium* | Wood | CID_439246 | 6783 | ChEMBL |
| *Pterocarpus marsupium* | Wood | CID_439246 | 7276 | ChEMBL |
| *Pterocarpus marsupium* | Wood | CID_439246 | 7299 | ChEMBL \| NPASS |
| *Pterocarpus marsupium* | Wood | CID_439246 | 7366 | ChEMBL \| NPASS |
| *Pterocarpus marsupium* | Wood | CID_439246 | 7398 | NPASS |
| *Pterocarpus marsupium* | Wood | CID_439246 | 7498 | BindingDB \| ChEMBL \| NPASS |
| *Pterocarpus marsupium* | Wood | CID_439246 | 759 | BindingDB \| ChEMBL \| NPASS |
| *Pterocarpus marsupium* | Wood | CID_439246 | 760 | BindingDB \| ChEMBL \| NPASS |
| *Pterocarpus marsupium* | Wood | CID_439246 | 762 | BindingDB \| ChEMBL \| NPASS |
| *Pterocarpus marsupium* | Wood | CID_439246 | 766 | BindingDB \| ChEMBL \| NPASS |
| *Pterocarpus marsupium* | Wood | CID_439246 | 771 | BindingDB \| ChEMBL \| NPASS |
| *Pterocarpus marsupium* | Wood | CID_439246 | 805 | ChEMBL |
| *Pterocarpus marsupium* | Wood | CID_439246 | 8654 | ChEMBL \| NPASS |
| *Pterocarpus marsupium* | Wood | CID_439246 | 873 | BindingDB \| ChEMBL \| NPASS |
| *Pterocarpus marsupium* | Wood | CID_439246 | 874 | ChEMBL \| NPASS |
| *Pterocarpus marsupium* | Wood | CID_439246 | 9429 | ChEMBL |
| *Pterocarpus marsupium* | Wood | CID_10494 | 10599 | ChEMBL \| NPASS |
| *Pterocarpus marsupium* | Wood | CID_10494 | 1066 | ChEMBL \| NPASS |
| *Pterocarpus marsupium* | Wood | CID_10494 | 1071 | ChEMBL \| NPASS |
| *Pterocarpus marsupium* | Wood | CID_10494 | 10951 | NPASS |
| *Pterocarpus marsupium* | Wood | CID_10494 | 11069 | NPASS |
| *Pterocarpus marsupium* | Wood | CID_10494 | 112398 | ChEMBL \| NPASS |
| *Pterocarpus marsupium* | Wood | CID_10494 | 112399 | ChEMBL \| NPASS |
| *Pterocarpus marsupium* | Wood | CID_10494 | 11343 | ChEMBL \| NPASS |
| *Pterocarpus marsupium* | Wood | CID_10494 | 151306 | BindingDB \| ChEMBL \| NPASS |
| *Pterocarpus marsupium* | Wood | CID_10494 | 1588 | ChEMBL \| NPASS |
| *Pterocarpus marsupium* | Wood | CID_10494 | 1845 | BindingDB \| ChEMBL \| NPASS |
| *Pterocarpus marsupium* | Wood | CID_10494 | 1956 | BindingDB \| ChEMBL \| NPASS |
| *Pterocarpus marsupium* | Wood | CID_10494 | 2152 | BindingDB \| ChEMBL \| NPASS |
| *Pterocarpus marsupium* | Wood | CID_10494 | 231 | NPASS |
| *Pterocarpus marsupium* | Wood | CID_10494 | 23408 | ChEMBL |
| *Pterocarpus marsupium* | Wood | CID_10494 | 240 | ChEMBL \| NPASS |
| *Pterocarpus marsupium* | Wood | CID_10494 | 247 | ChEMBL \| NPASS |
| *Pterocarpus marsupium* | Wood | CID_10494 | 2648 | NPASS |
| *Pterocarpus marsupium* | Wood | CID_10494 | 28234 | ChEMBL \| NPASS |
| *Pterocarpus marsupium* | Wood | CID_10494 | 2932 | ChEMBL |
| *Pterocarpus marsupium* | Wood | CID_10494 | 3294 | ChEMBL \| NPASS |
| *Pterocarpus marsupium* | Wood | CID_10494 | 3417 | ChEMBL \| NPASS |
| *Pterocarpus marsupium* | Wood | CID_10494 | 351 | ChEMBL \| NPASS |
| *Pterocarpus marsupium* | Wood | CID_10494 | 3978 | BindingDB \| NPASS |
| *Pterocarpus marsupium* | Wood | CID_10494 | 4780 | ChEMBL \| NPASS |
| *Pterocarpus marsupium* | Wood | CID_10494 | 4843 | BindingDB \| ChEMBL \| NPASS |
| *Pterocarpus marsupium* | Wood | CID_10494 | 51053 | NPASS |
| *Pterocarpus marsupium* | Wood | CID_10494 | 5111 | ChEMBL |
| *Pterocarpus marsupium* | Wood | CID_10494 | 51422 | ChEMBL |
| *Pterocarpus marsupium* | Wood | CID_10494 | 52 | ChEMBL \| NPASS |
| *Pterocarpus marsupium* | Wood | CID_10494 | 5243 | ChEMBL \| NPASS |
| *Pterocarpus marsupium* | Wood | CID_10494 | 5319 | BindingDB \| ChEMBL \| NPASS |
| *Pterocarpus marsupium* | Wood | CID_10494 | 5347 | NPASS |
| *Pterocarpus marsupium* | Wood | CID_10494 | 53632 | ChEMBL |
| *Pterocarpus marsupium* | Wood | CID_10494 | 5423 | BindingDB \| ChEMBL \| NPASS |
| *Pterocarpus marsupium* | Wood | CID_10494 | 54583 | BindingDB \| ChEMBL \| NPASS |
| *Pterocarpus marsupium* | Wood | CID_10494 | 54681 | ChEMBL \| NPASS |
| *Pterocarpus marsupium* | Wood | CID_10494 | 5562 | ChEMBL |
| *Pterocarpus marsupium* | Wood | CID_10494 | 5563 | ChEMBL |
| *Pterocarpus marsupium* | Wood | CID_10494 | 5564 | ChEMBL |
| *Pterocarpus marsupium* | Wood | CID_10494 | 5565 | ChEMBL |
| *Pterocarpus marsupium* | Wood | CID_10494 | 5571 | ChEMBL |
| *Pterocarpus marsupium* | Wood | CID_10494 | 55775 | ChEMBL \| NPASS |
| *Pterocarpus marsupium* | Wood | CID_10494 | 57016 | ChEMBL \| NPASS |
| *Pterocarpus marsupium* | Wood | CID_10494 | 5742 | NPASS |
| *Pterocarpus marsupium* | Wood | CID_10494 | 5743 | BindingDB \| ChEMBL \| NPASS |
| *Pterocarpus marsupium* | Wood | CID_10494 | 5770 | BindingDB \| ChEMBL \| NPASS |
| *Pterocarpus marsupium* | Wood | CID_10494 | 5771 | BindingDB \| ChEMBL \| NPASS |
| *Pterocarpus marsupium* | Wood | CID_10494 | 5777 | BindingDB \| ChEMBL \| NPASS |
| *Pterocarpus marsupium* | Wood | CID_10494 | 5781 | BindingDB \| ChEMBL \| NPASS |
| *Pterocarpus marsupium* | Wood | CID_10494 | 5786 | ChEMBL \| NPASS |
| *Pterocarpus marsupium* | Wood | CID_10494 | 5788 | BindingDB \| ChEMBL \| NPASS |
| *Pterocarpus marsupium* | Wood | CID_10494 | 5791 | ChEMBL \| NPASS |
| *Pterocarpus marsupium* | Wood | CID_10494 | 5792 | BindingDB \| ChEMBL \| NPASS |
| *Pterocarpus marsupium* | Wood | CID_10494 | 5970 | ChEMBL \| NPASS |
| *Pterocarpus marsupium* | Wood | CID_10494 | 6774 | ChEMBL \| NPASS |
| *Pterocarpus marsupium* | Wood | CID_10494 | 7150 | ChEMBL \| NPASS |
| *Pterocarpus marsupium* | Wood | CID_10494 | 7153 | ChEMBL \| NPASS |
| *Pterocarpus marsupium* | Wood | CID_10494 | 7428 | ChEMBL |
| *Pterocarpus marsupium* | Wood | CID_10494 | 7442 | ChEMBL \| NPASS |
| *Pterocarpus marsupium* | Wood | CID_10494 | 7874 | BindingDB \| ChEMBL |
| *Pterocarpus marsupium* | Wood | CID_10494 | 79915 | NPASS |
| *Pterocarpus marsupium* | Wood | CID_10494 | 994 | ChEMBL \| NPASS |
| *Pterocarpus marsupium* | Wood | CID_10494 | 9971 | NPASS |
| *Pterocarpus marsupium* | Wood | CID_91457 | 8989 | ChEMBL \| NPASS |
| *Pterocarpus marsupium* | Wood | CID_5281611 | 1557 | ChEMBL |
| *Pterocarpus marsupium* | Wood | CID_5281611 | 1559 | ChEMBL |
| *Pterocarpus marsupium* | Wood | CID_5281611 | 1588 | ChEMBL |
| *Pterocarpus marsupium* | Wood | CID_5281611 | 1719 | ChEMBL |
| *Pterocarpus marsupium* | Wood | CID_5281611 | 2203 | ChEMBL |
| *Pterocarpus marsupium* | Wood | CID_5281611 | 3028 | ChEMBL |
| *Pterocarpus marsupium* | Wood | CID_5281611 | 5292 | ChEMBL |
| *Pterocarpus marsupium* | Wood | CID_5281611 | 55775 | ChEMBL |
| *Pterocarpus marsupium* | Wood | CID_638278 | 10013 | ChEMBL |
| *Pterocarpus marsupium* | Wood | CID_638278 | 10018 | BindingDB |
| *Pterocarpus marsupium* | Wood | CID_638278 | 10599 | NPASS |
| *Pterocarpus marsupium* | Wood | CID_638278 | 10919 | NPASS |
| *Pterocarpus marsupium* | Wood | CID_638278 | 10951 | NPASS |
| *Pterocarpus marsupium* | Wood | CID_638278 | 11201 | NPASS |
| *Pterocarpus marsupium* | Wood | CID_638278 | 1139 | BindingDB \| ChEMBL \| NPASS |
| *Pterocarpus marsupium* | Wood | CID_638278 | 114548 | BindingDB \| ChEMBL \| NPASS |
| *Pterocarpus marsupium* | Wood | CID_638278 | 1543 | ChEMBL \| NPASS |
| *Pterocarpus marsupium* | Wood | CID_638278 | 1545 | ChEMBL \| NPASS |
| *Pterocarpus marsupium* | Wood | CID_638278 | 1557 | ChEMBL \| NPASS |
| *Pterocarpus marsupium* | Wood | CID_638278 | 1559 | NPASS |
| *Pterocarpus marsupium* | Wood | CID_638278 | 1576 | ChEMBL \| NPASS |
| *Pterocarpus marsupium* | Wood | CID_638278 | 1588 | BindingDB \| ChEMBL \| NPASS |
| *Pterocarpus marsupium* | Wood | CID_638278 | 1812 | NPASS |
| *Pterocarpus marsupium* | Wood | CID_638278 | 1956 | ChEMBL \| NPASS |
| *Pterocarpus marsupium* | Wood | CID_638278 | 2099 | ChEMBL \| NPASS |
| *Pterocarpus marsupium* | Wood | CID_638278 | 2100 | ChEMBL \| NPASS |
| *Pterocarpus marsupium* | Wood | CID_638278 | 2237 | NPASS |
| *Pterocarpus marsupium* | Wood | CID_638278 | 2260 | ChEMBL \| NPASS |
| *Pterocarpus marsupium* | Wood | CID_638278 | 2263 | ChEMBL \| NPASS |
| *Pterocarpus marsupium* | Wood | CID_638278 | 231 | ChEMBL \| NPASS |
| *Pterocarpus marsupium* | Wood | CID_638278 | 23621 | BindingDB \| ChEMBL \| NPASS |
| *Pterocarpus marsupium* | Wood | CID_638278 | 23683 | ChEMBL |
| *Pterocarpus marsupium* | Wood | CID_638278 | 2475 | NPASS |
| *Pterocarpus marsupium* | Wood | CID_638278 | 2629 | ChEMBL \| NPASS |
| *Pterocarpus marsupium* | Wood | CID_638278 | 2740 | NPASS |
| *Pterocarpus marsupium* | Wood | CID_638278 | 28234 | ChEMBL \| NPASS |
| *Pterocarpus marsupium* | Wood | CID_638278 | 3248 | NPASS |
| *Pterocarpus marsupium* | Wood | CID_638278 | 328 | NPASS |
| *Pterocarpus marsupium* | Wood | CID_638278 | 3292 | BindingDB \| ChEMBL \| NPASS |
| *Pterocarpus marsupium* | Wood | CID_638278 | 3294 | BindingDB \| ChEMBL \| NPASS |
| *Pterocarpus marsupium* | Wood | CID_638278 | 3309 | NPASS |
| *Pterocarpus marsupium* | Wood | CID_638278 | 3417 | ChEMBL \| NPASS |
| *Pterocarpus marsupium* | Wood | CID_638278 | 351 | BindingDB \| ChEMBL \| NPASS |
| *Pterocarpus marsupium* | Wood | CID_638278 | 3725 | ChEMBL |
| *Pterocarpus marsupium* | Wood | CID_638278 | 3757 | ChEMBL \| NPASS |
| *Pterocarpus marsupium* | Wood | CID_638278 | 3791 | ChEMBL \| NPASS |
| *Pterocarpus marsupium* | Wood | CID_638278 | 3815 | ChEMBL \| NPASS |
| *Pterocarpus marsupium* | Wood | CID_638278 | 387129 | NPASS |
| *Pterocarpus marsupium* | Wood | CID_638278 | 4000 | NPASS |
| *Pterocarpus marsupium* | Wood | CID_638278 | 4088 | NPASS |
| *Pterocarpus marsupium* | Wood | CID_638278 | 4137 | ChEMBL \| NPASS |
| *Pterocarpus marsupium* | Wood | CID_638278 | 4297 | NPASS |
| *Pterocarpus marsupium* | Wood | CID_638278 | 4312 | ChEMBL |
| *Pterocarpus marsupium* | Wood | CID_638278 | 4313 | ChEMBL |
| *Pterocarpus marsupium* | Wood | CID_638278 | 4314 | ChEMBL |
| *Pterocarpus marsupium* | Wood | CID_638278 | 4316 | BindingDB \| ChEMBL |
| *Pterocarpus marsupium* | Wood | CID_638278 | 4318 | BindingDB \| ChEMBL \| NPASS |
| *Pterocarpus marsupium* | Wood | CID_638278 | 4322 | BindingDB \| ChEMBL |
| *Pterocarpus marsupium* | Wood | CID_638278 | 4780 | ChEMBL \| NPASS |
| *Pterocarpus marsupium* | Wood | CID_638278 | 4790 | ChEMBL |
| *Pterocarpus marsupium* | Wood | CID_638278 | 4791 | ChEMBL |
| *Pterocarpus marsupium* | Wood | CID_638278 | 5034 | BindingDB \| ChEMBL |
| *Pterocarpus marsupium* | Wood | CID_638278 | 51053 | ChEMBL \| NPASS |
| *Pterocarpus marsupium* | Wood | CID_638278 | 51426 | NPASS |
| *Pterocarpus marsupium* | Wood | CID_638278 | 5300 | NPASS |
| *Pterocarpus marsupium* | Wood | CID_638278 | 5376 | NPASS |
| *Pterocarpus marsupium* | Wood | CID_638278 | 5468 | ChEMBL \| NPASS |
| *Pterocarpus marsupium* | Wood | CID_638278 | 54737 | NPASS |
| *Pterocarpus marsupium* | Wood | CID_638278 | 55775 | ChEMBL \| NPASS |
| *Pterocarpus marsupium* | Wood | CID_638278 | 5578 | ChEMBL |
| *Pterocarpus marsupium* | Wood | CID_638278 | 5579 | ChEMBL |
| *Pterocarpus marsupium* | Wood | CID_638278 | 5580 | ChEMBL |
| *Pterocarpus marsupium* | Wood | CID_638278 | 5581 | ChEMBL |
| *Pterocarpus marsupium* | Wood | CID_638278 | 5582 | ChEMBL |
| *Pterocarpus marsupium* | Wood | CID_638278 | 5583 | ChEMBL |
| *Pterocarpus marsupium* | Wood | CID_638278 | 5584 | ChEMBL |
| *Pterocarpus marsupium* | Wood | CID_638278 | 5587 | ChEMBL |
| *Pterocarpus marsupium* | Wood | CID_638278 | 5588 | ChEMBL |
| *Pterocarpus marsupium* | Wood | CID_638278 | 5590 | ChEMBL |
| *Pterocarpus marsupium* | Wood | CID_638278 | 5693 | BindingDB \| ChEMBL \| NPASS |
| *Pterocarpus marsupium* | Wood | CID_638278 | 5742 | NPASS |
| *Pterocarpus marsupium* | Wood | CID_638278 | 5743 | BindingDB |
| *Pterocarpus marsupium* | Wood | CID_638278 | 5770 | BindingDB \| ChEMBL \| NPASS |
| *Pterocarpus marsupium* | Wood | CID_638278 | 5970 | ChEMBL \| NPASS |
| *Pterocarpus marsupium* | Wood | CID_638278 | 6311 | NPASS |
| *Pterocarpus marsupium* | Wood | CID_638278 | 641 | NPASS |
| *Pterocarpus marsupium* | Wood | CID_638278 | 6595 | BindingDB \| NPASS |
| *Pterocarpus marsupium* | Wood | CID_638278 | 6622 | NPASS |
| *Pterocarpus marsupium* | Wood | CID_638278 | 6714 | ChEMBL \| NPASS |
| *Pterocarpus marsupium* | Wood | CID_638278 | 7157 | NPASS |
| *Pterocarpus marsupium* | Wood | CID_638278 | 7299 | ChEMBL \| NPASS |
| *Pterocarpus marsupium* | Wood | CID_638278 | 7398 | ChEMBL \| NPASS |
| *Pterocarpus marsupium* | Wood | CID_638278 | 7421 | ChEMBL \| NPASS |
| *Pterocarpus marsupium* | Wood | CID_638278 | 79915 | ChEMBL \| NPASS |
| *Pterocarpus marsupium* | Wood | CID_126 | 18 | BindingDB \| ChEMBL \| NPASS |
| *Pterocarpus marsupium* | Wood | CID_126 | 3620 | ChEMBL \| NPASS |
| *Pterocarpus marsupium* | Wood | CID_126 | 4353 | ChEMBL |
| *Pterocarpus marsupium* | Wood | CID_126 | 7915 | BindingDB \| ChEMBL \| NPASS |
| *Pterocarpus marsupium* | Wood | CID_114829 | 10013 | ChEMBL |
| *Pterocarpus marsupium* | Wood | CID_114829 | 1543 | ChEMBL \| NPASS |
| *Pterocarpus marsupium* | Wood | CID_114829 | 1545 | ChEMBL \| NPASS |
| *Pterocarpus marsupium* | Wood | CID_114829 | 1588 | BindingDB \| ChEMBL \| NPASS |
| *Pterocarpus marsupium* | Wood | CID_114829 | 2099 | ChEMBL |
| *Pterocarpus marsupium* | Wood | CID_114829 | 2100 | ChEMBL |
| *Pterocarpus marsupium* | Wood | CID_114829 | 3725 | ChEMBL \| NPASS |
| *Pterocarpus marsupium* | Wood | CID_114829 | 4312 | ChEMBL |
| *Pterocarpus marsupium* | Wood | CID_114829 | 4313 | ChEMBL |
| *Pterocarpus marsupium* | Wood | CID_114829 | 4314 | ChEMBL |
| *Pterocarpus marsupium* | Wood | CID_114829 | 4316 | ChEMBL |
| *Pterocarpus marsupium* | Wood | CID_114829 | 4318 | ChEMBL \| NPASS |
| *Pterocarpus marsupium* | Wood | CID_114829 | 4322 | ChEMBL |
| *Pterocarpus marsupium* | Wood | CID_114829 | 4780 | ChEMBL \| NPASS |
| *Pterocarpus marsupium* | Wood | CID_114829 | 5693 | BindingDB \| ChEMBL \| NPASS |
| *Pterocarpus marsupium* | Wood | CID_114829 | 5770 | BindingDB \| ChEMBL \| NPASS |
| *Pterocarpus marsupium* | Wood | CID_114829 | 5970 | ChEMBL \| NPASS |
| *Pterocarpus marsupium* | Wood | CID_114829 | 7299 | ChEMBL \| NPASS |
| *Pterocarpus marsupium* | Wood | CID_114829 | 7525 | BindingDB \| ChEMBL \| NPASS |
| *Pterocarpus marsupium* | Wood | CID_442410 | 1588 | ChEMBL \| NPASS |
| *Pterocarpus marsupium* | Wood | CID_5281805 | 5465 | ChEMBL \| NPASS |
| *Pterocarpus marsupium* | Wood | CID_5281805 | 5467 | ChEMBL \| NPASS |
| *Pterocarpus marsupium* | Wood | CID_5281805 | 5468 | BindingDB \| ChEMBL \| NPASS |
| *Pterocarpus marsupium* | Wood | CID_5282073 | 1576 | NPASS |
| *Pterocarpus marsupium* | Wood | CID_5282073 | 1588 | BindingDB \| ChEMBL \| NPASS |
| *Pterocarpus marsupium* | Wood | CID_5282073 | 216 | NPASS |
| *Pterocarpus marsupium* | Wood | CID_5282073 | 217 | ChEMBL |
| *Pterocarpus marsupium* | Wood | CID_5282073 | 2203 | ChEMBL \| NPASS |
| *Pterocarpus marsupium* | Wood | CID_5282073 | 240 | ChEMBL \| NPASS |
| *Pterocarpus marsupium* | Wood | CID_5282073 | 253430 | ChEMBL \| NPASS |
| *Pterocarpus marsupium* | Wood | CID_5282073 | 25797 | BindingDB \| ChEMBL \| NPASS |
| *Pterocarpus marsupium* | Wood | CID_5282073 | 2717 | NPASS |
| *Pterocarpus marsupium* | Wood | CID_5282073 | 3028 | ChEMBL \| NPASS |
| *Pterocarpus marsupium* | Wood | CID_5282073 | 3248 | NPASS |
| *Pterocarpus marsupium* | Wood | CID_5282073 | 390245 | NPASS |
| *Pterocarpus marsupium* | Wood | CID_5282073 | 3932 | NPASS |
| *Pterocarpus marsupium* | Wood | CID_5282073 | 4128 | ChEMBL |
| *Pterocarpus marsupium* | Wood | CID_5282073 | 4129 | ChEMBL \| NPASS |
| *Pterocarpus marsupium* | Wood | CID_5282073 | 4353 | BindingDB \| ChEMBL \| NPASS |
| *Pterocarpus marsupium* | Wood | CID_5282073 | 4759 | BindingDB \| NPASS |
| *Pterocarpus marsupium* | Wood | CID_5282073 | 4780 | ChEMBL \| NPASS |
| *Pterocarpus marsupium* | Wood | CID_5282073 | 4985 | ChEMBL \| NPASS |
| *Pterocarpus marsupium* | Wood | CID_5282073 | 4986 | ChEMBL \| NPASS |
| *Pterocarpus marsupium* | Wood | CID_5282073 | 4988 | ChEMBL \| NPASS |
| *Pterocarpus marsupium* | Wood | CID_5282073 | 51447 | ChEMBL \| NPASS |
| *Pterocarpus marsupium* | Wood | CID_5282073 | 5163 | ChEMBL \| NPASS |
| *Pterocarpus marsupium* | Wood | CID_5282073 | 5465 | ChEMBL \| NPASS |
| *Pterocarpus marsupium* | Wood | CID_5282073 | 5467 | ChEMBL \| NPASS |
| *Pterocarpus marsupium* | Wood | CID_5282073 | 5468 | ChEMBL \| NPASS |
| *Pterocarpus marsupium* | Wood | CID_5282073 | 55775 | NPASS |
| *Pterocarpus marsupium* | Wood | CID_5282073 | 5594 | NPASS |
| *Pterocarpus marsupium* | Wood | CID_5282073 | 5770 | ChEMBL \| NPASS |
| *Pterocarpus marsupium* | Wood | CID_5282073 | 6606 | NPASS |
| *Pterocarpus marsupium* | Wood | CID_638088 | 367 | NPASS |
| *Pterocarpus marsupium* | Wood | CID_638088 | 4780 | NPASS |
| *Pterocarpus marsupium* | Wood | CID_688857 | 5693 | BindingDB \| ChEMBL \| NPASS |
| *Pterocarpus marsupium* | Wood | CID_5281727 | 10013 | ChEMBL |
| *Pterocarpus marsupium* | Wood | CID_5281727 | 1543 | BindingDB \| ChEMBL \| NPASS |
| *Pterocarpus marsupium* | Wood | CID_5281727 | 1545 | BindingDB \| ChEMBL \| NPASS |
| *Pterocarpus marsupium* | Wood | CID_5281727 | 1588 | ChEMBL \| NPASS |
| *Pterocarpus marsupium* | Wood | CID_5281727 | 23435 | ChEMBL \| NPASS |
| *Pterocarpus marsupium* | Wood | CID_5281727 | 240 | BindingDB \| ChEMBL \| NPASS |
| *Pterocarpus marsupium* | Wood | CID_5281727 | 25 | BindingDB \| ChEMBL |
| *Pterocarpus marsupium* | Wood | CID_5281727 | 2548 | NPASS |
| *Pterocarpus marsupium* | Wood | CID_5281727 | 2671 | ChEMBL \| NPASS |
| *Pterocarpus marsupium* | Wood | CID_5281727 | 2740 | NPASS |
| *Pterocarpus marsupium* | Wood | CID_5281727 | 390245 | NPASS |
| *Pterocarpus marsupium* | Wood | CID_5281727 | 406991 | ChEMBL |
| *Pterocarpus marsupium* | Wood | CID_5281727 | 4137 | ChEMBL \| NPASS |
| *Pterocarpus marsupium* | Wood | CID_5281727 | 4297 | NPASS |
| *Pterocarpus marsupium* | Wood | CID_5281727 | 4780 | ChEMBL |
| *Pterocarpus marsupium* | Wood | CID_5281727 | 4835 | BindingDB \| ChEMBL \| NPASS |
| *Pterocarpus marsupium* | Wood | CID_5281727 | 4864 | NPASS |
| *Pterocarpus marsupium* | Wood | CID_5281727 | 51053 | NPASS |
| *Pterocarpus marsupium* | Wood | CID_5281727 | 5243 | ChEMBL \| NPASS |
| *Pterocarpus marsupium* | Wood | CID_5281727 | 5328 | ChEMBL \| NPASS |
| *Pterocarpus marsupium* | Wood | CID_5281727 | 55775 | NPASS |
| *Pterocarpus marsupium* | Wood | CID_5281727 | 5594 | NPASS |
| *Pterocarpus marsupium* | Wood | CID_5281727 | 5742 | ChEMBL \| NPASS |
| *Pterocarpus marsupium* | Wood | CID_5281727 | 5743 | BindingDB \| ChEMBL \| NPASS |
| *Pterocarpus marsupium* | Wood | CID_5281727 | 5879 | ChEMBL \| NPASS |
| *Pterocarpus marsupium* | Wood | CID_5281727 | 613 | ChEMBL \| NPASS |
| *Pterocarpus marsupium* | Wood | CID_5281727 | 672 | NPASS |
| *Pterocarpus marsupium* | Wood | CID_5281727 | 7276 | BindingDB \| ChEMBL |
| *Pterocarpus marsupium* | Wood | CID_5281727 | 7442 | BindingDB \| ChEMBL \| NPASS |
| *Pterocarpus marsupium* | Wood | CID_5281727 | 9367 | ChEMBL \| NPASS |
| *Pterocarpus marsupium* | Wood | CID_5281727 | 9817 | ChEMBL |
| *Commiphora wightii* | Plant exudate | CID_222284 | 10013 | ChEMBL |
| *Commiphora wightii* | Plant exudate | CID_222284 | 10599 | ChEMBL \| NPASS |
| *Commiphora wightii* | Plant exudate | CID_222284 | 1565 | ChEMBL \| NPASS |
| *Commiphora wightii* | Plant exudate | CID_222284 | 1576 | BindingDB \| ChEMBL \| NPASS |
| *Commiphora wightii* | Plant exudate | CID_222284 | 1803 | ChEMBL \| NPASS |
| *Commiphora wightii* | Plant exudate | CID_222284 | 2147 | ChEMBL \| NPASS |
| *Commiphora wightii* | Plant exudate | CID_222284 | 28234 | ChEMBL \| NPASS |
| *Commiphora wightii* | Plant exudate | CID_222284 | 2932 | ChEMBL |
| *Commiphora wightii* | Plant exudate | CID_222284 | 3417 | ChEMBL \| NPASS |
| *Commiphora wightii* | Plant exudate | CID_222284 | 51422 | ChEMBL |
| *Commiphora wightii* | Plant exudate | CID_222284 | 5243 | ChEMBL |
| *Commiphora wightii* | Plant exudate | CID_222284 | 53632 | ChEMBL |
| *Commiphora wightii* | Plant exudate | CID_222284 | 5423 | BindingDB \| ChEMBL \| NPASS |
| *Commiphora wightii* | Plant exudate | CID_222284 | 5562 | ChEMBL |
| *Commiphora wightii* | Plant exudate | CID_222284 | 5563 | ChEMBL |
| *Commiphora wightii* | Plant exudate | CID_222284 | 5564 | ChEMBL |
| *Commiphora wightii* | Plant exudate | CID_222284 | 5565 | ChEMBL |
| *Commiphora wightii* | Plant exudate | CID_222284 | 5571 | ChEMBL |
| *Commiphora wightii* | Plant exudate | CID_222284 | 7299 | ChEMBL \| NPASS |
| *Commiphora wightii* | Plant exudate | CID_5281515 | 10013 | ChEMBL |
| *Commiphora wightii* | Plant exudate | CID_5281515 | 10599 | ChEMBL \| NPASS |
| *Commiphora wightii* | Plant exudate | CID_5281515 | 1269 | BindingDB \| ChEMBL \| NPASS |
| *Commiphora wightii* | Plant exudate | CID_5281515 | 28234 | ChEMBL \| NPASS |
| *Commiphora wightii* | Plant exudate | CID_5281515 | 4297 | NPASS |
| *Commiphora wightii* | Plant exudate | CID_5281515 | 5465 | ChEMBL |
| *Commiphora wightii* | Plant exudate | CID_5997 | 10013 | ChEMBL |
| *Commiphora wightii* | Plant exudate | CID_5997 | 10062 | ChEMBL \| NPASS |
| *Commiphora wightii* | Plant exudate | CID_5997 | 10599 | ChEMBL \| NPASS |
| *Commiphora wightii* | Plant exudate | CID_5997 | 10858 | ChEMBL \| NPASS |
| *Commiphora wightii* | Plant exudate | CID_5997 | 1595 | NPASS |
| *Commiphora wightii* | Plant exudate | CID_5997 | 23762 | BindingDB \| ChEMBL |
| *Commiphora wightii* | Plant exudate | CID_5997 | 28234 | ChEMBL \| NPASS |
| *Commiphora wightii* | Plant exudate | CID_5997 | 2908 | NPASS |
| *Commiphora wightii* | Plant exudate | CID_5997 | 3417 | ChEMBL \| NPASS |
| *Commiphora wightii* | Plant exudate | CID_5997 | 367 | NPASS |
| *Commiphora wightii* | Plant exudate | CID_5997 | 4000 | NPASS |
| *Commiphora wightii* | Plant exudate | CID_5997 | 4780 | ChEMBL \| NPASS |
| *Commiphora wightii* | Plant exudate | CID_5997 | 5007 | ChEMBL |
| *Commiphora wightii* | Plant exudate | CID_5997 | 5243 | ChEMBL \| NPASS |
| *Commiphora wightii* | Plant exudate | CID_5997 | 5422 | BindingDB \| NPASS |
| *Commiphora wightii* | Plant exudate | CID_5997 | 5467 | NPASS |
| *Commiphora wightii* | Plant exudate | CID_5997 | 6097 | ChEMBL \| NPASS |
| *Commiphora wightii* | Plant exudate | CID_5997 | 6323 | ChEMBL |
| *Commiphora wightii* | Plant exudate | CID_5997 | 6326 | ChEMBL |
| *Commiphora wightii* | Plant exudate | CID_5997 | 6328 | ChEMBL \| NPASS |
| *Commiphora wightii* | Plant exudate | CID_5997 | 6622 | ChEMBL \| NPASS |
| *Commiphora wightii* | Plant exudate | CID_5997 | 6820 | ChEMBL |
| *Commiphora wightii* | Plant exudate | CID_5997 | 7155 | NPASS |
| *Commiphora wightii* | Plant exudate | CID_5997 | 9429 | ChEMBL |
| *Commiphora wightii* | Plant exudate | CID_5280794 | 10013 | ChEMBL |
| *Commiphora wightii* | Plant exudate | CID_5280794 | 10599 | ChEMBL \| NPASS |
| *Commiphora wightii* | Plant exudate | CID_5280794 | 28234 | ChEMBL \| NPASS |
| *Commiphora wightii* | Plant exudate | CID_5280794 | 3363 | ChEMBL \| NPASS |
| *Commiphora wightii* | Plant exudate | CID_5280794 | 3417 | ChEMBL \| NPASS |
| *Commiphora wightii* | Plant exudate | CID_5280794 | 5423 | BindingDB \| ChEMBL \| NPASS |
| *Commiphora wightii* | Plant exudate | CID_31253 | 2908 | NPASS |
| *Commiphora wightii* | Plant exudate | CID_31253 | 3315 | NPASS |
| *Commiphora wightii* | Plant exudate | CID_31253 | 3725 | NPASS |
| *Commiphora wightii* | Plant exudate | CID_31253 | 4790 | NPASS |
| *Commiphora wightii* | Plant exudate | CID_31253 | 5467 | NPASS |
| *Commiphora wightii* | Plant exudate | CID_31253 | 7253 | NPASS |
| *Commiphora wightii* | Plant exudate | CID_31253 | 9971 | NPASS |
| *Commiphora wightii* | Plant exudate | CID_5204 | 10951 | NPASS |
| *Commiphora wightii* | Plant exudate | CID_5204 | 11201 | NPASS |
| *Commiphora wightii* | Plant exudate | CID_5204 | 23435 | NPASS |
| *Commiphora wightii* | Plant exudate | CID_5204 | 2740 | NPASS |
| *Commiphora wightii* | Plant exudate | CID_5204 | 51053 | NPASS |
| *Commiphora wightii* | Plant exudate | CID_5204 | 55775 | NPASS |
| *Commiphora wightii* | Plant exudate | CID_5204 | 5745 | NPASS |
| *Commiphora wightii* | Plant exudate | CID_445858 | 10013 | ChEMBL |
| *Commiphora wightii* | Plant exudate | CID_445858 | 10014 | ChEMBL |
| *Commiphora wightii* | Plant exudate | CID_445858 | 10280 | ChEMBL |
| *Commiphora wightii* | Plant exudate | CID_445858 | 10599 | ChEMBL \| NPASS |
| *Commiphora wightii* | Plant exudate | CID_445858 | 10800 | ChEMBL |
| *Commiphora wightii* | Plant exudate | CID_445858 | 11238 | BindingDB \| ChEMBL \| NPASS |
| *Commiphora wightii* | Plant exudate | CID_445858 | 1128 | ChEMBL |
| *Commiphora wightii* | Plant exudate | CID_445858 | 1129 | ChEMBL |
| *Commiphora wightii* | Plant exudate | CID_445858 | 1131 | ChEMBL |
| *Commiphora wightii* | Plant exudate | CID_445858 | 1132 | ChEMBL |
| *Commiphora wightii* | Plant exudate | CID_445858 | 1133 | ChEMBL |
| *Commiphora wightii* | Plant exudate | CID_445858 | 1233 | ChEMBL |
| *Commiphora wightii* | Plant exudate | CID_445858 | 1234 | ChEMBL |
| *Commiphora wightii* | Plant exudate | CID_445858 | 1268 | ChEMBL |
| *Commiphora wightii* | Plant exudate | CID_445858 | 134 | ChEMBL |
| *Commiphora wightii* | Plant exudate | CID_445858 | 135 | ChEMBL |
| *Commiphora wightii* | Plant exudate | CID_445858 | 140 | ChEMBL |
| *Commiphora wightii* | Plant exudate | CID_445858 | 1432 | ChEMBL |
| *Commiphora wightii* | Plant exudate | CID_445858 | 1457 | ChEMBL |
| *Commiphora wightii* | Plant exudate | CID_445858 | 1459 | ChEMBL |
| *Commiphora wightii* | Plant exudate | CID_445858 | 146 | ChEMBL |
| *Commiphora wightii* | Plant exudate | CID_445858 | 1460 | ChEMBL |
| *Commiphora wightii* | Plant exudate | CID_445858 | 150 | ChEMBL |
| *Commiphora wightii* | Plant exudate | CID_445858 | 151 | ChEMBL |
| *Commiphora wightii* | Plant exudate | CID_445858 | 1511 | ChEMBL |
| *Commiphora wightii* | Plant exudate | CID_445858 | 152 | ChEMBL |
| *Commiphora wightii* | Plant exudate | CID_445858 | 153 | ChEMBL |
| *Commiphora wightii* | Plant exudate | CID_445858 | 154 | ChEMBL |
| *Commiphora wightii* | Plant exudate | CID_445858 | 1544 | ChEMBL |
| *Commiphora wightii* | Plant exudate | CID_445858 | 1548 | ChEMBL |
| *Commiphora wightii* | Plant exudate | CID_445858 | 155 | ChEMBL |
| *Commiphora wightii* | Plant exudate | CID_445858 | 1557 | ChEMBL |
| *Commiphora wightii* | Plant exudate | CID_445858 | 1559 | ChEMBL |
| *Commiphora wightii* | Plant exudate | CID_445858 | 1565 | ChEMBL |
| *Commiphora wightii* | Plant exudate | CID_445858 | 1571 | ChEMBL |
| *Commiphora wightii* | Plant exudate | CID_445858 | 1576 | ChEMBL |
| *Commiphora wightii* | Plant exudate | CID_445858 | 1812 | ChEMBL |
| *Commiphora wightii* | Plant exudate | CID_445858 | 1813 | ChEMBL |
| *Commiphora wightii* | Plant exudate | CID_445858 | 1814 | ChEMBL |
| *Commiphora wightii* | Plant exudate | CID_445858 | 1815 | ChEMBL |
| *Commiphora wightii* | Plant exudate | CID_445858 | 186 | ChEMBL |
| *Commiphora wightii* | Plant exudate | CID_445858 | 1909 | ChEMBL |
| *Commiphora wightii* | Plant exudate | CID_445858 | 1956 | ChEMBL |
| *Commiphora wightii* | Plant exudate | CID_445858 | 1991 | ChEMBL |
| *Commiphora wightii* | Plant exudate | CID_445858 | 2064 | ChEMBL |
| *Commiphora wightii* | Plant exudate | CID_445858 | 2099 | ChEMBL \| NPASS |
| *Commiphora wightii* | Plant exudate | CID_445858 | 2100 | ChEMBL |
| *Commiphora wightii* | Plant exudate | CID_445858 | 2260 | ChEMBL \| NPASS |
| *Commiphora wightii* | Plant exudate | CID_445858 | 2263 | ChEMBL \| NPASS |
| *Commiphora wightii* | Plant exudate | CID_445858 | 2321 | ChEMBL |
| *Commiphora wightii* | Plant exudate | CID_445858 | 23632 | BindingDB \| ChEMBL \| NPASS |
| *Commiphora wightii* | Plant exudate | CID_445858 | 2534 | ChEMBL |
| *Commiphora wightii* | Plant exudate | CID_445858 | 259285 | ChEMBL \| NPASS |
| *Commiphora wightii* | Plant exudate | CID_445858 | 28234 | ChEMBL \| NPASS |
| *Commiphora wightii* | Plant exudate | CID_445858 | 283106 | BindingDB \| ChEMBL |
| *Commiphora wightii* | Plant exudate | CID_445858 | 2908 | ChEMBL |
| *Commiphora wightii* | Plant exudate | CID_445858 | 3065 | ChEMBL |
| *Commiphora wightii* | Plant exudate | CID_445858 | 3066 | ChEMBL |
| *Commiphora wightii* | Plant exudate | CID_445858 | 3156 | ChEMBL |
| *Commiphora wightii* | Plant exudate | CID_445858 | 3269 | ChEMBL |
| *Commiphora wightii* | Plant exudate | CID_445858 | 3274 | ChEMBL |
| *Commiphora wightii* | Plant exudate | CID_445858 | 3356 | ChEMBL |
| *Commiphora wightii* | Plant exudate | CID_445858 | 3357 | ChEMBL |
| *Commiphora wightii* | Plant exudate | CID_445858 | 3358 | ChEMBL |
| *Commiphora wightii* | Plant exudate | CID_445858 | 3362 | ChEMBL |
| *Commiphora wightii* | Plant exudate | CID_445858 | 3363 | ChEMBL \| NPASS |
| *Commiphora wightii* | Plant exudate | CID_445858 | 3373 | BindingDB \| NPASS |
| *Commiphora wightii* | Plant exudate | CID_445858 | 338442 | ChEMBL \| NPASS |
| *Commiphora wightii* | Plant exudate | CID_445858 | 351 | BindingDB \| ChEMBL \| NPASS |
| *Commiphora wightii* | Plant exudate | CID_445858 | 3577 | ChEMBL |
| *Commiphora wightii* | Plant exudate | CID_445858 | 3579 | ChEMBL |
| *Commiphora wightii* | Plant exudate | CID_445858 | 3757 | ChEMBL |
| *Commiphora wightii* | Plant exudate | CID_445858 | 390245 | NPASS |
| *Commiphora wightii* | Plant exudate | CID_445858 | 3932 | ChEMBL |
| *Commiphora wightii* | Plant exudate | CID_445858 | 4128 | ChEMBL |
| *Commiphora wightii* | Plant exudate | CID_445858 | 4159 | ChEMBL |
| *Commiphora wightii* | Plant exudate | CID_445858 | 4160 | ChEMBL |
| *Commiphora wightii* | Plant exudate | CID_445858 | 4161 | ChEMBL |
| *Commiphora wightii* | Plant exudate | CID_445858 | 43 | BindingDB \| ChEMBL \| NPASS |
| *Commiphora wightii* | Plant exudate | CID_445858 | 4312 | ChEMBL |
| *Commiphora wightii* | Plant exudate | CID_445858 | 4318 | ChEMBL |
| *Commiphora wightii* | Plant exudate | CID_445858 | 4886 | ChEMBL |
| *Commiphora wightii* | Plant exudate | CID_445858 | 4887 | ChEMBL |
| *Commiphora wightii* | Plant exudate | CID_445858 | 4985 | ChEMBL |
| *Commiphora wightii* | Plant exudate | CID_445858 | 4986 | ChEMBL |
| *Commiphora wightii* | Plant exudate | CID_445858 | 4988 | ChEMBL |
| *Commiphora wightii* | Plant exudate | CID_445858 | 51053 | NPASS |
| *Commiphora wightii* | Plant exudate | CID_445858 | 51564 | ChEMBL |
| *Commiphora wightii* | Plant exudate | CID_445858 | 5315 | ChEMBL |
| *Commiphora wightii* | Plant exudate | CID_445858 | 54657 | ChEMBL \| NPASS |
| *Commiphora wightii* | Plant exudate | CID_445858 | 54658 | ChEMBL \| NPASS |
| *Commiphora wightii* | Plant exudate | CID_445858 | 552 | ChEMBL |
| *Commiphora wightii* | Plant exudate | CID_445858 | 5530 | ChEMBL |
| *Commiphora wightii* | Plant exudate | CID_445858 | 55775 | NPASS |
| *Commiphora wightii* | Plant exudate | CID_445858 | 5578 | ChEMBL |
| *Commiphora wightii* | Plant exudate | CID_445858 | 55869 | ChEMBL |
| *Commiphora wightii* | Plant exudate | CID_445858 | 5594 | ChEMBL |
| *Commiphora wightii* | Plant exudate | CID_445858 | 5595 | ChEMBL |
| *Commiphora wightii* | Plant exudate | CID_445858 | 5724 | ChEMBL |
| *Commiphora wightii* | Plant exudate | CID_445858 | 5742 | ChEMBL |
| *Commiphora wightii* | Plant exudate | CID_445858 | 5743 | ChEMBL |
| *Commiphora wightii* | Plant exudate | CID_445858 | 5745 | NPASS |
| *Commiphora wightii* | Plant exudate | CID_445858 | 5788 | ChEMBL |
| *Commiphora wightii* | Plant exudate | CID_445858 | 590 | BindingDB \| ChEMBL \| NPASS |
| *Commiphora wightii* | Plant exudate | CID_445858 | 624 | ChEMBL |
| *Commiphora wightii* | Plant exudate | CID_445858 | 6530 | ChEMBL |
| *Commiphora wightii* | Plant exudate | CID_445858 | 6531 | ChEMBL |
| *Commiphora wightii* | Plant exudate | CID_445858 | 6532 | ChEMBL |
| *Commiphora wightii* | Plant exudate | CID_445858 | 6622 | ChEMBL \| NPASS |
| *Commiphora wightii* | Plant exudate | CID_445858 | 6865 | ChEMBL |
| *Commiphora wightii* | Plant exudate | CID_445858 | 6869 | ChEMBL |
| *Commiphora wightii* | Plant exudate | CID_445858 | 6916 | ChEMBL |
| *Commiphora wightii* | Plant exudate | CID_445858 | 729230 | ChEMBL |
| *Commiphora wightii* | Plant exudate | CID_445858 | 7299 | ChEMBL \| NPASS |
| *Commiphora wightii* | Plant exudate | CID_445858 | 7366 | ChEMBL \| NPASS |
| *Commiphora wightii* | Plant exudate | CID_445858 | 7433 | ChEMBL |
| *Commiphora wightii* | Plant exudate | CID_445858 | 759 | BindingDB \| ChEMBL \| NPASS |
| *Commiphora wightii* | Plant exudate | CID_445858 | 760 | BindingDB \| ChEMBL \| NPASS |
| *Commiphora wightii* | Plant exudate | CID_445858 | 761 | BindingDB \| ChEMBL \| NPASS |
| *Commiphora wightii* | Plant exudate | CID_445858 | 762 | BindingDB \| ChEMBL \| NPASS |
| *Commiphora wightii* | Plant exudate | CID_445858 | 763 | BindingDB \| ChEMBL \| NPASS |
| *Commiphora wightii* | Plant exudate | CID_445858 | 765 | ChEMBL \| NPASS |
| *Commiphora wightii* | Plant exudate | CID_445858 | 766 | BindingDB \| ChEMBL \| NPASS |
| *Commiphora wightii* | Plant exudate | CID_445858 | 768 | BindingDB \| ChEMBL \| NPASS |
| *Commiphora wightii* | Plant exudate | CID_445858 | 771 | BindingDB \| ChEMBL \| NPASS |
| *Commiphora wightii* | Plant exudate | CID_445858 | 79885 | ChEMBL |
| *Commiphora wightii* | Plant exudate | CID_445858 | 799 | ChEMBL |
| *Commiphora wightii* | Plant exudate | CID_445858 | 834 | ChEMBL |
| *Commiphora wightii* | Plant exudate | CID_445858 | 83933 | ChEMBL \| NPASS |
| *Commiphora wightii* | Plant exudate | CID_445858 | 8654 | ChEMBL |
| *Commiphora wightii* | Plant exudate | CID_445858 | 8841 | ChEMBL |
| *Commiphora wightii* | Plant exudate | CID_445858 | 886 | ChEMBL |
| *Commiphora wightii* | Plant exudate | CID_445858 | 9734 | ChEMBL |
| *Commiphora wightii* | Plant exudate | CID_445858 | 9759 | ChEMBL |
| *Commiphora wightii* | Plant exudate | CID_5281384 | 5319 | ChEMBL \| NPASS |
| *Commiphora wightii* | Plant exudate | CID_92747 | 1586 | BindingDB \| ChEMBL \| NPASS |
| *Commiphora wightii* | Plant exudate | CID_92747 | 5241 | ChEMBL \| NPASS |
| *Commiphora wightii* | Plant exudate | CID_6439929 | 10062 | ChEMBL \| NPASS |
| *Commiphora wightii* | Plant exudate | CID_6439929 | 1576 | ChEMBL \| NPASS |
| *Commiphora wightii* | Plant exudate | CID_6439929 | 2099 | ChEMBL \| NPASS |
| *Commiphora wightii* | Plant exudate | CID_6439929 | 2100 | ChEMBL |
| *Commiphora wightii* | Plant exudate | CID_6439929 | 216 | NPASS |
| *Commiphora wightii* | Plant exudate | CID_6439929 | 2908 | ChEMBL \| NPASS |
| *Commiphora wightii* | Plant exudate | CID_6439929 | 3315 | NPASS |
| *Commiphora wightii* | Plant exudate | CID_6439929 | 3576 | NPASS |
| *Commiphora wightii* | Plant exudate | CID_6439929 | 367 | BindingDB \| ChEMBL \| NPASS |
| *Commiphora wightii* | Plant exudate | CID_6439929 | 3757 | NPASS |
| *Commiphora wightii* | Plant exudate | CID_6439929 | 4306 | BindingDB \| ChEMBL \| NPASS |
| *Commiphora wightii* | Plant exudate | CID_6439929 | 4780 | ChEMBL \| NPASS |
| *Commiphora wightii* | Plant exudate | CID_6439929 | 4790 | NPASS |
| *Commiphora wightii* | Plant exudate | CID_6439929 | 5241 | ChEMBL \| NPASS |
| *Commiphora wightii* | Plant exudate | CID_6439929 | 5465 | ChEMBL \| NPASS |
| *Commiphora wightii* | Plant exudate | CID_6439929 | 5467 | ChEMBL \| NPASS |
| *Commiphora wightii* | Plant exudate | CID_6439929 | 5468 | ChEMBL \| NPASS |
| *Commiphora wightii* | Plant exudate | CID_6439929 | 5594 | NPASS |
| *Commiphora wightii* | Plant exudate | CID_6439929 | 5914 | ChEMBL \| NPASS |
| *Commiphora wightii* | Plant exudate | CID_6439929 | 5915 | ChEMBL \| NPASS |
| *Commiphora wightii* | Plant exudate | CID_6439929 | 5916 | ChEMBL \| NPASS |
| *Commiphora wightii* | Plant exudate | CID_6439929 | 6097 | ChEMBL \| NPASS |
| *Commiphora wightii* | Plant exudate | CID_6439929 | 6256 | ChEMBL \| NPASS |
| *Commiphora wightii* | Plant exudate | CID_6439929 | 6606 | NPASS |
| *Commiphora wightii* | Plant exudate | CID_6439929 | 7067 | ChEMBL |
| *Commiphora wightii* | Plant exudate | CID_6439929 | 7068 | ChEMBL \| NPASS |
| *Commiphora wightii* | Plant exudate | CID_6439929 | 7157 | NPASS |
| *Commiphora wightii* | Plant exudate | CID_6439929 | 7253 | NPASS |
| *Commiphora wightii* | Plant exudate | CID_6439929 | 7376 | ChEMBL \| NPASS |
| *Commiphora wightii* | Plant exudate | CID_6439929 | 7421 | ChEMBL \| NPASS |
| *Commiphora wightii* | Plant exudate | CID_6439929 | 865 | NPASS |
| *Commiphora wightii* | Plant exudate | CID_6439929 | 8856 | ChEMBL \| NPASS |
| *Commiphora wightii* | Plant exudate | CID_6439929 | 9971 | BindingDB \| ChEMBL \| NPASS |
| *Commiphora wightii* | Plant exudate | CID_70695727 | 5465 | ChEMBL \| NPASS |
| *Commiphora wightii* | Plant exudate | CID_70695727 | 5467 | ChEMBL \| NPASS |
| *Commiphora wightii* | Plant exudate | CID_70695727 | 5468 | BindingDB \| ChEMBL \| NPASS |
| *Terminalia arjuna* | Bark | CID_222284 | 10013 | ChEMBL |
| *Terminalia arjuna* | Bark | CID_222284 | 10599 | ChEMBL \| NPASS |
| *Terminalia arjuna* | Bark | CID_222284 | 1565 | ChEMBL \| NPASS |
| *Terminalia arjuna* | Bark | CID_222284 | 1576 | BindingDB \| ChEMBL \| NPASS |
| *Terminalia arjuna* | Bark | CID_222284 | 1803 | ChEMBL \| NPASS |
| *Terminalia arjuna* | Bark | CID_222284 | 2147 | ChEMBL \| NPASS |
| *Terminalia arjuna* | Bark | CID_222284 | 28234 | ChEMBL \| NPASS |
| *Terminalia arjuna* | Bark | CID_222284 | 2932 | ChEMBL |
| *Terminalia arjuna* | Bark | CID_222284 | 3417 | ChEMBL \| NPASS |
| *Terminalia arjuna* | Bark | CID_222284 | 51422 | ChEMBL |
| *Terminalia arjuna* | Bark | CID_222284 | 5243 | ChEMBL |
| *Terminalia arjuna* | Bark | CID_222284 | 53632 | ChEMBL |
| *Terminalia arjuna* | Bark | CID_222284 | 5423 | BindingDB \| ChEMBL \| NPASS |
| *Terminalia arjuna* | Bark | CID_222284 | 5562 | ChEMBL |
| *Terminalia arjuna* | Bark | CID_222284 | 5563 | ChEMBL |
| *Terminalia arjuna* | Bark | CID_222284 | 5564 | ChEMBL |
| *Terminalia arjuna* | Bark | CID_222284 | 5565 | ChEMBL |
| *Terminalia arjuna* | Bark | CID_222284 | 5571 | ChEMBL |
| *Terminalia arjuna* | Bark | CID_222284 | 7299 | ChEMBL \| NPASS |
| *Terminalia arjuna* | Bark | CID_971 | 11309 | ChEMBL |
| *Terminalia arjuna* | Bark | CID_971 | 6714 | BindingDB |
| *Terminalia arjuna* | Bark | CID_91472 | 10599 | ChEMBL \| NPASS |
| *Terminalia arjuna* | Bark | CID_91472 | 151306 | BindingDB \| ChEMBL \| NPASS |
| *Terminalia arjuna* | Bark | CID_91472 | 28234 | ChEMBL \| NPASS |
| *Terminalia arjuna* | Bark | CID_91472 | 3417 | ChEMBL \| NPASS |
| *Terminalia arjuna* | Bark | CID_91472 | 9971 | NPASS |
| *Terminalia arjuna* | Bark | CID_65084 | 1432 | BindingDB |
| *Terminalia arjuna* | Bark | CID_65084 | 23621 | BindingDB \| ChEMBL \| NPASS |
| *Terminalia arjuna* | Bark | CID_65084 | 5602 | BindingDB |
| *Terminalia arjuna* | Bark | CID_65084 | 5742 | ChEMBL \| NPASS |
| *Terminalia arjuna* | Bark | CID_65084 | 5743 | ChEMBL \| NPASS |
| *Terminalia arjuna* | Bark | CID_65084 | 596 | ChEMBL \| NPASS |
| *Terminalia arjuna* | Bark | CID_65084 | 6622 | ChEMBL \| NPASS |
| *Terminalia arjuna* | Bark | CID_5281855 | 10013 | ChEMBL |
| *Terminalia arjuna* | Bark | CID_5281855 | 10073 | NPASS |
| *Terminalia arjuna* | Bark | CID_5281855 | 1017 | BindingDB \| ChEMBL |
| *Terminalia arjuna* | Bark | CID_5281855 | 1019 | BindingDB \| ChEMBL |
| *Terminalia arjuna* | Bark | CID_5281855 | 10411 | NPASS |
| *Terminalia arjuna* | Bark | CID_5281855 | 10498 | ChEMBL \| NPASS |
| *Terminalia arjuna* | Bark | CID_5281855 | 10587 | ChEMBL |
| *Terminalia arjuna* | Bark | CID_5281855 | 10599 | ChEMBL \| NPASS |
| *Terminalia arjuna* | Bark | CID_5281855 | 10733 | BindingDB \| ChEMBL \| NPASS |
| *Terminalia arjuna* | Bark | CID_5281855 | 10919 | ChEMBL \| NPASS |
| *Terminalia arjuna* | Bark | CID_5281855 | 10951 | NPASS |
| *Terminalia arjuna* | Bark | CID_5281855 | 11069 | NPASS |
| *Terminalia arjuna* | Bark | CID_5281855 | 11201 | ChEMBL \| NPASS |
| *Terminalia arjuna* | Bark | CID_5281855 | 11238 | BindingDB \| ChEMBL \| NPASS |
| *Terminalia arjuna* | Bark | CID_5281855 | 114112 | ChEMBL \| NPASS |
| *Terminalia arjuna* | Bark | CID_5281855 | 1326 | BindingDB \| ChEMBL \| NPASS |
| *Terminalia arjuna* | Bark | CID_5281855 | 1457 | BindingDB \| ChEMBL \| NPASS |
| *Terminalia arjuna* | Bark | CID_5281855 | 1459 | ChEMBL \| NPASS |
| *Terminalia arjuna* | Bark | CID_5281855 | 1460 | ChEMBL |
| *Terminalia arjuna* | Bark | CID_5281855 | 1557 | ChEMBL \| NPASS |
| *Terminalia arjuna* | Bark | CID_5281855 | 1559 | NPASS |
| *Terminalia arjuna* | Bark | CID_5281855 | 1636 | BindingDB \| NPASS |
| *Terminalia arjuna* | Bark | CID_5281855 | 1845 | BindingDB \| NPASS |
| *Terminalia arjuna* | Bark | CID_5281855 | 1859 | BindingDB \| ChEMBL \| NPASS |
| *Terminalia arjuna* | Bark | CID_5281855 | 1956 | BindingDB \| ChEMBL \| NPASS |
| *Terminalia arjuna* | Bark | CID_5281855 | 1981 | BindingDB |
| *Terminalia arjuna* | Bark | CID_5281855 | 1995 | ChEMBL \| NPASS |
| *Terminalia arjuna* | Bark | CID_5281855 | 2050 | BindingDB \| ChEMBL \| NPASS |
| *Terminalia arjuna* | Bark | CID_5281855 | 2053 | ChEMBL \| NPASS |
| *Terminalia arjuna* | Bark | CID_5281855 | 2064 | ChEMBL \| NPASS |
| *Terminalia arjuna* | Bark | CID_5281855 | 207 | ChEMBL \| NPASS |
| *Terminalia arjuna* | Bark | CID_5281855 | 2237 | ChEMBL \| NPASS |
| *Terminalia arjuna* | Bark | CID_5281855 | 231 | ChEMBL \| NPASS |
| *Terminalia arjuna* | Bark | CID_5281855 | 23192 | BindingDB \| ChEMBL \| NPASS |
| *Terminalia arjuna* | Bark | CID_5281855 | 2322 | ChEMBL \| NPASS |
| *Terminalia arjuna* | Bark | CID_5281855 | 2324 | BindingDB \| ChEMBL \| NPASS |
| *Terminalia arjuna* | Bark | CID_5281855 | 23405 | ChEMBL |
| *Terminalia arjuna* | Bark | CID_5281855 | 23621 | BindingDB \| ChEMBL \| NPASS |
| *Terminalia arjuna* | Bark | CID_5281855 | 23632 | BindingDB \| ChEMBL \| NPASS |
| *Terminalia arjuna* | Bark | CID_5281855 | 238 | BindingDB \| ChEMBL |
| *Terminalia arjuna* | Bark | CID_5281855 | 246 | NPASS |
| *Terminalia arjuna* | Bark | CID_5281855 | 247 | NPASS |
| *Terminalia arjuna* | Bark | CID_5281855 | 2548 | NPASS |
| *Terminalia arjuna* | Bark | CID_5281855 | 2590 | ChEMBL \| NPASS |
| *Terminalia arjuna* | Bark | CID_5281855 | 2597 | NPASS |
| *Terminalia arjuna* | Bark | CID_5281855 | 2648 | NPASS |
| *Terminalia arjuna* | Bark | CID_5281855 | 2744 | ChEMBL \| NPASS |
| *Terminalia arjuna* | Bark | CID_5281855 | 28234 | ChEMBL \| NPASS |
| *Terminalia arjuna* | Bark | CID_5281855 | 2859 | BindingDB \| ChEMBL \| NPASS |
| *Terminalia arjuna* | Bark | CID_5281855 | 2932 | BindingDB \| ChEMBL \| NPASS |
| *Terminalia arjuna* | Bark | CID_5281855 | 2936 | ChEMBL \| NPASS |
| *Terminalia arjuna* | Bark | CID_5281855 | 2950 | ChEMBL \| NPASS |
| *Terminalia arjuna* | Bark | CID_5281855 | 29994 | ChEMBL \| NPASS |
| *Terminalia arjuna* | Bark | CID_5281855 | 3028 | ChEMBL \| NPASS |
| *Terminalia arjuna* | Bark | CID_5281855 | 3043 | ChEMBL \| NPASS |
| *Terminalia arjuna* | Bark | CID_5281855 | 3248 | ChEMBL \| NPASS |
| *Terminalia arjuna* | Bark | CID_5281855 | 328 | ChEMBL \| NPASS |
| *Terminalia arjuna* | Bark | CID_5281855 | 3303 | BindingDB \| ChEMBL \| NPASS |
| *Terminalia arjuna* | Bark | CID_5281855 | 3309 | NPASS |
| *Terminalia arjuna* | Bark | CID_5281855 | 3312 | BindingDB \| ChEMBL \| NPASS |
| *Terminalia arjuna* | Bark | CID_5281855 | 3417 | NPASS |
| *Terminalia arjuna* | Bark | CID_5281855 | 3480 | BindingDB \| ChEMBL \| NPASS |
| *Terminalia arjuna* | Bark | CID_5281855 | 351 | BindingDB \| ChEMBL \| NPASS |
| *Terminalia arjuna* | Bark | CID_5281855 | 3551 | BindingDB \| ChEMBL \| NPASS |
| *Terminalia arjuna* | Bark | CID_5281855 | 3643 | BindingDB \| ChEMBL \| NPASS |
| *Terminalia arjuna* | Bark | CID_5281855 | 377677 | ChEMBL \| NPASS |
| *Terminalia arjuna* | Bark | CID_5281855 | 3791 | BindingDB \| ChEMBL \| NPASS |
| *Terminalia arjuna* | Bark | CID_5281855 | 3837 | NPASS |
| *Terminalia arjuna* | Bark | CID_5281855 | 390245 | ChEMBL \| NPASS |
| *Terminalia arjuna* | Bark | CID_5281855 | 4000 | NPASS |
| *Terminalia arjuna* | Bark | CID_5281855 | 4088 | BindingDB \| ChEMBL \| NPASS |
| *Terminalia arjuna* | Bark | CID_5281855 | 4128 | ChEMBL \| NPASS |
| *Terminalia arjuna* | Bark | CID_5281855 | 4137 | ChEMBL \| NPASS |
| *Terminalia arjuna* | Bark | CID_5281855 | 4154 | NPASS |
| *Terminalia arjuna* | Bark | CID_5281855 | 4170 | BindingDB \| ChEMBL \| NPASS |
| *Terminalia arjuna* | Bark | CID_5281855 | 4233 | BindingDB \| ChEMBL \| NPASS |
| *Terminalia arjuna* | Bark | CID_5281855 | 4282 | BindingDB \| ChEMBL \| NPASS |
| *Terminalia arjuna* | Bark | CID_5281855 | 4297 | NPASS |
| *Terminalia arjuna* | Bark | CID_5281855 | 43 | BindingDB |
| *Terminalia arjuna* | Bark | CID_5281855 | 472 | NPASS |
| *Terminalia arjuna* | Bark | CID_5281855 | 4780 | NPASS |
| *Terminalia arjuna* | Bark | CID_5281855 | 4869 | ChEMBL \| NPASS |
| *Terminalia arjuna* | Bark | CID_5281855 | 4907 | ChEMBL \| NPASS |
| *Terminalia arjuna* | Bark | CID_5281855 | 51053 | NPASS |
| *Terminalia arjuna* | Bark | CID_5281855 | 51426 | ChEMBL \| NPASS |
| *Terminalia arjuna* | Bark | CID_5281855 | 5159 | BindingDB \| ChEMBL \| NPASS |
| *Terminalia arjuna* | Bark | CID_5281855 | 5300 | NPASS |
| *Terminalia arjuna* | Bark | CID_5281855 | 5313 | ChEMBL |
| *Terminalia arjuna* | Bark | CID_5281855 | 5347 | BindingDB \| ChEMBL \| NPASS |
| *Terminalia arjuna* | Bark | CID_5281855 | 5423 | ChEMBL \| NPASS |
| *Terminalia arjuna* | Bark | CID_5281855 | 5429 | BindingDB \| ChEMBL \| NPASS |
| *Terminalia arjuna* | Bark | CID_5281855 | 5566 | BindingDB \| ChEMBL |
| *Terminalia arjuna* | Bark | CID_5281855 | 5567 | ChEMBL |
| *Terminalia arjuna* | Bark | CID_5281855 | 5568 | ChEMBL \| NPASS |
| *Terminalia arjuna* | Bark | CID_5281855 | 55775 | ChEMBL \| NPASS |
| *Terminalia arjuna* | Bark | CID_5281855 | 55867 | ChEMBL \| NPASS |
| *Terminalia arjuna* | Bark | CID_5281855 | 5594 | ChEMBL \| NPASS |
| *Terminalia arjuna* | Bark | CID_5281855 | 5747 | BindingDB \| ChEMBL \| NPASS |
| *Terminalia arjuna* | Bark | CID_5281855 | 5770 | BindingDB |
| *Terminalia arjuna* | Bark | CID_5281855 | 57787 | ChEMBL |
| *Terminalia arjuna* | Bark | CID_5281855 | 595 | ChEMBL \| NPASS |
| *Terminalia arjuna* | Bark | CID_5281855 | 5965 | ChEMBL \| NPASS |
| *Terminalia arjuna* | Bark | CID_5281855 | 5979 | BindingDB \| ChEMBL \| NPASS |
| *Terminalia arjuna* | Bark | CID_5281855 | 5999 | NPASS |
| *Terminalia arjuna* | Bark | CID_5281855 | 641 | ChEMBL \| NPASS |
| *Terminalia arjuna* | Bark | CID_5281855 | 6714 | ChEMBL \| NPASS |
| *Terminalia arjuna* | Bark | CID_5281855 | 673 | BindingDB \| ChEMBL \| NPASS |
| *Terminalia arjuna* | Bark | CID_5281855 | 6790 | BindingDB \| ChEMBL \| NPASS |
| *Terminalia arjuna* | Bark | CID_5281855 | 7010 | BindingDB \| ChEMBL \| NPASS |
| *Terminalia arjuna* | Bark | CID_5281855 | 7068 | NPASS |
| *Terminalia arjuna* | Bark | CID_5281855 | 7155 | ChEMBL \| NPASS |
| *Terminalia arjuna* | Bark | CID_5281855 | 7157 | NPASS |
| *Terminalia arjuna* | Bark | CID_5281855 | 7257 | ChEMBL \| NPASS |
| *Terminalia arjuna* | Bark | CID_5281855 | 7296 | ChEMBL |
| *Terminalia arjuna* | Bark | CID_5281855 | 7398 | NPASS |
| *Terminalia arjuna* | Bark | CID_5281855 | 7421 | ChEMBL \| NPASS |
| *Terminalia arjuna* | Bark | CID_5281855 | 759 | BindingDB \| ChEMBL \| NPASS |
| *Terminalia arjuna* | Bark | CID_5281855 | 760 | BindingDB \| ChEMBL \| NPASS |
| *Terminalia arjuna* | Bark | CID_5281855 | 761 | BindingDB \| ChEMBL \| NPASS |
| *Terminalia arjuna* | Bark | CID_5281855 | 762 | BindingDB \| ChEMBL \| NPASS |
| *Terminalia arjuna* | Bark | CID_5281855 | 763 | BindingDB \| ChEMBL \| NPASS |
| *Terminalia arjuna* | Bark | CID_5281855 | 765 | ChEMBL \| NPASS |
| *Terminalia arjuna* | Bark | CID_5281855 | 766 | BindingDB \| ChEMBL \| NPASS |
| *Terminalia arjuna* | Bark | CID_5281855 | 768 | BindingDB \| ChEMBL \| NPASS |
| *Terminalia arjuna* | Bark | CID_5281855 | 771 | BindingDB \| ChEMBL \| NPASS |
| *Terminalia arjuna* | Bark | CID_5281855 | 861 | ChEMBL |
| *Terminalia arjuna* | Bark | CID_5281855 | 865 | ChEMBL \| NPASS |
| *Terminalia arjuna* | Bark | CID_5281855 | 8877 | ChEMBL |
| *Terminalia arjuna* | Bark | CID_5281855 | 890 | ChEMBL |
| *Terminalia arjuna* | Bark | CID_5281855 | 8900 | ChEMBL \| NPASS |
| *Terminalia arjuna* | Bark | CID_5281855 | 9099 | ChEMBL \| NPASS |
| *Terminalia arjuna* | Bark | CID_5281855 | 9212 | BindingDB \| ChEMBL \| NPASS |
| *Terminalia arjuna* | Bark | CID_5281855 | 9356 | BindingDB \| ChEMBL \| NPASS |
| *Terminalia arjuna* | Bark | CID_5281855 | 9682 | ChEMBL \| NPASS |
| *Terminalia arjuna* | Bark | CID_5281855 | 9891 | BindingDB \| ChEMBL \| NPASS |
| *Terminalia arjuna* | Bark | CID_289 | 10280 | ChEMBL |
| *Terminalia arjuna* | Bark | CID_289 | 10800 | ChEMBL |
| *Terminalia arjuna* | Bark | CID_289 | 10919 | ChEMBL \| NPASS |
| *Terminalia arjuna* | Bark | CID_289 | 10951 | NPASS |
| *Terminalia arjuna* | Bark | CID_289 | 11201 | ChEMBL \| NPASS |
| *Terminalia arjuna* | Bark | CID_289 | 11238 | BindingDB \| ChEMBL \| NPASS |
| *Terminalia arjuna* | Bark | CID_289 | 1128 | ChEMBL |
| *Terminalia arjuna* | Bark | CID_289 | 1129 | ChEMBL |
| *Terminalia arjuna* | Bark | CID_289 | 1131 | ChEMBL |
| *Terminalia arjuna* | Bark | CID_289 | 1132 | ChEMBL |
| *Terminalia arjuna* | Bark | CID_289 | 1133 | ChEMBL |
| *Terminalia arjuna* | Bark | CID_289 | 1233 | ChEMBL |
| *Terminalia arjuna* | Bark | CID_289 | 1234 | ChEMBL |
| *Terminalia arjuna* | Bark | CID_289 | 1268 | ChEMBL |
| *Terminalia arjuna* | Bark | CID_289 | 134 | ChEMBL |
| *Terminalia arjuna* | Bark | CID_289 | 135 | ChEMBL |
| *Terminalia arjuna* | Bark | CID_289 | 140 | ChEMBL |
| *Terminalia arjuna* | Bark | CID_289 | 1432 | ChEMBL |
| *Terminalia arjuna* | Bark | CID_289 | 146 | ChEMBL |
| *Terminalia arjuna* | Bark | CID_289 | 150 | ChEMBL |
| *Terminalia arjuna* | Bark | CID_289 | 151 | ChEMBL |
| *Terminalia arjuna* | Bark | CID_289 | 1511 | ChEMBL |
| *Terminalia arjuna* | Bark | CID_289 | 152 | ChEMBL |
| *Terminalia arjuna* | Bark | CID_289 | 153 | ChEMBL |
| *Terminalia arjuna* | Bark | CID_289 | 154 | ChEMBL |
| *Terminalia arjuna* | Bark | CID_289 | 1544 | ChEMBL |
| *Terminalia arjuna* | Bark | CID_289 | 1548 | ChEMBL |
| *Terminalia arjuna* | Bark | CID_289 | 155 | ChEMBL |
| *Terminalia arjuna* | Bark | CID_289 | 1557 | ChEMBL |
| *Terminalia arjuna* | Bark | CID_289 | 1559 | ChEMBL |
| *Terminalia arjuna* | Bark | CID_289 | 1565 | ChEMBL |
| *Terminalia arjuna* | Bark | CID_289 | 1571 | ChEMBL |
| *Terminalia arjuna* | Bark | CID_289 | 1576 | ChEMBL \| NPASS |
| *Terminalia arjuna* | Bark | CID_289 | 1812 | ChEMBL |
| *Terminalia arjuna* | Bark | CID_289 | 1813 | ChEMBL |
| *Terminalia arjuna* | Bark | CID_289 | 1814 | ChEMBL |
| *Terminalia arjuna* | Bark | CID_289 | 1815 | ChEMBL |
| *Terminalia arjuna* | Bark | CID_289 | 186 | ChEMBL |
| *Terminalia arjuna* | Bark | CID_289 | 1909 | ChEMBL |
| *Terminalia arjuna* | Bark | CID_289 | 1956 | ChEMBL \| NPASS |
| *Terminalia arjuna* | Bark | CID_289 | 1991 | ChEMBL |
| *Terminalia arjuna* | Bark | CID_289 | 2064 | ChEMBL |
| *Terminalia arjuna* | Bark | CID_289 | 2099 | ChEMBL \| NPASS |
| *Terminalia arjuna* | Bark | CID_289 | 2100 | ChEMBL |
| *Terminalia arjuna* | Bark | CID_289 | 216 | ChEMBL \| NPASS |
| *Terminalia arjuna* | Bark | CID_289 | 2237 | ChEMBL \| NPASS |
| *Terminalia arjuna* | Bark | CID_289 | 2321 | ChEMBL |
| *Terminalia arjuna* | Bark | CID_289 | 23632 | BindingDB \| ChEMBL \| NPASS |
| *Terminalia arjuna* | Bark | CID_289 | 240 | ChEMBL \| NPASS |
| *Terminalia arjuna* | Bark | CID_289 | 246 | ChEMBL \| NPASS |
| *Terminalia arjuna* | Bark | CID_289 | 247 | NPASS |
| *Terminalia arjuna* | Bark | CID_289 | 2475 | ChEMBL \| NPASS |
| *Terminalia arjuna* | Bark | CID_289 | 2534 | ChEMBL \| NPASS |
| *Terminalia arjuna* | Bark | CID_289 | 25797 | BindingDB \| ChEMBL |
| *Terminalia arjuna* | Bark | CID_289 | 26013 | ChEMBL \| NPASS |
| *Terminalia arjuna* | Bark | CID_289 | 2744 | NPASS |
| *Terminalia arjuna* | Bark | CID_289 | 2908 | ChEMBL \| NPASS |
| *Terminalia arjuna* | Bark | CID_289 | 3028 | ChEMBL \| NPASS |
| *Terminalia arjuna* | Bark | CID_289 | 3091 | ChEMBL \| NPASS |
| *Terminalia arjuna* | Bark | CID_289 | 3156 | ChEMBL |
| *Terminalia arjuna* | Bark | CID_289 | 3248 | ChEMBL \| NPASS |
| *Terminalia arjuna* | Bark | CID_289 | 3269 | ChEMBL |
| *Terminalia arjuna* | Bark | CID_289 | 3274 | ChEMBL |
| *Terminalia arjuna* | Bark | CID_289 | 328 | ChEMBL \| NPASS |
| *Terminalia arjuna* | Bark | CID_289 | 3309 | NPASS |
| *Terminalia arjuna* | Bark | CID_289 | 3315 | NPASS |
| *Terminalia arjuna* | Bark | CID_289 | 3356 | ChEMBL |
| *Terminalia arjuna* | Bark | CID_289 | 3357 | ChEMBL |
| *Terminalia arjuna* | Bark | CID_289 | 3358 | ChEMBL |
| *Terminalia arjuna* | Bark | CID_289 | 3362 | ChEMBL |
| *Terminalia arjuna* | Bark | CID_289 | 3417 | NPASS |
| *Terminalia arjuna* | Bark | CID_289 | 3576 | NPASS |
| *Terminalia arjuna* | Bark | CID_289 | 3577 | ChEMBL |
| *Terminalia arjuna* | Bark | CID_289 | 3579 | ChEMBL |
| *Terminalia arjuna* | Bark | CID_289 | 3620 | ChEMBL \| NPASS |
| *Terminalia arjuna* | Bark | CID_289 | 367 | NPASS |
| *Terminalia arjuna* | Bark | CID_289 | 3725 | NPASS |
| *Terminalia arjuna* | Bark | CID_289 | 3757 | ChEMBL |
| *Terminalia arjuna* | Bark | CID_289 | 390245 | ChEMBL \| NPASS |
| *Terminalia arjuna* | Bark | CID_289 | 3932 | ChEMBL |
| *Terminalia arjuna* | Bark | CID_289 | 4000 | NPASS |
| *Terminalia arjuna* | Bark | CID_289 | 4128 | ChEMBL |
| *Terminalia arjuna* | Bark | CID_289 | 4137 | NPASS |
| *Terminalia arjuna* | Bark | CID_289 | 4159 | ChEMBL |
| *Terminalia arjuna* | Bark | CID_289 | 4160 | ChEMBL |
| *Terminalia arjuna* | Bark | CID_289 | 4161 | ChEMBL |
| *Terminalia arjuna* | Bark | CID_289 | 4193 | BindingDB \| ChEMBL \| NPASS |
| *Terminalia arjuna* | Bark | CID_289 | 43 | ChEMBL |
| *Terminalia arjuna* | Bark | CID_289 | 4312 | ChEMBL \| NPASS |
| *Terminalia arjuna* | Bark | CID_289 | 4313 | ChEMBL \| NPASS |
| *Terminalia arjuna* | Bark | CID_289 | 4314 | ChEMBL \| NPASS |
| *Terminalia arjuna* | Bark | CID_289 | 4317 | ChEMBL \| NPASS |
| *Terminalia arjuna* | Bark | CID_289 | 4318 | ChEMBL \| NPASS |
| *Terminalia arjuna* | Bark | CID_289 | 4780 | ChEMBL \| NPASS |
| *Terminalia arjuna* | Bark | CID_289 | 4790 | NPASS |
| *Terminalia arjuna* | Bark | CID_289 | 4864 | NPASS |
| *Terminalia arjuna* | Bark | CID_289 | 4886 | ChEMBL |
| *Terminalia arjuna* | Bark | CID_289 | 4887 | ChEMBL |
| *Terminalia arjuna* | Bark | CID_289 | 4985 | ChEMBL |
| *Terminalia arjuna* | Bark | CID_289 | 4986 | ChEMBL |
| *Terminalia arjuna* | Bark | CID_289 | 4988 | ChEMBL |
| *Terminalia arjuna* | Bark | CID_289 | 51053 | NPASS |
| *Terminalia arjuna* | Bark | CID_289 | 51426 | ChEMBL \| NPASS |
| *Terminalia arjuna* | Bark | CID_289 | 5290 | BindingDB \| ChEMBL \| NPASS |
| *Terminalia arjuna* | Bark | CID_289 | 5300 | NPASS |
| *Terminalia arjuna* | Bark | CID_289 | 5376 | NPASS |
| *Terminalia arjuna* | Bark | CID_289 | 5423 | ChEMBL \| NPASS |
| *Terminalia arjuna* | Bark | CID_289 | 5429 | NPASS |
| *Terminalia arjuna* | Bark | CID_289 | 5467 | NPASS |
| *Terminalia arjuna* | Bark | CID_289 | 5468 | NPASS |
| *Terminalia arjuna* | Bark | CID_289 | 54737 | ChEMBL \| NPASS |
| *Terminalia arjuna* | Bark | CID_289 | 552 | ChEMBL |
| *Terminalia arjuna* | Bark | CID_289 | 5530 | ChEMBL |
| *Terminalia arjuna* | Bark | CID_289 | 55775 | NPASS |
| *Terminalia arjuna* | Bark | CID_289 | 5578 | ChEMBL |
| *Terminalia arjuna* | Bark | CID_289 | 5594 | ChEMBL |
| *Terminalia arjuna* | Bark | CID_289 | 5595 | ChEMBL |
| *Terminalia arjuna* | Bark | CID_289 | 5724 | ChEMBL |
| *Terminalia arjuna* | Bark | CID_289 | 5742 | ChEMBL |
| *Terminalia arjuna* | Bark | CID_289 | 5743 | ChEMBL |
| *Terminalia arjuna* | Bark | CID_289 | 5745 | NPASS |
| *Terminalia arjuna* | Bark | CID_289 | 5788 | ChEMBL |
| *Terminalia arjuna* | Bark | CID_289 | 5965 | ChEMBL \| NPASS |
| *Terminalia arjuna* | Bark | CID_289 | 5999 | NPASS |
| *Terminalia arjuna* | Bark | CID_289 | 60489 | NPASS |
| *Terminalia arjuna* | Bark | CID_289 | 6097 | ChEMBL \| NPASS |
| *Terminalia arjuna* | Bark | CID_289 | 624 | ChEMBL |
| *Terminalia arjuna* | Bark | CID_289 | 6311 | NPASS |
| *Terminalia arjuna* | Bark | CID_289 | 641 | NPASS |
| *Terminalia arjuna* | Bark | CID_289 | 6530 | ChEMBL |
| *Terminalia arjuna* | Bark | CID_289 | 6531 | ChEMBL |
| *Terminalia arjuna* | Bark | CID_289 | 6532 | ChEMBL |
| *Terminalia arjuna* | Bark | CID_289 | 6606 | NPASS |
| *Terminalia arjuna* | Bark | CID_289 | 6865 | ChEMBL |
| *Terminalia arjuna* | Bark | CID_289 | 6869 | ChEMBL |
| *Terminalia arjuna* | Bark | CID_289 | 6916 | ChEMBL |
| *Terminalia arjuna* | Bark | CID_289 | 7066 | NPASS |
| *Terminalia arjuna* | Bark | CID_289 | 729230 | ChEMBL |
| *Terminalia arjuna* | Bark | CID_289 | 7433 | ChEMBL |
| *Terminalia arjuna* | Bark | CID_289 | 7498 | ChEMBL \| NPASS |
| *Terminalia arjuna* | Bark | CID_289 | 759 | BindingDB \| ChEMBL \| NPASS |
| *Terminalia arjuna* | Bark | CID_289 | 760 | BindingDB \| ChEMBL \| NPASS |
| *Terminalia arjuna* | Bark | CID_289 | 761 | BindingDB \| ChEMBL \| NPASS |
| *Terminalia arjuna* | Bark | CID_289 | 762 | BindingDB \| ChEMBL \| NPASS |
| *Terminalia arjuna* | Bark | CID_289 | 763 | BindingDB \| ChEMBL \| NPASS |
| *Terminalia arjuna* | Bark | CID_289 | 765 | NPASS |
| *Terminalia arjuna* | Bark | CID_289 | 766 | BindingDB \| NPASS |
| *Terminalia arjuna* | Bark | CID_289 | 768 | BindingDB \| NPASS |
| *Terminalia arjuna* | Bark | CID_289 | 771 | BindingDB \| ChEMBL \| NPASS |
| *Terminalia arjuna* | Bark | CID_289 | 799 | ChEMBL |
| *Terminalia arjuna* | Bark | CID_289 | 834 | ChEMBL |
| *Terminalia arjuna* | Bark | CID_289 | 8654 | ChEMBL |
| *Terminalia arjuna* | Bark | CID_289 | 886 | ChEMBL |
| *Terminalia arjuna* | Bark | CID_289 | 9367 | NPASS |
| *Terminalia arjuna* | Bark | CID_289 | 9682 | ChEMBL \| NPASS |
| *Terminalia arjuna* | Bark | CID_289 | 9971 | NPASS |
| *Terminalia arjuna* | Bark | CID_6251 | 10280 | ChEMBL |
| *Terminalia arjuna* | Bark | CID_6251 | 10800 | ChEMBL |
| *Terminalia arjuna* | Bark | CID_6251 | 11255 | ChEMBL |
| *Terminalia arjuna* | Bark | CID_6251 | 1128 | ChEMBL |
| *Terminalia arjuna* | Bark | CID_6251 | 1129 | ChEMBL |
| *Terminalia arjuna* | Bark | CID_6251 | 1131 | ChEMBL |
| *Terminalia arjuna* | Bark | CID_6251 | 1132 | ChEMBL |
| *Terminalia arjuna* | Bark | CID_6251 | 1133 | ChEMBL |
| *Terminalia arjuna* | Bark | CID_6251 | 1233 | ChEMBL |
| *Terminalia arjuna* | Bark | CID_6251 | 1234 | ChEMBL |
| *Terminalia arjuna* | Bark | CID_6251 | 1268 | ChEMBL |
| *Terminalia arjuna* | Bark | CID_6251 | 134 | ChEMBL |
| *Terminalia arjuna* | Bark | CID_6251 | 135 | ChEMBL |
| *Terminalia arjuna* | Bark | CID_6251 | 140 | ChEMBL |
| *Terminalia arjuna* | Bark | CID_6251 | 1432 | ChEMBL |
| *Terminalia arjuna* | Bark | CID_6251 | 146 | ChEMBL |
| *Terminalia arjuna* | Bark | CID_6251 | 148 | ChEMBL |
| *Terminalia arjuna* | Bark | CID_6251 | 150 | ChEMBL |
| *Terminalia arjuna* | Bark | CID_6251 | 151 | ChEMBL |
| *Terminalia arjuna* | Bark | CID_6251 | 1511 | ChEMBL |
| *Terminalia arjuna* | Bark | CID_6251 | 152 | ChEMBL |
| *Terminalia arjuna* | Bark | CID_6251 | 153 | ChEMBL |
| *Terminalia arjuna* | Bark | CID_6251 | 154 | ChEMBL |
| *Terminalia arjuna* | Bark | CID_6251 | 1544 | ChEMBL |
| *Terminalia arjuna* | Bark | CID_6251 | 1548 | ChEMBL |
| *Terminalia arjuna* | Bark | CID_6251 | 155 | ChEMBL |
| *Terminalia arjuna* | Bark | CID_6251 | 1557 | ChEMBL |
| *Terminalia arjuna* | Bark | CID_6251 | 1559 | ChEMBL |
| *Terminalia arjuna* | Bark | CID_6251 | 1565 | ChEMBL |
| *Terminalia arjuna* | Bark | CID_6251 | 1571 | ChEMBL |
| *Terminalia arjuna* | Bark | CID_6251 | 1576 | ChEMBL |
| *Terminalia arjuna* | Bark | CID_6251 | 1812 | ChEMBL |
| *Terminalia arjuna* | Bark | CID_6251 | 1813 | ChEMBL |
| *Terminalia arjuna* | Bark | CID_6251 | 1814 | ChEMBL |
| *Terminalia arjuna* | Bark | CID_6251 | 1815 | ChEMBL |
| *Terminalia arjuna* | Bark | CID_6251 | 186 | ChEMBL |
| *Terminalia arjuna* | Bark | CID_6251 | 1909 | ChEMBL |
| *Terminalia arjuna* | Bark | CID_6251 | 1956 | ChEMBL |
| *Terminalia arjuna* | Bark | CID_6251 | 1991 | ChEMBL |
| *Terminalia arjuna* | Bark | CID_6251 | 2064 | ChEMBL |
| *Terminalia arjuna* | Bark | CID_6251 | 2099 | ChEMBL \| NPASS |
| *Terminalia arjuna* | Bark | CID_6251 | 2100 | ChEMBL |
| *Terminalia arjuna* | Bark | CID_6251 | 2147 | ChEMBL |
| *Terminalia arjuna* | Bark | CID_6251 | 2321 | ChEMBL |
| *Terminalia arjuna* | Bark | CID_6251 | 23435 | NPASS |
| *Terminalia arjuna* | Bark | CID_6251 | 2534 | ChEMBL |
| *Terminalia arjuna* | Bark | CID_6251 | 2554 | ChEMBL |
| *Terminalia arjuna* | Bark | CID_6251 | 2555 | ChEMBL |
| *Terminalia arjuna* | Bark | CID_6251 | 2561 | ChEMBL |
| *Terminalia arjuna* | Bark | CID_6251 | 2566 | ChEMBL |
| *Terminalia arjuna* | Bark | CID_6251 | 26013 | NPASS |
| *Terminalia arjuna* | Bark | CID_6251 | 2908 | ChEMBL \| NPASS |
| *Terminalia arjuna* | Bark | CID_6251 | 3156 | ChEMBL |
| *Terminalia arjuna* | Bark | CID_6251 | 3269 | ChEMBL |
| *Terminalia arjuna* | Bark | CID_6251 | 3274 | ChEMBL |
| *Terminalia arjuna* | Bark | CID_6251 | 3350 | ChEMBL |
| *Terminalia arjuna* | Bark | CID_6251 | 3356 | ChEMBL |
| *Terminalia arjuna* | Bark | CID_6251 | 3357 | ChEMBL |
| *Terminalia arjuna* | Bark | CID_6251 | 3358 | ChEMBL |
| *Terminalia arjuna* | Bark | CID_6251 | 3362 | ChEMBL |
| *Terminalia arjuna* | Bark | CID_6251 | 3577 | ChEMBL |
| *Terminalia arjuna* | Bark | CID_6251 | 3579 | ChEMBL |
| *Terminalia arjuna* | Bark | CID_6251 | 367 | ChEMBL \| NPASS |
| *Terminalia arjuna* | Bark | CID_6251 | 3757 | ChEMBL |
| *Terminalia arjuna* | Bark | CID_6251 | 3791 | ChEMBL |
| *Terminalia arjuna* | Bark | CID_6251 | 3932 | ChEMBL |
| *Terminalia arjuna* | Bark | CID_6251 | 4000 | NPASS |
| *Terminalia arjuna* | Bark | CID_6251 | 4128 | ChEMBL |
| *Terminalia arjuna* | Bark | CID_6251 | 4159 | ChEMBL |
| *Terminalia arjuna* | Bark | CID_6251 | 4160 | ChEMBL |
| *Terminalia arjuna* | Bark | CID_6251 | 4161 | ChEMBL |
| *Terminalia arjuna* | Bark | CID_6251 | 43 | ChEMBL |
| *Terminalia arjuna* | Bark | CID_6251 | 4312 | ChEMBL |
| *Terminalia arjuna* | Bark | CID_6251 | 4318 | ChEMBL |
| *Terminalia arjuna* | Bark | CID_6251 | 4780 | NPASS |
| *Terminalia arjuna* | Bark | CID_6251 | 4886 | ChEMBL |
| *Terminalia arjuna* | Bark | CID_6251 | 4887 | ChEMBL |
| *Terminalia arjuna* | Bark | CID_6251 | 49 | ChEMBL \| NPASS |
| *Terminalia arjuna* | Bark | CID_6251 | 4985 | ChEMBL |
| *Terminalia arjuna* | Bark | CID_6251 | 4986 | ChEMBL |
| *Terminalia arjuna* | Bark | CID_6251 | 4988 | ChEMBL |
| *Terminalia arjuna* | Bark | CID_6251 | 51053 | NPASS |
| *Terminalia arjuna* | Bark | CID_6251 | 5139 | ChEMBL |
| *Terminalia arjuna* | Bark | CID_6251 | 5141 | ChEMBL |
| *Terminalia arjuna* | Bark | CID_6251 | 5241 | ChEMBL |
| *Terminalia arjuna* | Bark | CID_6251 | 5243 | ChEMBL \| NPASS |
| *Terminalia arjuna* | Bark | CID_6251 | 552 | ChEMBL |
| *Terminalia arjuna* | Bark | CID_6251 | 5530 | ChEMBL |
| *Terminalia arjuna* | Bark | CID_6251 | 5578 | ChEMBL |
| *Terminalia arjuna* | Bark | CID_6251 | 5594 | ChEMBL |
| *Terminalia arjuna* | Bark | CID_6251 | 5595 | ChEMBL |
| *Terminalia arjuna* | Bark | CID_6251 | 5724 | ChEMBL |
| *Terminalia arjuna* | Bark | CID_6251 | 5742 | ChEMBL |
| *Terminalia arjuna* | Bark | CID_6251 | 5743 | ChEMBL |
| *Terminalia arjuna* | Bark | CID_6251 | 5788 | ChEMBL |
| *Terminalia arjuna* | Bark | CID_6251 | 624 | ChEMBL |
| *Terminalia arjuna* | Bark | CID_6251 | 6530 | ChEMBL |
| *Terminalia arjuna* | Bark | CID_6251 | 6531 | ChEMBL |
| *Terminalia arjuna* | Bark | CID_6251 | 6532 | ChEMBL |
| *Terminalia arjuna* | Bark | CID_6251 | 6580 | ChEMBL \| NPASS |
| *Terminalia arjuna* | Bark | CID_6251 | 6865 | ChEMBL |
| *Terminalia arjuna* | Bark | CID_6251 | 6869 | ChEMBL |
| *Terminalia arjuna* | Bark | CID_6251 | 6915 | ChEMBL |
| *Terminalia arjuna* | Bark | CID_6251 | 6916 | ChEMBL |
| *Terminalia arjuna* | Bark | CID_6251 | 729230 | ChEMBL |
| *Terminalia arjuna* | Bark | CID_6251 | 7398 | NPASS |
| *Terminalia arjuna* | Bark | CID_6251 | 7433 | ChEMBL |
| *Terminalia arjuna* | Bark | CID_6251 | 760 | ChEMBL |
| *Terminalia arjuna* | Bark | CID_6251 | 799 | ChEMBL |
| *Terminalia arjuna* | Bark | CID_6251 | 834 | ChEMBL |
| *Terminalia arjuna* | Bark | CID_6251 | 8654 | ChEMBL |
| *Terminalia arjuna* | Bark | CID_6251 | 886 | ChEMBL |
| *Terminalia arjuna* | Bark | CID_6251 | 9429 | ChEMBL \| NPASS |
| *Terminalia arjuna* | Bark | CID_72277 | 10599 | ChEMBL \| NPASS |
| *Terminalia arjuna* | Bark | CID_72277 | 11309 | ChEMBL |
| *Terminalia arjuna* | Bark | CID_72277 | 118429 | NPASS |
| *Terminalia arjuna* | Bark | CID_72277 | 1268 | BindingDB \| ChEMBL \| NPASS |
| *Terminalia arjuna* | Bark | CID_72277 | 1269 | BindingDB \| NPASS |
| *Terminalia arjuna* | Bark | CID_72277 | 213 | ChEMBL |
| *Terminalia arjuna* | Bark | CID_72277 | 28234 | ChEMBL \| NPASS |
| *Terminalia arjuna* | Bark | CID_72277 | 4985 | ChEMBL \| NPASS |
| *Terminalia arjuna* | Bark | CID_72277 | 4986 | ChEMBL \| NPASS |
| *Terminalia arjuna* | Bark | CID_72277 | 4988 | ChEMBL \| NPASS |
| *Terminalia arjuna* | Bark | CID_72277 | 5226 | BindingDB \| NPASS |
| *Terminalia arjuna* | Bark | CID_72277 | 5693 | BindingDB \| NPASS |
| *Terminalia arjuna* | Bark | CID_72277 | 5893 | ChEMBL \| NPASS |
| *Terminalia arjuna* | Bark | CID_72277 | 596 | ChEMBL \| NPASS |
| *Terminalia arjuna* | Bark | CID_72277 | 65220 | BindingDB \| ChEMBL |
| *Terminalia arjuna* | Bark | CID_72277 | 6622 | ChEMBL \| NPASS |
| *Terminalia arjuna* | Bark | CID_72277 | 7257 | ChEMBL \| NPASS |
| *Terminalia arjuna* | Bark | CID_72277 | 7498 | BindingDB \| ChEMBL \| NPASS |
| *Terminalia arjuna* | Bark | CID_10494 | 10599 | ChEMBL \| NPASS |
| *Terminalia arjuna* | Bark | CID_10494 | 1066 | ChEMBL \| NPASS |
| *Terminalia arjuna* | Bark | CID_10494 | 1071 | ChEMBL \| NPASS |
| *Terminalia arjuna* | Bark | CID_10494 | 10951 | NPASS |
| *Terminalia arjuna* | Bark | CID_10494 | 11069 | NPASS |
| *Terminalia arjuna* | Bark | CID_10494 | 112398 | ChEMBL \| NPASS |
| *Terminalia arjuna* | Bark | CID_10494 | 112399 | ChEMBL \| NPASS |
| *Terminalia arjuna* | Bark | CID_10494 | 11343 | ChEMBL \| NPASS |
| *Terminalia arjuna* | Bark | CID_10494 | 151306 | BindingDB \| ChEMBL \| NPASS |
| *Terminalia arjuna* | Bark | CID_10494 | 1588 | ChEMBL \| NPASS |
| *Terminalia arjuna* | Bark | CID_10494 | 1845 | BindingDB \| ChEMBL \| NPASS |
| *Terminalia arjuna* | Bark | CID_10494 | 1956 | BindingDB \| ChEMBL \| NPASS |
| *Terminalia arjuna* | Bark | CID_10494 | 2152 | BindingDB \| ChEMBL \| NPASS |
| *Terminalia arjuna* | Bark | CID_10494 | 231 | NPASS |
| *Terminalia arjuna* | Bark | CID_10494 | 23408 | ChEMBL |
| *Terminalia arjuna* | Bark | CID_10494 | 240 | ChEMBL \| NPASS |
| *Terminalia arjuna* | Bark | CID_10494 | 247 | ChEMBL \| NPASS |
| *Terminalia arjuna* | Bark | CID_10494 | 2648 | NPASS |
| *Terminalia arjuna* | Bark | CID_10494 | 28234 | ChEMBL \| NPASS |
| *Terminalia arjuna* | Bark | CID_10494 | 2932 | ChEMBL |
| *Terminalia arjuna* | Bark | CID_10494 | 3294 | ChEMBL \| NPASS |
| *Terminalia arjuna* | Bark | CID_10494 | 3417 | ChEMBL \| NPASS |
| *Terminalia arjuna* | Bark | CID_10494 | 351 | ChEMBL \| NPASS |
| *Terminalia arjuna* | Bark | CID_10494 | 3978 | BindingDB \| NPASS |
| *Terminalia arjuna* | Bark | CID_10494 | 4780 | ChEMBL \| NPASS |
| *Terminalia arjuna* | Bark | CID_10494 | 4843 | BindingDB \| ChEMBL \| NPASS |
| *Terminalia arjuna* | Bark | CID_10494 | 51053 | NPASS |
| *Terminalia arjuna* | Bark | CID_10494 | 5111 | ChEMBL |
| *Terminalia arjuna* | Bark | CID_10494 | 51422 | ChEMBL |
| *Terminalia arjuna* | Bark | CID_10494 | 52 | ChEMBL \| NPASS |
| *Terminalia arjuna* | Bark | CID_10494 | 5243 | ChEMBL \| NPASS |
| *Terminalia arjuna* | Bark | CID_10494 | 5319 | BindingDB \| ChEMBL \| NPASS |
| *Terminalia arjuna* | Bark | CID_10494 | 5347 | NPASS |
| *Terminalia arjuna* | Bark | CID_10494 | 53632 | ChEMBL |
| *Terminalia arjuna* | Bark | CID_10494 | 5423 | BindingDB \| ChEMBL \| NPASS |
| *Terminalia arjuna* | Bark | CID_10494 | 54583 | BindingDB \| ChEMBL \| NPASS |
| *Terminalia arjuna* | Bark | CID_10494 | 54681 | ChEMBL \| NPASS |
| *Terminalia arjuna* | Bark | CID_10494 | 5562 | ChEMBL |
| *Terminalia arjuna* | Bark | CID_10494 | 5563 | ChEMBL |
| *Terminalia arjuna* | Bark | CID_10494 | 5564 | ChEMBL |
| *Terminalia arjuna* | Bark | CID_10494 | 5565 | ChEMBL |
| *Terminalia arjuna* | Bark | CID_10494 | 5571 | ChEMBL |
| *Terminalia arjuna* | Bark | CID_10494 | 55775 | ChEMBL \| NPASS |
| *Terminalia arjuna* | Bark | CID_10494 | 57016 | ChEMBL \| NPASS |
| *Terminalia arjuna* | Bark | CID_10494 | 5742 | NPASS |
| *Terminalia arjuna* | Bark | CID_10494 | 5743 | BindingDB \| ChEMBL \| NPASS |
| *Terminalia arjuna* | Bark | CID_10494 | 5770 | BindingDB \| ChEMBL \| NPASS |
| *Terminalia arjuna* | Bark | CID_10494 | 5771 | BindingDB \| ChEMBL \| NPASS |
| *Terminalia arjuna* | Bark | CID_10494 | 5777 | BindingDB \| ChEMBL \| NPASS |
| *Terminalia arjuna* | Bark | CID_10494 | 5781 | BindingDB \| ChEMBL \| NPASS |
| *Terminalia arjuna* | Bark | CID_10494 | 5786 | ChEMBL \| NPASS |
| *Terminalia arjuna* | Bark | CID_10494 | 5788 | BindingDB \| ChEMBL \| NPASS |
| *Terminalia arjuna* | Bark | CID_10494 | 5791 | ChEMBL \| NPASS |
| *Terminalia arjuna* | Bark | CID_10494 | 5792 | BindingDB \| ChEMBL \| NPASS |
| *Terminalia arjuna* | Bark | CID_10494 | 5970 | ChEMBL \| NPASS |
| *Terminalia arjuna* | Bark | CID_10494 | 6774 | ChEMBL \| NPASS |
| *Terminalia arjuna* | Bark | CID_10494 | 7150 | ChEMBL \| NPASS |
| *Terminalia arjuna* | Bark | CID_10494 | 7153 | ChEMBL \| NPASS |
| *Terminalia arjuna* | Bark | CID_10494 | 7428 | ChEMBL |
| *Terminalia arjuna* | Bark | CID_10494 | 7442 | ChEMBL \| NPASS |
| *Terminalia arjuna* | Bark | CID_10494 | 7874 | BindingDB \| ChEMBL |
| *Terminalia arjuna* | Bark | CID_10494 | 79915 | NPASS |
| *Terminalia arjuna* | Bark | CID_10494 | 994 | ChEMBL \| NPASS |
| *Terminalia arjuna* | Bark | CID_10494 | 9971 | NPASS |
| *Terminalia arjuna* | Bark | CID_73641 | 151306 | BindingDB \| ChEMBL \| NPASS |
| *Terminalia arjuna* | Bark | CID_73641 | 240 | BindingDB \| ChEMBL \| NPASS |
| *Terminalia arjuna* | Bark | CID_73641 | 247 | ChEMBL \| NPASS |
| *Terminalia arjuna* | Bark | CID_73641 | 5743 | BindingDB \| ChEMBL \| NPASS |
| *Terminalia arjuna* | Bark | CID_73641 | 9971 | NPASS |
| *Terminalia arjuna* | Bark | CID_14034812 | 4193 | ChEMBL |
| *Terminalia arjuna* | Bark | CID_14034812 | 7157 | ChEMBL |
| *Terminalia arjuna* | Bark | CID_5281605 | 10013 | ChEMBL |
| *Terminalia arjuna* | Bark | CID_5281605 | 100861540 | ChEMBL |
| *Terminalia arjuna* | Bark | CID_5281605 | 10411 | NPASS |
| *Terminalia arjuna* | Bark | CID_5281605 | 10599 | ChEMBL \| NPASS |
| *Terminalia arjuna* | Bark | CID_5281605 | 10769 | BindingDB |
| *Terminalia arjuna* | Bark | CID_5281605 | 11201 | NPASS |
| *Terminalia arjuna* | Bark | CID_5281605 | 11283 | ChEMBL |
| *Terminalia arjuna* | Bark | CID_5281605 | 113612 | ChEMBL |
| *Terminalia arjuna* | Bark | CID_5281605 | 120227 | ChEMBL |
| *Terminalia arjuna* | Bark | CID_5281605 | 1263 | BindingDB \| ChEMBL |
| *Terminalia arjuna* | Bark | CID_5281605 | 1385 | ChEMBL \| NPASS |
| *Terminalia arjuna* | Bark | CID_5281605 | 1457 | ChEMBL \| NPASS |
| *Terminalia arjuna* | Bark | CID_5281605 | 1459 | ChEMBL \| NPASS |
| *Terminalia arjuna* | Bark | CID_5281605 | 1460 | ChEMBL \| NPASS |
| *Terminalia arjuna* | Bark | CID_5281605 | 1543 | BindingDB \| ChEMBL \| NPASS |
| *Terminalia arjuna* | Bark | CID_5281605 | 1544 | ChEMBL \| NPASS |
| *Terminalia arjuna* | Bark | CID_5281605 | 1545 | BindingDB \| ChEMBL \| NPASS |
| *Terminalia arjuna* | Bark | CID_5281605 | 1548 | ChEMBL |
| *Terminalia arjuna* | Bark | CID_5281605 | 1549 | ChEMBL |
| *Terminalia arjuna* | Bark | CID_5281605 | 1553 | ChEMBL |
| *Terminalia arjuna* | Bark | CID_5281605 | 1555 | ChEMBL |
| *Terminalia arjuna* | Bark | CID_5281605 | 1557 | ChEMBL \| NPASS |
| *Terminalia arjuna* | Bark | CID_5281605 | 1558 | ChEMBL |
| *Terminalia arjuna* | Bark | CID_5281605 | 1559 | ChEMBL \| NPASS |
| *Terminalia arjuna* | Bark | CID_5281605 | 1562 | ChEMBL |
| *Terminalia arjuna* | Bark | CID_5281605 | 1565 | ChEMBL \| NPASS |
| *Terminalia arjuna* | Bark | CID_5281605 | 1571 | ChEMBL |
| *Terminalia arjuna* | Bark | CID_5281605 | 1572 | ChEMBL |
| *Terminalia arjuna* | Bark | CID_5281605 | 1573 | ChEMBL |
| *Terminalia arjuna* | Bark | CID_5281605 | 1576 | BindingDB \| ChEMBL \| NPASS |
| *Terminalia arjuna* | Bark | CID_5281605 | 1577 | ChEMBL |
| *Terminalia arjuna* | Bark | CID_5281605 | 1579 | ChEMBL |
| *Terminalia arjuna* | Bark | CID_5281605 | 1580 | ChEMBL |
| *Terminalia arjuna* | Bark | CID_5281605 | 1612 | BindingDB \| NPASS |
| *Terminalia arjuna* | Bark | CID_5281605 | 18 | BindingDB \| NPASS |
| *Terminalia arjuna* | Bark | CID_5281605 | 1803 | BindingDB \| NPASS |
| *Terminalia arjuna* | Bark | CID_5281605 | 207 | ChEMBL |
| *Terminalia arjuna* | Bark | CID_5281605 | 2099 | ChEMBL \| NPASS |
| *Terminalia arjuna* | Bark | CID_5281605 | 216 | ChEMBL \| NPASS |
| *Terminalia arjuna* | Bark | CID_5281605 | 2203 | ChEMBL \| NPASS |
| *Terminalia arjuna* | Bark | CID_5281605 | 2237 | ChEMBL \| NPASS |
| *Terminalia arjuna* | Bark | CID_5281605 | 23028 | BindingDB \| ChEMBL \| NPASS |
| *Terminalia arjuna* | Bark | CID_5281605 | 2322 | ChEMBL \| NPASS |
| *Terminalia arjuna* | Bark | CID_5281605 | 239 | BindingDB \| ChEMBL \| NPASS |
| *Terminalia arjuna* | Bark | CID_5281605 | 240 | BindingDB \| ChEMBL \| NPASS |
| *Terminalia arjuna* | Bark | CID_5281605 | 246 | BindingDB \| ChEMBL \| NPASS |
| *Terminalia arjuna* | Bark | CID_5281605 | 247 | BindingDB \| ChEMBL \| NPASS |
| *Terminalia arjuna* | Bark | CID_5281605 | 2548 | BindingDB \| NPASS |
| *Terminalia arjuna* | Bark | CID_5281605 | 2554 | ChEMBL |
| *Terminalia arjuna* | Bark | CID_5281605 | 2555 | ChEMBL |
| *Terminalia arjuna* | Bark | CID_5281605 | 2556 | ChEMBL |
| *Terminalia arjuna* | Bark | CID_5281605 | 2557 | ChEMBL |
| *Terminalia arjuna* | Bark | CID_5281605 | 2558 | ChEMBL |
| *Terminalia arjuna* | Bark | CID_5281605 | 2559 | ChEMBL \| NPASS |
| *Terminalia arjuna* | Bark | CID_5281605 | 2560 | ChEMBL |
| *Terminalia arjuna* | Bark | CID_5281605 | 2561 | ChEMBL |
| *Terminalia arjuna* | Bark | CID_5281605 | 2562 | ChEMBL |
| *Terminalia arjuna* | Bark | CID_5281605 | 2563 | ChEMBL |
| *Terminalia arjuna* | Bark | CID_5281605 | 2564 | ChEMBL |
| *Terminalia arjuna* | Bark | CID_5281605 | 2565 | ChEMBL |
| *Terminalia arjuna* | Bark | CID_5281605 | 2566 | ChEMBL |
| *Terminalia arjuna* | Bark | CID_5281605 | 2567 | ChEMBL |
| *Terminalia arjuna* | Bark | CID_5281605 | 2568 | ChEMBL |
| *Terminalia arjuna* | Bark | CID_5281605 | 25939 | ChEMBL |
| *Terminalia arjuna* | Bark | CID_5281605 | 2597 | BindingDB \| ChEMBL |
| *Terminalia arjuna* | Bark | CID_5281605 | 26013 | ChEMBL \| NPASS |
| *Terminalia arjuna* | Bark | CID_5281605 | 2629 | NPASS |
| *Terminalia arjuna* | Bark | CID_5281605 | 2739 | ChEMBL \| NPASS |
| *Terminalia arjuna* | Bark | CID_5281605 | 28234 | ChEMBL \| NPASS |
| *Terminalia arjuna* | Bark | CID_5281605 | 283106 | ChEMBL \| NPASS |
| *Terminalia arjuna* | Bark | CID_5281605 | 284541 | ChEMBL |
| *Terminalia arjuna* | Bark | CID_5281605 | 2859 | BindingDB \| ChEMBL \| NPASS |
| *Terminalia arjuna* | Bark | CID_5281605 | 2870 | BindingDB \| ChEMBL \| NPASS |
| *Terminalia arjuna* | Bark | CID_5281605 | 2908 | ChEMBL \| NPASS |
| *Terminalia arjuna* | Bark | CID_5281605 | 29785 | ChEMBL |
| *Terminalia arjuna* | Bark | CID_5281605 | 3028 | ChEMBL \| NPASS |
| *Terminalia arjuna* | Bark | CID_5281605 | 3043 | NPASS |
| *Terminalia arjuna* | Bark | CID_5281605 | 3248 | ChEMBL \| NPASS |
| *Terminalia arjuna* | Bark | CID_5281605 | 328 | ChEMBL \| NPASS |
| *Terminalia arjuna* | Bark | CID_5281605 | 3293 | BindingDB \| NPASS |
| *Terminalia arjuna* | Bark | CID_5281605 | 3363 | BindingDB \| NPASS |
| *Terminalia arjuna* | Bark | CID_5281605 | 367 | ChEMBL \| NPASS |
| *Terminalia arjuna* | Bark | CID_5281605 | 3725 | ChEMBL \| NPASS |
| *Terminalia arjuna* | Bark | CID_5281605 | 3741 | ChEMBL |
| *Terminalia arjuna* | Bark | CID_5281605 | 3757 | ChEMBL |
| *Terminalia arjuna* | Bark | CID_5281605 | 390245 | BindingDB \| ChEMBL \| NPASS |
| *Terminalia arjuna* | Bark | CID_5281605 | 4051 | ChEMBL |
| *Terminalia arjuna* | Bark | CID_5281605 | 4137 | ChEMBL \| NPASS |
| *Terminalia arjuna* | Bark | CID_5281605 | 4297 | NPASS |
| *Terminalia arjuna* | Bark | CID_5281605 | 4759 | BindingDB \| NPASS |
| *Terminalia arjuna* | Bark | CID_5281605 | 51053 | NPASS |
| *Terminalia arjuna* | Bark | CID_5281605 | 51422 | ChEMBL |
| *Terminalia arjuna* | Bark | CID_5281605 | 51426 | ChEMBL \| NPASS |
| *Terminalia arjuna* | Bark | CID_5281605 | 5241 | ChEMBL |
| *Terminalia arjuna* | Bark | CID_5281605 | 5243 | ChEMBL \| NPASS |
| *Terminalia arjuna* | Bark | CID_5281605 | 5292 | ChEMBL \| NPASS |
| *Terminalia arjuna* | Bark | CID_5281605 | 5300 | NPASS |
| *Terminalia arjuna* | Bark | CID_5281605 | 5347 | BindingDB \| ChEMBL |
| *Terminalia arjuna* | Bark | CID_5281605 | 53632 | ChEMBL |
| *Terminalia arjuna* | Bark | CID_5281605 | 54205 | ChEMBL \| NPASS |
| *Terminalia arjuna* | Bark | CID_5281605 | 5429 | BindingDB \| ChEMBL \| NPASS |
| *Terminalia arjuna* | Bark | CID_5281605 | 54583 | ChEMBL |
| *Terminalia arjuna* | Bark | CID_5281605 | 54658 | ChEMBL \| NPASS |
| *Terminalia arjuna* | Bark | CID_5281605 | 54905 | ChEMBL |
| *Terminalia arjuna* | Bark | CID_5281605 | 5550 | ChEMBL \| NPASS |
| *Terminalia arjuna* | Bark | CID_5281605 | 5562 | ChEMBL |
| *Terminalia arjuna* | Bark | CID_5281605 | 5563 | ChEMBL |
| *Terminalia arjuna* | Bark | CID_5281605 | 5564 | ChEMBL |
| *Terminalia arjuna* | Bark | CID_5281605 | 5565 | ChEMBL |
| *Terminalia arjuna* | Bark | CID_5281605 | 55662 | ChEMBL |
| *Terminalia arjuna* | Bark | CID_5281605 | 5571 | ChEMBL |
| *Terminalia arjuna* | Bark | CID_5281605 | 55775 | ChEMBL \| NPASS |
| *Terminalia arjuna* | Bark | CID_5281605 | 55879 | ChEMBL |
| *Terminalia arjuna* | Bark | CID_5281605 | 5644 | BindingDB \| ChEMBL |
| *Terminalia arjuna* | Bark | CID_5281605 | 5645 | ChEMBL |
| *Terminalia arjuna* | Bark | CID_5281605 | 5646 | ChEMBL \| NPASS |
| *Terminalia arjuna* | Bark | CID_5281605 | 5802 | ChEMBL \| NPASS |
| *Terminalia arjuna* | Bark | CID_5281605 | 5965 | ChEMBL \| NPASS |
| *Terminalia arjuna* | Bark | CID_5281605 | 6401 | ChEMBL \| NPASS |
| *Terminalia arjuna* | Bark | CID_5281605 | 6402 | ChEMBL \| NPASS |
| *Terminalia arjuna* | Bark | CID_5281605 | 6403 | ChEMBL \| NPASS |
| *Terminalia arjuna* | Bark | CID_5281605 | 641 | ChEMBL \| NPASS |
| *Terminalia arjuna* | Bark | CID_5281605 | 6476 | BindingDB \| ChEMBL \| NPASS |
| *Terminalia arjuna* | Bark | CID_5281605 | 64816 | ChEMBL |
| *Terminalia arjuna* | Bark | CID_5281605 | 6606 | NPASS |
| *Terminalia arjuna* | Bark | CID_5281605 | 6622 | ChEMBL \| NPASS |
| *Terminalia arjuna* | Bark | CID_5281605 | 672 | NPASS |
| *Terminalia arjuna* | Bark | CID_5281605 | 7068 | ChEMBL \| NPASS |
| *Terminalia arjuna* | Bark | CID_5281605 | 7157 | ChEMBL \| NPASS |
| *Terminalia arjuna* | Bark | CID_5281605 | 7366 | ChEMBL \| NPASS |
| *Terminalia arjuna* | Bark | CID_5281605 | 7486 | NPASS |
| *Terminalia arjuna* | Bark | CID_5281605 | 7498 | BindingDB \| ChEMBL \| NPASS |
| *Terminalia arjuna* | Bark | CID_5281605 | 7525 | BindingDB \| ChEMBL \| NPASS |
| *Terminalia arjuna* | Bark | CID_5281605 | 79915 | NPASS |
| *Terminalia arjuna* | Bark | CID_5281605 | 8529 | ChEMBL |
| *Terminalia arjuna* | Bark | CID_5281605 | 865 | NPASS |
| *Terminalia arjuna* | Bark | CID_5281605 | 891 | ChEMBL \| NPASS |
| *Terminalia arjuna* | Bark | CID_5281605 | 8972 | ChEMBL |
| *Terminalia arjuna* | Bark | CID_5281605 | 9099 | NPASS |
| *Terminalia arjuna* | Bark | CID_5281605 | 9212 | BindingDB |
| *Terminalia arjuna* | Bark | CID_5281605 | 9682 | NPASS |
| *Terminalia arjuna* | Bark | CID_5281605 | 983 | BindingDB \| ChEMBL |
| *Tectona grandis* | Wood | CID_222284 | 10013 | ChEMBL |
| *Tectona grandis* | Wood | CID_222284 | 10599 | ChEMBL \| NPASS |
| *Tectona grandis* | Wood | CID_222284 | 1565 | ChEMBL \| NPASS |
| *Tectona grandis* | Wood | CID_222284 | 1576 | BindingDB \| ChEMBL \| NPASS |
| *Tectona grandis* | Wood | CID_222284 | 1803 | ChEMBL \| NPASS |
| *Tectona grandis* | Wood | CID_222284 | 2147 | ChEMBL \| NPASS |
| *Tectona grandis* | Wood | CID_222284 | 28234 | ChEMBL \| NPASS |
| *Tectona grandis* | Wood | CID_222284 | 2932 | ChEMBL |
| *Tectona grandis* | Wood | CID_222284 | 3417 | ChEMBL \| NPASS |
| *Tectona grandis* | Wood | CID_222284 | 51422 | ChEMBL |
| *Tectona grandis* | Wood | CID_222284 | 5243 | ChEMBL |
| *Tectona grandis* | Wood | CID_222284 | 53632 | ChEMBL |
| *Tectona grandis* | Wood | CID_222284 | 5423 | BindingDB \| ChEMBL \| NPASS |
| *Tectona grandis* | Wood | CID_222284 | 5562 | ChEMBL |
| *Tectona grandis* | Wood | CID_222284 | 5563 | ChEMBL |
| *Tectona grandis* | Wood | CID_222284 | 5564 | ChEMBL |
| *Tectona grandis* | Wood | CID_222284 | 5565 | ChEMBL |
| *Tectona grandis* | Wood | CID_222284 | 5571 | ChEMBL |
| *Tectona grandis* | Wood | CID_222284 | 7299 | ChEMBL \| NPASS |
| *Tectona grandis* | Wood | CID_64971 | 10013 | ChEMBL |
| *Tectona grandis* | Wood | CID_64971 | 10054 | ChEMBL \| NPASS |
| *Tectona grandis* | Wood | CID_64971 | 10055 | BindingDB \| ChEMBL |
| *Tectona grandis* | Wood | CID_64971 | 10062 | ChEMBL \| NPASS |
| *Tectona grandis* | Wood | CID_64971 | 10599 | ChEMBL \| NPASS |
| *Tectona grandis* | Wood | CID_64971 | 1071 | ChEMBL \| NPASS |
| *Tectona grandis* | Wood | CID_64971 | 11069 | NPASS |
| *Tectona grandis* | Wood | CID_64971 | 112398 | ChEMBL \| NPASS |
| *Tectona grandis* | Wood | CID_64971 | 112399 | ChEMBL \| NPASS |
| *Tectona grandis* | Wood | CID_64971 | 151306 | BindingDB \| ChEMBL \| NPASS |
| *Tectona grandis* | Wood | CID_64971 | 1728 | ChEMBL \| NPASS |
| *Tectona grandis* | Wood | CID_64971 | 1803 | ChEMBL \| NPASS |
| *Tectona grandis* | Wood | CID_64971 | 200316 | NPASS |
| *Tectona grandis* | Wood | CID_64971 | 213 | BindingDB \| ChEMBL \| NPASS |
| *Tectona grandis* | Wood | CID_64971 | 231 | ChEMBL \| NPASS |
| *Tectona grandis* | Wood | CID_64971 | 2735 | BindingDB \| ChEMBL \| NPASS |
| *Tectona grandis* | Wood | CID_64971 | 2740 | NPASS |
| *Tectona grandis* | Wood | CID_64971 | 28234 | ChEMBL \| NPASS |
| *Tectona grandis* | Wood | CID_64971 | 2932 | ChEMBL |
| *Tectona grandis* | Wood | CID_64971 | 3417 | ChEMBL \| NPASS |
| *Tectona grandis* | Wood | CID_64971 | 4907 | ChEMBL |
| *Tectona grandis* | Wood | CID_64971 | 51053 | NPASS |
| *Tectona grandis* | Wood | CID_64971 | 51422 | ChEMBL |
| *Tectona grandis* | Wood | CID_64971 | 51426 | NPASS |
| *Tectona grandis* | Wood | CID_64971 | 53632 | ChEMBL |
| *Tectona grandis* | Wood | CID_64971 | 5423 | BindingDB \| ChEMBL \| NPASS |
| *Tectona grandis* | Wood | CID_64971 | 5429 | NPASS |
| *Tectona grandis* | Wood | CID_64971 | 54583 | BindingDB \| ChEMBL \| NPASS |
| *Tectona grandis* | Wood | CID_64971 | 54681 | ChEMBL \| NPASS |
| *Tectona grandis* | Wood | CID_64971 | 5562 | ChEMBL |
| *Tectona grandis* | Wood | CID_64971 | 5563 | ChEMBL |
| *Tectona grandis* | Wood | CID_64971 | 5564 | ChEMBL |
| *Tectona grandis* | Wood | CID_64971 | 5565 | ChEMBL |
| *Tectona grandis* | Wood | CID_64971 | 5571 | ChEMBL |
| *Tectona grandis* | Wood | CID_64971 | 55775 | ChEMBL \| NPASS |
| *Tectona grandis* | Wood | CID_64971 | 5579 | NPASS |
| *Tectona grandis* | Wood | CID_64971 | 5581 | NPASS |
| *Tectona grandis* | Wood | CID_64971 | 57016 | ChEMBL \| NPASS |
| *Tectona grandis* | Wood | CID_64971 | 5745 | NPASS |
| *Tectona grandis* | Wood | CID_64971 | 5770 | BindingDB \| ChEMBL |
| *Tectona grandis* | Wood | CID_64971 | 5970 | ChEMBL \| NPASS |
| *Tectona grandis* | Wood | CID_64971 | 5999 | NPASS |
| *Tectona grandis* | Wood | CID_64971 | 60489 | NPASS |
| *Tectona grandis* | Wood | CID_64971 | 7150 | ChEMBL |
| *Tectona grandis* | Wood | CID_64971 | 7376 | ChEMBL \| NPASS |
| *Tectona grandis* | Wood | CID_64971 | 9971 | NPASS |
| *Tectona grandis* | Wood | CID_3884 | 10599 | ChEMBL |
| *Tectona grandis* | Wood | CID_3884 | 1723 | ChEMBL |
| *Tectona grandis* | Wood | CID_3884 | 2739 | ChEMBL |
| *Tectona grandis* | Wood | CID_3884 | 28234 | ChEMBL |
| *Tectona grandis* | Wood | CID_3884 | 2859 | ChEMBL |
| *Tectona grandis* | Wood | CID_3884 | 4137 | ChEMBL |
| *Tectona grandis* | Wood | CID_3884 | 5315 | ChEMBL |
| *Tectona grandis* | Wood | CID_3884 | 6197 | ChEMBL |
| *Tectona grandis* | Wood | CID_3884 | 6606 | ChEMBL |
| *Tectona grandis* | Wood | CID_3884 | 7398 | ChEMBL |
| *Tectona grandis* | Wood | CID_3884 | 7421 | ChEMBL |
| *Tectona grandis* | Wood | CID_3884 | 80854 | ChEMBL |
| *Tectona grandis* | Wood | CID_3884 | 993 | ChEMBL |
| *Tectona grandis* | Wood | CID_3884 | 994 | ChEMBL |
| *Tectona grandis* | Wood | CID_6773 | 1991 | BindingDB \| ChEMBL \| NPASS |
| *Tectona grandis* | Wood | CID_6773 | 367 | NPASS |
| *Tectona grandis* | Wood | CID_6773 | 4780 | NPASS |
| *Tectona grandis* | Wood | CID_931 | 1544 | BindingDB \| NPASS |
| *Tectona grandis* | Wood | CID_931 | 1548 | BindingDB \| ChEMBL \| NPASS |
| *Tectona grandis* | Wood | CID_931 | 239 | NPASS |
| *Tectona grandis* | Wood | CID_931 | 367 | NPASS |
| *Tectona grandis* | Wood | CID_931 | 51053 | NPASS |
| *Tectona grandis* | Wood | CID_931 | 5745 | NPASS |
| *Tectona grandis* | Wood | CID_6780 | 10013 | ChEMBL |
| *Tectona grandis* | Wood | CID_6780 | 10599 | ChEMBL \| NPASS |
| *Tectona grandis* | Wood | CID_6780 | 10919 | ChEMBL \| NPASS |
| *Tectona grandis* | Wood | CID_6780 | 2099 | ChEMBL \| NPASS |
| *Tectona grandis* | Wood | CID_6780 | 2100 | ChEMBL |
| *Tectona grandis* | Wood | CID_6780 | 28234 | ChEMBL \| NPASS |
| *Tectona grandis* | Wood | CID_6780 | 4129 | ChEMBL |
| *Tectona grandis* | Wood | CID_6780 | 4297 | NPASS |
| *Tectona grandis* | Wood | CID_6780 | 4780 | NPASS |
| *Tectona grandis* | Wood | CID_6780 | 5166 | ChEMBL \| NPASS |
| *Tectona grandis* | Wood | CID_6780 | 54576 | ChEMBL |
| *Tectona grandis* | Wood | CID_6780 | 54658 | ChEMBL \| NPASS |
| *Tectona grandis* | Wood | CID_6780 | 6256 | ChEMBL \| NPASS |
| *Tectona grandis* | Wood | CID_6780 | 79813 | BindingDB \| ChEMBL \| NPASS |
| *Tectona grandis* | Wood | CID_6780 | 994 | BindingDB \| ChEMBL \| NPASS |
| *Tectona grandis* | Wood | CID_72734 | 1728 | ChEMBL \| NPASS |
| *Tectona grandis* | Wood | CID_72734 | 3620 | BindingDB \| ChEMBL \| NPASS |
| *Tectona grandis* | Wood | CID_72734 | 5447 | ChEMBL \| NPASS |
| *Tectona grandis* | Wood | CID_67030 | 4759 | BindingDB \| NPASS |
| *Tectona grandis* | Wood | CID_67030 | 7276 | ChEMBL |
| *Aegle marmelos* | Bark | CID_222284 | 10013 | ChEMBL |
| *Aegle marmelos* | Bark | CID_222284 | 10599 | ChEMBL \| NPASS |
| *Aegle marmelos* | Bark | CID_222284 | 1565 | ChEMBL \| NPASS |
| *Aegle marmelos* | Bark | CID_222284 | 1576 | BindingDB \| ChEMBL \| NPASS |
| *Aegle marmelos* | Bark | CID_222284 | 1803 | ChEMBL \| NPASS |
| *Aegle marmelos* | Bark | CID_222284 | 2147 | ChEMBL \| NPASS |
| *Aegle marmelos* | Bark | CID_222284 | 28234 | ChEMBL \| NPASS |
| *Aegle marmelos* | Bark | CID_222284 | 2932 | ChEMBL |
| *Aegle marmelos* | Bark | CID_222284 | 3417 | ChEMBL \| NPASS |
| *Aegle marmelos* | Bark | CID_222284 | 51422 | ChEMBL |
| *Aegle marmelos* | Bark | CID_222284 | 5243 | ChEMBL |
| *Aegle marmelos* | Bark | CID_222284 | 53632 | ChEMBL |
| *Aegle marmelos* | Bark | CID_222284 | 5423 | BindingDB \| ChEMBL \| NPASS |
| *Aegle marmelos* | Bark | CID_222284 | 5562 | ChEMBL |
| *Aegle marmelos* | Bark | CID_222284 | 5563 | ChEMBL |
| *Aegle marmelos* | Bark | CID_222284 | 5564 | ChEMBL |
| *Aegle marmelos* | Bark | CID_222284 | 5565 | ChEMBL |
| *Aegle marmelos* | Bark | CID_222284 | 5571 | ChEMBL |
| *Aegle marmelos* | Bark | CID_222284 | 7299 | ChEMBL \| NPASS |
| *Aegle marmelos* | Bark | CID_259846 | 1066 | ChEMBL \| NPASS |
| *Aegle marmelos* | Bark | CID_259846 | 387 | ChEMBL |
| *Aegle marmelos* | Bark | CID_259846 | 5770 | BindingDB \| ChEMBL \| NPASS |
| *Aegle marmelos* | Bark | CID_259846 | 5879 | ChEMBL |
| *Aegle marmelos* | Bark | CID_259846 | 7153 | ChEMBL \| NPASS |
| *Aegle marmelos* | Bark | CID_259846 | 8824 | BindingDB \| ChEMBL \| NPASS |
| *Aegle marmelos* | Bark | CID_259846 | 998 | ChEMBL |
| *Aegle marmelos* | Bark | CID_323 | 10013 | ChEMBL |
| *Aegle marmelos* | Bark | CID_323 | 10280 | ChEMBL |
| *Aegle marmelos* | Bark | CID_323 | 10599 | ChEMBL \| NPASS |
| *Aegle marmelos* | Bark | CID_323 | 10800 | ChEMBL |
| *Aegle marmelos* | Bark | CID_323 | 11201 | NPASS |
| *Aegle marmelos* | Bark | CID_323 | 11238 | BindingDB \| ChEMBL \| NPASS |
| *Aegle marmelos* | Bark | CID_323 | 1128 | ChEMBL |
| *Aegle marmelos* | Bark | CID_323 | 1129 | ChEMBL |
| *Aegle marmelos* | Bark | CID_323 | 11309 | ChEMBL \| NPASS |
| *Aegle marmelos* | Bark | CID_323 | 1131 | ChEMBL |
| *Aegle marmelos* | Bark | CID_323 | 1132 | ChEMBL |
| *Aegle marmelos* | Bark | CID_323 | 1133 | ChEMBL |
| *Aegle marmelos* | Bark | CID_323 | 1233 | ChEMBL |
| *Aegle marmelos* | Bark | CID_323 | 1234 | ChEMBL |
| *Aegle marmelos* | Bark | CID_323 | 1268 | ChEMBL |
| *Aegle marmelos* | Bark | CID_323 | 134 | ChEMBL |
| *Aegle marmelos* | Bark | CID_323 | 135 | ChEMBL |
| *Aegle marmelos* | Bark | CID_323 | 140 | ChEMBL |
| *Aegle marmelos* | Bark | CID_323 | 1432 | ChEMBL |
| *Aegle marmelos* | Bark | CID_323 | 146 | ChEMBL |
| *Aegle marmelos* | Bark | CID_323 | 150 | ChEMBL |
| *Aegle marmelos* | Bark | CID_323 | 1504 | ChEMBL \| NPASS |
| *Aegle marmelos* | Bark | CID_323 | 151 | ChEMBL |
| *Aegle marmelos* | Bark | CID_323 | 1511 | ChEMBL |
| *Aegle marmelos* | Bark | CID_323 | 152 | ChEMBL |
| *Aegle marmelos* | Bark | CID_323 | 153 | ChEMBL |
| *Aegle marmelos* | Bark | CID_323 | 154 | ChEMBL |
| *Aegle marmelos* | Bark | CID_323 | 1544 | ChEMBL |
| *Aegle marmelos* | Bark | CID_323 | 1548 | ChEMBL \| NPASS |
| *Aegle marmelos* | Bark | CID_323 | 155 | ChEMBL |
| *Aegle marmelos* | Bark | CID_323 | 1557 | ChEMBL |
| *Aegle marmelos* | Bark | CID_323 | 1559 | ChEMBL |
| *Aegle marmelos* | Bark | CID_323 | 1565 | ChEMBL |
| *Aegle marmelos* | Bark | CID_323 | 1571 | ChEMBL |
| *Aegle marmelos* | Bark | CID_323 | 1576 | ChEMBL |
| *Aegle marmelos* | Bark | CID_323 | 1812 | ChEMBL |
| *Aegle marmelos* | Bark | CID_323 | 1813 | ChEMBL |
| *Aegle marmelos* | Bark | CID_323 | 1814 | ChEMBL |
| *Aegle marmelos* | Bark | CID_323 | 1815 | ChEMBL |
| *Aegle marmelos* | Bark | CID_323 | 186 | ChEMBL |
| *Aegle marmelos* | Bark | CID_323 | 1909 | ChEMBL |
| *Aegle marmelos* | Bark | CID_323 | 1956 | ChEMBL |
| *Aegle marmelos* | Bark | CID_323 | 1991 | ChEMBL |
| *Aegle marmelos* | Bark | CID_323 | 2064 | ChEMBL |
| *Aegle marmelos* | Bark | CID_323 | 2099 | ChEMBL \| NPASS |
| *Aegle marmelos* | Bark | CID_323 | 2100 | ChEMBL |
| *Aegle marmelos* | Bark | CID_323 | 213 | ChEMBL \| NPASS |
| *Aegle marmelos* | Bark | CID_323 | 231 | NPASS |
| *Aegle marmelos* | Bark | CID_323 | 2321 | ChEMBL |
| *Aegle marmelos* | Bark | CID_323 | 23621 | BindingDB \| ChEMBL \| NPASS |
| *Aegle marmelos* | Bark | CID_323 | 23632 | BindingDB \| ChEMBL \| NPASS |
| *Aegle marmelos* | Bark | CID_323 | 2534 | ChEMBL |
| *Aegle marmelos* | Bark | CID_323 | 2581 | NPASS |
| *Aegle marmelos* | Bark | CID_323 | 2597 | BindingDB \| NPASS |
| *Aegle marmelos* | Bark | CID_323 | 28234 | ChEMBL \| NPASS |
| *Aegle marmelos* | Bark | CID_323 | 2908 | ChEMBL \| NPASS |
| *Aegle marmelos* | Bark | CID_323 | 3156 | ChEMBL |
| *Aegle marmelos* | Bark | CID_323 | 3269 | ChEMBL |
| *Aegle marmelos* | Bark | CID_323 | 3274 | ChEMBL |
| *Aegle marmelos* | Bark | CID_323 | 3356 | ChEMBL |
| *Aegle marmelos* | Bark | CID_323 | 3357 | ChEMBL |
| *Aegle marmelos* | Bark | CID_323 | 3358 | ChEMBL |
| *Aegle marmelos* | Bark | CID_323 | 3362 | ChEMBL |
| *Aegle marmelos* | Bark | CID_323 | 3577 | ChEMBL |
| *Aegle marmelos* | Bark | CID_323 | 3579 | ChEMBL |
| *Aegle marmelos* | Bark | CID_323 | 367 | NPASS |
| *Aegle marmelos* | Bark | CID_323 | 3757 | ChEMBL |
| *Aegle marmelos* | Bark | CID_323 | 377677 | NPASS |
| *Aegle marmelos* | Bark | CID_323 | 3932 | ChEMBL |
| *Aegle marmelos* | Bark | CID_323 | 4000 | NPASS |
| *Aegle marmelos* | Bark | CID_323 | 4128 | ChEMBL |
| *Aegle marmelos* | Bark | CID_323 | 4154 | NPASS |
| *Aegle marmelos* | Bark | CID_323 | 4159 | ChEMBL |
| *Aegle marmelos* | Bark | CID_323 | 4160 | ChEMBL |
| *Aegle marmelos* | Bark | CID_323 | 4161 | ChEMBL |
| *Aegle marmelos* | Bark | CID_323 | 4170 | BindingDB \| ChEMBL \| NPASS |
| *Aegle marmelos* | Bark | CID_323 | 43 | ChEMBL |
| *Aegle marmelos* | Bark | CID_323 | 4312 | ChEMBL |
| *Aegle marmelos* | Bark | CID_323 | 4318 | ChEMBL |
| *Aegle marmelos* | Bark | CID_323 | 4780 | ChEMBL \| NPASS |
| *Aegle marmelos* | Bark | CID_323 | 4886 | ChEMBL |
| *Aegle marmelos* | Bark | CID_323 | 4887 | ChEMBL |
| *Aegle marmelos* | Bark | CID_323 | 4985 | ChEMBL |
| *Aegle marmelos* | Bark | CID_323 | 4986 | ChEMBL |
| *Aegle marmelos* | Bark | CID_323 | 4988 | ChEMBL |
| *Aegle marmelos* | Bark | CID_323 | 552 | ChEMBL |
| *Aegle marmelos* | Bark | CID_323 | 5530 | ChEMBL |
| *Aegle marmelos* | Bark | CID_323 | 55775 | NPASS |
| *Aegle marmelos* | Bark | CID_323 | 5578 | ChEMBL |
| *Aegle marmelos* | Bark | CID_323 | 5594 | ChEMBL |
| *Aegle marmelos* | Bark | CID_323 | 5595 | ChEMBL |
| *Aegle marmelos* | Bark | CID_323 | 5644 | ChEMBL \| NPASS |
| *Aegle marmelos* | Bark | CID_323 | 5724 | ChEMBL |
| *Aegle marmelos* | Bark | CID_323 | 5742 | ChEMBL |
| *Aegle marmelos* | Bark | CID_323 | 5743 | ChEMBL \| NPASS |
| *Aegle marmelos* | Bark | CID_323 | 5788 | ChEMBL |
| *Aegle marmelos* | Bark | CID_323 | 624 | ChEMBL |
| *Aegle marmelos* | Bark | CID_323 | 6530 | ChEMBL |
| *Aegle marmelos* | Bark | CID_323 | 6531 | ChEMBL |
| *Aegle marmelos* | Bark | CID_323 | 6532 | ChEMBL |
| *Aegle marmelos* | Bark | CID_323 | 6652 | BindingDB \| NPASS |
| *Aegle marmelos* | Bark | CID_323 | 6865 | ChEMBL |
| *Aegle marmelos* | Bark | CID_323 | 6869 | ChEMBL |
| *Aegle marmelos* | Bark | CID_323 | 6916 | ChEMBL |
| *Aegle marmelos* | Bark | CID_323 | 7068 | NPASS |
| *Aegle marmelos* | Bark | CID_323 | 729230 | ChEMBL |
| *Aegle marmelos* | Bark | CID_323 | 7398 | NPASS |
| *Aegle marmelos* | Bark | CID_323 | 7433 | ChEMBL |
| *Aegle marmelos* | Bark | CID_323 | 7514 | BindingDB \| ChEMBL \| NPASS |
| *Aegle marmelos* | Bark | CID_323 | 759 | BindingDB \| ChEMBL \| NPASS |
| *Aegle marmelos* | Bark | CID_323 | 760 | BindingDB \| ChEMBL \| NPASS |
| *Aegle marmelos* | Bark | CID_323 | 761 | BindingDB \| ChEMBL \| NPASS |
| *Aegle marmelos* | Bark | CID_323 | 762 | BindingDB \| ChEMBL \| NPASS |
| *Aegle marmelos* | Bark | CID_323 | 763 | BindingDB \| ChEMBL \| NPASS |
| *Aegle marmelos* | Bark | CID_323 | 765 | ChEMBL \| NPASS |
| *Aegle marmelos* | Bark | CID_323 | 766 | BindingDB \| ChEMBL \| NPASS |
| *Aegle marmelos* | Bark | CID_323 | 768 | BindingDB \| ChEMBL \| NPASS |
| *Aegle marmelos* | Bark | CID_323 | 771 | BindingDB \| ChEMBL \| NPASS |
| *Aegle marmelos* | Bark | CID_323 | 799 | ChEMBL |
| *Aegle marmelos* | Bark | CID_323 | 834 | ChEMBL |
| *Aegle marmelos* | Bark | CID_323 | 8647 | NPASS |
| *Aegle marmelos* | Bark | CID_323 | 865 | NPASS |
| *Aegle marmelos* | Bark | CID_323 | 8654 | ChEMBL |
| *Aegle marmelos* | Bark | CID_323 | 886 | ChEMBL |
| *Aegle marmelos* | Bark | CID_323 | 9971 | NPASS |
| *Aegle marmelos* | Bark | CID_1550607 | 10919 | NPASS |
| *Aegle marmelos* | Bark | CID_1550607 | 10951 | NPASS |
| *Aegle marmelos* | Bark | CID_1550607 | 126328 | ChEMBL \| NPASS |
| *Aegle marmelos* | Bark | CID_1550607 | 1576 | NPASS |
| *Aegle marmelos* | Bark | CID_1550607 | 216 | NPASS |
| *Aegle marmelos* | Bark | CID_1550607 | 23435 | NPASS |
| *Aegle marmelos* | Bark | CID_1550607 | 23621 | BindingDB \| NPASS |
| *Aegle marmelos* | Bark | CID_1550607 | 246 | BindingDB \| ChEMBL \| NPASS |
| *Aegle marmelos* | Bark | CID_1550607 | 25915 | ChEMBL \| NPASS |
| *Aegle marmelos* | Bark | CID_1550607 | 26013 | NPASS |
| *Aegle marmelos* | Bark | CID_1550607 | 2629 | NPASS |
| *Aegle marmelos* | Bark | CID_1550607 | 2717 | NPASS |
| *Aegle marmelos* | Bark | CID_1550607 | 2740 | NPASS |
| *Aegle marmelos* | Bark | CID_1550607 | 29078 | ChEMBL \| NPASS |
| *Aegle marmelos* | Bark | CID_1550607 | 29994 | NPASS |
| *Aegle marmelos* | Bark | CID_1550607 | 3028 | ChEMBL \| NPASS |
| *Aegle marmelos* | Bark | CID_1550607 | 3248 | NPASS |
| *Aegle marmelos* | Bark | CID_1550607 | 374291 | ChEMBL \| NPASS |
| *Aegle marmelos* | Bark | CID_1550607 | 390245 | NPASS |
| *Aegle marmelos* | Bark | CID_1550607 | 4088 | NPASS |
| *Aegle marmelos* | Bark | CID_1550607 | 4137 | NPASS |
| *Aegle marmelos* | Bark | CID_1550607 | 4535 | ChEMBL \| NPASS |
| *Aegle marmelos* | Bark | CID_1550607 | 4536 | ChEMBL \| NPASS |
| *Aegle marmelos* | Bark | CID_1550607 | 4537 | ChEMBL \| NPASS |
| *Aegle marmelos* | Bark | CID_1550607 | 4538 | ChEMBL \| NPASS |
| *Aegle marmelos* | Bark | CID_1550607 | 4539 | ChEMBL \| NPASS |
| *Aegle marmelos* | Bark | CID_1550607 | 4540 | ChEMBL \| NPASS |
| *Aegle marmelos* | Bark | CID_1550607 | 4541 | ChEMBL \| NPASS |
| *Aegle marmelos* | Bark | CID_1550607 | 4543 | BindingDB \| ChEMBL \| NPASS |
| *Aegle marmelos* | Bark | CID_1550607 | 4694 | ChEMBL \| NPASS |
| *Aegle marmelos* | Bark | CID_1550607 | 4695 | ChEMBL \| NPASS |
| *Aegle marmelos* | Bark | CID_1550607 | 4696 | ChEMBL \| NPASS |
| *Aegle marmelos* | Bark | CID_1550607 | 4697 | ChEMBL \| NPASS |
| *Aegle marmelos* | Bark | CID_1550607 | 4698 | ChEMBL \| NPASS |
| *Aegle marmelos* | Bark | CID_1550607 | 4700 | ChEMBL \| NPASS |
| *Aegle marmelos* | Bark | CID_1550607 | 4701 | ChEMBL \| NPASS |
| *Aegle marmelos* | Bark | CID_1550607 | 4702 | ChEMBL \| NPASS |
| *Aegle marmelos* | Bark | CID_1550607 | 4704 | ChEMBL \| NPASS |
| *Aegle marmelos* | Bark | CID_1550607 | 4705 | ChEMBL \| NPASS |
| *Aegle marmelos* | Bark | CID_1550607 | 4706 | ChEMBL \| NPASS |
| *Aegle marmelos* | Bark | CID_1550607 | 4707 | ChEMBL \| NPASS |
| *Aegle marmelos* | Bark | CID_1550607 | 4708 | ChEMBL \| NPASS |
| *Aegle marmelos* | Bark | CID_1550607 | 4709 | ChEMBL \| NPASS |
| *Aegle marmelos* | Bark | CID_1550607 | 4710 | ChEMBL \| NPASS |
| *Aegle marmelos* | Bark | CID_1550607 | 4711 | ChEMBL \| NPASS |
| *Aegle marmelos* | Bark | CID_1550607 | 4712 | ChEMBL \| NPASS |
| *Aegle marmelos* | Bark | CID_1550607 | 4713 | ChEMBL \| NPASS |
| *Aegle marmelos* | Bark | CID_1550607 | 4714 | ChEMBL \| NPASS |
| *Aegle marmelos* | Bark | CID_1550607 | 4715 | ChEMBL \| NPASS |
| *Aegle marmelos* | Bark | CID_1550607 | 4716 | ChEMBL \| NPASS |
| *Aegle marmelos* | Bark | CID_1550607 | 4717 | ChEMBL \| NPASS |
| *Aegle marmelos* | Bark | CID_1550607 | 4718 | ChEMBL \| NPASS |
| *Aegle marmelos* | Bark | CID_1550607 | 4719 | ChEMBL \| NPASS |
| *Aegle marmelos* | Bark | CID_1550607 | 472 | NPASS |
| *Aegle marmelos* | Bark | CID_1550607 | 4720 | ChEMBL \| NPASS |
| *Aegle marmelos* | Bark | CID_1550607 | 4722 | ChEMBL \| NPASS |
| *Aegle marmelos* | Bark | CID_1550607 | 4723 | ChEMBL \| NPASS |
| *Aegle marmelos* | Bark | CID_1550607 | 4724 | ChEMBL \| NPASS |
| *Aegle marmelos* | Bark | CID_1550607 | 4725 | ChEMBL \| NPASS |
| *Aegle marmelos* | Bark | CID_1550607 | 4726 | ChEMBL \| NPASS |
| *Aegle marmelos* | Bark | CID_1550607 | 4728 | ChEMBL \| NPASS |
| *Aegle marmelos* | Bark | CID_1550607 | 4729 | ChEMBL \| NPASS |
| *Aegle marmelos* | Bark | CID_1550607 | 4731 | ChEMBL \| NPASS |
| *Aegle marmelos* | Bark | CID_1550607 | 51053 | NPASS |
| *Aegle marmelos* | Bark | CID_1550607 | 51079 | ChEMBL \| NPASS |
| *Aegle marmelos* | Bark | CID_1550607 | 51103 | ChEMBL \| NPASS |
| *Aegle marmelos* | Bark | CID_1550607 | 5429 | NPASS |
| *Aegle marmelos* | Bark | CID_1550607 | 54539 | ChEMBL \| NPASS |
| *Aegle marmelos* | Bark | CID_1550607 | 5468 | BindingDB \| ChEMBL |
| *Aegle marmelos* | Bark | CID_1550607 | 55775 | NPASS |
| *Aegle marmelos* | Bark | CID_1550607 | 5594 | NPASS |
| *Aegle marmelos* | Bark | CID_1550607 | 55967 | ChEMBL \| NPASS |
| *Aegle marmelos* | Bark | CID_1550607 | 56901 | ChEMBL \| NPASS |
| *Aegle marmelos* | Bark | CID_1550607 | 5745 | NPASS |
| *Aegle marmelos* | Bark | CID_1550607 | 6609 | NPASS |
| *Aegle marmelos* | Bark | CID_1550607 | 7398 | NPASS |
| *Aegle marmelos* | Bark | CID_1550607 | 865 | NPASS |
| *Aegle marmelos* | Bark | CID_1550607 | 8989 | ChEMBL |
| *Aegle marmelos* | Bark | CID_1550607 | 91942 | ChEMBL \| NPASS |
| *Aegle marmelos* | Bark | CID_1550607 | 9682 | NPASS |
| *Aegle marmelos* | Bark | CID_1550607 | 9971 | ChEMBL \| NPASS |
| *Aegle marmelos* | Bark | CID_334704 | 10599 | ChEMBL \| NPASS |
| *Aegle marmelos* | Bark | CID_334704 | 1588 | ChEMBL \| NPASS |
| *Aegle marmelos* | Bark | CID_334704 | 23621 | BindingDB \| ChEMBL \| NPASS |
| *Aegle marmelos* | Bark | CID_334704 | 28234 | ChEMBL \| NPASS |
| *Aegle marmelos* | Bark | CID_334704 | 43 | BindingDB \| ChEMBL \| NPASS |
| *Aegle marmelos* | Bark | CID_6450230 | 126328 | ChEMBL \| NPASS |
| *Aegle marmelos* | Bark | CID_6450230 | 25915 | ChEMBL \| NPASS |
| *Aegle marmelos* | Bark | CID_6450230 | 29078 | ChEMBL \| NPASS |
| *Aegle marmelos* | Bark | CID_6450230 | 374291 | ChEMBL \| NPASS |
| *Aegle marmelos* | Bark | CID_6450230 | 4535 | ChEMBL \| NPASS |
| *Aegle marmelos* | Bark | CID_6450230 | 4536 | ChEMBL \| NPASS |
| *Aegle marmelos* | Bark | CID_6450230 | 4537 | ChEMBL \| NPASS |
| *Aegle marmelos* | Bark | CID_6450230 | 4538 | ChEMBL \| NPASS |
| *Aegle marmelos* | Bark | CID_6450230 | 4539 | ChEMBL \| NPASS |
| *Aegle marmelos* | Bark | CID_6450230 | 4540 | ChEMBL \| NPASS |
| *Aegle marmelos* | Bark | CID_6450230 | 4541 | ChEMBL \| NPASS |
| *Aegle marmelos* | Bark | CID_6450230 | 4694 | ChEMBL \| NPASS |
| *Aegle marmelos* | Bark | CID_6450230 | 4695 | ChEMBL \| NPASS |
| *Aegle marmelos* | Bark | CID_6450230 | 4696 | ChEMBL \| NPASS |
| *Aegle marmelos* | Bark | CID_6450230 | 4697 | ChEMBL \| NPASS |
| *Aegle marmelos* | Bark | CID_6450230 | 4698 | ChEMBL \| NPASS |
| *Aegle marmelos* | Bark | CID_6450230 | 4700 | ChEMBL \| NPASS |
| *Aegle marmelos* | Bark | CID_6450230 | 4701 | ChEMBL \| NPASS |
| *Aegle marmelos* | Bark | CID_6450230 | 4702 | ChEMBL \| NPASS |
| *Aegle marmelos* | Bark | CID_6450230 | 4704 | ChEMBL \| NPASS |
| *Aegle marmelos* | Bark | CID_6450230 | 4705 | ChEMBL \| NPASS |
| *Aegle marmelos* | Bark | CID_6450230 | 4706 | ChEMBL \| NPASS |
| *Aegle marmelos* | Bark | CID_6450230 | 4707 | ChEMBL \| NPASS |
| *Aegle marmelos* | Bark | CID_6450230 | 4708 | ChEMBL \| NPASS |
| *Aegle marmelos* | Bark | CID_6450230 | 4709 | ChEMBL \| NPASS |
| *Aegle marmelos* | Bark | CID_6450230 | 4710 | ChEMBL \| NPASS |
| *Aegle marmelos* | Bark | CID_6450230 | 4711 | ChEMBL \| NPASS |
| *Aegle marmelos* | Bark | CID_6450230 | 4712 | ChEMBL \| NPASS |
| *Aegle marmelos* | Bark | CID_6450230 | 4713 | ChEMBL \| NPASS |
| *Aegle marmelos* | Bark | CID_6450230 | 4714 | ChEMBL \| NPASS |
| *Aegle marmelos* | Bark | CID_6450230 | 4715 | ChEMBL \| NPASS |
| *Aegle marmelos* | Bark | CID_6450230 | 4716 | ChEMBL \| NPASS |
| *Aegle marmelos* | Bark | CID_6450230 | 4717 | ChEMBL \| NPASS |
| *Aegle marmelos* | Bark | CID_6450230 | 4718 | ChEMBL \| NPASS |
| *Aegle marmelos* | Bark | CID_6450230 | 4719 | ChEMBL \| NPASS |
| *Aegle marmelos* | Bark | CID_6450230 | 4720 | ChEMBL \| NPASS |
| *Aegle marmelos* | Bark | CID_6450230 | 4722 | ChEMBL \| NPASS |
| *Aegle marmelos* | Bark | CID_6450230 | 4723 | ChEMBL \| NPASS |
| *Aegle marmelos* | Bark | CID_6450230 | 4724 | ChEMBL \| NPASS |
| *Aegle marmelos* | Bark | CID_6450230 | 4725 | ChEMBL \| NPASS |
| *Aegle marmelos* | Bark | CID_6450230 | 4726 | ChEMBL \| NPASS |
| *Aegle marmelos* | Bark | CID_6450230 | 4728 | ChEMBL \| NPASS |
| *Aegle marmelos* | Bark | CID_6450230 | 4729 | ChEMBL \| NPASS |
| *Aegle marmelos* | Bark | CID_6450230 | 4731 | ChEMBL \| NPASS |
| *Aegle marmelos* | Bark | CID_6450230 | 51079 | ChEMBL \| NPASS |
| *Aegle marmelos* | Bark | CID_6450230 | 51103 | ChEMBL \| NPASS |
| *Aegle marmelos* | Bark | CID_6450230 | 54539 | ChEMBL \| NPASS |
| *Aegle marmelos* | Bark | CID_6450230 | 55775 | NPASS |
| *Aegle marmelos* | Bark | CID_6450230 | 55967 | ChEMBL \| NPASS |
| *Aegle marmelos* | Bark | CID_6450230 | 56901 | ChEMBL \| NPASS |
| *Aegle marmelos* | Bark | CID_6450230 | 6609 | NPASS |
| *Aegle marmelos* | Bark | CID_6450230 | 7157 | NPASS |
| *Aegle marmelos* | Bark | CID_6450230 | 79915 | NPASS |
| *Aegle marmelos* | Bark | CID_6450230 | 91942 | ChEMBL \| NPASS |
| *Aegle marmelos* | Bark | CID_6760 | 10411 | NPASS |
| *Aegle marmelos* | Bark | CID_6760 | 1544 | ChEMBL \| NPASS |
| *Aegle marmelos* | Bark | CID_6760 | 1557 | ChEMBL \| NPASS |
| *Aegle marmelos* | Bark | CID_6760 | 1565 | NPASS |
| *Aegle marmelos* | Bark | CID_6760 | 1576 | ChEMBL \| NPASS |
| *Aegle marmelos* | Bark | CID_6760 | 216 | NPASS |
| *Aegle marmelos* | Bark | CID_6760 | 2648 | NPASS |
| *Aegle marmelos* | Bark | CID_6760 | 3028 | ChEMBL \| NPASS |
| *Aegle marmelos* | Bark | CID_6760 | 3248 | NPASS |
| *Aegle marmelos* | Bark | CID_6760 | 390245 | NPASS |
| *Aegle marmelos* | Bark | CID_6760 | 4000 | NPASS |
| *Aegle marmelos* | Bark | CID_6760 | 4154 | NPASS |
| *Aegle marmelos* | Bark | CID_6760 | 4297 | NPASS |
| *Aegle marmelos* | Bark | CID_6760 | 43 | BindingDB \| ChEMBL \| NPASS |
| *Aegle marmelos* | Bark | CID_6760 | 51053 | NPASS |
| *Aegle marmelos* | Bark | CID_6760 | 55775 | NPASS |
| *Aegle marmelos* | Bark | CID_6760 | 7157 | NPASS |
| *Aegle marmelos* | Bark | CID_6760 | 7253 | ChEMBL \| NPASS |
| *Aegle marmelos* | Bark | CID_6760 | 79915 | NPASS |
| *Aegle marmelos* | Bark | CID_5281426 | 10073 | NPASS |
| *Aegle marmelos* | Bark | CID_5281426 | 10257 | ChEMBL |
| *Aegle marmelos* | Bark | CID_5281426 | 10599 | ChEMBL \| NPASS |
| *Aegle marmelos* | Bark | CID_5281426 | 10941 | ChEMBL |
| *Aegle marmelos* | Bark | CID_5281426 | 10951 | NPASS |
| *Aegle marmelos* | Bark | CID_5281426 | 1244 | ChEMBL |
| *Aegle marmelos* | Bark | CID_5281426 | 216 | NPASS |
| *Aegle marmelos* | Bark | CID_5281426 | 2237 | NPASS |
| *Aegle marmelos* | Bark | CID_5281426 | 231 | ChEMBL \| NPASS |
| *Aegle marmelos* | Bark | CID_5281426 | 23621 | BindingDB \| ChEMBL \| NPASS |
| *Aegle marmelos* | Bark | CID_5281426 | 2548 | NPASS |
| *Aegle marmelos* | Bark | CID_5281426 | 2597 | BindingDB \| NPASS |
| *Aegle marmelos* | Bark | CID_5281426 | 2629 | NPASS |
| *Aegle marmelos* | Bark | CID_5281426 | 2648 | NPASS |
| *Aegle marmelos* | Bark | CID_5281426 | 2717 | NPASS |
| *Aegle marmelos* | Bark | CID_5281426 | 2744 | NPASS |
| *Aegle marmelos* | Bark | CID_5281426 | 28234 | ChEMBL \| NPASS |
| *Aegle marmelos* | Bark | CID_5281426 | 3028 | ChEMBL \| NPASS |
| *Aegle marmelos* | Bark | CID_5281426 | 3248 | NPASS |
| *Aegle marmelos* | Bark | CID_5281426 | 3292 | ChEMBL |
| *Aegle marmelos* | Bark | CID_5281426 | 351 | ChEMBL \| NPASS |
| *Aegle marmelos* | Bark | CID_5281426 | 3837 | NPASS |
| *Aegle marmelos* | Bark | CID_5281426 | 390245 | NPASS |
| *Aegle marmelos* | Bark | CID_5281426 | 412 | ChEMBL |
| *Aegle marmelos* | Bark | CID_5281426 | 4170 | BindingDB \| ChEMBL \| NPASS |
| *Aegle marmelos* | Bark | CID_5281426 | 4297 | NPASS |
| *Aegle marmelos* | Bark | CID_5281426 | 43 | BindingDB \| ChEMBL \| NPASS |
| *Aegle marmelos* | Bark | CID_5281426 | 4363 | ChEMBL |
| *Aegle marmelos* | Bark | CID_5281426 | 4780 | ChEMBL \| NPASS |
| *Aegle marmelos* | Bark | CID_5281426 | 51053 | NPASS |
| *Aegle marmelos* | Bark | CID_5281426 | 51752 | ChEMBL \| NPASS |
| *Aegle marmelos* | Bark | CID_5281426 | 54575 | ChEMBL \| NPASS |
| *Aegle marmelos* | Bark | CID_5281426 | 54576 | ChEMBL \| NPASS |
| *Aegle marmelos* | Bark | CID_5281426 | 54657 | ChEMBL \| NPASS |
| *Aegle marmelos* | Bark | CID_5281426 | 54658 | ChEMBL \| NPASS |
| *Aegle marmelos* | Bark | CID_5281426 | 54659 | ChEMBL \| NPASS |
| *Aegle marmelos* | Bark | CID_5281426 | 55775 | NPASS |
| *Aegle marmelos* | Bark | CID_5281426 | 5743 | ChEMBL \| NPASS |
| *Aegle marmelos* | Bark | CID_5281426 | 6311 | NPASS |
| *Aegle marmelos* | Bark | CID_5281426 | 6652 | BindingDB \| NPASS |
| *Aegle marmelos* | Bark | CID_5281426 | 6715 | ChEMBL \| NPASS |
| *Aegle marmelos* | Bark | CID_5281426 | 7299 | BindingDB \| NPASS |
| *Aegle marmelos* | Bark | CID_5281426 | 7366 | ChEMBL \| NPASS |
| *Aegle marmelos* | Bark | CID_5281426 | 7498 | BindingDB \| NPASS |
| *Aegle marmelos* | Bark | CID_5281426 | 759 | BindingDB \| ChEMBL \| NPASS |
| *Aegle marmelos* | Bark | CID_5281426 | 760 | BindingDB \| ChEMBL \| NPASS |
| *Aegle marmelos* | Bark | CID_5281426 | 768 | BindingDB \| ChEMBL \| NPASS |
| *Aegle marmelos* | Bark | CID_5281426 | 771 | BindingDB \| ChEMBL \| NPASS |
| *Aegle marmelos* | Bark | CID_5281426 | 834 | NPASS |
| *Aegle marmelos* | Bark | CID_5281426 | 840 | NPASS |
| *Aegle marmelos* | Bark | CID_5281426 | 8714 | ChEMBL |
| *Aegle marmelos* | Bark | CID_5281426 | 9429 | ChEMBL |
| *Aegle marmelos* | Bark | CID_10212 | 10599 | NPASS |
| *Aegle marmelos* | Bark | CID_10212 | 10919 | NPASS |
| *Aegle marmelos* | Bark | CID_10212 | 10951 | NPASS |
| *Aegle marmelos* | Bark | CID_10212 | 1543 | ChEMBL \| NPASS |
| *Aegle marmelos* | Bark | CID_10212 | 1545 | BindingDB \| ChEMBL \| NPASS |
| *Aegle marmelos* | Bark | CID_10212 | 1576 | ChEMBL \| NPASS |
| *Aegle marmelos* | Bark | CID_10212 | 1845 | BindingDB \| ChEMBL \| NPASS |
| *Aegle marmelos* | Bark | CID_10212 | 2139 | NPASS |
| *Aegle marmelos* | Bark | CID_10212 | 216 | NPASS |
| *Aegle marmelos* | Bark | CID_10212 | 23435 | NPASS |
| *Aegle marmelos* | Bark | CID_10212 | 23621 | BindingDB \| ChEMBL \| NPASS |
| *Aegle marmelos* | Bark | CID_10212 | 2548 | NPASS |
| *Aegle marmelos* | Bark | CID_10212 | 2554 | ChEMBL |
| *Aegle marmelos* | Bark | CID_10212 | 2561 | ChEMBL |
| *Aegle marmelos* | Bark | CID_10212 | 2566 | ChEMBL \| NPASS |
| *Aegle marmelos* | Bark | CID_10212 | 2740 | NPASS |
| *Aegle marmelos* | Bark | CID_10212 | 28234 | ChEMBL \| NPASS |
| *Aegle marmelos* | Bark | CID_10212 | 328 | NPASS |
| *Aegle marmelos* | Bark | CID_10212 | 3363 | ChEMBL \| NPASS |
| *Aegle marmelos* | Bark | CID_10212 | 3417 | ChEMBL \| NPASS |
| *Aegle marmelos* | Bark | CID_10212 | 5243 | ChEMBL \| NPASS |
| *Aegle marmelos* | Bark | CID_10212 | 5329 | ChEMBL |
| *Aegle marmelos* | Bark | CID_10212 | 5347 | ChEMBL \| NPASS |
| *Aegle marmelos* | Bark | CID_10212 | 5778 | BindingDB \| ChEMBL \| NPASS |
| *Aegle marmelos* | Bark | CID_10212 | 5999 | NPASS |
| *Aegle marmelos* | Bark | CID_10212 | 7398 | NPASS |
| *Aegle marmelos* | Bark | CID_10212 | 7421 | NPASS |
| *Aegle marmelos* | Bark | CID_10212 | 9682 | NPASS |
| *Aegle marmelos* | Bark | CID_107936 | 10919 | NPASS |
| *Aegle marmelos* | Bark | CID_107936 | 216 | NPASS |
| *Aegle marmelos* | Bark | CID_107936 | 2778 | NPASS |
| *Aegle marmelos* | Bark | CID_107936 | 3248 | NPASS |
| *Aegle marmelos* | Bark | CID_107936 | 390245 | NPASS |
| *Aegle marmelos* | Bark | CID_107936 | 4154 | NPASS |
| *Aegle marmelos* | Bark | CID_107936 | 43 | ChEMBL \| NPASS |
| *Aegle marmelos* | Bark | CID_107936 | 672 | ChEMBL \| NPASS |
| *Aegle marmelos* | Bark | CID_107936 | 79915 | NPASS |
| *Butea monosperma* | Seed | CID_222284 | 10013 | ChEMBL |
| *Butea monosperma* | Seed | CID_222284 | 10599 | ChEMBL \| NPASS |
| *Butea monosperma* | Seed | CID_222284 | 1565 | ChEMBL \| NPASS |
| *Butea monosperma* | Seed | CID_222284 | 1576 | BindingDB \| ChEMBL \| NPASS |
| *Butea monosperma* | Seed | CID_222284 | 1803 | ChEMBL \| NPASS |
| *Butea monosperma* | Seed | CID_222284 | 2147 | ChEMBL \| NPASS |
| *Butea monosperma* | Seed | CID_222284 | 28234 | ChEMBL \| NPASS |
| *Butea monosperma* | Seed | CID_222284 | 2932 | ChEMBL |
| *Butea monosperma* | Seed | CID_222284 | 3417 | ChEMBL \| NPASS |
| *Butea monosperma* | Seed | CID_222284 | 51422 | ChEMBL |
| *Butea monosperma* | Seed | CID_222284 | 5243 | ChEMBL |
| *Butea monosperma* | Seed | CID_222284 | 53632 | ChEMBL |
| *Butea monosperma* | Seed | CID_222284 | 5423 | BindingDB \| ChEMBL \| NPASS |
| *Butea monosperma* | Seed | CID_222284 | 5562 | ChEMBL |
| *Butea monosperma* | Seed | CID_222284 | 5563 | ChEMBL |
| *Butea monosperma* | Seed | CID_222284 | 5564 | ChEMBL |
| *Butea monosperma* | Seed | CID_222284 | 5565 | ChEMBL |
| *Butea monosperma* | Seed | CID_222284 | 5571 | ChEMBL |
| *Butea monosperma* | Seed | CID_222284 | 7299 | ChEMBL \| NPASS |
| *Butea monosperma* | Seed | CID_11005 | 10411 | NPASS |
| *Butea monosperma* | Seed | CID_11005 | 10951 | NPASS |
| *Butea monosperma* | Seed | CID_11005 | 1588 | ChEMBL \| NPASS |
| *Butea monosperma* | Seed | CID_11005 | 1759 | BindingDB \| ChEMBL \| NPASS |
| *Butea monosperma* | Seed | CID_11005 | 2237 | NPASS |
| *Butea monosperma* | Seed | CID_11005 | 23435 | NPASS |
| *Butea monosperma* | Seed | CID_11005 | 247 | NPASS |
| *Butea monosperma* | Seed | CID_11005 | 2648 | NPASS |
| *Butea monosperma* | Seed | CID_11005 | 2740 | NPASS |
| *Butea monosperma* | Seed | CID_11005 | 3315 | NPASS |
| *Butea monosperma* | Seed | CID_11005 | 367 | NPASS |
| *Butea monosperma* | Seed | CID_11005 | 51053 | NPASS |
| *Butea monosperma* | Seed | CID_11005 | 51548 | BindingDB \| ChEMBL |
| *Butea monosperma* | Seed | CID_11005 | 5347 | NPASS |
| *Butea monosperma* | Seed | CID_11005 | 53831 | ChEMBL \| NPASS |
| *Butea monosperma* | Seed | CID_11005 | 54575 | ChEMBL \| NPASS |
| *Butea monosperma* | Seed | CID_11005 | 54576 | ChEMBL \| NPASS |
| *Butea monosperma* | Seed | CID_11005 | 5465 | ChEMBL \| NPASS |
| *Butea monosperma* | Seed | CID_11005 | 5467 | ChEMBL \| NPASS |
| *Butea monosperma* | Seed | CID_11005 | 5468 | BindingDB \| ChEMBL \| NPASS |
| *Butea monosperma* | Seed | CID_11005 | 55775 | NPASS |
| *Butea monosperma* | Seed | CID_11005 | 6609 | NPASS |
| *Butea monosperma* | Seed | CID_11005 | 7097 | ChEMBL \| NPASS |
| *Butea monosperma* | Seed | CID_11005 | 7398 | NPASS |
| *Butea monosperma* | Seed | CID_11005 | 7421 | NPASS |
| *Butea monosperma* | Seed | CID_11005 | 8989 | ChEMBL \| NPASS |
| *Butea monosperma* | Seed | CID_11005 | 9971 | NPASS |
| *Butea monosperma* | Seed | CID_985 | 1588 | ChEMBL \| NPASS |
| *Butea monosperma* | Seed | CID_985 | 2099 | ChEMBL \| NPASS |
| *Butea monosperma* | Seed | CID_985 | 2167 | ChEMBL \| NPASS |
| *Butea monosperma* | Seed | CID_985 | 2169 | ChEMBL \| NPASS |
| *Butea monosperma* | Seed | CID_985 | 2170 | BindingDB \| ChEMBL \| NPASS |
| *Butea monosperma* | Seed | CID_985 | 2171 | ChEMBL \| NPASS |
| *Butea monosperma* | Seed | CID_985 | 2908 | NPASS |
| *Butea monosperma* | Seed | CID_985 | 367 | NPASS |
| *Butea monosperma* | Seed | CID_985 | 4780 | NPASS |
| *Butea monosperma* | Seed | CID_985 | 51053 | NPASS |
| *Butea monosperma* | Seed | CID_985 | 5243 | ChEMBL |
| *Butea monosperma* | Seed | CID_985 | 5465 | ChEMBL \| NPASS |
| *Butea monosperma* | Seed | CID_985 | 5467 | ChEMBL \| NPASS |
| *Butea monosperma* | Seed | CID_985 | 5468 | BindingDB \| ChEMBL \| NPASS |
| *Butea monosperma* | Seed | CID_985 | 5770 | BindingDB \| ChEMBL \| NPASS |
| *Butea monosperma* | Seed | CID_985 | 7004 | ChEMBL |
| *Butea monosperma* | Seed | CID_985 | 7097 | ChEMBL \| NPASS |
| *Butea monosperma* | Seed | CID_985 | 7398 | NPASS |
| *Butea monosperma* | Seed | CID_985 | 81285 | ChEMBL \| NPASS |
| *Butea monosperma* | Seed | CID_985 | 9971 | NPASS |
| *Butea monosperma* | Seed | CID_5281 | 11255 | ChEMBL |
| *Butea monosperma* | Seed | CID_5281 | 1128 | ChEMBL |
| *Butea monosperma* | Seed | CID_5281 | 1129 | ChEMBL |
| *Butea monosperma* | Seed | CID_5281 | 140 | ChEMBL |
| *Butea monosperma* | Seed | CID_5281 | 148 | ChEMBL |
| *Butea monosperma* | Seed | CID_5281 | 150 | ChEMBL |
| *Butea monosperma* | Seed | CID_5281 | 1588 | ChEMBL \| NPASS |
| *Butea monosperma* | Seed | CID_5281 | 1812 | ChEMBL |
| *Butea monosperma* | Seed | CID_5281 | 1814 | ChEMBL |
| *Butea monosperma* | Seed | CID_5281 | 2099 | ChEMBL |
| *Butea monosperma* | Seed | CID_5281 | 2147 | ChEMBL |
| *Butea monosperma* | Seed | CID_5281 | 2167 | ChEMBL \| NPASS |
| *Butea monosperma* | Seed | CID_5281 | 246 | NPASS |
| *Butea monosperma* | Seed | CID_5281 | 2554 | ChEMBL |
| *Butea monosperma* | Seed | CID_5281 | 2555 | ChEMBL |
| *Butea monosperma* | Seed | CID_5281 | 2561 | ChEMBL |
| *Butea monosperma* | Seed | CID_5281 | 2566 | ChEMBL |
| *Butea monosperma* | Seed | CID_5281 | 2908 | NPASS |
| *Butea monosperma* | Seed | CID_5281 | 3028 | ChEMBL \| NPASS |
| *Butea monosperma* | Seed | CID_5281 | 328 | NPASS |
| *Butea monosperma* | Seed | CID_5281 | 3350 | ChEMBL |
| *Butea monosperma* | Seed | CID_5281 | 3357 | ChEMBL |
| *Butea monosperma* | Seed | CID_5281 | 367 | ChEMBL |
| *Butea monosperma* | Seed | CID_5281 | 3757 | ChEMBL |
| *Butea monosperma* | Seed | CID_5281 | 3791 | ChEMBL |
| *Butea monosperma* | Seed | CID_5281 | 4128 | ChEMBL |
| *Butea monosperma* | Seed | CID_5281 | 4297 | NPASS |
| *Butea monosperma* | Seed | CID_5281 | 43 | ChEMBL |
| *Butea monosperma* | Seed | CID_5281 | 4790 | NPASS |
| *Butea monosperma* | Seed | CID_5281 | 4988 | ChEMBL |
| *Butea monosperma* | Seed | CID_5281 | 5139 | ChEMBL |
| *Butea monosperma* | Seed | CID_5281 | 5141 | ChEMBL |
| *Butea monosperma* | Seed | CID_5281 | 5241 | ChEMBL |
| *Butea monosperma* | Seed | CID_5281 | 5465 | ChEMBL \| NPASS |
| *Butea monosperma* | Seed | CID_5281 | 5467 | ChEMBL \| NPASS |
| *Butea monosperma* | Seed | CID_5281 | 5468 | BindingDB \| ChEMBL \| NPASS |
| *Butea monosperma* | Seed | CID_5281 | 55775 | NPASS |
| *Butea monosperma* | Seed | CID_5281 | 5742 | ChEMBL |
| *Butea monosperma* | Seed | CID_5281 | 5770 | BindingDB \| ChEMBL \| NPASS |
| *Butea monosperma* | Seed | CID_5281 | 6530 | ChEMBL |
| *Butea monosperma* | Seed | CID_5281 | 6531 | ChEMBL |
| *Butea monosperma* | Seed | CID_5281 | 6532 | ChEMBL |
| *Butea monosperma* | Seed | CID_5281 | 6915 | ChEMBL |
| *Butea monosperma* | Seed | CID_5281 | 7068 | NPASS |
| *Butea monosperma* | Seed | CID_5281 | 7253 | NPASS |
| *Butea monosperma* | Seed | CID_5281 | 7421 | NPASS |
| *Butea monosperma* | Seed | CID_445639 | 10018 | BindingDB |
| *Butea monosperma* | Seed | CID_445639 | 100861540 | ChEMBL |
| *Butea monosperma* | Seed | CID_445639 | 10411 | NPASS |
| *Butea monosperma* | Seed | CID_445639 | 10951 | NPASS |
| *Butea monosperma* | Seed | CID_445639 | 11069 | NPASS |
| *Butea monosperma* | Seed | CID_445639 | 11255 | ChEMBL |
| *Butea monosperma* | Seed | CID_445639 | 1128 | ChEMBL |
| *Butea monosperma* | Seed | CID_445639 | 11283 | ChEMBL |
| *Butea monosperma* | Seed | CID_445639 | 1129 | ChEMBL |
| *Butea monosperma* | Seed | CID_445639 | 113612 | ChEMBL |
| *Butea monosperma* | Seed | CID_445639 | 120227 | ChEMBL |
| *Butea monosperma* | Seed | CID_445639 | 122706 | ChEMBL \| NPASS |
| *Butea monosperma* | Seed | CID_445639 | 140 | ChEMBL |
| *Butea monosperma* | Seed | CID_445639 | 143471 | ChEMBL \| NPASS |
| *Butea monosperma* | Seed | CID_445639 | 148 | ChEMBL |
| *Butea monosperma* | Seed | CID_445639 | 150 | ChEMBL |
| *Butea monosperma* | Seed | CID_445639 | 1543 | ChEMBL |
| *Butea monosperma* | Seed | CID_445639 | 1544 | ChEMBL |
| *Butea monosperma* | Seed | CID_445639 | 1545 | ChEMBL |
| *Butea monosperma* | Seed | CID_445639 | 1548 | ChEMBL |
| *Butea monosperma* | Seed | CID_445639 | 1549 | ChEMBL |
| *Butea monosperma* | Seed | CID_445639 | 1553 | ChEMBL |
| *Butea monosperma* | Seed | CID_445639 | 1555 | ChEMBL |
| *Butea monosperma* | Seed | CID_445639 | 1557 | ChEMBL |
| *Butea monosperma* | Seed | CID_445639 | 1558 | ChEMBL |
| *Butea monosperma* | Seed | CID_445639 | 1559 | ChEMBL |
| *Butea monosperma* | Seed | CID_445639 | 1562 | ChEMBL |
| *Butea monosperma* | Seed | CID_445639 | 1565 | ChEMBL |
| *Butea monosperma* | Seed | CID_445639 | 1571 | ChEMBL |
| *Butea monosperma* | Seed | CID_445639 | 1572 | ChEMBL |
| *Butea monosperma* | Seed | CID_445639 | 1573 | ChEMBL |
| *Butea monosperma* | Seed | CID_445639 | 1576 | ChEMBL |
| *Butea monosperma* | Seed | CID_445639 | 1577 | ChEMBL |
| *Butea monosperma* | Seed | CID_445639 | 1579 | ChEMBL |
| *Butea monosperma* | Seed | CID_445639 | 1580 | ChEMBL |
| *Butea monosperma* | Seed | CID_445639 | 1588 | BindingDB \| ChEMBL \| NPASS |
| *Butea monosperma* | Seed | CID_445639 | 1812 | ChEMBL |
| *Butea monosperma* | Seed | CID_445639 | 1814 | ChEMBL |
| *Butea monosperma* | Seed | CID_445639 | 2052 | ChEMBL |
| *Butea monosperma* | Seed | CID_445639 | 2053 | ChEMBL |
| *Butea monosperma* | Seed | CID_445639 | 2099 | ChEMBL |
| *Butea monosperma* | Seed | CID_445639 | 2147 | ChEMBL |
| *Butea monosperma* | Seed | CID_445639 | 2152 | BindingDB \| ChEMBL \| NPASS |
| *Butea monosperma* | Seed | CID_445639 | 2155 | ChEMBL |
| *Butea monosperma* | Seed | CID_445639 | 2166 | ChEMBL \| NPASS |
| *Butea monosperma* | Seed | CID_445639 | 2167 | ChEMBL \| NPASS |
| *Butea monosperma* | Seed | CID_445639 | 2168 | BindingDB \| ChEMBL \| NPASS |
| *Butea monosperma* | Seed | CID_445639 | 2171 | ChEMBL \| NPASS |
| *Butea monosperma* | Seed | CID_445639 | 2554 | ChEMBL |
| *Butea monosperma* | Seed | CID_445639 | 2555 | ChEMBL |
| *Butea monosperma* | Seed | CID_445639 | 2561 | ChEMBL |
| *Butea monosperma* | Seed | CID_445639 | 2566 | ChEMBL |
| *Butea monosperma* | Seed | CID_445639 | 2648 | ChEMBL \| NPASS |
| *Butea monosperma* | Seed | CID_445639 | 284541 | ChEMBL |
| *Butea monosperma* | Seed | CID_445639 | 29785 | ChEMBL |
| *Butea monosperma* | Seed | CID_445639 | 3028 | ChEMBL \| NPASS |
| *Butea monosperma* | Seed | CID_445639 | 328 | NPASS |
| *Butea monosperma* | Seed | CID_445639 | 3315 | NPASS |
| *Butea monosperma* | Seed | CID_445639 | 3350 | ChEMBL |
| *Butea monosperma* | Seed | CID_445639 | 3357 | ChEMBL |
| *Butea monosperma* | Seed | CID_445639 | 367 | ChEMBL |
| *Butea monosperma* | Seed | CID_445639 | 3757 | ChEMBL \| NPASS |
| *Butea monosperma* | Seed | CID_445639 | 3791 | ChEMBL |
| *Butea monosperma* | Seed | CID_445639 | 4051 | ChEMBL |
| *Butea monosperma* | Seed | CID_445639 | 4128 | ChEMBL |
| *Butea monosperma* | Seed | CID_445639 | 4137 | NPASS |
| *Butea monosperma* | Seed | CID_445639 | 4297 | NPASS |
| *Butea monosperma* | Seed | CID_445639 | 43 | ChEMBL |
| *Butea monosperma* | Seed | CID_445639 | 472 | NPASS |
| *Butea monosperma* | Seed | CID_445639 | 4929 | BindingDB \| ChEMBL |
| *Butea monosperma* | Seed | CID_445639 | 4988 | ChEMBL |
| *Butea monosperma* | Seed | CID_445639 | 51053 | ChEMBL \| NPASS |
| *Butea monosperma* | Seed | CID_445639 | 5139 | ChEMBL |
| *Butea monosperma* | Seed | CID_445639 | 5141 | ChEMBL |
| *Butea monosperma* | Seed | CID_445639 | 51426 | NPASS |
| *Butea monosperma* | Seed | CID_445639 | 51548 | ChEMBL |
| *Butea monosperma* | Seed | CID_445639 | 5241 | ChEMBL |
| *Butea monosperma* | Seed | CID_445639 | 5347 | ChEMBL \| NPASS |
| *Butea monosperma* | Seed | CID_445639 | 5465 | ChEMBL \| NPASS |
| *Butea monosperma* | Seed | CID_445639 | 5467 | ChEMBL \| NPASS |
| *Butea monosperma* | Seed | CID_445639 | 5468 | BindingDB \| ChEMBL \| NPASS |
| *Butea monosperma* | Seed | CID_445639 | 54905 | ChEMBL |
| *Butea monosperma* | Seed | CID_445639 | 5578 | ChEMBL \| NPASS |
| *Butea monosperma* | Seed | CID_445639 | 5682 | ChEMBL \| NPASS |
| *Butea monosperma* | Seed | CID_445639 | 5683 | ChEMBL \| NPASS |
| *Butea monosperma* | Seed | CID_445639 | 5684 | ChEMBL \| NPASS |
| *Butea monosperma* | Seed | CID_445639 | 5685 | ChEMBL \| NPASS |
| *Butea monosperma* | Seed | CID_445639 | 5686 | ChEMBL \| NPASS |
| *Butea monosperma* | Seed | CID_445639 | 5687 | ChEMBL \| NPASS |
| *Butea monosperma* | Seed | CID_445639 | 5688 | ChEMBL \| NPASS |
| *Butea monosperma* | Seed | CID_445639 | 5689 | ChEMBL \| NPASS |
| *Butea monosperma* | Seed | CID_445639 | 5690 | ChEMBL \| NPASS |
| *Butea monosperma* | Seed | CID_445639 | 5691 | ChEMBL \| NPASS |
| *Butea monosperma* | Seed | CID_445639 | 5692 | ChEMBL \| NPASS |
| *Butea monosperma* | Seed | CID_445639 | 5693 | ChEMBL \| NPASS |
| *Butea monosperma* | Seed | CID_445639 | 5694 | ChEMBL \| NPASS |
| *Butea monosperma* | Seed | CID_445639 | 5695 | ChEMBL \| NPASS |
| *Butea monosperma* | Seed | CID_445639 | 5696 | ChEMBL \| NPASS |
| *Butea monosperma* | Seed | CID_445639 | 5698 | ChEMBL \| NPASS |
| *Butea monosperma* | Seed | CID_445639 | 5699 | ChEMBL \| NPASS |
| *Butea monosperma* | Seed | CID_445639 | 5742 | ChEMBL |
| *Butea monosperma* | Seed | CID_445639 | 5745 | NPASS |
| *Butea monosperma* | Seed | CID_445639 | 5770 | BindingDB \| ChEMBL \| NPASS |
| *Butea monosperma* | Seed | CID_445639 | 6097 | NPASS |
| *Butea monosperma* | Seed | CID_445639 | 641 | NPASS |
| *Butea monosperma* | Seed | CID_445639 | 64816 | ChEMBL |
| *Butea monosperma* | Seed | CID_445639 | 6530 | ChEMBL |
| *Butea monosperma* | Seed | CID_445639 | 6531 | ChEMBL |
| *Butea monosperma* | Seed | CID_445639 | 6532 | ChEMBL |
| *Butea monosperma* | Seed | CID_445639 | 6646 | BindingDB \| NPASS |
| *Butea monosperma* | Seed | CID_445639 | 6915 | ChEMBL |
| *Butea monosperma* | Seed | CID_445639 | 7015 | ChEMBL \| NPASS |
| *Butea monosperma* | Seed | CID_445639 | 7150 | ChEMBL \| NPASS |
| *Butea monosperma* | Seed | CID_445639 | 7398 | ChEMBL \| NPASS |
| *Butea monosperma* | Seed | CID_445639 | 79915 | NPASS |
| *Butea monosperma* | Seed | CID_445639 | 8529 | ChEMBL |
| *Butea monosperma* | Seed | CID_445639 | 9682 | ChEMBL \| NPASS |
| *Butea monosperma* | Seed | CID_445639 | 9971 | NPASS |
| *Butea monosperma* | Seed | CID_5280934 | 10073 | NPASS |
| *Butea monosperma* | Seed | CID_5280934 | 10919 | ChEMBL \| NPASS |
| *Butea monosperma* | Seed | CID_5280934 | 11201 | ChEMBL \| NPASS |
| *Butea monosperma* | Seed | CID_5280934 | 11255 | ChEMBL |
| *Butea monosperma* | Seed | CID_5280934 | 1128 | ChEMBL |
| *Butea monosperma* | Seed | CID_5280934 | 1129 | ChEMBL |
| *Butea monosperma* | Seed | CID_5280934 | 1131 | ChEMBL |
| *Butea monosperma* | Seed | CID_5280934 | 1137 | ChEMBL |
| *Butea monosperma* | Seed | CID_5280934 | 1268 | ChEMBL |
| *Butea monosperma* | Seed | CID_5280934 | 134 | ChEMBL |
| *Butea monosperma* | Seed | CID_5280934 | 135 | ChEMBL |
| *Butea monosperma* | Seed | CID_5280934 | 140 | ChEMBL |
| *Butea monosperma* | Seed | CID_5280934 | 148 | ChEMBL |
| *Butea monosperma* | Seed | CID_5280934 | 150 | ChEMBL |
| *Butea monosperma* | Seed | CID_5280934 | 151 | ChEMBL |
| *Butea monosperma* | Seed | CID_5280934 | 152 | ChEMBL |
| *Butea monosperma* | Seed | CID_5280934 | 153 | ChEMBL |
| *Butea monosperma* | Seed | CID_5280934 | 154 | ChEMBL |
| *Butea monosperma* | Seed | CID_5280934 | 1576 | ChEMBL \| NPASS |
| *Butea monosperma* | Seed | CID_5280934 | 1588 | BindingDB \| ChEMBL \| NPASS |
| *Butea monosperma* | Seed | CID_5280934 | 1812 | ChEMBL |
| *Butea monosperma* | Seed | CID_5280934 | 1813 | ChEMBL |
| *Butea monosperma* | Seed | CID_5280934 | 1814 | ChEMBL |
| *Butea monosperma* | Seed | CID_5280934 | 185 | ChEMBL |
| *Butea monosperma* | Seed | CID_5280934 | 1909 | ChEMBL |
| *Butea monosperma* | Seed | CID_5280934 | 200316 | NPASS |
| *Butea monosperma* | Seed | CID_5280934 | 2030 | ChEMBL |
| *Butea monosperma* | Seed | CID_5280934 | 2152 | BindingDB \| ChEMBL \| NPASS |
| *Butea monosperma* | Seed | CID_5280934 | 2155 | ChEMBL |
| *Butea monosperma* | Seed | CID_5280934 | 216 | ChEMBL \| NPASS |
| *Butea monosperma* | Seed | CID_5280934 | 23192 | BindingDB \| ChEMBL \| NPASS |
| *Butea monosperma* | Seed | CID_5280934 | 246 | ChEMBL \| NPASS |
| *Butea monosperma* | Seed | CID_5280934 | 247 | ChEMBL \| NPASS |
| *Butea monosperma* | Seed | CID_5280934 | 2554 | ChEMBL |
| *Butea monosperma* | Seed | CID_5280934 | 2555 | ChEMBL |
| *Butea monosperma* | Seed | CID_5280934 | 2561 | ChEMBL |
| *Butea monosperma* | Seed | CID_5280934 | 2566 | ChEMBL |
| *Butea monosperma* | Seed | CID_5280934 | 26013 | NPASS |
| *Butea monosperma* | Seed | CID_5280934 | 2648 | ChEMBL \| NPASS |
| *Butea monosperma* | Seed | CID_5280934 | 2693 | ChEMBL |
| *Butea monosperma* | Seed | CID_5280934 | 2744 | ChEMBL \| NPASS |
| *Butea monosperma* | Seed | CID_5280934 | 2862 | ChEMBL |
| *Butea monosperma* | Seed | CID_5280934 | 2864 | ChEMBL \| NPASS |
| *Butea monosperma* | Seed | CID_5280934 | 2908 | ChEMBL \| NPASS |
| *Butea monosperma* | Seed | CID_5280934 | 3028 | ChEMBL \| NPASS |
| *Butea monosperma* | Seed | CID_5280934 | 3248 | NPASS |
| *Butea monosperma* | Seed | CID_5280934 | 3269 | ChEMBL |
| *Butea monosperma* | Seed | CID_5280934 | 3274 | ChEMBL |
| *Butea monosperma* | Seed | CID_5280934 | 328 | NPASS |
| *Butea monosperma* | Seed | CID_5280934 | 3315 | NPASS |
| *Butea monosperma* | Seed | CID_5280934 | 3350 | ChEMBL |
| *Butea monosperma* | Seed | CID_5280934 | 3356 | ChEMBL |
| *Butea monosperma* | Seed | CID_5280934 | 3357 | ChEMBL |
| *Butea monosperma* | Seed | CID_5280934 | 3358 | ChEMBL |
| *Butea monosperma* | Seed | CID_5280934 | 3359 | ChEMBL |
| *Butea monosperma* | Seed | CID_5280934 | 338557 | BindingDB \| ChEMBL \| NPASS |
| *Butea monosperma* | Seed | CID_5280934 | 367 | NPASS |
| *Butea monosperma* | Seed | CID_5280934 | 3725 | NPASS |
| *Butea monosperma* | Seed | CID_5280934 | 3757 | ChEMBL |
| *Butea monosperma* | Seed | CID_5280934 | 3837 | NPASS |
| *Butea monosperma* | Seed | CID_5280934 | 390245 | NPASS |
| *Butea monosperma* | Seed | CID_5280934 | 4000 | NPASS |
| *Butea monosperma* | Seed | CID_5280934 | 4128 | ChEMBL |
| *Butea monosperma* | Seed | CID_5280934 | 4137 | ChEMBL \| NPASS |
| *Butea monosperma* | Seed | CID_5280934 | 4159 | ChEMBL |
| *Butea monosperma* | Seed | CID_5280934 | 4297 | NPASS |
| *Butea monosperma* | Seed | CID_5280934 | 4780 | ChEMBL \| NPASS |
| *Butea monosperma* | Seed | CID_5280934 | 4790 | NPASS |
| *Butea monosperma* | Seed | CID_5280934 | 4886 | ChEMBL |
| *Butea monosperma* | Seed | CID_5280934 | 4985 | ChEMBL |
| *Butea monosperma* | Seed | CID_5280934 | 4986 | ChEMBL |
| *Butea monosperma* | Seed | CID_5280934 | 4988 | ChEMBL |
| *Butea monosperma* | Seed | CID_5280934 | 51053 | NPASS |
| *Butea monosperma* | Seed | CID_5280934 | 5139 | ChEMBL |
| *Butea monosperma* | Seed | CID_5280934 | 51426 | ChEMBL \| NPASS |
| *Butea monosperma* | Seed | CID_5280934 | 5144 | ChEMBL |
| *Butea monosperma* | Seed | CID_5280934 | 5241 | ChEMBL |
| *Butea monosperma* | Seed | CID_5280934 | 5300 | NPASS |
| *Butea monosperma* | Seed | CID_5280934 | 5319 | BindingDB \| NPASS |
| *Butea monosperma* | Seed | CID_5280934 | 5347 | NPASS |
| *Butea monosperma* | Seed | CID_5280934 | 5429 | ChEMBL \| NPASS |
| *Butea monosperma* | Seed | CID_5280934 | 5465 | ChEMBL \| NPASS |
| *Butea monosperma* | Seed | CID_5280934 | 5467 | ChEMBL \| NPASS |
| *Butea monosperma* | Seed | CID_5280934 | 5468 | BindingDB \| ChEMBL \| NPASS |
| *Butea monosperma* | Seed | CID_5280934 | 552 | ChEMBL |
| *Butea monosperma* | Seed | CID_5280934 | 554 | ChEMBL |
| *Butea monosperma* | Seed | CID_5280934 | 55775 | ChEMBL \| NPASS |
| *Butea monosperma* | Seed | CID_5280934 | 5742 | ChEMBL \| NPASS |
| *Butea monosperma* | Seed | CID_5280934 | 5743 | ChEMBL \| NPASS |
| *Butea monosperma* | Seed | CID_5280934 | 5965 | ChEMBL \| NPASS |
| *Butea monosperma* | Seed | CID_5280934 | 5999 | NPASS |
| *Butea monosperma* | Seed | CID_5280934 | 60489 | NPASS |
| *Butea monosperma* | Seed | CID_5280934 | 624 | ChEMBL |
| *Butea monosperma* | Seed | CID_5280934 | 6256 | NPASS |
| *Butea monosperma* | Seed | CID_5280934 | 641 | NPASS |
| *Butea monosperma* | Seed | CID_5280934 | 6530 | ChEMBL |
| *Butea monosperma* | Seed | CID_5280934 | 6531 | ChEMBL |
| *Butea monosperma* | Seed | CID_5280934 | 6532 | ChEMBL |
| *Butea monosperma* | Seed | CID_5280934 | 6915 | ChEMBL |
| *Butea monosperma* | Seed | CID_5280934 | 7015 | ChEMBL \| NPASS |
| *Butea monosperma* | Seed | CID_5280934 | 7398 | NPASS |
| *Butea monosperma* | Seed | CID_5280934 | 7421 | ChEMBL \| NPASS |
| *Butea monosperma* | Seed | CID_5280934 | 7486 | ChEMBL \| NPASS |
| *Butea monosperma* | Seed | CID_5280934 | 865 | NPASS |
| *Butea monosperma* | Seed | CID_5280934 | 8856 | ChEMBL \| NPASS |
| *Butea monosperma* | Seed | CID_5280934 | 886 | ChEMBL |
| *Butea monosperma* | Seed | CID_5280934 | 887 | ChEMBL |
| *Butea monosperma* | Seed | CID_5280934 | 9971 | NPASS |
| *Butea monosperma* | Seed | CID_5280450 | 1576 | NPASS |
| *Butea monosperma* | Seed | CID_5280450 | 1588 | BindingDB \| ChEMBL \| NPASS |
| *Butea monosperma* | Seed | CID_5280450 | 1845 | BindingDB \| ChEMBL \| NPASS |
| *Butea monosperma* | Seed | CID_5280450 | 2099 | ChEMBL \| NPASS |
| *Butea monosperma* | Seed | CID_5280450 | 2152 | BindingDB \| ChEMBL \| NPASS |
| *Butea monosperma* | Seed | CID_5280450 | 2155 | ChEMBL |
| *Butea monosperma* | Seed | CID_5280450 | 216 | ChEMBL \| NPASS |
| *Butea monosperma* | Seed | CID_5280450 | 2167 | ChEMBL \| NPASS |
| *Butea monosperma* | Seed | CID_5280450 | 2170 | BindingDB \| ChEMBL \| NPASS |
| *Butea monosperma* | Seed | CID_5280450 | 246 | ChEMBL \| NPASS |
| *Butea monosperma* | Seed | CID_5280450 | 2648 | NPASS |
| *Butea monosperma* | Seed | CID_5280450 | 2769 | ChEMBL |
| *Butea monosperma* | Seed | CID_5280450 | 2864 | ChEMBL \| NPASS |
| *Butea monosperma* | Seed | CID_5280450 | 3028 | NPASS |
| *Butea monosperma* | Seed | CID_5280450 | 3248 | ChEMBL \| NPASS |
| *Butea monosperma* | Seed | CID_5280450 | 3315 | NPASS |
| *Butea monosperma* | Seed | CID_5280450 | 338557 | BindingDB \| ChEMBL \| NPASS |
| *Butea monosperma* | Seed | CID_5280450 | 367 | NPASS |
| *Butea monosperma* | Seed | CID_5280450 | 3725 | NPASS |
| *Butea monosperma* | Seed | CID_5280450 | 4000 | NPASS |
| *Butea monosperma* | Seed | CID_5280450 | 4780 | NPASS |
| *Butea monosperma* | Seed | CID_5280450 | 4790 | NPASS |
| *Butea monosperma* | Seed | CID_5280450 | 4929 | BindingDB \| ChEMBL |
| *Butea monosperma* | Seed | CID_5280450 | 51053 | NPASS |
| *Butea monosperma* | Seed | CID_5280450 | 51426 | NPASS |
| *Butea monosperma* | Seed | CID_5280450 | 51548 | ChEMBL |
| *Butea monosperma* | Seed | CID_5280450 | 5465 | ChEMBL \| NPASS |
| *Butea monosperma* | Seed | CID_5280450 | 5467 | ChEMBL \| NPASS |
| *Butea monosperma* | Seed | CID_5280450 | 5468 | BindingDB \| ChEMBL \| NPASS |
| *Butea monosperma* | Seed | CID_5280450 | 55775 | NPASS |
| *Butea monosperma* | Seed | CID_5280450 | 5742 | ChEMBL \| NPASS |
| *Butea monosperma* | Seed | CID_5280450 | 5778 | BindingDB \| ChEMBL \| NPASS |
| *Butea monosperma* | Seed | CID_5280450 | 6256 | NPASS |
| *Butea monosperma* | Seed | CID_5280450 | 6646 | BindingDB \| NPASS |
| *Butea monosperma* | Seed | CID_5280450 | 7015 | ChEMBL \| NPASS |
| *Butea monosperma* | Seed | CID_5280450 | 7253 | NPASS |
| *Butea monosperma* | Seed | CID_5280450 | 7398 | NPASS |
| *Butea monosperma* | Seed | CID_5280450 | 8644 | BindingDB \| ChEMBL \| NPASS |
| *Butea monosperma* | Seed | CID_5280450 | 9971 | NPASS |
| *Butea monosperma* | Seed | CID_5742590 | 10013 | ChEMBL |
| *Butea monosperma* | Seed | CID_5742590 | 2932 | ChEMBL |
| *Butea monosperma* | Seed | CID_5742590 | 3417 | ChEMBL \| NPASS |
| *Butea monosperma* | Seed | CID_5742590 | 51422 | ChEMBL |
| *Butea monosperma* | Seed | CID_5742590 | 53632 | ChEMBL |
| *Butea monosperma* | Seed | CID_5742590 | 5423 | BindingDB \| ChEMBL \| NPASS |
| *Butea monosperma* | Seed | CID_5742590 | 5562 | ChEMBL |
| *Butea monosperma* | Seed | CID_5742590 | 5563 | ChEMBL |
| *Butea monosperma* | Seed | CID_5742590 | 5564 | ChEMBL |
| *Butea monosperma* | Seed | CID_5742590 | 5565 | ChEMBL |
| *Butea monosperma* | Seed | CID_5742590 | 5571 | ChEMBL |
| *Butea monosperma* | Seed | CID_10467 | 5465 | ChEMBL \| NPASS |
| *Butea monosperma* | Seed | CID_10467 | 5467 | ChEMBL \| NPASS |
| *Butea monosperma* | Seed | CID_10467 | 5468 | BindingDB \| ChEMBL \| NPASS |
| *Butea monosperma* | Seed | CID_10467 | 5770 | BindingDB \| ChEMBL \| NPASS |
| *Butea monosperma* | Seed | CID_8215 | 10013 | ChEMBL |
| *Butea monosperma* | Seed | CID_8215 | 5465 | ChEMBL \| NPASS |
| *Butea monosperma* | Seed | CID_8215 | 5467 | ChEMBL \| NPASS |
| *Butea monosperma* | Seed | CID_8215 | 5468 | BindingDB \| ChEMBL \| NPASS |
| *Butea monosperma* | Seed | CID_73170 | 151306 | BindingDB \| ChEMBL \| NPASS |
| *Butea monosperma* | Seed | CID_73170 | 240 | BindingDB \| ChEMBL \| NPASS |
| *Butea monosperma* | Seed | CID_73170 | 247 | ChEMBL \| NPASS |
| *Butea monosperma* | Seed | CID_73170 | 3417 | ChEMBL \| NPASS |
| *Butea monosperma* | Seed | CID_73170 | 5743 | ChEMBL \| NPASS |
| *Butea monosperma* | Seed | CID_73170 | 5770 | BindingDB \| ChEMBL \| NPASS |
| *Butea monosperma* | Seed | CID_73170 | 7150 | NPASS |
| *Butea monosperma* | Seed | CID_73170 | 7153 | NPASS |
| *Butea monosperma* | Seed | CID_73170 | 9971 | NPASS |
| *Butea monosperma* | Seed | CID_2969 | 2908 | NPASS |
| *Butea monosperma* | Seed | CID_2969 | 3315 | NPASS |
| *Butea monosperma* | Seed | CID_2969 | 367 | NPASS |
| *Butea monosperma* | Seed | CID_2969 | 4780 | NPASS |
| *Butea monosperma* | Seed | CID_2969 | 51053 | NPASS |
| *Butea monosperma* | Seed | CID_2969 | 53831 | ChEMBL \| NPASS |
| *Butea monosperma* | Seed | CID_2969 | 54575 | ChEMBL \| NPASS |
| *Butea monosperma* | Seed | CID_2969 | 54576 | ChEMBL \| NPASS |
| *Butea monosperma* | Seed | CID_2969 | 5465 | ChEMBL \| NPASS |
| *Butea monosperma* | Seed | CID_2969 | 54659 | ChEMBL \| NPASS |
| *Butea monosperma* | Seed | CID_2969 | 5467 | ChEMBL \| NPASS |
| *Butea monosperma* | Seed | CID_2969 | 5468 | ChEMBL \| NPASS |
| *Butea monosperma* | Seed | CID_2969 | 55775 | NPASS |
| *Butea monosperma* | Seed | CID_2969 | 6097 | NPASS |
| *Butea monosperma* | Seed | CID_2969 | 7253 | ChEMBL \| NPASS |
| *Butea monosperma* | Seed | CID_379 | 216 | NPASS |
| *Butea monosperma* | Seed | CID_379 | 2908 | NPASS |
| *Butea monosperma* | Seed | CID_379 | 4000 | NPASS |
| *Butea monosperma* | Seed | CID_379 | 51053 | NPASS |
| *Butea monosperma* | Seed | CID_379 | 53831 | ChEMBL \| NPASS |
| *Butea monosperma* | Seed | CID_379 | 54575 | ChEMBL \| NPASS |
| *Butea monosperma* | Seed | CID_379 | 54576 | ChEMBL \| NPASS |
| *Butea monosperma* | Seed | CID_379 | 55867 | NPASS |
| *Butea monosperma* | Seed | CID_379 | 7253 | NPASS |
| *Butea monosperma* | Seed | CID_379 | 9356 | BindingDB \| ChEMBL \| NPASS |
| *Butea monosperma* | Seed | CID_379 | 9376 | BindingDB \| ChEMBL \| NPASS |
| *Butea monosperma* | Seed | CID_3893 | 10013 | ChEMBL |
| *Butea monosperma* | Seed | CID_3893 | 10951 | NPASS |
| *Butea monosperma* | Seed | CID_3893 | 11069 | NPASS |
| *Butea monosperma* | Seed | CID_3893 | 1571 | ChEMBL |
| *Butea monosperma* | Seed | CID_3893 | 2099 | NPASS |
| *Butea monosperma* | Seed | CID_3893 | 216 | NPASS |
| *Butea monosperma* | Seed | CID_3893 | 2740 | NPASS |
| *Butea monosperma* | Seed | CID_3893 | 2778 | NPASS |
| *Butea monosperma* | Seed | CID_3893 | 2864 | ChEMBL \| NPASS |
| *Butea monosperma* | Seed | CID_3893 | 338557 | ChEMBL \| NPASS |
| *Butea monosperma* | Seed | CID_3893 | 3576 | NPASS |
| *Butea monosperma* | Seed | CID_3893 | 367 | NPASS |
| *Butea monosperma* | Seed | CID_3893 | 4790 | NPASS |
| *Butea monosperma* | Seed | CID_3893 | 53831 | ChEMBL \| NPASS |
| *Butea monosperma* | Seed | CID_3893 | 54575 | ChEMBL \| NPASS |
| *Butea monosperma* | Seed | CID_3893 | 54576 | ChEMBL \| NPASS |
| *Butea monosperma* | Seed | CID_3893 | 5465 | ChEMBL \| NPASS |
| *Butea monosperma* | Seed | CID_3893 | 54659 | ChEMBL \| NPASS |
| *Butea monosperma* | Seed | CID_3893 | 5467 | ChEMBL \| NPASS |
| *Butea monosperma* | Seed | CID_3893 | 5468 | BindingDB \| ChEMBL \| NPASS |
| *Butea monosperma* | Seed | CID_3893 | 6097 | NPASS |
| *Butea monosperma* | Seed | CID_3893 | 6256 | NPASS |
| *Butea monosperma* | Seed | CID_3893 | 7068 | NPASS |
| *Butea monosperma* | Seed | CID_3893 | 7253 | NPASS |
| *Butea monosperma* | Seed | CID_3893 | 9971 | NPASS |
| *Diospyros malabarica* | Fruit | CID_222284 | 10013 | ChEMBL |
| *Diospyros malabarica* | Fruit | CID_222284 | 10599 | ChEMBL \| NPASS |
| *Diospyros malabarica* | Fruit | CID_222284 | 1565 | ChEMBL \| NPASS |
| *Diospyros malabarica* | Fruit | CID_222284 | 1576 | BindingDB \| ChEMBL \| NPASS |
| *Diospyros malabarica* | Fruit | CID_222284 | 1803 | ChEMBL \| NPASS |
| *Diospyros malabarica* | Fruit | CID_222284 | 2147 | ChEMBL \| NPASS |
| *Diospyros malabarica* | Fruit | CID_222284 | 28234 | ChEMBL \| NPASS |
| *Diospyros malabarica* | Fruit | CID_222284 | 2932 | ChEMBL |
| *Diospyros malabarica* | Fruit | CID_222284 | 3417 | ChEMBL \| NPASS |
| *Diospyros malabarica* | Fruit | CID_222284 | 51422 | ChEMBL |
| *Diospyros malabarica* | Fruit | CID_222284 | 5243 | ChEMBL |
| *Diospyros malabarica* | Fruit | CID_222284 | 53632 | ChEMBL |
| *Diospyros malabarica* | Fruit | CID_222284 | 5423 | BindingDB \| ChEMBL \| NPASS |
| *Diospyros malabarica* | Fruit | CID_222284 | 5562 | ChEMBL |
| *Diospyros malabarica* | Fruit | CID_222284 | 5563 | ChEMBL |
| *Diospyros malabarica* | Fruit | CID_222284 | 5564 | ChEMBL |
| *Diospyros malabarica* | Fruit | CID_222284 | 5565 | ChEMBL |
| *Diospyros malabarica* | Fruit | CID_222284 | 5571 | ChEMBL |
| *Diospyros malabarica* | Fruit | CID_222284 | 7299 | ChEMBL \| NPASS |
| *Diospyros malabarica* | Fruit | CID_5742590 | 10013 | ChEMBL |
| *Diospyros malabarica* | Fruit | CID_5742590 | 2932 | ChEMBL |
| *Diospyros malabarica* | Fruit | CID_5742590 | 3417 | ChEMBL \| NPASS |
| *Diospyros malabarica* | Fruit | CID_5742590 | 51422 | ChEMBL |
| *Diospyros malabarica* | Fruit | CID_5742590 | 53632 | ChEMBL |
| *Diospyros malabarica* | Fruit | CID_5742590 | 5423 | BindingDB \| ChEMBL \| NPASS |
| *Diospyros malabarica* | Fruit | CID_5742590 | 5562 | ChEMBL |
| *Diospyros malabarica* | Fruit | CID_5742590 | 5563 | ChEMBL |
| *Diospyros malabarica* | Fruit | CID_5742590 | 5564 | ChEMBL |
| *Diospyros malabarica* | Fruit | CID_5742590 | 5565 | ChEMBL |
| *Diospyros malabarica* | Fruit | CID_5742590 | 5571 | ChEMBL |
| *Diospyros malabarica* | Fruit | CID_73145 | 10599 | ChEMBL \| NPASS |
| *Diospyros malabarica* | Fruit | CID_73145 | 240 | BindingDB \| ChEMBL \| NPASS |
| *Diospyros malabarica* | Fruit | CID_73145 | 247 | BindingDB \| ChEMBL \| NPASS |
| *Diospyros malabarica* | Fruit | CID_73145 | 28234 | ChEMBL \| NPASS |
| *Diospyros malabarica* | Fruit | CID_73145 | 5743 | ChEMBL \| NPASS |
| *Diospyros malabarica* | Fruit | CID_73145 | 5770 | BindingDB \| ChEMBL \| NPASS |
| *Diospyros malabarica* | Fruit | CID_259846 | 1066 | ChEMBL \| NPASS |
| *Diospyros malabarica* | Fruit | CID_259846 | 387 | ChEMBL |
| *Diospyros malabarica* | Fruit | CID_259846 | 5770 | BindingDB \| ChEMBL \| NPASS |
| *Diospyros malabarica* | Fruit | CID_259846 | 5879 | ChEMBL |
| *Diospyros malabarica* | Fruit | CID_259846 | 7153 | ChEMBL \| NPASS |
| *Diospyros malabarica* | Fruit | CID_259846 | 8824 | BindingDB \| ChEMBL \| NPASS |
| *Diospyros malabarica* | Fruit | CID_259846 | 998 | ChEMBL |
| *Diospyros malabarica* | Fruit | CID_370 | 10013 | ChEMBL |
| *Diospyros malabarica* | Fruit | CID_370 | 10599 | ChEMBL \| NPASS |
| *Diospyros malabarica* | Fruit | CID_370 | 10919 | ChEMBL \| NPASS |
| *Diospyros malabarica* | Fruit | CID_370 | 11238 | BindingDB \| ChEMBL \| NPASS |
| *Diospyros malabarica* | Fruit | CID_370 | 118429 | NPASS |
| *Diospyros malabarica* | Fruit | CID_370 | 1588 | ChEMBL \| NPASS |
| *Diospyros malabarica* | Fruit | CID_370 | 1636 | BindingDB \| NPASS |
| *Diospyros malabarica* | Fruit | CID_370 | 1803 | BindingDB \| ChEMBL \| NPASS |
| *Diospyros malabarica* | Fruit | CID_370 | 2099 | ChEMBL \| NPASS |
| *Diospyros malabarica* | Fruit | CID_370 | 216 | ChEMBL \| NPASS |
| *Diospyros malabarica* | Fruit | CID_370 | 231 | NPASS |
| *Diospyros malabarica* | Fruit | CID_370 | 23632 | BindingDB \| ChEMBL \| NPASS |
| *Diospyros malabarica* | Fruit | CID_370 | 2526 | ChEMBL \| NPASS |
| *Diospyros malabarica* | Fruit | CID_370 | 2529 | ChEMBL \| NPASS |
| *Diospyros malabarica* | Fruit | CID_370 | 2590 | ChEMBL \| NPASS |
| *Diospyros malabarica* | Fruit | CID_370 | 259285 | ChEMBL \| NPASS |
| *Diospyros malabarica* | Fruit | CID_370 | 2597 | ChEMBL \| NPASS |
| *Diospyros malabarica* | Fruit | CID_370 | 2648 | NPASS |
| *Diospyros malabarica* | Fruit | CID_370 | 2744 | ChEMBL \| NPASS |
| *Diospyros malabarica* | Fruit | CID_370 | 28234 | ChEMBL \| NPASS |
| *Diospyros malabarica* | Fruit | CID_370 | 2908 | NPASS |
| *Diospyros malabarica* | Fruit | CID_370 | 3028 | ChEMBL \| NPASS |
| *Diospyros malabarica* | Fruit | CID_370 | 3248 | ChEMBL \| NPASS |
| *Diospyros malabarica* | Fruit | CID_370 | 328 | NPASS |
| *Diospyros malabarica* | Fruit | CID_370 | 3315 | NPASS |
| *Diospyros malabarica* | Fruit | CID_370 | 3417 | ChEMBL \| NPASS |
| *Diospyros malabarica* | Fruit | CID_370 | 351 | BindingDB \| ChEMBL |
| *Diospyros malabarica* | Fruit | CID_370 | 3576 | NPASS |
| *Diospyros malabarica* | Fruit | CID_370 | 367 | NPASS |
| *Diospyros malabarica* | Fruit | CID_370 | 3725 | NPASS |
| *Diospyros malabarica* | Fruit | CID_370 | 390245 | ChEMBL \| NPASS |
| *Diospyros malabarica* | Fruit | CID_370 | 4000 | NPASS |
| *Diospyros malabarica* | Fruit | CID_370 | 4137 | ChEMBL \| NPASS |
| *Diospyros malabarica* | Fruit | CID_370 | 4316 | BindingDB |
| *Diospyros malabarica* | Fruit | CID_370 | 4780 | ChEMBL \| NPASS |
| *Diospyros malabarica* | Fruit | CID_370 | 5054 | BindingDB |
| *Diospyros malabarica* | Fruit | CID_370 | 51426 | ChEMBL \| NPASS |
| *Diospyros malabarica* | Fruit | CID_370 | 54658 | ChEMBL \| NPASS |
| *Diospyros malabarica* | Fruit | CID_370 | 55775 | ChEMBL \| NPASS |
| *Diospyros malabarica* | Fruit | CID_370 | 5594 | NPASS |
| *Diospyros malabarica* | Fruit | CID_370 | 5693 | BindingDB \| NPASS |
| *Diospyros malabarica* | Fruit | CID_370 | 5743 | ChEMBL \| NPASS |
| *Diospyros malabarica* | Fruit | CID_370 | 6401 | BindingDB \| NPASS |
| *Diospyros malabarica* | Fruit | CID_370 | 6402 | BindingDB \| ChEMBL \| NPASS |
| *Diospyros malabarica* | Fruit | CID_370 | 6403 | BindingDB \| ChEMBL \| NPASS |
| *Diospyros malabarica* | Fruit | CID_370 | 6606 | NPASS |
| *Diospyros malabarica* | Fruit | CID_370 | 6622 | ChEMBL \| NPASS |
| *Diospyros malabarica* | Fruit | CID_370 | 6868 | ChEMBL \| NPASS |
| *Diospyros malabarica* | Fruit | CID_370 | 7157 | ChEMBL \| NPASS |
| *Diospyros malabarica* | Fruit | CID_370 | 7421 | ChEMBL \| NPASS |
| *Diospyros malabarica* | Fruit | CID_370 | 759 | BindingDB \| ChEMBL \| NPASS |
| *Diospyros malabarica* | Fruit | CID_370 | 760 | BindingDB \| ChEMBL \| NPASS |
| *Diospyros malabarica* | Fruit | CID_370 | 761 | BindingDB \| ChEMBL \| NPASS |
| *Diospyros malabarica* | Fruit | CID_370 | 762 | BindingDB \| ChEMBL \| NPASS |
| *Diospyros malabarica* | Fruit | CID_370 | 763 | BindingDB \| ChEMBL \| NPASS |
| *Diospyros malabarica* | Fruit | CID_370 | 765 | ChEMBL \| NPASS |
| *Diospyros malabarica* | Fruit | CID_370 | 766 | BindingDB \| ChEMBL \| NPASS |
| *Diospyros malabarica* | Fruit | CID_370 | 768 | BindingDB \| ChEMBL \| NPASS |
| *Diospyros malabarica* | Fruit | CID_370 | 771 | BindingDB \| ChEMBL \| NPASS |
| *Diospyros malabarica* | Fruit | CID_64971 | 10013 | ChEMBL |
| *Diospyros malabarica* | Fruit | CID_64971 | 10054 | ChEMBL \| NPASS |
| *Diospyros malabarica* | Fruit | CID_64971 | 10055 | BindingDB \| ChEMBL |
| *Diospyros malabarica* | Fruit | CID_64971 | 10062 | ChEMBL \| NPASS |
| *Diospyros malabarica* | Fruit | CID_64971 | 10599 | ChEMBL \| NPASS |
| *Diospyros malabarica* | Fruit | CID_64971 | 1071 | ChEMBL \| NPASS |
| *Diospyros malabarica* | Fruit | CID_64971 | 11069 | NPASS |
| *Diospyros malabarica* | Fruit | CID_64971 | 112398 | ChEMBL \| NPASS |
| *Diospyros malabarica* | Fruit | CID_64971 | 112399 | ChEMBL \| NPASS |
| *Diospyros malabarica* | Fruit | CID_64971 | 151306 | BindingDB \| ChEMBL \| NPASS |
| *Diospyros malabarica* | Fruit | CID_64971 | 1728 | ChEMBL \| NPASS |
| *Diospyros malabarica* | Fruit | CID_64971 | 1803 | ChEMBL \| NPASS |
| *Diospyros malabarica* | Fruit | CID_64971 | 200316 | NPASS |
| *Diospyros malabarica* | Fruit | CID_64971 | 213 | BindingDB \| ChEMBL \| NPASS |
| *Diospyros malabarica* | Fruit | CID_64971 | 231 | ChEMBL \| NPASS |
| *Diospyros malabarica* | Fruit | CID_64971 | 2735 | BindingDB \| ChEMBL \| NPASS |
| *Diospyros malabarica* | Fruit | CID_64971 | 2740 | NPASS |
| *Diospyros malabarica* | Fruit | CID_64971 | 28234 | ChEMBL \| NPASS |
| *Diospyros malabarica* | Fruit | CID_64971 | 2932 | ChEMBL |
| *Diospyros malabarica* | Fruit | CID_64971 | 3417 | ChEMBL \| NPASS |
| *Diospyros malabarica* | Fruit | CID_64971 | 4907 | ChEMBL |
| *Diospyros malabarica* | Fruit | CID_64971 | 51053 | NPASS |
| *Diospyros malabarica* | Fruit | CID_64971 | 51422 | ChEMBL |
| *Diospyros malabarica* | Fruit | CID_64971 | 51426 | NPASS |
| *Diospyros malabarica* | Fruit | CID_64971 | 53632 | ChEMBL |
| *Diospyros malabarica* | Fruit | CID_64971 | 5423 | BindingDB \| ChEMBL \| NPASS |
| *Diospyros malabarica* | Fruit | CID_64971 | 5429 | NPASS |
| *Diospyros malabarica* | Fruit | CID_64971 | 54583 | BindingDB \| ChEMBL \| NPASS |
| *Diospyros malabarica* | Fruit | CID_64971 | 54681 | ChEMBL \| NPASS |
| *Diospyros malabarica* | Fruit | CID_64971 | 5562 | ChEMBL |
| *Diospyros malabarica* | Fruit | CID_64971 | 5563 | ChEMBL |
| *Diospyros malabarica* | Fruit | CID_64971 | 5564 | ChEMBL |
| *Diospyros malabarica* | Fruit | CID_64971 | 5565 | ChEMBL |
| *Diospyros malabarica* | Fruit | CID_64971 | 5571 | ChEMBL |
| *Diospyros malabarica* | Fruit | CID_64971 | 55775 | ChEMBL \| NPASS |
| *Diospyros malabarica* | Fruit | CID_64971 | 5579 | NPASS |
| *Diospyros malabarica* | Fruit | CID_64971 | 5581 | NPASS |
| *Diospyros malabarica* | Fruit | CID_64971 | 57016 | ChEMBL \| NPASS |
| *Diospyros malabarica* | Fruit | CID_64971 | 5745 | NPASS |
| *Diospyros malabarica* | Fruit | CID_64971 | 5770 | BindingDB \| ChEMBL |
| *Diospyros malabarica* | Fruit | CID_64971 | 5970 | ChEMBL \| NPASS |
| *Diospyros malabarica* | Fruit | CID_64971 | 5999 | NPASS |
| *Diospyros malabarica* | Fruit | CID_64971 | 60489 | NPASS |
| *Diospyros malabarica* | Fruit | CID_64971 | 7150 | ChEMBL |
| *Diospyros malabarica* | Fruit | CID_64971 | 7376 | ChEMBL \| NPASS |
| *Diospyros malabarica* | Fruit | CID_64971 | 9971 | NPASS |
| *Diospyros malabarica* | Fruit | CID_72326 | 10013 | ChEMBL |
| *Diospyros malabarica* | Fruit | CID_72326 | 10599 | ChEMBL \| NPASS |
| *Diospyros malabarica* | Fruit | CID_72326 | 151306 | BindingDB \| ChEMBL \| NPASS |
| *Diospyros malabarica* | Fruit | CID_72326 | 28234 | ChEMBL \| NPASS |
| *Diospyros malabarica* | Fruit | CID_72326 | 387 | ChEMBL |
| *Diospyros malabarica* | Fruit | CID_72326 | 4907 | ChEMBL \| NPASS |
| *Diospyros malabarica* | Fruit | CID_72326 | 5579 | BindingDB \| NPASS |
| *Diospyros malabarica* | Fruit | CID_72326 | 5581 | BindingDB \| NPASS |
| *Diospyros malabarica* | Fruit | CID_72326 | 5743 | ChEMBL \| NPASS |
| *Diospyros malabarica* | Fruit | CID_72326 | 5879 | ChEMBL |
| *Diospyros malabarica* | Fruit | CID_72326 | 5970 | ChEMBL \| NPASS |
| *Diospyros malabarica* | Fruit | CID_72326 | 6554 | BindingDB |
| *Diospyros malabarica* | Fruit | CID_72326 | 7153 | ChEMBL \| NPASS |
| *Diospyros malabarica* | Fruit | CID_72326 | 9971 | NPASS |
| *Diospyros malabarica* | Fruit | CID_72326 | 998 | ChEMBL |
| *Berberis aristata* | Stem | CID_2353 | 10013 | ChEMBL |
| *Berberis aristata* | Stem | CID_2353 | 10599 | ChEMBL \| NPASS |
| *Berberis aristata* | Stem | CID_2353 | 11309 | ChEMBL \| NPASS |
| *Berberis aristata* | Stem | CID_2353 | 1544 | ChEMBL \| NPASS |
| *Berberis aristata* | Stem | CID_2353 | 1545 | ChEMBL \| NPASS |
| *Berberis aristata* | Stem | CID_2353 | 1565 | ChEMBL \| NPASS |
| *Berberis aristata* | Stem | CID_2353 | 1576 | NPASS |
| *Berberis aristata* | Stem | CID_2353 | 1646 | ChEMBL \| NPASS |
| *Berberis aristata* | Stem | CID_2353 | 216 | NPASS |
| *Berberis aristata* | Stem | CID_2353 | 23028 | ChEMBL |
| *Berberis aristata* | Stem | CID_2353 | 28234 | ChEMBL \| NPASS |
| *Berberis aristata* | Stem | CID_2353 | 3091 | NPASS |
| *Berberis aristata* | Stem | CID_2353 | 3576 | ChEMBL |
| *Berberis aristata* | Stem | CID_2353 | 3757 | ChEMBL \| NPASS |
| *Berberis aristata* | Stem | CID_2353 | 4137 | NPASS |
| *Berberis aristata* | Stem | CID_2353 | 4297 | NPASS |
| *Berberis aristata* | Stem | CID_2353 | 43 | ChEMBL \| NPASS |
| *Berberis aristata* | Stem | CID_2353 | 4513 | ChEMBL \| NPASS |
| *Berberis aristata* | Stem | CID_2353 | 51422 | ChEMBL |
| *Berberis aristata* | Stem | CID_2353 | 53632 | ChEMBL |
| *Berberis aristata* | Stem | CID_2353 | 5465 | ChEMBL |
| *Berberis aristata* | Stem | CID_2353 | 5467 | ChEMBL |
| *Berberis aristata* | Stem | CID_2353 | 5550 | NPASS |
| *Berberis aristata* | Stem | CID_2353 | 5562 | ChEMBL |
| *Berberis aristata* | Stem | CID_2353 | 5563 | ChEMBL |
| *Berberis aristata* | Stem | CID_2353 | 5564 | ChEMBL |
| *Berberis aristata* | Stem | CID_2353 | 5565 | ChEMBL |
| *Berberis aristata* | Stem | CID_2353 | 5571 | ChEMBL |
| *Berberis aristata* | Stem | CID_2353 | 5770 | ChEMBL \| NPASS |
| *Berberis aristata* | Stem | CID_2353 | 590 | ChEMBL \| NPASS |
| *Berberis aristata* | Stem | CID_2353 | 6606 | NPASS |
| *Berberis aristata* | Stem | CID_2353 | 7015 | ChEMBL \| NPASS |
| *Berberis aristata* | Stem | CID_2353 | 7253 | ChEMBL \| NPASS |
| *Berberis aristata* | Stem | CID_2353 | 8644 | ChEMBL \| NPASS |
| *Berberis aristata* | Stem | CID_2353 | 865 | NPASS |
| *Berberis aristata* | Stem | CID_2353 | 9536 | ChEMBL \| NPASS |
| *Berberis aristata* | Stem | CID_19009 | 10013 | ChEMBL |
| *Berberis aristata* | Stem | CID_19009 | 10599 | ChEMBL \| NPASS |
| *Berberis aristata* | Stem | CID_19009 | 1565 | NPASS |
| *Berberis aristata* | Stem | CID_19009 | 1576 | BindingDB |
| *Berberis aristata* | Stem | CID_19009 | 2033 | BindingDB \| ChEMBL \| NPASS |
| *Berberis aristata* | Stem | CID_19009 | 23028 | BindingDB \| ChEMBL |
| *Berberis aristata* | Stem | CID_19009 | 28234 | ChEMBL \| NPASS |
| *Berberis aristata* | Stem | CID_19009 | 3091 | NPASS |
| *Berberis aristata* | Stem | CID_19009 | 4000 | NPASS |
| *Berberis aristata* | Stem | CID_19009 | 43 | BindingDB \| ChEMBL \| NPASS |
| *Berberis aristata* | Stem | CID_19009 | 5550 | NPASS |
| *Berberis aristata* | Stem | CID_19009 | 5770 | NPASS |
| *Berberis aristata* | Stem | CID_19009 | 590 | BindingDB \| ChEMBL \| NPASS |
| *Berberis aristata* | Stem | CID_19009 | 7398 | NPASS |
| *Berberis aristata* | Stem | CID_72323 | 10919 | NPASS |
| *Berberis aristata* | Stem | CID_72323 | 11201 | NPASS |
| *Berberis aristata* | Stem | CID_72323 | 1576 | BindingDB \| ChEMBL \| NPASS |
| *Berberis aristata* | Stem | CID_72323 | 2237 | NPASS |
| *Berberis aristata* | Stem | CID_72323 | 23028 | BindingDB \| ChEMBL |
| *Berberis aristata* | Stem | CID_72323 | 2548 | NPASS |
| *Berberis aristata* | Stem | CID_72323 | 2629 | NPASS |
| *Berberis aristata* | Stem | CID_72323 | 2744 | NPASS |
| *Berberis aristata* | Stem | CID_72323 | 390245 | NPASS |
| *Berberis aristata* | Stem | CID_72323 | 4000 | NPASS |
| *Berberis aristata* | Stem | CID_72323 | 4137 | NPASS |
| *Berberis aristata* | Stem | CID_72323 | 4154 | NPASS |
| *Berberis aristata* | Stem | CID_72323 | 5423 | NPASS |
| *Berberis aristata* | Stem | CID_72323 | 55775 | NPASS |
| *Berberis aristata* | Stem | CID_72323 | 5893 | ChEMBL \| NPASS |
| *Berberis aristata* | Stem | CID_72323 | 6606 | NPASS |
| *Berberis aristata* | Stem | CID_11066 | 351 | NPASS |
| *Berberis aristata* | Stem | CID_11066 | 43 | BindingDB \| ChEMBL \| NPASS |
| *Berberis aristata* | Stem | CID_11066 | 5550 | NPASS |
| *Berberis aristata* | Stem | CID_11066 | 590 | BindingDB \| ChEMBL \| NPASS |
| *Berberis aristata* | Stem | CID_442333 | 3417 | NPASS |
| *Berberis aristata* | Stem | CID_442333 | 4088 | NPASS |
| *Berberis aristata* | Stem | CID_442333 | 55775 | ChEMBL \| NPASS |
| *Berberis aristata* | Stem | CID_442333 | 5893 | ChEMBL \| NPASS |
| *Berberis aristata* | Stem | CID_442333 | 6311 | ChEMBL \| NPASS |
| *Berberis aristata* | Stem | CID_442333 | 7398 | NPASS |
| *Berberis aristata* | Stem | CID_442333 | 9682 | NPASS |
| *Gymnema sylvestre* | Leaf | CID_11005 | 10411 | NPASS |
| *Gymnema sylvestre* | Leaf | CID_11005 | 10951 | NPASS |
| *Gymnema sylvestre* | Leaf | CID_11005 | 1588 | ChEMBL \| NPASS |
| *Gymnema sylvestre* | Leaf | CID_11005 | 1759 | BindingDB \| ChEMBL \| NPASS |
| *Gymnema sylvestre* | Leaf | CID_11005 | 2237 | NPASS |
| *Gymnema sylvestre* | Leaf | CID_11005 | 23435 | NPASS |
| *Gymnema sylvestre* | Leaf | CID_11005 | 247 | NPASS |
| *Gymnema sylvestre* | Leaf | CID_11005 | 2648 | NPASS |
| *Gymnema sylvestre* | Leaf | CID_11005 | 2740 | NPASS |
| *Gymnema sylvestre* | Leaf | CID_11005 | 3315 | NPASS |
| *Gymnema sylvestre* | Leaf | CID_11005 | 367 | NPASS |
| *Gymnema sylvestre* | Leaf | CID_11005 | 51053 | NPASS |
| *Gymnema sylvestre* | Leaf | CID_11005 | 51548 | BindingDB \| ChEMBL |
| *Gymnema sylvestre* | Leaf | CID_11005 | 5347 | NPASS |
| *Gymnema sylvestre* | Leaf | CID_11005 | 53831 | ChEMBL \| NPASS |
| *Gymnema sylvestre* | Leaf | CID_11005 | 54575 | ChEMBL \| NPASS |
| *Gymnema sylvestre* | Leaf | CID_11005 | 54576 | ChEMBL \| NPASS |
| *Gymnema sylvestre* | Leaf | CID_11005 | 5465 | ChEMBL \| NPASS |
| *Gymnema sylvestre* | Leaf | CID_11005 | 5467 | ChEMBL \| NPASS |
| *Gymnema sylvestre* | Leaf | CID_11005 | 5468 | BindingDB \| ChEMBL \| NPASS |
| *Gymnema sylvestre* | Leaf | CID_11005 | 55775 | NPASS |
| *Gymnema sylvestre* | Leaf | CID_11005 | 6609 | NPASS |
| *Gymnema sylvestre* | Leaf | CID_11005 | 7097 | ChEMBL \| NPASS |
| *Gymnema sylvestre* | Leaf | CID_11005 | 7398 | NPASS |
| *Gymnema sylvestre* | Leaf | CID_11005 | 7421 | NPASS |
| *Gymnema sylvestre* | Leaf | CID_11005 | 8989 | ChEMBL \| NPASS |
| *Gymnema sylvestre* | Leaf | CID_11005 | 9971 | NPASS |
| *Gymnema sylvestre* | Leaf | CID_985 | 1588 | ChEMBL \| NPASS |
| *Gymnema sylvestre* | Leaf | CID_985 | 2099 | ChEMBL \| NPASS |
| *Gymnema sylvestre* | Leaf | CID_985 | 2167 | ChEMBL \| NPASS |
| *Gymnema sylvestre* | Leaf | CID_985 | 2169 | ChEMBL \| NPASS |
| *Gymnema sylvestre* | Leaf | CID_985 | 2170 | BindingDB \| ChEMBL \| NPASS |
| *Gymnema sylvestre* | Leaf | CID_985 | 2171 | ChEMBL \| NPASS |
| *Gymnema sylvestre* | Leaf | CID_985 | 2908 | NPASS |
| *Gymnema sylvestre* | Leaf | CID_985 | 367 | NPASS |
| *Gymnema sylvestre* | Leaf | CID_985 | 4780 | NPASS |
| *Gymnema sylvestre* | Leaf | CID_985 | 51053 | NPASS |
| *Gymnema sylvestre* | Leaf | CID_985 | 5243 | ChEMBL |
| *Gymnema sylvestre* | Leaf | CID_985 | 5465 | ChEMBL \| NPASS |
| *Gymnema sylvestre* | Leaf | CID_985 | 5467 | ChEMBL \| NPASS |
| *Gymnema sylvestre* | Leaf | CID_985 | 5468 | BindingDB \| ChEMBL \| NPASS |
| *Gymnema sylvestre* | Leaf | CID_985 | 5770 | BindingDB \| ChEMBL \| NPASS |
| *Gymnema sylvestre* | Leaf | CID_985 | 7004 | ChEMBL |
| *Gymnema sylvestre* | Leaf | CID_985 | 7097 | ChEMBL \| NPASS |
| *Gymnema sylvestre* | Leaf | CID_985 | 7398 | NPASS |
| *Gymnema sylvestre* | Leaf | CID_985 | 81285 | ChEMBL \| NPASS |
| *Gymnema sylvestre* | Leaf | CID_985 | 9971 | NPASS |
| *Gymnema sylvestre* | Leaf | CID_5281 | 11255 | ChEMBL |
| *Gymnema sylvestre* | Leaf | CID_5281 | 1128 | ChEMBL |
| *Gymnema sylvestre* | Leaf | CID_5281 | 1129 | ChEMBL |
| *Gymnema sylvestre* | Leaf | CID_5281 | 140 | ChEMBL |
| *Gymnema sylvestre* | Leaf | CID_5281 | 148 | ChEMBL |
| *Gymnema sylvestre* | Leaf | CID_5281 | 150 | ChEMBL |
| *Gymnema sylvestre* | Leaf | CID_5281 | 1588 | ChEMBL \| NPASS |
| *Gymnema sylvestre* | Leaf | CID_5281 | 1812 | ChEMBL |
| *Gymnema sylvestre* | Leaf | CID_5281 | 1814 | ChEMBL |
| *Gymnema sylvestre* | Leaf | CID_5281 | 2099 | ChEMBL |
| *Gymnema sylvestre* | Leaf | CID_5281 | 2147 | ChEMBL |
| *Gymnema sylvestre* | Leaf | CID_5281 | 2167 | ChEMBL \| NPASS |
| *Gymnema sylvestre* | Leaf | CID_5281 | 246 | NPASS |
| *Gymnema sylvestre* | Leaf | CID_5281 | 2554 | ChEMBL |
| *Gymnema sylvestre* | Leaf | CID_5281 | 2555 | ChEMBL |
| *Gymnema sylvestre* | Leaf | CID_5281 | 2561 | ChEMBL |
| *Gymnema sylvestre* | Leaf | CID_5281 | 2566 | ChEMBL |
| *Gymnema sylvestre* | Leaf | CID_5281 | 2908 | NPASS |
| *Gymnema sylvestre* | Leaf | CID_5281 | 3028 | ChEMBL \| NPASS |
| *Gymnema sylvestre* | Leaf | CID_5281 | 328 | NPASS |
| *Gymnema sylvestre* | Leaf | CID_5281 | 3350 | ChEMBL |
| *Gymnema sylvestre* | Leaf | CID_5281 | 3357 | ChEMBL |
| *Gymnema sylvestre* | Leaf | CID_5281 | 367 | ChEMBL |
| *Gymnema sylvestre* | Leaf | CID_5281 | 3757 | ChEMBL |
| *Gymnema sylvestre* | Leaf | CID_5281 | 3791 | ChEMBL |
| *Gymnema sylvestre* | Leaf | CID_5281 | 4128 | ChEMBL |
| *Gymnema sylvestre* | Leaf | CID_5281 | 4297 | NPASS |
| *Gymnema sylvestre* | Leaf | CID_5281 | 43 | ChEMBL |
| *Gymnema sylvestre* | Leaf | CID_5281 | 4790 | NPASS |
| *Gymnema sylvestre* | Leaf | CID_5281 | 4988 | ChEMBL |
| *Gymnema sylvestre* | Leaf | CID_5281 | 5139 | ChEMBL |
| *Gymnema sylvestre* | Leaf | CID_5281 | 5141 | ChEMBL |
| *Gymnema sylvestre* | Leaf | CID_5281 | 5241 | ChEMBL |
| *Gymnema sylvestre* | Leaf | CID_5281 | 5465 | ChEMBL \| NPASS |
| *Gymnema sylvestre* | Leaf | CID_5281 | 5467 | ChEMBL \| NPASS |
| *Gymnema sylvestre* | Leaf | CID_5281 | 5468 | BindingDB \| ChEMBL \| NPASS |
| *Gymnema sylvestre* | Leaf | CID_5281 | 55775 | NPASS |
| *Gymnema sylvestre* | Leaf | CID_5281 | 5742 | ChEMBL |
| *Gymnema sylvestre* | Leaf | CID_5281 | 5770 | BindingDB \| ChEMBL \| NPASS |
| *Gymnema sylvestre* | Leaf | CID_5281 | 6530 | ChEMBL |
| *Gymnema sylvestre* | Leaf | CID_5281 | 6531 | ChEMBL |
| *Gymnema sylvestre* | Leaf | CID_5281 | 6532 | ChEMBL |
| *Gymnema sylvestre* | Leaf | CID_5281 | 6915 | ChEMBL |
| *Gymnema sylvestre* | Leaf | CID_5281 | 7068 | NPASS |
| *Gymnema sylvestre* | Leaf | CID_5281 | 7253 | NPASS |
| *Gymnema sylvestre* | Leaf | CID_5281 | 7421 | NPASS |
| *Gymnema sylvestre* | Leaf | CID_957 | 10951 | NPASS |
| *Gymnema sylvestre* | Leaf | CID_957 | 216 | NPASS |
| *Gymnema sylvestre* | Leaf | CID_957 | 2908 | NPASS |
| *Gymnema sylvestre* | Leaf | CID_957 | 3028 | NPASS |
| *Gymnema sylvestre* | Leaf | CID_957 | 328 | NPASS |
| *Gymnema sylvestre* | Leaf | CID_957 | 7068 | NPASS |
| *Gymnema sylvestre* | Leaf | CID_8193 | 4780 | NPASS |
| *Gymnema sylvestre* | Leaf | CID_8193 | 8989 | ChEMBL \| NPASS |
| *Gymnema sylvestre* | Leaf | CID_5280794 | 10013 | ChEMBL |
| *Gymnema sylvestre* | Leaf | CID_5280794 | 10599 | ChEMBL \| NPASS |
| *Gymnema sylvestre* | Leaf | CID_5280794 | 28234 | ChEMBL \| NPASS |
| *Gymnema sylvestre* | Leaf | CID_5280794 | 3363 | ChEMBL \| NPASS |
| *Gymnema sylvestre* | Leaf | CID_5280794 | 3417 | ChEMBL \| NPASS |
| *Gymnema sylvestre* | Leaf | CID_5280794 | 5423 | BindingDB \| ChEMBL \| NPASS |
| *Gymnema sylvestre* | Leaf | CID_73145 | 10599 | ChEMBL \| NPASS |
| *Gymnema sylvestre* | Leaf | CID_73145 | 240 | BindingDB \| ChEMBL \| NPASS |
| *Gymnema sylvestre* | Leaf | CID_73145 | 247 | BindingDB \| ChEMBL \| NPASS |
| *Gymnema sylvestre* | Leaf | CID_73145 | 28234 | ChEMBL \| NPASS |
| *Gymnema sylvestre* | Leaf | CID_73145 | 5743 | ChEMBL \| NPASS |
| *Gymnema sylvestre* | Leaf | CID_73145 | 5770 | BindingDB \| ChEMBL \| NPASS |
| *Gymnema sylvestre* | Leaf | CID_259846 | 1066 | ChEMBL \| NPASS |
| *Gymnema sylvestre* | Leaf | CID_259846 | 387 | ChEMBL |
| *Gymnema sylvestre* | Leaf | CID_259846 | 5770 | BindingDB \| ChEMBL \| NPASS |
| *Gymnema sylvestre* | Leaf | CID_259846 | 5879 | ChEMBL |
| *Gymnema sylvestre* | Leaf | CID_259846 | 7153 | ChEMBL \| NPASS |
| *Gymnema sylvestre* | Leaf | CID_259846 | 8824 | BindingDB \| ChEMBL \| NPASS |
| *Gymnema sylvestre* | Leaf | CID_259846 | 998 | ChEMBL |
| *Gymnema sylvestre* | Leaf | CID_305 | 10919 | NPASS |
| *Gymnema sylvestre* | Leaf | CID_305 | 11255 | ChEMBL |
| *Gymnema sylvestre* | Leaf | CID_305 | 1128 | ChEMBL |
| *Gymnema sylvestre* | Leaf | CID_305 | 1129 | ChEMBL |
| *Gymnema sylvestre* | Leaf | CID_305 | 1137 | BindingDB |
| *Gymnema sylvestre* | Leaf | CID_305 | 1139 | ChEMBL |
| *Gymnema sylvestre* | Leaf | CID_305 | 140 | ChEMBL |
| *Gymnema sylvestre* | Leaf | CID_305 | 148 | ChEMBL |
| *Gymnema sylvestre* | Leaf | CID_305 | 150 | ChEMBL |
| *Gymnema sylvestre* | Leaf | CID_305 | 1812 | ChEMBL |
| *Gymnema sylvestre* | Leaf | CID_305 | 1814 | ChEMBL |
| *Gymnema sylvestre* | Leaf | CID_305 | 2099 | ChEMBL |
| *Gymnema sylvestre* | Leaf | CID_305 | 2147 | ChEMBL |
| *Gymnema sylvestre* | Leaf | CID_305 | 3350 | ChEMBL |
| *Gymnema sylvestre* | Leaf | CID_305 | 3357 | ChEMBL |
| *Gymnema sylvestre* | Leaf | CID_305 | 367 | ChEMBL |
| *Gymnema sylvestre* | Leaf | CID_305 | 4128 | ChEMBL |
| *Gymnema sylvestre* | Leaf | CID_305 | 43 | BindingDB \| ChEMBL \| NPASS |
| *Gymnema sylvestre* | Leaf | CID_305 | 4988 | ChEMBL |
| *Gymnema sylvestre* | Leaf | CID_305 | 5139 | ChEMBL |
| *Gymnema sylvestre* | Leaf | CID_305 | 5141 | ChEMBL |
| *Gymnema sylvestre* | Leaf | CID_305 | 5241 | ChEMBL |
| *Gymnema sylvestre* | Leaf | CID_305 | 5742 | ChEMBL |
| *Gymnema sylvestre* | Leaf | CID_305 | 6530 | ChEMBL |
| *Gymnema sylvestre* | Leaf | CID_305 | 6531 | ChEMBL |
| *Gymnema sylvestre* | Leaf | CID_305 | 6532 | ChEMBL |
| *Gymnema sylvestre* | Leaf | CID_305 | 6580 | ChEMBL \| NPASS |
| *Gymnema sylvestre* | Leaf | CID_305 | 6582 | ChEMBL \| NPASS |
| *Gymnema sylvestre* | Leaf | CID_305 | 6583 | ChEMBL \| NPASS |
| *Gymnema sylvestre* | Leaf | CID_305 | 6584 | ChEMBL \| NPASS |
| *Gymnema sylvestre* | Leaf | CID_305 | 6915 | ChEMBL |
| *Gymnema sylvestre* | Leaf | CID_247 | 10919 | NPASS |
| *Gymnema sylvestre* | Leaf | CID_247 | 10951 | NPASS |
| *Gymnema sylvestre* | Leaf | CID_247 | 1559 | NPASS |
| *Gymnema sylvestre* | Leaf | CID_247 | 1565 | NPASS |
| *Gymnema sylvestre* | Leaf | CID_247 | 1576 | NPASS |
| *Gymnema sylvestre* | Leaf | CID_247 | 206358 | NPASS |
| *Gymnema sylvestre* | Leaf | CID_247 | 25939 | ChEMBL |
| *Gymnema sylvestre* | Leaf | CID_247 | 7253 | NPASS |
| *Gymnema sylvestre* | Leaf | CID_247 | 79915 | NPASS |
| *Gymnema sylvestre* | Leaf | CID_247 | 8647 | NPASS |
| *Gymnema sylvestre* | Leaf | CID_2758 | 10062 | ChEMBL \| NPASS |
| *Gymnema sylvestre* | Leaf | CID_2758 | 10280 | ChEMBL |
| *Gymnema sylvestre* | Leaf | CID_2758 | 10800 | ChEMBL |
| *Gymnema sylvestre* | Leaf | CID_2758 | 10951 | NPASS |
| *Gymnema sylvestre* | Leaf | CID_2758 | 1128 | ChEMBL |
| *Gymnema sylvestre* | Leaf | CID_2758 | 1129 | ChEMBL |
| *Gymnema sylvestre* | Leaf | CID_2758 | 1131 | ChEMBL |
| *Gymnema sylvestre* | Leaf | CID_2758 | 1132 | ChEMBL |
| *Gymnema sylvestre* | Leaf | CID_2758 | 1133 | ChEMBL |
| *Gymnema sylvestre* | Leaf | CID_2758 | 1233 | ChEMBL |
| *Gymnema sylvestre* | Leaf | CID_2758 | 1234 | ChEMBL |
| *Gymnema sylvestre* | Leaf | CID_2758 | 1268 | ChEMBL |
| *Gymnema sylvestre* | Leaf | CID_2758 | 134 | ChEMBL |
| *Gymnema sylvestre* | Leaf | CID_2758 | 135 | ChEMBL |
| *Gymnema sylvestre* | Leaf | CID_2758 | 140 | ChEMBL |
| *Gymnema sylvestre* | Leaf | CID_2758 | 1432 | ChEMBL |
| *Gymnema sylvestre* | Leaf | CID_2758 | 146 | ChEMBL |
| *Gymnema sylvestre* | Leaf | CID_2758 | 150 | ChEMBL |
| *Gymnema sylvestre* | Leaf | CID_2758 | 151 | ChEMBL |
| *Gymnema sylvestre* | Leaf | CID_2758 | 1511 | ChEMBL |
| *Gymnema sylvestre* | Leaf | CID_2758 | 152 | ChEMBL |
| *Gymnema sylvestre* | Leaf | CID_2758 | 153 | ChEMBL |
| *Gymnema sylvestre* | Leaf | CID_2758 | 154 | ChEMBL |
| *Gymnema sylvestre* | Leaf | CID_2758 | 1544 | ChEMBL |
| *Gymnema sylvestre* | Leaf | CID_2758 | 1548 | ChEMBL |
| *Gymnema sylvestre* | Leaf | CID_2758 | 155 | ChEMBL |
| *Gymnema sylvestre* | Leaf | CID_2758 | 1557 | ChEMBL |
| *Gymnema sylvestre* | Leaf | CID_2758 | 1559 | ChEMBL |
| *Gymnema sylvestre* | Leaf | CID_2758 | 1565 | ChEMBL |
| *Gymnema sylvestre* | Leaf | CID_2758 | 1571 | ChEMBL |
| *Gymnema sylvestre* | Leaf | CID_2758 | 1576 | ChEMBL |
| *Gymnema sylvestre* | Leaf | CID_2758 | 1812 | ChEMBL |
| *Gymnema sylvestre* | Leaf | CID_2758 | 1813 | ChEMBL |
| *Gymnema sylvestre* | Leaf | CID_2758 | 1814 | ChEMBL |
| *Gymnema sylvestre* | Leaf | CID_2758 | 1815 | ChEMBL |
| *Gymnema sylvestre* | Leaf | CID_2758 | 186 | ChEMBL |
| *Gymnema sylvestre* | Leaf | CID_2758 | 1909 | ChEMBL |
| *Gymnema sylvestre* | Leaf | CID_2758 | 1956 | ChEMBL |
| *Gymnema sylvestre* | Leaf | CID_2758 | 1991 | ChEMBL |
| *Gymnema sylvestre* | Leaf | CID_2758 | 2064 | ChEMBL |
| *Gymnema sylvestre* | Leaf | CID_2758 | 2099 | ChEMBL |
| *Gymnema sylvestre* | Leaf | CID_2758 | 2100 | ChEMBL |
| *Gymnema sylvestre* | Leaf | CID_2758 | 2321 | ChEMBL |
| *Gymnema sylvestre* | Leaf | CID_2758 | 2534 | ChEMBL |
| *Gymnema sylvestre* | Leaf | CID_2758 | 25939 | ChEMBL |
| *Gymnema sylvestre* | Leaf | CID_2758 | 2908 | ChEMBL |
| *Gymnema sylvestre* | Leaf | CID_2758 | 3156 | ChEMBL |
| *Gymnema sylvestre* | Leaf | CID_2758 | 3269 | ChEMBL |
| *Gymnema sylvestre* | Leaf | CID_2758 | 3274 | ChEMBL |
| *Gymnema sylvestre* | Leaf | CID_2758 | 328 | NPASS |
| *Gymnema sylvestre* | Leaf | CID_2758 | 3356 | ChEMBL |
| *Gymnema sylvestre* | Leaf | CID_2758 | 3357 | ChEMBL |
| *Gymnema sylvestre* | Leaf | CID_2758 | 3358 | ChEMBL |
| *Gymnema sylvestre* | Leaf | CID_2758 | 3359 | BindingDB \| NPASS |
| *Gymnema sylvestre* | Leaf | CID_2758 | 3362 | ChEMBL |
| *Gymnema sylvestre* | Leaf | CID_2758 | 3577 | ChEMBL |
| *Gymnema sylvestre* | Leaf | CID_2758 | 3579 | ChEMBL |
| *Gymnema sylvestre* | Leaf | CID_2758 | 367 | NPASS |
| *Gymnema sylvestre* | Leaf | CID_2758 | 3725 | NPASS |
| *Gymnema sylvestre* | Leaf | CID_2758 | 3757 | ChEMBL |
| *Gymnema sylvestre* | Leaf | CID_2758 | 3932 | ChEMBL |
| *Gymnema sylvestre* | Leaf | CID_2758 | 4000 | NPASS |
| *Gymnema sylvestre* | Leaf | CID_2758 | 4128 | ChEMBL |
| *Gymnema sylvestre* | Leaf | CID_2758 | 4137 | NPASS |
| *Gymnema sylvestre* | Leaf | CID_2758 | 4159 | ChEMBL |
| *Gymnema sylvestre* | Leaf | CID_2758 | 4160 | ChEMBL |
| *Gymnema sylvestre* | Leaf | CID_2758 | 4161 | ChEMBL |
| *Gymnema sylvestre* | Leaf | CID_2758 | 43 | ChEMBL |
| *Gymnema sylvestre* | Leaf | CID_2758 | 4312 | ChEMBL |
| *Gymnema sylvestre* | Leaf | CID_2758 | 4318 | ChEMBL |
| *Gymnema sylvestre* | Leaf | CID_2758 | 4886 | ChEMBL |
| *Gymnema sylvestre* | Leaf | CID_2758 | 4887 | ChEMBL |
| *Gymnema sylvestre* | Leaf | CID_2758 | 4985 | ChEMBL |
| *Gymnema sylvestre* | Leaf | CID_2758 | 4986 | ChEMBL |
| *Gymnema sylvestre* | Leaf | CID_2758 | 4988 | ChEMBL |
| *Gymnema sylvestre* | Leaf | CID_2758 | 51053 | NPASS |
| *Gymnema sylvestre* | Leaf | CID_2758 | 5467 | NPASS |
| *Gymnema sylvestre* | Leaf | CID_2758 | 552 | ChEMBL |
| *Gymnema sylvestre* | Leaf | CID_2758 | 5530 | ChEMBL |
| *Gymnema sylvestre* | Leaf | CID_2758 | 55775 | NPASS |
| *Gymnema sylvestre* | Leaf | CID_2758 | 5578 | ChEMBL |
| *Gymnema sylvestre* | Leaf | CID_2758 | 5594 | ChEMBL \| NPASS |
| *Gymnema sylvestre* | Leaf | CID_2758 | 5595 | ChEMBL |
| *Gymnema sylvestre* | Leaf | CID_2758 | 5724 | ChEMBL |
| *Gymnema sylvestre* | Leaf | CID_2758 | 5742 | ChEMBL |
| *Gymnema sylvestre* | Leaf | CID_2758 | 5743 | ChEMBL \| NPASS |
| *Gymnema sylvestre* | Leaf | CID_2758 | 5788 | ChEMBL |
| *Gymnema sylvestre* | Leaf | CID_2758 | 5965 | NPASS |
| *Gymnema sylvestre* | Leaf | CID_2758 | 624 | ChEMBL |
| *Gymnema sylvestre* | Leaf | CID_2758 | 6530 | ChEMBL |
| *Gymnema sylvestre* | Leaf | CID_2758 | 6531 | ChEMBL |
| *Gymnema sylvestre* | Leaf | CID_2758 | 6532 | ChEMBL |
| *Gymnema sylvestre* | Leaf | CID_2758 | 6865 | ChEMBL |
| *Gymnema sylvestre* | Leaf | CID_2758 | 6869 | ChEMBL |
| *Gymnema sylvestre* | Leaf | CID_2758 | 6916 | ChEMBL |
| *Gymnema sylvestre* | Leaf | CID_2758 | 7253 | NPASS |
| *Gymnema sylvestre* | Leaf | CID_2758 | 729230 | ChEMBL |
| *Gymnema sylvestre* | Leaf | CID_2758 | 7433 | ChEMBL |
| *Gymnema sylvestre* | Leaf | CID_2758 | 760 | ChEMBL |
| *Gymnema sylvestre* | Leaf | CID_2758 | 79054 | ChEMBL \| NPASS |
| *Gymnema sylvestre* | Leaf | CID_2758 | 799 | ChEMBL |
| *Gymnema sylvestre* | Leaf | CID_2758 | 834 | ChEMBL |
| *Gymnema sylvestre* | Leaf | CID_2758 | 8654 | ChEMBL |
| *Gymnema sylvestre* | Leaf | CID_2758 | 886 | ChEMBL |
| *Gymnema sylvestre* | Leaf | CID_6918391 | 10013 | ChEMBL |
| *Gymnema sylvestre* | Leaf | CID_6918391 | 5315 | ChEMBL |
| *Gymnema sylvestre* | Leaf | CID_439655 | 10013 | ChEMBL |
| *Gymnema sylvestre* | Leaf | CID_439655 | 10014 | ChEMBL |
| *Gymnema sylvestre* | Leaf | CID_439655 | 3065 | ChEMBL |
| *Gymnema sylvestre* | Leaf | CID_439655 | 3066 | ChEMBL |
| *Gymnema sylvestre* | Leaf | CID_439655 | 51053 | NPASS |
| *Gymnema sylvestre* | Leaf | CID_439655 | 51564 | ChEMBL |
| *Gymnema sylvestre* | Leaf | CID_439655 | 55869 | ChEMBL |
| *Gymnema sylvestre* | Leaf | CID_439655 | 7253 | NPASS |
| *Gymnema sylvestre* | Leaf | CID_439655 | 79885 | ChEMBL |
| *Gymnema sylvestre* | Leaf | CID_439655 | 83933 | ChEMBL \| NPASS |
| *Gymnema sylvestre* | Leaf | CID_439655 | 8841 | ChEMBL |
| *Gymnema sylvestre* | Leaf | CID_439655 | 9682 | NPASS |
| *Gymnema sylvestre* | Leaf | CID_439655 | 9734 | ChEMBL |
| *Gymnema sylvestre* | Leaf | CID_439655 | 9759 | ChEMBL |
| *Gymnema sylvestre* | Leaf | CID_13849 | 1588 | ChEMBL \| NPASS |
| *Gymnema sylvestre* | Leaf | CID_13849 | 367 | NPASS |
| *Gymnema sylvestre* | Leaf | CID_13849 | 5467 | NPASS |
| *Gymnema sylvestre* | Leaf | CID_12535 | 7068 | NPASS |
| *Gymnema sylvestre* | Leaf | CID_3314 | 10280 | ChEMBL |
| *Gymnema sylvestre* | Leaf | CID_3314 | 10599 | ChEMBL \| NPASS |
| *Gymnema sylvestre* | Leaf | CID_3314 | 10800 | ChEMBL |
| *Gymnema sylvestre* | Leaf | CID_3314 | 10941 | ChEMBL |
| *Gymnema sylvestre* | Leaf | CID_3314 | 1128 | ChEMBL |
| *Gymnema sylvestre* | Leaf | CID_3314 | 1129 | ChEMBL |
| *Gymnema sylvestre* | Leaf | CID_3314 | 1131 | ChEMBL |
| *Gymnema sylvestre* | Leaf | CID_3314 | 1132 | ChEMBL |
| *Gymnema sylvestre* | Leaf | CID_3314 | 1133 | ChEMBL |
| *Gymnema sylvestre* | Leaf | CID_3314 | 1233 | ChEMBL |
| *Gymnema sylvestre* | Leaf | CID_3314 | 1234 | ChEMBL |
| *Gymnema sylvestre* | Leaf | CID_3314 | 1268 | ChEMBL |
| *Gymnema sylvestre* | Leaf | CID_3314 | 134 | ChEMBL |
| *Gymnema sylvestre* | Leaf | CID_3314 | 135 | ChEMBL |
| *Gymnema sylvestre* | Leaf | CID_3314 | 140 | ChEMBL |
| *Gymnema sylvestre* | Leaf | CID_3314 | 1432 | ChEMBL |
| *Gymnema sylvestre* | Leaf | CID_3314 | 146 | ChEMBL |
| *Gymnema sylvestre* | Leaf | CID_3314 | 150 | ChEMBL |
| *Gymnema sylvestre* | Leaf | CID_3314 | 151 | ChEMBL |
| *Gymnema sylvestre* | Leaf | CID_3314 | 1511 | ChEMBL |
| *Gymnema sylvestre* | Leaf | CID_3314 | 152 | ChEMBL |
| *Gymnema sylvestre* | Leaf | CID_3314 | 153 | ChEMBL |
| *Gymnema sylvestre* | Leaf | CID_3314 | 154 | ChEMBL |
| *Gymnema sylvestre* | Leaf | CID_3314 | 1544 | ChEMBL |
| *Gymnema sylvestre* | Leaf | CID_3314 | 1548 | ChEMBL |
| *Gymnema sylvestre* | Leaf | CID_3314 | 155 | ChEMBL |
| *Gymnema sylvestre* | Leaf | CID_3314 | 1557 | ChEMBL |
| *Gymnema sylvestre* | Leaf | CID_3314 | 1559 | ChEMBL |
| *Gymnema sylvestre* | Leaf | CID_3314 | 1565 | ChEMBL |
| *Gymnema sylvestre* | Leaf | CID_3314 | 1571 | ChEMBL |
| *Gymnema sylvestre* | Leaf | CID_3314 | 1576 | ChEMBL \| NPASS |
| *Gymnema sylvestre* | Leaf | CID_3314 | 1812 | ChEMBL |
| *Gymnema sylvestre* | Leaf | CID_3314 | 1813 | ChEMBL |
| *Gymnema sylvestre* | Leaf | CID_3314 | 1814 | ChEMBL |
| *Gymnema sylvestre* | Leaf | CID_3314 | 1815 | ChEMBL |
| *Gymnema sylvestre* | Leaf | CID_3314 | 186 | ChEMBL |
| *Gymnema sylvestre* | Leaf | CID_3314 | 1909 | ChEMBL |
| *Gymnema sylvestre* | Leaf | CID_3314 | 1956 | ChEMBL |
| *Gymnema sylvestre* | Leaf | CID_3314 | 1991 | ChEMBL |
| *Gymnema sylvestre* | Leaf | CID_3314 | 2064 | ChEMBL |
| *Gymnema sylvestre* | Leaf | CID_3314 | 2099 | ChEMBL |
| *Gymnema sylvestre* | Leaf | CID_3314 | 2100 | ChEMBL |
| *Gymnema sylvestre* | Leaf | CID_3314 | 216 | ChEMBL \| NPASS |
| *Gymnema sylvestre* | Leaf | CID_3314 | 2321 | ChEMBL |
| *Gymnema sylvestre* | Leaf | CID_3314 | 240 | BindingDB \| ChEMBL \| NPASS |
| *Gymnema sylvestre* | Leaf | CID_3314 | 2534 | ChEMBL |
| *Gymnema sylvestre* | Leaf | CID_3314 | 26339 | ChEMBL \| NPASS |
| *Gymnema sylvestre* | Leaf | CID_3314 | 28234 | ChEMBL \| NPASS |
| *Gymnema sylvestre* | Leaf | CID_3314 | 2908 | ChEMBL |
| *Gymnema sylvestre* | Leaf | CID_3314 | 2990 | NPASS |
| *Gymnema sylvestre* | Leaf | CID_3314 | 3156 | ChEMBL |
| *Gymnema sylvestre* | Leaf | CID_3314 | 3269 | ChEMBL |
| *Gymnema sylvestre* | Leaf | CID_3314 | 3274 | ChEMBL |
| *Gymnema sylvestre* | Leaf | CID_3314 | 3356 | ChEMBL |
| *Gymnema sylvestre* | Leaf | CID_3314 | 3357 | ChEMBL |
| *Gymnema sylvestre* | Leaf | CID_3314 | 3358 | ChEMBL |
| *Gymnema sylvestre* | Leaf | CID_3314 | 3362 | ChEMBL |
| *Gymnema sylvestre* | Leaf | CID_3314 | 3576 | NPASS |
| *Gymnema sylvestre* | Leaf | CID_3314 | 3577 | ChEMBL |
| *Gymnema sylvestre* | Leaf | CID_3314 | 3579 | ChEMBL |
| *Gymnema sylvestre* | Leaf | CID_3314 | 367 | BindingDB \| ChEMBL \| NPASS |
| *Gymnema sylvestre* | Leaf | CID_3314 | 3725 | NPASS |
| *Gymnema sylvestre* | Leaf | CID_3314 | 3757 | ChEMBL |
| *Gymnema sylvestre* | Leaf | CID_3314 | 3932 | ChEMBL |
| *Gymnema sylvestre* | Leaf | CID_3314 | 4128 | ChEMBL |
| *Gymnema sylvestre* | Leaf | CID_3314 | 4159 | ChEMBL |
| *Gymnema sylvestre* | Leaf | CID_3314 | 4160 | ChEMBL |
| *Gymnema sylvestre* | Leaf | CID_3314 | 4161 | ChEMBL |
| *Gymnema sylvestre* | Leaf | CID_3314 | 43 | ChEMBL |
| *Gymnema sylvestre* | Leaf | CID_3314 | 4312 | ChEMBL |
| *Gymnema sylvestre* | Leaf | CID_3314 | 4318 | ChEMBL |
| *Gymnema sylvestre* | Leaf | CID_3314 | 4780 | NPASS |
| *Gymnema sylvestre* | Leaf | CID_3314 | 4886 | ChEMBL |
| *Gymnema sylvestre* | Leaf | CID_3314 | 4887 | ChEMBL |
| *Gymnema sylvestre* | Leaf | CID_3314 | 4985 | ChEMBL |
| *Gymnema sylvestre* | Leaf | CID_3314 | 4986 | ChEMBL |
| *Gymnema sylvestre* | Leaf | CID_3314 | 4988 | ChEMBL |
| *Gymnema sylvestre* | Leaf | CID_3314 | 51053 | NPASS |
| *Gymnema sylvestre* | Leaf | CID_3314 | 54575 | ChEMBL \| NPASS |
| *Gymnema sylvestre* | Leaf | CID_3314 | 54576 | ChEMBL \| NPASS |
| *Gymnema sylvestre* | Leaf | CID_3314 | 54578 | ChEMBL \| NPASS |
| *Gymnema sylvestre* | Leaf | CID_3314 | 54600 | ChEMBL \| NPASS |
| *Gymnema sylvestre* | Leaf | CID_3314 | 54657 | ChEMBL \| NPASS |
| *Gymnema sylvestre* | Leaf | CID_3314 | 54658 | ChEMBL \| NPASS |
| *Gymnema sylvestre* | Leaf | CID_3314 | 54659 | ChEMBL \| NPASS |
| *Gymnema sylvestre* | Leaf | CID_3314 | 552 | ChEMBL |
| *Gymnema sylvestre* | Leaf | CID_3314 | 5530 | ChEMBL |
| *Gymnema sylvestre* | Leaf | CID_3314 | 5578 | ChEMBL |
| *Gymnema sylvestre* | Leaf | CID_3314 | 5594 | ChEMBL |
| *Gymnema sylvestre* | Leaf | CID_3314 | 5595 | ChEMBL |
| *Gymnema sylvestre* | Leaf | CID_3314 | 5724 | ChEMBL |
| *Gymnema sylvestre* | Leaf | CID_3314 | 5742 | ChEMBL \| NPASS |
| *Gymnema sylvestre* | Leaf | CID_3314 | 5743 | BindingDB \| ChEMBL \| NPASS |
| *Gymnema sylvestre* | Leaf | CID_3314 | 5788 | ChEMBL |
| *Gymnema sylvestre* | Leaf | CID_3314 | 60489 | NPASS |
| *Gymnema sylvestre* | Leaf | CID_3314 | 624 | ChEMBL |
| *Gymnema sylvestre* | Leaf | CID_3314 | 6530 | ChEMBL |
| *Gymnema sylvestre* | Leaf | CID_3314 | 6531 | ChEMBL |
| *Gymnema sylvestre* | Leaf | CID_3314 | 6532 | ChEMBL |
| *Gymnema sylvestre* | Leaf | CID_3314 | 6865 | ChEMBL |
| *Gymnema sylvestre* | Leaf | CID_3314 | 6869 | ChEMBL |
| *Gymnema sylvestre* | Leaf | CID_3314 | 6916 | ChEMBL |
| *Gymnema sylvestre* | Leaf | CID_3314 | 729230 | ChEMBL |
| *Gymnema sylvestre* | Leaf | CID_3314 | 7366 | ChEMBL \| NPASS |
| *Gymnema sylvestre* | Leaf | CID_3314 | 7367 | ChEMBL \| NPASS |
| *Gymnema sylvestre* | Leaf | CID_3314 | 7433 | ChEMBL |
| *Gymnema sylvestre* | Leaf | CID_3314 | 760 | ChEMBL |
| *Gymnema sylvestre* | Leaf | CID_3314 | 775 | NPASS |
| *Gymnema sylvestre* | Leaf | CID_3314 | 79054 | ChEMBL |
| *Gymnema sylvestre* | Leaf | CID_3314 | 799 | ChEMBL |
| *Gymnema sylvestre* | Leaf | CID_3314 | 834 | ChEMBL |
| *Gymnema sylvestre* | Leaf | CID_3314 | 8654 | ChEMBL |
| *Gymnema sylvestre* | Leaf | CID_3314 | 886 | ChEMBL |
| *Gymnema sylvestre* | Leaf | CID_3314 | 8989 | ChEMBL \| NPASS |
| *Gymnema sylvestre* | Leaf | CID_7127 | 216 | NPASS |
| *Gymnema sylvestre* | Leaf | CID_7127 | 2648 | NPASS |
| *Gymnema sylvestre* | Leaf | CID_7127 | 2908 | NPASS |
| *Gymnema sylvestre* | Leaf | CID_7127 | 367 | NPASS |
| *Gymnema sylvestre* | Leaf | CID_7127 | 4088 | NPASS |
| *Gymnema sylvestre* | Leaf | CID_7127 | 4780 | ChEMBL \| NPASS |
| *Gymnema sylvestre* | Leaf | CID_7127 | 5465 | ChEMBL \| NPASS |
| *Gymnema sylvestre* | Leaf | CID_7127 | 5467 | ChEMBL \| NPASS |
| *Gymnema sylvestre* | Leaf | CID_7127 | 5468 | BindingDB \| ChEMBL \| NPASS |
| *Gymnema sylvestre* | Leaf | CID_7127 | 7068 | NPASS |
| *Gymnema sylvestre* | Leaf | CID_7127 | 79915 | NPASS |
| *Gymnema sylvestre* | Leaf | CID_7127 | 9971 | NPASS |
| *Gymnema sylvestre* | Leaf | CID_2682 | 10013 | ChEMBL |
| *Gymnema sylvestre* | Leaf | CID_2682 | 328 | NPASS |
| *Gymnema sylvestre* | Leaf | CID_2682 | 4000 | NPASS |
| *Gymnema sylvestre* | Leaf | CID_2682 | 4297 | NPASS |
| *Gymnema sylvestre* | Leaf | CID_2682 | 4780 | NPASS |
| *Gymnema sylvestre* | Leaf | CID_2682 | 7398 | NPASS |
| *Gymnema sylvestre* | Leaf | CID_7410 | 2932 | BindingDB \| ChEMBL \| NPASS |
| *Gymnema sylvestre* | Leaf | CID_7410 | 6093 | ChEMBL \| NPASS |
| *Gymnema sylvestre* | Leaf | CID_7410 | 6095 | ChEMBL |
| *Gymnema sylvestre* | Leaf | CID_7410 | 6096 | ChEMBL \| NPASS |
| *Gymnema sylvestre* | Leaf | CID_7410 | 6097 | ChEMBL \| NPASS |
| *Gymnema sylvestre* | Leaf | CID_61303 | 8989 | ChEMBL \| NPASS |
| *Gymnema sylvestre* | Leaf | CID_8221 | 10013 | ChEMBL |
| *Gymnema sylvestre* | Leaf | CID_8221 | 328 | NPASS |
| *Gymnema sylvestre* | Leaf | CID_8221 | 4000 | NPASS |
| *Gymnema sylvestre* | Leaf | CID_8221 | 51053 | NPASS |
| *Gymnema sylvestre* | Leaf | CID_5280435 | 5770 | BindingDB \| ChEMBL \| NPASS |
| *Gymnema sylvestre* | Leaf | CID_5280435 | 5771 | ChEMBL \| NPASS |
| *Gymnema sylvestre* | Leaf | CID_5280435 | 5777 | ChEMBL \| NPASS |
| *Gymnema sylvestre* | Leaf | CID_5280435 | 5781 | ChEMBL \| NPASS |
| *Gymnema sylvestre* | Leaf | CID_5280435 | 5792 | ChEMBL \| NPASS |
| *Gymnema sylvestre* | Leaf | CID_5280435 | 994 | ChEMBL \| NPASS |
| *Gymnema sylvestre* | Leaf | CID_8181 | 1268 | ChEMBL \| NPASS |
| *Gymnema sylvestre* | Leaf | CID_8181 | 1269 | ChEMBL \| NPASS |
| *Gymnema sylvestre* | Leaf | CID_8181 | 8989 | ChEMBL \| NPASS |
| *Gymnema sylvestre* | Leaf | CID_119 | 10013 | ChEMBL |
| *Gymnema sylvestre* | Leaf | CID_119 | 10014 | ChEMBL |
| *Gymnema sylvestre* | Leaf | CID_119 | 10599 | ChEMBL |
| *Gymnema sylvestre* | Leaf | CID_119 | 10919 | ChEMBL |
| *Gymnema sylvestre* | Leaf | CID_119 | 117247 | ChEMBL |
| *Gymnema sylvestre* | Leaf | CID_119 | 18 | ChEMBL |
| *Gymnema sylvestre* | Leaf | CID_119 | 200959 | BindingDB \| ChEMBL |
| *Gymnema sylvestre* | Leaf | CID_119 | 206358 | BindingDB |
| *Gymnema sylvestre* | Leaf | CID_119 | 2550 | BindingDB \| ChEMBL |
| *Gymnema sylvestre* | Leaf | CID_119 | 2554 | BindingDB \| ChEMBL |
| *Gymnema sylvestre* | Leaf | CID_119 | 2555 | ChEMBL |
| *Gymnema sylvestre* | Leaf | CID_119 | 2556 | BindingDB \| ChEMBL |
| *Gymnema sylvestre* | Leaf | CID_119 | 2557 | BindingDB \| ChEMBL |
| *Gymnema sylvestre* | Leaf | CID_119 | 2558 | BindingDB \| ChEMBL |
| *Gymnema sylvestre* | Leaf | CID_119 | 2559 | BindingDB \| ChEMBL |
| *Gymnema sylvestre* | Leaf | CID_119 | 2560 | ChEMBL |
| *Gymnema sylvestre* | Leaf | CID_119 | 2561 | ChEMBL |
| *Gymnema sylvestre* | Leaf | CID_119 | 2562 | ChEMBL |
| *Gymnema sylvestre* | Leaf | CID_119 | 2563 | ChEMBL |
| *Gymnema sylvestre* | Leaf | CID_119 | 2564 | ChEMBL |
| *Gymnema sylvestre* | Leaf | CID_119 | 2565 | ChEMBL |
| *Gymnema sylvestre* | Leaf | CID_119 | 2566 | ChEMBL |
| *Gymnema sylvestre* | Leaf | CID_119 | 2567 | ChEMBL |
| *Gymnema sylvestre* | Leaf | CID_119 | 2568 | ChEMBL |
| *Gymnema sylvestre* | Leaf | CID_119 | 2569 | BindingDB \| ChEMBL |
| *Gymnema sylvestre* | Leaf | CID_119 | 2570 | ChEMBL |
| *Gymnema sylvestre* | Leaf | CID_119 | 25939 | ChEMBL |
| *Gymnema sylvestre* | Leaf | CID_119 | 28234 | ChEMBL |
| *Gymnema sylvestre* | Leaf | CID_119 | 3065 | BindingDB \| ChEMBL |
| *Gymnema sylvestre* | Leaf | CID_119 | 3066 | ChEMBL |
| *Gymnema sylvestre* | Leaf | CID_119 | 51206 | ChEMBL |
| *Gymnema sylvestre* | Leaf | CID_119 | 51564 | ChEMBL |
| *Gymnema sylvestre* | Leaf | CID_119 | 55869 | ChEMBL |
| *Gymnema sylvestre* | Leaf | CID_119 | 55879 | ChEMBL |
| *Gymnema sylvestre* | Leaf | CID_119 | 6529 | BindingDB \| ChEMBL |
| *Gymnema sylvestre* | Leaf | CID_119 | 6538 | BindingDB \| ChEMBL |
| *Gymnema sylvestre* | Leaf | CID_119 | 6539 | BindingDB \| ChEMBL |
| *Gymnema sylvestre* | Leaf | CID_119 | 6540 | BindingDB \| ChEMBL |
| *Gymnema sylvestre* | Leaf | CID_119 | 79885 | ChEMBL |
| *Gymnema sylvestre* | Leaf | CID_119 | 83933 | ChEMBL |
| *Gymnema sylvestre* | Leaf | CID_119 | 8841 | ChEMBL |
| *Gymnema sylvestre* | Leaf | CID_119 | 9568 | ChEMBL |
| *Gymnema sylvestre* | Leaf | CID_119 | 9734 | ChEMBL |
| *Gymnema sylvestre* | Leaf | CID_119 | 9759 | ChEMBL |
| *Gymnema sylvestre* | Leaf | CID_264 | 10013 | BindingDB \| ChEMBL \| NPASS |
| *Gymnema sylvestre* | Leaf | CID_264 | 10014 | BindingDB \| ChEMBL \| NPASS |
| *Gymnema sylvestre* | Leaf | CID_264 | 2865 | BindingDB \| ChEMBL \| NPASS |
| *Gymnema sylvestre* | Leaf | CID_264 | 3065 | BindingDB \| ChEMBL \| NPASS |
| *Gymnema sylvestre* | Leaf | CID_264 | 3066 | BindingDB \| ChEMBL \| NPASS |
| *Gymnema sylvestre* | Leaf | CID_264 | 390245 | BindingDB |
| *Gymnema sylvestre* | Leaf | CID_264 | 4790 | NPASS |
| *Gymnema sylvestre* | Leaf | CID_264 | 51181 | ChEMBL \| NPASS |
| *Gymnema sylvestre* | Leaf | CID_264 | 51564 | BindingDB \| ChEMBL \| NPASS |
| *Gymnema sylvestre* | Leaf | CID_264 | 5315 | ChEMBL |
| *Gymnema sylvestre* | Leaf | CID_264 | 54575 | ChEMBL \| NPASS |
| *Gymnema sylvestre* | Leaf | CID_264 | 54576 | ChEMBL \| NPASS |
| *Gymnema sylvestre* | Leaf | CID_264 | 55869 | BindingDB \| ChEMBL \| NPASS |
| *Gymnema sylvestre* | Leaf | CID_264 | 79885 | ChEMBL |
| *Gymnema sylvestre* | Leaf | CID_264 | 83933 | ChEMBL \| NPASS |
| *Gymnema sylvestre* | Leaf | CID_264 | 84929 | ChEMBL \| NPASS |
| *Gymnema sylvestre* | Leaf | CID_264 | 8841 | BindingDB \| ChEMBL \| NPASS |
| *Gymnema sylvestre* | Leaf | CID_264 | 9734 | BindingDB \| ChEMBL \| NPASS |
| *Gymnema sylvestre* | Leaf | CID_264 | 9759 | BindingDB \| ChEMBL \| NPASS |
| *Gymnema sylvestre* | Leaf | CID_445858 | 10013 | ChEMBL |
| *Gymnema sylvestre* | Leaf | CID_445858 | 10014 | ChEMBL |
| *Gymnema sylvestre* | Leaf | CID_445858 | 10280 | ChEMBL |
| *Gymnema sylvestre* | Leaf | CID_445858 | 10599 | ChEMBL \| NPASS |
| *Gymnema sylvestre* | Leaf | CID_445858 | 10800 | ChEMBL |
| *Gymnema sylvestre* | Leaf | CID_445858 | 11238 | BindingDB \| ChEMBL \| NPASS |
| *Gymnema sylvestre* | Leaf | CID_445858 | 1128 | ChEMBL |
| *Gymnema sylvestre* | Leaf | CID_445858 | 1129 | ChEMBL |
| *Gymnema sylvestre* | Leaf | CID_445858 | 1131 | ChEMBL |
| *Gymnema sylvestre* | Leaf | CID_445858 | 1132 | ChEMBL |
| *Gymnema sylvestre* | Leaf | CID_445858 | 1133 | ChEMBL |
| *Gymnema sylvestre* | Leaf | CID_445858 | 1233 | ChEMBL |
| *Gymnema sylvestre* | Leaf | CID_445858 | 1234 | ChEMBL |
| *Gymnema sylvestre* | Leaf | CID_445858 | 1268 | ChEMBL |
| *Gymnema sylvestre* | Leaf | CID_445858 | 134 | ChEMBL |
| *Gymnema sylvestre* | Leaf | CID_445858 | 135 | ChEMBL |
| *Gymnema sylvestre* | Leaf | CID_445858 | 140 | ChEMBL |
| *Gymnema sylvestre* | Leaf | CID_445858 | 1432 | ChEMBL |
| *Gymnema sylvestre* | Leaf | CID_445858 | 1457 | ChEMBL |
| *Gymnema sylvestre* | Leaf | CID_445858 | 1459 | ChEMBL |
| *Gymnema sylvestre* | Leaf | CID_445858 | 146 | ChEMBL |
| *Gymnema sylvestre* | Leaf | CID_445858 | 1460 | ChEMBL |
| *Gymnema sylvestre* | Leaf | CID_445858 | 150 | ChEMBL |
| *Gymnema sylvestre* | Leaf | CID_445858 | 151 | ChEMBL |
| *Gymnema sylvestre* | Leaf | CID_445858 | 1511 | ChEMBL |
| *Gymnema sylvestre* | Leaf | CID_445858 | 152 | ChEMBL |
| *Gymnema sylvestre* | Leaf | CID_445858 | 153 | ChEMBL |
| *Gymnema sylvestre* | Leaf | CID_445858 | 154 | ChEMBL |
| *Gymnema sylvestre* | Leaf | CID_445858 | 1544 | ChEMBL |
| *Gymnema sylvestre* | Leaf | CID_445858 | 1548 | ChEMBL |
| *Gymnema sylvestre* | Leaf | CID_445858 | 155 | ChEMBL |
| *Gymnema sylvestre* | Leaf | CID_445858 | 1557 | ChEMBL |
| *Gymnema sylvestre* | Leaf | CID_445858 | 1559 | ChEMBL |
| *Gymnema sylvestre* | Leaf | CID_445858 | 1565 | ChEMBL |
| *Gymnema sylvestre* | Leaf | CID_445858 | 1571 | ChEMBL |
| *Gymnema sylvestre* | Leaf | CID_445858 | 1576 | ChEMBL |
| *Gymnema sylvestre* | Leaf | CID_445858 | 1812 | ChEMBL |
| *Gymnema sylvestre* | Leaf | CID_445858 | 1813 | ChEMBL |
| *Gymnema sylvestre* | Leaf | CID_445858 | 1814 | ChEMBL |
| *Gymnema sylvestre* | Leaf | CID_445858 | 1815 | ChEMBL |
| *Gymnema sylvestre* | Leaf | CID_445858 | 186 | ChEMBL |
| *Gymnema sylvestre* | Leaf | CID_445858 | 1909 | ChEMBL |
| *Gymnema sylvestre* | Leaf | CID_445858 | 1956 | ChEMBL |
| *Gymnema sylvestre* | Leaf | CID_445858 | 1991 | ChEMBL |
| *Gymnema sylvestre* | Leaf | CID_445858 | 2064 | ChEMBL |
| *Gymnema sylvestre* | Leaf | CID_445858 | 2099 | ChEMBL \| NPASS |
| *Gymnema sylvestre* | Leaf | CID_445858 | 2100 | ChEMBL |
| *Gymnema sylvestre* | Leaf | CID_445858 | 2260 | ChEMBL \| NPASS |
| *Gymnema sylvestre* | Leaf | CID_445858 | 2263 | ChEMBL \| NPASS |
| *Gymnema sylvestre* | Leaf | CID_445858 | 2321 | ChEMBL |
| *Gymnema sylvestre* | Leaf | CID_445858 | 23632 | BindingDB \| ChEMBL \| NPASS |
| *Gymnema sylvestre* | Leaf | CID_445858 | 2534 | ChEMBL |
| *Gymnema sylvestre* | Leaf | CID_445858 | 259285 | ChEMBL \| NPASS |
| *Gymnema sylvestre* | Leaf | CID_445858 | 28234 | ChEMBL \| NPASS |
| *Gymnema sylvestre* | Leaf | CID_445858 | 283106 | BindingDB \| ChEMBL |
| *Gymnema sylvestre* | Leaf | CID_445858 | 2908 | ChEMBL |
| *Gymnema sylvestre* | Leaf | CID_445858 | 3065 | ChEMBL |
| *Gymnema sylvestre* | Leaf | CID_445858 | 3066 | ChEMBL |
| *Gymnema sylvestre* | Leaf | CID_445858 | 3156 | ChEMBL |
| *Gymnema sylvestre* | Leaf | CID_445858 | 3269 | ChEMBL |
| *Gymnema sylvestre* | Leaf | CID_445858 | 3274 | ChEMBL |
| *Gymnema sylvestre* | Leaf | CID_445858 | 3356 | ChEMBL |
| *Gymnema sylvestre* | Leaf | CID_445858 | 3357 | ChEMBL |
| *Gymnema sylvestre* | Leaf | CID_445858 | 3358 | ChEMBL |
| *Gymnema sylvestre* | Leaf | CID_445858 | 3362 | ChEMBL |
| *Gymnema sylvestre* | Leaf | CID_445858 | 3363 | ChEMBL \| NPASS |
| *Gymnema sylvestre* | Leaf | CID_445858 | 3373 | BindingDB \| NPASS |
| *Gymnema sylvestre* | Leaf | CID_445858 | 338442 | ChEMBL \| NPASS |
| *Gymnema sylvestre* | Leaf | CID_445858 | 351 | BindingDB \| ChEMBL \| NPASS |
| *Gymnema sylvestre* | Leaf | CID_445858 | 3577 | ChEMBL |
| *Gymnema sylvestre* | Leaf | CID_445858 | 3579 | ChEMBL |
| *Gymnema sylvestre* | Leaf | CID_445858 | 3757 | ChEMBL |
| *Gymnema sylvestre* | Leaf | CID_445858 | 390245 | NPASS |
| *Gymnema sylvestre* | Leaf | CID_445858 | 3932 | ChEMBL |
| *Gymnema sylvestre* | Leaf | CID_445858 | 4128 | ChEMBL |
| *Gymnema sylvestre* | Leaf | CID_445858 | 4159 | ChEMBL |
| *Gymnema sylvestre* | Leaf | CID_445858 | 4160 | ChEMBL |
| *Gymnema sylvestre* | Leaf | CID_445858 | 4161 | ChEMBL |
| *Gymnema sylvestre* | Leaf | CID_445858 | 43 | BindingDB \| ChEMBL \| NPASS |
| *Gymnema sylvestre* | Leaf | CID_445858 | 4312 | ChEMBL |
| *Gymnema sylvestre* | Leaf | CID_445858 | 4318 | ChEMBL |
| *Gymnema sylvestre* | Leaf | CID_445858 | 4886 | ChEMBL |
| *Gymnema sylvestre* | Leaf | CID_445858 | 4887 | ChEMBL |
| *Gymnema sylvestre* | Leaf | CID_445858 | 4985 | ChEMBL |
| *Gymnema sylvestre* | Leaf | CID_445858 | 4986 | ChEMBL |
| *Gymnema sylvestre* | Leaf | CID_445858 | 4988 | ChEMBL |
| *Gymnema sylvestre* | Leaf | CID_445858 | 51053 | NPASS |
| *Gymnema sylvestre* | Leaf | CID_445858 | 51564 | ChEMBL |
| *Gymnema sylvestre* | Leaf | CID_445858 | 5315 | ChEMBL |
| *Gymnema sylvestre* | Leaf | CID_445858 | 54657 | ChEMBL \| NPASS |
| *Gymnema sylvestre* | Leaf | CID_445858 | 54658 | ChEMBL \| NPASS |
| *Gymnema sylvestre* | Leaf | CID_445858 | 552 | ChEMBL |
| *Gymnema sylvestre* | Leaf | CID_445858 | 5530 | ChEMBL |
| *Gymnema sylvestre* | Leaf | CID_445858 | 55775 | NPASS |
| *Gymnema sylvestre* | Leaf | CID_445858 | 5578 | ChEMBL |
| *Gymnema sylvestre* | Leaf | CID_445858 | 55869 | ChEMBL |
| *Gymnema sylvestre* | Leaf | CID_445858 | 5594 | ChEMBL |
| *Gymnema sylvestre* | Leaf | CID_445858 | 5595 | ChEMBL |
| *Gymnema sylvestre* | Leaf | CID_445858 | 5724 | ChEMBL |
| *Gymnema sylvestre* | Leaf | CID_445858 | 5742 | ChEMBL |
| *Gymnema sylvestre* | Leaf | CID_445858 | 5743 | ChEMBL |
| *Gymnema sylvestre* | Leaf | CID_445858 | 5745 | NPASS |
| *Gymnema sylvestre* | Leaf | CID_445858 | 5788 | ChEMBL |
| *Gymnema sylvestre* | Leaf | CID_445858 | 590 | BindingDB \| ChEMBL \| NPASS |
| *Gymnema sylvestre* | Leaf | CID_445858 | 624 | ChEMBL |
| *Gymnema sylvestre* | Leaf | CID_445858 | 6530 | ChEMBL |
| *Gymnema sylvestre* | Leaf | CID_445858 | 6531 | ChEMBL |
| *Gymnema sylvestre* | Leaf | CID_445858 | 6532 | ChEMBL |
| *Gymnema sylvestre* | Leaf | CID_445858 | 6622 | ChEMBL \| NPASS |
| *Gymnema sylvestre* | Leaf | CID_445858 | 6865 | ChEMBL |
| *Gymnema sylvestre* | Leaf | CID_445858 | 6869 | ChEMBL |
| *Gymnema sylvestre* | Leaf | CID_445858 | 6916 | ChEMBL |
| *Gymnema sylvestre* | Leaf | CID_445858 | 729230 | ChEMBL |
| *Gymnema sylvestre* | Leaf | CID_445858 | 7299 | ChEMBL \| NPASS |
| *Gymnema sylvestre* | Leaf | CID_445858 | 7366 | ChEMBL \| NPASS |
| *Gymnema sylvestre* | Leaf | CID_445858 | 7433 | ChEMBL |
| *Gymnema sylvestre* | Leaf | CID_445858 | 759 | BindingDB \| ChEMBL \| NPASS |
| *Gymnema sylvestre* | Leaf | CID_445858 | 760 | BindingDB \| ChEMBL \| NPASS |
| *Gymnema sylvestre* | Leaf | CID_445858 | 761 | BindingDB \| ChEMBL \| NPASS |
| *Gymnema sylvestre* | Leaf | CID_445858 | 762 | BindingDB \| ChEMBL \| NPASS |
| *Gymnema sylvestre* | Leaf | CID_445858 | 763 | BindingDB \| ChEMBL \| NPASS |
| *Gymnema sylvestre* | Leaf | CID_445858 | 765 | ChEMBL \| NPASS |
| *Gymnema sylvestre* | Leaf | CID_445858 | 766 | BindingDB \| ChEMBL \| NPASS |
| *Gymnema sylvestre* | Leaf | CID_445858 | 768 | BindingDB \| ChEMBL \| NPASS |
| *Gymnema sylvestre* | Leaf | CID_445858 | 771 | BindingDB \| ChEMBL \| NPASS |
| *Gymnema sylvestre* | Leaf | CID_445858 | 79885 | ChEMBL |
| *Gymnema sylvestre* | Leaf | CID_445858 | 799 | ChEMBL |
| *Gymnema sylvestre* | Leaf | CID_445858 | 834 | ChEMBL |
| *Gymnema sylvestre* | Leaf | CID_445858 | 83933 | ChEMBL \| NPASS |
| *Gymnema sylvestre* | Leaf | CID_445858 | 8654 | ChEMBL |
| *Gymnema sylvestre* | Leaf | CID_445858 | 8841 | ChEMBL |
| *Gymnema sylvestre* | Leaf | CID_445858 | 886 | ChEMBL |
| *Gymnema sylvestre* | Leaf | CID_445858 | 9734 | ChEMBL |
| *Gymnema sylvestre* | Leaf | CID_445858 | 9759 | ChEMBL |
| *Gymnema sylvestre* | Leaf | CID_8209 | 5467 | NPASS |
| *Gymnema sylvestre* | Leaf | CID_8209 | 9971 | NPASS |
| *Gymnema sylvestre* | Leaf | CID_798 | 1020 | ChEMBL |
| *Gymnema sylvestre* | Leaf | CID_798 | 132 | BindingDB \| ChEMBL \| NPASS |
| *Gymnema sylvestre* | Leaf | CID_798 | 2931 | ChEMBL |
| *Gymnema sylvestre* | Leaf | CID_798 | 2932 | ChEMBL \| NPASS |
| *Gymnema sylvestre* | Leaf | CID_798 | 3383 | ChEMBL |
| *Gymnema sylvestre* | Leaf | CID_798 | 3683 | BindingDB \| ChEMBL |
| *Gymnema sylvestre* | Leaf | CID_798 | 3689 | ChEMBL \| NPASS |
| *Gymnema sylvestre* | Leaf | CID_798 | 3725 | NPASS |
| *Gymnema sylvestre* | Leaf | CID_798 | 759 | BindingDB \| ChEMBL \| NPASS |
| *Gymnema sylvestre* | Leaf | CID_798 | 760 | BindingDB \| ChEMBL \| NPASS |
| *Gymnema sylvestre* | Leaf | CID_798 | 762 | BindingDB \| ChEMBL \| NPASS |
| *Gymnema sylvestre* | Leaf | CID_798 | 765 | ChEMBL \| NPASS |
| *Gymnema sylvestre* | Leaf | CID_798 | 8851 | ChEMBL \| NPASS |
| *Gymnema sylvestre* | Leaf | CID_798 | 8989 | ChEMBL \| NPASS |
| *Gymnema sylvestre* | Leaf | CID_12397 | 367 | NPASS |
| *Gymnema sylvestre* | Leaf | CID_892 | 10013 | ChEMBL |
| *Gymnema sylvestre* | Leaf | CID_892 | 11255 | ChEMBL |
| *Gymnema sylvestre* | Leaf | CID_892 | 1128 | ChEMBL |
| *Gymnema sylvestre* | Leaf | CID_892 | 1129 | ChEMBL |
| *Gymnema sylvestre* | Leaf | CID_892 | 140 | ChEMBL |
| *Gymnema sylvestre* | Leaf | CID_892 | 148 | ChEMBL |
| *Gymnema sylvestre* | Leaf | CID_892 | 150 | ChEMBL |
| *Gymnema sylvestre* | Leaf | CID_892 | 1812 | ChEMBL |
| *Gymnema sylvestre* | Leaf | CID_892 | 1814 | ChEMBL |
| *Gymnema sylvestre* | Leaf | CID_892 | 2099 | ChEMBL |
| *Gymnema sylvestre* | Leaf | CID_892 | 2147 | ChEMBL |
| *Gymnema sylvestre* | Leaf | CID_892 | 3350 | ChEMBL |
| *Gymnema sylvestre* | Leaf | CID_892 | 3357 | ChEMBL |
| *Gymnema sylvestre* | Leaf | CID_892 | 351 | ChEMBL |
| *Gymnema sylvestre* | Leaf | CID_892 | 367 | ChEMBL |
| *Gymnema sylvestre* | Leaf | CID_892 | 3757 | ChEMBL |
| *Gymnema sylvestre* | Leaf | CID_892 | 3791 | ChEMBL |
| *Gymnema sylvestre* | Leaf | CID_892 | 4128 | ChEMBL |
| *Gymnema sylvestre* | Leaf | CID_892 | 43 | ChEMBL |
| *Gymnema sylvestre* | Leaf | CID_892 | 49 | ChEMBL |
| *Gymnema sylvestre* | Leaf | CID_892 | 4988 | ChEMBL |
| *Gymnema sylvestre* | Leaf | CID_892 | 5139 | ChEMBL |
| *Gymnema sylvestre* | Leaf | CID_892 | 5141 | ChEMBL |
| *Gymnema sylvestre* | Leaf | CID_892 | 5241 | ChEMBL |
| *Gymnema sylvestre* | Leaf | CID_892 | 55775 | ChEMBL |
| *Gymnema sylvestre* | Leaf | CID_892 | 5742 | ChEMBL |
| *Gymnema sylvestre* | Leaf | CID_892 | 6530 | ChEMBL |
| *Gymnema sylvestre* | Leaf | CID_892 | 6531 | ChEMBL |
| *Gymnema sylvestre* | Leaf | CID_892 | 6532 | ChEMBL |
| *Gymnema sylvestre* | Leaf | CID_892 | 6915 | ChEMBL |
| *Gymnema sylvestre* | Leaf | CID_785 | 10013 | ChEMBL |
| *Gymnema sylvestre* | Leaf | CID_785 | 10599 | ChEMBL \| NPASS |
| *Gymnema sylvestre* | Leaf | CID_785 | 10919 | ChEMBL \| NPASS |
| *Gymnema sylvestre* | Leaf | CID_785 | 10951 | NPASS |
| *Gymnema sylvestre* | Leaf | CID_785 | 11238 | BindingDB \| ChEMBL \| NPASS |
| *Gymnema sylvestre* | Leaf | CID_785 | 1576 | NPASS |
| *Gymnema sylvestre* | Leaf | CID_785 | 196 | NPASS |
| *Gymnema sylvestre* | Leaf | CID_785 | 2099 | NPASS |
| *Gymnema sylvestre* | Leaf | CID_785 | 216 | NPASS |
| *Gymnema sylvestre* | Leaf | CID_785 | 2237 | NPASS |
| *Gymnema sylvestre* | Leaf | CID_785 | 23632 | BindingDB \| ChEMBL \| NPASS |
| *Gymnema sylvestre* | Leaf | CID_785 | 246 | NPASS |
| *Gymnema sylvestre* | Leaf | CID_785 | 2475 | NPASS |
| *Gymnema sylvestre* | Leaf | CID_785 | 2744 | NPASS |
| *Gymnema sylvestre* | Leaf | CID_785 | 28234 | ChEMBL \| NPASS |
| *Gymnema sylvestre* | Leaf | CID_785 | 2908 | NPASS |
| *Gymnema sylvestre* | Leaf | CID_785 | 3028 | ChEMBL \| NPASS |
| *Gymnema sylvestre* | Leaf | CID_785 | 3043 | NPASS |
| *Gymnema sylvestre* | Leaf | CID_785 | 3091 | ChEMBL \| NPASS |
| *Gymnema sylvestre* | Leaf | CID_785 | 3248 | ChEMBL \| NPASS |
| *Gymnema sylvestre* | Leaf | CID_785 | 3315 | NPASS |
| *Gymnema sylvestre* | Leaf | CID_785 | 3576 | ChEMBL \| NPASS |
| *Gymnema sylvestre* | Leaf | CID_785 | 367 | NPASS |
| *Gymnema sylvestre* | Leaf | CID_785 | 3725 | NPASS |
| *Gymnema sylvestre* | Leaf | CID_785 | 390245 | NPASS |
| *Gymnema sylvestre* | Leaf | CID_785 | 4000 | NPASS |
| *Gymnema sylvestre* | Leaf | CID_785 | 410 | NPASS |
| *Gymnema sylvestre* | Leaf | CID_785 | 4137 | NPASS |
| *Gymnema sylvestre* | Leaf | CID_785 | 4297 | NPASS |
| *Gymnema sylvestre* | Leaf | CID_785 | 4780 | ChEMBL \| NPASS |
| *Gymnema sylvestre* | Leaf | CID_785 | 4790 | ChEMBL \| NPASS |
| *Gymnema sylvestre* | Leaf | CID_785 | 4864 | NPASS |
| *Gymnema sylvestre* | Leaf | CID_785 | 51053 | NPASS |
| *Gymnema sylvestre* | Leaf | CID_785 | 5300 | NPASS |
| *Gymnema sylvestre* | Leaf | CID_785 | 5376 | NPASS |
| *Gymnema sylvestre* | Leaf | CID_785 | 54575 | ChEMBL \| NPASS |
| *Gymnema sylvestre* | Leaf | CID_785 | 54577 | ChEMBL \| NPASS |
| *Gymnema sylvestre* | Leaf | CID_785 | 5467 | NPASS |
| *Gymnema sylvestre* | Leaf | CID_785 | 54737 | NPASS |
| *Gymnema sylvestre* | Leaf | CID_785 | 55775 | ChEMBL \| NPASS |
| *Gymnema sylvestre* | Leaf | CID_785 | 5743 | ChEMBL \| NPASS |
| *Gymnema sylvestre* | Leaf | CID_785 | 5965 | NPASS |
| *Gymnema sylvestre* | Leaf | CID_785 | 5999 | NPASS |
| *Gymnema sylvestre* | Leaf | CID_785 | 6097 | ChEMBL \| NPASS |
| *Gymnema sylvestre* | Leaf | CID_785 | 6256 | NPASS |
| *Gymnema sylvestre* | Leaf | CID_785 | 6311 | NPASS |
| *Gymnema sylvestre* | Leaf | CID_785 | 641 | NPASS |
| *Gymnema sylvestre* | Leaf | CID_785 | 7066 | ChEMBL \| NPASS |
| *Gymnema sylvestre* | Leaf | CID_785 | 7068 | NPASS |
| *Gymnema sylvestre* | Leaf | CID_785 | 7299 | BindingDB \| ChEMBL \| NPASS |
| *Gymnema sylvestre* | Leaf | CID_785 | 759 | BindingDB \| ChEMBL \| NPASS |
| *Gymnema sylvestre* | Leaf | CID_785 | 760 | BindingDB \| ChEMBL \| NPASS |
| *Gymnema sylvestre* | Leaf | CID_785 | 761 | BindingDB \| ChEMBL \| NPASS |
| *Gymnema sylvestre* | Leaf | CID_785 | 762 | BindingDB \| ChEMBL \| NPASS |
| *Gymnema sylvestre* | Leaf | CID_785 | 763 | BindingDB \| ChEMBL \| NPASS |
| *Gymnema sylvestre* | Leaf | CID_785 | 765 | NPASS |
| *Gymnema sylvestre* | Leaf | CID_785 | 766 | BindingDB \| NPASS |
| *Gymnema sylvestre* | Leaf | CID_785 | 768 | BindingDB \| ChEMBL \| NPASS |
| *Gymnema sylvestre* | Leaf | CID_785 | 771 | BindingDB \| ChEMBL \| NPASS |
| *Gymnema sylvestre* | Leaf | CID_785 | 79915 | NPASS |
| *Gymnema sylvestre* | Leaf | CID_785 | 9367 | NPASS |
| *Gymnema sylvestre* | Leaf | CID_785 | 9971 | NPASS |
| *Gymnema sylvestre* | Leaf | CID_9015 | 10013 | ChEMBL |
| *Gymnema sylvestre* | Leaf | CID_9015 | 10941 | ChEMBL |
| *Gymnema sylvestre* | Leaf | CID_9015 | 3315 | NPASS |
| *Gymnema sylvestre* | Leaf | CID_9015 | 3417 | NPASS |
| *Gymnema sylvestre* | Leaf | CID_9015 | 4780 | NPASS |
| *Gymnema sylvestre* | Leaf | CID_9015 | 5467 | NPASS |
| *Gymnema sylvestre* | Leaf | CID_9015 | 55775 | NPASS |
| *Gymnema sylvestre* | Leaf | CID_9015 | 6097 | NPASS |
| *Gymnema sylvestre* | Leaf | CID_12101 | 4780 | NPASS |
| *Enicostema axillare* | Whole Plant | CID_11005 | 10411 | NPASS |
| *Enicostema axillare* | Whole Plant | CID_11005 | 10951 | NPASS |
| *Enicostema axillare* | Whole Plant | CID_11005 | 1588 | ChEMBL \| NPASS |
| *Enicostema axillare* | Whole Plant | CID_11005 | 1759 | BindingDB \| ChEMBL \| NPASS |
| *Enicostema axillare* | Whole Plant | CID_11005 | 2237 | NPASS |
| *Enicostema axillare* | Whole Plant | CID_11005 | 23435 | NPASS |
| *Enicostema axillare* | Whole Plant | CID_11005 | 247 | NPASS |
| *Enicostema axillare* | Whole Plant | CID_11005 | 2648 | NPASS |
| *Enicostema axillare* | Whole Plant | CID_11005 | 2740 | NPASS |
| *Enicostema axillare* | Whole Plant | CID_11005 | 3315 | NPASS |
| *Enicostema axillare* | Whole Plant | CID_11005 | 367 | NPASS |
| *Enicostema axillare* | Whole Plant | CID_11005 | 51053 | NPASS |
| *Enicostema axillare* | Whole Plant | CID_11005 | 51548 | BindingDB \| ChEMBL |
| *Enicostema axillare* | Whole Plant | CID_11005 | 5347 | NPASS |
| *Enicostema axillare* | Whole Plant | CID_11005 | 53831 | ChEMBL \| NPASS |
| *Enicostema axillare* | Whole Plant | CID_11005 | 54575 | ChEMBL \| NPASS |
| *Enicostema axillare* | Whole Plant | CID_11005 | 54576 | ChEMBL \| NPASS |
| *Enicostema axillare* | Whole Plant | CID_11005 | 5465 | ChEMBL \| NPASS |
| *Enicostema axillare* | Whole Plant | CID_11005 | 5467 | ChEMBL \| NPASS |
| *Enicostema axillare* | Whole Plant | CID_11005 | 5468 | BindingDB \| ChEMBL \| NPASS |
| *Enicostema axillare* | Whole Plant | CID_11005 | 55775 | NPASS |
| *Enicostema axillare* | Whole Plant | CID_11005 | 6609 | NPASS |
| *Enicostema axillare* | Whole Plant | CID_11005 | 7097 | ChEMBL \| NPASS |
| *Enicostema axillare* | Whole Plant | CID_11005 | 7398 | NPASS |
| *Enicostema axillare* | Whole Plant | CID_11005 | 7421 | NPASS |
| *Enicostema axillare* | Whole Plant | CID_11005 | 8989 | ChEMBL \| NPASS |
| *Enicostema axillare* | Whole Plant | CID_11005 | 9971 | NPASS |
| *Enicostema axillare* | Whole Plant | CID_162350 | 2548 | BindingDB \| ChEMBL \| NPASS |
| *Enicostema axillare* | Whole Plant | CID_162350 | 2778 | NPASS |
| *Enicostema axillare* | Whole Plant | CID_162350 | 3417 | NPASS |
| *Enicostema axillare* | Whole Plant | CID_162350 | 55775 | NPASS |
| *Enicostema axillare* | Whole Plant | CID_445639 | 10018 | BindingDB |
| *Enicostema axillare* | Whole Plant | CID_445639 | 100861540 | ChEMBL |
| *Enicostema axillare* | Whole Plant | CID_445639 | 10411 | NPASS |
| *Enicostema axillare* | Whole Plant | CID_445639 | 10951 | NPASS |
| *Enicostema axillare* | Whole Plant | CID_445639 | 11069 | NPASS |
| *Enicostema axillare* | Whole Plant | CID_445639 | 11255 | ChEMBL |
| *Enicostema axillare* | Whole Plant | CID_445639 | 1128 | ChEMBL |
| *Enicostema axillare* | Whole Plant | CID_445639 | 11283 | ChEMBL |
| *Enicostema axillare* | Whole Plant | CID_445639 | 1129 | ChEMBL |
| *Enicostema axillare* | Whole Plant | CID_445639 | 113612 | ChEMBL |
| *Enicostema axillare* | Whole Plant | CID_445639 | 120227 | ChEMBL |
| *Enicostema axillare* | Whole Plant | CID_445639 | 122706 | ChEMBL \| NPASS |
| *Enicostema axillare* | Whole Plant | CID_445639 | 140 | ChEMBL |
| *Enicostema axillare* | Whole Plant | CID_445639 | 143471 | ChEMBL \| NPASS |
| *Enicostema axillare* | Whole Plant | CID_445639 | 148 | ChEMBL |
| *Enicostema axillare* | Whole Plant | CID_445639 | 150 | ChEMBL |
| *Enicostema axillare* | Whole Plant | CID_445639 | 1543 | ChEMBL |
| *Enicostema axillare* | Whole Plant | CID_445639 | 1544 | ChEMBL |
| *Enicostema axillare* | Whole Plant | CID_445639 | 1545 | ChEMBL |
| *Enicostema axillare* | Whole Plant | CID_445639 | 1548 | ChEMBL |
| *Enicostema axillare* | Whole Plant | CID_445639 | 1549 | ChEMBL |
| *Enicostema axillare* | Whole Plant | CID_445639 | 1553 | ChEMBL |
| *Enicostema axillare* | Whole Plant | CID_445639 | 1555 | ChEMBL |
| *Enicostema axillare* | Whole Plant | CID_445639 | 1557 | ChEMBL |
| *Enicostema axillare* | Whole Plant | CID_445639 | 1558 | ChEMBL |
| *Enicostema axillare* | Whole Plant | CID_445639 | 1559 | ChEMBL |
| *Enicostema axillare* | Whole Plant | CID_445639 | 1562 | ChEMBL |
| *Enicostema axillare* | Whole Plant | CID_445639 | 1565 | ChEMBL |
| *Enicostema axillare* | Whole Plant | CID_445639 | 1571 | ChEMBL |
| *Enicostema axillare* | Whole Plant | CID_445639 | 1572 | ChEMBL |
| *Enicostema axillare* | Whole Plant | CID_445639 | 1573 | ChEMBL |
| *Enicostema axillare* | Whole Plant | CID_445639 | 1576 | ChEMBL |
| *Enicostema axillare* | Whole Plant | CID_445639 | 1577 | ChEMBL |
| *Enicostema axillare* | Whole Plant | CID_445639 | 1579 | ChEMBL |
| *Enicostema axillare* | Whole Plant | CID_445639 | 1580 | ChEMBL |
| *Enicostema axillare* | Whole Plant | CID_445639 | 1588 | BindingDB \| ChEMBL \| NPASS |
| *Enicostema axillare* | Whole Plant | CID_445639 | 1812 | ChEMBL |
| *Enicostema axillare* | Whole Plant | CID_445639 | 1814 | ChEMBL |
| *Enicostema axillare* | Whole Plant | CID_445639 | 2052 | ChEMBL |
| *Enicostema axillare* | Whole Plant | CID_445639 | 2053 | ChEMBL |
| *Enicostema axillare* | Whole Plant | CID_445639 | 2099 | ChEMBL |
| *Enicostema axillare* | Whole Plant | CID_445639 | 2147 | ChEMBL |
| *Enicostema axillare* | Whole Plant | CID_445639 | 2152 | BindingDB \| ChEMBL \| NPASS |
| *Enicostema axillare* | Whole Plant | CID_445639 | 2155 | ChEMBL |
| *Enicostema axillare* | Whole Plant | CID_445639 | 2166 | ChEMBL \| NPASS |
| *Enicostema axillare* | Whole Plant | CID_445639 | 2167 | ChEMBL \| NPASS |
| *Enicostema axillare* | Whole Plant | CID_445639 | 2168 | BindingDB \| ChEMBL \| NPASS |
| *Enicostema axillare* | Whole Plant | CID_445639 | 2171 | ChEMBL \| NPASS |
| *Enicostema axillare* | Whole Plant | CID_445639 | 2554 | ChEMBL |
| *Enicostema axillare* | Whole Plant | CID_445639 | 2555 | ChEMBL |
| *Enicostema axillare* | Whole Plant | CID_445639 | 2561 | ChEMBL |
| *Enicostema axillare* | Whole Plant | CID_445639 | 2566 | ChEMBL |
| *Enicostema axillare* | Whole Plant | CID_445639 | 2648 | ChEMBL \| NPASS |
| *Enicostema axillare* | Whole Plant | CID_445639 | 284541 | ChEMBL |
| *Enicostema axillare* | Whole Plant | CID_445639 | 29785 | ChEMBL |
| *Enicostema axillare* | Whole Plant | CID_445639 | 3028 | ChEMBL \| NPASS |
| *Enicostema axillare* | Whole Plant | CID_445639 | 328 | NPASS |
| *Enicostema axillare* | Whole Plant | CID_445639 | 3315 | NPASS |
| *Enicostema axillare* | Whole Plant | CID_445639 | 3350 | ChEMBL |
| *Enicostema axillare* | Whole Plant | CID_445639 | 3357 | ChEMBL |
| *Enicostema axillare* | Whole Plant | CID_445639 | 367 | ChEMBL |
| *Enicostema axillare* | Whole Plant | CID_445639 | 3757 | ChEMBL \| NPASS |
| *Enicostema axillare* | Whole Plant | CID_445639 | 3791 | ChEMBL |
| *Enicostema axillare* | Whole Plant | CID_445639 | 4051 | ChEMBL |
| *Enicostema axillare* | Whole Plant | CID_445639 | 4128 | ChEMBL |
| *Enicostema axillare* | Whole Plant | CID_445639 | 4137 | NPASS |
| *Enicostema axillare* | Whole Plant | CID_445639 | 4297 | NPASS |
| *Enicostema axillare* | Whole Plant | CID_445639 | 43 | ChEMBL |
| *Enicostema axillare* | Whole Plant | CID_445639 | 472 | NPASS |
| *Enicostema axillare* | Whole Plant | CID_445639 | 4929 | BindingDB \| ChEMBL |
| *Enicostema axillare* | Whole Plant | CID_445639 | 4988 | ChEMBL |
| *Enicostema axillare* | Whole Plant | CID_445639 | 51053 | ChEMBL \| NPASS |
| *Enicostema axillare* | Whole Plant | CID_445639 | 5139 | ChEMBL |
| *Enicostema axillare* | Whole Plant | CID_445639 | 5141 | ChEMBL |
| *Enicostema axillare* | Whole Plant | CID_445639 | 51426 | NPASS |
| *Enicostema axillare* | Whole Plant | CID_445639 | 51548 | ChEMBL |
| *Enicostema axillare* | Whole Plant | CID_445639 | 5241 | ChEMBL |
| *Enicostema axillare* | Whole Plant | CID_445639 | 5347 | ChEMBL \| NPASS |
| *Enicostema axillare* | Whole Plant | CID_445639 | 5465 | ChEMBL \| NPASS |
| *Enicostema axillare* | Whole Plant | CID_445639 | 5467 | ChEMBL \| NPASS |
| *Enicostema axillare* | Whole Plant | CID_445639 | 5468 | BindingDB \| ChEMBL \| NPASS |
| *Enicostema axillare* | Whole Plant | CID_445639 | 54905 | ChEMBL |
| *Enicostema axillare* | Whole Plant | CID_445639 | 5578 | ChEMBL \| NPASS |
| *Enicostema axillare* | Whole Plant | CID_445639 | 5682 | ChEMBL \| NPASS |
| *Enicostema axillare* | Whole Plant | CID_445639 | 5683 | ChEMBL \| NPASS |
| *Enicostema axillare* | Whole Plant | CID_445639 | 5684 | ChEMBL \| NPASS |
| *Enicostema axillare* | Whole Plant | CID_445639 | 5685 | ChEMBL \| NPASS |
| *Enicostema axillare* | Whole Plant | CID_445639 | 5686 | ChEMBL \| NPASS |
| *Enicostema axillare* | Whole Plant | CID_445639 | 5687 | ChEMBL \| NPASS |
| *Enicostema axillare* | Whole Plant | CID_445639 | 5688 | ChEMBL \| NPASS |
| *Enicostema axillare* | Whole Plant | CID_445639 | 5689 | ChEMBL \| NPASS |
| *Enicostema axillare* | Whole Plant | CID_445639 | 5690 | ChEMBL \| NPASS |
| *Enicostema axillare* | Whole Plant | CID_445639 | 5691 | ChEMBL \| NPASS |
| *Enicostema axillare* | Whole Plant | CID_445639 | 5692 | ChEMBL \| NPASS |
| *Enicostema axillare* | Whole Plant | CID_445639 | 5693 | ChEMBL \| NPASS |
| *Enicostema axillare* | Whole Plant | CID_445639 | 5694 | ChEMBL \| NPASS |
| *Enicostema axillare* | Whole Plant | CID_445639 | 5695 | ChEMBL \| NPASS |
| *Enicostema axillare* | Whole Plant | CID_445639 | 5696 | ChEMBL \| NPASS |
| *Enicostema axillare* | Whole Plant | CID_445639 | 5698 | ChEMBL \| NPASS |
| *Enicostema axillare* | Whole Plant | CID_445639 | 5699 | ChEMBL \| NPASS |
| *Enicostema axillare* | Whole Plant | CID_445639 | 5742 | ChEMBL |
| *Enicostema axillare* | Whole Plant | CID_445639 | 5745 | NPASS |
| *Enicostema axillare* | Whole Plant | CID_445639 | 5770 | BindingDB \| ChEMBL \| NPASS |
| *Enicostema axillare* | Whole Plant | CID_445639 | 6097 | NPASS |
| *Enicostema axillare* | Whole Plant | CID_445639 | 641 | NPASS |
| *Enicostema axillare* | Whole Plant | CID_445639 | 64816 | ChEMBL |
| *Enicostema axillare* | Whole Plant | CID_445639 | 6530 | ChEMBL |
| *Enicostema axillare* | Whole Plant | CID_445639 | 6531 | ChEMBL |
| *Enicostema axillare* | Whole Plant | CID_445639 | 6532 | ChEMBL |
| *Enicostema axillare* | Whole Plant | CID_445639 | 6646 | BindingDB \| NPASS |
| *Enicostema axillare* | Whole Plant | CID_445639 | 6915 | ChEMBL |
| *Enicostema axillare* | Whole Plant | CID_445639 | 7015 | ChEMBL \| NPASS |
| *Enicostema axillare* | Whole Plant | CID_445639 | 7150 | ChEMBL \| NPASS |
| *Enicostema axillare* | Whole Plant | CID_445639 | 7398 | ChEMBL \| NPASS |
| *Enicostema axillare* | Whole Plant | CID_445639 | 79915 | NPASS |
| *Enicostema axillare* | Whole Plant | CID_445639 | 8529 | ChEMBL |
| *Enicostema axillare* | Whole Plant | CID_445639 | 9682 | ChEMBL \| NPASS |
| *Enicostema axillare* | Whole Plant | CID_445639 | 9971 | NPASS |
| *Enicostema axillare* | Whole Plant | CID_5280443 | 10013 | ChEMBL |
| *Enicostema axillare* | Whole Plant | CID_5280443 | 10072 | BindingDB |
| *Enicostema axillare* | Whole Plant | CID_5280443 | 100861540 | ChEMBL |
| *Enicostema axillare* | Whole Plant | CID_5280443 | 1017 | ChEMBL \| NPASS |
| *Enicostema axillare* | Whole Plant | CID_5280443 | 1020 | ChEMBL |
| *Enicostema axillare* | Whole Plant | CID_5280443 | 1021 | BindingDB \| ChEMBL \| NPASS |
| *Enicostema axillare* | Whole Plant | CID_5280443 | 10298 | ChEMBL \| NPASS |
| *Enicostema axillare* | Whole Plant | CID_5280443 | 10494 | ChEMBL \| NPASS |
| *Enicostema axillare* | Whole Plant | CID_5280443 | 10599 | ChEMBL \| NPASS |
| *Enicostema axillare* | Whole Plant | CID_5280443 | 10645 | ChEMBL \| NPASS |
| *Enicostema axillare* | Whole Plant | CID_5280443 | 106821730 | ChEMBL |
| *Enicostema axillare* | Whole Plant | CID_5280443 | 10733 | ChEMBL \| NPASS |
| *Enicostema axillare* | Whole Plant | CID_5280443 | 10783 | ChEMBL \| NPASS |
| *Enicostema axillare* | Whole Plant | CID_5280443 | 1080 | BindingDB \| ChEMBL \| NPASS |
| *Enicostema axillare* | Whole Plant | CID_5280443 | 11040 | ChEMBL \| NPASS |
| *Enicostema axillare* | Whole Plant | CID_5280443 | 1111 | ChEMBL \| NPASS |
| *Enicostema axillare* | Whole Plant | CID_5280443 | 11183 | ChEMBL \| NPASS |
| *Enicostema axillare* | Whole Plant | CID_5280443 | 11200 | ChEMBL \| NPASS |
| *Enicostema axillare* | Whole Plant | CID_5280443 | 11201 | NPASS |
| *Enicostema axillare* | Whole Plant | CID_5280443 | 11283 | ChEMBL |
| *Enicostema axillare* | Whole Plant | CID_5280443 | 11309 | ChEMBL |
| *Enicostema axillare* | Whole Plant | CID_5280443 | 11329 | ChEMBL \| NPASS |
| *Enicostema axillare* | Whole Plant | CID_5280443 | 113612 | ChEMBL |
| *Enicostema axillare* | Whole Plant | CID_5280443 | 114548 | BindingDB \| ChEMBL \| NPASS |
| *Enicostema axillare* | Whole Plant | CID_5280443 | 1195 | ChEMBL \| NPASS |
| *Enicostema axillare* | Whole Plant | CID_5280443 | 1196 | ChEMBL \| NPASS |
| *Enicostema axillare* | Whole Plant | CID_5280443 | 1198 | ChEMBL \| NPASS |
| *Enicostema axillare* | Whole Plant | CID_5280443 | 120227 | ChEMBL |
| *Enicostema axillare* | Whole Plant | CID_5280443 | 1244 | ChEMBL \| NPASS |
| *Enicostema axillare* | Whole Plant | CID_5280443 | 140 | ChEMBL \| NPASS |
| *Enicostema axillare* | Whole Plant | CID_5280443 | 142 | ChEMBL \| NPASS |
| *Enicostema axillare* | Whole Plant | CID_5280443 | 1432 | BindingDB \| ChEMBL \| NPASS |
| *Enicostema axillare* | Whole Plant | CID_5280443 | 1445 | ChEMBL \| NPASS |
| *Enicostema axillare* | Whole Plant | CID_5280443 | 1452 | ChEMBL \| NPASS |
| *Enicostema axillare* | Whole Plant | CID_5280443 | 1453 | ChEMBL |
| *Enicostema axillare* | Whole Plant | CID_5280443 | 1454 | ChEMBL |
| *Enicostema axillare* | Whole Plant | CID_5280443 | 1455 | ChEMBL \| NPASS |
| *Enicostema axillare* | Whole Plant | CID_5280443 | 1456 | ChEMBL \| NPASS |
| *Enicostema axillare* | Whole Plant | CID_5280443 | 1457 | BindingDB \| ChEMBL \| NPASS |
| *Enicostema axillare* | Whole Plant | CID_5280443 | 1459 | ChEMBL \| NPASS |
| *Enicostema axillare* | Whole Plant | CID_5280443 | 1460 | ChEMBL \| NPASS |
| *Enicostema axillare* | Whole Plant | CID_5280443 | 1536 | ChEMBL |
| *Enicostema axillare* | Whole Plant | CID_5280443 | 1543 | BindingDB \| ChEMBL \| NPASS |
| *Enicostema axillare* | Whole Plant | CID_5280443 | 1544 | BindingDB \| ChEMBL \| NPASS |
| *Enicostema axillare* | Whole Plant | CID_5280443 | 1545 | BindingDB \| ChEMBL \| NPASS |
| *Enicostema axillare* | Whole Plant | CID_5280443 | 1548 | ChEMBL |
| *Enicostema axillare* | Whole Plant | CID_5280443 | 1549 | ChEMBL |
| *Enicostema axillare* | Whole Plant | CID_5280443 | 1553 | ChEMBL |
| *Enicostema axillare* | Whole Plant | CID_5280443 | 1555 | ChEMBL |
| *Enicostema axillare* | Whole Plant | CID_5280443 | 1557 | ChEMBL \| NPASS |
| *Enicostema axillare* | Whole Plant | CID_5280443 | 1558 | ChEMBL |
| *Enicostema axillare* | Whole Plant | CID_5280443 | 1559 | ChEMBL \| NPASS |
| *Enicostema axillare* | Whole Plant | CID_5280443 | 1562 | ChEMBL |
| *Enicostema axillare* | Whole Plant | CID_5280443 | 1565 | ChEMBL \| NPASS |
| *Enicostema axillare* | Whole Plant | CID_5280443 | 1571 | ChEMBL |
| *Enicostema axillare* | Whole Plant | CID_5280443 | 1572 | ChEMBL |
| *Enicostema axillare* | Whole Plant | CID_5280443 | 1573 | ChEMBL |
| *Enicostema axillare* | Whole Plant | CID_5280443 | 1576 | ChEMBL \| NPASS |
| *Enicostema axillare* | Whole Plant | CID_5280443 | 1577 | ChEMBL |
| *Enicostema axillare* | Whole Plant | CID_5280443 | 1579 | ChEMBL |
| *Enicostema axillare* | Whole Plant | CID_5280443 | 1580 | ChEMBL |
| *Enicostema axillare* | Whole Plant | CID_5280443 | 1588 | BindingDB \| ChEMBL \| NPASS |
| *Enicostema axillare* | Whole Plant | CID_5280443 | 1612 | BindingDB \| ChEMBL \| NPASS |
| *Enicostema axillare* | Whole Plant | CID_5280443 | 1613 | ChEMBL \| NPASS |
| *Enicostema axillare* | Whole Plant | CID_5280443 | 1760 | ChEMBL \| NPASS |
| *Enicostema axillare* | Whole Plant | CID_5280443 | 1803 | BindingDB \| ChEMBL \| NPASS |
| *Enicostema axillare* | Whole Plant | CID_5280443 | 1859 | ChEMBL \| NPASS |
| *Enicostema axillare* | Whole Plant | CID_5280443 | 1956 | BindingDB \| ChEMBL \| NPASS |
| *Enicostema axillare* | Whole Plant | CID_5280443 | 196 | NPASS |
| *Enicostema axillare* | Whole Plant | CID_5280443 | 1991 | BindingDB \| ChEMBL \| NPASS |
| *Enicostema axillare* | Whole Plant | CID_5280443 | 207 | ChEMBL \| NPASS |
| *Enicostema axillare* | Whole Plant | CID_5280443 | 2099 | ChEMBL \| NPASS |
| *Enicostema axillare* | Whole Plant | CID_5280443 | 2100 | ChEMBL \| NPASS |
| *Enicostema axillare* | Whole Plant | CID_5280443 | 216 | ChEMBL \| NPASS |
| *Enicostema axillare* | Whole Plant | CID_5280443 | 2203 | ChEMBL \| NPASS |
| *Enicostema axillare* | Whole Plant | CID_5280443 | 2237 | NPASS |
| *Enicostema axillare* | Whole Plant | CID_5280443 | 2268 | ChEMBL \| NPASS |
| *Enicostema axillare* | Whole Plant | CID_5280443 | 23043 | ChEMBL \| NPASS |
| *Enicostema axillare* | Whole Plant | CID_5280443 | 231 | ChEMBL \| NPASS |
| *Enicostema axillare* | Whole Plant | CID_5280443 | 2322 | ChEMBL \| NPASS |
| *Enicostema axillare* | Whole Plant | CID_5280443 | 23621 | BindingDB \| ChEMBL \| NPASS |
| *Enicostema axillare* | Whole Plant | CID_5280443 | 240 | ChEMBL \| NPASS |
| *Enicostema axillare* | Whole Plant | CID_5280443 | 246 | ChEMBL \| NPASS |
| *Enicostema axillare* | Whole Plant | CID_5280443 | 253430 | ChEMBL \| NPASS |
| *Enicostema axillare* | Whole Plant | CID_5280443 | 2554 | ChEMBL |
| *Enicostema axillare* | Whole Plant | CID_5280443 | 2555 | ChEMBL |
| *Enicostema axillare* | Whole Plant | CID_5280443 | 2556 | ChEMBL |
| *Enicostema axillare* | Whole Plant | CID_5280443 | 2557 | ChEMBL |
| *Enicostema axillare* | Whole Plant | CID_5280443 | 2558 | ChEMBL |
| *Enicostema axillare* | Whole Plant | CID_5280443 | 2559 | ChEMBL \| NPASS |
| *Enicostema axillare* | Whole Plant | CID_5280443 | 2560 | ChEMBL |
| *Enicostema axillare* | Whole Plant | CID_5280443 | 2561 | ChEMBL |
| *Enicostema axillare* | Whole Plant | CID_5280443 | 2562 | ChEMBL |
| *Enicostema axillare* | Whole Plant | CID_5280443 | 2563 | ChEMBL |
| *Enicostema axillare* | Whole Plant | CID_5280443 | 2564 | ChEMBL |
| *Enicostema axillare* | Whole Plant | CID_5280443 | 2565 | ChEMBL |
| *Enicostema axillare* | Whole Plant | CID_5280443 | 2566 | ChEMBL |
| *Enicostema axillare* | Whole Plant | CID_5280443 | 2567 | ChEMBL |
| *Enicostema axillare* | Whole Plant | CID_5280443 | 2568 | ChEMBL |
| *Enicostema axillare* | Whole Plant | CID_5280443 | 25797 | ChEMBL \| NPASS |
| *Enicostema axillare* | Whole Plant | CID_5280443 | 2629 | NPASS |
| *Enicostema axillare* | Whole Plant | CID_5280443 | 2744 | NPASS |
| *Enicostema axillare* | Whole Plant | CID_5280443 | 2770 | NPASS |
| *Enicostema axillare* | Whole Plant | CID_5280443 | 28234 | ChEMBL \| NPASS |
| *Enicostema axillare* | Whole Plant | CID_5280443 | 283106 | BindingDB \| ChEMBL \| NPASS |
| *Enicostema axillare* | Whole Plant | CID_5280443 | 284541 | ChEMBL |
| *Enicostema axillare* | Whole Plant | CID_5280443 | 2872 | BindingDB \| ChEMBL \| NPASS |
| *Enicostema axillare* | Whole Plant | CID_5280443 | 2908 | ChEMBL \| NPASS |
| *Enicostema axillare* | Whole Plant | CID_5280443 | 2931 | ChEMBL |
| *Enicostema axillare* | Whole Plant | CID_5280443 | 2932 | BindingDB \| ChEMBL \| NPASS |
| *Enicostema axillare* | Whole Plant | CID_5280443 | 29785 | ChEMBL |
| *Enicostema axillare* | Whole Plant | CID_5280443 | 3028 | ChEMBL \| NPASS |
| *Enicostema axillare* | Whole Plant | CID_5280443 | 3043 | NPASS |
| *Enicostema axillare* | Whole Plant | CID_5280443 | 3091 | ChEMBL \| NPASS |
| *Enicostema axillare* | Whole Plant | CID_5280443 | 3248 | ChEMBL \| NPASS |
| *Enicostema axillare* | Whole Plant | CID_5280443 | 328 | ChEMBL \| NPASS |
| *Enicostema axillare* | Whole Plant | CID_5280443 | 3292 | BindingDB \| ChEMBL \| NPASS |
| *Enicostema axillare* | Whole Plant | CID_5280443 | 3293 | BindingDB |
| *Enicostema axillare* | Whole Plant | CID_5280443 | 3294 | BindingDB \| ChEMBL \| NPASS |
| *Enicostema axillare* | Whole Plant | CID_5280443 | 3309 | NPASS |
| *Enicostema axillare* | Whole Plant | CID_5280443 | 3315 | NPASS |
| *Enicostema axillare* | Whole Plant | CID_5280443 | 3417 | ChEMBL \| NPASS |
| *Enicostema axillare* | Whole Plant | CID_5280443 | 351 | BindingDB \| ChEMBL \| NPASS |
| *Enicostema axillare* | Whole Plant | CID_5280443 | 3576 | NPASS |
| *Enicostema axillare* | Whole Plant | CID_5280443 | 3643 | ChEMBL \| NPASS |
| *Enicostema axillare* | Whole Plant | CID_5280443 | 367 | BindingDB \| ChEMBL \| NPASS |
| *Enicostema axillare* | Whole Plant | CID_5280443 | 3716 | ChEMBL \| NPASS |
| *Enicostema axillare* | Whole Plant | CID_5280443 | 3725 | NPASS |
| *Enicostema axillare* | Whole Plant | CID_5280443 | 3778 | ChEMBL \| NPASS |
| *Enicostema axillare* | Whole Plant | CID_5280443 | 387129 | ChEMBL \| NPASS |
| *Enicostema axillare* | Whole Plant | CID_5280443 | 390245 | ChEMBL \| NPASS |
| *Enicostema axillare* | Whole Plant | CID_5280443 | 3932 | ChEMBL \| NPASS |
| *Enicostema axillare* | Whole Plant | CID_5280443 | 4000 | NPASS |
| *Enicostema axillare* | Whole Plant | CID_5280443 | 4051 | ChEMBL |
| *Enicostema axillare* | Whole Plant | CID_5280443 | 4067 | ChEMBL \| NPASS |
| *Enicostema axillare* | Whole Plant | CID_5280443 | 4088 | NPASS |
| *Enicostema axillare* | Whole Plant | CID_5280443 | 410 | NPASS |
| *Enicostema axillare* | Whole Plant | CID_5280443 | 4128 | ChEMBL \| NPASS |
| *Enicostema axillare* | Whole Plant | CID_5280443 | 4129 | ChEMBL \| NPASS |
| *Enicostema axillare* | Whole Plant | CID_5280443 | 4137 | ChEMBL \| NPASS |
| *Enicostema axillare* | Whole Plant | CID_5280443 | 415116 | ChEMBL \| NPASS |
| *Enicostema axillare* | Whole Plant | CID_5280443 | 4217 | ChEMBL \| NPASS |
| *Enicostema axillare* | Whole Plant | CID_5280443 | 4297 | NPASS |
| *Enicostema axillare* | Whole Plant | CID_5280443 | 43 | BindingDB \| ChEMBL \| NPASS |
| *Enicostema axillare* | Whole Plant | CID_5280443 | 4353 | BindingDB \| ChEMBL \| NPASS |
| *Enicostema axillare* | Whole Plant | CID_5280443 | 4363 | ChEMBL \| NPASS |
| *Enicostema axillare* | Whole Plant | CID_5280443 | 4582 | ChEMBL \| NPASS |
| *Enicostema axillare* | Whole Plant | CID_5280443 | 4653 | ChEMBL \| NPASS |
| *Enicostema axillare* | Whole Plant | CID_5280443 | 4751 | ChEMBL \| NPASS |
| *Enicostema axillare* | Whole Plant | CID_5280443 | 4759 | BindingDB \| NPASS |
| *Enicostema axillare* | Whole Plant | CID_5280443 | 4780 | ChEMBL \| NPASS |
| *Enicostema axillare* | Whole Plant | CID_5280443 | 4790 | ChEMBL \| NPASS |
| *Enicostema axillare* | Whole Plant | CID_5280443 | 4791 | ChEMBL |
| *Enicostema axillare* | Whole Plant | CID_5280443 | 4864 | NPASS |
| *Enicostema axillare* | Whole Plant | CID_5280443 | 4985 | ChEMBL \| NPASS |
| *Enicostema axillare* | Whole Plant | CID_5280443 | 4986 | ChEMBL \| NPASS |
| *Enicostema axillare* | Whole Plant | CID_5280443 | 4988 | ChEMBL \| NPASS |
| *Enicostema axillare* | Whole Plant | CID_5280443 | 50507 | BindingDB \| ChEMBL \| NPASS |
| *Enicostema axillare* | Whole Plant | CID_5280443 | 51053 | NPASS |
| *Enicostema axillare* | Whole Plant | CID_5280443 | 51231 | ChEMBL \| NPASS |
| *Enicostema axillare* | Whole Plant | CID_5280443 | 5127 | ChEMBL \| NPASS |
| *Enicostema axillare* | Whole Plant | CID_5280443 | 51426 | ChEMBL \| NPASS |
| *Enicostema axillare* | Whole Plant | CID_5280443 | 51447 | ChEMBL \| NPASS |
| *Enicostema axillare* | Whole Plant | CID_5280443 | 5163 | ChEMBL \| NPASS |
| *Enicostema axillare* | Whole Plant | CID_5280443 | 5170 | ChEMBL \| NPASS |
| *Enicostema axillare* | Whole Plant | CID_5280443 | 51765 | ChEMBL \| NPASS |
| *Enicostema axillare* | Whole Plant | CID_5280443 | 5243 | BindingDB \| ChEMBL \| NPASS |
| *Enicostema axillare* | Whole Plant | CID_5280443 | 5292 | ChEMBL \| NPASS |
| *Enicostema axillare* | Whole Plant | CID_5280443 | 5300 | NPASS |
| *Enicostema axillare* | Whole Plant | CID_5280443 | 5315 | ChEMBL \| NPASS |
| *Enicostema axillare* | Whole Plant | CID_5280443 | 5347 | ChEMBL \| NPASS |
| *Enicostema axillare* | Whole Plant | CID_5280443 | 5376 | NPASS |
| *Enicostema axillare* | Whole Plant | CID_5280443 | 53944 | ChEMBL \| NPASS |
| *Enicostema axillare* | Whole Plant | CID_5280443 | 5406 | ChEMBL |
| *Enicostema axillare* | Whole Plant | CID_5280443 | 5429 | BindingDB \| ChEMBL \| NPASS |
| *Enicostema axillare* | Whole Plant | CID_5280443 | 5444 | ChEMBL \| NPASS |
| *Enicostema axillare* | Whole Plant | CID_5280443 | 54575 | ChEMBL \| NPASS |
| *Enicostema axillare* | Whole Plant | CID_5280443 | 54576 | ChEMBL \| NPASS |
| *Enicostema axillare* | Whole Plant | CID_5280443 | 5465 | ChEMBL \| NPASS |
| *Enicostema axillare* | Whole Plant | CID_5280443 | 54657 | ChEMBL \| NPASS |
| *Enicostema axillare* | Whole Plant | CID_5280443 | 54658 | ChEMBL \| NPASS |
| *Enicostema axillare* | Whole Plant | CID_5280443 | 54659 | ChEMBL \| NPASS |
| *Enicostema axillare* | Whole Plant | CID_5280443 | 5467 | ChEMBL \| NPASS |
| *Enicostema axillare* | Whole Plant | CID_5280443 | 5468 | ChEMBL \| NPASS |
| *Enicostema axillare* | Whole Plant | CID_5280443 | 54737 | NPASS |
| *Enicostema axillare* | Whole Plant | CID_5280443 | 54905 | ChEMBL |
| *Enicostema axillare* | Whole Plant | CID_5280443 | 5563 | ChEMBL \| NPASS |
| *Enicostema axillare* | Whole Plant | CID_5280443 | 5566 | ChEMBL \| NPASS |
| *Enicostema axillare* | Whole Plant | CID_5280443 | 5567 | ChEMBL |
| *Enicostema axillare* | Whole Plant | CID_5280443 | 5568 | ChEMBL \| NPASS |
| *Enicostema axillare* | Whole Plant | CID_5280443 | 55775 | ChEMBL \| NPASS |
| *Enicostema axillare* | Whole Plant | CID_5280443 | 5578 | ChEMBL \| NPASS |
| *Enicostema axillare* | Whole Plant | CID_5280443 | 55781 | ChEMBL \| NPASS |
| *Enicostema axillare* | Whole Plant | CID_5280443 | 55872 | ChEMBL \| NPASS |
| *Enicostema axillare* | Whole Plant | CID_5280443 | 55879 | ChEMBL |
| *Enicostema axillare* | Whole Plant | CID_5280443 | 5594 | ChEMBL \| NPASS |
| *Enicostema axillare* | Whole Plant | CID_5280443 | 5595 | ChEMBL \| NPASS |
| *Enicostema axillare* | Whole Plant | CID_5280443 | 5597 | ChEMBL \| NPASS |
| *Enicostema axillare* | Whole Plant | CID_5280443 | 5599 | ChEMBL |
| *Enicostema axillare* | Whole Plant | CID_5280443 | 5600 | ChEMBL \| NPASS |
| *Enicostema axillare* | Whole Plant | CID_5280443 | 5601 | ChEMBL \| NPASS |
| *Enicostema axillare* | Whole Plant | CID_5280443 | 5602 | BindingDB \| ChEMBL |
| *Enicostema axillare* | Whole Plant | CID_5280443 | 5603 | ChEMBL \| NPASS |
| *Enicostema axillare* | Whole Plant | CID_5280443 | 5604 | ChEMBL \| NPASS |
| *Enicostema axillare* | Whole Plant | CID_5280443 | 5605 | ChEMBL \| NPASS |
| *Enicostema axillare* | Whole Plant | CID_5280443 | 5608 | ChEMBL \| NPASS |
| *Enicostema axillare* | Whole Plant | CID_5280443 | 56924 | NPASS |
| *Enicostema axillare* | Whole Plant | CID_5280443 | 5693 | BindingDB \| ChEMBL \| NPASS |
| *Enicostema axillare* | Whole Plant | CID_5280443 | 57016 | ChEMBL \| NPASS |
| *Enicostema axillare* | Whole Plant | CID_5280443 | 57118 | ChEMBL \| NPASS |
| *Enicostema axillare* | Whole Plant | CID_5280443 | 57144 | ChEMBL \| NPASS |
| *Enicostema axillare* | Whole Plant | CID_5280443 | 57172 | ChEMBL \| NPASS |
| *Enicostema axillare* | Whole Plant | CID_5280443 | 5742 | ChEMBL \| NPASS |
| *Enicostema axillare* | Whole Plant | CID_5280443 | 5743 | BindingDB \| ChEMBL \| NPASS |
| *Enicostema axillare* | Whole Plant | CID_5280443 | 5745 | NPASS |
| *Enicostema axillare* | Whole Plant | CID_5280443 | 5770 | BindingDB \| ChEMBL \| NPASS |
| *Enicostema axillare* | Whole Plant | CID_5280443 | 590 | BindingDB \| ChEMBL \| NPASS |
| *Enicostema axillare* | Whole Plant | CID_5280443 | 5970 | ChEMBL |
| *Enicostema axillare* | Whole Plant | CID_5280443 | 5999 | NPASS |
| *Enicostema axillare* | Whole Plant | CID_5280443 | 6002 | NPASS |
| *Enicostema axillare* | Whole Plant | CID_5280443 | 6097 | NPASS |
| *Enicostema axillare* | Whole Plant | CID_5280443 | 6195 | ChEMBL \| NPASS |
| *Enicostema axillare* | Whole Plant | CID_5280443 | 6197 | ChEMBL \| NPASS |
| *Enicostema axillare* | Whole Plant | CID_5280443 | 6198 | ChEMBL \| NPASS |
| *Enicostema axillare* | Whole Plant | CID_5280443 | 6256 | NPASS |
| *Enicostema axillare* | Whole Plant | CID_5280443 | 6300 | ChEMBL \| NPASS |
| *Enicostema axillare* | Whole Plant | CID_5280443 | 6311 | NPASS |
| *Enicostema axillare* | Whole Plant | CID_5280443 | 6446 | ChEMBL \| NPASS |
| *Enicostema axillare* | Whole Plant | CID_5280443 | 64816 | ChEMBL |
| *Enicostema axillare* | Whole Plant | CID_5280443 | 6495 | ChEMBL |
| *Enicostema axillare* | Whole Plant | CID_5280443 | 6580 | BindingDB \| ChEMBL |
| *Enicostema axillare* | Whole Plant | CID_5280443 | 6582 | BindingDB \| ChEMBL |
| *Enicostema axillare* | Whole Plant | CID_5280443 | 6606 | ChEMBL \| NPASS |
| *Enicostema axillare* | Whole Plant | CID_5280443 | 672 | ChEMBL \| NPASS |
| *Enicostema axillare* | Whole Plant | CID_5280443 | 6789 | ChEMBL \| NPASS |
| *Enicostema axillare* | Whole Plant | CID_5280443 | 6793 | ChEMBL \| NPASS |
| *Enicostema axillare* | Whole Plant | CID_5280443 | 6850 | BindingDB \| ChEMBL \| NPASS |
| *Enicostema axillare* | Whole Plant | CID_5280443 | 7066 | NPASS |
| *Enicostema axillare* | Whole Plant | CID_5280443 | 7068 | NPASS |
| *Enicostema axillare* | Whole Plant | CID_5280443 | 7153 | ChEMBL \| NPASS |
| *Enicostema axillare* | Whole Plant | CID_5280443 | 7155 | ChEMBL \| NPASS |
| *Enicostema axillare* | Whole Plant | CID_5280443 | 7157 | NPASS |
| *Enicostema axillare* | Whole Plant | CID_5280443 | 7224 | ChEMBL \| NPASS |
| *Enicostema axillare* | Whole Plant | CID_5280443 | 7253 | ChEMBL \| NPASS |
| *Enicostema axillare* | Whole Plant | CID_5280443 | 7276 | ChEMBL \| NPASS |
| *Enicostema axillare* | Whole Plant | CID_5280443 | 7366 | ChEMBL \| NPASS |
| *Enicostema axillare* | Whole Plant | CID_5280443 | 7398 | NPASS |
| *Enicostema axillare* | Whole Plant | CID_5280443 | 7421 | NPASS |
| *Enicostema axillare* | Whole Plant | CID_5280443 | 7443 | ChEMBL \| NPASS |
| *Enicostema axillare* | Whole Plant | CID_5280443 | 7444 | ChEMBL \| NPASS |
| *Enicostema axillare* | Whole Plant | CID_5280443 | 7498 | BindingDB \| ChEMBL \| NPASS |
| *Enicostema axillare* | Whole Plant | CID_5280443 | 759 | BindingDB |
| *Enicostema axillare* | Whole Plant | CID_5280443 | 760 | BindingDB |
| *Enicostema axillare* | Whole Plant | CID_5280443 | 762 | BindingDB |
| *Enicostema axillare* | Whole Plant | CID_5280443 | 79915 | NPASS |
| *Enicostema axillare* | Whole Plant | CID_5280443 | 80351 | ChEMBL \| NPASS |
| *Enicostema axillare* | Whole Plant | CID_5280443 | 814 | ChEMBL \| NPASS |
| *Enicostema axillare* | Whole Plant | CID_5280443 | 815 | ChEMBL \| NPASS |
| *Enicostema axillare* | Whole Plant | CID_5280443 | 816 | ChEMBL \| NPASS |
| *Enicostema axillare* | Whole Plant | CID_5280443 | 817 | ChEMBL \| NPASS |
| *Enicostema axillare* | Whole Plant | CID_5280443 | 818 | ChEMBL \| NPASS |
| *Enicostema axillare* | Whole Plant | CID_5280443 | 8529 | ChEMBL |
| *Enicostema axillare* | Whole Plant | CID_5280443 | 85417 | ChEMBL |
| *Enicostema axillare* | Whole Plant | CID_5280443 | 8550 | ChEMBL \| NPASS |
| *Enicostema axillare* | Whole Plant | CID_5280443 | 8576 | ChEMBL \| NPASS |
| *Enicostema axillare* | Whole Plant | CID_5280443 | 865 | NPASS |
| *Enicostema axillare* | Whole Plant | CID_5280443 | 8654 | ChEMBL \| NPASS |
| *Enicostema axillare* | Whole Plant | CID_5280443 | 8658 | ChEMBL \| NPASS |
| *Enicostema axillare* | Whole Plant | CID_5280443 | 8814 | ChEMBL \| NPASS |
| *Enicostema axillare* | Whole Plant | CID_5280443 | 8851 | ChEMBL \| NPASS |
| *Enicostema axillare* | Whole Plant | CID_5280443 | 891 | ChEMBL |
| *Enicostema axillare* | Whole Plant | CID_5280443 | 9099 | NPASS |
| *Enicostema axillare* | Whole Plant | CID_5280443 | 9133 | ChEMBL \| NPASS |
| *Enicostema axillare* | Whole Plant | CID_5280443 | 9212 | BindingDB |
| *Enicostema axillare* | Whole Plant | CID_5280443 | 9252 | ChEMBL \| NPASS |
| *Enicostema axillare* | Whole Plant | CID_5280443 | 9261 | ChEMBL \| NPASS |
| *Enicostema axillare* | Whole Plant | CID_5280443 | 9263 | ChEMBL \| NPASS |
| *Enicostema axillare* | Whole Plant | CID_5280443 | 9429 | BindingDB \| ChEMBL \| NPASS |
| *Enicostema axillare* | Whole Plant | CID_5280443 | 9475 | ChEMBL \| NPASS |
| *Enicostema axillare* | Whole Plant | CID_5280443 | 9748 | ChEMBL \| NPASS |
| *Enicostema axillare* | Whole Plant | CID_5280443 | 983 | ChEMBL |
| *Enicostema axillare* | Whole Plant | CID_5280443 | 9943 | ChEMBL \| NPASS |
| *Enicostema axillare* | Whole Plant | CID_5280443 | 9971 | ChEMBL \| NPASS |
| *Enicostema axillare* | Whole Plant | CID_5281 | 11255 | ChEMBL |
| *Enicostema axillare* | Whole Plant | CID_5281 | 1128 | ChEMBL |
| *Enicostema axillare* | Whole Plant | CID_5281 | 1129 | ChEMBL |
| *Enicostema axillare* | Whole Plant | CID_5281 | 140 | ChEMBL |
| *Enicostema axillare* | Whole Plant | CID_5281 | 148 | ChEMBL |
| *Enicostema axillare* | Whole Plant | CID_5281 | 150 | ChEMBL |
| *Enicostema axillare* | Whole Plant | CID_5281 | 1588 | ChEMBL \| NPASS |
| *Enicostema axillare* | Whole Plant | CID_5281 | 1812 | ChEMBL |
| *Enicostema axillare* | Whole Plant | CID_5281 | 1814 | ChEMBL |
| *Enicostema axillare* | Whole Plant | CID_5281 | 2099 | ChEMBL |
| *Enicostema axillare* | Whole Plant | CID_5281 | 2147 | ChEMBL |
| *Enicostema axillare* | Whole Plant | CID_5281 | 2167 | ChEMBL \| NPASS |
| *Enicostema axillare* | Whole Plant | CID_5281 | 246 | NPASS |
| *Enicostema axillare* | Whole Plant | CID_5281 | 2554 | ChEMBL |
| *Enicostema axillare* | Whole Plant | CID_5281 | 2555 | ChEMBL |
| *Enicostema axillare* | Whole Plant | CID_5281 | 2561 | ChEMBL |
| *Enicostema axillare* | Whole Plant | CID_5281 | 2566 | ChEMBL |
| *Enicostema axillare* | Whole Plant | CID_5281 | 2908 | NPASS |
| *Enicostema axillare* | Whole Plant | CID_5281 | 3028 | ChEMBL \| NPASS |
| *Enicostema axillare* | Whole Plant | CID_5281 | 328 | NPASS |
| *Enicostema axillare* | Whole Plant | CID_5281 | 3350 | ChEMBL |
| *Enicostema axillare* | Whole Plant | CID_5281 | 3357 | ChEMBL |
| *Enicostema axillare* | Whole Plant | CID_5281 | 367 | ChEMBL |
| *Enicostema axillare* | Whole Plant | CID_5281 | 3757 | ChEMBL |
| *Enicostema axillare* | Whole Plant | CID_5281 | 3791 | ChEMBL |
| *Enicostema axillare* | Whole Plant | CID_5281 | 4128 | ChEMBL |
| *Enicostema axillare* | Whole Plant | CID_5281 | 4297 | NPASS |
| *Enicostema axillare* | Whole Plant | CID_5281 | 43 | ChEMBL |
| *Enicostema axillare* | Whole Plant | CID_5281 | 4790 | NPASS |
| *Enicostema axillare* | Whole Plant | CID_5281 | 4988 | ChEMBL |
| *Enicostema axillare* | Whole Plant | CID_5281 | 5139 | ChEMBL |
| *Enicostema axillare* | Whole Plant | CID_5281 | 5141 | ChEMBL |
| *Enicostema axillare* | Whole Plant | CID_5281 | 5241 | ChEMBL |
| *Enicostema axillare* | Whole Plant | CID_5281 | 5465 | ChEMBL \| NPASS |
| *Enicostema axillare* | Whole Plant | CID_5281 | 5467 | ChEMBL \| NPASS |
| *Enicostema axillare* | Whole Plant | CID_5281 | 5468 | BindingDB \| ChEMBL \| NPASS |
| *Enicostema axillare* | Whole Plant | CID_5281 | 55775 | NPASS |
| *Enicostema axillare* | Whole Plant | CID_5281 | 5742 | ChEMBL |
| *Enicostema axillare* | Whole Plant | CID_5281 | 5770 | BindingDB \| ChEMBL \| NPASS |
| *Enicostema axillare* | Whole Plant | CID_5281 | 6530 | ChEMBL |
| *Enicostema axillare* | Whole Plant | CID_5281 | 6531 | ChEMBL |
| *Enicostema axillare* | Whole Plant | CID_5281 | 6532 | ChEMBL |
| *Enicostema axillare* | Whole Plant | CID_5281 | 6915 | ChEMBL |
| *Enicostema axillare* | Whole Plant | CID_5281 | 7068 | NPASS |
| *Enicostema axillare* | Whole Plant | CID_5281 | 7253 | NPASS |
| *Enicostema axillare* | Whole Plant | CID_5281 | 7421 | NPASS |
| *Enicostema axillare* | Whole Plant | CID_5281617 | 10599 | ChEMBL \| NPASS |
| *Enicostema axillare* | Whole Plant | CID_5281617 | 1543 | BindingDB \| ChEMBL \| NPASS |
| *Enicostema axillare* | Whole Plant | CID_5281617 | 1545 | BindingDB \| ChEMBL \| NPASS |
| *Enicostema axillare* | Whole Plant | CID_5281617 | 2322 | ChEMBL |
| *Enicostema axillare* | Whole Plant | CID_5281617 | 253430 | ChEMBL \| NPASS |
| *Enicostema axillare* | Whole Plant | CID_5281617 | 276 | ChEMBL |
| *Enicostema axillare* | Whole Plant | CID_5281617 | 28234 | ChEMBL \| NPASS |
| *Enicostema axillare* | Whole Plant | CID_5281617 | 2872 | BindingDB \| ChEMBL \| NPASS |
| *Enicostema axillare* | Whole Plant | CID_5281617 | 3932 | NPASS |
| *Enicostema axillare* | Whole Plant | CID_5281617 | 4759 | BindingDB \| NPASS |
| *Enicostema axillare* | Whole Plant | CID_5281617 | 51447 | ChEMBL \| NPASS |
| *Enicostema axillare* | Whole Plant | CID_5281617 | 5163 | ChEMBL \| NPASS |
| *Enicostema axillare* | Whole Plant | CID_72326 | 10013 | ChEMBL |
| *Enicostema axillare* | Whole Plant | CID_72326 | 10599 | ChEMBL \| NPASS |
| *Enicostema axillare* | Whole Plant | CID_72326 | 151306 | BindingDB \| ChEMBL \| NPASS |
| *Enicostema axillare* | Whole Plant | CID_72326 | 28234 | ChEMBL \| NPASS |
| *Enicostema axillare* | Whole Plant | CID_72326 | 387 | ChEMBL |
| *Enicostema axillare* | Whole Plant | CID_72326 | 4907 | ChEMBL \| NPASS |
| *Enicostema axillare* | Whole Plant | CID_72326 | 5579 | BindingDB \| NPASS |
| *Enicostema axillare* | Whole Plant | CID_72326 | 5581 | BindingDB \| NPASS |
| *Enicostema axillare* | Whole Plant | CID_72326 | 5743 | ChEMBL \| NPASS |
| *Enicostema axillare* | Whole Plant | CID_72326 | 5879 | ChEMBL |
| *Enicostema axillare* | Whole Plant | CID_72326 | 5970 | ChEMBL \| NPASS |
| *Enicostema axillare* | Whole Plant | CID_72326 | 6554 | BindingDB |
| *Enicostema axillare* | Whole Plant | CID_72326 | 7153 | ChEMBL \| NPASS |
| *Enicostema axillare* | Whole Plant | CID_72326 | 9971 | NPASS |
| *Enicostema axillare* | Whole Plant | CID_72326 | 998 | ChEMBL |
| *Tecomella undulata* | Bark | CID_222284 | 10013 | ChEMBL |
| *Tecomella undulata* | Bark | CID_222284 | 10599 | ChEMBL \| NPASS |
| *Tecomella undulata* | Bark | CID_222284 | 1565 | ChEMBL \| NPASS |
| *Tecomella undulata* | Bark | CID_222284 | 1576 | BindingDB \| ChEMBL \| NPASS |
| *Tecomella undulata* | Bark | CID_222284 | 1803 | ChEMBL \| NPASS |
| *Tecomella undulata* | Bark | CID_222284 | 2147 | ChEMBL \| NPASS |
| *Tecomella undulata* | Bark | CID_222284 | 28234 | ChEMBL \| NPASS |
| *Tecomella undulata* | Bark | CID_222284 | 2932 | ChEMBL |
| *Tecomella undulata* | Bark | CID_222284 | 3417 | ChEMBL \| NPASS |
| *Tecomella undulata* | Bark | CID_222284 | 51422 | ChEMBL |
| *Tecomella undulata* | Bark | CID_222284 | 5243 | ChEMBL |
| *Tecomella undulata* | Bark | CID_222284 | 53632 | ChEMBL |
| *Tecomella undulata* | Bark | CID_222284 | 5423 | BindingDB \| ChEMBL \| NPASS |
| *Tecomella undulata* | Bark | CID_222284 | 5562 | ChEMBL |
| *Tecomella undulata* | Bark | CID_222284 | 5563 | ChEMBL |
| *Tecomella undulata* | Bark | CID_222284 | 5564 | ChEMBL |
| *Tecomella undulata* | Bark | CID_222284 | 5565 | ChEMBL |
| *Tecomella undulata* | Bark | CID_222284 | 5571 | ChEMBL |
| *Tecomella undulata* | Bark | CID_222284 | 7299 | ChEMBL \| NPASS |
| *Tecomella undulata* | Bark | CID_5742590 | 10013 | ChEMBL |
| *Tecomella undulata* | Bark | CID_5742590 | 2932 | ChEMBL |
| *Tecomella undulata* | Bark | CID_5742590 | 3417 | ChEMBL \| NPASS |
| *Tecomella undulata* | Bark | CID_5742590 | 51422 | ChEMBL |
| *Tecomella undulata* | Bark | CID_5742590 | 53632 | ChEMBL |
| *Tecomella undulata* | Bark | CID_5742590 | 5423 | BindingDB \| ChEMBL \| NPASS |
| *Tecomella undulata* | Bark | CID_5742590 | 5562 | ChEMBL |
| *Tecomella undulata* | Bark | CID_5742590 | 5563 | ChEMBL |
| *Tecomella undulata* | Bark | CID_5742590 | 5564 | ChEMBL |
| *Tecomella undulata* | Bark | CID_5742590 | 5565 | ChEMBL |
| *Tecomella undulata* | Bark | CID_5742590 | 5571 | ChEMBL |
| *Tecomella undulata* | Bark | CID_68406 | 10013 | ChEMBL |
| *Tecomella undulata* | Bark | CID_12535 | 7068 | NPASS |
| *Tecomella undulata* | Bark | CID_3884 | 10599 | ChEMBL |
| *Tecomella undulata* | Bark | CID_3884 | 1723 | ChEMBL |
| *Tecomella undulata* | Bark | CID_3884 | 2739 | ChEMBL |
| *Tecomella undulata* | Bark | CID_3884 | 28234 | ChEMBL |
| *Tecomella undulata* | Bark | CID_3884 | 2859 | ChEMBL |
| *Tecomella undulata* | Bark | CID_3884 | 4137 | ChEMBL |
| *Tecomella undulata* | Bark | CID_3884 | 5315 | ChEMBL |
| *Tecomella undulata* | Bark | CID_3884 | 6197 | ChEMBL |
| *Tecomella undulata* | Bark | CID_3884 | 6606 | ChEMBL |
| *Tecomella undulata* | Bark | CID_3884 | 7398 | ChEMBL |
| *Tecomella undulata* | Bark | CID_3884 | 7421 | ChEMBL |
| *Tecomella undulata* | Bark | CID_3884 | 80854 | ChEMBL |
| *Tecomella undulata* | Bark | CID_3884 | 993 | ChEMBL |
| *Tecomella undulata* | Bark | CID_3884 | 994 | ChEMBL |
| *Tecomella undulata* | Bark | CID_445858 | 10013 | ChEMBL |
| *Tecomella undulata* | Bark | CID_445858 | 10014 | ChEMBL |
| *Tecomella undulata* | Bark | CID_445858 | 10280 | ChEMBL |
| *Tecomella undulata* | Bark | CID_445858 | 10599 | ChEMBL \| NPASS |
| *Tecomella undulata* | Bark | CID_445858 | 10800 | ChEMBL |
| *Tecomella undulata* | Bark | CID_445858 | 11238 | BindingDB \| ChEMBL \| NPASS |
| *Tecomella undulata* | Bark | CID_445858 | 1128 | ChEMBL |
| *Tecomella undulata* | Bark | CID_445858 | 1129 | ChEMBL |
| *Tecomella undulata* | Bark | CID_445858 | 1131 | ChEMBL |
| *Tecomella undulata* | Bark | CID_445858 | 1132 | ChEMBL |
| *Tecomella undulata* | Bark | CID_445858 | 1133 | ChEMBL |
| *Tecomella undulata* | Bark | CID_445858 | 1233 | ChEMBL |
| *Tecomella undulata* | Bark | CID_445858 | 1234 | ChEMBL |
| *Tecomella undulata* | Bark | CID_445858 | 1268 | ChEMBL |
| *Tecomella undulata* | Bark | CID_445858 | 134 | ChEMBL |
| *Tecomella undulata* | Bark | CID_445858 | 135 | ChEMBL |
| *Tecomella undulata* | Bark | CID_445858 | 140 | ChEMBL |
| *Tecomella undulata* | Bark | CID_445858 | 1432 | ChEMBL |
| *Tecomella undulata* | Bark | CID_445858 | 1457 | ChEMBL |
| *Tecomella undulata* | Bark | CID_445858 | 1459 | ChEMBL |
| *Tecomella undulata* | Bark | CID_445858 | 146 | ChEMBL |
| *Tecomella undulata* | Bark | CID_445858 | 1460 | ChEMBL |
| *Tecomella undulata* | Bark | CID_445858 | 150 | ChEMBL |
| *Tecomella undulata* | Bark | CID_445858 | 151 | ChEMBL |
| *Tecomella undulata* | Bark | CID_445858 | 1511 | ChEMBL |
| *Tecomella undulata* | Bark | CID_445858 | 152 | ChEMBL |
| *Tecomella undulata* | Bark | CID_445858 | 153 | ChEMBL |
| *Tecomella undulata* | Bark | CID_445858 | 154 | ChEMBL |
| *Tecomella undulata* | Bark | CID_445858 | 1544 | ChEMBL |
| *Tecomella undulata* | Bark | CID_445858 | 1548 | ChEMBL |
| *Tecomella undulata* | Bark | CID_445858 | 155 | ChEMBL |
| *Tecomella undulata* | Bark | CID_445858 | 1557 | ChEMBL |
| *Tecomella undulata* | Bark | CID_445858 | 1559 | ChEMBL |
| *Tecomella undulata* | Bark | CID_445858 | 1565 | ChEMBL |
| *Tecomella undulata* | Bark | CID_445858 | 1571 | ChEMBL |
| *Tecomella undulata* | Bark | CID_445858 | 1576 | ChEMBL |
| *Tecomella undulata* | Bark | CID_445858 | 1812 | ChEMBL |
| *Tecomella undulata* | Bark | CID_445858 | 1813 | ChEMBL |
| *Tecomella undulata* | Bark | CID_445858 | 1814 | ChEMBL |
| *Tecomella undulata* | Bark | CID_445858 | 1815 | ChEMBL |
| *Tecomella undulata* | Bark | CID_445858 | 186 | ChEMBL |
| *Tecomella undulata* | Bark | CID_445858 | 1909 | ChEMBL |
| *Tecomella undulata* | Bark | CID_445858 | 1956 | ChEMBL |
| *Tecomella undulata* | Bark | CID_445858 | 1991 | ChEMBL |
| *Tecomella undulata* | Bark | CID_445858 | 2064 | ChEMBL |
| *Tecomella undulata* | Bark | CID_445858 | 2099 | ChEMBL \| NPASS |
| *Tecomella undulata* | Bark | CID_445858 | 2100 | ChEMBL |
| *Tecomella undulata* | Bark | CID_445858 | 2260 | ChEMBL \| NPASS |
| *Tecomella undulata* | Bark | CID_445858 | 2263 | ChEMBL \| NPASS |
| *Tecomella undulata* | Bark | CID_445858 | 2321 | ChEMBL |
| *Tecomella undulata* | Bark | CID_445858 | 23632 | BindingDB \| ChEMBL \| NPASS |
| *Tecomella undulata* | Bark | CID_445858 | 2534 | ChEMBL |
| *Tecomella undulata* | Bark | CID_445858 | 259285 | ChEMBL \| NPASS |
| *Tecomella undulata* | Bark | CID_445858 | 28234 | ChEMBL \| NPASS |
| *Tecomella undulata* | Bark | CID_445858 | 283106 | BindingDB \| ChEMBL |
| *Tecomella undulata* | Bark | CID_445858 | 2908 | ChEMBL |
| *Tecomella undulata* | Bark | CID_445858 | 3065 | ChEMBL |
| *Tecomella undulata* | Bark | CID_445858 | 3066 | ChEMBL |
| *Tecomella undulata* | Bark | CID_445858 | 3156 | ChEMBL |
| *Tecomella undulata* | Bark | CID_445858 | 3269 | ChEMBL |
| *Tecomella undulata* | Bark | CID_445858 | 3274 | ChEMBL |
| *Tecomella undulata* | Bark | CID_445858 | 3356 | ChEMBL |
| *Tecomella undulata* | Bark | CID_445858 | 3357 | ChEMBL |
| *Tecomella undulata* | Bark | CID_445858 | 3358 | ChEMBL |
| *Tecomella undulata* | Bark | CID_445858 | 3362 | ChEMBL |
| *Tecomella undulata* | Bark | CID_445858 | 3363 | ChEMBL \| NPASS |
| *Tecomella undulata* | Bark | CID_445858 | 3373 | BindingDB \| NPASS |
| *Tecomella undulata* | Bark | CID_445858 | 338442 | ChEMBL \| NPASS |
| *Tecomella undulata* | Bark | CID_445858 | 351 | BindingDB \| ChEMBL \| NPASS |
| *Tecomella undulata* | Bark | CID_445858 | 3577 | ChEMBL |
| *Tecomella undulata* | Bark | CID_445858 | 3579 | ChEMBL |
| *Tecomella undulata* | Bark | CID_445858 | 3757 | ChEMBL |
| *Tecomella undulata* | Bark | CID_445858 | 390245 | NPASS |
| *Tecomella undulata* | Bark | CID_445858 | 3932 | ChEMBL |
| *Tecomella undulata* | Bark | CID_445858 | 4128 | ChEMBL |
| *Tecomella undulata* | Bark | CID_445858 | 4159 | ChEMBL |
| *Tecomella undulata* | Bark | CID_445858 | 4160 | ChEMBL |
| *Tecomella undulata* | Bark | CID_445858 | 4161 | ChEMBL |
| *Tecomella undulata* | Bark | CID_445858 | 43 | BindingDB \| ChEMBL \| NPASS |
| *Tecomella undulata* | Bark | CID_445858 | 4312 | ChEMBL |
| *Tecomella undulata* | Bark | CID_445858 | 4318 | ChEMBL |
| *Tecomella undulata* | Bark | CID_445858 | 4886 | ChEMBL |
| *Tecomella undulata* | Bark | CID_445858 | 4887 | ChEMBL |
| *Tecomella undulata* | Bark | CID_445858 | 4985 | ChEMBL |
| *Tecomella undulata* | Bark | CID_445858 | 4986 | ChEMBL |
| *Tecomella undulata* | Bark | CID_445858 | 4988 | ChEMBL |
| *Tecomella undulata* | Bark | CID_445858 | 51053 | NPASS |
| *Tecomella undulata* | Bark | CID_445858 | 51564 | ChEMBL |
| *Tecomella undulata* | Bark | CID_445858 | 5315 | ChEMBL |
| *Tecomella undulata* | Bark | CID_445858 | 54657 | ChEMBL \| NPASS |
| *Tecomella undulata* | Bark | CID_445858 | 54658 | ChEMBL \| NPASS |
| *Tecomella undulata* | Bark | CID_445858 | 552 | ChEMBL |
| *Tecomella undulata* | Bark | CID_445858 | 5530 | ChEMBL |
| *Tecomella undulata* | Bark | CID_445858 | 55775 | NPASS |
| *Tecomella undulata* | Bark | CID_445858 | 5578 | ChEMBL |
| *Tecomella undulata* | Bark | CID_445858 | 55869 | ChEMBL |
| *Tecomella undulata* | Bark | CID_445858 | 5594 | ChEMBL |
| *Tecomella undulata* | Bark | CID_445858 | 5595 | ChEMBL |
| *Tecomella undulata* | Bark | CID_445858 | 5724 | ChEMBL |
| *Tecomella undulata* | Bark | CID_445858 | 5742 | ChEMBL |
| *Tecomella undulata* | Bark | CID_445858 | 5743 | ChEMBL |
| *Tecomella undulata* | Bark | CID_445858 | 5745 | NPASS |
| *Tecomella undulata* | Bark | CID_445858 | 5788 | ChEMBL |
| *Tecomella undulata* | Bark | CID_445858 | 590 | BindingDB \| ChEMBL \| NPASS |
| *Tecomella undulata* | Bark | CID_445858 | 624 | ChEMBL |
| *Tecomella undulata* | Bark | CID_445858 | 6530 | ChEMBL |
| *Tecomella undulata* | Bark | CID_445858 | 6531 | ChEMBL |
| *Tecomella undulata* | Bark | CID_445858 | 6532 | ChEMBL |
| *Tecomella undulata* | Bark | CID_445858 | 6622 | ChEMBL \| NPASS |
| *Tecomella undulata* | Bark | CID_445858 | 6865 | ChEMBL |
| *Tecomella undulata* | Bark | CID_445858 | 6869 | ChEMBL |
| *Tecomella undulata* | Bark | CID_445858 | 6916 | ChEMBL |
| *Tecomella undulata* | Bark | CID_445858 | 729230 | ChEMBL |
| *Tecomella undulata* | Bark | CID_445858 | 7299 | ChEMBL \| NPASS |
| *Tecomella undulata* | Bark | CID_445858 | 7366 | ChEMBL \| NPASS |
| *Tecomella undulata* | Bark | CID_445858 | 7433 | ChEMBL |
| *Tecomella undulata* | Bark | CID_445858 | 759 | BindingDB \| ChEMBL \| NPASS |
| *Tecomella undulata* | Bark | CID_445858 | 760 | BindingDB \| ChEMBL \| NPASS |
| *Tecomella undulata* | Bark | CID_445858 | 761 | BindingDB \| ChEMBL \| NPASS |
| *Tecomella undulata* | Bark | CID_445858 | 762 | BindingDB \| ChEMBL \| NPASS |
| *Tecomella undulata* | Bark | CID_445858 | 763 | BindingDB \| ChEMBL \| NPASS |
| *Tecomella undulata* | Bark | CID_445858 | 765 | ChEMBL \| NPASS |
| *Tecomella undulata* | Bark | CID_445858 | 766 | BindingDB \| ChEMBL \| NPASS |
| *Tecomella undulata* | Bark | CID_445858 | 768 | BindingDB \| ChEMBL \| NPASS |
| *Tecomella undulata* | Bark | CID_445858 | 771 | BindingDB \| ChEMBL \| NPASS |
| *Tecomella undulata* | Bark | CID_445858 | 79885 | ChEMBL |
| *Tecomella undulata* | Bark | CID_445858 | 799 | ChEMBL |
| *Tecomella undulata* | Bark | CID_445858 | 834 | ChEMBL |
| *Tecomella undulata* | Bark | CID_445858 | 83933 | ChEMBL \| NPASS |
| *Tecomella undulata* | Bark | CID_445858 | 8654 | ChEMBL |
| *Tecomella undulata* | Bark | CID_445858 | 8841 | ChEMBL |
| *Tecomella undulata* | Bark | CID_445858 | 886 | ChEMBL |
| *Tecomella undulata* | Bark | CID_445858 | 9734 | ChEMBL |
| *Tecomella undulata* | Bark | CID_445858 | 9759 | ChEMBL |
| *Tecomella undulata* | Bark | CID_7121 | 10599 | ChEMBL \| NPASS |
| *Tecomella undulata* | Bark | CID_7121 | 23632 | BindingDB \| ChEMBL \| NPASS |
| *Tecomella undulata* | Bark | CID_7121 | 259285 | ChEMBL \| NPASS |
| *Tecomella undulata* | Bark | CID_7121 | 28234 | ChEMBL \| NPASS |
| *Tecomella undulata* | Bark | CID_7121 | 4088 | NPASS |
| *Tecomella undulata* | Bark | CID_7121 | 7172 | ChEMBL \| NPASS |
| *Tecomella undulata* | Bark | CID_7121 | 759 | BindingDB \| ChEMBL \| NPASS |
| *Tecomella undulata* | Bark | CID_7121 | 760 | BindingDB \| ChEMBL \| NPASS |
| *Tecomella undulata* | Bark | CID_7121 | 762 | BindingDB \| ChEMBL \| NPASS |
| *Tecomella undulata* | Bark | CID_7121 | 765 | ChEMBL \| NPASS |
| *Tecomella undulata* | Bark | CID_7121 | 766 | BindingDB \| ChEMBL \| NPASS |
| *Tecomella undulata* | Bark | CID_7121 | 768 | BindingDB \| ChEMBL \| NPASS |
| *Tecomella undulata* | Bark | CID_7121 | 771 | BindingDB \| ChEMBL \| NPASS |

**Table S6: Curated gene set associated with type 2 diabetes mellitus.** Genes were compiled from multiple resources. The table includes genes that are either known targets of FDA-approved antidiabetic drugs or supported by at least two of the four considered databases. For each gene, the Entrez gene ID, gene symbol, evidence source(s), and approved drug information (where applicable) are reported.

| Entrez Gene ID | Gene symbol | Approved drug target | Drug names | Evidence source |
|---|---|---|---|---|
| 2740 | GLP1R | Yes | Pramlintide \| Semaglutide \| Tirzepatide | DrugBank (TTD) \| DisGeNET \| DISEASES \| GeneCards \| CTD |
| 3767 | KCNJ11 | Yes | Glimepiride \| Glipizide \| Gliquidone \| Glyburide \| Nateglinide \| Repaglinide \| Tolazamide | DrugBank (Drug Central \| Santos et al., Drug Central) \| DisGeNET \| DISEASES \| GeneCards \| CTD |
| 5468 | PPARG | Yes | Pioglitazone | DrugBank (Santos et al., Drug Central \| TTD) \| DisGeNET \| DISEASES \| GeneCards \| CTD |
| 1803 | DPP4 | Yes | Alogliptin \| Linagliptin \| Saxagliptin \| Sitagliptin \| Vildagliptin | DrugBank (Drug Central \| Santos et al., Drug Central \| TTD) \| DisGeNET \| DISEASES \| GeneCards |
| 3643 | INSR | Yes | Metformin | DrugBank (TTD) \| DisGeNET \| DISEASES \| GeneCards |
| 6833 | ABCC8 | Yes | Glimepiride \| Glipizide \| Gliquidone \| Glyburide \| Nateglinide \| Repaglinide \| Tolazamide | DrugBank (Drug Central \| Santos et al., Drug Central) \| DisGeNET \| GeneCards \| CTD |
| 4535 | MT-ND1\|ND1 | Yes | Metformin | DrugBank (Santos et al., Drug Central) \| DisGeNET \| GeneCards |
| 6524 | SLC5A2 | Yes | Bexagliflozin \| Canagliflozin \| Dapagliflozin \| Empagliflozin \| Ertugliflozin \| Sotagliflozin | DrugBank (Drug Central \| Santos et al., Drug Central \| TTD) \| DISEASES \| GeneCards |
| 6523 | SLC5A1 | Yes | Sotagliflozin | DrugBank (Drug Central) \| DISEASES |
| 8972 | MGAM | Yes | Miglitol | DrugBank (Santos et al., Drug Central) \| DISEASES |
| 10060 | ABCC9 | Yes | Glimepiride | DrugBank (Santos et al., Drug Central) |
| 126328 | NDUFA11 | Yes | Metformin | DrugBank (Santos et al., Drug Central) |
| 2548 | GAA | Yes | Miglitol | DrugBank (Santos et al., Drug Central) |
| 25915 | NDUFAF3 | Yes | Metformin | DrugBank (Santos et al., Drug Central) |
| 29078 | NDUFAF4 | Yes | Metformin | DrugBank (Santos et al., Drug Central) |
| 32 | ACACB | Yes | Metformin | DrugBank (TTD) |
| 374291 | NDUFS7 | Yes | Metformin | DrugBank (Santos et al., Drug Central) |
| 4536 | MT-ND2 | Yes | Metformin | DrugBank (Santos et al., Drug Central) |
| 4537 | MT-ND3 | Yes | Metformin | DrugBank (Santos et al., Drug Central) |
| 4538 | MT-ND4 | Yes | Metformin | DrugBank (Santos et al., Drug Central) |
| 4539 | MT-ND4L | Yes | Metformin | DrugBank (Santos et al., Drug Central) |
| 4540 | MT-ND5 | Yes | Metformin | DrugBank (Santos et al., Drug Central) |
| 4541 | MT-ND6 | Yes | Metformin | DrugBank (Santos et al., Drug Central) |
| 4694 | NDUFA1 | Yes | Metformin | DrugBank (Santos et al., Drug Central) |
| 4695 | NDUFA2 | Yes | Metformin | DrugBank (Santos et al., Drug Central) |
| 4696 | NDUFA3 | Yes | Metformin | DrugBank (Santos et al., Drug Central) |
| 4697 | COXFA4 | Yes | Metformin | DrugBank (Santos et al., Drug Central) |
| 4698 | NDUFA5 | Yes | Metformin | DrugBank (Santos et al., Drug Central) |
| 4700 | NDUFA6 | Yes | Metformin | DrugBank (Santos et al., Drug Central) |
| 4701 | NDUFA7 | Yes | Metformin | DrugBank (Santos et al., Drug Central) |
| 4702 | NDUFA8 | Yes | Metformin | DrugBank (Santos et al., Drug Central) |
| 4704 | NDUFA9 | Yes | Metformin | DrugBank (Santos et al., Drug Central) |
| 4705 | NDUFA10 | Yes | Metformin | DrugBank (Santos et al., Drug Central) |
| 4706 | NDUFAB1 | Yes | Metformin | DrugBank (Santos et al., Drug Central) |
| 4707 | NDUFB1 | Yes | Metformin | DrugBank (Santos et al., Drug Central) |
| 4708 | NDUFB2 | Yes | Metformin | DrugBank (Santos et al., Drug Central) |
| 4709 | NDUFB3 | Yes | Metformin | DrugBank (Santos et al., Drug Central) |
| 4710 | NDUFB4 | Yes | Metformin | DrugBank (Santos et al., Drug Central) |
| 4711 | NDUFB5 | Yes | Metformin | DrugBank (Santos et al., Drug Central) |
| 4712 | NDUFB6 | Yes | Metformin | DrugBank (Santos et al., Drug Central) |
| 4713 | NDUFB7 | Yes | Metformin | DrugBank (Santos et al., Drug Central) |
| 4714 | NDUFB8 | Yes | Metformin | DrugBank (Santos et al., Drug Central) |
| 4715 | NDUFB9 | Yes | Metformin | DrugBank (Santos et al., Drug Central) |
| 4716 | NDUFB10 | Yes | Metformin | DrugBank (Santos et al., Drug Central) |
| 4717 | NDUFC1 | Yes | Metformin | DrugBank (Santos et al., Drug Central) |
| 4718 | NDUFC2 | Yes | Metformin | DrugBank (Santos et al., Drug Central) |
| 4719 | NDUFS1 | Yes | Metformin | DrugBank (Santos et al., Drug Central) |
| 4720 | NDUFS2 | Yes | Metformin | DrugBank (Santos et al., Drug Central) |
| 4722 | NDUFS3 | Yes | Metformin | DrugBank (Santos et al., Drug Central) |
| 4723 | NDUFV1 | Yes | Metformin | DrugBank (Santos et al., Drug Central) |
| 4724 | NDUFS4 | Yes | Metformin | DrugBank (Santos et al., Drug Central) |
| 4725 | NDUFS5 | Yes | Metformin | DrugBank (Santos et al., Drug Central) |
| 4726 | NDUFS6 | Yes | Metformin | DrugBank (Santos et al., Drug Central) |
| 4728 | NDUFS8 | Yes | Metformin | DrugBank (Santos et al., Drug Central) |
| 4729 | NDUFV2 | Yes | Metformin | DrugBank (Santos et al., Drug Central) |
| 4731 | NDUFV3 | Yes | Metformin | DrugBank (Santos et al., Drug Central) |
| 51079 | NDUFA13 | Yes | Metformin | DrugBank (Santos et al., Drug Central) |
| 51103 | NDUFAF1 | Yes | Metformin | DrugBank (Santos et al., Drug Central) |
| 54539 | NDUFB11 | Yes | Metformin | DrugBank (Santos et al., Drug Central) |
| 55244 | SLC47A1 | Yes | Metformin | DrugBank (TTD) |
| 55967 | NDUFA12 | Yes | Metformin | DrugBank (Santos et al., Drug Central) |
| 56901 | COXFA4L2 | Yes | Metformin | DrugBank (Santos et al., Drug Central) |
| 91942 | NDUFAF2 | Yes | Metformin | DrugBank (Santos et al., Drug Central) |
| 10891 | PPARGC1A | No | - | DisGeNET \| DISEASES \| GeneCards \| CTD |
| 169026 | SLC30A8 | No | - | DisGeNET \| DISEASES \| GeneCards \| CTD |
| 23411 | SIRT1 | No | - | DisGeNET \| DISEASES \| GeneCards \| CTD |
| 26291 | FGF21 | No | - | DisGeNET \| DISEASES \| GeneCards \| CTD |
| 2641 | GCG | No | - | DisGeNET \| DISEASES \| GeneCards \| CTD |
| 2645 | GCK | No | - | DisGeNET \| DISEASES \| GeneCards \| CTD |
| 3375 | IAPP | No | - | DisGeNET \| DISEASES \| GeneCards \| CTD |
| 3383 | ICAM1 | No | - | DisGeNET \| DISEASES \| GeneCards \| CTD |
| 3569 | IL6 | No | - | DisGeNET \| DISEASES \| GeneCards \| CTD |
| 3630 | INS | No | - | DisGeNET \| DISEASES \| GeneCards \| CTD |
| 3667 | IRS1 | No | - | DisGeNET \| DISEASES \| GeneCards \| CTD |
| 3952 | LEP | No | - | DisGeNET \| DISEASES \| GeneCards \| CTD |
| 3953 | LEPR | No | - | DisGeNET \| DISEASES \| GeneCards \| CTD |
| 4846 | NOS3 | No | - | DisGeNET \| DISEASES \| GeneCards \| CTD |
| 5465 | PPARA | No | - | DisGeNET \| DISEASES \| GeneCards \| CTD |
| 54901 | CDKAL1 | No | - | DisGeNET \| DISEASES \| GeneCards \| CTD |
| 56729 | RETN | No | - | DisGeNET \| DISEASES \| GeneCards \| CTD |
| 5770 | PTPN1 | No | - | DisGeNET \| DISEASES \| GeneCards \| CTD |
| 6462 | SHBG | No | - | DisGeNET \| DISEASES \| GeneCards \| CTD |
| 6514 | SLC2A2 | No | - | DisGeNET \| DISEASES \| GeneCards \| CTD |
| 6517 | SLC2A4 | No | - | DisGeNET \| DISEASES \| GeneCards \| CTD |
| 6934 | TCF7L2 | No | - | DisGeNET \| DISEASES \| GeneCards \| CTD |
| 7124 | TNF | No | - | DisGeNET \| DISEASES \| GeneCards \| CTD |
| 7351 | UCP2 | No | - | DisGeNET \| DISEASES \| GeneCards \| CTD |
| 79068 | FTO | No | - | DisGeNET \| DISEASES \| GeneCards \| CTD |
| 8660 | IRS2 | No | - | DisGeNET \| DISEASES \| GeneCards \| CTD |
| 9370 | ADIPOQ | No | - | DisGeNET \| DISEASES \| GeneCards \| CTD |
| 948 | CD36 | No | - | DisGeNET \| DISEASES \| GeneCards \| CTD |
| 10644 | IGF2BP2 | No | - | DisGeNET \| GeneCards \| CTD |
| 11132 | CAPN10 | No | - | DisGeNET \| GeneCards \| CTD |
| 1636 | ACE | No | - | DisGeNET \| DISEASES \| GeneCards |
| 1906 | EDN1 | No | - | DisGeNET \| DISEASES \| CTD |
| 207 | AKT1 | No | - | DisGeNET \| DISEASES \| CTD |
| 208 | AKT2 | No | - | DisGeNET \| GeneCards \| CTD |
| 221895 | JAZF1 | No | - | DisGeNET \| GeneCards \| CTD |
| 2646 | GCKR | No | - | DisGeNET \| GeneCards \| CTD |
| 2820 | GPD2 | No | - | DisGeNET \| GeneCards \| CTD |
| 3087 | HHEX | No | - | DisGeNET \| GeneCards \| CTD |
| 3159 | HMGA1 | No | - | DisGeNET \| GeneCards \| CTD |
| 3172 | HNF4A | No | - | DisGeNET \| GeneCards \| CTD |
| 338 | APOB | No | - | DisGeNET \| DISEASES \| GeneCards |
| 3479 | IGF1 | No | - | DisGeNET \| DISEASES \| GeneCards |
| 3553 | IL1B | No | - | DisGeNET \| DISEASES \| GeneCards |
| 3586 | IL10 | No | - | DisGeNET \| DISEASES \| GeneCards |
| 3651 | PDX1 | No | - | DisGeNET \| GeneCards \| CTD |
| 3784 | KCNQ1 | No | - | DisGeNET \| GeneCards \| CTD |
| 3990 | LIPC | No | - | DisGeNET \| GeneCards \| CTD |
| 4023 | LPL | No | - | DisGeNET \| DISEASES \| GeneCards |
| 4544 | MTNR1B | No | - | DisGeNET \| GeneCards \| CTD |
| 4760 | NEUROD1 | No | - | DisGeNET \| GeneCards \| CTD |
| 4790 | NFKB1 | No | - | DisGeNET \| DISEASES \| CTD |
| 4853 | NOTCH2 | No | - | DisGeNET \| GeneCards \| CTD |
| 5078 | PAX4 | No | - | DisGeNET \| GeneCards \| CTD |
| 5167 | ENPP1 | No | - | DisGeNET \| GeneCards \| CTD |
| 51738 | GHRL | No | - | DisGeNET \| DISEASES \| GeneCards |
| 5294 | PIK3CG | No | - | DisGeNET \| DISEASES \| CTD |
| 5506 | PPP1R3A | No | - | DisGeNET \| GeneCards \| CTD |
| 55600 | ITLN1 | No | - | DisGeNET \| GeneCards \| CTD |
| 56999 | ADAMTS9 | No | - | DisGeNET \| GeneCards \| CTD |
| 5950 | RBP4 | No | - | DisGeNET \| DISEASES \| GeneCards |
| 63892 | THADA | No | - | DisGeNET \| GeneCards \| CTD |
| 6648 | SOD2\|SOD2-2 | No | - | DisGeNET \| DISEASES \| CTD |
| 6720 | SREBF1 | No | - | DisGeNET \| DISEASES \| GeneCards |
| 6774 | STAT3 | No | - | DisGeNET \| DISEASES \| GeneCards |
| 6927 | HNF1A | No | - | DisGeNET \| GeneCards \| CTD |
| 6928 | HNF1B | No | - | DisGeNET \| GeneCards \| CTD |
| 7099 | TLR4 | No | - | DisGeNET \| DISEASES \| GeneCards |
| 7350 | UCP1 | No | - | DisGeNET \| DISEASES \| GeneCards |
| 7466 | WFS1 | No | - | DisGeNET \| GeneCards \| CTD |
| 79602 | ADIPOR2 | No | - | DisGeNET \| DISEASES \| GeneCards |
| 836 | CASP3 | No | - | DisGeNET \| DISEASES \| CTD |
| 847 | CAT | No | - | DisGeNET \| GeneCards \| CTD |
| 9479 | MAPK8IP1 | No | - | DisGeNET \| GeneCards \| CTD |
| 10018 | BCL2L11 | No | - | DisGeNET \| CTD |
| 10239 | AP3S2 | No | - | DisGeNET \| CTD |
| 10296 | MAEA | No | - | DisGeNET \| CTD |
| 1036 | CDO1 | No | - | DisGeNET \| CTD |
| 10363 | HMG20A | No | - | DisGeNET \| CTD |
| 111 | ADCY5 | No | - | DisGeNET \| CTD |
| 11317 | RBPJL | No | - | DisGeNET \| GeneCards |
| 116150 | NUS1 | No | - | DisGeNET \| CTD |
| 129787 | TMEM18 | No | - | DisGeNET \| CTD |
| 132332 | SMIM43 | No | - | DisGeNET \| CTD |
| 136259 | KLF14 | No | - | DisGeNET \| CTD |
| 1363 | CPE | No | - | DisGeNET \| GeneCards |
| 1374 | CPT1A | No | - | DisGeNET \| CTD |
| 1401 | CRP | No | - | DISEASES \| GeneCards |
| 145741 | C2CD4A | No | - | DisGeNET \| CTD |
| 150 | ADRA2A | No | - | DisGeNET \| GeneCards |
| 1535 | CYBA | No | - | DisGeNET \| CTD |
| 1544 | CYP1A2 | No | - | DisGeNET \| CTD |
| 155 | ADRB3 | No | - | DisGeNET \| GeneCards |
| 157855 | KCNU1 | No | - | DisGeNET \| CTD |
| 169792 | GLIS3 | No | - | DisGeNET \| CTD |
| 177 | AGER | No | - | DisGeNET \| GeneCards |
| 181 | AGRP | No | - | DisGeNET \| DISEASES |
| 1889 | ECE1 | No | - | DisGeNET \| CTD |
| 19 | ABCA1 | No | - | DisGeNET \| GeneCards |
| 1909 | EDNRA | No | - | DisGeNET \| CTD |
| 1910 | EDNRB | No | - | DisGeNET \| CTD |
| 1956 | EGFR | No | - | DisGeNET \| CTD |
| 197 | AHSG | No | - | DisGeNET \| GeneCards |
| 2113 | ETS1 | No | - | DisGeNET \| CTD |
| 213 | ALB | No | - | DISEASES \| GeneCards |
| 2167 | FABP4 | No | - | DISEASES \| GeneCards |
| 22808 | MRAS | No | - | DisGeNET \| CTD |
| 22866 | CNKSR2 | No | - | DisGeNET \| CTD |
| 23338 | JADE2 | No | - | DisGeNET \| CTD |
| 2475 | MTOR | No | - | DisGeNET \| DISEASES |
| 252995 | FNDC5 | No | - | DISEASES \| GeneCards |
| 26053 | AUTS2 | No | - | DisGeNET \| CTD |
| 26122 | EPC2 | No | - | DisGeNET \| CTD |
| 2642 | GCGR | No | - | DisGeNET \| GeneCards |
| 2695 | GIP | No | - | DisGeNET \| DISEASES |
| 2729 | GCLC | No | - | DisGeNET \| CTD |
| 2730 | GCLM | No | - | DisGeNET \| CTD |
| 2784 | GNB3 | No | - | DisGeNET \| CTD |
| 2813 | GP2 | No | - | DisGeNET \| CTD |
| 283450 | HECTD4 | No | - | DisGeNET \| CTD |
| 283455 | KSR2 | No | - | DisGeNET \| CTD |
| 2864 | FFAR1 | No | - | DisGeNET \| GeneCards |
| 2876 | GPX1 | No | - | DisGeNET \| CTD |
| 2878 | GPX3 | No | - | DisGeNET \| DISEASES |
| 2888 | GRB14 | No | - | DisGeNET \| CTD |
| 2944 | GSTM1 | No | - | DisGeNET \| CTD |
| 3039 | HBA1 | No | - | DisGeNET \| CTD |
| 3077 | HFE | No | - | DisGeNET \| GeneCards |
| 3098 | HK1 | No | - | DisGeNET \| CTD |
| 3156 | HMGCR | No | - | DisGeNET \| DISEASES |
| 3162 | HMOX1 | No | - | DisGeNET \| CTD |
| 3170 | FOXA2 | No | - | DisGeNET \| GeneCards |
| 3240 | HP | No | - | DisGeNET \| CTD |
| 3263 | HPX | No | - | DisGeNET \| CTD |
| 335 | APOA1 | No | - | DISEASES \| GeneCards |
| 3397 | ID1 | No | - | DisGeNET \| CTD |
| 3416 | IDE | No | - | DisGeNET \| GeneCards |
| 345 | APOC3 | No | - | DISEASES \| GeneCards |
| 3458 | IFNG | No | - | DisGeNET \| DISEASES |
| 348 | APOE | No | - | DISEASES \| GeneCards |
| 3516 | RBPJ | No | - | DisGeNET \| GeneCards |
| 355 | FAS | No | - | DisGeNET \| CTD |
| 3597 | IL13RA1 | No | - | DisGeNET \| CTD |
| 3636 | INPPL1 | No | - | DisGeNET \| CTD |
| 3672 | ITGA1 | No | - | DisGeNET \| CTD |
| 388125 | C2CD4B | No | - | DisGeNET \| CTD |
| 4000 | LMNA | No | - | DisGeNET \| GeneCards |
| 4090 | SMAD5 | No | - | DisGeNET \| CTD |
| 4143 | MAT1A | No | - | DisGeNET \| CTD |
| 4160 | MC4R | No | - | DISEASES \| GeneCards |
| 4524 | MTHFR | No | - | DisGeNET \| GeneCards |
| 467 | ATF3 | No | - | DisGeNET \| CTD |
| 4773 | NFATC2 | No | - | DisGeNET \| CTD |
| 4825 | NKX6-1 | No | - | DisGeNET \| CTD |
| 4843 | NOS2 | No | - | DisGeNET \| CTD |
| 4852 | NPY | No | - | DisGeNET \| DISEASES |
| 488 | ATP2A2 | No | - | DisGeNET \| CTD |
| 489 | ATP2A3 | No | - | DisGeNET \| CTD |
| 493856 | CISD2 | No | - | DisGeNET \| CTD |
| 4968 | OGG1 | No | - | DisGeNET \| CTD |
| 5054 | SERPINE1 | No | - | DISEASES \| GeneCards |
| 5066 | PAM | No | - | DisGeNET \| CTD |
| 50674 | NEUROG3 | No | - | DisGeNET \| GeneCards |
| 5080 | PAX6 | No | - | DisGeNET \| CTD |
| 51094 | ADIPOR1 | No | - | DISEASES \| GeneCards |
| 5126 | PCSK2 | No | - | DisGeNET \| CTD |
| 51530 | ZC3HC1 | No | - | DisGeNET \| CTD |
| 5184 | PEPD | No | - | DisGeNET \| CTD |
| 5346 | PLIN1 | No | - | DISEASES \| GeneCards |
| 5444 | PON1 | No | - | DisGeNET \| GeneCards |
| 5467 | PPARD | No | - | DisGeNET \| DISEASES |
| 5563 | PRKAA2 | No | - | DisGeNET \| GeneCards |
| 5579 | PRKCB | No | - | DisGeNET \| CTD |
| 5629 | PROX1 | No | - | DisGeNET \| CTD |
| 57818 | G6PC2 | No | - | DISEASES \| GeneCards |
| 581 | BAX | No | - | DisGeNET \| CTD |
| 59338 | PLEKHA1 | No | - | DisGeNET \| CTD |
| 596 | BCL2 | No | - | DisGeNET \| CTD |
| 598 | BCL2L1 | No | - | DisGeNET \| CTD |
| 6049 | RNF6 | No | - | DisGeNET \| CTD |
| 60685 | ZFAND3 | No | - | DisGeNET \| CTD |
| 6277 | S100A6 | No | - | DisGeNET \| CTD |
| 6344 | SCTR | No | - | DisGeNET \| CTD |
| 6347 | CCL2 | No | - | DISEASES \| GeneCards |
| 635 | BHMT | No | - | DisGeNET \| CTD |
| 6424 | SFRP4 | No | - | DisGeNET \| CTD |
| 6480 | ST6GAL1 | No | - | DisGeNET \| CTD |
| 64897 | C12orf43 | No | - | DisGeNET \| GeneCards |
| 6506 | SLC1A2 | No | - | DisGeNET \| CTD |
| 6513 | SLC2A1 | No | - | DisGeNET \| CTD |
| 6581 | SLC22A3 | No | - | DisGeNET \| CTD |
| 6616 | SNAP25 | No | - | DisGeNET \| CTD |
| 6647 | SOD1 | No | - | DisGeNET \| CTD |
| 673 | BRAF | No | - | DisGeNET \| CTD |
| 7040 | TGFB1 | No | - | DISEASES \| GeneCards |
| 7076 | TIMP1 | No | - | DisGeNET \| CTD |
| 7132 | TNFRSF1A | No | - | DisGeNET \| CTD |
| 7133 | TNFRSF1B | No | - | DisGeNET \| CTD |
| 718 | C3 | No | - | DisGeNET \| CTD |
| 723961 | INS-IGF2 | No | - | DisGeNET \| GeneCards |
| 7325 | UBE2E2 | No | - | DisGeNET \| CTD |
| 7352 | UCP3 | No | - | DisGeNET \| GeneCards |
| 7486 | WRN | No | - | DisGeNET \| GeneCards |
| 80212 | CCDC92 | No | - | DisGeNET \| CTD |
| 80339 | PNPLA3 | No | - | DisGeNET \| DISEASES |
| 80790 | CMIP | No | - | DisGeNET \| CTD |
| 83795 | KCNK16 | No | - | DisGeNET \| CTD |
| 841 | CASP8 | No | - | DisGeNET \| CTD |
| 84196 | USP48 | No | - | DisGeNET \| CTD |
| 8462 | KLF11 | No | - | DisGeNET \| GeneCards |
| 8527 | DGKD | No | - | DisGeNET \| CTD |
| 875 | CBS | No | - | DisGeNET \| CTD |
| 894 | CCND2 | No | - | DisGeNET \| CTD |
| 9365 | KL | No | - | DisGeNET \| CTD |
| 9559 | VPS26A | No | - | DisGeNET \| CTD |
| 9861 | PSMD6 | No | - | DisGeNET \| CTD |
| 9882 | TBC1D4 | No | - | DISEASES \| GeneCards |

**Table S7: Curated gene set associated with obesity.** Genes were compiled from multiple resources. The table includes genes that are either known targets of FDA-approved antiobesity drugs or supported by at least two of the four considered databases. For each gene, the Entrez gene ID, gene symbol, evidence source(s), and approved drug information (where applicable) are reported.

| Entrez Gene ID | Gene symbol | Approved drug target | Drug Names | Evidence source |
|---|---|---|---|---|
| 4160 | MC4R | Yes | Setmelanotide | DrugBank (TTD) \| DisGeNET \| DISEASES \| GeneCards \| CTD |
| 2194 | FASN | Yes | Orlistat | DrugBank (Santos et al., Drug Central) \| DisGeNET \| DISEASES \| CTD |
| 5406 | PNLIP | Yes | Orlistat | DrugBank (Santos et al., Drug Central \| TTD) \| DISEASES \| GeneCards |
| 6531 | SLC6A3 | Yes | Amphetamine \| Diethylpropion \| Metamfetamine \| Phendimetrazine | DrugBank (Drug Central \| Santos et al., Drug Central) \| DisGeNET |
| 134864 | TAAR1 | Yes | Amphetamine | DrugBank (Santos et al., Drug Central) |
| 147 | ADRA1B | Yes | Phendimetrazine | DrugBank (TTD) |
| 56144 | PCDHA4 | Yes | Orlistat | DrugBank (TTD) |
| 6530 | SLC6A2 | Yes | Amphetamine \| Benzphetamine \| Diethylpropion \| Phendimetrazine \| Phentermine | DrugBank (Santos et al., Drug Central \| TTD) |
| 6571 | SLC18A2 | Yes | Benzphetamine | DrugBank (Santos et al., Drug Central) |
| 8513 | LIPF | Yes | Orlistat | DrugBank (Santos et al., Drug Central) |
| 5468 | PPARG | No | - | DisGeNET \| DISEASES \| GeneCards \| CTD |
| 2740 | GLP1R | No | - | DisGeNET \| DISEASES \| GeneCards |
| 32 | ACACB | No | - | DisGeNET \| CTD |
| 3643 | INSR | No | - | DISEASES \| GeneCards |
| 10135 | NAMPT | No | - | DisGeNET \| DISEASES \| GeneCards \| CTD |
| 109 | ADCY3 | No | - | DisGeNET \| DISEASES \| GeneCards \| CTD |
| 1268 | CNR1 | No | - | DisGeNET \| DISEASES \| GeneCards \| CTD |
| 1363 | CPE | No | - | DisGeNET \| DISEASES \| GeneCards \| CTD |
| 1401 | CRP | No | - | DisGeNET \| DISEASES \| GeneCards \| CTD |
| 155 | ADRB3 | No | - | DisGeNET \| DISEASES \| GeneCards \| CTD |
| 181 | AGRP | No | - | DisGeNET \| DISEASES \| GeneCards \| CTD |
| 2641 | GCG | No | - | DisGeNET \| DISEASES \| GeneCards \| CTD |
| 27125 | AFF4 | No | - | DisGeNET \| DISEASES \| GeneCards \| CTD |
| 2778 | GNAS | No | - | DisGeNET \| DISEASES \| GeneCards \| CTD |
| 348 | APOE | No | - | DisGeNET \| DISEASES \| GeneCards \| CTD |
| 3569 | IL6 | No | - | DisGeNET \| DISEASES \| GeneCards \| CTD |
| 3630 | INS | No | - | DisGeNET \| DISEASES \| GeneCards \| CTD |
| 3667 | IRS1 | No | - | DisGeNET \| DISEASES \| GeneCards \| CTD |
| 3952 | LEP | No | - | DisGeNET \| DISEASES \| GeneCards \| CTD |
| 3953 | LEPR | No | - | DisGeNET \| DISEASES \| GeneCards \| CTD |
| 4023 | LPL | No | - | DisGeNET \| DISEASES \| GeneCards \| CTD |
| 4915 | NTRK2 | No | - | DisGeNET \| DISEASES \| GeneCards \| CTD |
| 5054 | SERPINE1 | No | - | DisGeNET \| DISEASES \| GeneCards \| CTD |
| 5122 | PCSK1 | No | - | DisGeNET \| DISEASES \| GeneCards \| CTD |
| 5167 | ENPP1 | No | - | DisGeNET \| DISEASES \| GeneCards \| CTD |
| 51738 | GHRL | No | - | DisGeNET \| DISEASES \| GeneCards \| CTD |
| 5346 | PLIN1 | No | - | DisGeNET \| DISEASES \| GeneCards \| CTD |
| 5443 | POMC | No | - | DisGeNET \| DISEASES \| GeneCards \| CTD |
| 56623 | INPP5E | No | - | DisGeNET \| DISEASES \| GeneCards \| CTD |
| 5697 | PYY | No | - | DisGeNET \| DISEASES \| GeneCards \| CTD |
| 5770 | PTPN1 | No | - | DisGeNET \| DISEASES \| GeneCards \| CTD |
| 6347 | CCL2 | No | - | DisGeNET \| DISEASES \| GeneCards \| CTD |
| 6720 | SREBF1 | No | - | DisGeNET \| DISEASES \| GeneCards \| CTD |
| 7124 | TNF | No | - | DisGeNET \| DISEASES \| GeneCards \| CTD |
| 7350 | UCP1 | No | - | DisGeNET \| DISEASES \| GeneCards \| CTD |
| 7351 | UCP2 | No | - | DisGeNET \| DISEASES \| GeneCards \| CTD |
| 7352 | UCP3 | No | - | DisGeNET \| DISEASES \| GeneCards \| CTD |
| 79068 | FTO | No | - | DisGeNET \| DISEASES \| GeneCards \| CTD |
| 8431 | NR0B2 | No | - | DisGeNET \| DISEASES \| GeneCards \| CTD |
| 9370 | ADIPOQ | No | - | DisGeNET \| DISEASES \| GeneCards \| CTD |
| 9607 | CARTPT | No | - | DisGeNET \| DISEASES \| GeneCards \| CTD |
| 1050 | CEBPA | No | - | DisGeNET \| DISEASES \| CTD |
| 11254 | SLC6A14 | No | - | DisGeNET \| DISEASES \| GeneCards |
| 1149 | CIDEA | No | - | DisGeNET \| DISEASES \| CTD |
| 133522 | PPARGC1B | No | - | DisGeNET \| GeneCards \| CTD |
| 154 | ADRB2 | No | - | DisGeNET \| GeneCards \| CTD |
| 207 | AKT1 | No | - | DisGeNET \| DISEASES \| CTD |
| 2099 | ESR1 | No | - | DisGeNET \| DISEASES \| CTD |
| 23411 | SIRT1 | No | - | DisGeNET \| DISEASES \| CTD |
| 25970 | SH2B1 | No | - | DisGeNET \| GeneCards \| CTD |
| 26291 | FGF21 | No | - | DisGeNET \| DISEASES \| CTD |
| 2688 | GH1 | No | - | DisGeNET \| GeneCards \| CTD |
| 3060 | HCRT | No | - | DisGeNET \| DISEASES \| CTD |
| 3290 | HSD11B1 | No | - | DisGeNET \| GeneCards \| CTD |
| 3383 | ICAM1 | No | - | DisGeNET \| DISEASES \| CTD |
| 364 | AQP7 | No | - | DisGeNET \| GeneCards \| CTD |
| 3949 | LDLR | No | - | DisGeNET \| DISEASES \| CTD |
| 5465 | PPARA | No | - | DisGeNET \| DISEASES \| CTD |
| 5467 | PPARD | No | - | DisGeNET \| DISEASES \| CTD |
| 5573 | PRKAR1A | No | - | DisGeNET \| GeneCards \| CTD |
| 56729 | RETN | No | - | DisGeNET \| DISEASES \| GeneCards |
| 583 | BBS2 | No | - | DisGeNET \| DISEASES \| GeneCards |
| 585 | BBS4 | No | - | DisGeNET \| DISEASES \| GeneCards |
| 627 | BDNF | No | - | DisGeNET \| DISEASES \| GeneCards |
| 6648 | SOD2\|SOD2-2 | No | - | DisGeNET \| DISEASES \| CTD |
| 6774 | STAT3 | No | - | DisGeNET \| DISEASES \| GeneCards |
| 8195 | MKKS | No | - | DisGeNET \| DISEASES \| GeneCards |
| 9672 | SDC3 | No | - | DisGeNET \| GeneCards \| CTD |
| 10059 | DNM1L | No | - | DisGeNET \| GeneCards |
| 10411 | RAPGEF3 | No | - | DisGeNET \| CTD |
| 1066 | CES1 | No | - | DisGeNET \| CTD |
| 10726 | NUDC | No | - | DisGeNET \| GeneCards |
| 10743 | RAI1 | No | - | DisGeNET \| CTD |
| 10806 | SDCCAG8 | No | - | DISEASES \| GeneCards |
| 10840 | ALDH1L1 | No | - | DisGeNET \| CTD |
| 10891 | PPARGC1A | No | - | DISEASES \| GeneCards |
| 112609 | MRAP2 | No | - | DISEASES \| GeneCards |
| 114294 | LACTB | No | - | DisGeNET \| CTD |
| 123016 | TTC8 | No | - | DISEASES \| GeneCards |
| 129787 | TMEM18 | No | - | DisGeNET \| CTD |
| 129880 | BBS5 | No | - | DISEASES \| GeneCards |
| 1312 | COMT | No | - | DisGeNET \| GeneCards |
| 132789 | GNPDA2 | No | - | DisGeNET \| CTD |
| 1350 | COX7C | No | - | DisGeNET \| CTD |
| 1361 | CPB2 | No | - | DisGeNET \| CTD |
| 142 | PARP1 | No | - | DisGeNET \| CTD |
| 1431 | CS | No | - | DisGeNET \| CTD |
| 1489 | CTF1 | No | - | DisGeNET \| CTD |
| 151742 | PPM1L | No | - | DisGeNET \| CTD |
| 1520 | CTSS | No | - | DisGeNET \| CTD |
| 1528 | CYB5A | No | - | DisGeNET \| CTD |
| 153 | ADRB1 | No | - | DisGeNET \| CTD |
| 1545 | CYP1B1 | No | - | DisGeNET \| CTD |
| 1571 | CYP2E1 | No | - | DisGeNET \| CTD |
| 157657 | C8orf37\|CFAP418 | No | - | DISEASES \| GeneCards |
| 157680 | VPS13B | No | - | DISEASES \| GeneCards |
| 1636 | ACE | No | - | DISEASES \| GeneCards |
| 1649 | DDIT3 | No | - | DisGeNET \| CTD |
| 166379 | BBS12 | No | - | DISEASES \| GeneCards |
| 169841 | ZNF169 | No | - | DisGeNET \| CTD |
| 1728 | NQO1 | No | - | DisGeNET \| CTD |
| 1788 | DNMT3A | No | - | DisGeNET \| GeneCards |
| 1806 | DPYD | No | - | DisGeNET \| CTD |
| 183 | AGT | No | - | DisGeNET \| DISEASES |
| 1892 | ECHS1 | No | - | DisGeNET \| CTD |
| 1947 | EFNB1 | No | - | DisGeNET \| CTD |
| 196 | AHR | No | - | DisGeNET \| CTD |
| 2033 | EP300 | No | - | DisGeNET \| CTD |
| 2110 | ETFDH | No | - | DisGeNET \| CTD |
| 2147 | F2 | No | - | DisGeNET \| CTD |
| 2166 | FAAH | No | - | DisGeNET \| CTD |
| 2180 | ACSL1 | No | - | DisGeNET \| CTD |
| 23011 | RAB21 | No | - | DisGeNET \| CTD |
| 2309 | FOXO3 | No | - | DisGeNET \| CTD |
| 23216 | TBC1D1 | No | - | DisGeNET \| CTD |
| 23217 | ZFR2 | No | - | DisGeNET \| CTD |
| 23410 | SIRT3 | No | - | DisGeNET \| CTD |
| 2353 | FOS | No | - | DisGeNET \| CTD |
| 23788 | MTCH2 | No | - | DisGeNET \| CTD |
| 2512 | FTL | No | - | DisGeNET \| CTD |
| 252995 | FNDC5 | No | - | DISEASES \| GeneCards |
| 257194 | NEGR1 | No | - | DisGeNET \| CTD |
| 2673 | GFPT1 | No | - | DisGeNET \| CTD |
| 2695 | GIP | No | - | DisGeNET \| DISEASES |
| 27241 | BBS9 | No | - | DISEASES \| GeneCards |
| 2731 | GLDC | No | - | DisGeNET \| GeneCards |
| 2752 | GLUL | No | - | DisGeNET \| CTD |
| 2784 | GNB3 | No | - | DisGeNET \| CTD |
| 283455 | KSR2 | No | - | DisGeNET \| GeneCards |
| 2840 | GPR17 | No | - | DisGeNET \| CTD |
| 2876 | GPX1 | No | - | DisGeNET \| CTD |
| 2878 | GPX3 | No | - | DisGeNET \| CTD |
| 2908 | NR3C1 | No | - | DisGeNET \| GeneCards |
| 3033 | HADH | No | - | DisGeNET \| CTD |
| 3099 | HK2 | No | - | DisGeNET \| CTD |
| 3148 | HMGB2 | No | - | DisGeNET \| CTD |
| 3162 | HMOX1 | No | - | DisGeNET \| CTD |
| 3215 | HOXB5 | No | - | DisGeNET \| CTD |
| 3291 | HSD11B2 | No | - | DisGeNET \| GeneCards |
| 3309 | HSPA5 | No | - | DisGeNET \| CTD |
| 3356 | HTR2A | No | - | DisGeNET \| CTD |
| 3358 | HTR2C | No | - | DisGeNET \| CTD |
| 338 | APOB | No | - | DISEASES \| GeneCards |
| 34 | ACADM | No | - | DisGeNET \| CTD |
| 3479 | IGF1 | No | - | DISEASES \| GeneCards |
| 3481 | IGF2 | No | - | DisGeNET \| CTD |
| 3485 | IGFBP2 | No | - | DisGeNET \| CTD |
| 3553 | IL1B | No | - | DISEASES \| GeneCards |
| 3574 | IL7 | No | - | DisGeNET \| CTD |
| 3620 | IDO1 | No | - | DisGeNET \| CTD |
| 3636 | INPPL1 | No | - | DisGeNET \| GeneCards |
| 3684 | ITGAM | No | - | DisGeNET \| CTD |
| 3778 | KCNMA1 | No | - | DisGeNET \| CTD |
| 3929 | LBP | No | - | DisGeNET \| CTD |
| 412 | STS | No | - | DisGeNET \| CTD |
| 4159 | MC3R | No | - | DisGeNET \| GeneCards |
| 4199 | ME1 | No | - | DisGeNET \| CTD |
| 43 | ACHE | No | - | DisGeNET \| CTD |
| 4318 | MMP9 | No | - | DisGeNET \| CTD |
| 4329 | ALDH6A1 | No | - | DisGeNET \| CTD |
| 434 | ASIP | No | - | DisGeNET \| DISEASES |
| 4360 | MRC1 | No | - | DisGeNET \| CTD |
| 4547 | MTTP | No | - | DisGeNET \| GeneCards |
| 4594 | MMUT | No | - | DisGeNET \| CTD |
| 463 | ZFHX3 | No | - | DisGeNET \| CTD |
| 47 | ACLY | No | - | DisGeNET \| CTD |
| 4808 | NHLH2 | No | - | DisGeNET \| CTD |
| 4852 | NPY | No | - | DISEASES \| GeneCards |
| 4864 | NPC1 | No | - | DisGeNET \| CTD |
| 4886 | NPY1R | No | - | DisGeNET \| CTD |
| 4889 | NPY5R | No | - | DisGeNET \| CTD |
| 4968 | OGG1 | No | - | DisGeNET \| CTD |
| 51043 | ZBTB7B | No | - | DisGeNET \| CTD |
| 5105 | PCK1 | No | - | DisGeNET \| CTD |
| 51094 | ADIPOR1 | No | - | DISEASES \| GeneCards |
| 51129 | ANGPTL4 | No | - | DisGeNET \| CTD |
| 51181 | DCXR | No | - | DisGeNET \| CTD |
| 5209 | PFKFB3 | No | - | DisGeNET \| CTD |
| 5290 | PIK3CA | No | - | DISEASES \| GeneCards |
| 5367 | PMCH | No | - | DisGeNET \| CTD |
| 54 | ACP5 | No | - | DisGeNET \| CTD |
| 54205 | CYCS | No | - | DisGeNET \| CTD |
| 54496 | PRMT7 | No | - | DisGeNET \| GeneCards |
| 54551 | MAGEL2 | No | - | DISEASES \| GeneCards |
| 54585 | LZTFL1 | No | - | DISEASES \| GeneCards |
| 54901 | CDKAL1 | No | - | DisGeNET \| GeneCards |
| 55212 | BBS7 | No | - | DISEASES \| GeneCards |
| 55277 | FGGY | No | - | DisGeNET \| CTD |
| 55690 | PACS1 | No | - | DisGeNET \| CTD |
| 5577 | PRKAR2B | No | - | DisGeNET \| CTD |
| 5583 | PRKCH | No | - | DisGeNET \| CTD |
| 56603 | CYP26B1 | No | - | DisGeNET \| CTD |
| 57104 | PNPLA2 | No | - | DISEASES \| GeneCards |
| 5743 | PTGS2 | No | - | DisGeNET \| CTD |
| 57498 | KIDINS220 | No | - | DISEASES \| GeneCards |
| 582 | BBS1 | No | - | DISEASES \| GeneCards |
| 590 | BCHE | No | - | DisGeNET \| CTD |
| 5919 | RARRES2 | No | - | DISEASES \| GeneCards |
| 6303 | SAT1 | No | - | DisGeNET \| CTD |
| 6319 | SCD | No | - | DisGeNET \| DISEASES |
| 6462 | SHBG | No | - | DISEASES \| GeneCards |
| 64756 | ATPAF1 | No | - | DisGeNET \| CTD |
| 6492 | SIM1 | No | - | DisGeNET \| GeneCards |
| 6517 | SLC2A4 | No | - | DISEASES \| GeneCards |
| 6580 | SLC22A1 | No | - | DisGeNET \| CTD |
| 6581 | SLC22A3 | No | - | DisGeNET \| CTD |
| 6582 | SLC22A2 | No | - | DisGeNET \| CTD |
| 6647 | SOD1 | No | - | DisGeNET \| CTD |
| 7018 | TF | No | - | DisGeNET \| CTD |
| 7037 | TFRC | No | - | DisGeNET \| CTD |
| 7133 | TNFRSF1B | No | - | DisGeNET \| CTD |
| 7275 | TUB | No | - | DISEASES \| GeneCards |
| 7385 | UQCRC2 | No | - | DisGeNET \| CTD |
| 7425 | VGF | No | - | DisGeNET \| CTD |
| 7436 | VLDLR | No | - | DisGeNET \| CTD |
| 7442 | TRPV1 | No | - | DisGeNET \| CTD |
| 761 | CA3 | No | - | DisGeNET \| CTD |
| 7840 | ALMS1 | No | - | DISEASES \| GeneCards |
| 79047 | KCTD15 | No | - | DisGeNET \| CTD |
| 79661 | NEIL1 | No | - | DisGeNET \| CTD |
| 79738 | BBS10 | No | - | DISEASES \| GeneCards |
| 80184 | CEP290 | No | - | DISEASES \| GeneCards |
| 8165 | AKAP1 | No | - | DisGeNET \| CTD |
| 834 | CASP1 | No | - | DisGeNET \| CTD |
| 84100 | ARL 6.00\|ARL6 | No | - | DISEASES \| GeneCards |
| 84872 | ZC3H10 | No | - | DisGeNET \| CTD |
| 84984 | CEP19 | No | - | DISEASES \| GeneCards |
| 8522 | GAS7 | No | - | DisGeNET \| CTD |
| 8694 | DGAT1 | No | - | DISEASES \| GeneCards |
| 8800 | PEX11A | No | - | DisGeNET \| CTD |
| 8801 | SUCLG2 | No | - | DisGeNET \| CTD |
| 8803 | SUCLA2 | No | - | DisGeNET \| CTD |
| 8856 | NR1I2 | No | - | DisGeNET \| CTD |
| 89780 | WNT3A | No | - | DisGeNET \| CTD |
| 91147 | TMEM67 | No | - | DISEASES \| GeneCards |
| 9149 | DYRK1B | No | - | DISEASES \| GeneCards |
| 9194 | SLC16A7 | No | - | DisGeNET \| CTD |
| 92482 | BBIP1 | No | - | DISEASES \| GeneCards |
| 9332 | CD163 | No | - | DisGeNET \| CTD |
| 955 | ENTPD6 | No | - | DisGeNET \| CTD |
| 958 | CD40 | No | - | DisGeNET \| CTD |
| 968 | CD68 | No | - | DisGeNET \| CTD |
| 9970 | NR1I3 | No | - | DisGeNET \| CTD |

**Table S8: Distribution of phytochemicals across the selected Single Herbal Drugs (SHDs).** This table summarizes the occurrence frequency of phytochemicals across the selected SHDs, highlighting overlap of phytochemicals across multiple SHDs. For each phytochemical, the number of SHDs in which it is present is reported, along with the names of the corresponding plants.

| **Chemical name** | **External chemical identifier** | **Number of SHDs containing the phytochemical** | **Plant name** |
|---|---|---|---|
| Beta-sitosterol | CID_222284 | 7 | *Aegle marmelos \| Butea monosperma \| Commiphora wightii \| Diospyros malabarica \| Tecomella undulata \| Tectona grandis \| Terminalia arjuna* |
| Lupeol | CID_259846 | 4 | *Aegle marmelos \| Diospyros malabarica \| Gymnema sylvestre \| Pterocarpus marsupium* |
| Myristic acid | CID_11005 | 3 | *Butea monosperma \| Enicostema axillare \| Gymnema sylvestre* |
| n-Heptacosane | CID_11636 | 3 | *Enicostema axillare \| Gymnema sylvestre \| Tecomella undulata* |
| Octadecanoic acid | CID_5281 | 3 | *Butea monosperma \| Enicostema axillare \| Gymnema sylvestre* |
| n-Nonacosane | CID_12409 | 3 | *Enicostema axillare \| Gymnema sylvestre \| Tecomella undulata* |
| Ferulate | CID_445858 | 3 | *Commiphora wightii \| Gymnema sylvestre \| Tecomella undulata* |
| Beta-sitosterol glucoside | CID_5742590 | 3 | *Butea monosperma \| Diospyros malabarica \| Tecomella undulata* |
| Beta-Sitosterol-beta-D-glucoside | CID_12309055 | 3 | *Butea monosperma \| Diospyros malabarica \| Tecomella undulata* |
| Lapachol | CID_3884 | 2 | *Tecomella undulata \| Tectona grandis* |
| Dehydrotectol | CID_3037329 | 2 | *Tecomella undulata \| Tectona grandis* |
| Betulin | CID_72326 | 2 | *Diospyros malabarica \| Enicostema axillare* |
| n-Triacontanol | CID_68972 | 2 | *Commiphora wightii \| Tecomella undulata* |
| n-Hexacosanol | CID_68171 | 2 | *Diospyros malabarica \| Enicostema axillare* |
| Palmitic acid | CID_985 | 2 | *Butea monosperma \| Gymnema sylvestre* |
| Hentriacontane | CID_12410 | 2 | *Gymnema sylvestre \| Terminalia arjuna* |
| n-Triacontane | CID_12535 | 2 | *Gymnema sylvestre \| Tecomella undulata* |
| Oleic acid | CID_445639 | 2 | *Butea monosperma \| Enicostema axillare* |
| Oleanolic acid | CID_10494 | 2 | *Pterocarpus marsupium \| Terminalia arjuna* |
| Betulinic acid | CID_64971 | 2 | *Diospyros malabarica \| Tectona grandis* |
| Beta amyrin | CID_73145 | 2 | *Diospyros malabarica \| Gymnema sylvestre* |
| Stigmasterol | CID_5280794 | 2 | *Commiphora wightii \| Gymnema sylvestre* |
| Lignoceric acid | CID_11197 | 1 | *Butea monosperma* |
| Oxyberberine | CID_11066 | 1 | *Berberis aristata* |
| Gentiocrucine | CAS_22108-77-6 | 1 | *Enicostema axillare* |
| Aleurilic acid | CID_10790 | 1 | *Butea monosperma* |
| Garbanzol | CID_442410 | 1 | *Pterocarpus marsupium* |
| Hexahydrofarnesyl acetone | CID_10408 | 1 | *Gymnema sylvestre* |
| Conduritol A | CID_10290861 | 1 | *Gymnema sylvestre* |
| Arjunic acid | CID_15385516 | 1 | *Terminalia arjuna* |
| Aurapten | CID_1550607 | 1 | *Aegle marmelos* |
| Gymnestrogenin | CID_15560302 | 1 | *Gymnema sylvestre* |
| Kalashine | CID_156697 | 1 | *Berberis aristata* |
| Ketones | CID_114522 | 1 | *Gymnema sylvestre* |
| Liquiritigenin | CID_114829 | 1 | *Pterocarpus marsupium* |
| Coreopsin | CID_12303942 | 1 | *Butea monosperma* |
| Alpha-camphorene | CID_101750 | 1 | *Commiphora wightii* |
| Chitraline | CAS_77754-91-7 | 1 | *Berberis aristata* |
| Arjunolone | CAS_82178-34-5 | 1 | *Terminalia arjuna* |
| Cerasidin | CID_14034812 | 1 | *Terminalia arjuna* |
| Pongaflavone | CID_14033983 | 1 | *Diospyros malabarica* |
| Pentadecanoic acid | CID_13849 | 1 | *Gymnema sylvestre* |
| 1-Hexadecanol | CID_2682 | 1 | *Gymnema sylvestre* |
| Isomonospermoside | CID_42607822 | 1 | *Butea monosperma* |
| Lauric acid | CID_3893 | 1 | *Butea monosperma* |
| Monospermoside | CID_42607524 | 1 | *Butea monosperma* |
| Capric acid | CID_2969 | 1 | *Butea monosperma* |
| Anthraquinone-2-carboxaldehyde | CID_344310 | 1 | *Tectona grandis* |
| Gentianine | CID_354616 | 1 | *Enicostema axillare* |
| Tecomaquinone I | CID_3574508 | 1 | *Tectona grandis* |
| Caprylic acid | CID_379 | 1 | *Butea monosperma* |
| 1 - O - methyl pakistanine | CID_181478 | 1 | *Berberis aristata* |
| Myrcene | CID_31253 | 1 | *Commiphora wightii* |
| Coumarins | CID_323 | 1 | *Aegle marmelos* |
| Eugenol | CID_3314 | 1 | *Gymnema sylvestre* |
| Oxyacanthine | CID_442333 | 1 | *Berberis aristata* |
| E-guggulsterone | CID_6439929 | 1 | *Commiphora wightii* |
| Germacrene A | CID_9548706 | 1 | *Gymnema sylvestre* |
| Mukulol | CID_90472510 | 1 | *Commiphora wightii* |
| Allylcembrol | CID_5368823 | 1 | *Commiphora wightii* |
| Betaine | CID_247 | 1 | *Gymnema sylvestre* |
| 20 alpha hydroxy-4-pregnen-3-one | CID_161109 | 1 | *Commiphora wightii* |
| Tartaric acid | CID_439655 | 1 | *Gymnema sylvestre* |
| Z-guggulsterone | CID_122173119 | 1 | *Commiphora wightii* |
| 20 beta hydroxy -4-pregnen-3-one | CID_92747 | 1 | *Commiphora wightii* |
| Choline | CID_305 | 1 | *Gymnema sylvestre* |
| Catechol | CID_289 | 1 | *Terminalia arjuna* |
| Glucuronic acid | CID_94715 | 1 | *Gymnema sylvestre* |
| Undulatosides A | CID_5321494 | 1 | *Tecomella undulata* |
| Pterostilbene | CID_5281727 | 1 | *Pterocarpus marsupium* |
| Pseudobaptigenin | CID_5281805 | 1 | *Pterocarpus marsupium* |
| Apigenin | CID_5280443 | 1 | *Enicostema axillare* |
| 5-hydroxyalpachol | CID_5318245 | 1 | *Tectona grandis* |
| n-Heneicosanoic acid | CID_16898 | 1 | *Butea monosperma* |
| Erythrocentaurin | CID_191120 | 1 | *Enicostema axillare* |
| Propterol | CID_185124 | 1 | *Pterocarpus marsupium* |
| Palmatine | CID_19009 | 1 | *Berberis aristata* |
| 5-deoxykaempferol | CID_5281611 | 1 | *Pterocarpus marsupium* |
| Genkwanin | CID_5281617 | 1 | *Enicostema axillare* |
| Ellagic acid | CID_5281855 | 1 | *Terminalia arjuna* |
| Umbelliferone | CID_5281426 | 1 | *Aegle marmelos* |
| Cembrene A | CID_5281384 | 1 | *Commiphora wightii* |
| Baicalein | CID_5281605 | 1 | *Terminalia arjuna* |
| Enicoflavin | CID_5281564 | 1 | *Enicostema axillare* |
| Berberine | CID_2353 | 1 | *Berberis aristata* |
| Pakistanine | CID_193239 | 1 | *Berberis aristata* |
| Myristyl oleate | CID_5365034 | 1 | *Terminalia arjuna* |
| Dehydro-alpha-lapachone | CID_72734 | 1 | *Tectona grandis* |
| Angelic acid | CID_643915 | 1 | *Gymnema sylvestre* |
| Marmin | CID_6450230 | 1 | *Aegle marmelos* |
| Cholesterol | CID_5997 | 1 | *Commiphora wightii* |
| 2-Pentadecanone | CID_61303 | 1 | *Gymnema sylvestre* |
| Stilbenes | CID_638088 | 1 | *Pterocarpus marsupium* |
| Isoliquiritigenin | CID_638278 | 1 | *Pterocarpus marsupium* |
| Marsupol | CID_6437266 | 1 | *Pterocarpus marsupium* |
| Pluviatilol | CID_70695727 | 1 | *Commiphora wightii* |
| Veratric acid | CID_7121 | 1 | *Tecomella undulata* |
| Methyl eugenol | CID_7127 | 1 | *Gymnema sylvestre* |
| Peregrinol | CID_7092583 | 1 | *Diospyros malabarica* |
| Palasonin | CID_198727 | 1 | *Butea monosperma* |
| Octadecanol | CID_8221 | 1 | *Gymnema sylvestre* |
| Hydroquinone | CID_785 | 1 | *Gymnema sylvestre* |
| Methyl palmitate | CID_8181 | 1 | *Gymnema sylvestre* |
| n-Octacosanol | CID_68406 | 1 | *Tecomella undulata* |
| Indole | CID_798 | 1 | *Gymnema sylvestre* |
| Acetophenone | CID_7410 | 1 | *Gymnema sylvestre* |
| Dodecanol | CID_8193 | 1 | *Gymnema sylvestre* |
| Tetradecanol | CID_8209 | 1 | *Gymnema sylvestre* |
| Jatrorrhizine | CID_72323 | 1 | *Berberis aristata* |
| Tectoquinone | CID_6773 | 1 | *Tectona grandis* |
| Anthraquinones | CID_6780 | 1 | *Tectona grandis* |
| Behenic acid | CID_8215 | 1 | *Butea monosperma* |
| Anthraquinone-2-carboxylic acid | CID_67030 | 1 | *Tectona grandis* |
| Skimmianine | CID_6760 | 1 | *Aegle marmelos* |
| Taraxerone | CID_92785 | 1 | *Diospyros malabarica* |
| Oxalic acid | CID_971 | 1 | *Terminalia arjuna* |
| Octanol | CID_957 | 1 | *Gymnema sylvestre* |
| Karachine | CID_630739 | 1 | *Berberis aristata* |
| Inositol | CID_892 | 1 | *Gymnema sylvestre* |
| Marsformosanone | CID_22296838 | 1 | *Diospyros malabarica* |
| Lysophosphatidylcholine | CID_5311264 | 1 | *Butea monosperma* |
| Terminic acid | CID_132568257 | 1 | *Terminalia arjuna* |
| Isovitexin | CID_162350 | 1 | *Enicostema axillare* |
| Guggulsterol III | CID_101297675 | 1 | *Commiphora wightii* |
| Guggulsterol I | CID_101297673 | 1 | *Commiphora wightii* |
| Guggulsterol II | CID_101297674 | 1 | *Commiphora wightii* |
| Imperatorin | CID_10212 | 1 | *Aegle marmelos* |
| Tetracosane | CID_12592 | 1 | *Gymnema sylvestre* |
| P-Hydroxybenzaldehyde | CID_126 | 1 | *Pterocarpus marsupium* |
| Hexacosane | CID_12407 | 1 | *Diospyros malabarica* |
| Pentadecanol | CID_12397 | 1 | *Gymnema sylvestre* |
| Swertisin | CID_124034 | 1 | *Enicostema axillare* |
| Pentacosane | CID_12406 | 1 | *Gymnema sylvestre* |
| Isocoreopsin | CID_12309899 | 1 | *Butea monosperma* |
| Tritriacontane | CID_12411 | 1 | *Gymnema sylvestre* |
| Pentatriacontane | CID_12413 | 1 | *Gymnema sylvestre* |
| Marsupsin | CID_134369 | 1 | *Pterocarpus marsupium* |
| Ethyl palmitate | CID_12366 | 1 | *Gymnema sylvestre* |
| Pterosupin | CID_133775 | 1 | *Pterocarpus marsupium* |
| Tetradecadiene | CID_6365430 | 1 | *Gymnema sylvestre* |
| Phosphatidylethanolamine | CID_102515444 | 1 | *Butea monosperma* |
| Naphthalene | CID_931 | 1 | *Tectona grandis* |
| p-Guaiacol | CID_9015 | 1 | *Gymnema sylvestre* |
| 1,8-Cineole | CID_2758 | 1 | *Gymnema sylvestre* |
| b-Elemene | CID_6918391 | 1 | *Gymnema sylvestre* |
| Butyric acid | CID_264 | 1 | *Gymnema sylvestre* |
| Cembrene | CID_11747713 | 1 | *Commiphora wightii* |
| Sulfurein | CID_10071442 | 1 | *Butea monosperma* |
| Naringenin | CID_439246 | 1 | *Pterocarpus marsupium* |
| 7,4'-dihydroxyflavone | CID_5282073 | 1 | *Pterocarpus marsupium* |
| 2-Dodecenol | CID_5352845 | 1 | *Gymnema sylvestre* |
| Arachidic acid | CID_10467 | 1 | *Butea monosperma* |
| Propterol B | CHEMSPIDER_10306372 | 1 | *Pterocarpus marsupium* |
| 7-hydroxyflavanone | CID_688857 | 1 | *Pterocarpus marsupium* |
| Beta eudesmol | CID_91457 | 1 | *Pterocarpus marsupium* |
| Fagarine | CID_107936 | 1 | *Aegle marmelos* |
| Alpha-amyrin | CID_73170 | 1 | *Butea monosperma* |
| Arjunolic acid | CID_73641 | 1 | *Terminalia arjuna* |
| Marmesin | CID_334704 | 1 | *Aegle marmelos* |
| Friedelin | CID_91472 | 1 | *Terminalia arjuna* |
| Mannitol | CID_6251 | 1 | *Terminalia arjuna* |
| (+)-Gallocatechol | CID_65084 | 1 | *Terminalia arjuna* |
| Epigallocatechol | CID_72277 | 1 | *Terminalia arjuna* |
| D-quercitol | CID_441437 | 1 | *Gymnema sylvestre* |
| Leucodelphinidin | CID_440835 | 1 | *Terminalia arjuna* |
| m-Ethyl phenol | CID_12101 | 1 | *Gymnema sylvestre* |
| Gamma-butyric acid | CID_119 | 1 | *Gymnema sylvestre* |
| Gallic acid | CID_370 | 1 | *Diospyros malabarica* |
| Aegelinol | CID_600671 | 1 | *Aegle marmelos* |
| Tomentosic acid | CID_622032 | 1 | *Terminalia arjuna* |
| Arjungenin | CID_12444386 | 1 | *Terminalia arjuna* |
| Phytol | CID_5280435 | 1 | *Gymnema sylvestre* |
| Linolenic acid | CID_5280934 | 1 | *Butea monosperma* |
| 9,12,15-Octadecatrienal | CID_5283384 | 1 | *Gymnema sylvestre* |
| 7-hydroxy-5,4'-dimethoxy-8-methylisoflavone- | CID_44257328 | 1 | *Pterocarpus marsupium* |
| Delta-lactone | CID_49831545 | 1 | *Butea monosperma* |
| Triterpenoids | CID_71597391 | 1 | *Diospyros malabarica* |
| 15-hydroxypentacosanoic acid | CID_5312784 | 1 | *Butea monosperma* |
| 1-hydroxy-5-methoxy-2-methyl-9,10-anthraqu | CID_13970503 | 1 | *Tectona grandis* |
| 9,10-dimethoxy-2-methyl-1,4-anthraquinone | CID_14283285 | 1 | *Tectona grandis* |
| Caryophyllene | CID_5281515 | 1 | *Commiphora wightii* |
| Sesamin | CID_5204 | 1 | *Commiphora wightii* |
| (-)-Epicatechin | CID_72276 | 1 | *Pterocarpus marsupium* |
| Ethyl octadec-9-enoate | CID_8123 | 1 | *Gymnema sylvestre* |
| Galacturonic acid | CID_439215 | 1 | *Gymnema sylvestre* |
| Linoleic acid | CID_5280450 | 1 | *Butea monosperma* |
| Formic Acid | CID_284 | 1 | *Gymnema sylvestre* |
| Tecomin | | 1 | *Tecomella undulata* |

**Table S9: Statistical overlap between SHD-associated target sets and disease-associated gene sets.** For each SHD and disease condition, the table reports the size of the SHD target set, the size of the disease-associated gene set, the number of overlapping genes, and the corresponding statistical significance. *p*-values were calculated using a right-tailed hypergeometric test to assess whether observed overlaps were greater than expected by chance.

| **Plant name** | **Disease** | **SHD target set size** | **Disease gene set size** | **Overlap size** | ***p*-value** |
|---|---|---|---|---|---|
| *Aegle marmelos* | Diabetes | 268 | 279 | 65 | 4.91E-62 |
| *Berberis aristata* | Diabetes | 55 | 279 | 7 | 1.22E-05 |
| *Butea monosperma* | Diabetes | 226 | 279 | 16 | 1.58E-07 |
| *Commiphora wightii* | Diabetes | 195 | 279 | 15 | 1.32E-07 |
| *Diospyros malabarica* | Diabetes | 111 | 279 | 9 | 2.94E-05 |
| *Enicostema axillare* | Diabetes | 360 | 279 | 19 | 1.04E-06 |
| *Gymnema sylvestre* | Diabetes | 292 | 279 | 17 | 1.03E-06 |
| *Pterocarpus marsupium* | Diabetes | 237 | 279 | 22 | 3.64E-12 |
| *Tecomella undulata* | Diabetes | 161 | 279 | 9 | 0.0005020986237 |
| *Tectona grandis* | Diabetes | 83 | 279 | 7 | 0.0001782975477 |
| *Terminalia arjuna* | Diabetes | 388 | 279 | 26 | 6.30E-11 |
| *Aegle marmelos* | Obesity | 268 | 253 | 28 | 1.29E-17 |
| *Berberis aristata* | Obesity | 55 | 253 | 7 | 6.46E-06 |
| *Butea monosperma* | Obesity | 226 | 253 | 25 | 1.98E-16 |
| *Commiphora wightii* | Obesity | 195 | 253 | 26 | 4.31E-19 |
| *Diospyros malabarica* | Obesity | 111 | 253 | 11 | 1.97E-07 |
| *Enicostema axillare* | Obesity | 360 | 253 | 28 | 2.61E-14 |
| *Gymnema sylvestre* | Obesity | 292 | 253 | 34 | 9.77E-23 |
| *Pterocarpus marsupium* | Obesity | 237 | 253 | 23 | 5.70E-14 |
| *Tecomella undulata* | Obesity | 161 | 253 | 21 | 1.95E-15 |
| *Tectona grandis* | Obesity | 83 | 253 | 6 | 0.0007076191123 |
| *Terminalia arjuna* | Obesity | 388 | 253 | 37 | 1.24E-21 |

**Table S10. SHD – phytochemical – target – FDA-approved drug – disease associations linking single herbal drugs to therapeutically relevant drugs.** This table presents single herbal drugs (SHDs), their constituent phytochemicals, associated molecular targets, and corresponding FDA-approved antidiabetic drugs, highlighting target-level overlaps between SHDs and clinically approved therapeutics. Note that for metformin, multiple genes associated with mitochondrial complex I (e.g., NDUF and MT-ND family genes) are grouped and represented as MT-I in the Sankey plot in the main text. Genes associated with mitochondrial complex IV (e.g., COX family genes) are represented as MT-IV.

| SHD | Phytochemical | Gene Symbol | Mimicked Drug | Disease |
|---|---|---|---|---|
| *Butea monosperma* | Beta-sitosterol | DPP4 | Sitagliptin | Type 2 Diabetes |
| *Tecomella undulata* | Beta-sitosterol | DPP4 | Sitagliptin | Type 2 Diabetes |
| *Commiphora wightii* | Beta-sitosterol | DPP4 | Sitagliptin | Type 2 Diabetes |
| *Tectona grandis* | Beta-sitosterol | DPP4 | Sitagliptin | Type 2 Diabetes |
| *Aegle marmelos* | Beta-sitosterol | DPP4 | Sitagliptin | Type 2 Diabetes |
| *Terminalia arjuna* | Beta-sitosterol | DPP4 | Sitagliptin | Type 2 Diabetes |
| *Diospyros malabarica* | Beta-sitosterol | DPP4 | Sitagliptin | Type 2 Diabetes |
| *Enicostema axillare* | Apigenin | DPP4 | Sitagliptin | Type 2 Diabetes |
| *Diospyros malabarica* | Gallic acid | DPP4 | Sitagliptin | Type 2 Diabetes |
| *Tectona grandis* | Betulinic acid | DPP4 | Sitagliptin | Type 2 Diabetes |
| *Diospyros malabarica* | Betulinic acid | DPP4 | Sitagliptin | Type 2 Diabetes |
| *Terminalia arjuna* | Baicalein | DPP4 | Sitagliptin | Type 2 Diabetes |
| *Pterocarpus marsupium* | Naringenin | DPP4 | Sitagliptin | Type 2 Diabetes |
| *Butea monosperma* | Beta-sitosterol | DPP4 | Vildagliptin | Type 2 Diabetes |
| *Tecomella undulata* | Beta-sitosterol | DPP4 | Vildagliptin | Type 2 Diabetes |
| *Commiphora wightii* | Beta-sitosterol | DPP4 | Vildagliptin | Type 2 Diabetes |
| *Tectona grandis* | Beta-sitosterol | DPP4 | Vildagliptin | Type 2 Diabetes |
| *Aegle marmelos* | Beta-sitosterol | DPP4 | Vildagliptin | Type 2 Diabetes |
| *Terminalia arjuna* | Beta-sitosterol | DPP4 | Vildagliptin | Type 2 Diabetes |
| *Diospyros malabarica* | Beta-sitosterol | DPP4 | Vildagliptin | Type 2 Diabetes |
| *Enicostema axillare* | Apigenin | DPP4 | Vildagliptin | Type 2 Diabetes |
| *Diospyros malabarica* | Gallic acid | DPP4 | Vildagliptin | Type 2 Diabetes |
| *Tectona grandis* | Betulinic acid | DPP4 | Vildagliptin | Type 2 Diabetes |
| *Diospyros malabarica* | Betulinic acid | DPP4 | Vildagliptin | Type 2 Diabetes |
| *Terminalia arjuna* | Baicalein | DPP4 | Vildagliptin | Type 2 Diabetes |
| *Pterocarpus marsupium* | Naringenin | DPP4 | Vildagliptin | Type 2 Diabetes |
| *Terminalia arjuna* | Baicalein | MGAM | Miglitol | Type 2 Diabetes |
| *Enicostema axillare* | Isovitexin | GAA | Miglitol | Type 2 Diabetes |
| *Terminalia arjuna* | Baicalein | GAA | Miglitol | Type 2 Diabetes |
| *Pterocarpus marsupium* | (-)-Epicatechin | GAA | Miglitol | Type 2 Diabetes |
| *Berberis aristata* | Jatrorrhizine | GAA | Miglitol | Type 2 Diabetes |
| *Terminalia arjuna* | Ellagic acid | GAA | Miglitol | Type 2 Diabetes |
| *Aegle marmelos* | Umbelliferone | GAA | Miglitol | Type 2 Diabetes |
| *Aegle marmelos* | Imperatorin | GAA | Miglitol | Type 2 Diabetes |
| *Pterocarpus marsupium* | Pterostilbene | GAA | Miglitol | Type 2 Diabetes |
| *Aegle marmelos* | Imperatorin | GLP1R | Semaglutide | Type 2 Diabetes |
| *Butea monosperma* | Lauric acid | GLP1R | Semaglutide | Type 2 Diabetes |
| *Pterocarpus marsupium* | Pterostilbene | GLP1R | Semaglutide | Type 2 Diabetes |
| *Commiphora wightii* | Sesamin | GLP1R | Semaglutide | Type 2 Diabetes |
| *Aegle marmelos* | Aurapten | GLP1R | Semaglutide | Type 2 Diabetes |
| *Gymnema sylvestre* | Myristic acid | GLP1R | Semaglutide | Type 2 Diabetes |
| *Butea monosperma* | Myristic acid | GLP1R | Semaglutide | Type 2 Diabetes |
| *Enicostema axillare* | Myristic acid | GLP1R | Semaglutide | Type 2 Diabetes |
| *Tectona grandis* | Betulinic acid | GLP1R | Semaglutide | Type 2 Diabetes |
| *Diospyros malabarica* | Betulinic acid | GLP1R | Semaglutide | Type 2 Diabetes |
| *Pterocarpus marsupium* | Isoliquiritigenin | GLP1R | Semaglutide | Type 2 Diabetes |
| *Aegle marmelos* | Imperatorin | GLP1R | Tirzepatide | Type 2 Diabetes |
| *Butea monosperma* | Lauric acid | GLP1R | Tirzepatide | Type 2 Diabetes |
| *Pterocarpus marsupium* | Pterostilbene | GLP1R | Tirzepatide | Type 2 Diabetes |
| *Commiphora wightii* | Sesamin | GLP1R | Tirzepatide | Type 2 Diabetes |
| *Aegle marmelos* | Aurapten | GLP1R | Tirzepatide | Type 2 Diabetes |
| *Gymnema sylvestre* | Myristic acid | GLP1R | Tirzepatide | Type 2 Diabetes |
| *Butea monosperma* | Myristic acid | GLP1R | Tirzepatide | Type 2 Diabetes |
| *Enicostema axillare* | Myristic acid | GLP1R | Tirzepatide | Type 2 Diabetes |
| *Tectona grandis* | Betulinic acid | GLP1R | Tirzepatide | Type 2 Diabetes |
| *Diospyros malabarica* | Betulinic acid | GLP1R | Tirzepatide | Type 2 Diabetes |
| *Pterocarpus marsupium* | Isoliquiritigenin | GLP1R | Tirzepatide | Type 2 Diabetes |
| *Butea monosperma* | Beta-sitosterol | DPP4 | Saxagliptin | Type 2 Diabetes |
| *Tecomella undulata* | Beta-sitosterol | DPP4 | Saxagliptin | Type 2 Diabetes |
| *Commiphora wightii* | Beta-sitosterol | DPP4 | Saxagliptin | Type 2 Diabetes |
| *Tectona grandis* | Beta-sitosterol | DPP4 | Saxagliptin | Type 2 Diabetes |
| *Aegle marmelos* | Beta-sitosterol | DPP4 | Saxagliptin | Type 2 Diabetes |
| *Terminalia arjuna* | Beta-sitosterol | DPP4 | Saxagliptin | Type 2 Diabetes |
| *Diospyros malabarica* | Beta-sitosterol | DPP4 | Saxagliptin | Type 2 Diabetes |
| *Enicostema axillare* | Apigenin | DPP4 | Saxagliptin | Type 2 Diabetes |
| *Diospyros malabarica* | Gallic acid | DPP4 | Saxagliptin | Type 2 Diabetes |
| *Tectona grandis* | Betulinic acid | DPP4 | Saxagliptin | Type 2 Diabetes |
| *Diospyros malabarica* | Betulinic acid | DPP4 | Saxagliptin | Type 2 Diabetes |
| *Terminalia arjuna* | Baicalein | DPP4 | Saxagliptin | Type 2 Diabetes |
| *Pterocarpus marsupium* | Naringenin | DPP4 | Saxagliptin | Type 2 Diabetes |
| *Butea monosperma* | Beta-sitosterol | DPP4 | Alogliptin | Type 2 Diabetes |
| *Tecomella undulata* | Beta-sitosterol | DPP4 | Alogliptin | Type 2 Diabetes |
| *Commiphora wightii* | Beta-sitosterol | DPP4 | Alogliptin | Type 2 Diabetes |
| *Tectona grandis* | Beta-sitosterol | DPP4 | Alogliptin | Type 2 Diabetes |
| *Aegle marmelos* | Beta-sitosterol | DPP4 | Alogliptin | Type 2 Diabetes |
| *Terminalia arjuna* | Beta-sitosterol | DPP4 | Alogliptin | Type 2 Diabetes |
| *Diospyros malabarica* | Beta-sitosterol | DPP4 | Alogliptin | Type 2 Diabetes |
| *Enicostema axillare* | Apigenin | DPP4 | Alogliptin | Type 2 Diabetes |
| *Diospyros malabarica* | Gallic acid | DPP4 | Alogliptin | Type 2 Diabetes |
| *Tectona grandis* | Betulinic acid | DPP4 | Alogliptin | Type 2 Diabetes |
| *Diospyros malabarica* | Betulinic acid | DPP4 | Alogliptin | Type 2 Diabetes |
| *Terminalia arjuna* | Baicalein | DPP4 | Alogliptin | Type 2 Diabetes |
| *Pterocarpus marsupium* | Naringenin | DPP4 | Alogliptin | Type 2 Diabetes |
| *Aegle marmelos* | Marmin | NDUFA12 | Metformin | Type 2 Diabetes |
| *Aegle marmelos* | Aurapten | NDUFA12 | Metformin | Type 2 Diabetes |
| *Aegle marmelos* | Marmin | NDUFAF3 | Metformin | Type 2 Diabetes |
| *Aegle marmelos* | Aurapten | NDUFAF3 | Metformin | Type 2 Diabetes |
| *Aegle marmelos* | Marmin | NDUFAF1 | Metformin | Type 2 Diabetes |
| *Aegle marmelos* | Aurapten | NDUFAF1 | Metformin | Type 2 Diabetes |
| *Aegle marmelos* | Aurapten | NDUFS8 | Metformin | Type 2 Diabetes |
| *Aegle marmelos* | Marmin | NDUFS8 | Metformin | Type 2 Diabetes |
| *Aegle marmelos* | Aurapten | COXFA4 | Metformin | Type 2 Diabetes |
| *Aegle marmelos* | Marmin | COXFA4 | Metformin | Type 2 Diabetes |
| *Aegle marmelos* | Aurapten | NDUFAB1 | Metformin | Type 2 Diabetes |
| *Aegle marmelos* | Marmin | NDUFAB1 | Metformin | Type 2 Diabetes |
| *Aegle marmelos* | Aurapten | NDUFA1 | Metformin | Type 2 Diabetes |
| *Aegle marmelos* | Marmin | NDUFA1 | Metformin | Type 2 Diabetes |
| *Aegle marmelos* | Aurapten | NDUFS4 | Metformin | Type 2 Diabetes |
| *Aegle marmelos* | Marmin | NDUFS4 | Metformin | Type 2 Diabetes |
| *Aegle marmelos* | Marmin | NDUFB5 | Metformin | Type 2 Diabetes |
| *Aegle marmelos* | Aurapten | NDUFB5 | Metformin | Type 2 Diabetes |
| *Aegle marmelos* | Marmin | NDUFB3 | Metformin | Type 2 Diabetes |
| *Aegle marmelos* | Aurapten | NDUFB3 | Metformin | Type 2 Diabetes |
| *Aegle marmelos* | Marmin | NDUFC1 | Metformin | Type 2 Diabetes |
| *Aegle marmelos* | Aurapten | NDUFC1 | Metformin | Type 2 Diabetes |
| *Aegle marmelos* | Marmin | NDUFA2 | Metformin | Type 2 Diabetes |
| *Aegle marmelos* | Aurapten | NDUFA2 | Metformin | Type 2 Diabetes |
| *Aegle marmelos* | Aurapten | NDUFS5 | Metformin | Type 2 Diabetes |
| *Aegle marmelos* | Marmin | NDUFS5 | Metformin | Type 2 Diabetes |
| *Aegle marmelos* | Marmin | NDUFS7 | Metformin | Type 2 Diabetes |
| *Aegle marmelos* | Aurapten | NDUFS7 | Metformin | Type 2 Diabetes |
| *Aegle marmelos* | Aurapten | NDUFS2 | Metformin | Type 2 Diabetes |
| *Aegle marmelos* | Marmin | NDUFS2 | Metformin | Type 2 Diabetes |
| *Aegle marmelos* | Aurapten | NDUFS6 | Metformin | Type 2 Diabetes |
| *Aegle marmelos* | Marmin | NDUFS6 | Metformin | Type 2 Diabetes |
| *Aegle marmelos* | Marmin | NDUFB1 | Metformin | Type 2 Diabetes |
| *Aegle marmelos* | Aurapten | NDUFB1 | Metformin | Type 2 Diabetes |
| *Aegle marmelos* | Aurapten | NDUFS3 | Metformin | Type 2 Diabetes |
| *Aegle marmelos* | Marmin | NDUFS3 | Metformin | Type 2 Diabetes |
| *Aegle marmelos* | Marmin | NDUFB6 | Metformin | Type 2 Diabetes |
| *Aegle marmelos* | Aurapten | NDUFB6 | Metformin | Type 2 Diabetes |
| *Aegle marmelos* | Aurapten | NDUFA3 | Metformin | Type 2 Diabetes |
| *Aegle marmelos* | Marmin | NDUFA3 | Metformin | Type 2 Diabetes |
| *Aegle marmelos* | Aurapten | NDUFB4 | Metformin | Type 2 Diabetes |
| *Aegle marmelos* | Marmin | NDUFB4 | Metformin | Type 2 Diabetes |
| *Aegle marmelos* | Marmin | NDUFB8 | Metformin | Type 2 Diabetes |
| *Aegle marmelos* | Aurapten | NDUFB8 | Metformin | Type 2 Diabetes |
| *Aegle marmelos* | Aurapten | NDUFB2 | Metformin | Type 2 Diabetes |
| *Aegle marmelos* | Marmin | NDUFB2 | Metformin | Type 2 Diabetes |
| *Aegle marmelos* | Marmin | NDUFA7 | Metformin | Type 2 Diabetes |
| *Aegle marmelos* | Aurapten | NDUFA7 | Metformin | Type 2 Diabetes |
| *Aegle marmelos* | Marmin | NDUFC2 | Metformin | Type 2 Diabetes |
| *Aegle marmelos* | Aurapten | NDUFC2 | Metformin | Type 2 Diabetes |
| *Aegle marmelos* | Aurapten | NDUFA10 | Metformin | Type 2 Diabetes |
| *Aegle marmelos* | Marmin | NDUFA10 | Metformin | Type 2 Diabetes |
| *Aegle marmelos* | Aurapten | NDUFB10 | Metformin | Type 2 Diabetes |
| *Aegle marmelos* | Marmin | NDUFB10 | Metformin | Type 2 Diabetes |
| *Aegle marmelos* | Aurapten | MT-ND1 | Metformin | Type 2 Diabetes |
| *Aegle marmelos* | Marmin | MT-ND1 | Metformin | Type 2 Diabetes |
| *Aegle marmelos* | Aurapten | MT-ND2 | Metformin | Type 2 Diabetes |
| *Aegle marmelos* | Marmin | MT-ND2 | Metformin | Type 2 Diabetes |
| *Aegle marmelos* | Aurapten | MT-ND3 | Metformin | Type 2 Diabetes |
| *Aegle marmelos* | Marmin | MT-ND3 | Metformin | Type 2 Diabetes |
| *Aegle marmelos* | Aurapten | MT-ND4L | Metformin | Type 2 Diabetes |
| *Aegle marmelos* | Marmin | MT-ND4L | Metformin | Type 2 Diabetes |
| *Aegle marmelos* | Marmin | MT-ND4 | Metformin | Type 2 Diabetes |
| *Aegle marmelos* | Aurapten | MT-ND4 | Metformin | Type 2 Diabetes |
| *Aegle marmelos* | Aurapten | NDUFA11 | Metformin | Type 2 Diabetes |
| *Aegle marmelos* | Marmin | NDUFA11 | Metformin | Type 2 Diabetes |
| *Terminalia arjuna* | Ellagic acid | INSR | Metformin | Type 2 Diabetes |
| *Enicostema axillare* | Apigenin | INSR | Metformin | Type 2 Diabetes |
| *Aegle marmelos* | Marmin | NDUFA6 | Metformin | Type 2 Diabetes |
| *Aegle marmelos* | Aurapten | NDUFA6 | Metformin | Type 2 Diabetes |
| *Aegle marmelos* | Marmin | NDUFA8 | Metformin | Type 2 Diabetes |
| *Aegle marmelos* | Aurapten | NDUFA8 | Metformin | Type 2 Diabetes |
| *Aegle marmelos* | Marmin | COXFA4L2 | Metformin | Type 2 Diabetes |
| *Aegle marmelos* | Aurapten | COXFA4L2 | Metformin | Type 2 Diabetes |
| *Aegle marmelos* | Aurapten | NDUFAF4 | Metformin | Type 2 Diabetes |
| *Aegle marmelos* | Marmin | NDUFAF4 | Metformin | Type 2 Diabetes |
| *Aegle marmelos* | Aurapten | NDUFA5 | Metformin | Type 2 Diabetes |
| *Aegle marmelos* | Marmin | NDUFA5 | Metformin | Type 2 Diabetes |
| *Aegle marmelos* | Marmin | NDUFA13 | Metformin | Type 2 Diabetes |
| *Aegle marmelos* | Aurapten | NDUFA13 | Metformin | Type 2 Diabetes |
| *Aegle marmelos* | Aurapten | NDUFA9 | Metformin | Type 2 Diabetes |
| *Aegle marmelos* | Marmin | NDUFA9 | Metformin | Type 2 Diabetes |
| *Aegle marmelos* | Marmin | NDUFV3 | Metformin | Type 2 Diabetes |
| *Aegle marmelos* | Aurapten | NDUFV3 | Metformin | Type 2 Diabetes |
| *Aegle marmelos* | Marmin | NDUFB11 | Metformin | Type 2 Diabetes |
| *Aegle marmelos* | Aurapten | NDUFB11 | Metformin | Type 2 Diabetes |
| *Aegle marmelos* | Aurapten | NDUFAF2 | Metformin | Type 2 Diabetes |
| *Aegle marmelos* | Marmin | NDUFAF2 | Metformin | Type 2 Diabetes |
| *Aegle marmelos* | Aurapten | NDUFS1 | Metformin | Type 2 Diabetes |
| *Aegle marmelos* | Marmin | NDUFS1 | Metformin | Type 2 Diabetes |
| *Aegle marmelos* | Aurapten | MT-ND5 | Metformin | Type 2 Diabetes |
| *Aegle marmelos* | Marmin | MT-ND5 | Metformin | Type 2 Diabetes |
| *Aegle marmelos* | Aurapten | MT-ND6 | Metformin | Type 2 Diabetes |
| *Aegle marmelos* | Marmin | MT-ND6 | Metformin | Type 2 Diabetes |
| *Aegle marmelos* | Aurapten | NDUFV2 | Metformin | Type 2 Diabetes |
| *Aegle marmelos* | Marmin | NDUFV2 | Metformin | Type 2 Diabetes |
| *Aegle marmelos* | Marmin | NDUFV1 | Metformin | Type 2 Diabetes |
| *Aegle marmelos* | Aurapten | NDUFV1 | Metformin | Type 2 Diabetes |
| *Aegle marmelos* | Marmin | NDUFB7 | Metformin | Type 2 Diabetes |
| *Aegle marmelos* | Aurapten | NDUFB7 | Metformin | Type 2 Diabetes |
| *Aegle marmelos* | Aurapten | NDUFB9 | Metformin | Type 2 Diabetes |
| *Aegle marmelos* | Marmin | NDUFB9 | Metformin | Type 2 Diabetes |
| *Aegle marmelos* | Imperatorin | GLP1R | Pramlintide | Type 2 Diabetes |
| *Butea monosperma* | Lauric acid | GLP1R | Pramlintide | Type 2 Diabetes |
| *Pterocarpus marsupium* | Pterostilbene | GLP1R | Pramlintide | Type 2 Diabetes |
| *Commiphora wightii* | Sesamin | GLP1R | Pramlintide | Type 2 Diabetes |
| *Aegle marmelos* | Aurapten | GLP1R | Pramlintide | Type 2 Diabetes |
| *Gymnema sylvestre* | Myristic acid | GLP1R | Pramlintide | Type 2 Diabetes |
| *Butea monosperma* | Myristic acid | GLP1R | Pramlintide | Type 2 Diabetes |
| *Enicostema axillare* | Myristic acid | GLP1R | Pramlintide | Type 2 Diabetes |
| *Tectona grandis* | Betulinic acid | GLP1R | Pramlintide | Type 2 Diabetes |
| *Diospyros malabarica* | Betulinic acid | GLP1R | Pramlintide | Type 2 Diabetes |
| *Pterocarpus marsupium* | Isoliquiritigenin | GLP1R | Pramlintide | Type 2 Diabetes |
| *Terminalia arjuna* | Catechol | PPARG | Pioglitazone | Type 2 Diabetes |
| *Gymnema sylvestre* | Methyl eugenol | PPARG | Pioglitazone | Type 2 Diabetes |
| *Commiphora wightii* | E-guggulsterone | PPARG | Pioglitazone | Type 2 Diabetes |
| *Enicostema axillare* | Octadecanoic acid | PPARG | Pioglitazone | Type 2 Diabetes |
| *Butea monosperma* | Octadecanoic acid | PPARG | Pioglitazone | Type 2 Diabetes |
| *Gymnema sylvestre* | Octadecanoic acid | PPARG | Pioglitazone | Type 2 Diabetes |
| *Pterocarpus marsupium* | Pseudobaptigenin | PPARG | Pioglitazone | Type 2 Diabetes |
| *Butea monosperma* | Capric acid | PPARG | Pioglitazone | Type 2 Diabetes |
| *Butea monosperma* | Behenic acid | PPARG | Pioglitazone | Type 2 Diabetes |
| *Butea monosperma* | Palmitic acid | PPARG | Pioglitazone | Type 2 Diabetes |
| *Gymnema sylvestre* | Palmitic acid | PPARG | Pioglitazone | Type 2 Diabetes |
| *Butea monosperma* | Linoleic acid | PPARG | Pioglitazone | Type 2 Diabetes |
| *Enicostema axillare* | Apigenin | PPARG | Pioglitazone | Type 2 Diabetes |
| *Pterocarpus marsupium* | 7,4'-dihydroxyflavone | PPARG | Pioglitazone | Type 2 Diabetes |
| *Aegle marmelos* | Aurapten | PPARG | Pioglitazone | Type 2 Diabetes |
| *Butea monosperma* | Lauric acid | PPARG | Pioglitazone | Type 2 Diabetes |
| *Gymnema sylvestre* | Myristic acid | PPARG | Pioglitazone | Type 2 Diabetes |
| *Butea monosperma* | Myristic acid | PPARG | Pioglitazone | Type 2 Diabetes |
| *Enicostema axillare* | Myristic acid | PPARG | Pioglitazone | Type 2 Diabetes |
| *Butea monosperma* | Arachidic acid | PPARG | Pioglitazone | Type 2 Diabetes |
| *Commiphora wightii* | Pluviatilol | PPARG | Pioglitazone | Type 2 Diabetes |
| *Pterocarpus marsupium* | Naringenin | PPARG | Pioglitazone | Type 2 Diabetes |
| *Butea monosperma* | Linolenic acid | PPARG | Pioglitazone | Type 2 Diabetes |
| *Pterocarpus marsupium* | Isoliquiritigenin | PPARG | Pioglitazone | Type 2 Diabetes |
| *Enicostema axillare* | Oleic acid | PPARG | Pioglitazone | Type 2 Diabetes |
| *Butea monosperma* | Oleic acid | PPARG | Pioglitazone | Type 2 Diabetes |
| *Butea monosperma* | Beta-sitosterol | DPP4 | Linagliptin | Type 2 Diabetes |
| *Tecomella undulata* | Beta-sitosterol | DPP4 | Linagliptin | Type 2 Diabetes |
| *Commiphora wightii* | Beta-sitosterol | DPP4 | Linagliptin | Type 2 Diabetes |
| *Tectona grandis* | Beta-sitosterol | DPP4 | Linagliptin | Type 2 Diabetes |
| *Aegle marmelos* | Beta-sitosterol | DPP4 | Linagliptin | Type 2 Diabetes |
| *Terminalia arjuna* | Beta-sitosterol | DPP4 | Linagliptin | Type 2 Diabetes |
| *Diospyros malabarica* | Beta-sitosterol | DPP4 | Linagliptin | Type 2 Diabetes |
| *Enicostema axillare* | Apigenin | DPP4 | Linagliptin | Type 2 Diabetes |
| *Diospyros malabarica* | Gallic acid | DPP4 | Linagliptin | Type 2 Diabetes |
| *Tectona grandis* | Betulinic acid | DPP4 | Linagliptin | Type 2 Diabetes |
| *Diospyros malabarica* | Betulinic acid | DPP4 | Linagliptin | Type 2 Diabetes |
| *Terminalia arjuna* | Baicalein | DPP4 | Linagliptin | Type 2 Diabetes |
| *Pterocarpus marsupium* | Naringenin | DPP4 | Linagliptin | Type 2 Diabetes |
| *Terminalia arjuna* | Catechol | SLC6A3 | Diethylpropion | Obesity |
| *Gymnema sylvestre* | Choline | SLC6A3 | Diethylpropion | Obesity |
| *Butea monosperma* | Linolenic acid | SLC6A3 | Diethylpropion | Obesity |
| *Aegle marmelos* | Coumarins | SLC6A3 | Diethylpropion | Obesity |
| *Gymnema sylvestre* | Inositol | SLC6A3 | Diethylpropion | Obesity |
| *Gymnema sylvestre* | Eugenol | SLC6A3 | Diethylpropion | Obesity |
| *Enicostema axillare* | Oleic acid | SLC6A3 | Diethylpropion | Obesity |
| *Butea monosperma* | Oleic acid | SLC6A3 | Diethylpropion | Obesity |
| *Terminalia arjuna* | Mannitol | SLC6A3 | Diethylpropion | Obesity |
| *Commiphora wightii* | Ferulate | SLC6A3 | Diethylpropion | Obesity |
| *Gymnema sylvestre* | Ferulate | SLC6A3 | Diethylpropion | Obesity |
| *Tecomella undulata* | Ferulate | SLC6A3 | Diethylpropion | Obesity |
| *Gymnema sylvestre* | 1,8-Cineole | SLC6A3 | Diethylpropion | Obesity |
| *Enicostema axillare* | Octadecanoic acid | SLC6A3 | Diethylpropion | Obesity |
| *Butea monosperma* | Octadecanoic acid | SLC6A3 | Diethylpropion | Obesity |
| *Gymnema sylvestre* | Octadecanoic acid | SLC6A3 | Diethylpropion | Obesity |
| *Gymnema sylvestre* | Eugenol | SLC6A2 | Diethylpropion | Obesity |
| *Commiphora wightii* | Ferulate | SLC6A2 | Diethylpropion | Obesity |
| *Gymnema sylvestre* | Ferulate | SLC6A2 | Diethylpropion | Obesity |
| *Tecomella undulata* | Ferulate | SLC6A2 | Diethylpropion | Obesity |
| *Gymnema sylvestre* | 1,8-Cineole | SLC6A2 | Diethylpropion | Obesity |
| *Aegle marmelos* | Coumarins | SLC6A2 | Diethylpropion | Obesity |
| *Enicostema axillare* | Octadecanoic acid | SLC6A2 | Diethylpropion | Obesity |
| *Butea monosperma* | Octadecanoic acid | SLC6A2 | Diethylpropion | Obesity |
| *Gymnema sylvestre* | Octadecanoic acid | SLC6A2 | Diethylpropion | Obesity |
| *Terminalia arjuna* | Catechol | SLC6A2 | Diethylpropion | Obesity |
| *Gymnema sylvestre* | Choline | SLC6A2 | Diethylpropion | Obesity |
| *Terminalia arjuna* | Mannitol | SLC6A2 | Diethylpropion | Obesity |
| *Butea monosperma* | Linolenic acid | SLC6A2 | Diethylpropion | Obesity |
| *Enicostema axillare* | Oleic acid | SLC6A2 | Diethylpropion | Obesity |
| *Butea monosperma* | Oleic acid | SLC6A2 | Diethylpropion | Obesity |
| *Gymnema sylvestre* | Inositol | SLC6A2 | Diethylpropion | Obesity |
| *Terminalia arjuna* | Catechol | SLC6A3 | Phendimetrazine | Obesity |
| *Gymnema sylvestre* | Choline | SLC6A3 | Phendimetrazine | Obesity |
| *Butea monosperma* | Linolenic acid | SLC6A3 | Phendimetrazine | Obesity |
| *Aegle marmelos* | Coumarins | SLC6A3 | Phendimetrazine | Obesity |
| *Gymnema sylvestre* | Inositol | SLC6A3 | Phendimetrazine | Obesity |
| *Gymnema sylvestre* | Eugenol | SLC6A3 | Phendimetrazine | Obesity |
| *Enicostema axillare* | Oleic acid | SLC6A3 | Phendimetrazine | Obesity |
| *Butea monosperma* | Oleic acid | SLC6A3 | Phendimetrazine | Obesity |
| *Terminalia arjuna* | Mannitol | SLC6A3 | Phendimetrazine | Obesity |
| *Commiphora wightii* | Ferulate | SLC6A3 | Phendimetrazine | Obesity |
| *Gymnema sylvestre* | Ferulate | SLC6A3 | Phendimetrazine | Obesity |
| *Tecomella undulata* | Ferulate | SLC6A3 | Phendimetrazine | Obesity |
| *Gymnema sylvestre* | 1,8-Cineole | SLC6A3 | Phendimetrazine | Obesity |
| *Enicostema axillare* | Octadecanoic acid | SLC6A3 | Phendimetrazine | Obesity |
| *Butea monosperma* | Octadecanoic acid | SLC6A3 | Phendimetrazine | Obesity |
| *Gymnema sylvestre* | Octadecanoic acid | SLC6A3 | Phendimetrazine | Obesity |
| *Gymnema sylvestre* | Eugenol | SLC6A2 | Phendimetrazine | Obesity |
| *Commiphora wightii* | Ferulate | SLC6A2 | Phendimetrazine | Obesity |
| *Gymnema sylvestre* | Ferulate | SLC6A2 | Phendimetrazine | Obesity |
| *Tecomella undulata* | Ferulate | SLC6A2 | Phendimetrazine | Obesity |
| *Gymnema sylvestre* | 1,8-Cineole | SLC6A2 | Phendimetrazine | Obesity |
| *Aegle marmelos* | Coumarins | SLC6A2 | Phendimetrazine | Obesity |
| *Enicostema axillare* | Octadecanoic acid | SLC6A2 | Phendimetrazine | Obesity |
| *Butea monosperma* | Octadecanoic acid | SLC6A2 | Phendimetrazine | Obesity |
| *Gymnema sylvestre* | Octadecanoic acid | SLC6A2 | Phendimetrazine | Obesity |
| *Terminalia arjuna* | Catechol | SLC6A2 | Phendimetrazine | Obesity |
| *Gymnema sylvestre* | Choline | SLC6A2 | Phendimetrazine | Obesity |
| *Terminalia arjuna* | Mannitol | SLC6A2 | Phendimetrazine | Obesity |
| *Butea monosperma* | Linolenic acid | SLC6A2 | Phendimetrazine | Obesity |
| *Enicostema axillare* | Oleic acid | SLC6A2 | Phendimetrazine | Obesity |
| *Butea monosperma* | Oleic acid | SLC6A2 | Phendimetrazine | Obesity |
| *Gymnema sylvestre* | Inositol | SLC6A2 | Phendimetrazine | Obesity |
| *Terminalia arjuna* | Catechol | SLC6A3 | Metamfetamine | Obesity |
| *Gymnema sylvestre* | Choline | SLC6A3 | Metamfetamine | Obesity |
| *Butea monosperma* | Linolenic acid | SLC6A3 | Metamfetamine | Obesity |
| *Aegle marmelos* | Coumarins | SLC6A3 | Metamfetamine | Obesity |
| *Gymnema sylvestre* | Inositol | SLC6A3 | Metamfetamine | Obesity |
| *Gymnema sylvestre* | Eugenol | SLC6A3 | Metamfetamine | Obesity |
| *Enicostema axillare* | Oleic acid | SLC6A3 | Metamfetamine | Obesity |
| *Butea monosperma* | Oleic acid | SLC6A3 | Metamfetamine | Obesity |
| *Terminalia arjuna* | Mannitol | SLC6A3 | Metamfetamine | Obesity |
| *Commiphora wightii* | Ferulate | SLC6A3 | Metamfetamine | Obesity |
| *Gymnema sylvestre* | Ferulate | SLC6A3 | Metamfetamine | Obesity |
| *Tecomella undulata* | Ferulate | SLC6A3 | Metamfetamine | Obesity |
| *Gymnema sylvestre* | 1,8-Cineole | SLC6A3 | Metamfetamine | Obesity |
| *Enicostema axillare* | Octadecanoic acid | SLC6A3 | Metamfetamine | Obesity |
| *Butea monosperma* | Octadecanoic acid | SLC6A3 | Metamfetamine | Obesity |
| *Gymnema sylvestre* | Octadecanoic acid | SLC6A3 | Metamfetamine | Obesity |
| *Gymnema sylvestre* | Eugenol | SLC6A2 | Amphetamine | Obesity |
| *Commiphora wightii* | Ferulate | SLC6A2 | Amphetamine | Obesity |
| *Gymnema sylvestre* | Ferulate | SLC6A2 | Amphetamine | Obesity |
| *Tecomella undulata* | Ferulate | SLC6A2 | Amphetamine | Obesity |
| *Gymnema sylvestre* | 1,8-Cineole | SLC6A2 | Amphetamine | Obesity |
| *Aegle marmelos* | Coumarins | SLC6A2 | Amphetamine | Obesity |
| *Enicostema axillare* | Octadecanoic acid | SLC6A2 | Amphetamine | Obesity |
| *Butea monosperma* | Octadecanoic acid | SLC6A2 | Amphetamine | Obesity |
| *Gymnema sylvestre* | Octadecanoic acid | SLC6A2 | Amphetamine | Obesity |
| *Terminalia arjuna* | Catechol | SLC6A2 | Amphetamine | Obesity |
| *Gymnema sylvestre* | Choline | SLC6A2 | Amphetamine | Obesity |
| *Terminalia arjuna* | Mannitol | SLC6A2 | Amphetamine | Obesity |
| *Butea monosperma* | Linolenic acid | SLC6A2 | Amphetamine | Obesity |
| *Enicostema axillare* | Oleic acid | SLC6A2 | Amphetamine | Obesity |
| *Butea monosperma* | Oleic acid | SLC6A2 | Amphetamine | Obesity |
| *Gymnema sylvestre* | Inositol | SLC6A2 | Amphetamine | Obesity |
| *Terminalia arjuna* | Catechol | SLC6A3 | Amphetamine | Obesity |
| *Gymnema sylvestre* | Choline | SLC6A3 | Amphetamine | Obesity |
| *Butea monosperma* | Linolenic acid | SLC6A3 | Amphetamine | Obesity |
| *Aegle marmelos* | Coumarins | SLC6A3 | Amphetamine | Obesity |
| *Gymnema sylvestre* | Inositol | SLC6A3 | Amphetamine | Obesity |
| *Gymnema sylvestre* | Eugenol | SLC6A3 | Amphetamine | Obesity |
| *Enicostema axillare* | Oleic acid | SLC6A3 | Amphetamine | Obesity |
| *Butea monosperma* | Oleic acid | SLC6A3 | Amphetamine | Obesity |
| *Terminalia arjuna* | Mannitol | SLC6A3 | Amphetamine | Obesity |
| *Commiphora wightii* | Ferulate | SLC6A3 | Amphetamine | Obesity |
| *Gymnema sylvestre* | Ferulate | SLC6A3 | Amphetamine | Obesity |
| *Tecomella undulata* | Ferulate | SLC6A3 | Amphetamine | Obesity |
| *Gymnema sylvestre* | 1,8-Cineole | SLC6A3 | Amphetamine | Obesity |
| *Enicostema axillare* | Octadecanoic acid | SLC6A3 | Amphetamine | Obesity |
| *Butea monosperma* | Octadecanoic acid | SLC6A3 | Amphetamine | Obesity |
| *Gymnema sylvestre* | Octadecanoic acid | SLC6A3 | Amphetamine | Obesity |
| *Gymnema sylvestre* | Eugenol | MC4R | Setmelanotide | Obesity |
| *Aegle marmelos* | Coumarins | MC4R | Setmelanotide | Obesity |
| *Terminalia arjuna* | Catechol | MC4R | Setmelanotide | Obesity |
| *Terminalia arjuna* | Mannitol | MC4R | Setmelanotide | Obesity |
| *Commiphora wightii* | Ferulate | MC4R | Setmelanotide | Obesity |
| *Gymnema sylvestre* | Ferulate | MC4R | Setmelanotide | Obesity |
| *Tecomella undulata* | Ferulate | MC4R | Setmelanotide | Obesity |
| *Gymnema sylvestre* | 1,8-Cineole | MC4R | Setmelanotide | Obesity |
| *Pterocarpus marsupium* | (-)-Epicatechin | FASN | Orlistat | Obesity |
| *Enicostema axillare* | Apigenin | PNLIP | Orlistat | Obesity |
| *Gymnema sylvestre* | Eugenol | SLC6A2 | Benzphetamine | Obesity |
| *Commiphora wightii* | Ferulate | SLC6A2 | Benzphetamine | Obesity |
| *Gymnema sylvestre* | Ferulate | SLC6A2 | Benzphetamine | Obesity |
| *Tecomella undulata* | Ferulate | SLC6A2 | Benzphetamine | Obesity |
| *Gymnema sylvestre* | 1,8-Cineole | SLC6A2 | Benzphetamine | Obesity |
| *Aegle marmelos* | Coumarins | SLC6A2 | Benzphetamine | Obesity |
| *Enicostema axillare* | Octadecanoic acid | SLC6A2 | Benzphetamine | Obesity |
| *Butea monosperma* | Octadecanoic acid | SLC6A2 | Benzphetamine | Obesity |
| *Gymnema sylvestre* | Octadecanoic acid | SLC6A2 | Benzphetamine | Obesity |
| *Terminalia arjuna* | Catechol | SLC6A2 | Benzphetamine | Obesity |
| *Gymnema sylvestre* | Choline | SLC6A2 | Benzphetamine | Obesity |
| *Terminalia arjuna* | Mannitol | SLC6A2 | Benzphetamine | Obesity |
| *Butea monosperma* | Linolenic acid | SLC6A2 | Benzphetamine | Obesity |
| *Enicostema axillare* | Oleic acid | SLC6A2 | Benzphetamine | Obesity |
| *Butea monosperma* | Oleic acid | SLC6A2 | Benzphetamine | Obesity |
| *Gymnema sylvestre* | Inositol | SLC6A2 | Benzphetamine | Obesity |
| *Gymnema sylvestre* | Eugenol | SLC6A2 | Phentermine | Obesity |
| *Commiphora wightii* | Ferulate | SLC6A2 | Phentermine | Obesity |
| *Gymnema sylvestre* | Ferulate | SLC6A2 | Phentermine | Obesity |
| *Tecomella undulata* | Ferulate | SLC6A2 | Phentermine | Obesity |
| *Gymnema sylvestre* | 1,8-Cineole | SLC6A2 | Phentermine | Obesity |
| *Aegle marmelos* | Coumarins | SLC6A2 | Phentermine | Obesity |
| *Enicostema axillare* | Octadecanoic acid | SLC6A2 | Phentermine | Obesity |
| *Butea monosperma* | Octadecanoic acid | SLC6A2 | Phentermine | Obesity |
| *Gymnema sylvestre* | Octadecanoic acid | SLC6A2 | Phentermine | Obesity |
| *Terminalia arjuna* | Catechol | SLC6A2 | Phentermine | Obesity |
| *Gymnema sylvestre* | Choline | SLC6A2 | Phentermine | Obesity |
| *Terminalia arjuna* | Mannitol | SLC6A2 | Phentermine | Obesity |
| *Butea monosperma* | Linolenic acid | SLC6A2 | Phentermine | Obesity |
| *Enicostema axillare* | Oleic acid | SLC6A2 | Phentermine | Obesity |
| *Butea monosperma* | Oleic acid | SLC6A2 | Phentermine | Obesity |
| *Gymnema sylvestre* | Inositol | SLC6A2 | Phentermine | Obesity |

**Table S11: Identification of consensus hub proteins across SHD–disease combinations using network pharmacology analysis.** Each sub-table corresponds to a specific SHD–disease pair and reports the top ten proteins identified by five different network centrality measures. Consensus hub proteins denote proteins that are ranked among the top ten by all five centrality measures.

| *Aegle marmelos* - Obesity | | | | | | | | | | | |
|---|---|---|---|---|---|---|---|---|---|---|---|
| **MCC** | | **MNC** | | **Degree** | | **Closeness centrality** | | **Betweenness centrality** | | **Consensus hub proteins** | |
| **Gene symbol** | **Entrez Gene ID** | **Gene symbol** | **Entrez Gene ID** | **Gene symbol** | **Entrez Gene ID** | **Gene symbol** | **Entrez Gene ID** | **Gene symbol** | **Entrez Gene ID** | **Gene symbol** | **Entrez Gene ID** |
| ADRB1 | 153 | ADRB1 | 153 | ADRB1 | 153 | ADRB1 | 153 | ADRB1 | 153 | ADRB1 | 153 |
| ADRB3 | 155 | ADRB3 | 155 | PTGS2 | 5743 | PTGS2 | 5743 | PTGS2 | 5743 | PTGS2 | 5743 |
| PTGS2 | 5743 | PTGS2 | 5743 | PPARG | 5468 | PPARG | 5468 | PPARG | 5468 | PPARG | 5468 |
| PPARG | 5468 | PPARG | 5468 | CNR1 | 1268 | CNR1 | 1268 | CNR1 | 1268 | GNAS | 2778 |
| MMP9 | 4318 | CNR1 | 1268 | ESR1 | 2099 | ESR1 | 2099 | CYP2E1 | 1571 | ADRB2 | 154 |
| GNAS | 2778 | GNAS | 2778 | GNAS | 2778 | GNAS | 2778 | ESR1 | 2099 | GLP1R | 2740 |
| NR3C1 | 2908 | MC4R | 4160 | NR3C1 | 2908 | NR3C1 | 2908 | GNAS | 2778 | | |
| HTR2C | 3358 | HTR2C | 3358 | HTR2C | 3358 | HTR2C | 3358 | NR3C1 | 2908 | | |
| ADRB2 | 154 | ADRB2 | 154 | ADRB2 | 154 | ADRB2 | 154 | ADRB2 | 154 | | |
| GLP1R | 2740 | GLP1R | 2740 | GLP1R | 2740 | GLP1R | 2740 | GLP1R | 2740 | | |

| *Butea monosperma* - Obesity | | | | | | | | | | | |
|---|---|---|---|---|---|---|---|---|---|---|---|
| **MCC** | | **MNC** | | **Degree** | | **Closeness centrality** | | **Betweenness centrality** | | **Consensus hub proteins** | |
| **Gene symbol** | **Entrez Gene ID** | **Gene symbol** | **Entrez Gene ID** | **Gene symbol** | **Entrez Gene ID** | **Gene symbol** | **Entrez Gene ID** | **Gene symbol** | **Entrez Gene ID** | **Gene symbol** | **Entrez Gene ID** |
| NR3C1 | 2908 | CNR1 | 1268 | NR3C1 | 2908 | NR3C1 | 2908 | NR3C1 | 2908 | CNR1 | 1268 |
| CNR1 | 1268 | HTR2C | 3358 | CNR1 | 1268 | CNR1 | 1268 | CNR1 | 1268 | ADRB2 | 154 |
| ADRB2 | 154 | ADRB2 | 154 | ADRB2 | 154 | ADRB2 | 154 | ADRB2 | 154 | ESR1 | 2099 |
| FAAH | 2166 | ADRB1 | 153 | ADRB1 | 153 | ADRB1 | 153 | ESR1 | 2099 | PTGS2 | 5743 |
| ESR1 | 2099 | ESR1 | 2099 | ESR1 | 2099 | ESR1 | 2099 | ADRB1 | 153 | PPARG | 5468 |
| GNAS | 2778 | GNAS | 2778 | GNAS | 2778 | GNAS | 2778 | PTGS2 | 5743 | PPARA | 5465 |
| PTGS2 | 5743 | PTGS2 | 5743 | PTGS2 | 5743 | PTGS2 | 5743 | PPARG | 5468 | GLP1R | 2740 |
| PPARG | 5468 | PPARG | 5468 | PPARG | 5468 | PPARG | 5468 | PPARA | 5465 | | |
| PPARA | 5465 | PPARA | 5465 | PPARA | 5465 | PPARA | 5465 | SLC6A3 | 6531 | | |
| GLP1R | 2740 | GLP1R | 2740 | GLP1R | 2740 | GLP1R | 2740 | GLP1R | 2740 | | |

| *Commiphora wightii* - Obesity | | | | | | | | | | | |
|---|---|---|---|---|---|---|---|---|---|---|---|
| **MCC** | | **MNC** | | **Degree** | | **Closeness centrality** | | **Betweenness centrality** | | **Consensus hub proteins** | |
| **Gene symbol** | **Entrez Gene ID** | **Gene symbol** | **Entrez Gene ID** | **Gene symbol** | **Entrez Gene ID** | **Gene symbol** | **Entrez Gene ID** | **Gene symbol** | **Entrez Gene ID** | **Gene symbol** | **Entrez Gene ID** |
| PPARD | 5467 | MMP9 | 4318 | MMP9 | 4318 | MMP9 | 4318 | SLC6A3 | 6531 | GLP1R | 2740 |
| MMP9 | 4318 | PPARD | 5467 | SLC6A3 | 6531 | SLC6A3 | 6531 | NR3C1 | 2908 | ADRB2 | 154 |
| NR3C1 | 2908 | GLP1R | 2740 | NR3C1 | 2908 | NR3C1 | 2908 | GLP1R | 2740 | PPARG | 5468 |
| GLP1R | 2740 | HTR2C | 3358 | GLP1R | 2740 | GLP1R | 2740 | HTR2C | 3358 | PPARA | 5465 |
| ADRB2 | 154 | ADRB3 | 155 | ADRB3 | 155 | ADRB3 | 155 | ADRB2 | 154 | PTGS2 | 5743 |
| CNR1 | 1268 | ADRB2 | 154 | ADRB2 | 154 | ADRB2 | 154 | CNR1 | 1268 | | |
| ESR1 | 2099 | ADRB1 | 153 | CNR1 | 1268 | CNR1 | 1268 | PPARG | 5468 | | |
| PPARG | 5468 | PPARG | 5468 | PPARG | 5468 | PPARG | 5468 | PPARA | 5465 | | |
| PPARA | 5465 | PTGS2 | 5743 | PPARA | 5465 | PPARA | 5465 | PTGS2 | 5743 | | |
| PTGS2 | 5743 | PPARA | 5465 | PTGS2 | 5743 | PTGS2 | 5743 | BCHE | 590 | | |

| *Enicostema axillare* - Obesity | | | | | | | | | | | |
|---|---|---|---|---|---|---|---|---|---|---|---|
| **MCC** | | **MNC** | | **Degree** | | **Closeness centrality** | | **Betweenness centrality** | | **Consensus hub proteins** | |
| **Gene symbol** | **Entrez Gene ID** | **Gene symbol** | **Entrez Gene ID** | **Gene symbol** | **Entrez Gene ID** | **Gene symbol** | **Entrez Gene ID** | **Gene symbol** | **Entrez Gene ID** | **Gene symbol** | **Entrez Gene ID** |
| PTGS2 | 5743 | PTGS2 | 5743 | PTGS2 | 5743 | PTGS2 | 5743 | PTGS2 | 5743 | PTGS2 | 5743 |
| ESR1 | 2099 | ESR1 | 2099 | ESR1 | 2099 | ESR1 | 2099 | ESR1 | 2099 | ESR1 | 2099 |
| AKT1 | 207 | AKT1 | 207 | AKT1 | 207 | AKT1 | 207 | AKT1 | 207 | AKT1 | 207 |
| PPARG | 5468 | PPARG | 5468 | PPARG | 5468 | PPARG | 5468 | SLC6A3 | 6531 | PPARG | 5468 |
| PARP1 | 142 | PARP1 | 142 | PARP1 | 142 | PARP1 | 142 | SLC6A2 | 6530 | PARP1 | 142 |
| INSR | 3643 | INSR | 3643 | INSR | 3643 | INSR | 3643 | PPARG | 5468 | PPARA | 5465 |
| HSPA5 | 3309 | HSPA5 | 3309 | HSPA5 | 3309 | HSPA5 | 3309 | PARP1 | 142 | | |
| PPARA | 5465 | GLP1R | 2740 | GLP1R | 2740 | PPARA | 5465 | BCHE | 590 | | |
| PTPN1 | 5770 | PPARA | 5465 | PPARA | 5465 | NR3C1 | 2908 | PPARA | 5465 | | |
| NR3C1 | 2908 | NR3C1 | 2908 | NR3C1 | 2908 | CYP2E1 | 1571 | CYP2E1 | 1571 | | |

| *Gymnema sylvestre* - Obesity | | | | | | | | | | | |
|---|---|---|---|---|---|---|---|---|---|---|---|
| **MCC** | | **MNC** | | **Degree** | | **Closeness centrality** | | **Betweenness centrality** | | **Consensus hub proteins** | |
| **Gene symbol** | **Entrez Gene ID** | **Gene symbol** | **Entrez Gene ID** | **Gene symbol** | **Entrez Gene ID** | **Gene symbol** | **Entrez Gene ID** | **Gene symbol** | **Entrez Gene ID** | **Gene symbol** | **Entrez Gene ID** |
| PTGS2 | 5743 | PTGS2 | 5743 | PTGS2 | 5743 | PTGS2 | 5743 | PTGS2 | 5743 | PTGS2 | 5743 |
| MMP9 | 4318 | GLP1R | 2740 | GLP1R | 2740 | MMP9 | 4318 | SLC6A3 | 6531 | GLP1R | 2740 |
| GLP1R | 2740 | MMP9 | 4318 | MMP9 | 4318 | GLP1R | 2740 | GLP1R | 2740 | ESR1 | 2099 |
| NR3C1 | 2908 | NR3C1 | 2908 | NR3C1 | 2908 | NR3C1 | 2908 | ESR1 | 2099 | PPARG | 5468 |
| CASP1 | 834 | ESR1 | 2099 | ESR1 | 2099 | ESR1 | 2099 | PPARG | 5468 | PPARA | 5465 |
| ESR1 | 2099 | PPARG | 5468 | PPARG | 5468 | PPARG | 5468 | PPARA | 5465 | ADRB2 | 154 |
| PPARG | 5468 | PPARA | 5465 | PPARA | 5465 | PPARA | 5465 | ADRB2 | 154 | ICAM1 | 3383 |
| PPARA | 5465 | ADRB2 | 154 | ADRB2 | 154 | ADRB2 | 154 | MC4R | 4160 | | |
| ADRB2 | 154 | ICAM1 | 3383 | ICAM1 | 3383 | ICAM1 | 3383 | CYP2E1 | 1571 | | |
| ICAM1 | 3383 | ADRB1 | 153 | ADRB1 | 153 | ADRB1 | 153 | ICAM1 | 3383 | | |

| *Pterocarpus marsupium* - Obesity | | | | | | | | | | | |
|---|---|---|---|---|---|---|---|---|---|---|---|
| **MCC** | | **MNC** | | **Degree** | | **Closeness centrality** | | **Betweenness centrality** | | **Consensus hub proteins** | |
| **Gene symbol** | **Entrez Gene ID** | **Gene symbol** | **Entrez Gene ID** | **Gene symbol** | **Entrez Gene ID** | **Gene symbol** | **Entrez Gene ID** | **Gene symbol** | **Entrez Gene ID** | **Gene symbol** | **Entrez Gene ID** |
| CYCS | 54205 | CYCS | 54205 | CYCS | 54205 | CYCS | 54205 | BCHE | 590 | CYCS | 54205 |
| PPARA | 5465 | PPARA | 5465 | PPARA | 5465 | PPARA | 5465 | GLP1R | 2740 | PPARA | 5465 |
| PPARD | 5467 | PPARD | 5467 | PPARD | 5467 | PPARD | 5467 | CYCS | 54205 | ESR1 | 2099 |
| ESR1 | 2099 | ESR1 | 2099 | ESR1 | 2099 | ESR1 | 2099 | PPARA | 5465 | MMP9 | 4318 |
| MMP9 | 4318 | MMP9 | 4318 | MMP9 | 4318 | MMP9 | 4318 | ESR1 | 2099 | STAT3 | 6774 |
| FASN | 2194 | FASN | 2194 | FASN | 2194 | FASN | 2194 | MMP9 | 4318 | PPARG | 5468 |
| STAT3 | 6774 | STAT3 | 6774 | STAT3 | 6774 | STAT3 | 6774 | STAT3 | 6774 | PTGS2 | 5743 |
| PPARG | 5468 | PPARG | 5468 | PPARG | 5468 | PPARG | 5468 | F2 | 2147 | | |
| HSPA5 | 3309 | HSPA5 | 3309 | HSPA5 | 3309 | HSPA5 | 3309 | PPARG | 5468 | | |
| PTGS2 | 5743 | PTGS2 | 5743 | PTGS2 | 5743 | PTGS2 | 5743 | PTGS2 | 5743 | | |

| *Terminalia arjuna* - Obesity | | | | | | | | | | | |
|---|---|---|---|---|---|---|---|---|---|---|---|
| **MCC** | | **MNC** | | **Degree** | | **Closeness centrality** | | **Betweenness centrality** | | **Consensus hub proteins** | |
| **Gene symbol** | **Entrez Gene ID** | **Gene symbol** | **Entrez Gene ID** | **Gene symbol** | **Entrez Gene ID** | **Gene symbol** | **Entrez Gene ID** | **Gene symbol** | **Entrez Gene ID** | **Gene symbol** | **Entrez Gene ID** |
| MMP9 | 4318 | ESR1 | 2099 | ESR1 | 2099 | ESR1 | 2099 | ESR1 | 2099 | ESR1 | 2099 |
| ESR1 | 2099 | MMP9 | 4318 | MMP9 | 4318 | MMP9 | 4318 | AKT1 | 207 | AKT1 | 207 |
| AKT1 | 207 | AKT1 | 207 | AKT1 | 207 | AKT1 | 207 | NR3C1 | 2908 | NR3C1 | 2908 |
| NR3C1 | 2908 | NR3C1 | 2908 | NR3C1 | 2908 | NR3C1 | 2908 | CYP2E1 | 1571 | PPARG | 5468 |
| PPARG | 5468 | PPARG | 5468 | PPARG | 5468 | PPARG | 5468 | PPARG | 5468 | ACE | 1636 |
| ACE | 1636 | ACE | 1636 | ACE | 1636 | ACE | 1636 | ACE | 1636 | PTGS2 | 5743 |
| HSPA5 | 3309 | ADRB2 | 154 | ADRB2 | 154 | ADRB2 | 154 | MC4R | 4160 | | |
| PTGS2 | 5743 | PTGS2 | 5743 | PTGS2 | 5743 | PTGS2 | 5743 | ADRB2 | 154 | | |
| STAT3 | 6774 | STAT3 | 6774 | STAT3 | 6774 | STAT3 | 6774 | SLC6A3 | 6531 | | |
| CYCS | 54205 | CYCS | 54205 | CYCS | 54205 | CYCS | 54205 | PTGS2 | 5743 | | |

| *Tecomella undulata* - Obesity | | | | | | | | | | | |
|---|---|---|---|---|---|---|---|---|---|---|---|
| **MCC** | | **MNC** | | **Degree** | | **Closeness centrality** | | **Betweenness centrality** | | **Consensus hub proteins** | |
| **Gene symbol** | **Entrez Gene ID** | **Gene symbol** | **Entrez Gene ID** | **Gene symbol** | **Entrez Gene ID** | **Gene symbol** | **Entrez Gene ID** | **Gene symbol** | **Entrez Gene ID** | **Gene symbol** | **Entrez Gene ID** |
| MMP9 | 4318 | MMP9 | 4318 | MMP9 | 4318 | MMP9 | 4318 | NR3C1 | 2908 | NR3C1 | 2908 |
| NR3C1 | 2908 | NR3C1 | 2908 | NR3C1 | 2908 | NR3C1 | 2908 | BCHE | 590 | HTR2C | 3358 |
| HTR2A | 3356 | HTR2A | 3356 | HTR2A | 3356 | HTR2A | 3356 | HTR2C | 3358 | ADRB2 | 154 |
| HTR2C | 3358 | HTR2C | 3358 | HTR2C | 3358 | HTR2C | 3358 | ADRB2 | 154 | SLC6A3 | 6531 |
| ADRB2 | 154 | ADRB2 | 154 | ADRB2 | 154 | ADRB2 | 154 | SLC6A3 | 6531 | ADRB1 | 153 |
| SLC6A3 | 6531 | SLC6A3 | 6531 | SLC6A3 | 6531 | SLC6A3 | 6531 | MC3R | 4159 | PTGS2 | 5743 |
| ADRB1 | 153 | ADRB1 | 153 | ADRB1 | 153 | ADRB1 | 153 | ADRB1 | 153 | SLC6A2 | 6530 |
| PTGS2 | 5743 | PTGS2 | 5743 | PTGS2 | 5743 | PTGS2 | 5743 | PTGS2 | 5743 | | |
| ESR1 | 2099 | ADRB3 | 155 | SLC6A2 | 6530 | SLC6A2 | 6530 | NPY1R | 4886 | | |
| SLC6A2 | 6530 | SLC6A2 | 6530 | CNR1 | 1268 | CNR1 | 1268 | SLC6A2 | 6530 | | |

| *Aegle marmelos* - Type 2 diabetes | | | | | | | | | | | |
|---|---|---|---|---|---|---|---|---|---|---|---|
| **MCC** | | **MNC** | | **Degree** | | **Closeness centrality** | | **Betweenness centrality** | | **Consensus hub proteins** | |
| **Gene symbol** | **Entrez Gene ID** | **Gene symbol** | **Entrez Gene ID** | **Gene symbol** | **Entrez Gene ID** | **Gene symbol** | **Entrez Gene ID** | **Gene symbol** | **Entrez Gene ID** | **Gene symbol** | **Entrez Gene ID** |
| NDUFA5 | 4698 | NDUFA5 | 4698 | NDUFA5 | 4698 | NDUFA5 | 4698 | NDUFA5 | 4698 | NDUFA5 | 4698 |
| NDUFC2 | 4718 | NDUFC2 | 4718 | NDUFC2 | 4718 | NDUFC2 | 4718 | NDUFA9 | 4704 | NDUFA9 | 4704 |
| NDUFA13 | 51079 | NDUFA13 | 51079 | NDUFA13 | 51079 | NDUFA13 | 51079 | PTPN1 | 5770 | | |
| NDUFA9 | 4704 | NDUFA9 | 4704 | NDUFA9 | 4704 | NDUFA9 | 4704 | PPARG | 5468 | | |
| NDUFB5 | 4711 | NDUFB5 | 4711 | NDUFB5 | 4711 | NDUFB5 | 4711 | ALB | 213 | | |
| NDUFC1 | 4717 | NDUFC1 | 4717 | NDUFC1 | 4717 | NDUFC1 | 4717 | HMGCR | 3156 | | |
| NDUFV2 | 4729 | NDUFV2 | 4729 | NDUFV2 | 4729 | NDUFV2 | 4729 | EGFR | 1956 | | |
| NDUFS1 | 4719 | NDUFS1 | 4719 | NDUFS1 | 4719 | NDUFS1 | 4719 | DPP4 | 1803 | | |
| NDUFAB1 | 4706 | NDUFAB1 | 4706 | NDUFAB1 | 4706 | NDUFAB1 | 4706 | NDUFA4L2 | 56901 | | |
| NDUFB3 | 4709 | NDUFB3 | 4709 | NDUFB3 | 4709 | NDUFB3 | 4709 | GLP1R | 2740 | | |

| *Butea monosperma* - Type 2 diabetes | | | | | | | | | | | |
|---|---|---|---|---|---|---|---|---|---|---|---|
| **MCC** | | **MNC** | | **Degree** | | **Closeness centrality** | | **Betweenness centrality** | | **Consensus hub proteins** | |
| **Gene symbol** | **Entrez Gene ID** | **Gene symbol** | **Entrez Gene ID** | **Gene symbol** | **Entrez Gene ID** | **Gene symbol** | **Entrez Gene ID** | **Gene symbol** | **Entrez Gene ID** | **Gene symbol** | **Entrez Gene ID** |
| FFAR1 | 2864 | FFAR1 | 2864 | FFAR1 | 2864 | FFAR1 | 2864 | PPARG | 5468 | PPARG | 5468 |
| PPARG | 5468 | PPARG | 5468 | PPARG | 5468 | PPARG | 5468 | CYP1A2 | 1544 | PPARD | 5467 |
| PPARD | 5467 | CYP1A2 | 1544 | PPARD | 5467 | PPARD | 5467 | PPARD | 5467 | PPARA | 5465 |
| PPARA | 5465 | PPARD | 5467 | PPARA | 5465 | PPARA | 5465 | EDNRA | 1909 | DPP4 | 1803 |
| LMNA | 4000 | PPARA | 5465 | LMNA | 4000 | LMNA | 4000 | PPARA | 5465 | NFKB1 | 4790 |
| PTPN1 | 5770 | PTPN1 | 5770 | PTPN1 | 5770 | PTPN1 | 5770 | WRN | 7486 | | |
| DPP4 | 1803 | DPP4 | 1803 | DPP4 | 1803 | DPP4 | 1803 | LMNA | 4000 | | |
| FABP4 | 2167 | FABP4 | 2167 | FABP4 | 2167 | FABP4 | 2167 | BCL2L11 | 10018 | | |
| GLP1R | 2740 | GLP1R | 2740 | GLP1R | 2740 | GLP1R | 2740 | DPP4 | 1803 | | |
| NFKB1 | 4790 | NFKB1 | 4790 | NFKB1 | 4790 | NFKB1 | 4790 | NFKB1 | 4790 | | |

| *Commiphora wightii* - Type 2 diabetes | | | | | | | | | | | |
|---|---|---|---|---|---|---|---|---|---|---|---|
| **MCC** | | **MNC** | | **Degree** | | **Closeness centrality** | | **Betweenness centrality** | | **Consensus hub proteins** | |
| **Gene symbol** | **Entrez Gene ID** | **Gene symbol** | **Entrez Gene ID** | **Gene symbol** | **Entrez Gene ID** | **Gene symbol** | **Entrez Gene ID** | **Gene symbol** | **Entrez Gene ID** | **Gene symbol** | **Entrez Gene ID** |
| PPARD | 5467 | PPARD | 5467 | PPARD | 5467 | PPARG | 5468 | PPARD | 5467 | PPARD | 5467 |
| PPARA | 5465 | PPARA | 5465 | PPARA | 5465 | PPARD | 5467 | PPARA | 5465 | PPARA | 5465 |
| ADRB3 | 155 | ADRB3 | 155 | ADRB3 | 155 | PPARA | 5465 | ADRB3 | 155 | ADRB3 | 155 |
| GLP1R | 2740 | GLP1R | 2740 | GLP1R | 2740 | NFKB1 | 4790 | GLP1R | 2740 | GLP1R | 2740 |
| HMGCR | 3156 | HMGCR | 3156 | HMGCR | 3156 | MC4R | 4160 | HMGCR | 3156 | HMGCR | 3156 |
| MC4R | 4160 | MC4R | 4160 | MC4R | 4160 | HMGCR | 3156 | DPP4 | 1803 | DPP4 | 1803 |
| DPP4 | 1803 | DPP4 | 1803 | DPP4 | 1803 | GLP1R | 2740 | NFKB1 | 4790 | NFKB1 | 4790 |
| NFKB1 | 4790 | NFKB1 | 4790 | NFKB1 | 4790 | EGFR | 1956 | LMNA | 4000 | PPARG | 5468 |
| PPARG | 5468 | PPARG | 5468 | PPARG | 5468 | DPP4 | 1803 | PPARG | 5468 | EGFR | 1956 |
| EGFR | 1956 | EGFR | 1956 | EGFR | 1956 | ADRB3 | 155 | EGFR | 1956 | | |

| *Enicostema axillare* - Type 2 diabetes | | | | | | | | | | | |
|---|---|---|---|---|---|---|---|---|---|---|---|
| **MCC** | | **MNC** | | **Degree** | | **Closeness centrality** | | **Betweenness centrality** | | **Consensus hub proteins** | |
| **Gene symbol** | **Entrez Gene ID** | **Gene symbol** | **Entrez Gene ID** | **Gene symbol** | **Entrez Gene ID** | **Gene symbol** | **Entrez Gene ID** | **Gene symbol** | **Entrez Gene ID** | **Gene symbol** | **Entrez Gene ID** |
| PTPN1 | 5770 | INSR | 3643 | PRKCB | 5579 | PRKCB | 5579 | PON1 | 5444 | PTPN1 | 5770 |
| INSR | 3643 | PTPN1 | 5770 | INSR | 3643 | INSR | 3643 | PRKCB | 5579 | INSR | 3643 |
| DPP4 | 1803 | DPP4 | 1803 | PTPN1 | 5770 | PTPN1 | 5770 | INSR | 3643 | NFKB1 | 4790 |
| NFKB1 | 4790 | LMNA | 4000 | DPP4 | 1803 | DPP4 | 1803 | PTPN1 | 5770 | AKT1 | 207 |
| AKT1 | 207 | NFKB1 | 4790 | NFKB1 | 4790 | NFKB1 | 4790 | LMNA | 4000 | PPARG | 5468 |
| PPARG | 5468 | AKT1 | 207 | AKT1 | 207 | AKT1 | 207 | NFKB1 | 4790 | EGFR | 1956 |
| EGFR | 1956 | PPARG | 5468 | EGFR | 1956 | EGFR | 1956 | AKT1 | 207 | PPARA | 5465 |
| PPARA | 5465 | EGFR | 1956 | PPARG | 5468 | PPARG | 5468 | EGFR | 1956 | | |
| PPARD | 5467 | PPARA | 5465 | PPARA | 5465 | PPARA | 5465 | PPARG | 5468 | | |
| GLP1R | 2740 | PPARD | 5467 | PPARD | 5467 | PPARD | 5467 | PPARA | 5465 | | |

| ***Gymnema sylvestre*** **- Type 2 diabetes** | | | | | | | | | | | |
|---|---|---|---|---|---|---|---|---|---|---|---|
| **MCC** | | **MNC** | | **Degree** | | **Closeness centrality** | | **Betweenness centrality** | | **Consensus hub proteins** | |
| **Gene symbol** | **Entrez Gene ID** | **Gene symbol** | **Entrez Gene ID** | **Gene symbol** | **Entrez Gene ID** | **Gene symbol** | **Entrez Gene ID** | **Gene symbol** | **Entrez Gene ID** | **Gene symbol** | **Entrez Gene ID** |
| ICAM1 | 3383 | ICAM1 | 3383 | ICAM1 | 3383 | ICAM1 | 3383 | EGFR | 1956 | EGFR | 1956 |
| EGFR | 1956 | EGFR | 1956 | EGFR | 1956 | EGFR | 1956 | MTOR | 2475 | MTOR | 2475 |
| MTOR | 2475 | MTOR | 2475 | MTOR | 2475 | MTOR | 2475 | GLP1R | 2740 | PPARD | 5467 |
| PPARD | 5467 | PPARD | 5467 | PPARD | 5467 | PPARD | 5467 | PPARD | 5467 | PPARA | 5465 |
| PPARA | 5465 | PPARA | 5465 | PPARA | 5465 | PPARA | 5465 | PPARA | 5465 | ADRB3 | 155 |
| ADRB3 | 155 | ADRB3 | 155 | ADRB3 | 155 | ADRB3 | 155 | ADRB3 | 155 | HMGCR | 3156 |
| HMGCR | 3156 | HMGCR | 3156 | HMGCR | 3156 | HMGCR | 3156 | HMGCR | 3156 | FABP4 | 2167 |
| FABP4 | 2167 | FABP4 | 2167 | FABP4 | 2167 | FABP4 | 2167 | FABP4 | 2167 | PPARG | 5468 |
| PPARG | 5468 | PPARG | 5468 | PPARG | 5468 | PPARG | 5468 | PPARG | 5468 | NFKB1 | 4790 |
| NFKB1 | 4790 | NFKB1 | 4790 | NFKB1 | 4790 | NFKB1 | 4790 | NFKB1 | 4790 | | |

| ***Pterocarpus marsupium*** **-Type 2 Diabetes** | | | | | | | | | | | |
|---|---|---|---|---|---|---|---|---|---|---|---|
| **MCC** | | **MNC** | | **Degree** | | **Closeness centrality** | | **Betweenness centrality** | | **Consensus hub proteins** | |
| **Gene symbol** | **Entrez Gene ID** | **Gene symbol** | **Entrez Gene ID** | **Gene symbol** | **Entrez Gene ID** | **Gene symbol** | **Entrez Gene ID** | **Gene symbol** | **Entrez Gene ID** | **Gene symbol** | **Entrez Gene ID** |
| NFKB1 | 4790 | NFKB1 | 4790 | NFKB1 | 4790 | NFKB1 | 4790 | NFKB1 | 4790 | NFKB1 | 4790 |
| PPARD | 5467 | PPARD | 5467 | PPARD | 5467 | PPARD | 5467 | MTOR | 2475 | MTOR | 2475 |
| MTOR | 2475 | MTOR | 2475 | MTOR | 2475 | MTOR | 2475 | NOS3 | 4846 | NOS3 | 4846 |
| NOS3 | 4846 | NOS3 | 4846 | NOS3 | 4846 | NOS3 | 4846 | ALB | 213 | ALB | 213 |
| ALB | 213 | ALB | 213 | ALB | 213 | ALB | 213 | BCL2 | 596 | BCL2 | 596 |
| BCL2 | 596 | BCL2 | 596 | BCL2 | 596 | BCL2 | 596 | EGFR | 1956 | EGFR | 1956 |
| EGFR | 1956 | EGFR | 1956 | EGFR | 1956 | EGFR | 1956 | STAT3 | 6774 | STAT3 | 6774 |
| STAT3 | 6774 | STAT3 | 6774 | STAT3 | 6774 | STAT3 | 6774 | PON1 | 5444 | PPARA | 5465 |
| PPARA | 5465 | PPARA | 5465 | PPARA | 5465 | PPARA | 5465 | PPARA | 5465 | PPARG | 5468 |
| PPARG | 5468 | PPARG | 5468 | PPARG | 5468 | PPARG | 5468 | PPARG | 5468 | | |

| ***Terminalia arjuna*** **- Type 2 diabetes** | | | | | | | | | | | |
|---|---|---|---|---|---|---|---|---|---|---|---|
| **MCC** | | **MNC** | | **Degree** | | **Closeness centrality** | | **Betweenness centrality** | | **Consensus hub proteins** | |
| **Gene symbol** | **Entrez Gene ID** | **Gene symbol** | **Entrez Gene ID** | **Gene symbol** | **Entrez Gene ID** | **Gene symbol** | **Entrez Gene ID** | **Gene symbol** | **Entrez Gene ID** | **Gene symbol** | **Entrez Gene ID** |
| PPARG | 5468 | PPARG | 5468 | PPARG | 5468 | PPARG | 5468 | PPARG | 5468 | PPARG | 5468 |
| ALB | 213 | ALB | 213 | ALB | 213 | ALB | 213 | ALB | 213 | ALB | 213 |
| STAT3 | 6774 | STAT3 | 6774 | STAT3 | 6774 | STAT3 | 6774 | STAT3 | 6774 | STAT3 | 6774 |
| ACE | 1636 | ACE | 1636 | ACE | 1636 | ACE | 1636 | MGAM | 8972 | ACE | 1636 |
| AKT1 | 207 | AKT1 | 207 | AKT1 | 207 | AKT1 | 207 | ACE | 1636 | AKT1 | 207 |
| EGFR | 1956 | EGFR | 1956 | EGFR | 1956 | EGFR | 1956 | AKT1 | 207 | BCL2 | 596 |
| HMGCR | 3156 | HMGCR | 3156 | HMGCR | 3156 | HMGCR | 3156 | BCL2 | 596 | MTOR | 2475 |
| BCL2 | 596 | BCL2 | 596 | BCL2 | 596 | BCL2 | 596 | DPP4 | 1803 | | |
| MTOR | 2475 | MTOR | 2475 | MTOR | 2475 | MTOR | 2475 | MTOR | 2475 | | |
| NFKB1 | 4790 | NFKB1 | 4790 | NFKB1 | 4790 | NFKB1 | 4790 | LMNA | 4000 | | |

**Table S12: Recurrence of consensus hub proteins across SHD–Type 2 diabetes networks from network pharmacology analyses.** The table lists consensus hub proteins identified in one or multiple SHD–Type 2 diabetes networks, where each listed protein represents a hub that was consistently ranked among the top ten nodes by all five applied network centrality measures. For each protein, the table reports the number of SHDs in which it appeared as a consensus hub, along with the corresponding plant (SHD) name(s).

| Gene Symbol | Entrez Gene ID | Number of SHDs | Plant name(s) |
|---|---|---|---|
| PPARG | 5468 | 6 | *Pterocarpus marsupium* \| *Commiphora wightii* \| *Terminalia arjuna* \| *Butea monosperma* \| *Gymnema sylvestre* \| *Enicostema axillare* |
| PPARA | 5465 | 5 | *Pterocarpus marsupium* \| *Commiphora wightii* \| *Butea monosperma* \| *Gymnema sylvestra* \| *Enicostema axillare* |
| NFKB1 | 4790 | 5 | *Pterocarpus marsupium* \| *Commiphora wightii* \| *Butea monosperma* \| *Gymnema sylvestre* \| *Enicostema axillare* |
| EGFR | 1956 | 4 | *Pterocarpus marsupium* \| *Commiphora wightii* \| *Gymnema sylvestre* \| *Enicostema axillare* |
| MTOR | 2475 | 3 | *Pterocarpus marsupium* \| *Terminalia arjuna* \| *Gymnema sylvestre* |
| PPARD | 5467 | 3 | *Commiphora wightii* \| *Butea monosperma* \| *Gymnema sylvestre* |
| AKT1 | 207 | 2 | *Terminalia arjuna* \| *Enicostema axillare* |
| STAT3 | 6774 | 2 | *Pterocarpus marsupium* \| *Terminalia arjuna* |
| ALB | 213 | 2 | *Pterocarpus marsupium* \| *Terminalia arjuna* |
| BCL2 | 596 | 2 | *Pterocarpus marsupium* \| *Terminalia arjuna* |
| ADRB3 | 155 | 2 | *Commiphora wightii* \| *Gymnema sylvestre* |
| HMGCR | 3156 | 2 | *Commiphora wightii* \| *Gymnema sylvestre* |
| DPP4 | 1803 | 2 | *Commiphora wightii* \| *Butea monosperma* |
| GLP1R | 2740 | 1 | *Commiphora wightii* |
| ACE | 1636 | 1 | *Terminalia arjuna* |
| NOS3 | 4846 | 1 | *Pterocarpus marsupium* |
| NDUFA5 | 4698 | 1 | *Aegle marmelos* |
| NDUFA9 | 4704 | 1 | *Aegle marmelos* |
| FABP4 | 2167 | 1 | *Gymnema sylvestre* |
| PTPN1 | 5770 | 1 | *Enicostema axillare* |
| INSR | 3643 | 1 | *Enicostema axillare* |

**Table S13: Recurrence of consensus hub proteins across SHD–Obesity networks from network pharmacology analyses.** The table lists consensus hub proteins identified in one or multiple SHD–Obesity networks, where each listed protein represents a hub that was consistently ranked among the top ten nodes by all five applied network centrality measures. For each protein, the table reports the number of SHDs in which it appeared as a consensus hub, along with the corresponding plant (SHD) name(s).

| Gene Symbol | Entrez Gene ID | Number of SHDs | Plant name(s) |
|---|---|---|---|
| PTGS2 | 5743 | 8 | *Pterocarpus marsupium* \| *Commiphora wightii* \| *Terminalia arjuna* \| *Aegle marmelos* \| *Butea monosperma* \| *Gymnema sylvestre* \| *Enicostema axillare* \| *Tecomella undulata* |
| PPARG | 5468 | 7 | *Pterocarpus marsupium* \| *Commiphora wightii* \| *Terminalia arjuna* \| *Aegle marmelos* \| *Butea monosperma* \| *Gymnema sylvestre* \| *Enicostema axillare* |
| PPARA | 5465 | 5 | *Pterocarpus marsupium* \| *Commiphora wightii* \| *Butea monosperma* \| *Gymnema sylvestre* \| *Enicostema axillare* |
| ESR1 | 2099 | 5 | *Pterocarpus marsupium* \| *Terminalia arjuna* \| *Butea monosperma* \| *Gymnema sylvestre* \| *Enicostema axillare* |
| ADRB2 | 154 | 5 | *Commiphora wightii* \| *Aegle marmelos* \| *Butea monosperma* \| *Gymnema sylvestre* \| *Tecomella undulata* |
| GLP1R | 2740 | 4 | *Commiphora wightii* \| *Aegle marmelos*\| *Butea monosperma* \| *Gymnema sylvestre* |
| AKT1 | 207 | 2 | *Terminalia arjuna* \| *Enicostema axillare* |
| NR3C1 | 2908 | 2 | *Terminalia arjuna* \| *Tecomella undulata* |
| ADRB1 | 153 | 2 | *Aegle marmelos* \| *Tecomella undulata* |
| STAT3 | 6774 | 1 | *Pterocarpus marsupium* |
| ACE | 1636 | 1 | *Terminalia arjuna* |
| CYCS | 54205 | 1 | *Pterocarpus marsupium* |
| MMP9 | 4318 | 1 | *Pterocarpus marsupium* |
| GNAS | 2778 | 1 | *Aegle marmelos* |
| CNR1 | 1268 | 1 | *Butea monosperma* |
| ICAM1 | 3383 | 1 | *Gymnema sylvestre* |
| PARP1 | 142 | 1 | *Enicostema axillare* |
| HTR2C | 3358 | 1 | *Tecomella undulata* |
| SLC6A3 | 6531 | 1 | *Tecomella undulata* |
| SLC6A2 | 6530 | 1 | *Tecomella undulata* |

**Table S14: SHD-wise phytochemical pairs and their multi-level similarity profiles.** For each single herbal drug (SHD), phytochemical pairs (each of which targets at least one disease-associated protein) are reported along with their structural similarity (Tanimoto coefficient), target-level Jaccard similarity, and pathway-level Jaccard similarity, while restricting to disease-associated gene sets and pathways. Within each SHD, chemical pairs with low target overlap (target-level similarity < 0.25) and low chemical similarity (Tanimoto coefficient < 0.5) are highlighted. PubChem identifiers and chemical names are provided for each compound. Missing values in the pathway-level similarity column indicate that pathway-level similarity could not be computed because disease-overlapping targets for at least one chemical were not represented among the union of the top 20 enriched disease pathways for type 2 diabetes and obesity.

| Plant name | Chemical 1 | Chemical 1 name | Chemical 2 | Chemical 2 name | Tanimoto coefficient | Target-level similarity | Pathway-level similarity |
|---|---|---|---|---|---|---|---|
| *Aegle marmelos* | CID_259846 | Lupeol | CID_323 | Coumarins | 0 | 0 | |
| | CID_259846 | Lupeol | CID_5281426 | Umbelliferone | 0.008547008547 | 0 | |
| | CID_107936 | Fagarine | CID_259846 | Lupeol | 0.01526717557 | 0 | |
| | CID_222284 | Beta-sitosterol | CID_323 | Coumarins | 0.01769911504 | 0 | 0.375 |
| | CID_259846 | Lupeol | CID_6760 | Skimmianine | 0.02222222222 | 0 | |
| | CID_222284 | Beta-sitosterol | CID_5281426 | Umbelliferone | 0.02631578947 | 0 | 0 |
| | CID_10212 | Imperatorin | CID_259846 | Lupeol | 0.02919708029 | 0 | |
| | CID_107936 | Fagarine | CID_222284 | Beta-sitosterol | 0.03125 | 0 | 0 |
| | CID_1550607 | Aurapten | CID_259846 | Lupeol | 0.03546099291 | 0 | |
| | CID_10212 | Imperatorin | CID_222284 | Beta-sitosterol | 0.03703703704 | 0 | 0.25 |
| | CID_222284 | Beta-sitosterol | CID_6760 | Skimmianine | 0.03787878788 | 0 | 0 |
| | CID_259846 | Lupeol | CID_334704 | Marmesin | 0.03875968992 | 0 | |
| | CID_259846 | Lupeol | CID_6450230 | Marmin | 0.04166666667 | 0 | |
| | CID_1550607 | Aurapten | CID_222284 | Beta-sitosterol | 0.05839416058 | 0 | 0.25 |
| | CID_222284 | Beta-sitosterol | CID_334704 | Marmesin | 0.064 | 0 | |
| | CID_222284 | Beta-sitosterol | CID_6450230 | Marmin | 0.07971014493 | 0 | 0.1666666667 |
| | CID_6450230 | Marmin | CID_6760 | Skimmianine | 0.1538461538 | 0 | 0 |
| | CID_1550607 | Aurapten | CID_6760 | Skimmianine | 0.16 | 0 | 0.1428571429 |
| | CID_107936 | Fagarine | CID_6450230 | Marmin | 0.1616161616 | 0 | 0 |
| | CID_107936 | Fagarine | CID_1550607 | Aurapten | 0.1684210526 | 0 | 0.1428571429 |
| | CID_10212 | Imperatorin | CID_6760 | Skimmianine | 0.2333333333 | 0 | 0.5 |
| | CID_10212 | Imperatorin | CID_107936 | Fagarine | 0.2619047619 | 0 | 0.5 |
| | CID_222284 | Beta-sitosterol | CID_259846 | Lupeol | 0.2651515152 | 0 | |
| | CID_1550607 | Aurapten | CID_334704 | Marmesin | 0.2873563218 | 0 | |
| | CID_10212 | Imperatorin | CID_334704 | Marmesin | 0.2891566265 | 0 | |
| | CID_10212 | Imperatorin | CID_323 | Coumarins | 0.2898550725 | 0 | 0.25 |
| | CID_323 | Coumarins | CID_6450230 | Marmin | 0.3243243243 | 0 | 0.3333333333 |
| | CID_1550607 | Aurapten | CID_323 | Coumarins | 0.3428571429 | 0 | 0.6666666667 |
| | CID_334704 | Marmesin | CID_6450230 | Marmin | 0.3488372093 | 0 | |
| | CID_10212 | Imperatorin | CID_6450230 | Marmin | 0.3666666667 | 0 | 0 |
| | CID_5281426 | Umbelliferone | CID_6450230 | Marmin | 0.4084507042 | 0 | 0.2 |
| | CID_1550607 | Aurapten | CID_5281426 | Umbelliferone | 0.4117647059 | 0 | 0.2857142857 |
| | CID_10212 | Imperatorin | CID_1550607 | Aurapten | 0.4337349398 | 0.01851851852 | 0.2857142857 |
| | CID_107936 | Fagarine | CID_323 | Coumarins | 0.1911764706 | 0.03703703704 | 0.125 |
| | CID_323 | Coumarins | CID_334704 | Marmesin | 0.3225806452 | 0.03846153846 | |
| | CID_323 | Coumarins | CID_6760 | Skimmianine | 0.1466666667 | 0.1111111111 | 0.125 |
| | CID_5281426 | Umbelliferone | CID_6760 | Skimmianine | 0.1578947368 | 0.125 | 0 |
| | CID_10212 | Imperatorin | CID_5281426 | Umbelliferone | 0.3188405797 | 0.1428571429 | 0 |
| | CID_107936 | Fagarine | CID_5281426 | Umbelliferone | 0.1857142857 | 0.1666666667 | 0 |
| | CID_334704 | Marmesin | CID_5281426 | Umbelliferone | 0.4 | 0.2 | |
| | CID_323 | Coumarins | CID_5281426 | Umbelliferone | 0.5714285714 | 0.1071428571 | 0.25 |
| | CID_107936 | Fagarine | CID_6760 | Skimmianine | 0.6885245902 | 0.2 | 1 |
| | CID_334704 | Marmesin | CID_6760 | Skimmianine | 0.1555555556 | 0.25 | |
| | CID_107936 | Fagarine | CID_334704 | Marmesin | 0.1647058824 | 0.5 | |
| | CID_1550607 | Aurapten | CID_6450230 | Marmin | 0.6623376623 | 0.9615384615 | 0.5714285714 |
| *Berberis aristata* | CID_11066 | Oxyberberine | CID_72323 | Jatrorrhizine | 0.3904761905 | 0 | |
| | CID_2353 | Berberine | CID_72323 | Jatrorrhizine | 0.6744186047 | 0 | |
| | CID_19009 | Palmatine | CID_72323 | Jatrorrhizine | 0.8125 | 0.1666666667 | |
| | CID_11066 | Oxyberberine | CID_2353 | Berberine | 0.6086956522 | 0.25 | |
| | CID_19009 | Palmatine | CID_2353 | Berberine | 0.6516853933 | 0.3 | 0.5 |
| | CID_11066 | Oxyberberine | CID_19009 | Palmatine | 0.3796296296 | 0.4 | |
| | CID_73170 | Alpha-amyrin | CID_8215 | Behenic acid | 0.01324503311 | 0 | 0.1111111111 |
| | CID_5280450 | Linoleic acid | CID_73170 | Alpha-amyrin | 0.01438848921 | 0 | 0.2222222222 |
| | CID_11005 | Myristic acid | CID_73170 | Alpha-amyrin | 0.0157480315 | 0 | 0.1 |
| | CID_3893 | Lauric acid | CID_73170 | Alpha-amyrin | 0.01652892562 | 0 | 0.3 |
| | CID_2969 | Capric acid | CID_73170 | Alpha-amyrin | 0.01739130435 | 0 | 0.1 |
| | CID_379 | Caprylic acid | CID_73170 | Alpha-amyrin | 0.01834862385 | 0 | 0 |
| | CID_5742590 | Beta-sitosterol glucoside | CID_8215 | Behenic acid | 0.04 | 0 | 0.5714285714 |
| | CID_222284 | Beta-sitosterol | CID_8215 | Behenic acid | 0.04109589041 | 0 | 0.5 |
| | CID_10467 | Arachidic acid | CID_5742590 | Beta-sitosterol glucoside | 0.04142011834 | 0 | 0.5714285714 |
| | CID_10467 | Arachidic acid | CID_222284 | Beta-sitosterol | 0.04285714286 | 0 | 0.5 |
| | CID_445639 | Oleic acid | CID_5742590 | Beta-sitosterol glucoside | 0.04294478528 | 0 | 0.3636363636 |
| | CID_5280934 | Linolenic acid | CID_5742590 | Beta-sitosterol glucoside | 0.04294478528 | 0 | 0.3636363636 |
| | CID_5281 | Octadecanoic acid | CID_5742590 | Beta-sitosterol glucoside | 0.04294478528 | 0 | 0.4 |
| | CID_5280450 | Linoleic acid | CID_5742590 | Beta-sitosterol glucoside | 0.04294478528 | 0 | 0.5 |
| | CID_5742590 | Beta-sitosterol glucoside | CID_985 | Palmitic acid | 0.04458598726 | 0 | 0.4444444444 |
| | CID_222284 | Beta-sitosterol | CID_5280450 | Linoleic acid | 0.0447761194 | 0 | 0.4444444444 |
| | CID_222284 | Beta-sitosterol | CID_5280934 | Linolenic acid | 0.0447761194 | 0 | 0.4545454545 |
| | CID_11005 | Myristic acid | CID_5742590 | Beta-sitosterol glucoside | 0.04635761589 | 0 | 0.5 |
| | CID_222284 | Beta-sitosterol | CID_985 | Palmitic acid | 0.046875 | 0 | 0.5555555556 |
| | CID_3893 | Lauric acid | CID_5742590 | Beta-sitosterol glucoside | 0.04827586207 | 0 | 0.4 |
| | CID_11005 | Myristic acid | CID_222284 | Beta-sitosterol | 0.04918032787 | 0 | 0.625 |
| | CID_2969 | Capric acid | CID_5742590 | Beta-sitosterol glucoside | 0.05035971223 | 0 | 0.5 |
| | CID_222284 | Beta-sitosterol | CID_3893 | Lauric acid | 0.05172413793 | 0 | 0.5 |
| | CID_379 | Caprylic acid | CID_5742590 | Beta-sitosterol glucoside | 0.05263157895 | 0 | 0 |
| | CID_222284 | Beta-sitosterol | CID_2969 | Capric acid | 0.05454545455 | 0 | 0.625 |
| | CID_222284 | Beta-sitosterol | CID_379 | Caprylic acid | 0.05769230769 | 0 | 0.2 |
| | CID_5742590 | Beta-sitosterol glucoside | CID_73170 | Alpha-amyrin | 0.23125 | 0 | 0.1666666667 |
| | CID_222284 | Beta-sitosterol | CID_73170 | Alpha-amyrin | 0.2748091603 | 0 | 0.1428571429 |
| | CID_379 | Caprylic acid | CID_8215 | Behenic acid | 0.3913043478 | 0 | 0 |
| | CID_10467 | Arachidic acid | CID_379 | Caprylic acid | 0.4285714286 | 0 | 0 |
| | CID_379 | Caprylic acid | CID_445639 | Oleic acid | 0.4736842105 | 0 | 0.09090909091 |
| | CID_5280934 | Linolenic acid | CID_73170 | Alpha-amyrin | 0.01438848921 | 0.04545454545 | 0.2727272727 |
| | CID_222284 | Beta-sitosterol | CID_445639 | Oleic acid | 0.0447761194 | 0.05263157895 | 0.4545454545 |
| | CID_445639 | Oleic acid | CID_73170 | Alpha-amyrin | 0.01438848921 | 0.05555555556 | 0.2727272727 |
| | CID_222284 | Beta-sitosterol | CID_5281 | Octadecanoic acid | 0.0447761194 | 0.06666666667 | 0.5 |
| | CID_5281 | Octadecanoic acid | CID_73170 | Alpha-amyrin | 0.01438848921 | 0.07142857143 | 0.1818181818 |
| | CID_379 | Caprylic acid | CID_5281 | Octadecanoic acid | 0.4736842105 | 0.07142857143 | 0.1 |
| | CID_379 | Caprylic acid | CID_5280934 | Linolenic acid | 0.4 | 0.09523809524 | 0.09090909091 |
| | CID_379 | Caprylic acid | CID_5280450 | Linoleic acid | 0.4736842105 | 0.1111111111 | 0 |
| | CID_73170 | Alpha-amyrin | CID_985 | Palmitic acid | 0.01503759398 | 0.125 | 0.2 |
| | CID_11005 | Myristic acid | CID_5280934 | Linolenic acid | 0.4166666667 | 0.1304347826 | 0.7272727273 |
| | CID_10467 | Arachidic acid | CID_5280934 | Linolenic acid | 0.3333333333 | 0.1363636364 | 0.6363636364 |
| | CID_5280934 | Linolenic acid | CID_8215 | Behenic acid | 0.3125 | 0.1428571429 | 0.6363636364 |
| | CID_5280934 | Linolenic acid | CID_985 | Palmitic acid | 0.3846153846 | 0.1666666667 | 0.8181818182 |
| | CID_2969 | Capric acid | CID_5280934 | Linolenic acid | 0.4516129032 | 0.1904761905 | 0.7272727273 |
| ***Butea monosperma*** | CID_10467 | Arachidic acid | CID_73170 | Alpha-amyrin | 0.01379310345 | 0.2 | 0.1111111111 |
| | CID_3893 | Lauric acid | CID_5280934 | Linolenic acid | 0.4545454545 | 0.2 | 0.9090909091 |
| | CID_11005 | Myristic acid | CID_379 | Caprylic acid | 0.6 | 0 | 0.125 |
| | CID_379 | Caprylic acid | CID_3893 | Lauric acid | 0.6923076923 | 0 | 0.1 |
| | CID_379 | Caprylic acid | CID_985 | Palmitic acid | 0.5294117647 | 0.125 | 0.1111111111 |
| | CID_2969 | Capric acid | CID_445639 | Oleic acid | 0.5789473684 | 0.1666666667 | 0.7272727273 |
| | CID_445639 | Oleic acid | CID_8215 | Behenic acid | 0.5555555556 | 0.1764705882 | 0.6363636364 |
| | CID_445639 | Oleic acid | CID_5280934 | Linolenic acid | 0.5 | 0.1875 | 1 |
| | CID_11005 | Myristic acid | CID_5281 | Octadecanoic acid | 0.7894736842 | 0.2 | 0.8 |
| | CID_2969 | Capric acid | CID_379 | Caprylic acid | 0.8181818182 | 0.2 | 0.125 |
| | CID_11005 | Myristic acid | CID_445639 | Oleic acid | 0.7288135593 | 0.2222222222 | 0.7272727273 |
| | CID_5281 | Octadecanoic acid | CID_8215 | Behenic acid | 0.8260869565 | 0.2307692308 | 0.7 |
| | CID_10467 | Arachidic acid | CID_445639 | Oleic acid | 0.6 | 0.2352941176 | 0.6363636364 |
| | CID_3893 | Lauric acid | CID_445639 | Oleic acid | 0.6842105263 | 0.2380952381 | 0.9090909091 |
| | CID_445639 | Oleic acid | CID_5280450 | Linoleic acid | 0.7538461538 | 0.25 | 0.7272727273 |
| | CID_5280450 | Linoleic acid | CID_5280934 | Linolenic acid | 0.6764705882 | 0.2608695652 | 0.7272727273 |
| | CID_3893 | Lauric acid | CID_5281 | Octadecanoic acid | 0.6842105263 | 0.2941176471 | 0.8181818182 |
| | CID_11005 | Myristic acid | CID_5280450 | Linoleic acid | 0.5692307692 | 0.3 | 0.7777777778 |
| | CID_10467 | Arachidic acid | CID_3893 | Lauric acid | 0.619047619 | 0.3 | 0.7 |
| | CID_2969 | Capric acid | CID_3893 | Lauric acid | 0.8461538462 | 0.3 | 0.8 |
| | CID_5280934 | Linolenic acid | CID_5281 | Octadecanoic acid | 0.3571428571 | 0.3076923077 | 0.9090909091 |
| | CID_2969 | Capric acid | CID_5281 | Octadecanoic acid | 0.5789473684 | 0.3076923077 | 0.8 |
| | CID_10467 | Arachidic acid | CID_5281 | Octadecanoic acid | 0.9047619048 | 0.3076923077 | 0.7 |
| | CID_10467 | Arachidic acid | CID_5280450 | Linoleic acid | 0.4457831325 | 0.3333333333 | 0.875 |
| | CID_2969 | Capric acid | CID_5280450 | Linoleic acid | 0.5517241379 | 0.3333333333 | 0.7777777778 |
| | CID_3893 | Lauric acid | CID_8215 | Behenic acid | 0.5652173913 | 0.3333333333 | 0.7 |
| | CID_222284 | Beta-sitosterol | CID_5742590 | Beta-sitosterol glucoside | 0.6752136752 | 0.3333333333 | 0.8 |
| | CID_445639 | Oleic acid | CID_985 | Palmitic acid | 0.7142857143 | 0.3333333333 | 0.8181818182 |
| | CID_3893 | Lauric acid | CID_985 | Palmitic acid | 0.7647058824 | 0.3333333333 | 0.9 |
| | CID_11005 | Myristic acid | CID_985 | Palmitic acid | 0.8823529412 | 0.3333333333 | 0.8888888889 |
| | CID_5280450 | Linoleic acid | CID_8215 | Behenic acid | 0.4157303371 | 0.375 | 0.875 |
| | CID_5280450 | Linoleic acid | CID_5281 | Octadecanoic acid | 0.4805194805 | 0.4 | 0.8 |
| | CID_11005 | Myristic acid | CID_3893 | Lauric acid | 0.8666666667 | 0.4 | 0.8 |
| | CID_8215 | Behenic acid | CID_985 | Palmitic acid | 0.7391304348 | 0.4285714286 | 0.7777777778 |
| | CID_5280450 | Linoleic acid | CID_985 | Palmitic acid | 0.5211267606 | 0.5 | 0.8888888889 |
| | CID_10467 | Arachidic acid | CID_11005 | Myristic acid | 0.7142857143 | 0.5 | 0.875 |
| | CID_11005 | Myristic acid | CID_2969 | Capric acid | 0.7333333333 | 0.5 | 1 |
| | CID_5281 | Octadecanoic acid | CID_985 | Palmitic acid | 0.8947368421 | 0.5384615385 | 0.9 |
| | CID_3893 | Lauric acid | CID_5280450 | Linoleic acid | 0.5737704918 | 0.5454545455 | 0.8 |
| | CID_2969 | Capric acid | CID_985 | Palmitic acid | 0.6470588235 | 0.5714285714 | 0.8888888889 |
| | CID_10467 | Arachidic acid | CID_985 | Palmitic acid | 0.8095238095 | 0.5714285714 | 0.7777777778 |
| | CID_445639 | Oleic acid | CID_5281 | Octadecanoic acid | 0.652173913 | 0.5789473684 | 0.9090909091 |
| | CID_10467 | Arachidic acid | CID_2969 | Capric acid | 0.5238095238 | 0.6 | 0.875 |
| | CID_11005 | Myristic acid | CID_8215 | Behenic acid | 0.652173913 | 0.6 | 0.875 |
| | CID_2969 | Capric acid | CID_8215 | Behenic acid | 0.4782608696 | 0.75 | 0.875 |
| | CID_10467 | Arachidic acid | CID_8215 | Behenic acid | 0.9130434783 | 0.75 | 1 |
| ***Commiphora wightii*** | CID_31253 | Myrcene | CID_5204 | Sesamin | 0 | 0 | 0.2857142857 |
| | CID_222284 | Beta-sitosterol | CID_445858 | Ferulate | 0.03418803419 | 0 | 0.25 |
| | CID_222284 | Beta-sitosterol | CID_31253 | Myrcene | 0.0380952381 | 0 | 0.7142857143 |
| | CID_31253 | Myrcene | CID_5281515 | Caryophyllene | 0.046875 | 0 | 0.5714285714 |
| | CID_445858 | Ferulate | CID_5281515 | Caryophyllene | 0.05333333333 | 0 | 0.2857142857 |
| | CID_222284 | Beta-sitosterol | CID_5204 | Sesamin | 0.07333333333 | 0 | 0.1666666667 |
| | CID_5204 | Sesamin | CID_5997 | Cholesterol | 0.07586206897 | 0 | 0.5 |
| | CID_222284 | Beta-sitosterol | CID_70695727 | Pluviatilol | 0.08163265306 | 0 | 0.5714285714 |
| | CID_5204 | Sesamin | CID_5281515 | Caryophyllene | 0.08181818182 | 0 | 0.2 |
| | CID_445858 | Ferulate | CID_5204 | Sesamin | 0.08411214953 | 0 | 0.4 |
| | CID_5204 | Sesamin | CID_6439929 | E-guggulsterone | 0.1007751938 | 0 | 0.2857142857 |
| | CID_222284 | Beta-sitosterol | CID_5281515 | Caryophyllene | 0.1698113208 | 0 | 0.5 |
| | CID_5281515 | Caryophyllene | CID_5997 | Cholesterol | 0.1782178218 | 0 | 0 |
| | CID_445858 | Ferulate | CID_70695727 | Pluviatilol | 0.2258064516 | 0 | 0.375 |
| | CID_222284 | Beta-sitosterol | CID_6439929 | E-guggulsterone | 0.2894736842 | 0 | 0.7142857143 |
| | CID_445858 | Ferulate | CID_5997 | Cholesterol | 0.03571428571 | 0.03703703704 | 0.2 |
| | CID_31253 | Myrcene | CID_445858 | Ferulate | 0.06666666667 | 0.03703703704 | 0.5 |
| | CID_445858 | Ferulate | CID_6439929 | E-guggulsterone | 0.07368421053 | 0.06666666667 | 0.5 |
| | CID_5281515 | Caryophyllene | CID_6439929 | E-guggulsterone | 0.2352941176 | 0.1428571429 | 0.5714285714 |
| | CID_31253 | Myrcene | CID_70695727 | Pluviatilol | 0.009900990099 | 0.2 | 0.8571428571 |
| | CID_5997 | Cholesterol | CID_70695727 | Pluviatilol | 0.08450704225 | 0.2 | 0 |
| | CID_5204 | Sesamin | CID_70695727 | Pluviatilol | 0.6923076923 | 0 | 0.1428571429 |
| | CID_222284 | Beta-sitosterol | CID_5997 | Cholesterol | 0.808988764 | 0 | 0.2 |
| | CID_5997 | Cholesterol | CID_6439929 | E-guggulsterone | 0.3027522936 | 0.25 | 0.1428571429 |
| | CID_5281515 | Caryophyllene | CID_70695727 | Pluviatilol | 0.08333333333 | 0.3333333333 | 0.6666666667 |
| | CID_31253 | Myrcene | CID_6439929 | E-guggulsterone | 0.03448275862 | 0.4285714286 | 1 |
| | CID_6439929 | E-guggulsterone | CID_70695727 | Pluviatilol | 0.1023622047 | 0.4285714286 | 0.8571428571 |
| | CID_31253 | Myrcene | CID_5997 | Cholesterol | 0.04 | 0.5 | 0.1428571429 |
| ***Diospyros malabarica*** | CID_259846 | Lupeol | CID_370 | Gallic acid | 0.01769911504 | 0 | 0 |
| | CID_370 | Gallic acid | CID_5742590 | Beta-sitosterol glucoside | 0.04347826087 | 0 | 0 |
| | CID_5742590 | Beta-sitosterol glucoside | CID_73145 | Beta amyrin | 0.2085889571 | 0 | 0.25 |
| | CID_259846 | Lupeol | CID_5742590 | Beta-sitosterol glucoside | 0.2236024845 | 0 | 0.25 |
| | CID_5742590 | Beta-sitosterol glucoside | CID_72326 | Betulin | 0.2422360248 | 0 | 0.1 |
| | CID_222284 | Beta-sitosterol | CID_73145 | Beta amyrin | 0.2556390977 | 0 | 0.25 |
| | CID_222284 | Beta-sitosterol | CID_259846 | Lupeol | 0.2651515152 | 0 | 0.25 |
| | CID_222284 | Beta-sitosterol | CID_72326 | Betulin | 0.2686567164 | 0 | 0.1 |
| | CID_370 | Gallic acid | CID_64971 | Betulinic acid | 0.05263157895 | 0.07142857143 | 0.2 |
| | CID_222284 | Beta-sitosterol | CID_370 | Gallic acid | 0.02702702703 | 0.1 | 0 |
| | CID_370 | Gallic acid | CID_72326 | Betulin | 0.02608695652 | 0.1111111111 | 0.6666666667 |
| | CID_370 | Gallic acid | CID_73145 | Beta amyrin | 0.02678571429 | 0.1111111111 | 0 |
| | CID_64971 | Betulinic acid | CID_73145 | Beta amyrin | 0.4416666667 | 0.125 | 0.2 |
| | CID_5742590 | Beta-sitosterol glucoside | CID_64971 | Betulinic acid | 0.2242424242 | 0.1428571429 | 0.8 |
| | CID_259846 | Lupeol | CID_72326 | Betulin | 0.8191489362 | 0 | 0.25 |
| | CID_64971 | Betulinic acid | CID_72326 | Betulin | 0.7959183673 | 0.125 | 0.3 |
| | CID_259846 | Lupeol | CID_64971 | Betulinic acid | 0.8020833333 | 0.125 | 0.2 |
| | CID_222284 | Beta-sitosterol | CID_64971 | Betulinic acid | 0.2554744526 | 0.25 | 0.8 |
| | CID_72326 | Betulin | CID_73145 | Beta amyrin | 0.4491525424 | 0.3333333333 | 0.25 |
| | CID_259846 | Lupeol | CID_73145 | Beta amyrin | 0.5135135135 | 0.3333333333 | 1 |
| | CID_222284 | Beta-sitosterol | CID_5742590 | Beta-sitosterol glucoside | 0.6752136752 | 0.3333333333 | 1 |
| ***Enicostema axillare*** | CID_5281617 | Genkwanin | CID_72326 | Betulin | 0.02097902098 | 0 | |
| | CID_162350 | Isovitexin | CID_445639 | Oleic acid | 0.02158273381 | 0 | 0.2 |
| | CID_162350 | Isovitexin | CID_5281 | Octadecanoic acid | 0.02158273381 | 0 | 0.25 |
| | CID_11005 | Myristic acid | CID_162350 | Isovitexin | 0.02362204724 | 0 | 0.3333333333 |
| | CID_5281 | Octadecanoic acid | CID_5281617 | Genkwanin | 0.02654867257 | 0 | |
| | CID_445639 | Oleic acid | CID_72326 | Betulin | 0.02857142857 | 0 | 0.1666666667 |
| | CID_5281 | Octadecanoic acid | CID_72326 | Betulin | 0.02857142857 | 0 | 0.2 |
| | CID_11005 | Myristic acid | CID_5281617 | Genkwanin | 0.0297029703 | 0 | |
| | CID_11005 | Myristic acid | CID_72326 | Betulin | 0.03125 | 0 | 0 |
| | CID_162350 | Isovitexin | CID_72326 | Betulin | 0.06172839506 | 0 | 0.25 |
| | CID_162350 | Isovitexin | CID_5281617 | Genkwanin | 0.4693877551 | 0 | |
| | CID_5280443 | Apigenin | CID_72326 | Betulin | 0.01418439716 | 0.03448275862 | 0.3333333333 |
| | CID_445639 | Oleic acid | CID_5281617 | Genkwanin | 0.02654867257 | 0.05882352941 | |
| | CID_11005 | Myristic acid | CID_5280443 | Apigenin | 0.0202020202 | 0.1 | 0.4 |
| | CID_5280443 | Apigenin | CID_5281 | Octadecanoic acid | 0.01801801802 | 0.2424242424 | 0.6 |
| | CID_162350 | Isovitexin | CID_5280443 | Apigenin | 0.5161290323 | 0 | 0.1666666667 |
| | CID_5280443 | Apigenin | CID_5281617 | Genkwanin | 0.7692307692 | 0.03571428571 | |
| | CID_11005 | Myristic acid | CID_5281 | Octadecanoic acid | 0.7894736842 | 0.2 | 0.6666666667 |
| | CID_11005 | Myristic acid | CID_445639 | Oleic acid | 0.7288135593 | 0.2222222222 | 0.5 |
| | CID_445639 | Oleic acid | CID_5280443 | Apigenin | 0.01801801802 | 0.25 | 0.8 |
| | CID_445639 | Oleic acid | CID_5281 | Octadecanoic acid | 0.652173913 | 0.5789473684 | 0.4 |
| | CID_305 | Choline | CID_798 | Indole | 0 | 0 | 0 |
| | CID_798 | Indole | CID_8181 | Methyl palmitate | 0 | 0 | 0 |
| | CID_798 | Indole | CID_892 | Inositol | 0 | 0 | 0 |
| | CID_798 | Indole | CID_957 | Octanol | 0 | 0 | 0 |
| | CID_2758 | 1,8-Cineole | CID_798 | Indole | 0 | 0 | 0.2 |
| | CID_5281 | Octadecanoic acid | CID_798 | Indole | 0 | 0 | 0.25 |
| | CID_7127 | Methyl eugenol | CID_892 | Inositol | 0 | 0 | 0.25 |
| | CID_11005 | Myristic acid | CID_798 | Indole | 0 | 0 | 0.3333333333 |
| | CID_798 | Indole | CID_985 | Palmitic acid | 0 | 0 | 0.3333333333 |
| | CID_8181 | Methyl palmitate | CID_892 | Inositol | 0 | 0 | 0.3333333333 |
| | CID_264 | Butyric acid | CID_798 | Indole | 0 | 0 | 0.5 |
| | CID_785 | Hydroquinone | CID_8181 | Methyl palmitate | 0 | 0 | 0.5 |
| | CID_13849 | Pentadecanoic acid | CID_798 | Indole | 0 | 0 | |
| | CID_259846 | Lupeol | CID_798 | Indole | 0 | 0 | |
| | CID_2682 | 1-Hexadecanol | CID_798 | Indole | 0 | 0 | |
| | CID_5280435 | Phytol | CID_798 | Indole | 0 | 0 | |
| | CID_798 | Indole | CID_8209 | Tetradecanol | 0 | 0 | |
| | CID_798 | Indole | CID_8221 | Octadecanol | 0 | 0 | |
| | CID_259846 | Lupeol | CID_785 | Hydroquinone | 0.009523809524 | 0 | |
| | CID_5280435 | Phytol | CID_892 | Inositol | 0.01162790698 | 0 | |
| | CID_8221 | Octadecanol | CID_892 | Inositol | 0.0119047619 | 0 | |
| | CID_2682 | 1-Hexadecanol | CID_892 | Inositol | 0.01282051282 | 0 | |
| | CID_5280435 | Phytol | CID_785 | Hydroquinone | 0.01282051282 | 0 | |
| | CID_13849 | Pentadecanoic acid | CID_892 | Inositol | 0.01298701299 | 0 | |
| | CID_11005 | Myristic acid | CID_892 | Inositol | 0.01351351351 | 0 | 0.25 |
| | CID_8209 | Tetradecanol | CID_892 | Inositol | 0.01388888889 | 0 | |
| | CID_259846 | Lupeol | CID_8221 | Octadecanol | 0.01459854015 | 0 | |
| | CID_73145 | Beta amyrin | CID_8221 | Octadecanol | 0.01459854015 | 0 | |
| | CID_73145 | Beta amyrin | CID_8181 | Methyl palmitate | 0.01470588235 | 0 | 0.5 |
| | CID_259846 | Lupeol | CID_2682 | 1-Hexadecanol | 0.01526717557 | 0 | |
| | CID_2682 | 1-Hexadecanol | CID_73145 | Beta amyrin | 0.01526717557 | 0 | |
| | CID_13849 | Pentadecanoic acid | CID_73145 | Beta amyrin | 0.01538461538 | 0 | |
| | CID_11005 | Myristic acid | CID_73145 | Beta amyrin | 0.0157480315 | 0 | 0 |
| | CID_259846 | Lupeol | CID_8209 | Tetradecanol | 0.016 | 0 | |
| | CID_73145 | Beta amyrin | CID_8209 | Tetradecanol | 0.016 | 0 | |
| | CID_73145 | Beta amyrin | CID_798 | Indole | 0.01834862385 | 0 | 0 |
| | CID_892 | Inositol | CID_957 | Octanol | 0.01851851852 | 0 | 0.5 |
| | CID_892 | Inositol | CID_9015 | p-Guaiacol | 0.01851851852 | 0 | |
| | CID_73145 | Beta amyrin | CID_957 | Octanol | 0.01869158879 | 0 | 0 |
| | CID_259846 | Lupeol | CID_9015 | p-Guaiacol | 0.01869158879 | 0 | |
| | CID_259846 | Lupeol | CID_957 | Octanol | 0.01869158879 | 0 | |
| | CID_264 | Butyric acid | CID_73145 | Beta amyrin | 0.0206185567 | 0 | 0 |
| | CID_445858 | Ferulate | CID_8221 | Octadecanol | 0.02197802198 | 0 | |
| | CID_259846 | Lupeol | CID_8181 | Methyl palmitate | 0.02222222222 | 0 | |
| | CID_7127 | Methyl eugenol | CID_8221 | Octadecanol | 0.02247191011 | 0 | |
| | CID_264 | Butyric acid | CID_892 | Inositol | 0.02272727273 | 0 | 0 |
| | CID_264 | Butyric acid | CID_2758 | 1,8-Cineole | 0.02272727273 | 0 | 0.4 |
| | CID_13849 | Pentadecanoic acid | CID_259846 | Lupeol | 0.02325581395 | 0 | |
| | CID_2682 | 1-Hexadecanol | CID_445858 | Ferulate | 0.02352941176 | 0 | |
| | CID_11005 | Myristic acid | CID_259846 | Lupeol | 0.02380952381 | 0 | |
| | CID_2682 | 1-Hexadecanol | CID_7127 | Methyl eugenol | 0.02409638554 | 0 | |
| | CID_5280435 | Phytol | CID_9015 | p-Guaiacol | 0.025 | 0 | |
| | CID_259846 | Lupeol | CID_445858 | Ferulate | 0.02521008403 | 0 | |
| | CID_445858 | Ferulate | CID_8209 | Tetradecanol | 0.0253164557 | 0 | |
| | CID_259846 | Lupeol | CID_7127 | Methyl eugenol | 0.02564102564 | 0 | |
| | CID_8221 | Octadecanol | CID_9015 | p-Guaiacol | 0.02564102564 | 0 | |
| | CID_259846 | Lupeol | CID_3314 | Eugenol | 0.02631578947 | 0 | |
| | CID_2682 | 1-Hexadecanol | CID_9015 | p-Guaiacol | 0.02777777778 | 0 | |
| | CID_259846 | Lupeol | CID_264 | Butyric acid | 0.03125 | 0 | |
| | CID_7127 | Methyl eugenol | CID_73145 | Beta amyrin | 0.03448275862 | 0 | 0 |
| | CID_3314 | Eugenol | CID_8221 | Octadecanol | 0.03529411765 | 0 | |
| | CID_2758 | 1,8-Cineole | CID_5280435 | Phytol | 0.03571428571 | 0 | |
| | CID_2682 | 1-Hexadecanol | CID_3314 | Eugenol | 0.03797468354 | 0 | |
| | CID_73145 | Beta amyrin | CID_9015 | p-Guaiacol | 0.0380952381 | 0 | |
| | CID_13849 | Pentadecanoic acid | CID_3314 | Eugenol | 0.03846153846 | 0 | |
| | CID_8181 | Methyl palmitate | CID_9015 | p-Guaiacol | 0.03947368421 | 0 | |
| | CID_11005 | Myristic acid | CID_3314 | Eugenol | 0.04 | 0 | 0.6 |
| | CID_264 | Butyric acid | CID_7127 | Methyl eugenol | 0.04081632653 | 0 | 0.6666666667 |
| | CID_3314 | Eugenol | CID_8209 | Tetradecanol | 0.04109589041 | 0 | |
| | CID_305 | Choline | CID_73145 | Beta amyrin | 0.0412371134 | 0 | 0 |
| | CID_259846 | Lupeol | CID_305 | Choline | 0.0412371134 | 0 | |
| | CID_9015 | p-Guaiacol | CID_957 | Octanol | 0.04166666667 | 0 | |
| | CID_445858 | Ferulate | CID_5280435 | Phytol | 0.04395604396 | 0 | |
| | CID_5280435 | Phytol | CID_7127 | Methyl eugenol | 0.04494382022 | 0 | |
| | CID_3314 | Eugenol | CID_5280435 | Phytol | 0.04651162791 | 0 | |
| | CID_305 | Choline | CID_9015 | p-Guaiacol | 0.05 | 0 | |
| | CID_264 | Butyric acid | CID_9015 | p-Guaiacol | 0.05263157895 | 0 | |
| | CID_73145 | Beta amyrin | CID_892 | Inositol | 0.05555555556 | 0 | 0 |
| | CID_7127 | Methyl eugenol | CID_8181 | Methyl palmitate | 0.05882352941 | 0 | 0.25 |
| | CID_305 | Choline | CID_8181 | Methyl palmitate | 0.05970149254 | 0 | 0.3333333333 |
| | CID_305 | Choline | CID_7127 | Methyl eugenol | 0.06 | 0 | 0.25 |
| | CID_13849 | Pentadecanoic acid | CID_305 | Choline | 0.06557377049 | 0 | |
| | CID_264 | Butyric acid | CID_3314 | Eugenol | 0.06666666667 | 0 | 0.4 |
| | CID_11005 | Myristic acid | CID_305 | Choline | 0.06896551724 | 0 | 0.25 |
| | CID_13849 | Pentadecanoic acid | CID_445858 | Ferulate | 0.075 | 0 | |
| | CID_259846 | Lupeol | CID_892 | Inositol | 0.07547169811 | 0 | |
| | CID_11005 | Myristic acid | CID_445858 | Ferulate | 0.07792207792 | 0 | 0.6 |
| | CID_264 | Butyric acid | CID_5280435 | Phytol | 0.09090909091 | 0 | |
| | CID_305 | Choline | CID_8221 | Octadecanol | 0.09090909091 | 0 | |
| | CID_264 | Butyric acid | CID_8221 | Octadecanol | 0.09375 | 0 | |
| | CID_2682 | 1-Hexadecanol | CID_305 | Choline | 0.1 | 0 | |
| | CID_264 | Butyric acid | CID_2682 | 1-Hexadecanol | 0.1034482759 | 0 | |
| | CID_305 | Choline | CID_5280435 | Phytol | 0.1044776119 | 0 | |
| | CID_305 | Choline | CID_8209 | Tetradecanol | 0.1111111111 | 0 | |
| | CID_264 | Butyric acid | CID_8209 | Tetradecanol | 0.1153846154 | 0 | |
| | CID_445858 | Ferulate | CID_798 | Indole | 0.1206896552 | 0 | 0.2 |
| | CID_7127 | Methyl eugenol | CID_798 | Indole | 0.125 | 0 | 0.3333333333 |
| | CID_264 | Butyric acid | CID_445858 | Ferulate | 0.1276595745 | 0 | 0.4 |
| | CID_3314 | Eugenol | CID_798 | Indole | 0.1320754717 | 0 | 0.2 |
| | CID_264 | Butyric acid | CID_305 | Choline | 0.1428571429 | 0 | 0 |
| | CID_264 | Butyric acid | CID_8181 | Methyl palmitate | 0.15 | 0 | 0 |
| | CID_305 | Choline | CID_957 | Octanol | 0.1666666667 | 0 | 0.5 |
| | CID_5280435 | Phytol | CID_8221 | Octadecanol | 0.1666666667 | 0 | |
| | CID_259846 | Lupeol | CID_2758 | 1,8-Cineole | 0.175257732 | 0 | |
| | CID_264 | Butyric acid | CID_957 | Octanol | 0.1764705882 | 0 | 0 |
| | CID_2682 | 1-Hexadecanol | CID_5280435 | Phytol | 0.1777777778 | 0 | |
| | CID_5280435 | Phytol | CID_8181 | Methyl palmitate | 0.1808510638 | 0 | |
| | CID_5280435 | Phytol | CID_957 | Octanol | 0.1884057971 | 0 | |
| | CID_5280435 | Phytol | CID_8209 | Tetradecanol | 0.1904761905 | 0 | |
| | CID_13849 | Pentadecanoic acid | CID_5280435 | Phytol | 0.1931818182 | 0 | |
| | CID_11005 | Myristic acid | CID_5280435 | Phytol | 0.2 | 0 | |
| | CID_798 | Indole | CID_9015 | p-Guaiacol | 0.2093023256 | 0 | |
| | CID_785 | Hydroquinone | CID_798 | Indole | 0.225 | 0 | 0.25 |
| | CID_264 | Butyric acid | CID_985 | Palmitic acid | 0.2452830189 | 0 | 0.6666666667 |
| | CID_13849 | Pentadecanoic acid | CID_264 | Butyric acid | 0.26 | 0 | |
| ***Gymnema sylvestre*** | CID_445858 | Ferulate | CID_9015 | p-Guaiacol | 0.26 | 0 | |
| | CID_11005 | Myristic acid | CID_264 | Butyric acid | 0.2765957447 | 0 | 0.6666666667 |
| | CID_3314 | Eugenol | CID_9015 | p-Guaiacol | 0.2888888889 | 0 | |
| | CID_8181 | Methyl palmitate | CID_957 | Octanol | 0.3389830508 | 0 | 0.5 |
| | CID_13849 | Pentadecanoic acid | CID_957 | Octanol | 0.4038461538 | 0 | |
| | CID_11005 | Myristic acid | CID_957 | Octanol | 0.4285714286 | 0 | 0.3333333333 |
| | CID_8221 | Octadecanol | CID_957 | Octanol | 0.4545454545 | 0 | |
| | CID_11005 | Myristic acid | CID_2758 | 1,8-Cineole | 0.01351351351 | 0.03448275862 | 0.6 |
| | CID_445858 | Ferulate | CID_7127 | Methyl eugenol | 0.275862069 | 0.03571428571 | 0.6 |
| | CID_445858 | Ferulate | CID_73145 | Beta amyrin | 0.03389830508 | 0.03846153846 | 0.2 |
| | CID_2758 | 1,8-Cineole | CID_73145 | Beta amyrin | 0.175257732 | 0.03846153846 | 0.2 |
| | CID_2758 | 1,8-Cineole | CID_8221 | Octadecanol | 0.0119047619 | 0.04 | |
| | CID_2682 | 1-Hexadecanol | CID_2758 | 1,8-Cineole | 0.01282051282 | 0.04 | |
| | CID_13849 | Pentadecanoic acid | CID_2758 | 1,8-Cineole | 0.01298701299 | 0.04 | |
| | CID_2758 | 1,8-Cineole | CID_8209 | Tetradecanol | 0.01388888889 | 0.04 | |
| | CID_2758 | 1,8-Cineole | CID_957 | Octanol | 0.01851851852 | 0.04 | 0.2 |
| | CID_2758 | 1,8-Cineole | CID_9015 | p-Guaiacol | 0.01851851852 | 0.04 | |
| | CID_2758 | 1,8-Cineole | CID_8181 | Methyl palmitate | 0.0243902439 | 0.04 | 0.4 |
| | CID_445858 | Ferulate | CID_957 | Octanol | 0.03278688525 | 0.04 | 0.2 |

| | | | | | | |
|---|---|---|---|---|---|---|
| CID_445858 | Ferulate | CID_8181 | Methyl palmitate | 0.06976744186 | 0.04 | 0.4 |
| CID_3314 | Eugenol | CID_73145 | Beta amyrin | 0.03539823009 | 0.04166666667 | 0.2 |
| CID_3314 | Eugenol | CID_8181 | Methyl palmitate | 0.04819277108 | 0.04347826087 | 0.4 |
| CID_3314 | Eugenol | CID_957 | Octanol | 0.05454545455 | 0.04347826087 | 0.2 |
| CID_305 | Choline | CID_785 | Hydroquinone | 0.02631578947 | 0.05882352941 | 0.2 |
| CID_785 | Hydroquinone | CID_892 | Inositol | 0.04 | 0.06666666667 | 0.2 |
| CID_445858 | Ferulate | CID_985 | Palmitic acid | 0.07228915663 | 0.06666666667 | 0.6 |
| CID_5281 | Octadecanoic acid | CID_73145 | Beta amyrin | 0.01438848921 | 0.07142857143 | 0 |
| CID_11005 | Myristic acid | CID_785 | Hydroquinone | 0.01515151515 | 0.07142857143 | 0.75 |
| CID_259846 | Lupeol | CID_5281 | Octadecanoic acid | 0.02173913043 | 0.07142857143 | |
| CID_3314 | Eugenol | CID_985 | Palmitic acid | 0.03703703704 | 0.07142857143 | 0.6 |
| CID_305 | Choline | CID_985 | Palmitic acid | 0.0625 | 0.07142857143 | 0.25 |
| CID_264 | Butyric acid | CID_5281 | Octadecanoic acid | 0.2203389831 | 0.07142857143 | 0.5 |
| CID_2758 | 1,8-Cineole | CID_7127 | Methyl eugenol | 0.03125 | 0.07407407407 | 0.6 |
| CID_5281 | Octadecanoic acid | CID_9015 | p-Guaiacol | 0.025 | 0.07692307692 | |
| CID_5280435 | Phytol | CID_5281 | Octadecanoic acid | 0.175257732 | 0.07692307692 | |
| CID_5281 | Octadecanoic acid | CID_957 | Octanol | 0.3442622951 | 0.07692307692 | 0.25 |
| CID_892 | Inositol | CID_985 | Palmitic acid | 0.0125 | 0.08333333333 | 0.25 |
| CID_264 | Butyric acid | CID_785 | Hydroquinone | 0.02777777778 | 0.09090909091 | 0.5 |
| CID_73145 | Beta amyrin | CID_785 | Hydroquinone | 0.02912621359 | 0.09090909091 | 0.25 |
| CID_785 | Hydroquinone | CID_8221 | Octadecanol | 0.01315789474 | 0.1 | |
| CID_2682 | 1-Hexadecanol | CID_785 | Hydroquinone | 0.01428571429 | 0.1 | |
| CID_13849 | Pentadecanoic acid | CID_785 | Hydroquinone | 0.01449275362 | 0.1 | |
| CID_785 | Hydroquinone | CID_8209 | Tetradecanol | 0.015625 | 0.1 | |
| CID_785 | Hydroquinone | CID_957 | Octanol | 0.02173913043 | 0.1 | 0.25 |
| CID_3314 | Eugenol | CID_785 | Hydroquinone | 0.1956521739 | 0.1 | 0.8 |
| CID_2758 | 1,8-Cineole | CID_985 | Palmitic acid | 0.0125 | 0.1034482759 | 0.6 |
| CID_73145 | Beta amyrin | CID_985 | Palmitic acid | 0.01503759398 | 0.125 | 0 |
| CID_259846 | Lupeol | CID_985 | Palmitic acid | 0.02272727273 | 0.125 | |
| CID_445858 | Ferulate | CID_785 | Hydroquinone | 0.2 | 0.1290322581 | 0.8 |
| CID_9015 | p-Guaiacol | CID_985 | Palmitic acid | 0.02702702703 | 0.1428571429 | |
| CID_5280435 | Phytol | CID_985 | Palmitic acid | 0.1868131868 | 0.1428571429 | |
| CID_957 | Octanol | CID_985 | Palmitic acid | 0.3818181818 | 0.1428571429 | 0.3333333333 |
| CID_2758 | 1,8-Cineole | CID_785 | Hydroquinone | 0 | 0.1666666667 | 0.8 |
| CID_7127 | Methyl eugenol | CID_785 | Hydroquinone | 0.137254902 | 0.1666666667 | 0.75 |
| CID_305 | Choline | CID_445858 | Ferulate | 0.03773584906 | 0.1785714286 | 0.4 |
| CID_2758 | 1,8-Cineole | CID_305 | Choline | 0.06818181818 | 0.1785714286 | 0.4 |
| CID_445858 | Ferulate | CID_5281 | Octadecanoic acid | 0.06741573034 | 0.1875 | 0.8 |
| CID_2758 | 1,8-Cineole | CID_892 | Inositol | 0.01694915254 | 0.1923076923 | 0.4 |
| CID_445858 | Ferulate | CID_892 | Inositol | 0.0303030303 | 0.1923076923 | 0.4 |
| CID_305 | Choline | CID_3314 | Eugenol | 0.06382978723 | 0.1923076923 | 0.4 |
| CID_11005 | Myristic acid | CID_9015 | p-Guaiacol | 0.02941176471 | 0.2 | |
| CID_3314 | Eugenol | CID_5281 | Octadecanoic acid | 0.03448275862 | 0.2 | 0.8 |
| CID_3314 | Eugenol | CID_892 | Inositol | 0.01612903226 | 0.2083333333 | 0.4 |
| CID_5281 | Octadecanoic acid | CID_785 | Hydroquinone | 0.01282051282 | 0.2105263158 | 0.6 |
| CID_785 | Hydroquinone | CID_985 | Palmitic acid | 0.01388888889 | 0.2142857143 | 0.75 |
| CID_2758 | 1,8-Cineole | CID_5281 | Octadecanoic acid | 0.01162790698 | 0.2258064516 | 0.8 |
| CID_2682 | 1-Hexadecanol | CID_957 | Octanol | 0.5102040816 | 0 | |
| CID_11005 | Myristic acid | CID_8221 | Octadecanol | 0.5625 | 0 | |
| CID_8209 | Tetradecanol | CID_957 | Octanol | 0.5813953488 | 0 | |
| CID_8181 | Methyl palmitate | CID_8221 | Octadecanol | 0.6029411765 | 0 | |
| CID_13849 | Pentadecanoic acid | CID_8221 | Octadecanol | 0.609375 | 0 | |
| CID_11005 | Myristic acid | CID_2682 | 1-Hexadecanol | 0.6206896552 | 0 | |
| CID_8181 | Methyl palmitate | CID_8209 | Tetradecanol | 0.6440677966 | 0 | |
| CID_8221 | Octadecanol | CID_985 | Palmitic acid | 0.65625 | 0 | |
| CID_2682 | 1-Hexadecanol | CID_8181 | Methyl palmitate | 0.6612903226 | 0 | |
| CID_13849 | Pentadecanoic acid | CID_2682 | 1-Hexadecanol | 0.6724137931 | 0 | |
| CID_11005 | Myristic acid | CID_8181 | Methyl palmitate | 0.6779661017 | 0 | 0.25 |
| CID_5281 | Octadecanoic acid | CID_8181 | Methyl palmitate | 0.7076923077 | 0 | 0.2 |
| CID_2682 | 1-Hexadecanol | CID_985 | Palmitic acid | 0.724137931 | 0 | |
| CID_13849 | Pentadecanoic acid | CID_8181 | Methyl palmitate | 0.7288135593 | 0 | |
| CID_2682 | 1-Hexadecanol | CID_5281 | Octadecanoic acid | 0.737704918 | 0 | |
| CID_5281 | Octadecanoic acid | CID_8221 | Octadecanol | 0.75 | 0 | |
| CID_8181 | Methyl palmitate | CID_985 | Palmitic acid | 0.7796610169 | 0 | 0.25 |
| CID_8209 | Tetradecanol | CID_8221 | Octadecanol | 0.7818181818 | 0 | |
| CID_2682 | 1-Hexadecanol | CID_8209 | Tetradecanol | 0.8775510204 | 0 | |
| CID_3314 | Eugenol | CID_7127 | Methyl eugenol | 0.6428571429 | 0.03846153846 | 0.6 |
| CID_5281 | Octadecanoic acid | CID_8209 | Tetradecanol | 0.6393442623 | 0.07692307692 | |
| CID_13849 | Pentadecanoic acid | CID_5281 | Octadecanoic acid | 0.8421052632 | 0.07692307692 | |
| CID_785 | Hydroquinone | CID_9015 | p-Guaiacol | 0.5161290323 | 0.1 | |
| CID_8209 | Tetradecanol | CID_985 | Palmitic acid | 0.7090909091 | 0.1428571429 | |

| | | | | | | | |
|---|---|---|---|---|---|---|---|
| | CID_13849 | Pentadecanoic acid | CID_985 | Palmitic acid | 0.9411764706 | 0.1428571429 | |
| | CID_11005 | Myristic acid | CID_8209 | Tetradecanol | 0.6923076923 | 0.2 | |
| | CID_11005 | Myristic acid | CID_5281 | Octadecanoic acid | 0.7894736842 | 0.2 | 0.75 |
| | CID_11005 | Myristic acid | CID_13849 | Pentadecanoic acid | 0.9375 | 0.2 | |
| | CID_13849 | Pentadecanoic acid | CID_7127 | Methyl eugenol | 0.0243902439 | 0.25 | |
| | CID_7127 | Methyl eugenol | CID_8209 | Tetradecanol | 0.02597402597 | 0.25 | |
| | CID_7127 | Methyl eugenol | CID_957 | Octanol | 0.03389830508 | 0.25 | 0.3333333333 |
| | CID_7127 | Methyl eugenol | CID_9015 | p-Guaiacol | 0.22 | 0.25 | |
| | CID_5281 | Octadecanoic acid | CID_7127 | Methyl eugenol | 0.02197802198 | 0.3076923077 | 0.75 |
| | CID_259846 | Lupeol | CID_73145 | Beta amyrin | 0.5135135135 | 0.3333333333 | |
| | CID_11005 | Myristic acid | CID_985 | Palmitic acid | 0.8823529412 | 0.3333333333 | 1 |
| | CID_305 | Choline | CID_5281 | Octadecanoic acid | 0.05714285714 | 0.4 | 0.5 |
| | CID_5281 | Octadecanoic acid | CID_892 | Inositol | 0.01162790698 | 0.4615384615 | 0.5 |
| | CID_11005 | Myristic acid | CID_7127 | Methyl eugenol | 0.0253164557 | 0.5 | 1 |
| | CID_5280435 | Phytol | CID_73145 | Beta amyrin | 0.04444444444 | 0.5 | |
| | CID_259846 | Lupeol | CID_5280435 | Phytol | 0.06015037594 | 0.5 | |
| | CID_5281 | Octadecanoic acid | CID_985 | Palmitic acid | 0.8947368421 | 0.5384615385 | 0.75 |
| | CID_7127 | Methyl eugenol | CID_985 | Palmitic acid | 0.02352941176 | 0.5714285714 | 1 |
| | CID_305 | Choline | CID_892 | Inositol | 0.02173913043 | 0.75 | 1 |
| | CID_2758 | 1,8-Cineole | CID_445858 | Ferulate | 0.01492537313 | 0.8518518519 | 1 |
| | CID_2758 | 1,8-Cineole | CID_3314 | Eugenol | 0.01612903226 | 0.92 | 1 |
| | CID_3314 | Eugenol | CID_445858 | Ferulate | 0.4489795918 | 0.92 | 1 |
| | CID_13849 | Pentadecanoic acid | CID_9015 | p-Guaiacol | 0.02816901408 | 1 | |
| | CID_8209 | Tetradecanol | CID_9015 | p-Guaiacol | 0.0303030303 | 1 | |
| | CID_13849 | Pentadecanoic acid | CID_8209 | Tetradecanol | 0.75 | 1 | |
| | CID_2682 | 1-Hexadecanol | CID_8221 | Octadecanol | 0.8909090909 | 1 | |
| | CID_126 | P-Hydroxybenzaldehyde | CID_259846 | Lupeol | 0.009259259259 | 0 | |
| | CID_259846 | Lupeol | CID_5281805 | Pseudobaptigenin | 0.01398601399 | 0 | 0.2 |
| | CID_259846 | Lupeol | CID_5281727 | Pterostilbene | 0.02222222222 | 0 | |
| | CID_10494 | Oleanolic acid | CID_5281805 | Pseudobaptigenin | 0.03448275862 | 0 | 0.4444444444 |
| | CID_10494 | Oleanolic acid | CID_126 | P-Hydroxybenzaldehyde | 0.03636363636 | 0 | |
| | CID_259846 | Lupeol | CID_72276 | (-)-Epicatechin | 0.03649635036 | 0 | 0.25 |
| | CID_10494 | Oleanolic acid | CID_72276 | (-)-Epicatechin | 0.05755395683 | 0 | 0.8888888889 |
| | CID_126 | P-Hydroxybenzaldehyde | CID_72276 | (-)-Epicatechin | 0.1527777778 | 0 | |
| | CID_126 | P-Hydroxybenzaldehyde | CID_5281805 | Pseudobaptigenin | 0.2112676056 | 0 | |
| | CID_5281805 | Pseudobaptigenin | CID_72276 | (-)-Epicatechin | 0.2268041237 | 0 | 0.5 |
| | CID_5281727 | Pterostilbene | CID_5281805 | Pseudobaptigenin | 0.2365591398 | 0 | |
| | CID_5282073 | 7,4'-dihydroxyflavone | CID_72276 | (-)-Epicatechin | 0.2444444444 | 0 | 0.625 |
| | CID_126 | P-Hydroxybenzaldehyde | CID_439246 | Naringenin | 0.265625 | 0 | |
| | CID_114829 | Liquiritigenin | CID_126 | P-Hydroxybenzaldehyde | 0.2741935484 | 0 | |
| | CID_126 | P-Hydroxybenzaldehyde | CID_5282073 | 7,4'-dihydroxyflavone | 0.2741935484 | 0 | |
| | CID_5281727 | Pterostilbene | CID_5282073 | 7,4'-dihydroxyflavone | 0.2857142857 | 0 | |
| | CID_114829 | Liquiritigenin | CID_72276 | (-)-Epicatechin | 0.3333333333 | 0 | 0.5 |
| | CID_114829 | Liquiritigenin | CID_5281805 | Pseudobaptigenin | 0.3372093023 | 0 | 0.3333333333 |
| | CID_126 | P-Hydroxybenzaldehyde | CID_5281727 | Pterostilbene | 0.3389830508 | 0 | |
| | CID_126 | P-Hydroxybenzaldehyde | CID_638278 | Isoliquiritigenin | 0.3684210526 | 0 | |
| | CID_439246 | Naringenin | CID_72276 | (-)-Epicatechin | 0.4430379747 | 0 | 0.625 |
| | CID_638278 | Isoliquiritigenin | CID_72276 | (-)-Epicatechin | 0.2471910112 | 0.04761904762 | 0.8888888889 |
| ***Pterocarpus marsupium*** | CID_5281805 | Pseudobaptigenin | CID_638278 | Isoliquiritigenin | 0.2527472527 | 0.05882352941 | 0.4444444444 |
| | CID_259846 | Lupeol | CID_638278 | Isoliquiritigenin | 0.01481481481 | 0.0625 | 0.2222222222 |
| | CID_259846 | Lupeol | CID_439246 | Naringenin | 0.02189781022 | 0.0625 | 0.4 |
| | CID_5281727 | Pterostilbene | CID_72276 | (-)-Epicatechin | 0.1914893617 | 0.08333333333 | |
| | CID_10494 | Oleanolic acid | CID_5282073 | 7,4'-dihydroxyflavone | 0.03623188406 | 0.09090909091 | 0.5555555556 |
| | CID_10494 | Oleanolic acid | CID_114829 | Liquiritigenin | 0.05147058824 | 0.09090909091 | 0.4444444444 |
| | CID_10494 | Oleanolic acid | CID_439246 | Naringenin | 0.05072463768 | 0.09523809524 | 0.5555555556 |
| | CID_439246 | Naringenin | CID_5281727 | Pterostilbene | 0.2643678161 | 0.1052631579 | |
| | CID_114829 | Liquiritigenin | CID_5281727 | Pterostilbene | 0.2705882353 | 0.1111111111 | |
| | CID_5282073 | 7,4'-dihydroxyflavone | CID_638278 | Isoliquiritigenin | 0.4078947368 | 0.1176470588 | 0.5555555556 |
| | CID_114829 | Liquiritigenin | CID_5282073 | 7,4'-dihydroxyflavone | 0.4594594595 | 0.1428571429 | 0.5 |
| | CID_10494 | Oleanolic acid | CID_638278 | Isoliquiritigenin | 0.05185185185 | 0.15 | 1 |
| | CID_10494 | Oleanolic acid | CID_5281727 | Pterostilbene | 0.03623188406 | 0.1666666667 | |
| | CID_5281727 | Pterostilbene | CID_638278 | Isoliquiritigenin | 0.3896103896 | 0.1666666667 | |
| | CID_259846 | Lupeol | CID_5282073 | 7,4'-dihydroxyflavone | 0.007299270073 | 0.2 | 0.4 |
| | CID_114829 | Liquiritigenin | CID_259846 | Lupeol | 0.02222222222 | 0.2 | 0.5 |
| | CID_439246 | Naringenin | CID_5281805 | Pseudobaptigenin | 0.3146067416 | 0.2 | 0.8 |
| | CID_439246 | Naringenin | CID_638278 | Isoliquiritigenin | 0.3797468354 | 0.2 | 0.5555555556 |
| | CID_114829 | Liquiritigenin | CID_439246 | Naringenin | 0.6666666667 | 0.1875 | 0.5 |
| | CID_10494 | Oleanolic acid | CID_259846 | Lupeol | 0.4786324786 | 0.25 | 0.2222222222 |
| | CID_439246 | Naringenin | CID_5282073 | 7,4'-dihydroxyflavone | 0.3924050633 | 0.2666666667 | 1 |
| | CID_114829 | Liquiritigenin | CID_638278 | Isoliquiritigenin | 0.4078947368 | 0.2666666667 | 0.4444444444 |
| | CID_5281805 | Pseudobaptigenin | CID_5282073 | 7,4'-dihydroxyflavone | 0.4197530864 | 0.75 | 0.8 |
| ***Tecomella undulata*** | CID_222284 | Beta-sitosterol | CID_445858 | Ferulate | 0.03418803419 | 0 | 0.2857142857 |
| | CID_445858 | Ferulate | CID_5742590 | Beta-sitosterol glucoside | 0.04137931034 | 0 | 0.1428571429 |
| | CID_222284 | Beta-sitosterol | CID_5742590 | Beta-sitosterol glucoside | 0.6752136752 | 0.3333333333 | 0.8 |
| ***Tectona grandis*** | CID_64971 | Betulinic acid | CID_931 | Naphthalene | 0 | 0 | |
| | CID_64971 | Betulinic acid | CID_6780 | Anthraquinones | 0.007462686567 | 0 | |
| | CID_222284 | Beta-sitosterol | CID_6780 | Anthraquinones | 0.0157480315 | 0 | |
| | CID_222284 | Beta-sitosterol | CID_931 | Naphthalene | 0.01801801802 | 0 | |
| | CID_222284 | Beta-sitosterol | CID_72734 | Dehydro-alpha-lapachone | 0.05555555556 | 0 | |
| | CID_72734 | Dehydro-alpha-lapachone | CID_931 | Naphthalene | 0.2121212121 | 0 | |
| | CID_6780 | Anthraquinones | CID_931 | Naphthalene | 0.4074074074 | 0 | |
| | CID_6780 | Anthraquinones | CID_72734 | Dehydro-alpha-lapachone | 0.4545454545 | 0 | |
| | CID_64971 | Betulinic acid | CID_72734 | Dehydro-alpha-lapachone | 0.04511278195 | 0.125 | |
| | CID_222284 | Beta-sitosterol | CID_64971 | Betulinic acid | 0.2554744526 | 0.25 | 0.8888888889 |
| ***Terminalia arjuna*** | CID_222284 | Beta-sitosterol | CID_5281855 | Ellagic acid | 0.02142857143 | 0 | 0.4444444444 |
| | CID_222284 | Beta-sitosterol | CID_289 | Catechol | 0.02941176471 | 0 | 0.5 |
| | CID_5281605 | Baicalein | CID_73641 | Arjunolic acid | 0.04166666667 | 0 | 0 |
| | CID_5281855 | Ellagic acid | CID_73641 | Arjunolic acid | 0.04761904762 | 0 | 0 |
| | CID_10494 | Oleanolic acid | CID_72277 | Epigallocatechol | 0.05673758865 | 0 | 0.5555555556 |
| | CID_222284 | Beta-sitosterol | CID_65084 | (+)-Gallocatechol | 0.05925925926 | 0 | 0 |
| | CID_222284 | Beta-sitosterol | CID_72277 | Epigallocatechol | 0.05925925926 | 0 | 0.1111111111 |
| | CID_72277 | Epigallocatechol | CID_73641 | Arjunolic acid | 0.07692307692 | 0 | 0.2 |
| | CID_222284 | Beta-sitosterol | CID_73641 | Arjunolic acid | 0.2377622378 | 0 | 0 |
| | CID_5281605 | Baicalein | CID_65084 | (+)-Gallocatechol | 0.2608695652 | 0 | 0.3 |
| | CID_5281605 | Baicalein | CID_72277 | Epigallocatechol | 0.2608695652 | 0 | 0.4 |
| | CID_5281855 | Ellagic acid | CID_72277 | Epigallocatechol | 0.2765957447 | 0 | 0.3 |
| | CID_5281855 | Ellagic acid | CID_65084 | (+)-Gallocatechol | 0.2765957447 | 0 | 0.3333333333 |
| | CID_289 | Catechol | CID_72277 | Epigallocatechol | 0.1714285714 | 0.02857142857 | 0.5 |
| | CID_289 | Catechol | CID_65084 | (+)-Gallocatechol | 0.1714285714 | 0.02941176471 | 0.4 |
| | CID_289 | Catechol | CID_73641 | Arjunolic acid | 0.03571428571 | 0.0303030303 | 0.1 |
| | CID_222284 | Beta-sitosterol | CID_6251 | Mannitol | 0.05607476636 | 0.03571428571 | 0.2222222222 |
| | CID_6251 | Mannitol | CID_72277 | Epigallocatechol | 0.07142857143 | 0.03571428571 | 0.5714285714 |
| | CID_6251 | Mannitol | CID_65084 | (+)-Gallocatechol | 0.07142857143 | 0.03703703704 | 0.4285714286 |
| | CID_6251 | Mannitol | CID_73641 | Arjunolic acid | 0.05084745763 | 0.03846153846 | 0.1666666667 |
| | CID_10494 | Oleanolic acid | CID_5281605 | Baicalein | 0.03571428571 | 0.05 | 0.8 |
| | CID_10494 | Oleanolic acid | CID_289 | Catechol | 0.03738317757 | 0.05128205128 | 0.9 |
| | CID_10494 | Oleanolic acid | CID_6251 | Mannitol | 0.01709401709 | 0.0625 | 0.5 |
| | CID_5281855 | Ellagic acid | CID_6251 | Mannitol | 0.04651162791 | 0.08571428571 | 0.4 |
| | CID_289 | Catechol | CID_5281605 | Baicalein | 0.2786885246 | 0.09523809524 | 0.9 |
| | CID_10494 | Oleanolic acid | CID_222284 | Beta-sitosterol | 0.2463768116 | 0.1 | 0.5555555556 |
| | CID_10494 | Oleanolic acid | CID_5281855 | Ellagic acid | 0.03472222222 | 0.1111111111 | 0.7 |
| | CID_10494 | Oleanolic acid | CID_65084 | (+)-Gallocatechol | 0.05673758865 | 0.1111111111 | 0.4444444444 |
| | CID_5281605 | Baicalein | CID_6251 | Mannitol | 0.03614457831 | 0.1142857143 | 0.5 |
| | CID_289 | Catechol | CID_5281855 | Ellagic acid | 0.1388888889 | 0.125 | 0.8 |
| | CID_5281605 | Baicalein | CID_5281855 | Ellagic acid | 0.3033707865 | 0.1363636364 | 0.8888888889 |
| | CID_222284 | Beta-sitosterol | CID_5281605 | Baicalein | 0.02205882353 | 0.1428571429 | 0.5555555556 |
| | CID_10494 | Oleanolic acid | CID_73641 | Arjunolic acid | 0.7102803738 | 0.125 | 0.1111111111 |
| | CID_65084 | (+)-Gallocatechol | CID_72277 | Epigallocatechol | 1 | 0.25 | 0.8 |
| | CID_65084 | (+)-Gallocatechol | CID_73641 | Arjunolic acid | 0.07692307692 | 0.5 | 0.25 |
| | CID_289 | Catechol | CID_6251 | Mannitol | 0.04 | 0.6857142857 | 0.6 |

**Table S15: Docking-based binding energies (kcal/mol) of SHD-derived phytochemicals docked to DPP4 (PDB: 1X70).** Compounds were retained only if they interacted with at least 4 of 7 key DPP4 pocket residues. The biological reference ligand (Sitagliptin) is shown in bold.

| External chemical identifier | Chemical name | Docking-based Binding Energy (kcal/mol) |
|---|---|---|
| CAS_77754-91-7 | Chitraline | -9.756 |
| CID_162350 | Isovitexin | -9.699 |
| CID_193239 | Pakistanine | -9.278 |
| CID_101297673 | Guggulsterol I | -9.059 |
| CID_10071442 | Sulfurein | -8.976 |
| CID_156697 | Kalashine | -8.903 |
| CID_42607822 | Isomonospermoside | -8.898 |
| CID_3574508 | Tecomaquinone I | -8.893 |
| **CID_4369359** | **Sitagliptin** | **-8.871** |
| CID_181478 | 1 - O - methyl pakistanine | -8.871 |
| CID_12303942 | Coreopsin | -8.767 |
| CID_442333 | Oxyacanthine | -8.71 |
| CID_101297674 | Guggulsterol II | -8.488 |
| CID_133775 | Pterosupin | -8.441 |
| CID_5742590 | Beta-sitosterol glucoside | -8.427 |
| CID_67030 | Anthraquinone-2-carboxylic acid | -8.316 |
| CID_3037329 | Dehydrotectol | -8.304 |
| CID_101297675 | Guggulsterol III | -8.282 |
| CID_70695727 | Pluviatilol | -8.252 |
| CID_5282073 | 7,4'-dihydroxyflavone | -8.183 |
| CID_5281611 | 5-deoxykaempferol | -8.176 |
| CID_5281805 | Pseudobaptigenin | -8.146 |
| CID_6439929 | E-guggulsterone | -8.06 |
| CID_5280794 | Stigmasterol | -8.054 |
| CID_12309899 | Isocoreopsin | -8.049 |
| CID_5204 | Sesamin | -8.045 |
| CID_442410 | Garbanzol | -8.009 |
| CID_12309055 | Beta-Sitosterol-beta-D-glucoside | -7.952 |
| CID_440835 | Leucodelphinidin | -7.878 |
| CID_5997 | Cholesterol | -7.859 |
| CID_2353 | Berberine | -7.845 |
| CID_11066 | Oxyberberine | -7.783 |
| CID_65084 | (+)-Gallocatechol | -7.753 |
| CID_124034 | Swertisin | -7.744 |
| CID_1550607 | Aurapten | -7.714 |
| CID_42607524 | Monospermoside | -7.671 |
| CID_5321494 | Undulatosides A | -7.648 |
| CID_72277 | Epigallocatechol | -7.618 |
| CID_134369 | Marsupsin | -7.615 |
| CID_72276 | (-)-Epicatechin | -7.51 |
| CID_344310 | Anthraquinone-2-carboxaldehyde | -7.502 |
| CID_14033983 | Pongaflavone | -7.497 |
| CID_71597391 | Triterpenoids | -7.482 |
| CID_72734 | Dehydro-alpha-lapachone | -7.366 |
| CID_10494 | Oleanolic acid | -7.345 |
| CID_161109 | 20 alpha hydroxy-4-pregnen-3-one | -7.323 |
| CID_5281617 | Genkwanin | -7.305 |
| CID_92747 | 20 beta hydroxy -4-pregnen-3-one | -7.289 |
| CAS_82178-34-5 | Arjunolone | -7.212 |
| CID_638278 | Isoliquiritigenin | -7.202 |
| CID_5318245 | 5-hydroxyalpachol | -7.192 |
| CID_6773 | Tectoquinone | -7.167 |
| CHEMSPIDER_10306372 | Propterol B | -7.139 |
| CID_638088 | Stilbenes | -7.129 |
| CID_72326 | Betulin | -7.115 |
| CID_6450230 | Marmin | -7.105 |
| CID_15385516 | Arjunic acid | -7.084 |
| CID_630739 | Karachine | -7.072 |
| CID_15560302 | Gymnestrogenin | -7.043 |
| CID_3884 | Lapachol | -7.029 |
| CID_259846 | Lupeol | -7.012 |
| CID_334704 | Marmesin | -7.004 |
| CID_44257328 | 7-hydroxy-5,4'-dimethoxy-8-methylisoflavone-7-rhamnoside | -7 |
| CID_72323 | Jatrorrhizine | -6.987 |
| CID_10212 | Imperatorin | -6.961 |
| CID_73641 | Arjunolic acid | -6.946 |
| CID_13970503 | 1-hydroxy-5-methoxy-2-methyl-9,10-anthraquinone | -6.941 |
| CID_5281727 | Pterostilbene | -6.939 |
| CID_73145 | Beta amyrin | -6.925 |
| CID_600671 | Aegelinol | -6.912 |
| CID_622032 | Tomentosic acid | -6.849 |
| CID_6780 | Anthraquinones | -6.833 |
| | Tecomin | -6.733 |
| CID_101750 | Alpha-camphorene | -6.702 |
| CID_12444386 | Arjungenin | -6.686 |
| CID_91472 | Friedelin | -6.666 |
| CID_19009 | Palmatine | -6.624 |
| CID_122173119 | Z-guggulsterone | -6.623 |
| CID_185124 | Propterol | -6.564 |
| CID_132568257 | Terminic acid | -6.555 |
| CID_91457 | Beta Eudesmol | -6.292 |
| CID_14283285 | 9,10-dimethoxy-2-methyl-1,4-anthraquinone | -6.259 |
| CID_49831545 | Delta-lactone | -6.258 |
| CID_7092583 | Peregrinol | -6.174 |
| CID_5368823 | Allylcembrol | -6.126 |
| CID_102515444 | Phosphatidylethanolamine | -6.125 |
| CID_14034812 | Cerasidin | -6.023 |
| CID_68406 | n-Octacosanol | -5.843 |
| CID_5281384 | Cembrene A | -5.842 |
| CID_5281564 | Enicoflavin | -5.833 |
| CID_114522 | Ketones | -5.825 |
| CID_11197 | Lignoceric acid | -5.816 |
| CID_12410 | Hentriacontane | -5.813 |
| CID_107936 | Fagarine | -5.809 |
| CID_985 | Palmitic acid | -5.765 |
| CID_5280934 | Linolenic acid | -5.755 |
| CID_11005 | Myristic acid | -5.751 |
| CID_68972 | n-Triacontanol | -5.745 |
| CID_12409 | n-Nonacosane | -5.74 |
| CID_439215 | Galacturonic acid | -5.727 |
| CID_8215 | Behenic acid | -5.718 |
| CID_191120 | Erythrocentaurin | -5.687 |
| CID_11747713 | Cembrene | -5.625 |
| CID_90472510 | Mukulol | -5.622 |
| CID_5280435 | Phytol | -5.611 |
| CID_5312784 | 15-hydroxypentacosanoic acid | -5.595 |
| CID_12411 | Tritriacontane | -5.521 |
| CID_5281515 | Caryophyllene | -5.492 |
| CID_5365034 | Myristyl oleate | -5.492 |
| CID_94715 | Glucuronic acid | -5.49 |
| CID_10467 | Arachidic acid | -5.474 |
| CID_5352845 | 2-Dodecenol | -5.382 |
| CID_445858 | Ferulate | -5.375 |
| CID_8123 | Ethyl octadec-9-enoate | -5.37 |
| CID_441437 | D-quercitol | -5.325 |
| CID_12397 | Pentadecanol | -5.3 |
| CID_5283384 | 9,12,15-Octadecatrienal | -5.295 |
| CID_12406 | Pentacosane | -5.286 |
| CID_16898 | n-Heneicosanoic acid | -5.227 |
| CID_10290861 | Conduritol A | -5.189 |
| CID_10408 | Hexahydrofarnesyl acetone | -5.173 |
| CID_12413 | Pentatriacontane | -5.141 |
| CID_6437266 | Parsupol | -5.126 |
| CID_439655 | Tartaric acid | -5.116 |
| CID_31253 | Myrcene | -5.096 |
| CID_8193 | Dodecanol | -5.092 |
| CID_3893 | Lauric acid | -5.086 |
| CID_9548706 | Germacrene A | -5.076 |
| CID_892 | Inositol | -5.074 |
| CID_68171 | N-hexacosanol | -5.05 |
| CID_8209 | Tetradecanol | -5.042 |
| CID_2682 | 1-Hexadecanol | -4.987 |
| CID_8221 | Octadecanol | -4.976 |
| CID_6251 | Mannitol | -4.965 |
| CID_12366 | Ethyl palmitate | -4.889 |
| CID_6918391 | b-Elemene | -4.818 |
| CID_61303 | 2-Pentadecanone | -4.734 |
| CID_13849 | Pentadecanoic acid | -4.685 |
| CID_8181 | Methyl palmitate | -4.599 |
| CID_6365430 | Tetradecadiene | -4.594 |
| CID_379 | Caprylic acid | -4.525 |
| CID_798 | Indole | -4.52 |
| CID_2969 | Capric acid | -4.519 |
| CID_957 | Octanol | -4.41 |
| CID_119 | Gamma-butyric acid | -4.256 |

**Table S16: Docking-based binding energies (kcal/mol) of SHD-derived phytochemicals docked to PPARG (PDB: 5Y2O).** Compounds were retained only if they interacted with at least 8 of 13 key PPARG pocket residues. The biological reference ligand (Pioglitazone) is shown in bold.

| External chemical identifier | Chemical name | Docking-based Binding Energy (kcal/mol) |
|---|---|---|
| CID_10071442 | Sulfurein | -8.844 |
| **CID_4829** | **Pioglitazone** | **-8.842** |
| CID_5204 | Sesamin | -8.708 |
| CID_133775 | Pterosupin | -8.242 |
| CID_6450230 | Marmin | -7.966 |
| CID_42607524 | Monospermoside | -7.791 |
| CID_5742590 | Beta-sitosterol glucoside | -7.71 |
| CID_12309899 | Isocoreopsin | -7.375 |
| CID_185124 | Propterol | -7.143 |
| CHEMSPIDER_10306372 | Propterol B | -7.029 |
| CID_12411 | Tritriacontane | -6.932 |
| CID_638088 | Stilbenes | -6.872 |
| CID_5281617 | Genkwanin | -6.852 |
| CID_10212 | Imperatorin | -6.812 |
| CID_5283384 | 9,12,15-Octadecatrienal | -6.779 |
| CID_334704 | Marmesin | -6.772 |
| CID_5280794 | Stigmasterol | -6.721 |
| CID_222284 | Beta-sitosterol | -6.685 |
| CID_5312784 | 15-hydroxypentacosanoic acid | -6.489 |
| CID_12409 | n-Nonacosane | -6.471 |
| CID_102515444 | Phosphatidylethanolamine | -6.442 |
| CID_101750 | Alpha-camphorene | -6.418 |
| CID_12592 | Tetracosane | -6.379 |
| CID_600671 | Aegelinol | -6.373 |
| CID_114522 | Ketones | -6.322 |
| CID_11636 | n-Heptacosane | -6.269 |
| CID_8123 | Ethyl octadec-9-enoate | -6.262 |
| CID_10408 | Hexahydrofarnesyl acetone | -6.254 |
| CID_14034812 | Cerasidin | -6.213 |
| CID_8181 | Methyl palmitate | -6.078 |
| CID_11197 | Lignoceric acid | -6.047 |
| CID_12410 | Hentriacontane | -6.046 |
| CID_5280435 | Phytol | -6.026 |
| CID_5365034 | Myristyl oleate | -5.938 |
| CID_13849 | Pentadecanoic acid | -5.882 |
| CID_61303 | 2-Pentadecanone | -5.87 |
| CID_6365430 | Tetradecadiene | -5.861 |
| CID_323 | Coumarins | -5.856 |
| CID_12366 | Ethyl palmitate | -5.804 |
| CID_49831545 | Delta-lactone | -5.783 |
| CID_68171 | N-hexacosanol | -5.777 |
| CID_8209 | Tetradecanol | -5.768 |
| CID_5281426 | Umbelliferone | -5.684 |
| CID_68972 | n-Triacontanol | -5.605 |
| CID_12413 | Pentatriacontane | -5.578 |
| CID_2682 | 1-Hexadecanol | -5.498 |
| CID_8221 | Octadecanol | -5.494 |
| CAS_22108-77-6 | Gentiocrucine | -5.434 |
| CID_439655 | Tartaric acid | -5.431 |
| CID_5311264 | Lysophosphatidylcholine | -5.417 |
| CID_31253 | Myrcene | -5.355 |
| CID_12397 | Pentadecanol | -5.336 |
| CID_5352845 | 2-Dodecenol | -5.333 |
| CID_12406 | Pentacosane | -5.329 |
| CID_12535 | n-Triacontane | -5.309 |
| CID_8193 | Dodecanol | -5.255 |
| CID_94715 | Glucuronic acid | -5.236 |
| CID_3314 | Eugenol | -5.179 |
| CID_10790 | Aleurilic acid | -5.106 |
| CID_12407 | Hexacosane | -4.981 |
| CID_7410 | Acetophenone | -4.902 |
| CID_7121 | Veratric acid | -4.828 |
| CID_379 | Caprylic acid | -4.805 |
| CID_798 | Indole | -4.786 |
| CID_6251 | Mannitol | -4.741 |
| CID_68406 | n-Octacosanol | -4.736 |
| CID_12101 | m-Ethyl phenol | -4.735 |
| CID_785 | Hydroquinone | -4.692 |
| CID_957 | Octanol | -4.586 |
| CID_9015 | p-Guaiacol | -4.249 |
| CID_119 | Gamma-butyric acid | -4.204 |
| CID_643915 | Angelic acid | -3.813 |
| CID_264 | Butyric acid | -3.588 |